\documentclass[final,3p,times]{elsarticle}
\usepackage{simplewick,rotating}
\usepackage{setspace}
\usepackage[colorlinks,citecolor=red,urlcolor=black,bookmarks=false,hypertexnames=true]{hyperref} 
\usepackage{nicefrac,tikz,tensor,amsmath,amssymb,amsfonts,mathtools,leftidx,subfigure,color,here,booktabs,emp,bm,multirow,pdf14,tikz-feynman}
\usepackage{bm,braket}
\usepackage{glossaries}
\usepackage{nomencl}
\usepackage{scalefnt}
\usepackage{pifont}
\usepackage{gensymb}
\usepackage[utf8]{inputenc}

\usepackage{afterpage}

\usepackage[nameinlink]{cleveref}

\usepackage{framed}
\usepackage{multicol}
 
\makenomenclature

\usetikzlibrary{calc,intersections,through,backgrounds,fit,arrows,positioning}

\definecolor{cadmiumgreen}{rgb}{0.0, 0.42, 0.24}
\definecolor{electricviolet}{rgb}{0.56, 0.0, 1.0}
\definecolor{indianred}{rgb}{0.86, 0.08, 0.24}
\definecolor{royalblue}{rgb}{0.25, 0.41, 0.88}
\definecolor{darkorange}{rgb}{1.0, 0.55, 0}
\definecolor{mediumseagreen}{rgb}{0.24, 0.70, 0.44}
\definecolor{purple}{rgb}{0.5, 0, 0.5}
\definecolor{cyan3}{rgb}{0, 0.80, 0.80}

\newcommand*\pentagofill[1][]{\tikz[#1]{
        \draw [line width=0.05em,fill]
      (18:0.34em) -- (90:0.34em) -- (162:0.34em) -- (234:0.34em)
      -- (306:0.34em) -- cycle;
}}
\newcommand*\hexagofill[1][]{\tikz[#1]{
    \draw [line width=0.05em,fill]
      (30:0.34em) -- (90:0.34em) -- (150:0.34em) -- (210:0.34em)
      -- (270:0.34em) -- (330:0.34em) -- cycle;
}}

\newcommand{\MeV}{\, \text{MeV}}

\newcommand{\symboldiamond}[1][black]{{\color{#1}\ding{117}}}
\newcommand{\symboltriangle}[1][black]{{\color{#1}\ding{115}}}
\newcommand{\symboltriangleup}[1][black]{{\color{#1}\scalefont{0.9}{\raisebox{1.5ex}{\begin{turn}{180}$\blacktriangledown$\end{turn}}}}}

\newcommand{\symbolbox}[1][black]{{\color{#1}\scalefont{0.75}$\blacksquare$}}
\newcommand{\symbolcircle}[1][black]{{\color{#1}\scalefont{0.75}\ding{108}}}

\newcommand{\symboldiamondsym}[1][black]{{\color{#1}\scalefont{0.75}\raisebox{-.2ex}{\begin{turn}{45}$\blacksquare$\end{turn}}}}

\newcommand{\symbolstar}[1][black]{{\color{#1}\raisebox{0.2ex}{$\bigstar$}}}

\newcommand{\bluecircle}{{\scalefont{0.9}\symbolcircle[royalblue]}}

\newcommand{\greentriangleup}{{\scalefont{0.9}\symboltriangleup[mediumseagreen]}}

\newcommand{\orangediamond}{{\scalefont{0.9}{\scalefont{0.8}\symboldiamondsym[darkorange]}}}

\newcommand{\redsquare}{{\scalefont{0.9}\symbolbox[indianred]}}

\newcommand{\orangestar}{{\scalefont{0.9}{\scalefont{0.8}\symbolstar[darkorange]}}}

\definecolor{FGViolet}{rgb}{0.61,0.32,0.61}
\definecolor{FGDarkBlue}{rgb}{0,0,0.6}
\definecolor{FGBlue}{rgb}{0,0,0.8}
\definecolor{FGLightBlue}{rgb}{0.2, 0.6, 0.8}
\definecolor{FGGreen}{rgb}{0.2,0.7,0.2}
\definecolor{FGLightGreen}{rgb}{0.4,1,0.4}
\definecolor{FGYellow}{rgb}{1,0.95,0}
\definecolor{FGOrange}{rgb}{0.95,0.5,0.1}
\definecolor{FGRed}{rgb}{0.8,0,0}
\definecolor{FGWhite}{rgb}{1,1,1}
\definecolor{FGLightGray}{rgb}{0.8,0.8,0.8}
\definecolor{FGGray}{rgb}{0.5,0.5,0.5}
\definecolor{FGDarkGray}{rgb}{0.3,0.3,0.3}
\definecolor{FGBlack}{rgb}{0,0,0}

\newcommand{\be}{\begin{equation}}
\newcommand{\ee}{\end{equation}}
\newcommand{\ba}{\begin{align}}
\newcommand{\ea}{\end{align}}
\newcommand{\lm}{\Lambda}
\newcommand{\kf}{k_{\rm F}}

\newcommand{\vnn}{V_{\rm NN}}
\newcommand{\vtn}{V_{\rm 3N}}

\newcommand{\Trel}{T_{\rm rel}}
\newcommand{\clebschG}[6]{\mathcal{C}_{#1 #2 #3 #4}^{#5 #6}}  

\newcommand{\mev}{\, \text{MeV}}

\newcommand{\shortminus}{\scalebox{0.75}[1.0]{$-$}}

\newcommand{\openone}{\leavevmode\hbox{\small1\normalsize\kern-.33em1}}

\newcommand{\fmiq}{\, \text{fm}^{-3}}
\newcommand{\la}{\langle}
\newcommand{\ra}{\rangle}
\newcommand{\tl}{\tilde{L}}
\newcommand{\tlp}{\tilde{L'}}
\newcommand{\tj}{\tilde{J}}

\definecolor{graphicbackground}{rgb}{0.96,0.96,0.8}

\tikzset{%
   base/.style = {draw=black, minimum width=3cm, minimum height=1cm,
                    align=center, on chain},
 myarrows/.style = {-stealth, thick},
    }

\newcommand{\SolidBond}[6]%
{ \begin{pgfonlayer}{background}
        \colorlet{InColor}{#4}
        \colorlet{OutColor}{#5}
        \foreach \I in {#6,...,1}
        {   \pgfmathsetlengthmacro{\r}{#3/#6*\I}
            \pgfmathsetmacro{\C}{sqrt(1-\r*\r/#3/#3)*100}
            \draw[latex-][InColor!\C!OutColor, line width=\r] (#1) -- (#2);
        }
    \end{pgfonlayer}
}

\newcommand{\DashedBond}[7]%
{ \begin{pgfonlayer}{background}
        \colorlet{InColor}{#4}
        \colorlet{OutColor}{#5}
        \foreach \I in {#6,...,1}
        {   \pgfmathsetlengthmacro{\r}{#3/#6*\I}
            \pgfmathsetmacro{\C}{sqrt(1-\r*\r/#3/#3)*100}
            \draw[latex-][InColor!\C!OutColor, line width=\r, dashed] (#1) -- (#2);
        }
    \end{pgfonlayer}
}

\begin{document}

\title{Three-Nucleon Forces:\\Implementation and Applications to Atomic Nuclei and Dense Matter}
\author[add1,add2]{Kai Hebeler}
\ead{kai.hebeler@physik.tu-darmstadt.de}
\address[add1]{Technische Universit\"at Darmstadt, 64289 Darmstadt, Germany}
\address[add2]{ExtreMe Matter Institute EMMI, GSI Helmholtzzentrum f\"ur Schwerionenforschung GmbH, 64291 Darmstadt, Germany}

\begin{abstract} 
Recent advances in nuclear structure theory have significantly enlarged the
accessible part of the nuclear landscape via \textit{ab initio} many-body
calculations. These developments open new ways for microscopic studies of
light, medium-mass and heavy nuclei as well as nuclear matter and represent an
important step toward a systematic and comprehensive understanding of atomic
nuclei across the nuclear chart. While remarkable agreement has been found
between different many-body methods for a given nuclear Hamiltonian, the
comparison with experiment and the understanding of theoretical uncertainties
are still important open questions. The observed discrepancies to experiment
indicate deficiencies in presently used nuclear interactions and operators.
Chiral effective field theory (EFT) allows to systematically derive
contributions to nucleon-nucleon (NN), three-nucleon (3N) and higher-body
interactions including estimates of theoretical uncertainties. While the
treatment of NN interactions in many-body calculations is well established,
the calculation of 3N interactions and their incorporation in \textit{ab
initio} frameworks is still a frontier.

This work reviews in detail recent and current developments on the derivation
and implementation of improved 3N interactions and provides a comprehensive
introduction to fundamental methods for their practical calculation and
representation. We further give an overview of novel and established methods
that facilitate the inclusion and treatment of 3N interactions in \textit{ab
initio} nuclear structure frameworks and present a selection of the latest
calculations of atomic nuclei as well as nuclear matter based on
state-of-the-art nuclear NN and 3N interactions derived within chiral EFT.
Finally, we discuss ongoing efforts, open questions and future directions.
\end{abstract}

\maketitle


\tableofcontents

\clearpage 

\section*{List of Symbols and Abbreviations}
\begin{table}[h!]
\begin{tabular}{lp{8.1cm}}
$\hbar = c = 1 = \hbar c = 197.326$ MeV fm & unit system \\ 
\\[-3mm]
\hline \hline
\\[-3mm]
\multicolumn{2}{l}{\textit{Momenta (see Table~\ref{tab:Jacobi_momenta_crosstable} for details):}} \\
\\[-3mm]
$\hat{\mathbf{a}} = \mathbf{a}/|\mathbf{a}| = \mathbf{a}/a$ & unit vector \\
$\mathbf{k}_i$ & single-particle momentum of particle $i$\\
$\mathbf{k} (\mathbf{k}')$ & initial (final) state momentum \\
$\mathbf{Q}_i = \mathbf{k}'_i - \mathbf{k}_i$ & momentum transfer \\
$\mathbf{p}, \mathbf{q}$ & Jacobi momenta \\
\\[-3mm]
\hline \hline
\\[-3mm]
\multicolumn{2}{l}{\textit{Angular momentum coupling coefficients (using the conventions of Ref.~\cite{Vars88Gulag}):}} \\
\\[-3mm]
$\mathcal{C}_{j_1 m_1 j_2 m_2}^{j_3 m_3} = \bigl< j_1 m_1 j_2 m_2 | (j_1 j_2) j_3 m_3 \bigr>$ & Clebsch-Gordan coefficients \\
\\[-3mm]
$\left\{ \begin{array}{ccc}
j_1 & j_2 & j_3 \\ 
j_4 & j_5 & j_6 
\end{array} \right\}
$ & $6j$ symbols \\
\\[-3mm]
$\left\{ \begin{array}{ccc}
j_1 & j_2 & j_3 \\ 
j_4 & j_5 & j_6 \\
j_7 & j_8 & j_9 \\
\end{array} \right\}
$ & $9j$ symbols \\
\\[-3mm]
$Y_{l m} (\hat{\mathbf{a}}) = \bigl< \hat{\mathbf{a}} | l m \bigr>, Y_{l m}^* (\hat{\mathbf{a}}) = \bigl< l m |  \hat{\mathbf{a}} \bigr> = (-1)^m Y_{l -m} (\hat{\mathbf{a}})$ & spherical harmonics \\
\\[-3mm]
$\mathcal{Y}_{l_1 l_2}^{l_3 m_3} (\hat{\mathbf{a}},\hat{\mathbf{b}}) = \sum_{m_1, m_2} \mathcal{C}_{l_1 m_1 l_2 m_2}^{l_3 m_3}  Y_{l_1 m_1} (\hat{\mathbf{a}}) Y_{l_2 m_2} (\hat{\mathbf{b}})$
& coupled spherical harmonics \\
\\[-3mm]
\hline \hline
\\[-3mm]
\multicolumn{2}{l}{\textit{Three-body partial-wave states:}} \\
\\[-3mm]
$\bigl| p q \alpha \bigr> = \bigl| p q; \bigl[ (L S) J (l s) j \bigr] \mathcal{J} (T t) \mathcal{T} \bigr>$ & states in $Jj$-coupling scheme \\
\\[-3mm]
$\bigl| p q \beta \bigr> = \bigl| p q; \bigl[ (L l) \mathcal{L} (S s) \mathcal{S} \bigr] \mathcal{J} (T t) \mathcal{T} \bigr>$ & states in $LS$-coupling scheme \\
\\[-3mm]
$L$,$S$,$J$,$T$ & two-body quantum numbers: relative orbital angular momentum, spin, total relative angular momentum and isospin of the particles with Jacobi momentum $p$ \\
\\[-3mm]
$l$,$s=\tfrac{1}{2}$,$j$,$t=\tfrac{1}{2}$ & orbital angular momentum, spin, total angular momentum and isospin of the particle with Jacobi momentum $q$ \\
\\[-3mm]
$\mathcal{L}$,$\mathcal{S}$,$\mathcal{J}$,$\mathcal{T}$ & three-body quantum numbers: total orbital angular momentum, spin, angular momentum and isospin\\
\\[-3mm]
\hline \hline
\\[-3mm]
\end{tabular}
\begin{tabular}{lp{5cm}|lp{5cm}}
\multicolumn{4}{l}{\textit{Abbreviations:}} \\
\\[-3mm]
NN & nucleon-nucleon & 3N & three-nucleon \\
EFT & effective field theory & LEC & low-energy coupling \\
LO & leading order & NLO & next-to-LO \\
N$^2$LO & next-to-next-to-LO & N$^3$LO & next-to-next-to-next-to-LO \\
MS & momentum space  & CS & coordinate space \\
PNM & pure neutron matter & SNM & symmetric nuclear matter \\
RG & renormalization group & SRG & similarity renormalization group \\
IM-SRG & in-medium SRG & MR-IM-SRG & multi-reference IM-SRG \\
NCSM & no-core shell model & IT-NCSM & importance-truncated NSCM \\
CC & coupled cluster & QMC & quantum Monte Carlo \\
MBPT & many-body perturbation theory & BMBPT & Bogoliubov-MBPT\\
SCGF & self-consistent Green's function & GSCGF & Gorkov-SCGF \\
HO & harmonic oscillator &  QCD & quantum chromodynamics 
\end{tabular}
\end{table}

\addcontentsline{toc}{section}{Symbols and Notation}

\clearpage

\section{Introduction and Overview}
\label{sec:Intro}

\begin{figure}[b!]
\centering 
\tikzstyle{b} = [rectangle, draw, fill=blue!20, text width=15em, text centered, rounded corners, minimum height=2em, thick]

\begin{tikzpicture}

  \tikzset{>=latex}
  
  \node [b,fill=blue!40] (observables) at (0,6) {Nuclear structure and reaction observables};
  \node [b,fill=black!20] (LQCD) at (0,3) {Lattice QCD} edge [->,line width=0.3mm] (observables);
  \node [b,fill=red!40] (QCD) at (0,0) {Quantum chromodynamics} edge [->,line width=0.3mm] (LQCD);
  
  \node [b,fill=blue!40] (observables2) at (9,6) {Nuclear structure and reaction observables};
  \node [b,fill=green!40] (frameworks2) at (9,4.5) {\textit{Ab initio} many-body frameworks} edge [->,line width=0.3mm] (observables2);
  \node [b,fill=yellow!40] (RG2) at (9,3) {Renormalization group methods} edge [->,line width=0.3mm] (frameworks2);
  \node [b,fill=orange!40] (chiral2) at (9,1.5) {Chiral effective field theory\\nuclear interactions} edge [->,line width=0.3mm] (RG2);
  \node [b,fill=red!40] (QCD2) at (9,0) {Quantum chromodynamics} edge [->,line width=0.3mm] (chiral2);
  
\end{tikzpicture}
\caption{Two different paths from quantum chromodynamics (QCD), the
fundamental theory of the strong interaction, to low-energy nuclear structure
observables. The left panel shows the direct approach from first principles,
by explicitly simulating the quark-gluon dynamics on a discrete lattice and
extracting single- and multi-nucleon observables from such calculations. The
right panel shows the path based on low-energy effective degrees of freedom,
i.e., neutrons, protons and pions. The nuclear interaction between neutrons and
protons is constrained by the symmetries and the symmetry-breaking patterns
of QCD. The resulting chiral EFT interactions represent the microscopic input
for \textit{ab initio} many-body frameworks.}
\label{fig:QCD_to_observables}
\end{figure}
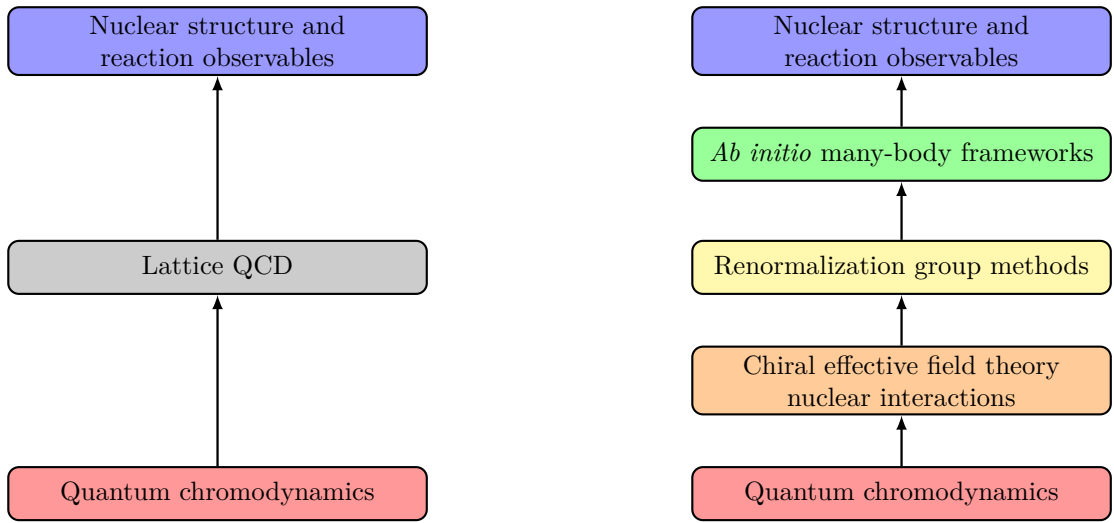

One of the central goals of \textit{ab initio} nuclear theory is the microscopic
understanding of the structure of atomic nuclei and dense matter starting from
the properties of the fundamental degrees of freedom and their interactions.
According to our present understanding, at the most fundamental level the
strong interaction is governed by the quark-gluon dynamics described by
quantum chromodynamics (QCD). However, even though it is now possible to
correctly predict properties of single-nucleon states via Lattice
QCD~\cite{Durr08lighth,Fodo12RMP,Yang18protmass}, the accurate and realistic
description of multi-nucleon systems directly based on QCD has so far remained
an elusive
goal~\cite{Bean11PPNP,Taka11HALQCD,Siny12HALQCD,NPLQCD13lightnuc,Bric14JPG,Orgi15LatticeQCD,Berk17LQCD,Detm19LatticeQCD,Irit19HALQCD,Dris19QCDnuclear}.
An alternative path from the principles of QCD to low-energy nuclear structure
observables consists in employing low-energy effective theories and quantum
many-body methods (see Figure~\ref{fig:QCD_to_observables}). The fundamental
principle of any effective low-energy description is based on the fact that
details at small distance scales are not resolved when a physical system is
probed at low energies. This makes it possible to introduce low-energy degrees
of freedom that encapsulate the complex high-energy dynamics.

The radius of an average atomic nucleus is of the order of a few femtometers
($1$ fm $= 10^{-15}$m). Hence, the uncertainty principle implies typical
momenta in nuclei to be on the order of the pion mass $m_{\pi} \approx 135$ MeV.
At this scale the relevant degrees of freedom are not quarks and gluons but
colorless hadrons, like neutrons, protons and pions. Chiral effective field
theory (EFT) allows to systematically derive contributions to interactions
between nucleons within a given power counting expansion
scheme~\cite{Epel09RMP,Mach11PR,Hamm19Rev}, which are parametrized in terms of
long-range pion exchange contributions and short-range couplings (see
Section~\ref{sec:chiral_EFT}). These contributions include nucleon-nucleon
(NN), three-nucleon (3N) and even higher-body interactions. The presence of
three- and higher-body nuclear interactions is a natural consequence of the
composite nature of nucleons since for any non-elementary particle its
constituents can be distorted by the presence of external forces. This
phenomenon is well known in classical systems consisting of extended objects
interacting via gravitational or electromagnetic interactions. For example,
the orbit of a satellite around the earth is affected by the location of the
moon due to the induced tides on earth, which in turn affect the gravitational
force between earth and the satellite. If earth, moon and the satellite are
all parametrized as point particles, such an effect can only be described via
three-body forces. In the case of the gravitational force such effects are
typically rather small. In contrast, for nuclear systems 3N contributions play
a central role for, e.g., the shell structure of atomic nuclei, the bulk
properties of nuclear matter and the evolution of systems toward the limits
of stability (see Section~\ref{sec:applications}).

The interactions derived within chiral EFT represent the fundamental
microscopic input for \textit{ab initio} many-body frameworks. In the following we
consider all frameworks as ``\textit{ab initio}'' which only use free-space nuclear
interactions as basic input and can be systematically improved such that in
the limit of infinite basis size and infinite order in the many-body expansion
one can in principle recover results of exact calculations (see
Table~\ref{tab:many_body_frameworks}). In Figure~\ref{fig:Heiko_nuclear_chart}
we illustrate the part of the nuclear chart studied within such approaches in
the years 2009, 2012, 2015 and 2018~\cite{Herg16PR,Herg19privcom}. Evidently,
in recent years there has been a dramatic increase in the scope of such \textit{ab
initio} frameworks. These advances were driven by developments in different
sectors (see also the right panel of Figure~\ref{fig:QCD_to_observables}):

\begin{figure}[t]
\centering
\includegraphics[width=0.45\textwidth]{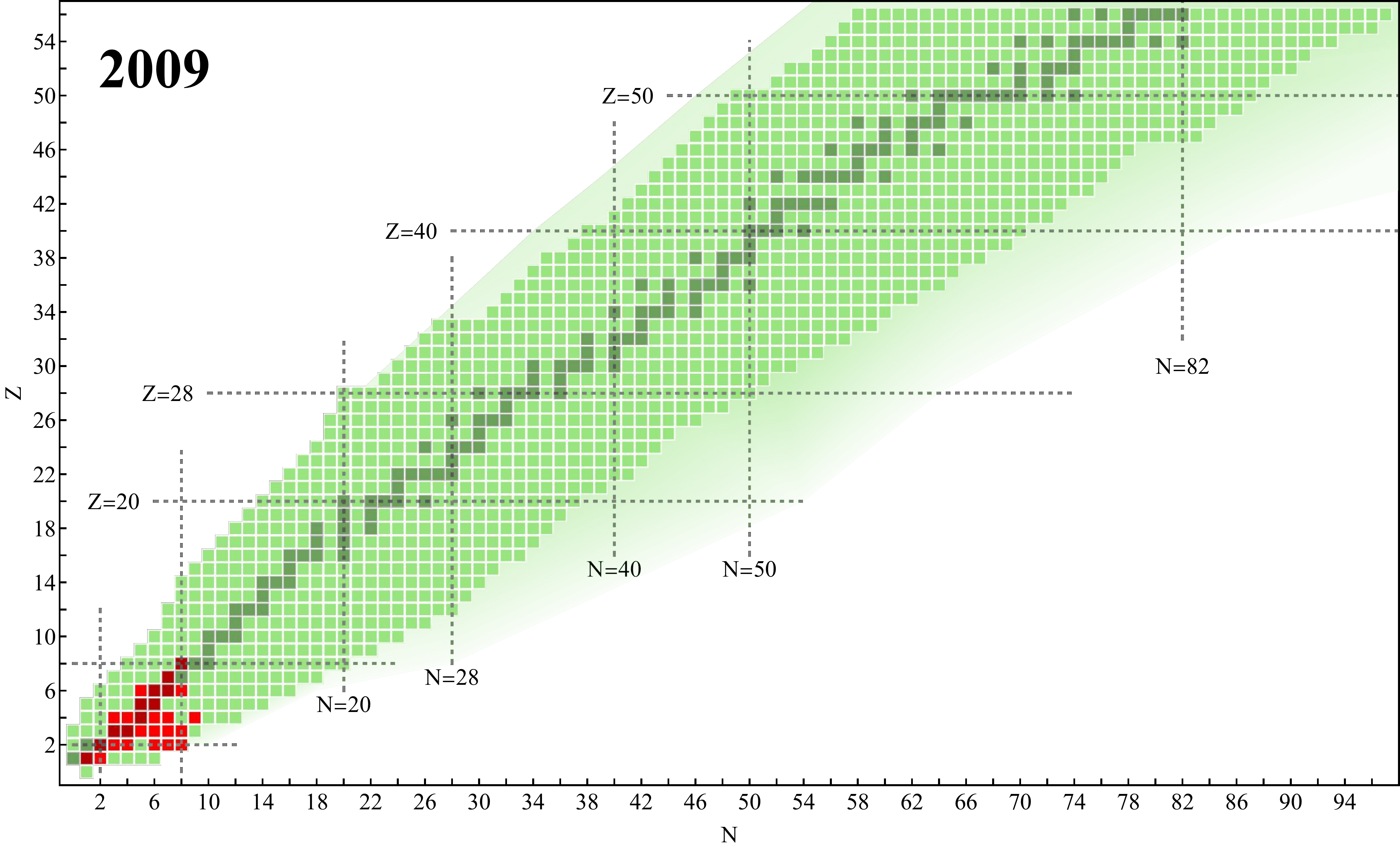}
\hspace{0cm}
\includegraphics[width=0.45\textwidth]{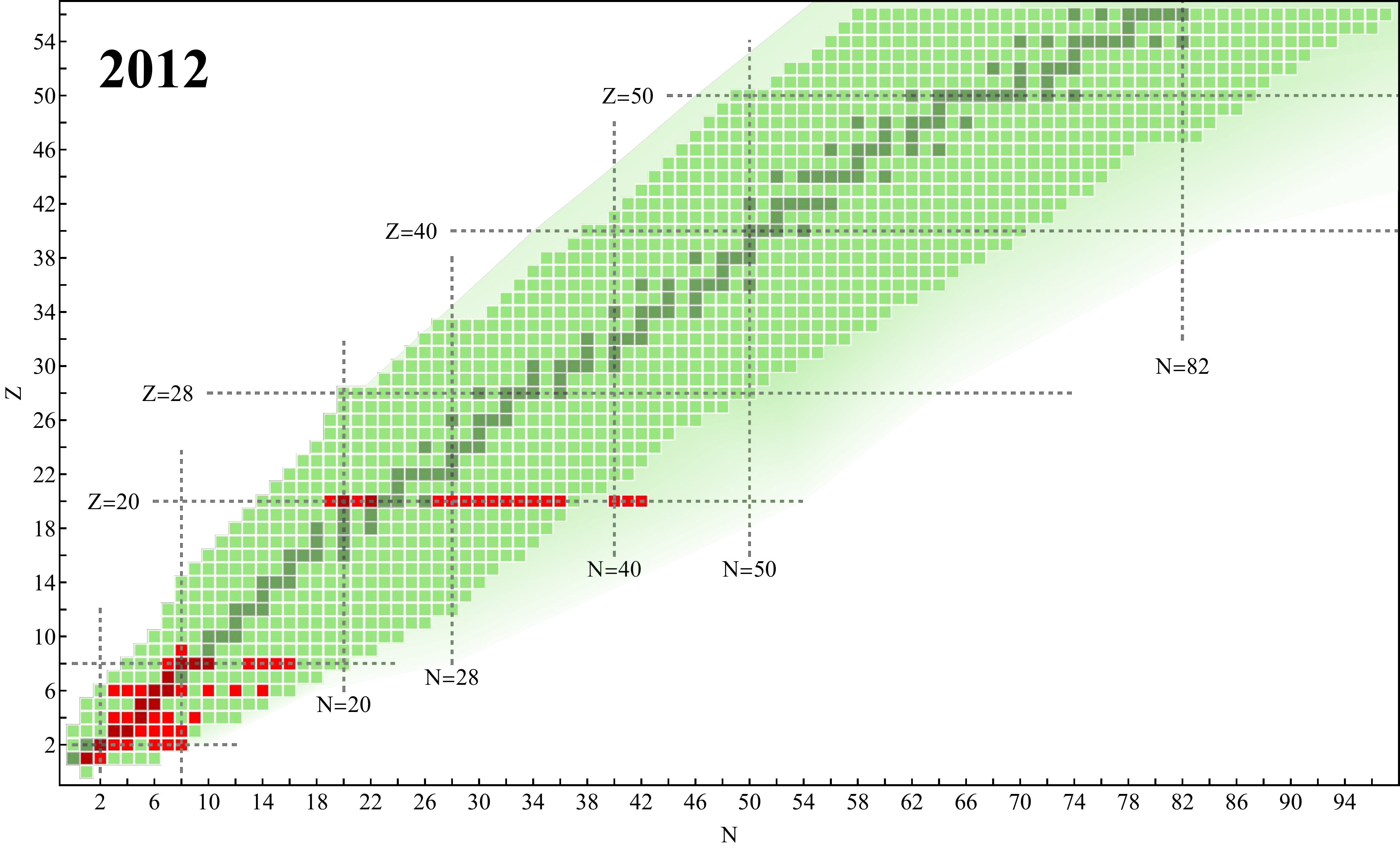}\\
\vspace{0.05cm}
\includegraphics[width=0.45\textwidth]{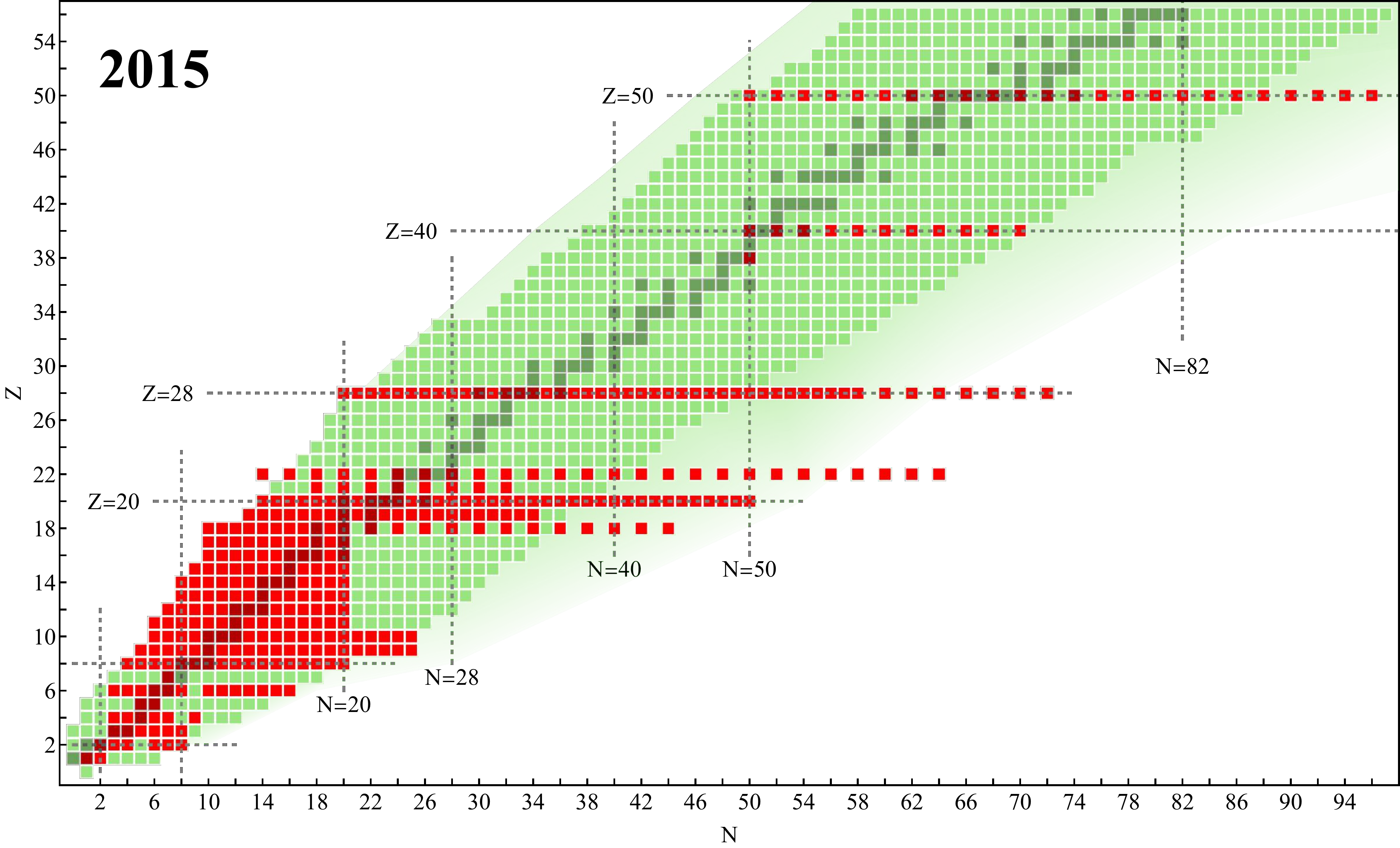}
\hspace{0cm}
\includegraphics[width=0.45\textwidth]{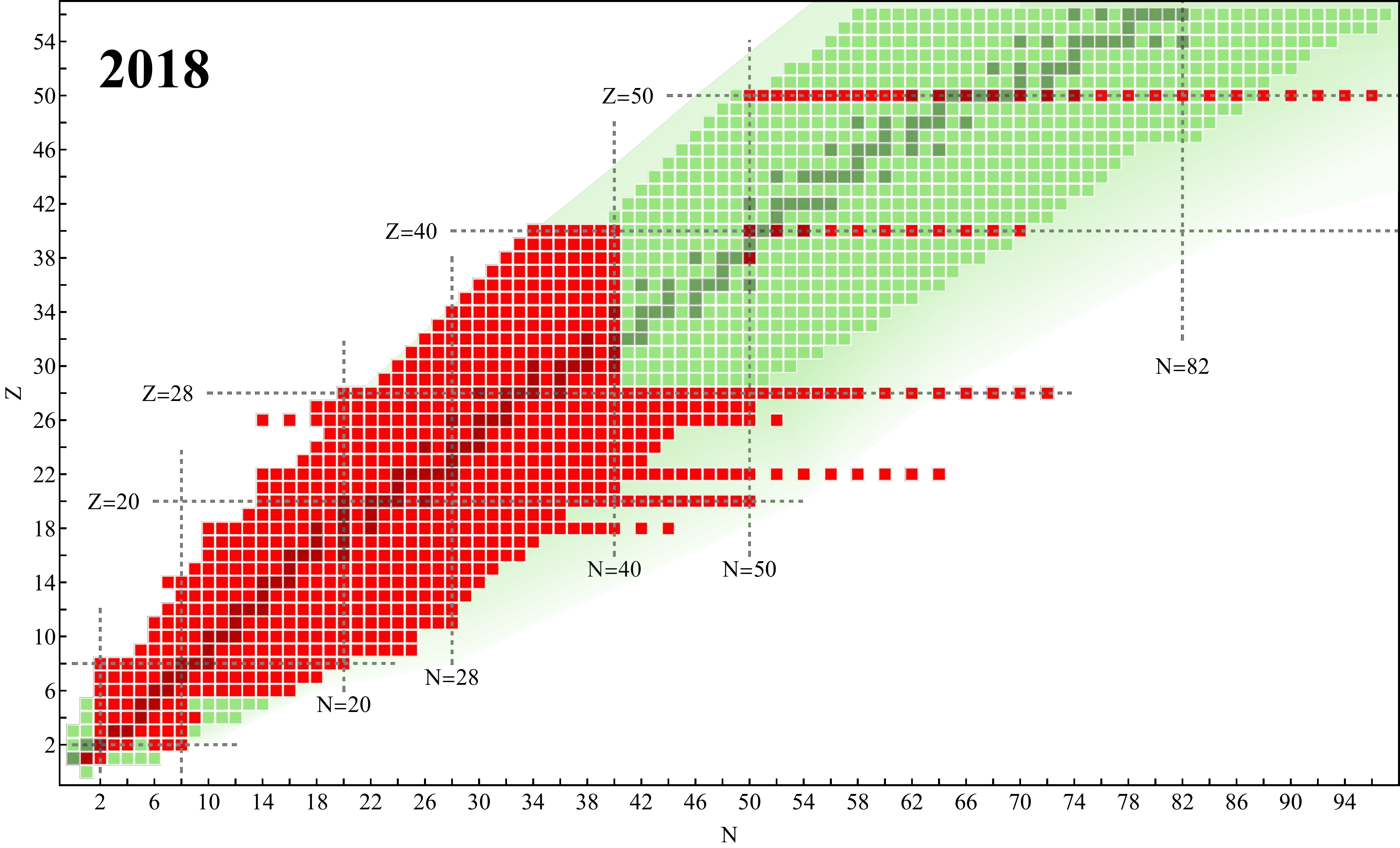}
\caption{Illustration of the scope of \textit{ab initio} many-body calculations of
nuclei from the year 2009 (upper left) to 2018 (lower right) using NN and 3N
interactions. Studied nuclei are highlighted in red. The figures include
calculations for which converged results with respect to the basis size have
been achieved. Credits to Heiko Hergert for providing the figures, see also
Ref.~\cite{Herg16PR}.}
\label{fig:Heiko_nuclear_chart}
\end{figure}

\textit{1. Nuclear interactions:} The predictive power of nuclear many-body
calculations is naturally limited by the quality of the employed nuclear
interactions. Even results of virtually exact calculations can only be as good
as the underlying input. During the recent years several new families of
interactions have been derived at different orders in the chiral expansion.
These efforts include the exploration of various different regularization
schemes and fitting strategies for the short-range low-energy couplings as
well as the investigation of different methods for estimating uncertainties
due to neglected higher-order terms of the chiral expansion. Most of these
investigations are still ongoing. In Section~\ref{sec:chiral_EFT} we give a
more detailed overview of these developments.

\textit{2. Renormalization Group methods:} Renormalization Group (RG)
techniques allow to systematically decouple low- and high-momentum components
of nuclear interactions while preserving low-energy
observables~\cite{Bogn10PPNP}. Such RG transformations to lower resolution
scales help to reduce the scheme dependence of nuclear interactions, offer new
tools to assess many-body uncertainties by studying residual resolution scale
dependencies, and, most importantly for practical calculations, can
dramatically improve the convergence of many-body calculations and generate
much less correlated wave functions. This improved perturbativeness is in
particular key for all many-body frameworks based on harmonic oscillator basis
expansions (see Table~\ref{tab:many_body_frameworks}). In fact, many of the
calculations illustrated in Figure~\ref{fig:Heiko_nuclear_chart} have only
been possible thanks to the use of low-resolution interactions obtained within
the Similarity Renormalization Group (SRG)~\cite{Bogn07SRG,Jurg08SRG3N1D} (see
also Section~\ref{sec:SRG}).

\textit{3. Improved and novel many-body frameworks:} Advances in many-body
theory have led to various new improved frameworks that allow to study
nuclear matter as well as nuclei in different regions of the nuclear chart. In
Table~\ref{tab:many_body_frameworks} we give an overview of currently used and
actively developed many-body frameworks for atomic nuclei and nuclear matter.
Each of the listed methods has its own benefits and drawbacks, determined by
factors like the required computational resources, the scaling behavior as a
function of mass number, the nature and level of many-body truncations, the
flexibility regarding which nuclear interactions can be used, as well as the
accessibility of different observables of closed- and/or open-shell systems.
The availability of various different methods offers new possibilities for
cross-benchmarks and for studying in detail the validity of different
many-body truncations for a given Hamiltonian. In particular, given that
various frameworks generally can differ quite substantially in terms of
computational cost, it is now possible to test novel nuclear interactions in
an efficient and reliable way by using computationally inexpensive methods and
by validating the results for a few systems using more precise and costly
methods. Furthermore there are active ongoing efforts to combine the
advantages of different methods by developing mixed and hybrid methods (see,
e.g., Refs.~\cite{Tich17NCSM-MCPT,Gebr17IMNCSM}).

\textit{4. Increased computational resources:} Finally, advances in the
hardware equipment of present high-performance computer clusters contributed
to the increased scope of present many-body calculations. These developments
include improved floating point performance as well as the availability of
large-memory machines, which are in particular key for handling matrix elements
of 3N interactions (see Section~\ref{sec:3N_incorporation}).

\begin{table}[t!]
\centering
\begin{tabular}{llll}
\multicolumn{4}{l}{\textit{atomic nuclei}} \\
\hline \hline
\textbf{method} & \textbf{type/representation} & \textbf{mass$^*$} & \\
\hline
Faddeev(-Yakubovsky) equations & momentum/coordinate space basis & 3-5 &
\cite{Gloe83QMFewbod,Gloe95cont,Stad91Faddeev,Nogg00Fadd4N,Epel05EGMN3LO,Delt064Nscatt,Vivi164N,Laza175N,Epel18SCS3N}\\
hyperspherical harmonics (HH) & momentum/coordinate space basis & 3-6 &
\cite{Kiev08HH,Barn01HH,Vivi05HH,Marc12mucap,Bacc13dipole}\\ no-core shell
model (NCSM) & harmonic oscillator configuration basis & $\lesssim 12$ &
\cite{Barr13PPNP,Navr99NCSM,Navr09NCSM,Navr07A1013,Roth11SRG,Roth14SRG3N,Wirt14hypernuc,Vora19NCSMcont}
\\ quantum Monte Carlo (QMC)  & eucl. time prop. in coordinate basis &
$\lesssim 16$ &
\cite{Carl15RMP,Pede17Lect,Lynn19QMCreview,Geze13QMCchi,Geze14long,Lynn16QMC3N,Lona17A16loc,Piar17LightNucl,Lona18mediummass}
\\ importance-truncated NCSM (IT-NCSM) & harmonic oscillator configuration
basis & $\lesssim 25$ &
\cite{Roth09ImTr,Roth07ITNCSM,Roth12NCSMCC3N,Gebr16MR,Wirt17hypernuc} \\
lattice EFT & eucl. time prop. on a discrete lattice & $\lesssim 28$ &
\cite{Lee09PPNP,Lee16Lect,Epel09LEFT,Epel09Lattice3N,Epel10Lattice,Epel11Hoyle,Epel12Hoyle,Epel12mquark,Rupa13capture,Lahd13LEFT,Epel14O16,Elha154Hescatt,Elha16quphase,Elha17LEFT}
\\ valence-shell diagonalization & harmonic oscillator configuration basis &
$\lesssim 100$ &
\cite{Lise08valshell,Stro19Shellmodel, Wien13Nat, Holt14Ca, Bogn14SM,Jans14SM, Jans16SM,Stro17ENO,Sun18CCshell}
\\ in-medium SRG (IM-SRG) & harmonic oscillator configuration basis &
$\lesssim 100$ &
\cite{Herg16PR,Herg17PS,Herg17lect,Tsuk11IMSRG,Tsuk12SM,Herg13MR,Herg14MR,Morr15Magnus,Stro16TNO,Parz17EOMIMSRG,Gebr17IMNCSM,Morr17Tin}
\\ coupled cluster (CC) & harmonic oscillator configuration basis & $\lesssim
100$ &
\cite{Hage14RPP,Kowa04CC,Hage07CC3N,Hage08spherCCSD,Hage12Ox3N,Roth12NCSMCC3N,Bind13expl3NCCSD,Bind13expl3NLCCSD(T),Hage12Ca3N,Bind14CCheavy,Sign14BogCC,Ruiz16Calcium,Hage16NatPhys,Birk17dipole,Morr17Tin}
\\ self-consistent Green's function (SCGF) & harmonic oscillator configuration
basis & $\lesssim 100$ &
\cite{Dick04PPNP,Barb16Lect,Soma11GGFform,Soma13GGF2N,Carb13SCGF3B,Cipo13Ox,Soma14GGF2N3N,Cipo14Oisotopes}\\
many-body perturbation theory (MBPT) & harmonic oscillator configuration basis
& $\lesssim 100$ & \cite{Tich16HFMBPT,Hu2016MBPT,Tich17NCSM-MCPT,Tich18BMBPT,Demo19PT,Tich20MBPT,Demo20MBPT} \vspace{0.1cm} \\
\multicolumn{4}{l}{\parbox{16cm}{$^*$\footnotesize{The given mass number limits include
constraints from limitations regarding the storage of three-body matrix
elements, which currently prevent converged results for systems beyond
the mass number regime $A \approx 100$.}}} \\ \\
\multicolumn{4}{p{10cm}}{\textit{infinite nuclear matter}} \\
\hline \hline
\textbf{method} & \textbf{type/representation} & & \\
\hline
many-body perturbation theory & momentum basis & & \cite{negele1995quantum,Hebe10nmatt,Hebe11fits,Tews13N3LO,Holt13PPNP,Cora14nmat,Well15therm,Dris16asym,Holt16eos3pt,Dris17MCshort} \\ 
self-consistent Green's function & momentum basis & & \cite{Carb13SCGF3B,Carb13nm,Carb14SCGFdd} \\
diagrammatic hole-line expansion & momentum basis & & \cite{Day67Brueckner,Li12Brueckner,Samm15numat,Hu16matter,Logo16matter} \\
coupled cluster & momentum basis on a discrete lattice & & \cite{Hage14RPP,Baar13CCinf,Hage14ccnm,Ekst13optNN,Ekst15sat,Ekst17deltasat} \\
quantum Monte Carlo & eucl. time prop. in coordinate space & & \cite{Lynn19QMCreview,Gand07AFDMCsnm,Rogg14QMC,Wlaz14QMC,Tews16QMCPNM} \\
lattice EFT & eucl. time prop. on a discrete lattice & & \cite{Lu19matter}
\end{tabular}
\caption{Summary of different microscopic many-body frameworks for
atomic nuclei and matter including selected developments and applications
based on NN and 3N interactions derived within chiral effective field theory.}
\label{tab:many_body_frameworks}
\end{table}

We stress that Figure~\ref{fig:Heiko_nuclear_chart} illustrates only the scope
of converged many-body calculations without quantifying the degree of
agreement with experimental data. Typically, results based on present nuclear
interactions can lead to good agreement with experiment for specific
observables (like, e.g., ground-state energies, excited states or charge
radii) in a restricted area of the nuclear chart (see
Section~\ref{sec:applications} for a detailed discussion). The agreement,
however, generally depends on details of the nuclear forces and on the
observables that have been included in the fitting process of the interaction
(see Section~\ref{sec:chiral_EFT}). A typical example is shown in
Figure~\ref{fig:Tichai_compare}, where we compare theoretical results for the
ground-state energies and two-neutron separation energies of the oxygen (left
panel), calcium (middle panel) and nickel (right panel) isotopic chain with
experimental data~\cite{Tich18BMBPT}. All the calculations have been performed
based on the same SRG-evolved NN plus 3N interaction~\cite{Gazi08lec}. While
the results of the theoretical calculations show excellent agreement with
experimental values for the oxygen chain (left panel), the agreement
deteriorates significantly for isotopic chains of heavier elements. However,
it is remarkable that the agreement between the predictions of different
many-body methods is excellent for all shown nuclei, given that the many-body
truncations are quite different in the various approaches. Based on these
observations we can draw the following general conclusions:

\begin{figure}[t]
\includegraphics[width=0.92\textwidth]{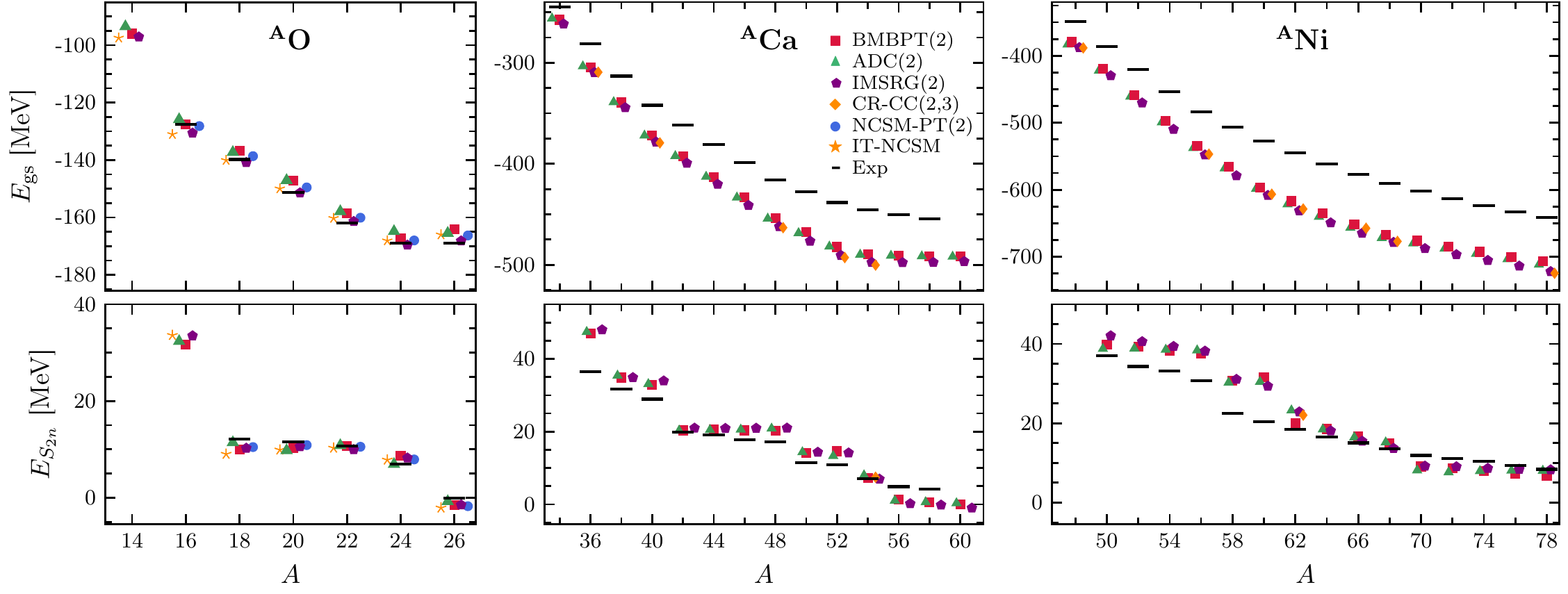}
\caption{Ground-state energies (top) and two-neutron separation energies
(bottom) for nuclei of the oxygen, calcium and nickel isotopic chains. Shown
are results of second-order BMBPT (\redsquare), perturbatively-improved
no-core shell model (NCSM-PT) (\bluecircle), IT-NCSM (\orangestar),
Gorkov-SCGF (GSCGF-ADC(2)) (\greentriangleup), MR-IM-SRG
(\pentagofill[purple]) and coupled cluster (CR-CC)
(\orangediamond). Black lines show the experimental values~\cite{Wang17AME16}.\\
\textit{Source:} Figure taken from Ref.~\cite{Tich18BMBPT}.}
\label{fig:Tichai_compare}
\end{figure}

\begin{itemize}
\item[a)] The agreement between the results of state-of-the-art many-body
frameworks is excellent for a given low-resolution Hamiltonian. This implies
that the many-body uncertainties are small for such interactions and possible
disagreement with experimental results can be mainly attributed to
deficiencies of currently used interactions and operators.

\item[b)] The agreement between theoretical predictions and experimental results
sensitively depends on the studied observable, the system under investigation
and the details of the employed nuclear interactions. Presently, there exist
no nuclear interactions that exhibit a systematic convergence pattern in the
chiral expansion and are able to correctly predict simultaneously different
few- and many-body observables of nuclei over the full range of mass number
within theoretical uncertainties.
\end{itemize}

Clearly, these findings emphasize the urgent need for improved nuclear
interactions which are able to correctly reproduce empirical properties of
known atomic nuclei as well as nuclear matter and can be applied in a reliable
and controlled way to systems in still unknown territory of the nuclear chart.
While the implementation of NN interactions in a form suitable for application
in many-body frameworks is straightforward and well established, the
calculation of 3N interactions is much more intricate and still involves major
challenges. This is true for both the conceptual development as well as their
practical implementation. The present work provides a comprehensive
introduction into methods for the practical implementation of 3N interactions
and their incorporation in microscopic many-body calculations. Even though the
techniques presented in this work can be applied to arbitrary 3N interactions,
in this work we will mainly focus on interactions derived within chiral
effective field theory (EFT).

Specifically, in Section~\ref{sec:chiral_EFT} we review recent efforts to
derive NN plus 3N interactions within chiral EFT with a particular focus on 3N
interactions and summarize different strategies to fit the unknown LECs. In
Section~\ref{sect:3NF_representation} we discuss in detail how 3N interactions
can be practically represented and computed in a form suitable for \textit{ab initio}
many-body frameworks. In addition, we summarize different currently employed
regularization schemes and illustrate the degree of scheme dependence of these
interactions. In Section~\ref{sec:3N_incorporation} we present novel methods
that help to simplify the incorporation and treatment of 3N interactions in
practical calculations. In particular, we focus here on the SRG and normal
ordering. In Section~\ref{sec:applications} we give a selection of
state-of-the-art calculations of atomic nuclei and dense matter based on the
most recent chiral EFT interactions. Finally, in Section~\ref{sec:outlook} we
conclude and give an outlook.

\clearpage
\section{Nuclear interactions and chiral effective field theory}
\label{sec:chiral_EFT}

In this section we review recent developments and the current status of
deriving NN, 3N and higher-body forces based on EFT principles. Compared to
more phenomenological approaches EFT provides a framework that allows to
derive contributions more systematically in a low-energy expansion scheme and
to estimate theoretical uncertainties due to neglected higher-order terms. In
this work we will not discuss the underlying concepts of chiral EFT, but
rather focus on providing an overview of recent developments and strategies to
derive novel and improved NN and 3N interactions for state-of-the-art
calculations of atomic nuclei and nuclear matter, with a particular focus on
3N interactions. For details on the underlying concepts of chiral EFT to
nuclear forces we refer the reader to
Refs.~\cite{Epel09RMP,Epel10primer,Mach11PR,Hamm19Rev} and references therein.
We also note that all ``high precision'' chiral interactions developed so far
and discussed below are based on the power counting scheme originally
suggested by Weinberg~\cite{Wein90NFch,Wein91chNp}. There are ongoing
discussions regarding alternative EFT expansion schemes~\cite{Hamm19Rev} which
involve the promotion of short-range couplings to lower orders in the chiral
expansion for improved RG invariance (see, e.g.,
Refs.~\cite{Kapl98pcount,Kapl98pionless,Nogg05renorm,Birs05powcount,Long12powercount,Vald14currents,Long16powercount,Vald16powercountsingl,Vald16powercount,Hamm19Rev})
and the perturbative many-body treatment of all interaction contributions
beyond leading order in the chiral expansion (see, e.g.,
Ref.~\cite{Dris19renorm}). However, so far no interactions suitable for
practical nuclear structure calculations have been developed within these
alternative schemes. This is currently work in progress. We stress that the
techniques discussed in Sections~\ref{sect:3NF_representation}
and~\ref{sec:3N_incorporation} on the calculation and treatment of 3N
interactions for many-body calculations are general and can be applied to
interactions derived in any power counting scheme and also to phenomenological
3N interactions.

\subsection{Chiral expansion of nuclear forces}
\label{sec:chiral_expansion}

Chiral EFT represents a systematic framework to describe the strong
interaction at momentum scales of the order of the pion mass $Q \approx m_\pi$,
constrained by the symmetries and symmetry-breaking patterns of QCD. In chiral
EFT the interaction between nucleons is parametrized in terms of long-range
pion-exchange interactions and short-range contact interactions. These
contributions to the nuclear interactions can be organized in a systematic
expansion in powers of $Q/\Lambda_{\rm b}$, where $\Lambda_{\rm b}$ denotes
the breakdown scale of the EFT. The expansion of any effective theory is based
on a separation of scales. Chiral EFT in particular exploits the mass
separation between the lightest meson, the pion with $m_{\pi} \approx 135$ MeV,
and the next lightest meson, the $\rho$ meson with $m_{\rho} \approx 770$ MeV.
This mass separation is rooted in the chiral symmetry breaking pattern of QCD
which requires the existence of an unnaturally light Goldstone boson in the
form of the pion. Since only pion contributions are treated explicitly in
chiral EFT while the contributions from heavier mesons are captured implicitly
in terms of short-range interactions, the breakdown scale is expected to be
below the $\rho$ mass and usually chosen to be around $\Lambda_{\rm b} \approx
500$ MeV. This results in an effective expansion parameter $Q/\Lambda_{\rm b}
\approx \tfrac{1}{3}$.

The values of the short-range interactions cannot be determined within the EFT
but need to be fixed either based on the underlying theory, i.e., QCD, or by
fitting them to observables. In the case of chiral EFT the short-range
couplings entering the NN interactions are usually fit to elastic two-nucleon
scattering cross sections and properties of the deuteron. In the future it might
be possible to connect nuclear forces directly to the underlying theory
through Lattice QCD~\cite{Bean11PPNP,Bric14JPG}. It is important, however, to
keep in mind that nuclear forces are generally non-observable, and in
particular the values of the short-range couplings are strongly scale and
scheme dependent, which can be made manifest in RG
frameworks~\cite{Bogn10PPNP} (see Section~\ref{sec:SRG}).

\begin{figure}[t]
\centering
\includegraphics[width=0.95\textwidth]{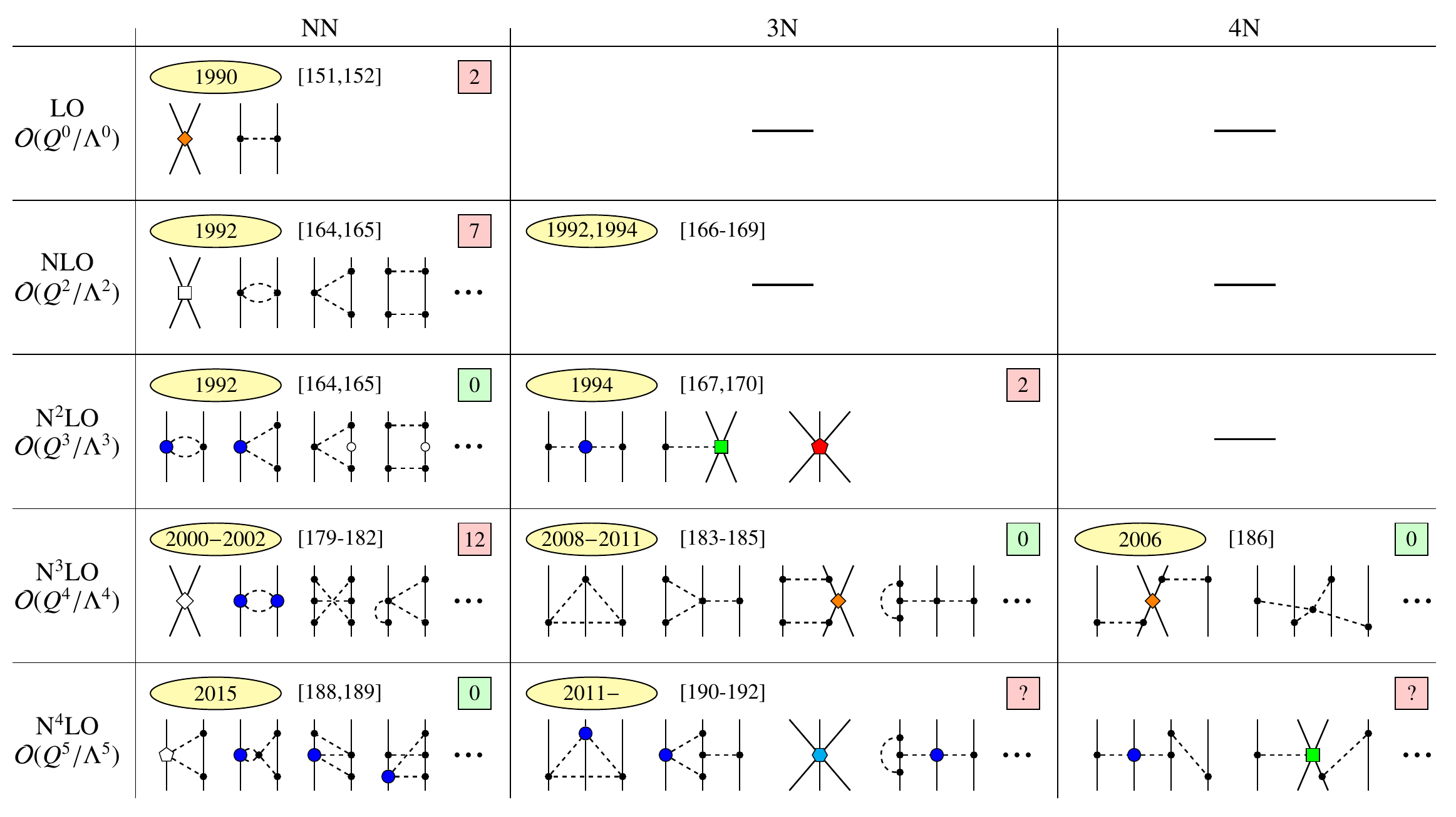}
\caption{Contributions to NN, 3N and 4N interactions in chiral
EFT at different orders in the chiral expansion within Weinberg's power
counting without explicit $\Delta$ degrees of freedom for intermediate states.
Solid and dashed lines denote nucleon and pion propagators, respectively. In
each panel we give the years when the terms were first derived, with
corresponding references and the number of new couplings (in naive dimensional
analysis) in the upper right corner. The low-energy coupling constants that
enter the 3N interactions are highlighted by vertices of different colors and
shapes. The following couplings show up first at N$^2$LO: the long-range
pion-nucleon couplings $c_i$ (\symbolcircle[FGBlue]), the intermediate-range
coupling $c_D$ (\symbolbox[green]) and the short-range 3N coupling $c_E$
($\pentagofill[red]$). The 3N and 4N interactions at N$^3$LO are
parameter-free in the sense that they only depend on short-range couplings
that appear already at lower orders, like, e.g., the leading-order NN
couplings $C_S$ and $C_T$ (\symboldiamond[orange]). The derivation of the 3N
interactions at N$^4$LO is still work in progress, in particular the
short-range contributions (formally indicated by the couplings
$\hexagofill[cyan]$). Higher-order couplings that only enter NN interactions
up to the shown orders are indicated by white vertices. See main text for
details.}
\label{fig:chiral_EFT_deltaless_table}
\end{figure}

\begin{figure}
\centering
\begin{minipage}[c]{0.3\textwidth}
\centering
\begin{tikzpicture} 
\begin{feynman}
\vertex (a) at (0,0) {}; 
\vertex (b) at (1.2,0) {};
\vertex (c) at (2.4,0) {};
\vertex [dot] (d) at (0,1) {}; 
\vertex [blob, /tikz/minimum size=7pt,fill=blue] (e) at (1.2,1) {};
\vertex [dot] (f) at (2.4,1) {};
\vertex (g) at (0,2) {}; 
\vertex (h) at (1.2,2) {};
\vertex (i) at (2.4,2) {};
\vertex (j) at (1.2,-0.3) {$V_{\text{3N}}^{2\pi} (c_1, c_3, c_4)$};
\diagram* {
(a) -- [fermion, line width=0.25mm] (d) -- [fermion, line width=0.25mm] (g);
(b) -- [fermion, line width=0.25mm] (e) -- [fermion, line width=0.25mm] (h);
(c) -- [fermion, line width=0.25mm] (f) -- [fermion, line width=0.25mm] (i);
(e) -- [scalar, edge label=$\pi$] (d);
(f) -- [scalar, edge label=$\pi$] (e);
};
\end{feynman}
\end{tikzpicture}
\end{minipage}
\hspace{0.3cm}
\begin{minipage}[c]{0.3\textwidth}
\centering
\begin{tikzpicture} 
\begin{feynman}
\vertex (a) at (0,0) {}; 
\vertex (b) at (1.2,0) {};
\vertex (c) at (2.4,0) {};
\vertex [dot] (d) at (0,1) {}; 
\vertex [blob, /tikz/minimum size=6pt,fill=green, shape=rectangle] (e) at (1.8,1) {};
\vertex (f) at (0,2) {}; 
\vertex (g) at (1.2,2) {};
\vertex (h) at (2.4,2) {};
\vertex (i) at (1.2,-0.3) {$V_{\text{3N}}^{1\pi} (c_D)$};
\diagram* {
(a) -- [fermion, line width=0.25mm] (d) -- [fermion, line width=0.25mm] (f);
(b) -- [fermion, line width=0.25mm] (e) -- [fermion, line width=0.25mm] (g);
(c) -- [fermion, line width=0.25mm] (e) -- [fermion, line width=0.25mm] (h);
(e) -- [scalar, edge label=$\pi$] (d);
};
\end{feynman}
\end{tikzpicture}
\end{minipage}
\hspace{0.1cm}
\begin{minipage}[c]{0.25\textwidth}
\centering
\begin{tikzpicture} 
\begin{feynman}
\vertex (a) at (0,0) {}; 
\vertex (b) at (0.8,0) {};
\vertex (c) at (1.6,0) {};
\vertex [blob, /tikz/minimum size=8pt,fill=red, regular polygon,regular polygon sides=5] (d) at (0.8,1) {};
\vertex (e) at (0,2) {}; 
\vertex (f) at (0.8,2) {};
\vertex (g) at (1.6,2) {};
\vertex (h) at (0.8,-0.3) {$V_{\text{3N}}^{\text{contact}} (c_E)$};
\diagram* {
(a) -- [fermion, line width=0.25mm] (d) -- [fermion, line width=0.25mm] (e);
(b) -- [fermion, line width=0.25mm] (d) -- [fermion, line width=0.25mm] (f);
(c) -- [fermion, line width=0.25mm] (d) -- [fermion, line width=0.25mm] (g);
};
\end{feynman}
\end{tikzpicture}
\end{minipage}
\caption{3N interactions at N$^2$LO in the chiral expansion. The long-range
pion-nucleon couplings $c_i$ (\symbolcircle[FGBlue]) also enter the NN interactions at
this and higher orders and can hence be constrained by NN observables and
$\pi$N scattering data. The short-range couplings $c_D$ (\symbolbox[green]) and
$c_E$ ($\pentagofill[red]$) need to be fixed in three- or higher-body systems.}
\label{fig:3N_N2LO_diags}
\end{figure}
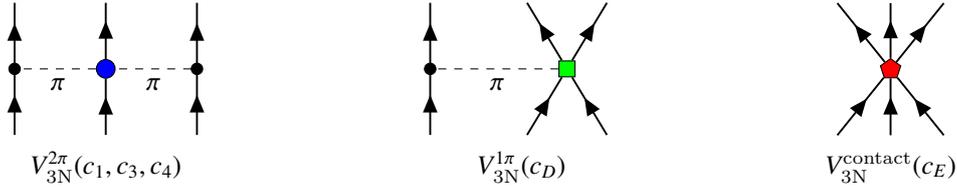

In Figure~\ref{fig:chiral_EFT_deltaless_table} we show the contributions to
NN, 3N and 4N interactions from one- or multi-pion exchanges, which govern the
long- and intermediate-range forces, as well as from short-range contact
interactions at different orders in the chiral expansion. The short-range
couplings are fit to low-energy data and thus capture all short-range effects
relevant at low energies. The leading-order (LO) contributions were derived by
Weinberg~\cite{Wein90NFch,Wein91chNp}. These seminal works represent the
foundation of the chiral expansion of nuclear forces that is still being used
today. At LO the NN interaction consists of a long-range pion exchange
interaction, which corresponds to the empirically well-known Yukawa
interaction~\cite{Yuka35mesonex} and two short-range interactions parametrized
by the coupling $C_S$ and $C_T$. In the following years this framework was
extended to higher orders in the expansion. At next-to-leading order (NLO) and
next-to-next-to-leading order (N$^2$LO) $2\pi$ exchange interactions and
higher-order short-range couplings contribute to NN
forces~\cite{Ordo92NNN2LO,Ordo93NN}\footnote{We note that there exist
different schemes to count contributions from relativistic corrections,
indicated by the white circles in Figure~\ref{fig:chiral_EFT_deltaless_table}
(see, e.g., Ref.~\cite{Epel09RMP}).}. For 3N interactions it was shown that
contributions at NLO cancel
exactly~\cite{Wein923body,Kolc94fewbody,Yang863N,Coon863N}, while the first
nonvanishing contributions to 3N interactions appear at
N$^2$LO~\cite{Kolc94fewbody,Epel02fewbody}. The 3N interactions at this order
include long-range two-pion exchange interactions $V_{\text{3N}}^{2\pi}$, an
intermediate-range one-pion plus contact interaction $V_{\text{3N}}^{1\pi}$
and a pure contact interaction $V_{\text{3N}}^{\text{contact}}$ (see
Figure~\ref{fig:3N_N2LO_diags}). Since these 3N interactions play a central
role in this work as well as in many recent and current nuclear structure
studies, we discuss them in more detail now. The interactions are given by the
following expressions~\cite{Epel02fewbody}:
\begin{equation}
V^{2\pi}_{\text{3N}} = \frac{1}{2} \, \biggl( \frac{g_A}{2 f_\pi} \biggr)^2 
\sum\limits_{i \neq j \neq k} 
\frac{({\bm \sigma}_i \cdot {\bf Q}_i) ({\bm \sigma}_j \cdot 
{\bf Q}_j)}{(Q_i^2 + m_\pi^2) (Q_j^2 + m_\pi^2)} \: F_{ijk}^{\alpha\beta} \,
\tau_i^\alpha \, \tau_j^\beta \,,
\label{eq:Vc}
\end{equation}
where ${\bf Q}_i = {\bf k}'_i - {\bf k}_i$ denotes the momentum transfers,
i.e., the difference between the initial and final single-particle momenta
($\mathbf{k}_i$ and $\mathbf{k}'_i$ respectively), with $i, j$ and $k=1,2,3$,
$\sigma_i^a$ ($\tau_i^a$) the Cartesian component $a$ of the spin (isospin)
operators of particle $i$ and
\begin{equation}
F_{ijk}^{\alpha\beta} = \delta^{\alpha\beta} \biggl[ - \frac{4 c_1 
m_\pi^2}{f_\pi^2} + \frac{2 c_3}{f_\pi^2} \: {\bf Q}_i \cdot {\bf Q}_j
\biggr] + \sum_\gamma \, \frac{c_4}{f_\pi^2} \: \epsilon^{\alpha\beta\gamma}
\: \tau_k^\gamma \: {\bm \sigma}_k \cdot ( {\bf Q}_i \times {\bf Q}_j)
\,. \label{eq:Vc2}
\end{equation}
The $1 \pi$-exchange and contact interactions are given respectively by
\begin{align}
V^{1\pi}_{\text{3N}} = - \frac{g_A}{8 f_\pi^2} \, \frac{c_D}{f_\pi^2 \lm_\chi}
\: \sum\limits_{i \neq j \neq k} 
\frac{{\bm \sigma}_j \cdot {\bf Q}_j}{Q_j^2 + m_\pi^2} \: ({\bm \tau}_i
\cdot {\bm \tau}_j) \, ({\bm \sigma}_i \cdot {\bf Q}_j) \,, \quad V_{\text{3N}}^{\text{contact}} = \frac{c_E}{2 f_\pi^4 \lm_\chi} \: \sum\limits_{j \neq k} ({\bm \tau}_j
\cdot {\bm \tau}_k) \,. \label{VE}
\end{align}
In Ref.~\cite{Epel02fewbody} the values  $g_A = 1.29$, $f_\pi = 92.4 \mev$,
$m_\pi = 138 \mev$ and $\lm_\chi = 700
\mev$ were chosen. Similarly to the LO one-pion exchange interactions,
the long-range two-pion exchange contribution $V^{2\pi}_{\text{3N}}$ resembles
features of previously developed phenomenological 3N
forces~\cite{Fuji573N,Coon783N,Coel843N}. However, we stress that in contrast
to these interactions, the $V^{2\pi}_{\text{3N}}$ interactions formally do not
contain any new parameters since the subleading pion-nucleon couplings $c_1$,
$c_3$ and $c_4$, which characterize the strength of $V^{2\pi}_{\text{3N}}$
(see Figure~\ref{fig:3N_N2LO_diags}), also appear in the NN interaction at
N$^2$LO (see Figure~\ref{fig:chiral_EFT_deltaless_table}) and play also a key
role in $\pi$-nucleon ($\pi$N) scattering. In fact, currently the most robust
extraction of the values of these long-range couplings was achieved based on
the Roy-Steiner-equation analysis of $\pi$N
scattering~\cite{Hofe15piNchiral,Hofe15PhysRep,Siem16RoySt}. This demonstrates that
contributions to NN and 3N interactions as well as terms determining the
pion-nucleon scattering dynamics are treated on equal footing in chiral EFT,
in contrast to most phenomenological approaches. The 3N interactions
$V_{\text{3N}}^{1\pi}$ and $V_{\text{3N}}^{\text{contact}}$ on the other hand
depend on two low-energy couplings $c_D$ and $c_E$ that encode pion
interactions with short-range NN pairs and short-range three-body physics,
respectively~\cite{Kolc94fewbody,Epel02fewbody}. These genuine three-body
couplings do not appear in NN interactions and hence need to be fixed in few-
or many-body systems (see Section~\ref{sec:3N_fits}).

\begin{figure}[b]
\centering
\includegraphics[width=0.9\textwidth]{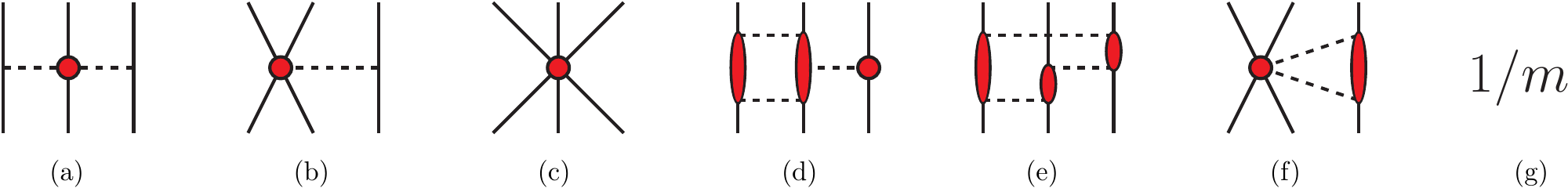}
\caption{Different topologies contributing to chiral 3N interactions up to
  N$^3$LO (and N$^4$LO). Nucleons and pions are represented by solid and
  dashed lines, respectively.  The vertices denote the amplitudes of the
  corresponding interaction. Specifically, the individual diagrams are: (a)
  2$\pi$ exchange, (b) 1$\pi$-contact, (c) pure contact, (d) 2$\pi$-1$\pi$
  exchange, (e) ring contributions, (f) 2$\pi$-contact and (g) relativistic
  corrections.\\
  \textit{Source:} Figure taken from Ref.~\cite{Hebe15N3LOpw}.}
  \label{fig:topologies}
\end{figure}

Even though nuclear forces are not observable, there are natural sizes of two-
and many-body-force contributions that are made manifest in the EFT power
counting (see Figure~\ref{fig:chiral_EFT_deltaless_table}) and which explain
the phenomenological hierarchy of contributions from NN and many-body forces
to observables, i.e., schematically $V_{\text{NN}} > V_{\text{3N}} >
V_{\text{4N}} $~\cite{Epel09RMP,Mach11PR}. Although it might be tempting to
neglect contributions from 3N interactions in cases when calculations based on
only NN forces already provide a good description of experimental data (see,
e.g., Ref.~\cite{Ekst13optNN}), EFT power counting dictates the inclusion of
all many-body forces up to a given order. In fact, explicit calculations show
that 3N forces typically provide important contributions in nuclei and
matter~\cite{Hebe15ARNPS} (see also Sections~\ref{sec:3N_fits}
and~\ref{sec:applications}).

The evaluation of the contributions to NN interactions at
next-to-next-to-next-to-leading-order (N$^3$LO) is quite involved as they
include two-loop pion contributions, three-pion exchange contributions as well
as relativistic
corrections~\cite{Kais993pi,Kais003pigA,Kais012pi2loop,Kais012pirel}. The 3N
interactions at this order also include many new structures as shown in
Figure~\ref{fig:topologies}, but are predicted in a parameter-free way since
they only depend on the leading NN contact interactions proportional to the
LECs $C_S$ and $C_T$~\cite{Bern083Nlong,Bern113Nshort,Ishi07N3LO} (see the 2$\pi$-contact
contributions (f) and the relativistic corrections (g) in
Figure~\ref{fig:topologies}). In addition, the first nonvanishing
contributions to 4N interactions appear at this order~\cite{Epel064N}, which
are also predicted in a parameter-free way. Remarkably, for systems consisting
of only neutrons, the N$^2$LO 3N interactions $V_{\text{3N}}^{1\pi}$ and
$V_{\text{3N}}^{\text{contact}}$ do not contribute for unregularized or
nonlocally regularized interactions \cite{Hebe10nmatt} (see
Section~\ref{sec:3N_regularization} for details). Therefore, chiral EFT
predicts all three-neutron and four-neutron forces up to N$^3$LO.

Due to the involved analytical structure of the 3N interactions at N$^3$LO the
implementation has only been completed relatively
recently~\cite{Gola14n3lo,Hebe15N3LOpw,Dris17MCshort}. However, the
development and implementation of optimized regularization schemes is still
work in progress (see Section~\ref{sec:3N_regularization}). The practical
calculation of all topologies shown in Figure~\ref{fig:topologies} in a form
suitable for few- and many-body calculations is a nontrivial task but is key
for improved many-body calculations and a systematic investigation of chiral
power counting in the 3N sector. Due to the large amounts of required
computational resources using traditional methods, the available 3N matrix
elements were at first restricted to small basis spaces~\cite{Gola14n3lo},
insufficient for converged many-body calculations. In
Section~\ref{sect:3NF_representation} we discuss in detail a more efficient
method for decomposing 3N forces in a momentum partial-wave
basis~\cite{Hebe15N3LOpw}. The new framework makes explicit use of the fact
that all unregularized contributions to chiral 3N forces are either local,
i.e., they depend only on momentum transfers, or they contain only polynomial
nonlocal terms (see Section~\ref{sec:PWD_3NF_local}). These new developments
allow to calculate matrix elements of all N$^3$LO 3N contributions for large
basis spaces, opening the way to \textit{ab initio} studies of nuclei and nucleonic
matter.

Recently the derivation of the NN contributions has been extended to
N$^4$LO~\cite{Ente17EMn4lo,Epel14SCSprl}. The corresponding 3N contributions
at this order have been worked out for the long- and intermediate-range
parts~\cite{Girl113N4Ncont,Kreb123Nlong,Kreb133Ninterm}, while the derivation
of all short-range parts is still work in progress. In contrast to the N$^3$LO
contributions new short-range couplings appear at this order and need to be
fixed based on properties of few- or many-body systems, in addition to the
couplings $c_D$ and $c_E$. The 4N contributions at N$^4$LO have not been
worked out yet.

\begin{figure}[t]
\centering
\includegraphics[width=0.7\textwidth]{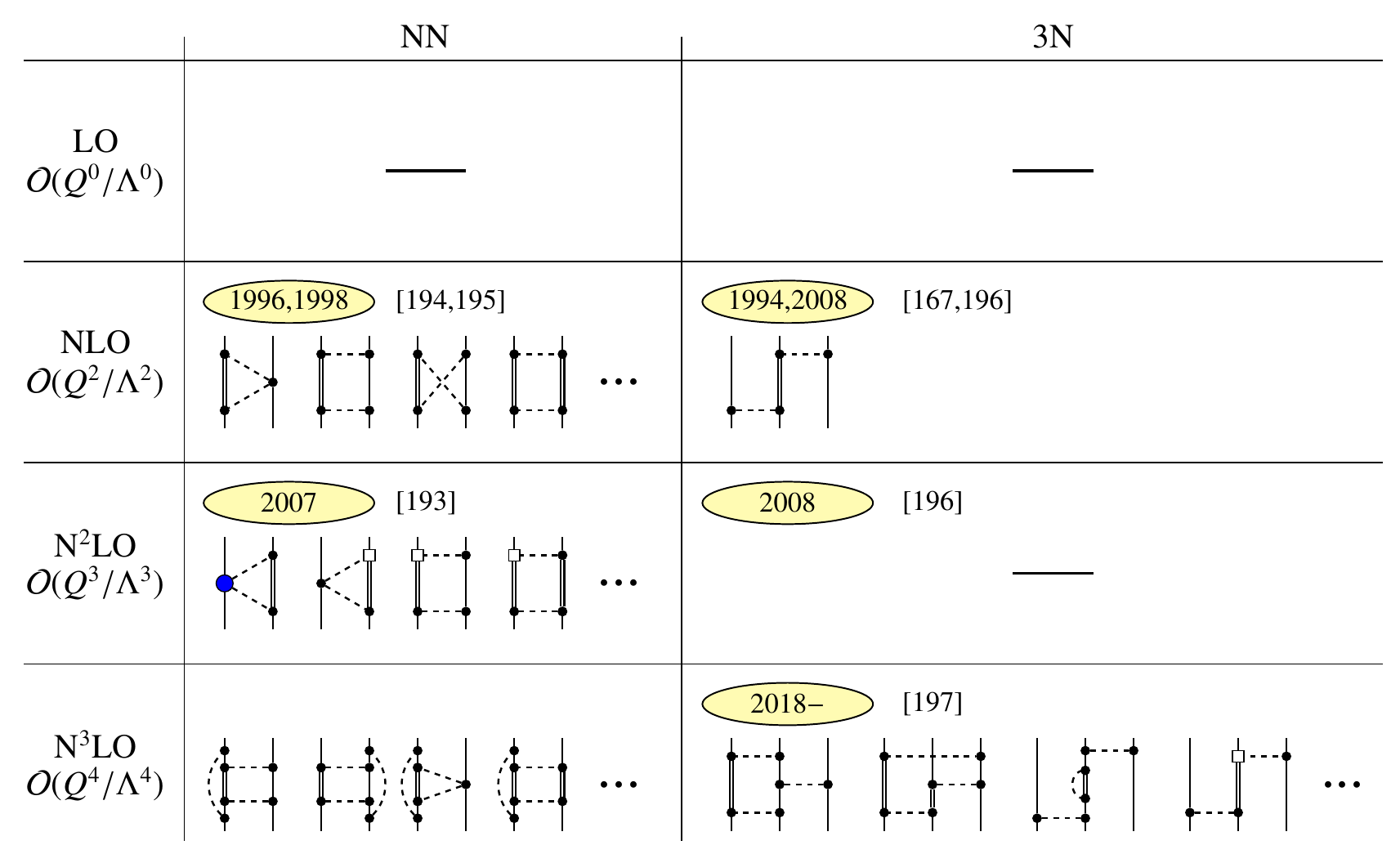}
\caption{Additional diagrams contributing to NN and 3N interactions
at different orders in $\Delta$-full chiral EFT compared to the terms shown in
Figure~\ref{fig:chiral_EFT_deltaless_table}. Double solid lines denote
intermediate states of the $\Delta$ resonance. We use the same notation as
in Figure~\ref{fig:chiral_EFT_deltaless_table}.}
\label{fig:chiral_EFT_deltafull_table}
\end{figure}


All the discussed developments above were performed in the so-called
$\Delta$-less chiral EFT formulation. Since it is known that the $\Delta$
resonance of the nucleon at the excitation energy of $\Delta = m_{\Delta} -
m_{N} \approx 300$ MeV plays an important role in nucleon scattering, it is
argued that the chiral EFT expansion might be more efficient and exhibit a
faster converging pattern by treating the $\Delta$ explicitly in intermediate
states rather than implicitly in low-energy couplings. By introducing this new
degree of freedom an additional small energy scale appears in the EFT
expansion and the power counting has to be adjusted. As a consequence
additional diagrams with new couplings appear at different orders, while a
specific types of diagrams in the $\Delta$-less formulation get promoted to
lower orders (see Figure~\ref{fig:chiral_EFT_deltafull_table}). In particular,
if Ref.~\cite{Kreb07Deltas} it was shown that the contributions from
intermediate $\Delta$ excitations expanded in powers of $1/\Delta$ can be
absorbed via a shift in the couplings in $\Delta$-less EFT. The $\Delta$
contributions in the subleading pion-nucleon couplings take the following
form: $c_3^{\Delta} = - 2 c_4^{\Delta} =- 4 h_A^2/(9 \Delta)$, with the $\pi N
\Delta$ axial coupling $h_A$. Comparing the values of these shifts with the
typical numerical values of the $c_i$ couplings in $\Delta$-less EFT
demonstrates that indeed a significant part of the $c_i$ contributions can be
attributed to the $\Delta$ resonance~\cite{Kreb07Deltas}.

The first additional contributions to NN and 3N interactions in $\Delta$-full
theory show up at NLO~\cite{Ordo95NN,Kais98Delta,Kolc94fewbody,Epel08Delta3N}.
These diagrams can all be interpreted as promoted diagrams at N$^2$LO and
N$^3$LO in Figure~\ref{fig:chiral_EFT_deltaless_table} by replacing the
subleading couplings $c_i$ by intermediate $\Delta$ resonance states. The fact
that 3N interactions already contribute at NLO in this EFT indicates that the
effects of 3N interactions might be formally underestimated in $\Delta$-less
theory and that 3N contributions get enhanced due to unnaturally large LEC
values and might also lead to a slower convergence of the expansion series. At
N$^2$LO various new NN contributions appear~\cite{Kreb07Deltas} while no new
3N contributions are generated~\cite{Epel08Delta3N}. This implies that up to
N$^2$LO the 3N interaction topologies are identical to those of $\Delta$-less
EFT, apart from a reshuffling of some contributions to NLO. Finally, the
derivation of contributions at N$^3$LO is still work in progress. In the case
of 3N interactions the long-range contributions have been
developed~\cite{Kreb183NDelta} while the intermediate-range and short-range
parts are still under development.

In the next subsections we will give an overview of recent developments of new
interactions within these two chiral EFT formulations for many-body
calculations. An important open question concerns how to fit the NN and 3N
LECs up to a given order. In particular, given that several low-energy
couplings appear on equal footing in NN and many-body forces, it may be
beneficial to fit several different two- and few-body observables
simultaneously within theoretical uncertainties or to also include information
beyond few-nucleon systems in the fits. In the following we will summarize
recent explorations of these different strategies and briefly discuss
advantages and disadvantages of each approach.

\subsection{A brief history of chiral EFT NN interactions}

The first ``high-precision'' NN interactions derived within chiral EFT
including contributions up to N$^3$LO were constructed in the years 2002 to
2005 in Refs.~\cite{Epel02NNint,Epel05EGMN3LO} and~\cite{Ente03EMN3LO}. The
low-energy couplings of these two families of interactions were fitted to
neutron-proton and proton-proton scattering phase shifts and the resulting
interactions were provided in a partial-wave decomposed form, suitable for
applications in many-body frameworks. Even though both interactions are
derived within the same power counting framework they differ in several ways,
particularly in the choice of regularization (see discussion in
Ref.~\cite{Epel05EGMN3LO} for details). The accuracy of the reproduction of NN
scattering phase shifts of the interaction presented in
Ref.~\cite{Ente03EMN3LO} was comparable to the best available phenomenological
NN interactions at this time up to laboratory energies of about 290 MeV. These
features made this force the interaction of choice for many-body calculations
in the following years.

It took about 10 years until the next generation of NN interactions was
developed. First, the NN interaction ``N$^2$LO$_{\text{opt}}$'' was
constructed~\cite{Ekst13optNN} by fitting the LECs up to N$^2$LO to phase
shifts up to 125 MeV using the automated derivative-free POUNDERS
method~\cite{Kort10edf} for the $\chi^2$ minimization. First calculations of
few-body systems based on only NN interactions showed remarkable agreement
with experimental results, suggesting that the missing 3N contributions might
be small. However, explicit calculations including 3N interactions showed that
they provide significant contributions~\cite{Navr19privcom}. Within the same
year another novel chiral EFT NN interaction up to N$^2$LO was
developed~\cite{Geze13QMCchi,Geze14long}. In contrast to the previous chiral
EFT interactions this interaction was local, including the choice for the
regulator (see also Section~\ref{sec:PWD_3NF_local}), which made this force
particularly suitable for Quantum Monte Carlo applications~\cite{Carl15RMP}.
Following these developments, another ``minimally nonlocal'' interaction was
developed in 2015~\cite{Piar14Deltas} and later reduced to a fully local
potential in 2016 by discarding the nonlocal terms~\cite{Piar16DeltaNuc}.
These forces contain contributions from the $\Delta$ degree of freedom to
2$\pi$ exchange interactions up to N$^2$LO.

In 2014 a new NN interaction including contributions up to N$^3$LO was
presented~\cite{Epel15improved}. One new feature of this interaction compared
to its predecessor published in Ref.~\cite{Epel05EGMN3LO} was the use of a
local coordinate-space regulator (see Section~\ref{sec:3N_regularization}) for
the long-range parts of the interaction, similar to the interactions of
Refs.~\cite{Geze13QMCchi,Geze14long}. It was argued that this choice of
regulator reduces finite cutoff artifacts and preserves the analytic structure
of the scattering amplitude (see Ref.~\cite{Epel15improved} for details). In
addition, a novel method for quantifying theoretical uncertainties due to the
truncation of the chiral expansion was proposed. The same year contributions
at N$^4$LO were included for the first time~\cite{Epel14SCSprl}, while it was
shown that long-range parameter-free terms at this order lead to a significant
improvement of the reproduction of scattering phase shifts. The interactions
were made available for different cutoff scales and at different orders in the
chiral expansion, allowing for a more systematic study of the theoretical
uncertainties of many-body observables. In 2017 a second set of NN
interactions up to N$^4$LO was presented~\cite{Ente17EMn4lo}. Similarly to the
potential of Ref.~\cite{Epel14SCSprl} a whole set of interactions with
different regularization cutoff scales and chiral orders were provided. On the
other hand, a nonlocal momentum regulator was used like in
Ref.~\cite{Ente03EMN3LO}. Finally, in 2018 a local momentum-space regularized
N$^4$LO interaction was presented~\cite{Rein17semilocal}. This new interaction
combines the advantages of the previously developed interaction of
Ref.~\cite{Epel15improved} with simplified calculations of many-body forces
and current operators in the novel regularization scheme (see
Section~\ref{sec:3N_regularization} for details). In addition, it was
demonstrated that the removal of redundant short-range couplings leads to
simplified fits of scattering observables and softer interactions. Finally, in
Ref.~\cite{Ekst17deltasat} a nuclear NN interaction (plus 3N interaction) with
explicit contributions from the $\Delta$ isobar was constructed. Interactions
at different orders up to N$^2$LO were fitted to NN scattering phase shifts
and by using the pion-nucleon LECs from the Roy-Steiner analysis of $\pi$N
scattering phase shifts~\cite{Hofe15piNchiral,Hofe15PhysRep}. Calculations of
heavier nuclei as well as matter based on these interactions at N$^2$LO, plus
corresponding 3N interactions (see Section~\ref{sec:sep_3N_fits_delta}),
showed remarkable agreement with experimental results, which might be an
indication for an improved and more natural convergence of the chiral
expansion (see Section~\ref{sec:chiral_expansion}).

\subsection{Chiral EFT three-nucleon interactions and fitting strategies}
\label{sec:3N_fits}

\begin{figure}[t]
\centering
\includegraphics[width=0.45\textwidth]{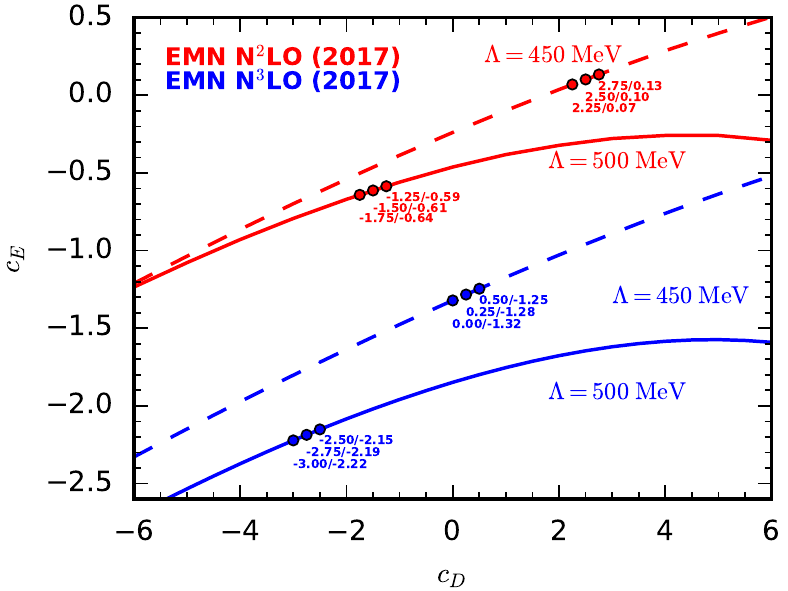}
\caption{Three-nucleon couplings $c_D$ and $c_E$, 
fit to the $^3$H binding energy using the NN potentials of
Ref.~\cite{Ente17EMn4lo} with $\Lambda = 450\MeV$~(dashed) and $\Lambda =
500\MeV$~(solid line) at N$^2$LO~(red) and N$^3$LO~(blue), combined with
consistent 3N interactions at these orders using $\Lambda =
\Lambda_{\text{NN,\,3N}}$.\\
\textit{Source:} Figure taken from Ref.~\cite{Dris17MCshort}.}
\label{fig:cd_ce}
\end{figure}

Parallel to the developments of new NN interactions as outlined in the
previous section first steps toward simultaneous fits of NN and 3N
interactions up to a given order in the chiral expansion were
achieved~\cite{Carl15sim,Ekst15sat}. While the determination of the
3N-interaction LECs is effectively a two-parameter problem (determination of
$c_D$ and $c_E$), a simultaneous fit of all LECs in NN and 3N interactions is
obviously of much higher complexity due to the large number of parameters. We
will first discuss different strategies that have been pursued in recent years
to determine the values for $c_D$ and $c_E$ for a given NN interaction based
on few-body observables (see also Ref.~\cite{Tews20ideas}) and then discuss
the simultaneous construction of NN and 3N interactions.

\subsubsection{Fits of 3N interactions based on fixed NN interactions ($\Delta$-less)}
\label{sec:sep_3N_fits}

One of the most natural observables for constraining the LECs of 3N
interactions is the binding energy of 3N systems. In fact, for most 3N
interactions developed in recent years the three-body ground-state energy has
been used as one of the fitting observables. Calculations based on only NN
interactions typically lead to a reasonable agreement with the experimental
ground-state energies of $^3$H and $^3$He ($E_{^3\text{H}} = -8.482$ MeV and
$E_{^3\text{He}} = -7.781$ MeV~\cite{Wang17AME16}), with a total net effect of
3N interactions typically of the order of 1 MeV or less. Due to subtleties
connected to the precise treatment and infrared regularization of the Coulomb
interaction for calculations of $^3$He, usually the 3N couplings are fitted to
the ground-state energy of $^3$H. Such fits provide a relation between the two
couplings, i.e., formally a function of the form $c_E (c_D)$. Figure
\ref{fig:cd_ce} shows an example of such relations for interactions at N$^2$LO
and N$^3$LO for the recently developed NN interactions of
Ref.~\cite{Ente17EMn4lo}. It can be argued if the perfect reproduction of the
experimental ground-state energy is useful, given that the interactions
contain inherent uncertainties due to truncation effects of neglected higher
order terms of the chiral expansion. It might be more natural and meaningful
to take into account these uncertainties at a given order in the chiral
expansion by fitting the LECs to some range around the experimental value.
Such a fit would result in a correlation band between $c_D$ and $c_E$ instead
of a correlation line. Such a strategy would allow to investigate to what
degree fits to different observables are consistent at a given order in the
chiral expansion. Work along these directions is currently in progress. For
the full determination of both 3N couplings, $c_D$ and $c_E$, a second
few-body observable is needed. Ideally both observables should be as
uncorrelated as possible. Examples of quantities that exhibit significant
correlations are the binding energy of $^3$He/$^3$H and $^4$He (``Tjon
line''~\cite{Tjon75tjonline,Plat04tjon}) or between $^3$He/$^3$H binding
energy and the neutron-deuteron (nd) scattering length (``Phillips
line''~\cite{Phil68line}). The correlation manifests itself in the form of a
mild sensitivity of the results for the correlated observables as a function
of the remaining coupling constant.

\begin{figure}[t]
\centering
\includegraphics[width=0.4\textwidth]{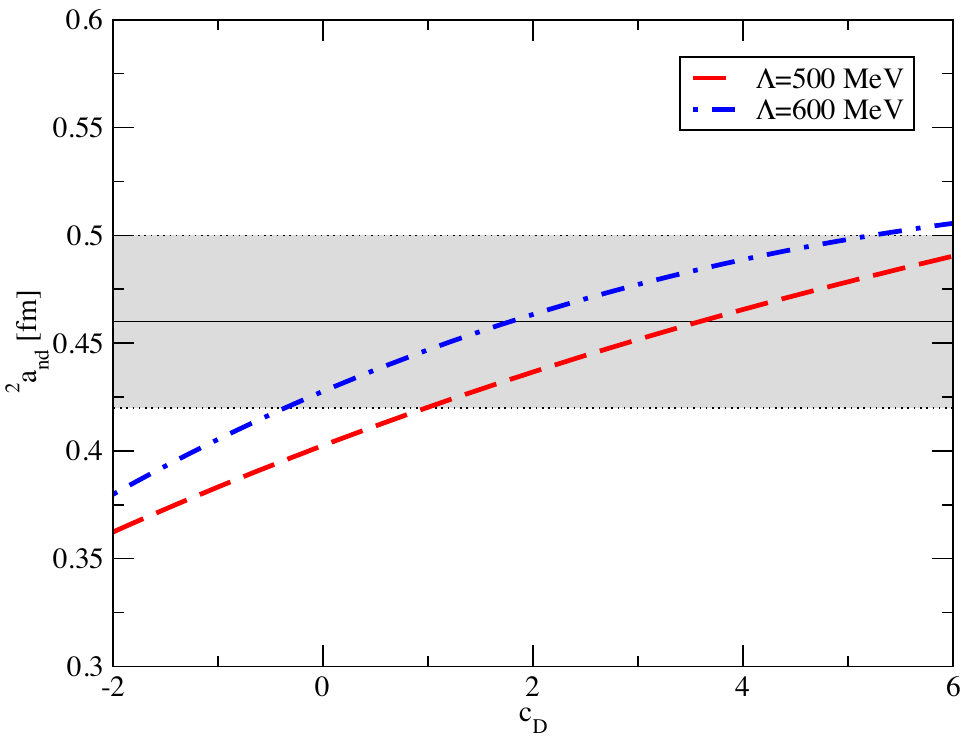}
\hspace{1cm}
\vspace{-0.2cm}
\includegraphics[width=0.45\textwidth]{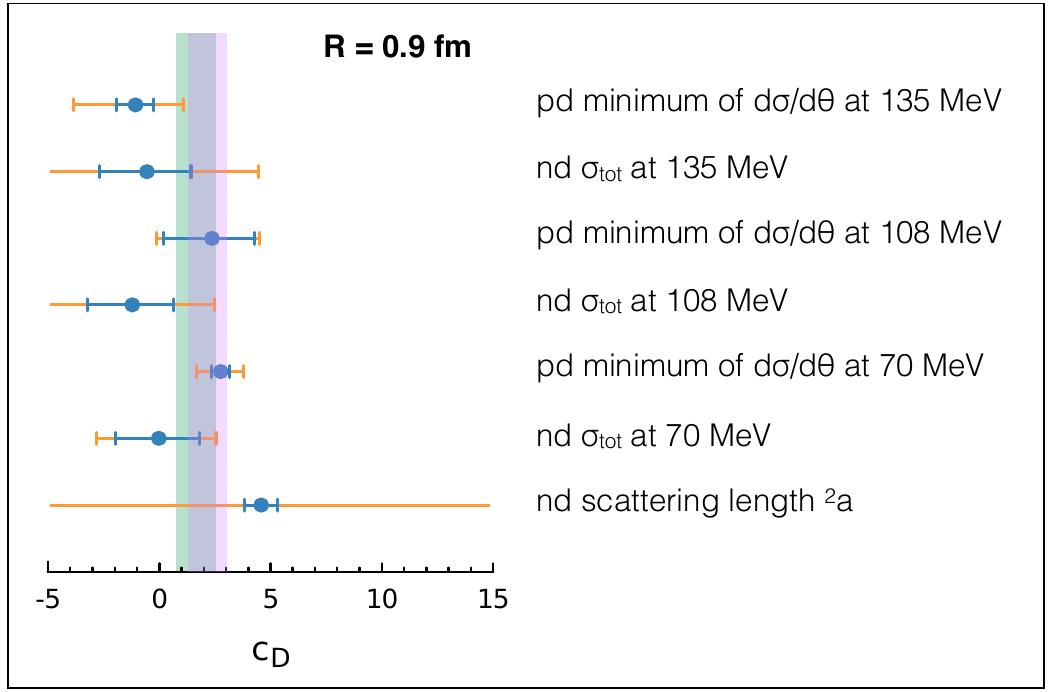}
\vspace{0.2cm}
\caption{Left: Neutron-deuteron scattering length $^2$a$_{nd}$
as function of the LEC $c_D$. The relation between $c_D$ and $c_E$ has been
determined by fits of the $^3$H binding energy based on the NN interactions of
Ref.~\cite{Epel02NNint} for two different cutoff values. The shaded band
indicates the uncertainty of the experimental scattering length. Right: Determination of the LEC $c_D$
from the differential cross section in elastic pd scattering, total nd cross
section and the nd doublet scattering length $^2$a for the cutoff choice of $R
= 0.9 \: \text{fm}$ of the interactions at N$^2$LO of Ref.~\cite{Epel18SCS3N}.
The smaller (blue) error bars correspond to the experimental uncertainty while
the larger (orange) error bars also take into account the theoretical
uncertainty estimated as described in Ref.~\cite{Epel15improved}. The pink
(green) bands show the results from a combined fit to all observables (to
observables up to $E = 108 \mev$).\\
\textit{Source:} Left figure taken from Ref.~\cite{Epel02fewbody} and right figure adapted from Ref.~\cite{Epel18SCS3N}.
}
\label{fig:cd_ce_ndscattering_length}
\end{figure}

In the left panel of Figure~\ref{fig:cd_ce_ndscattering_length} we show as an
example the results for the nd-scattering length as a function of the LEC
$c_D$ after the $c_E$ coupling has been fixed to the experimental $^3$H
binding energy for each value of $c_D$~\cite{Epel02fewbody}. The results show
that the nd scattering length is rather insensitive to $c_D$ since the
theoretical results are compatible with the experimental constraints over a
wide range of $c_D$ values. In addition, minor variations of the cutoff,
regularization schemes and inclusion of higher-order terms tend to lead to
significant changes in the values of the extracted LECs. This makes it hard or
even impossible to extract tight and robust constraints on this LEC from a fit
to the scattering length alone. One possibility to improve such a fit is to
include additional scattering observables and perform a global fit. An example
of such an analysis is shown in the right panel of
Figure~\ref{fig:cd_ce_ndscattering_length}. The figure shows the constraints
on $c_D$ resulting from the reproduction of the proton-deuteron differential
cross section data at $E = 70$ and $135$ MeV based on interactions at N$^2$LO
of Refs.~\cite{Epel14SCSprl,Epel15improved} (see Ref.~\cite{Epel18SCS3N} for
details). Such a more global analysis allows to improve the significance and
robustness of a fit based on three-body scattering observables.

Another three-body scattering observable sensitive to 3N interaction
contributions is connected to the long-standing ``$A_y$
puzzle''~\cite{Gloe95cont,Shim95Ay}. This puzzle refers to the observed large
discrepancy between theoretical predictions and experimental measurements of a
particular polarization observable, the so called vector analyzing power, in
elastic nucleon-deuteron scattering in the region of its maximum around the
center-of-mass angle $\Theta_{cm} \approx 120 \degree$ and for incoming
nucleon energies below $E
\approx 20$ MeV~\cite{Gloe95cont,Wita003N,Kiev01Ay}. So far, no satisfactory
resolution of this puzzle has been found. However, it should be emphasized
that the low-energy vector analysing power is a fine-tuned observable which is
very sensitive to changes in $^3P_j$ NN force components~\cite{Gola14n3lo}.
Thus, it is not obvious if the observed discrepancies can be mainly attributed
to deficiencies of presently used NN interaction or to genuine three-body
effects (see also Figure~\ref{fig:nd_scattering}).

\begin{figure}[t]
\centering
\includegraphics[width=0.45\textwidth]{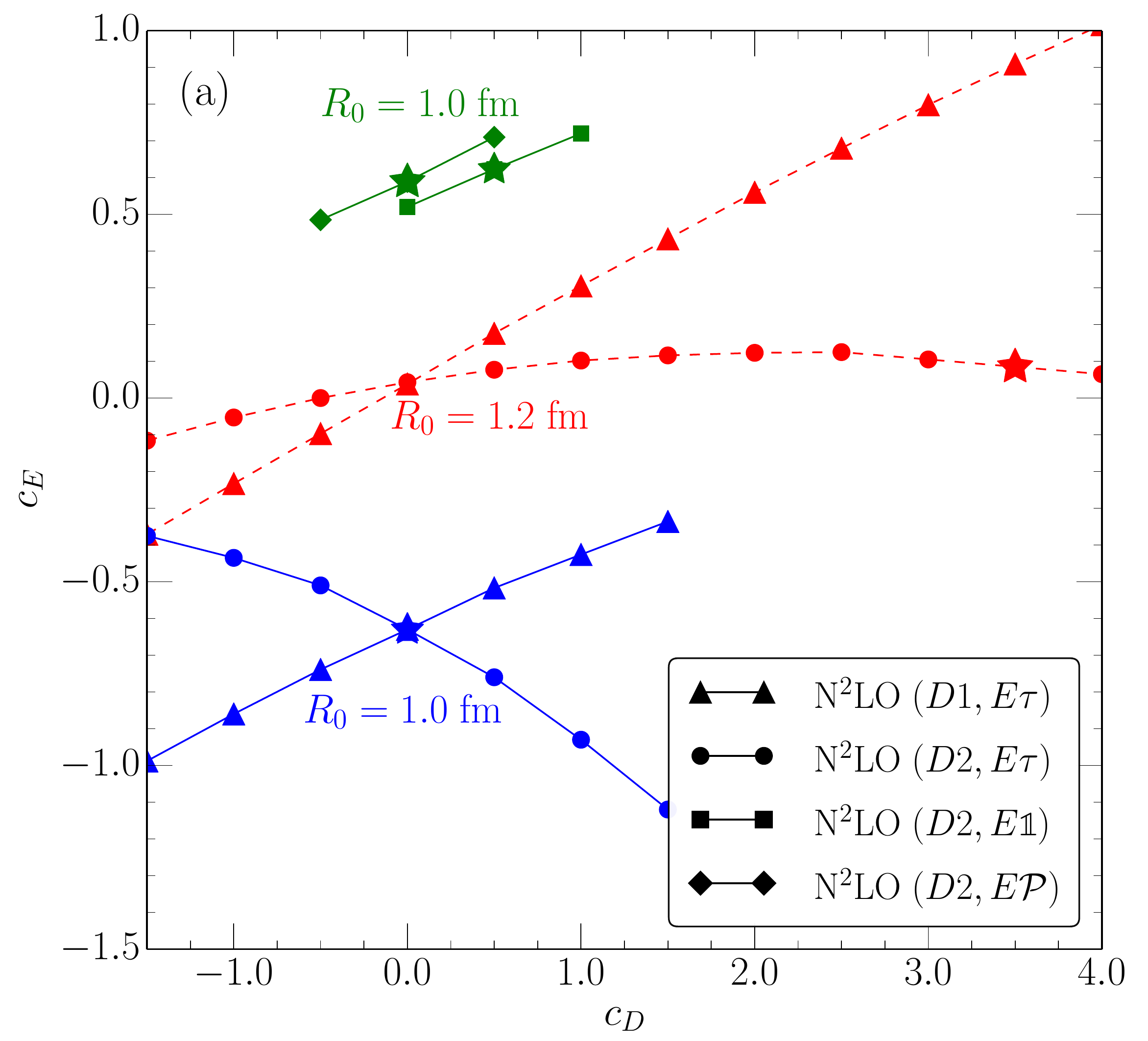}
\hspace{0.5cm}
\includegraphics[width=0.45\textwidth]{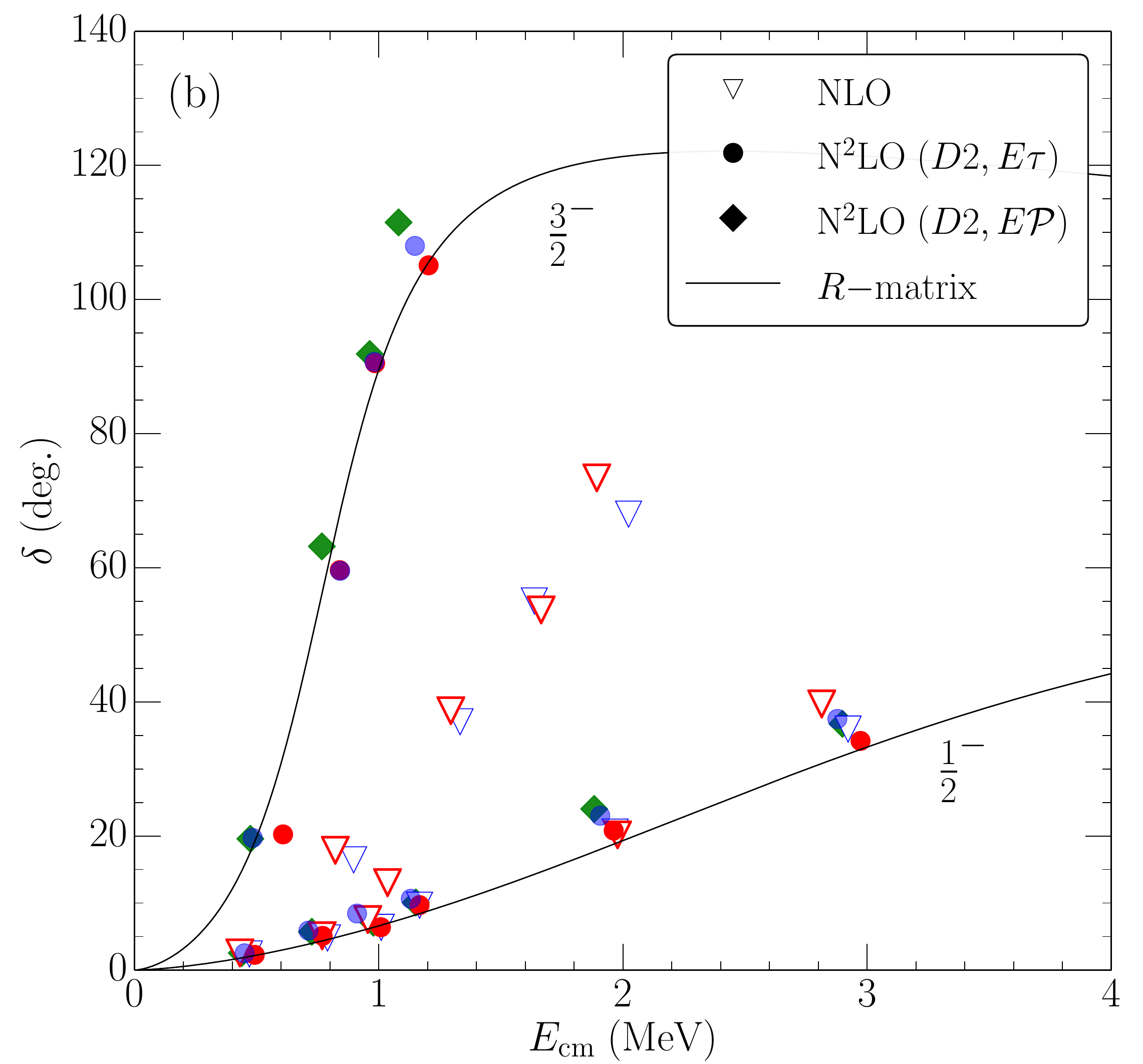}
\caption{
Left: Coupling values for $c_E$ and $c_D$ obtained by fits to the
$^4$He binding energy using different operator forms and different cutoff
values $R_0$ (see Ref.~\cite{Lynn16QMC3N} for details). The stars in the left
panel indicate the values of $c_D$ and $c_E$ that simultaneously fit the
experimental binding energy of $^4$He and the n$\alpha$ $P$-wave phase shifts
(see right panel). Right: $P$-wave n$\alpha$ elastic scattering phase
shifts compared with an $R$-matrix analysis of experimental data. The same
color coding as in the left panel has been used. Colors correspond to the
left panel.\\
\textit{Source:} Figures taken from Ref.~\cite{Lynn16QMC3N}. }
\label{fig:QMC_3N_fits}
\end{figure}

In Figure~\ref{fig:QMC_3N_fits} we show as another example the results of 3N
interaction fits using quantum Monte-Carlo methods to the binding energy of
$^4$He and the spin-orbit splitting in the n$\alpha$ $P$-wave phase shifts,
i.e., a five-body scattering observable~\cite{Lynn16QMC3N}. These calculations
are based on the NN interactions presented in
Refs.~\cite{Geze13QMCchi,Geze14long} and use a purely local coordinate-space
regularization scheme (see Section~\ref{sec:3N_regularization} for details).
These calculations demonstrate that the employed NN and 3N interactions
derived from chiral EFT up to N$^2$LO are capable of correctly predicting
n$\alpha$ scattering phase shifts and properties of light nuclei within
theoretical uncertainties. The inclusion of n$\alpha$ scattering phase shifts
in the fitting process was triggered by the inability of previous
phenomenological 3N interactions like the Urbana IX
interaction~\cite{Pudl95QMC} to correctly describe the spin-orbit splitting in
neutron-rich systems, which in turn motivated the inclusion of three-pion
exchange diagrams in the Illinois 3N models~\cite{Piep01Ill3N}. Since the
n$\alpha$ scattering phase shifts are sensitive to three-neutron forces, this
strategy might constrain this part of the 3N interactions and lead to a better
agreement in neutron-rich systems. In addition, the results of
Ref.~\cite{Lynn16QMC3N} investigated the impact of the Fierz-ambiguity on
observables when using local regulators (see
Section~\ref{sec:local_coordinate} and also Ref.~\cite{Lov113NNM}) and showed
that the fitted NN plus 3N interactions lead to pure neutron matter results in
good agreement with other works~\cite{Hebe15ARNPS}.

\begin{figure}[t]
\centering
\includegraphics[width=0.45\textwidth]{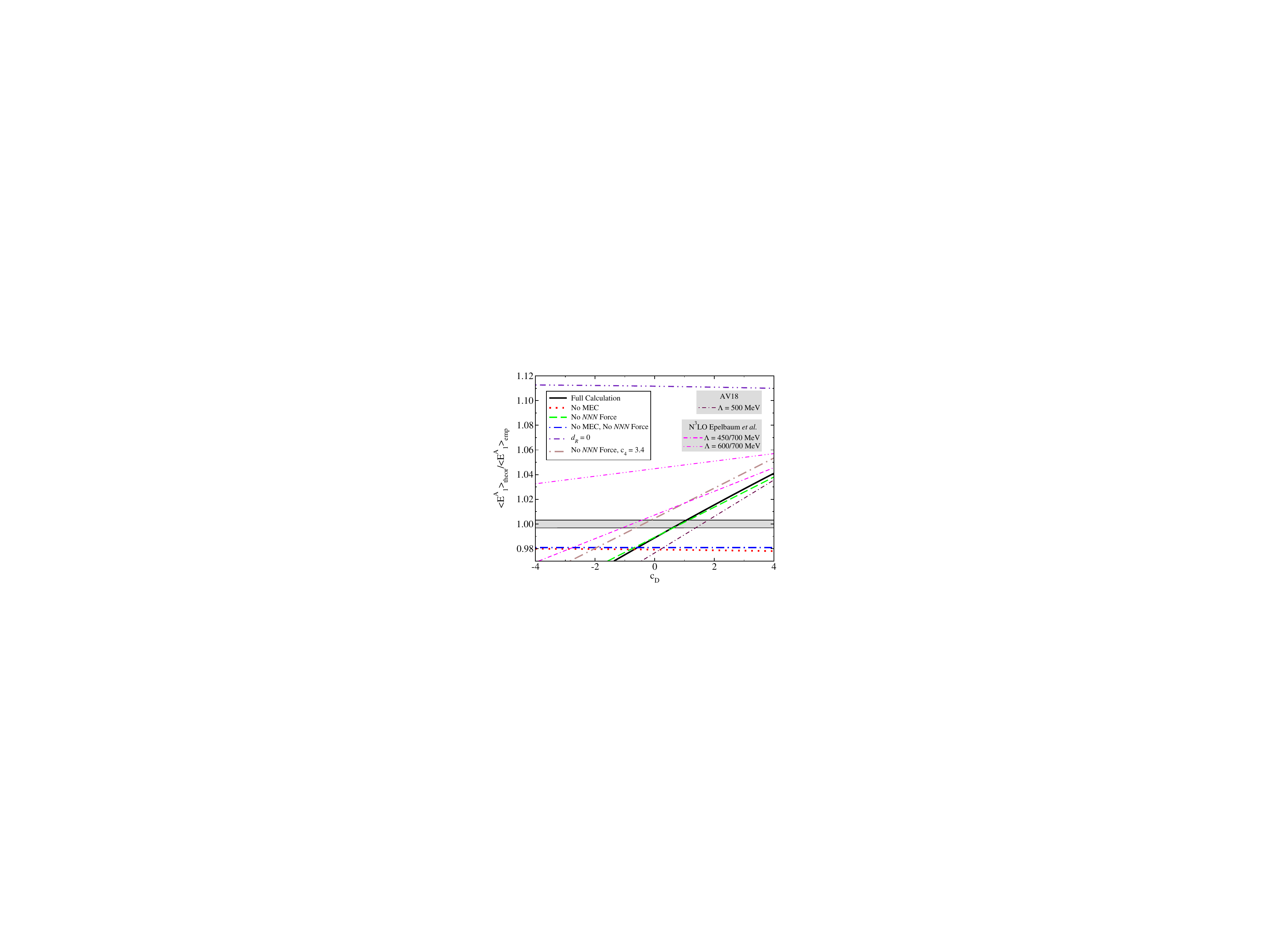}
\hspace{0.5cm}
\includegraphics[width=0.45\textwidth]{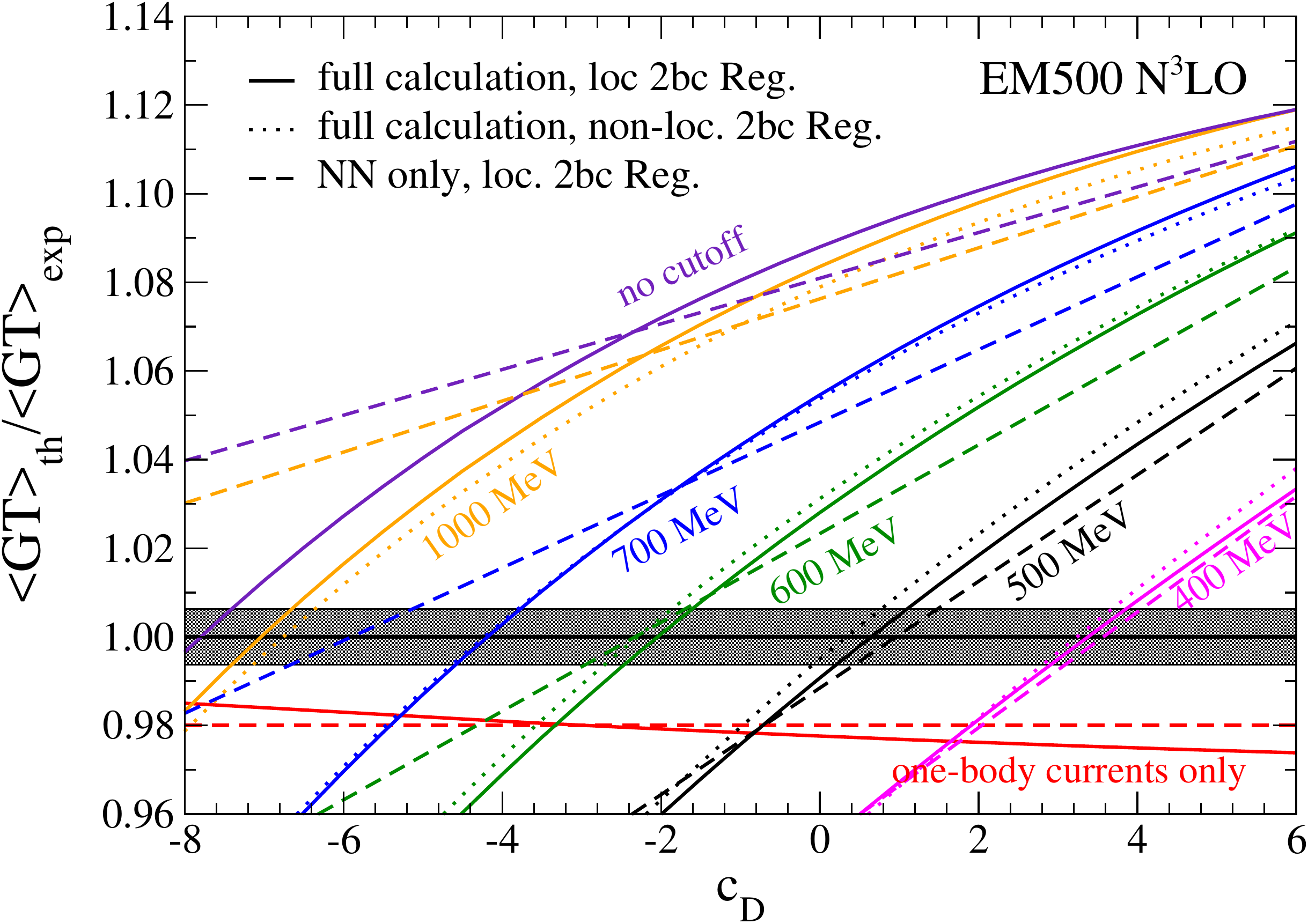}
\caption{Ratio of calculated and experimental Gamow-Teller
matrix elements as a function of $c_D$, while the width of the shaded bands
denotes the $2\sigma$ experimental uncertainty. Left: Results based on
the NN interaction of Refs.~\cite{Wiri95AV18,Ente03EMN3LO,Epel05EGMN3LO} plus
local 3N interactions~\cite{Navr07local3N}. Right: Results for different cutoff values and
regulators in the two-body currents, using the NN interaction of
Ref.~\cite{Ente03EMN3LO}. The solid (dotted) lines show results for nuclear
wave functions including contributions from 3N forces at N$^2$LO for a local
and a nonlocal regulator in the two-body currents.\\
\textit{Source:} Left figure taken from Ref.~\cite{Gazi08lec} and right figure taken from Ref.~\cite{Klos17triton}.}
\label{fig:interactions_fit_3Hdecay}
\end{figure}

For comparison to the results shown in
Figure~\ref{fig:cd_ce_ndscattering_length} we show in
Figure~\ref{fig:interactions_fit_3Hdecay} 3N fits based on observables that
are less correlated, in this case the $^3$H binding energy and the $\beta$
decay half-life of $^3$H. The latter observable was first suggested in
Ref.~\cite{Gard06weak3nf} in the year 2006 as a suitable observable to
constrain 3N forces, while it was first implemented in Ref.~\cite{Gazi08lec}
three years later. Such fits based on electroweak reactions take advantage of
one of the key strengths of chiral EFT, that nuclear interactions and nuclear
currents are derived from the same Lagrangian and hence contain the same
low-energy couplings. In particular, for the calculations in
Ref.~\cite{Gazi08lec} and later in
Refs.~\cite{Marc12mucap,Baro16beta,Klos17triton} the dependence of the axial
two-body current on the 3N coupling $c_D$ was exploited\footnote{We note that
in all original publications an incorrect factor of $-4$ was included in the
coefficient multiplying the coupling $c_D$ in the nuclear
current~\cite{Schi17privcom}. This error was corrected later in all those
references.}. This dependence leads a strong sensitivity in the predictions of
the $^3$H half-life, which can be measured quite
accurately~\cite{Akul053Hhalf}. The two panels of
Figure~\ref{fig:interactions_fit_3Hdecay} show the ratio of the theoretical
and experimental values of the reduced Gamow-Teller transition matrix elements
for different interactions as a function of $c_D$ (see also
Section~\ref{sec:beta_decay}). The shaded region indicates the experimental
uncertainty of the transition matrix element (see
Refs.~\cite{Gazi08lec,Klos17triton} for details). Of course, the specific
values of $c_D$ that are consistent with the experimental value sensitively
depend on the employed NN interactions. However, it is generally true that the
sensitivity of the $^3$H half-life on the 3N coupling is much stronger than
for more correlated observables like in
Figure~\ref{fig:cd_ce_ndscattering_length} and hence in principle allows for
tighter constraints on this coupling.

However, we note that contributions from the leading one-body currents already
contribute about $98 \%$ of the total transition strength for the shown NN
interaction in Figure~\ref{fig:interactions_fit_3Hdecay}, which implies that
the 3N coupling $c_D$ is effectively fitted to a $2 \%$ discrepancy to
experimental values. In addition, such fits based on electroweak currents
involve additional sources of uncertainty that are not present in pure nuclear
structure calculations. These uncertainties concern the way the nuclear
current operators are regularized. The right panel of
Figure~\ref{fig:interactions_fit_3Hdecay} shows that a variation in the
current cutoff scale can lead to a strong variation of the extracted $c_D$
values. These results indicate that the uncertainties of the LECs due to the
regulator dependence of the nuclear forces and currents can lead to
significant uncertainties for nuclear structure observables. The observed
dependence may also be related to possible inconsistencies in the power
counting in the currents due to the non-trivial enhancement of some
contributions~\cite{Vald14currents}. In order to systematically reduce these
uncertainties more detailed studies are required to find a consistent way of
regularizing nuclear interactions and currents ensuring also the validity of
the continuity equation. In Ref.~\cite{Kreb16axial} it was shown that the
currents and interactions indeed fulfill the continuity equation at the
operator level, i.e., for infinite cutoffs. Generalizing this analysis to
regularized matrix elements will provide additional nontrivial constraints for
a consistent way of regularizing electroweak currents.

In Ref.~\cite{Hebe11fits} another set of few-body observables was used to fit
the 3N interactions, the $^3$H binding energy and the charge radius of $^4$He.
The main focus of this work was to explore properties of symmetric nuclear
matter (see also Section~\ref{sec:applications_fits}) based on NN plus 3N
interactions which are only constrained by two- and few-body physics. The
historical route to heavy nuclei is through infinite nuclear matter, a
theoretical uniform limit that first turns off the Coulomb interaction, which
otherwise drives heavier stable nuclei toward an imbalance of neutrons over
protons and eventually, instability. However, predicting nuclear matter based
on microscopic nuclear forces has proved to be an elusive target for a long
time. In particular, few-body fits have not sufficiently constrained 3N
interactions around saturation density such that nuclear matter calculations
are predictive. Nuclear matter saturation is very delicate, with the binding
energy resulting from cancellations of much larger potential and kinetic
energy contributions. When a quantitative reproduction of empirical saturation
properties was obtained, it was imposed by hand through the adjustment of
short-range three-body forces (see, e.g., Refs.~\cite{Akma98EOS,Leje00EOS3N}).

\begin{figure}[t]
\centering
\includegraphics[width=0.45\textwidth]{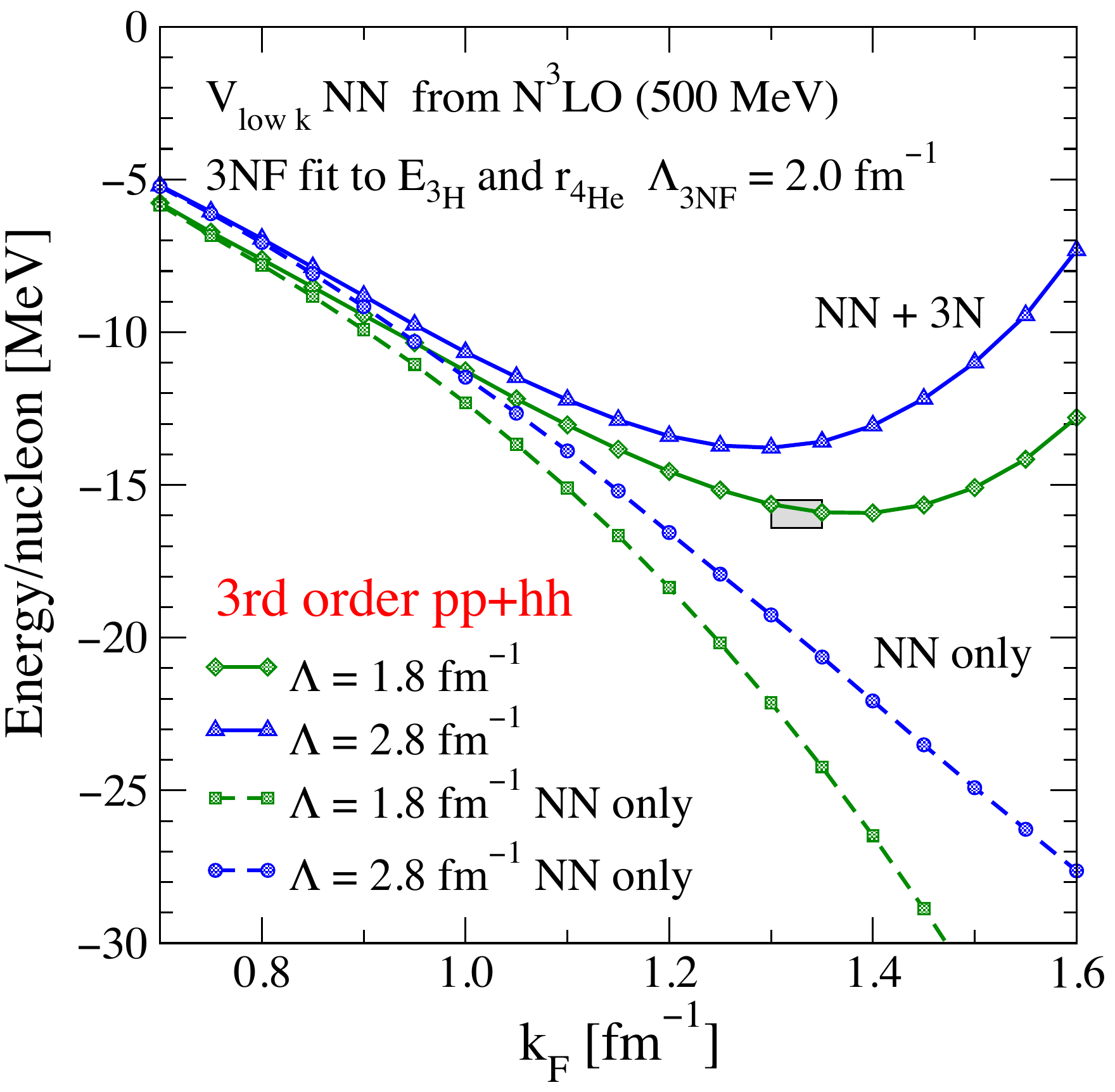}
\hspace{1cm}
\includegraphics[width=0.45\textwidth]{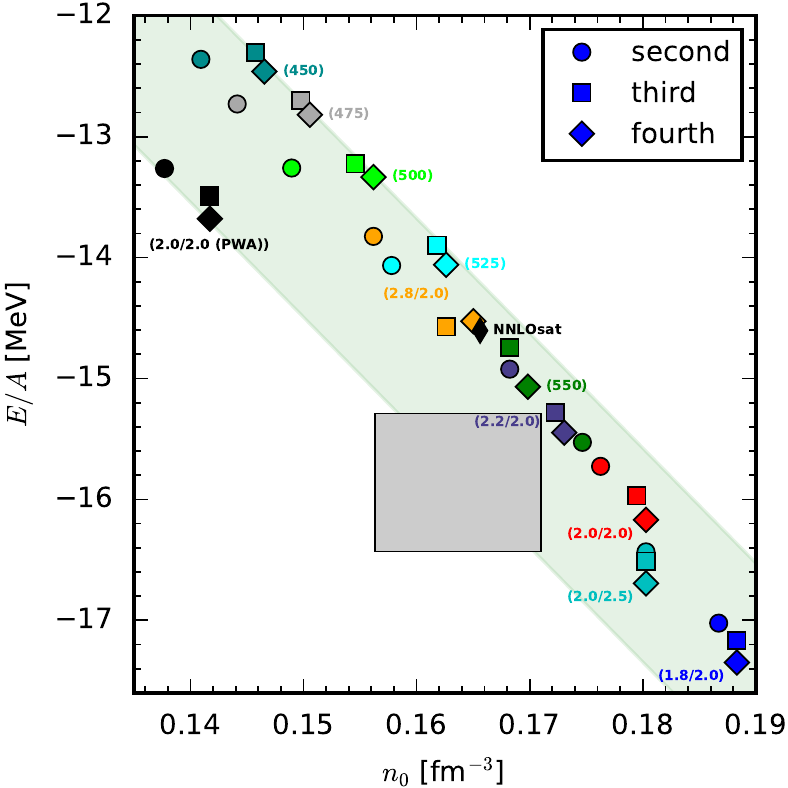}
\caption{Left: Symmetric nuclear matter energy per particle as a function of
the Fermi momentum $\kf$ including contributions up to third-order
particle-particle/hole-hole correlations, based on evolved N$^3$LO NN
potentials (NN-only, dashed lines) and NN+3N contributions, fit to
$E_{\rm^3H}$ and $r_{\rm^4He}$ (solid lines). Theoretical uncertainties are
estimated by the cutoff variation $\Lambda=1.8 - 2.8$ fm$^{-1}$. Right: Saturation density $n_0$ and
saturation energy $E/A$ for the interactions of Ref.~\cite{Hebe11fits}
(labeled by the NN resolution scale and 3N cutoff scales
$\lambda_{\text{SRG}}/\Lambda_{\text{3N}}$, see also
Table~\ref{tab:3Nmagicfits}), the interactions of Ref.~\cite{Carl15sim}, given
by the cutoff scale of the NN and 3N interactions, and the interaction
$\text{N$^2$LO}_{\text{sat}}$ of Ref.~\cite{Ekst15sat} (black diamond).\\
\textit{Source:} Left figure taken from Ref.~\cite{Hebe11fits} and right figure taken from Ref.~\cite{Dris17MCshort}. }
\label{fig:interactions_fit_4He}
\end{figure}

\begin{figure}[b!]
\centering
\includegraphics[width=0.45\textwidth]{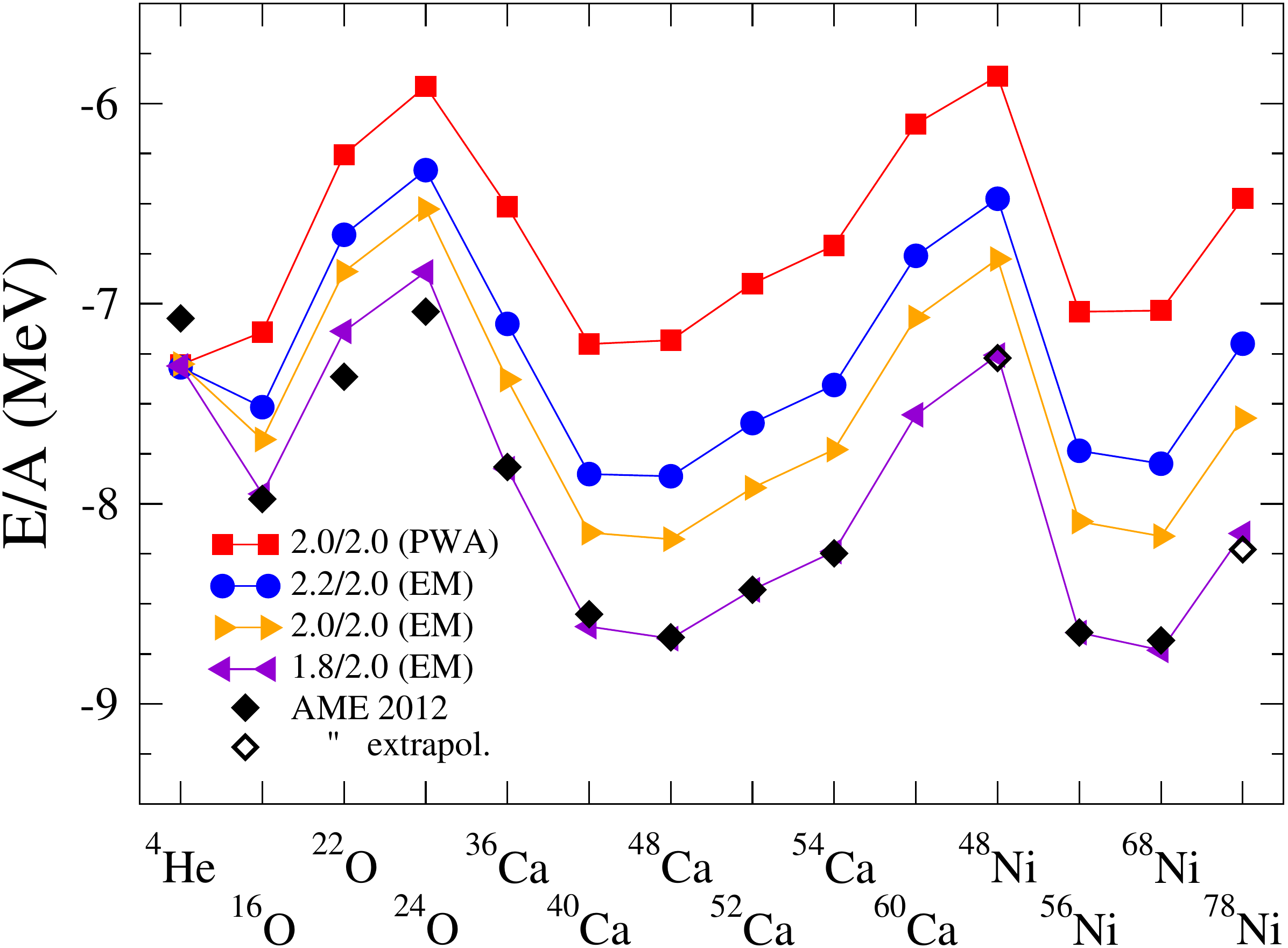}
\hspace{0.5cm}
\includegraphics[width=0.45\textwidth]{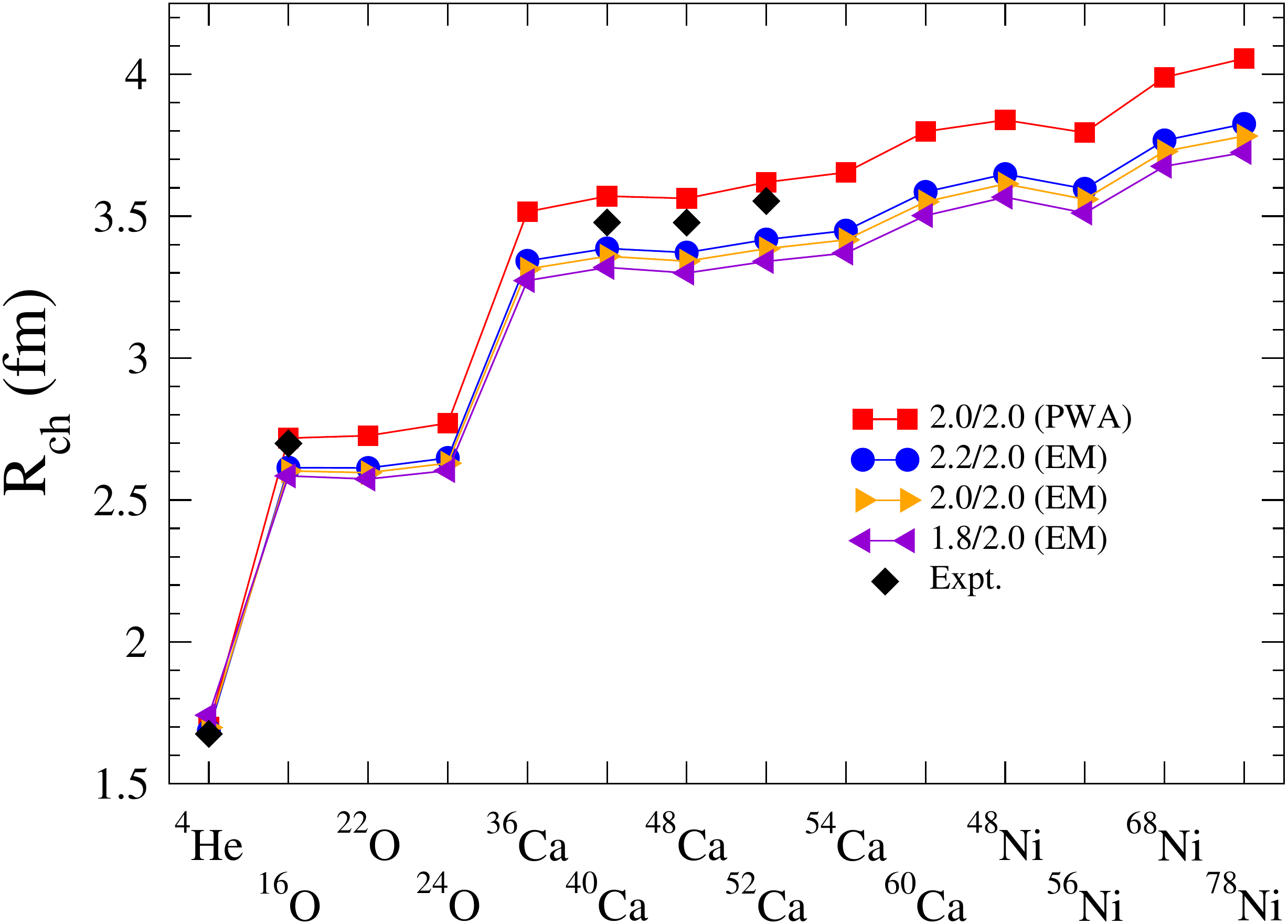}
\caption{Systematics of the energy per nucleon (left)
$E/A$ and charge radii (right) of closed-shell nuclei from $^{4}$He to
$^{78}$Ni calculated with the IM-SRG for four resolution scales given in the
legend. The energy results are compared against experimental ground-state
energies from the atomic mass evaluation (AME) 2012~\cite{Wang17AME16}
(extrapolated for $^{48, 78}$Ni), while the results for the radii are compared
against experimental charge radii~\cite{Ange13rch} where available.\\
\textit{Source:} Figures taken from Ref.~\cite{Simo17SatFinNuc}.
\label{fig:magic_interactions_matter_nuclei}
}
\end{figure}

It is not unnatural to expect a correlation between theoretical predictions
for heavier nuclei and nuclear matter. Naively, one might be tempted to expect
that nuclear interactions that predict a saturation point in good agreement
with the empirical saturation point $E/A = -16 \mev$ and $n_{0} = 0.16 \:
\text{fm}^{-3}$ should also lead to a reasonably realistic descriptions of
heavier nuclei. In Ref.~\cite{Hebe11fits} the nuclear matter many-body
calculations were simplified by an RG evolution of the NN interaction to lower
resolution scales~\cite{Bogn10PPNP} (see also Section~\ref{sec:SRG}). Given
that at the time of that work no consistent evolution of 3N interactions in
momentum space was possible yet, a hybrid approach for the determination of
the 3N interactions was employed instead. While the NN interaction of
Ref.~\cite{Ente03EMN3LO} was SRG-evolved to lower resolution scales
$\lambda_{\text{SRG}}$, the 3N interactions were fitted to the $^3$H binding
energy and $^4$He charge radius at each resolution scale with a fixed cutoff
scale $\Lambda_{\text{3N}}$. This strategy is implicitly based on the
assumption that the $c_i$ coefficients of the long-range two-pion-exchange
part are not modified by the RG and the N$^2$LO 3N interactions serve as a
truncated ``basis'' for low-momentum 3N interactions. In
Section~\ref{sec:applications_fits} we compare this approach to calculations
based on consistently-evolved NN plus 3N interactions and present the new
results based on the framework presented in Section~\ref{sec:SRG}.

The left panel of Figure~\ref{fig:interactions_fit_4He} shows the results for
symmetric nuclear matter energy per particle as a function of the Fermi
momentum $\kf$ with $n = 2 k_{\text{F}}^3/(3 \pi^2)$ and in particular
illustrates the role and importance of 3N interactions for saturation when
using low-resolution NN interactions. While calculations based on only NN
interactions do not exhibit saturation in the shown density region (dashed
lines), the inclusion of contributions from 3N interactions lead to saturation
properties in reasonable agreement with the empirical region (gray rectangle)
even though the 3N interactions have been fit to only few-body systems. The
right panel shows the detailed results for the saturation points of different
NN plus 3N interactions at different orders in the many-body expansion (see
Ref.~\cite{Dris17MCshort} for details), while the results based on the forces
derived in Ref.~\cite{Hebe11fits} are indicated by the labels
$\lambda_{\text{SRG}}/\Lambda_{\text{3N}}$. Evidently, there is a pronounced linear
correlation between the density and energy similar to the ``Coester
line''~\cite{Coes70nuclmatt}. In contrast to the original Coester line with NN
potentials only, however, the green band encompassing all shown theoretical
saturation points overlaps with the empirical saturation region because of the
inclusion of 3N forces.

Furthermore, a systematic trend toward higher saturation densities and larger
binding energies was found with decreasing NN resolution scale
$\lambda_{\text{SRG}}$. This trend translates in a systematic way to the
ground-state energies and radii of finite nuclei over a wide mass range, from
$^4$He to much heavier nuclei up to $^{78}$Ni as shown in
Figure~\ref{fig:magic_interactions_matter_nuclei}. Remarkably, all calculated
ground-state energies based on the ``1.8/2.0'' interaction are in very good
agreement with experiment, except for the neutron-rich oxygen isotopes
$^{22,24}$O. The other three shown interactions follow the same pattern but
are shifted by as much as $1.5$ MeV/A in the case of the ``2.0/2.0 (PWA)''
interaction (see Ref.~\cite{Hebe11fits} for details). The experimental charge
radii are enclosed by the ``2.2/2.0'' and ``2.0/2.0 (PWA)'' results, but the
trend observed for the closed-shell nuclei studied in detail already above
appears to hold at least up to $^{78}$Ni. That is, radii with ``1.8/2.0'' to
``2.2/2.0'' are too small, but ``2.0/2.0 (PWA)'' gives slightly too large
radii. As in the case of ground-state energies, the radius systematics is
similar for all Hamiltonians, with mainly only a constant shift for the
different interactions. This behavior for the ground-state energy and charge
radii is clearly reminiscent of the Coester-like line for the saturation
points of the four Hamiltonians considered, as shown in the right panel of
Figure~\ref{fig:interactions_fit_4He}. However, the reason why in particular
interaction ``1.8/2.0'' leads to such an excellent agreement with experimental
ground-state energies remains an open question. Nevertheless, thanks to these
promising results for heavier nuclei this set of interactions has been used
quite intensively in recent years in \textit{ab initio} studies of medium-mass
nuclei. We will present a selection of these results in more detail in
Section~\ref{sec:applications}.

\begin{figure}[t]
\begin{center}
\includegraphics[page=1,scale=1.02,clip]{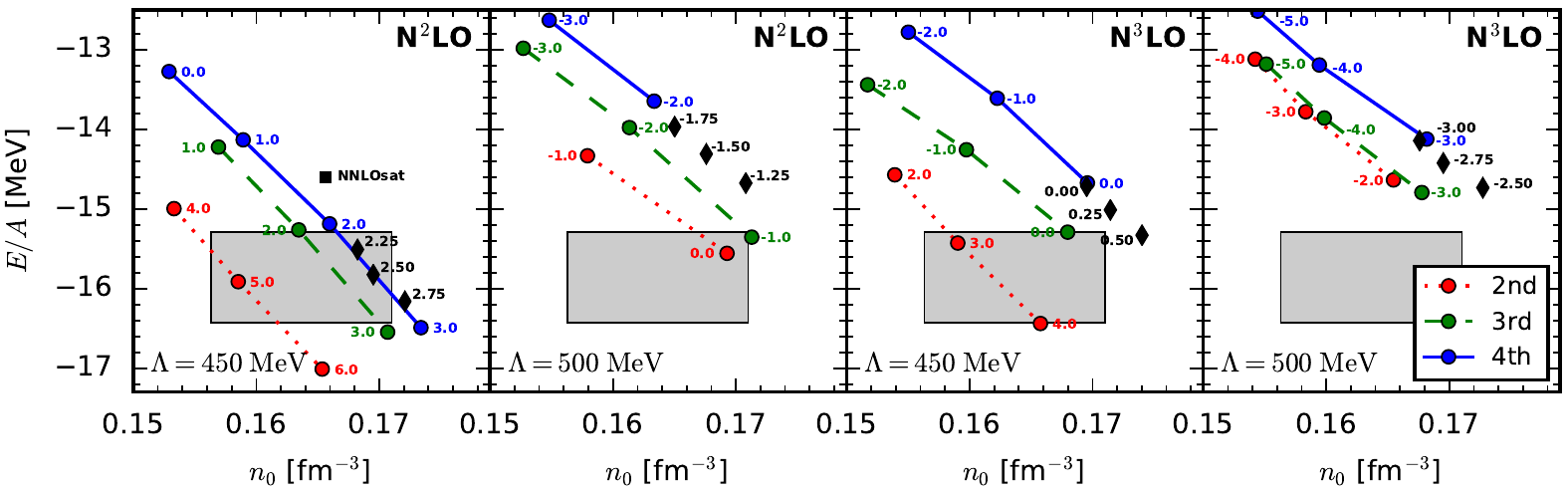}
\end{center}
\caption{Saturation density and energy of symmetric
nuclear matter at 2nd, 3rd and 4th order in MBPT for the NN and 3N
interactions at N$^2$LO and N$^3$LO. The points indicate different values of
$c_D$, while the red-dotted, green-dashed, and blue-solid lines correspond to
calculations at different orders. The left (right) two panels are for N$^2$LO
(N$^3$LO) interactions with $\Lambda = 450$ and $500
\MeV$~\cite{Ente17EMn4lo}. The diamonds in each panel represent the
calculations with the best simultaneous reproduction of both saturation density
and energy at fourth order.\\
\textit{Source:} Figure taken from Ref.~\cite{Dris17MCshort}.}
\label{fig:coester_fits}
\end{figure}

The results discussed above highlight the importance of realistic saturation
properties of infinite matter for nuclear forces, even though a deeper and
more quantitative understanding of the connection between properties of matter
and finite nuclei is still lacking. This suggests that it might be useful to
include information about saturation properties in the construction of the
interactions. However, the explicit incorporation of nuclear matter properties
in the fit process of nuclear forces has not been achieved until
recently~\cite{Dris17MCshort}. In Figure~\ref{fig:coester_fits} we show
results for the saturation point based on NN interactions of
Ref.~\cite{Ente17EMn4lo} at N$^2$LO and N$^3$LO as a function of the LEC $c_D$
(annotated numbers of the data points), while the relation between $c_D$ and
$c_E$ was determined via the $^3$H binding energy (see
Figure~\ref{fig:cd_ce}). Note that for such fits to nuclear matter properties
effectively two LECs are fitted to three observables, $E_{^3\text{H}}$, the
saturation energy $E(n_0)/A$, and the saturation density $n_0$. Hence it is \textit{a
priori} not obvious that a reasonable simultaneous reproduction of all
observables can be achieved. Remarkably, for the shown cases in
Figure~\ref{fig:coester_fits} a reasonable reproduction can be achieved for
all four interactions. These best fits are indicated by the black diamonds in
each panel. In Section~\ref{sec:applications} we will present first results
for finite nuclei based on these interactions.

Finally, properties of light and medium-mass nuclei have also been
investigated based on chiral NN and 3N interactions using Lattice EFT
methods~\cite{Epel09LEFT,Lahd13LEFT,Elha154Hescatt,Epel14O16}. For these
studies the relation between the couplings $c_D$ and $c_E$ was fixed via the
$^3$H binding energy~\cite{Epel09Lattice3N}, while the value for $c_D$ could
only be constrained to the regime $c_D \sim \mathcal{O} (1)$ using the
spin-doublet nucleon-deuteron scattering phase shifts~\cite{Epel10Lattice}.
For the practical calculations, like those of the Hoyle state in
$^{12}C$~\cite{Epel12Hoyle}, the authors set $c_D = 0$, while they studied the
sensitivity of the results to changes $c_D \pm 1$. Generally, the effects of
this modification were found to be minor~\cite{Lee19privcom}. In
Ref.~\cite{Lahd13LEFT} results for nuclei up to $^{28}\text{Si}$ were
presented based on NN and 3N interactions up to N$^2$LO and using a low cutoff
scale. For heavier systems, $A \ge 16$, a significant overbinding was observed
which was attributed to too attractive 3N interactions at this order and for
the employed low cutoff scale, which in turn leads to the formation of alpha
clusters on single lattice points. The addition of a single empirical
repulsive 4N interaction allowed to fix this overbinding effect. This
additional interaction was argued to mimic the effects of higher-order
interactions that should become particularly important when using low
resolution interactions, similar to those generated by SRG transformations
(see Section~\ref{sec:SRG}).

\subsubsection{Fits of $\Delta$-full 3N interactions}
\label{sec:sep_3N_fits_delta}
\begin{figure}[t]
\centering
\includegraphics[width=0.4\textwidth]{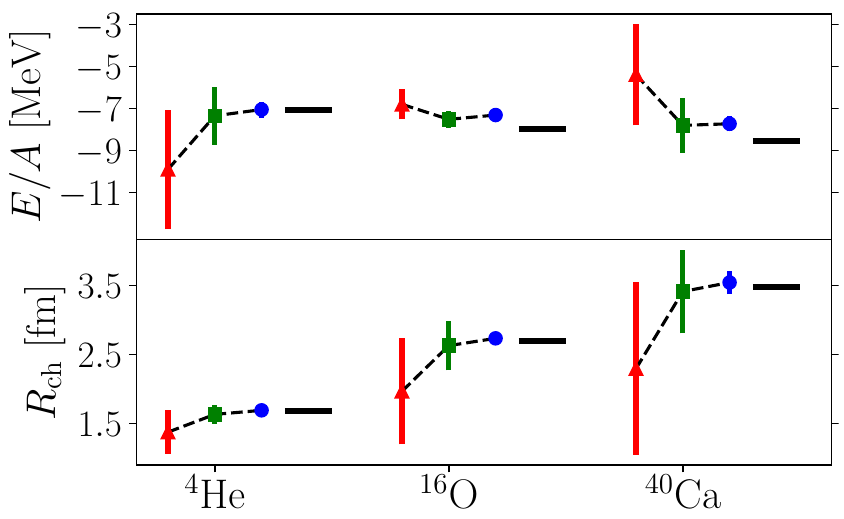}
\hspace{0.5cm}
\includegraphics[width=0.55\textwidth]{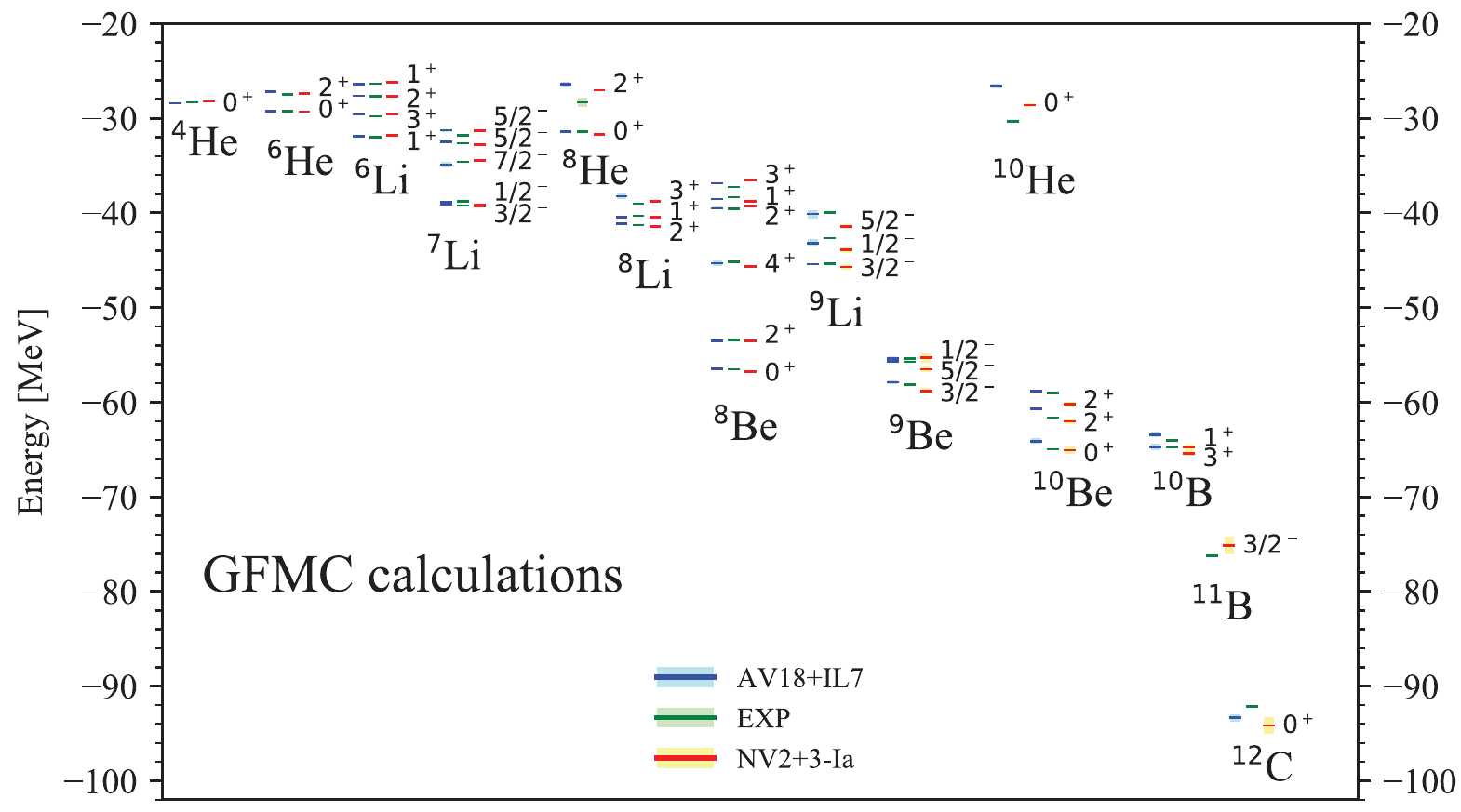}
\caption{Left: Ground-state energy per nucleon and charge radii of
selected nuclei based on the $\Delta$-full interactions of
Ref.~\cite{Ekst17deltasat} at LO (red triangle), NLO (green square), and N$^2$LO
(blue circle). The black bars show the experimental data. Error bars are estimated from the
order-by-order EFT truncation errors following Ref.~\cite{Epel15improved}
(see also Section\ref{sec:EFTtruncation}). Right: Energy spectra of selected nuclei
obtained with the interaction of Ref.~\cite{Piar17LightNucl} compared to
experimental data~\cite{Audi03AME03}.\\
\textit{Source:} Left figure taken from Ref.~\cite{Ekst17deltasat} and right figure taken from
Ref.~\cite{Piar17LightNucl}.}
\label{fig:fits_Deltafull}
\end{figure}

In Ref.~\cite{Ekst17deltasat} NN and 3N interactions were constructed with and
without explicit inclusion of the $\Delta$-isobar degree of freedom up to
N$^2$LO, following exactly the same fitting protocol for both formulations.
The 3N LECs were fitted to the binding energy and point-proton radius of
$^4$He, respectively. Based on these interactions, properties of medium-mass
nuclei and also matter were investigated, which in particular allowed to study
in detail the differences between $\Delta$-less and $\Delta$-full nuclear
interactions. While the results for radii and binding energies turn out to be
in remarkable agreement with experiment based on the $\Delta$-full interaction
(see left panel of Figure~\ref{fig:fits_Deltafull}), the $\Delta$-less
interactions produce nuclei that are not bound with respect to breakup into
$\alpha$ particles. In addition, the saturation properties of symmetric
nuclear matter are also significantly improved for the $\Delta$-full interaction.

In Ref.~\cite{Piar17LightNucl} 3N interactions within $\Delta$-full chiral EFT
were fitted to the ground-state energy of $^3$H and the central value of the
neutron-deuteron scattering length $^2a_{nd}$ based on the NN interactions
presented in Ref.~\cite{Piar16DeltaNuc}. Even though these two observables
exhibit a strong correlation (see Section~\ref{sec:sep_3N_fits}) these fits
lead to a good reproduction of ground state states and excited states of light
nuclei up to $A=12$ (see right panel of Figure~\ref{fig:fits_Deltafull}). The
agreement with experiment is of the same quality as calculations based on the
phenomenological Argonne v18 interaction, augmented with 3N interactions that
were fit to observables of nuclei beyond $A=3$. In Ref.~\cite{Logo16matter}
the 3N LECs $c_D$ and $c_E$ were fitted to properties of symmetric nuclear
matter based on the same local NN interaction of Ref.~\cite{Piar16DeltaNuc}.
The obtained values for the LECs differ quite significantly from the values
found in Ref.~\cite{Piar17LightNucl}. Some of the difference can most likely
be attributed to the approximate treatment of the angular dependence of the
momentum transfer in the local regulator during the incorporation of the 3N
interactions in the many-body calculations via normal-ordering (see
discussions in Section~\ref{sect:normal_ordering_matter} and also
Ref.~\cite{Logo2019consistent}).

\subsubsection{Simultaneous fits of NN and 3N interactions}
\label{sec:simul_NN_3N_fits}

\begin{figure}[t]
\includegraphics[width=0.38\textwidth]{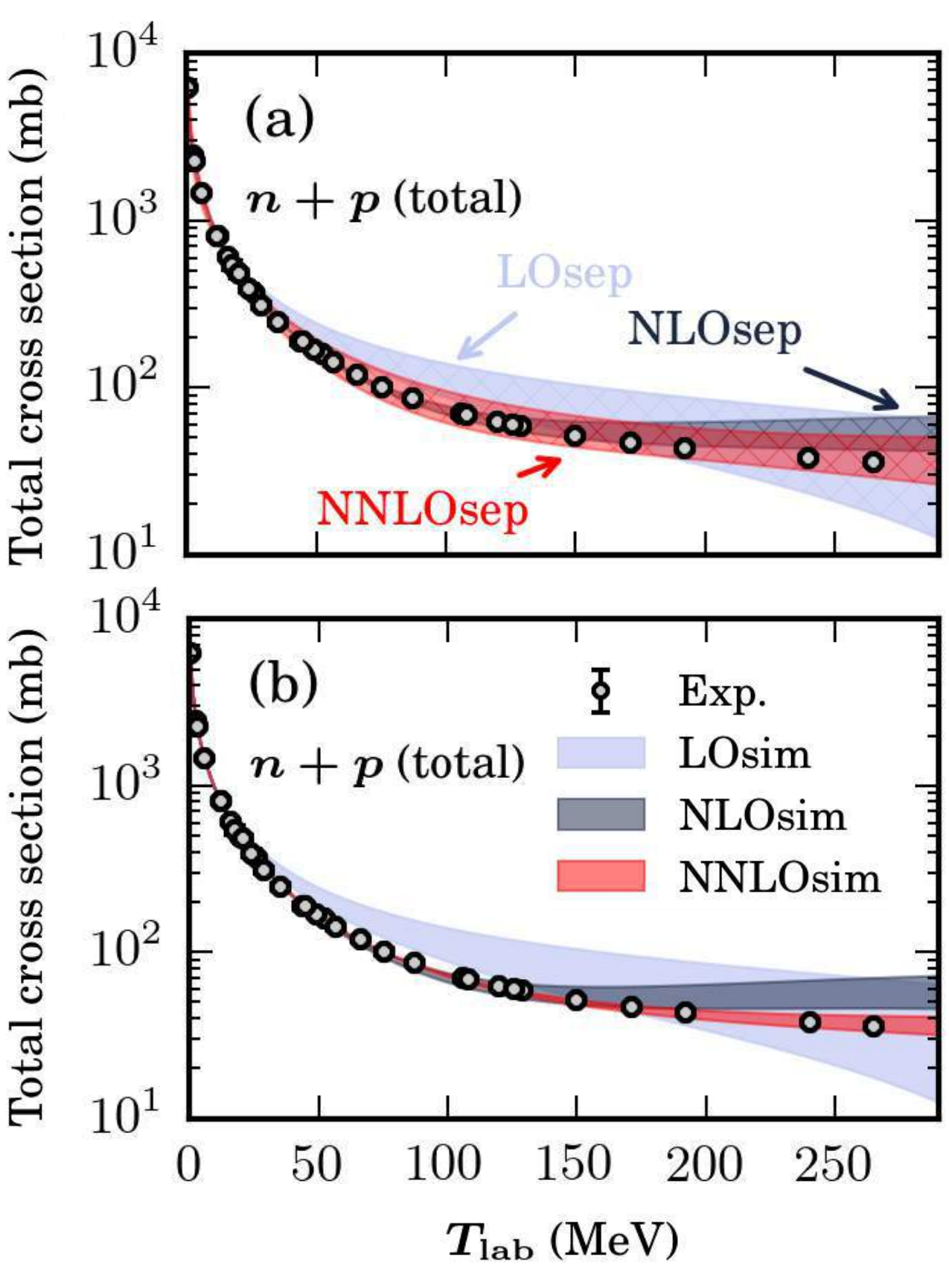}
\hspace{0.5cm}
\includegraphics[width=0.55\textwidth]{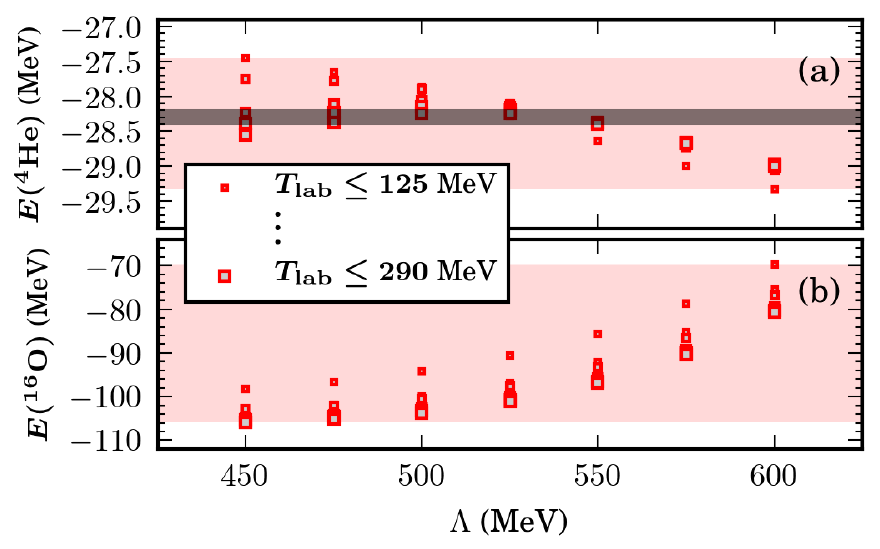}
\caption{Left: Comparison between neutron-proton total cross section for
the sequentially optimized interactions ((a), upper panel) and the
simultaneously-optimized interactions ((b), lower panel) at LO, NLO and
N$^2$LO. The bands show the total uncertainties, including statistical and
model uncertainties. Right: Ground-state energies of (a) $^4$He and (b)
$^{16}$O using different optimization parameters for the maximal kinetic
energy $T_{\text{lab}}$ and regularization cutoff scales $\Lambda$ for the
interaction N$^2$LO$_{\text{sim}}$. For reference, the experimental binding energies are
$E(^4\text{He}) \approx -28.3 \text{MeV}$, indicated by the gray band in the
upper panel and $E(^{16}\text{O}) \approx -127$ MeV, which is not visible on
the shown scale.\\
\textit{Source:} Figures adapted from Ref.~\cite{Carl15sim}.}
\label{fig:N2LO_sim}
\end{figure}

For all fits discussed in the previous section, the determination of the 3N
coupling constants amounts to an optimization problem in a two-dimensional
parameter space since all LECs in the NN interaction have been fixed using
two-nucleon observables plus pion-nucleon data. Furthermore, in cases when
only two few-body observables are used, the fitting problem of the 3N
interactions has by construction a unique solution if theoretical and
experimental uncertainties are neglected. If more observables are considered
(as, e.g., in the right panel of Figure~\ref{fig:cd_ce_ndscattering_length})
some kind of $\chi^2$ minimization can be employed to find the optimal values
for the LECs $c_D$ and $c_E$.

However, within chiral EFT all contributions to NN and 3N interactions are
derived on equal footing at each order in the chiral expansion and thus depend
on the same LECs. Hence it can be argued that it might be more natural and
consistent to determine the LECs of NN and 3N interactions as well as the
underlying couplings of $\pi$N scattering simultaneously instead of
sequentially up to a given order. Of course, such a simultaneous fit
is much more challenging since the number of couplings and hence the dimension
of the parameter space for the optimization process obviously increases
significantly. For example, the NN interactions derived within Weinberg's
power counting scheme (see Section~\ref{sec:chiral_expansion}) in naive
dimensional analysis contain 2,7,0,12,0 new unknown short-range couplings at
LO, NLO, N$^2$LO, N$^3$LO and N$^4$LO respectively~\cite{Rein17semilocal} (see
also Figure~\ref{fig:chiral_EFT_deltaless_table}), while 3N interactions
contain 2 new couplings at N$^2$LO and the terms at N$^3$LO are predicted in a
parameter free way\footnote{Note that in Ref.~\cite{Rein17semilocal} it was
shown that the number of LECs in NN interaction terms at N$^3$LO can be reduced
by three due to the presence of redundant couplings.}. The precise number of
new 3N LECs at N$^4$LO is still unknown. That means the parameter space
dimension involving 3N couplings increases from 2 to 11 at N$^2$LO and from 2
to 23 at N$^3$LO for simultaneous fits of NN and 3N interactions, which
requires powerful and efficient few-body and optimization frameworks.

In Ref.~\cite{Carl15sim} a first automated optimization framework for NN and
3N interactions was presented using scattering and bound-state observables in
the pion-nucleon, nucleon-nucleon, and few-nucleon sectors. The framework
allows to perform parameter optimizations in large spaces, study correlations
between the different parameters and perform a statistical analysis of
systematic uncertainties. Furthermore it is possible to in- and exclude
observables and adapt the weighting of scattering observables at different
energies in a straightforward and flexible way. The framework was used to
construct a set of NN plus 3N interactions at different orders in the chiral
expansion and using different regularization scales. In particular, the
effects of a simultaneous fit of NN and 3N couplings compared to a sequential
treatment was investigated. The left panel of Figure~\ref{fig:N2LO_sim} shows
the uncertainty bands for the total neutron-proton cross sections at different
orders for these two approaches. While the results for the sequential fit
(upper panel) show no clear sign of convergence with increasing chiral order,
the simultaneous fits lead to a reproduction of the scattering observables
with systematically reduced uncertainty bands toward higher orders. The right
panel of the figure shows the predictions for the ground-state energies of
$^4$He and $^{16}$O based on the derived potentials at N$^2$LO for different
cutoff scales and different values of $T_{\text{lab}}^{\text{max}}$, which
controls the upper energy limit of the included NN scattering observables in
the fit (see Ref.~\cite{Carl15sim} for details). While the energy variation
for all considered potentials is only about 2 MeV for $^4$He, it increases
significantly to about 35 MeV for $^{16}$O. In addition, the obtained energy
range for $^{16}$O does not include the experimental value $E \approx 127$ MeV
(not visible in the shown figure). These results raise some fundamental
questions concerning the size of theoretical uncertainties and the predictive
power for many-body observables when the underlying nuclear interactions are
only constrained by two- and few-nucleon observables.

\begin{figure}[t]
\includegraphics[width=0.45\textwidth]{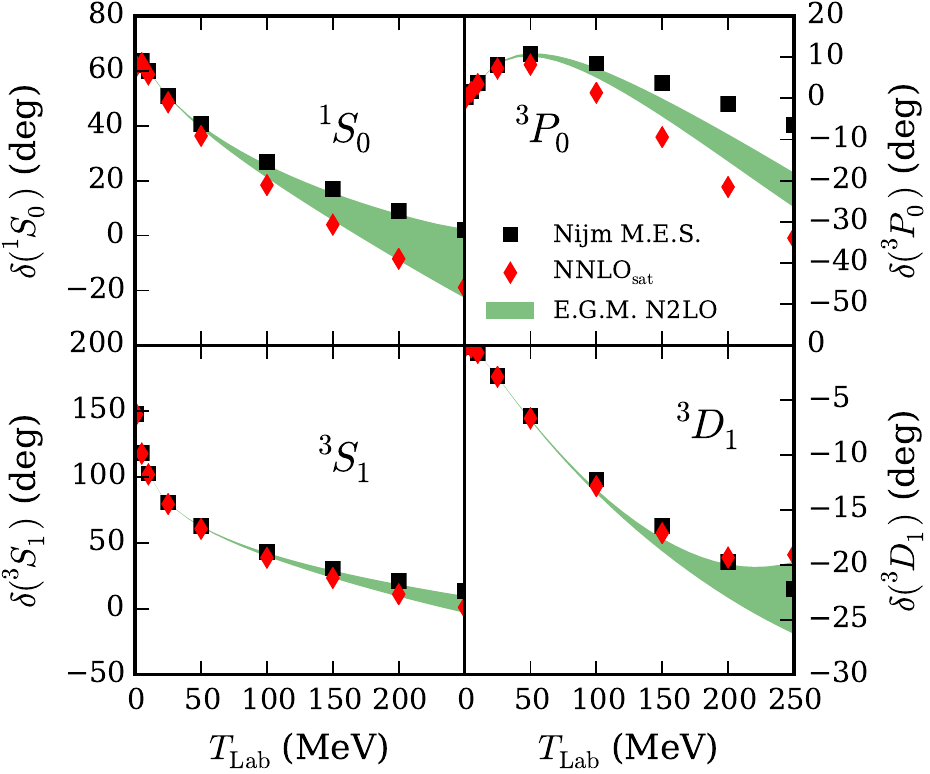}
\hspace{0.5cm}
\includegraphics[width=0.45\textwidth]{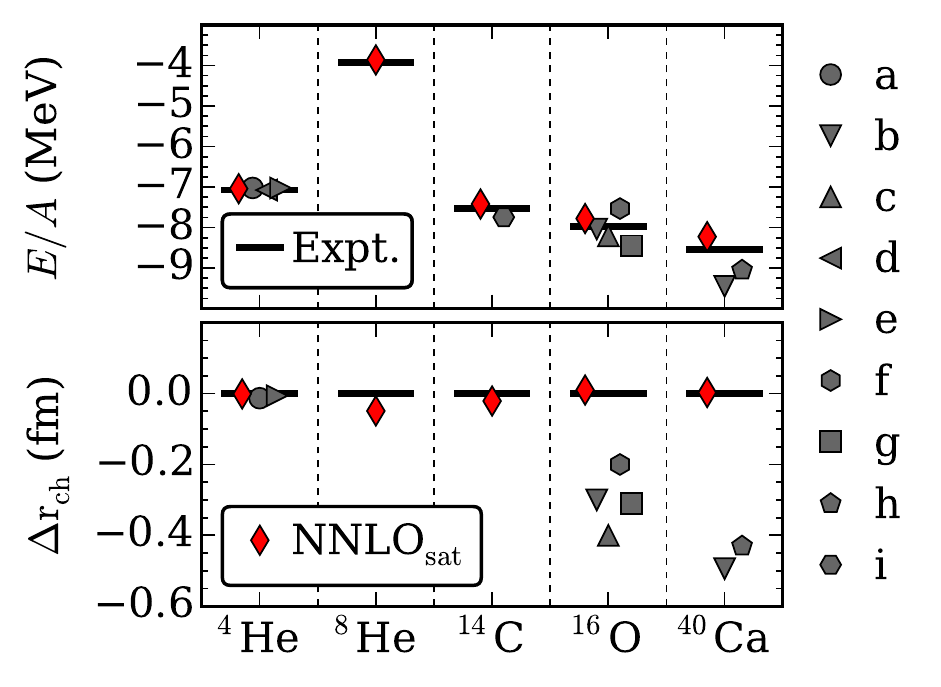}\\
\caption{Left: Neutron-proton scattering phase shifts obtained from
the interaction $\text{N$^2$LO}_{\text{sat}}$ (red diamonds) compared to the
Nijmegen phase shift analysis (black squares). The green bands shows the results
of the N$^2$LO interactions of Ref.~\cite{Epel04EGMN2LO} for comparison.
Right: Ground-state energies per nucleon (top panel), and differences between
computed and experimental charge radii (bottom panel) for selected
nuclei computed with different chiral interactions.\\
\textit{Source:} Figures taken from Ref.~\cite{Ekst15sat}.}
\label{fig:N2LO_sat_nuclei}
\end{figure}

The traditional paradigm for the construction of NN and 3N interactions has
been to determine short-range couplings in the lightest systems in which they
contribute. According to this approach the 3N couplings $c_D$ and $c_E$ should
be determined based on three-body observables. This approach has several
advantages:
\begin{itemize}
\item[1.] Uncertainties of the many-body calculations are minimized since
few-body systems can be solved exactly up to numerical uncertainties.
\item[2.] Effects from four- and even higher-body forces can be cleanly
disentangled as they do not contribute in three-body systems.
\item[3.] For few-body systems it is possible to include nuclear structure as well as
scattering observables in the fitting process.
\end{itemize}
However, this strategy also has practical disadvantages, in particular for
applications to medium-mass or even heavier systems. For example, for
interactions such as those used in Figures~\ref{fig:interactions_fit_4He},
~\ref{fig:magic_interactions_matter_nuclei} and~\ref{fig:N2LO_sim}, the
calculation of many-body systems like $^{16}$O involves a significant
extrapolation in particle number from the systems that have been used to fit
the underlying interactions and the system under investigation. It is
\textit{a priori} not obvious how sensitive many-body observables are to small
changes in interactions when fitted only to few-body systems. This question is
particularly relevant since fits of LECs up to a given chiral order generally
involve inherent uncertainties due to truncation effects of the chiral
expansion. In order to avoid such an extrapolation, it might be more efficient
and stable to include information about heavier systems in the construction of
the interactions. By using properties of the heaviest nuclei of interest as
anchor points the extrapolation problem can be effectively changed into an
interpolation problem, which mathematically is typically much better behaved.

A first interaction, ``$\text{N$^2$LO}_{\text{sat}}$'', following this strategy
was presented in Ref.~\cite{Ekst15sat}. For the practical fit of the NN and 3N
LECs the automated optimization framework of Ref.~\cite{Carl15sim} was
employed. Again, all LECs were fitted simultaneously while binding energies
and charge radii of $^3$H, $^{3,4}$He, $^{14}$C, and $^{16,22,24,25}$O were
included in the optimization. The implementation of such an optimization
including many-body calculations requires significant computational
optimizations in order to make such calculations feasible. The maximum energy
for the NN scattering observables had to be limited to 35 MeV in order find a
reasonable simultaneous reproduction of experimental scattering phase shifts
as well as many-body observables. The detailed theoretical results for
scattering phase shifts and many-body observables are shown in
Figure~\ref{fig:N2LO_sat_nuclei}. We note that the results for $^{40}$Ca are
also in good agreement with experiment even though the fit included only
information up to oxygen. In addition, the saturation properties of symmetric
nuclear matter based on $\text{N$^2$LO}_{\text{sat}}$ are in reasonable agreement
with the empirical constraints (see Figure~\ref{fig:interactions_fit_4He}).

\begin{figure}[t]
\centering
\includegraphics[width=0.43\textwidth]{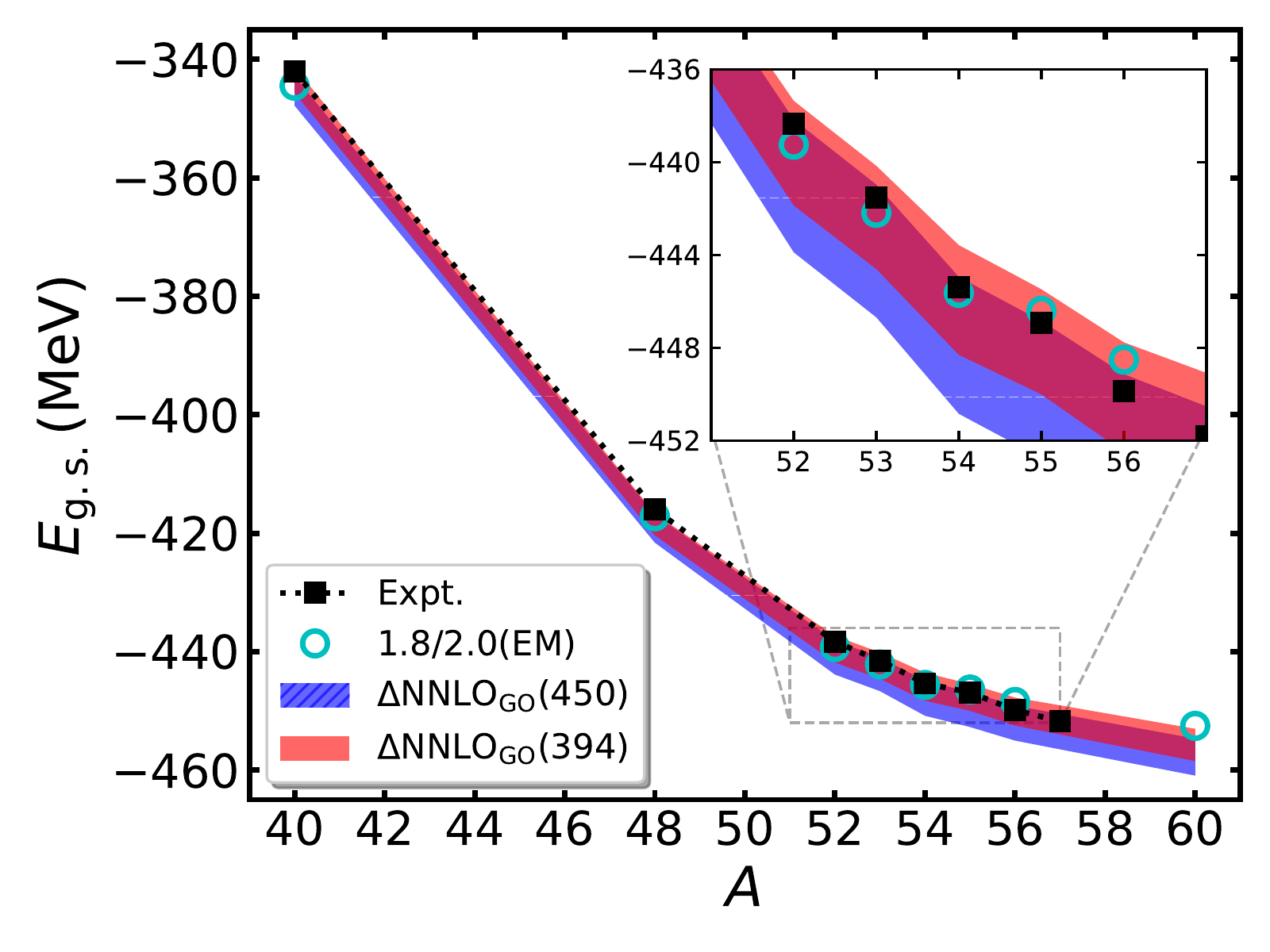} ~
\includegraphics[width=0.43\textwidth]{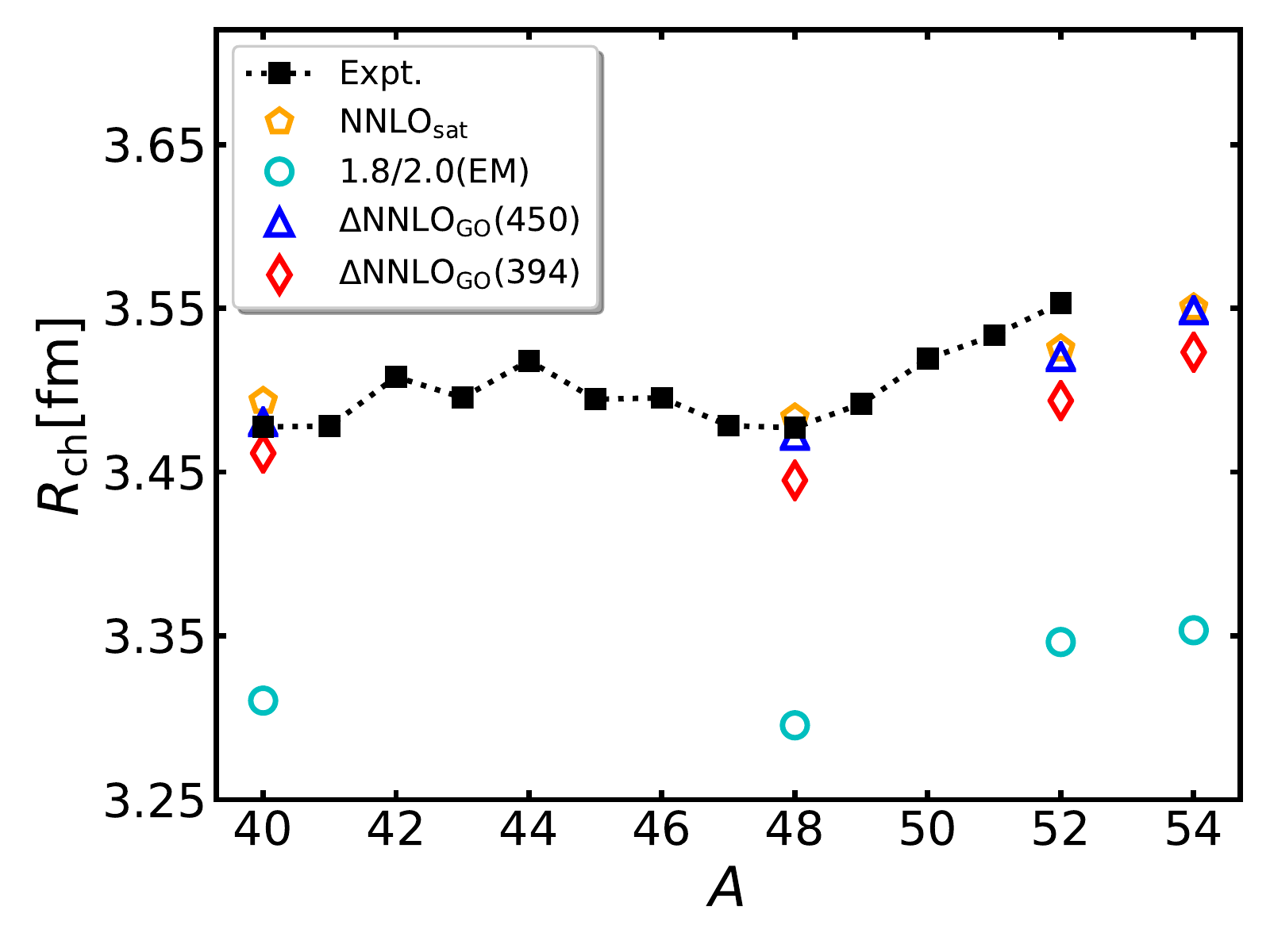}
\caption{
Left: Ground-state energies of calcium isotopes based on the interactions
``$\text{$\Delta$N$^2$LO}_{\text{GO}}$'' and ``$1.8/2.0$'' compared to
experimental values~\cite{Mich18Cameasure}. Right: Charge radii of calcium
isotopes based on the same interactions plus ``$\text{N$^2$LO}_{\text{sat}}$''
compared to experimental values~\cite{Ruiz16Calcium}.\\
\textit{Source:} Figures taken from Ref.~\cite{Jian20N2LOgo}.
}
\label{fig:N2LOgo}
\end{figure}

Very recently, the novel interactions ``$\text{$\Delta$N$^2$LO}_{\text{GO}}$''
were presented in Ref.~\cite{Jian20N2LOgo}. In contrast to
``$\text{N$^2$LO}_{\text{sat}}$'' these interactions are based on a chiral EFT
formulation with explict $\Delta$ degrees. The LECs of the NN and 3N
interactions were fitted simultaneously to the NN scattering data, bound state
properties of $A=2, 3$ and $4$ nucleon systems as well as nuclear matter,
while the long-range pion-nucleon couplings were taken from the Roy-Steiner
analysis of Refs.~\cite{Hofe15piNchiral,Hofe15PhysRep}. The new interactions
exhibit a significantly improved reproduction of experimental NN scattering
data up to laboratory energies of about 125 MeV compared to the interaction
``$\text{N$^2$LO}_{\text{sat}}$''. Furthermore, first results for medium-mass
and heavy nuclei up to $A=132$ indicate that the new interactions are able to
reproduce experimental ground-state energies to a similar degree as the
``1.8/2.0'' interaction of Ref.~\cite{Hebe11fits}, but shows a significantly
improved description of experimental charge radii (see
Figure~\ref{fig:N2LOgo}). However, still, they also fail to reproduce the
rapid increase of the charge radii toward neutron-rich isotopes (see also
Ref.~\cite{Ruiz16Calcium}). These promising first results could be an
indication that interactions based on a $\Delta$-full EFT formulation might be
capable of correctly reproducing different observables of nuclei over a wide
range of the nuclear chart already at order N$^2$LO. However, further
investigations are needed before definite conclusions can be drawn.

\subsubsection{Bayesian parameter estimation}
\label{sec:Bayes_parameter}

\begin{figure}[t]
\centering
\includegraphics[width=0.60\textwidth]{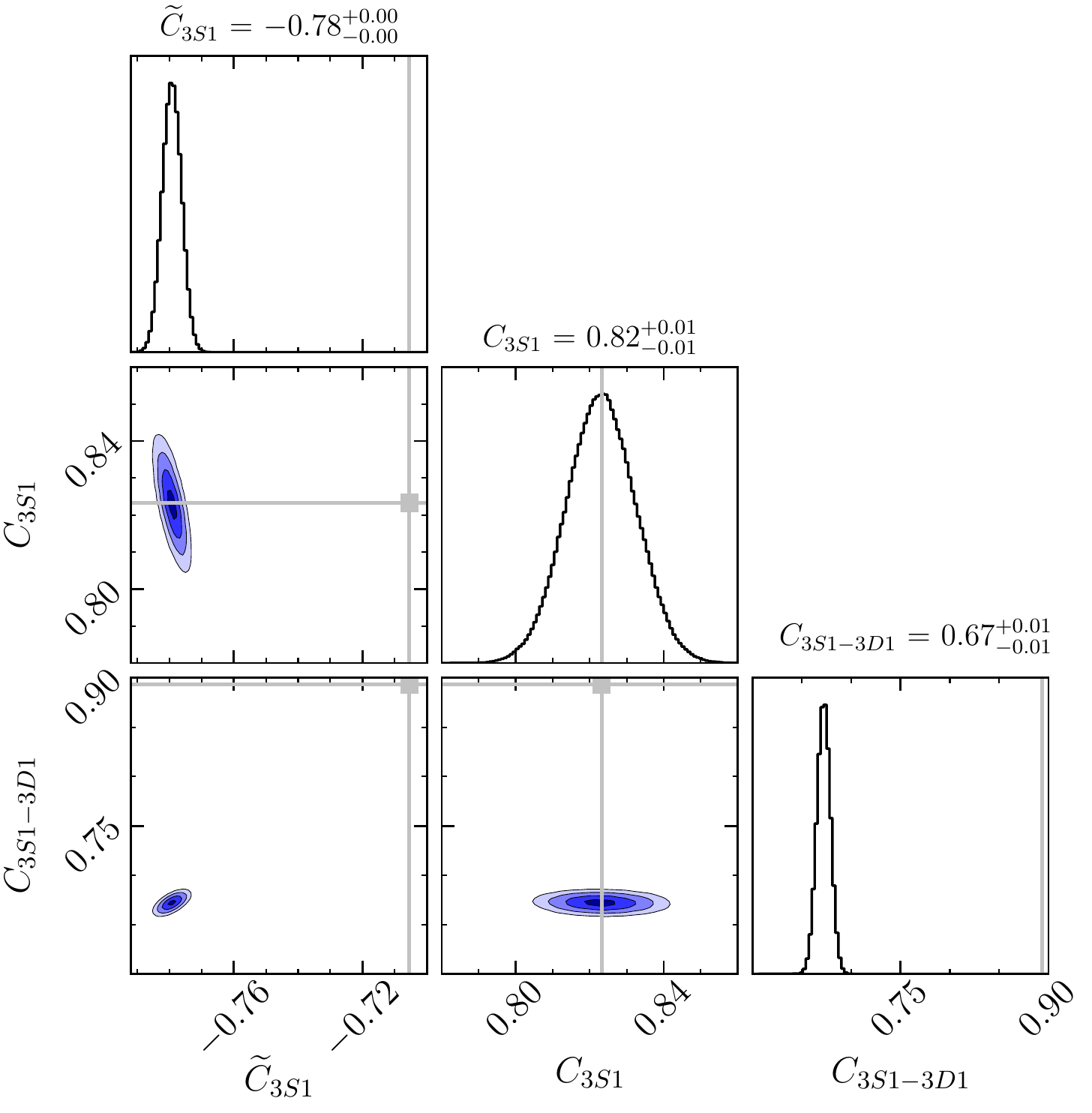}~%
\includegraphics[width=0.37\textwidth]{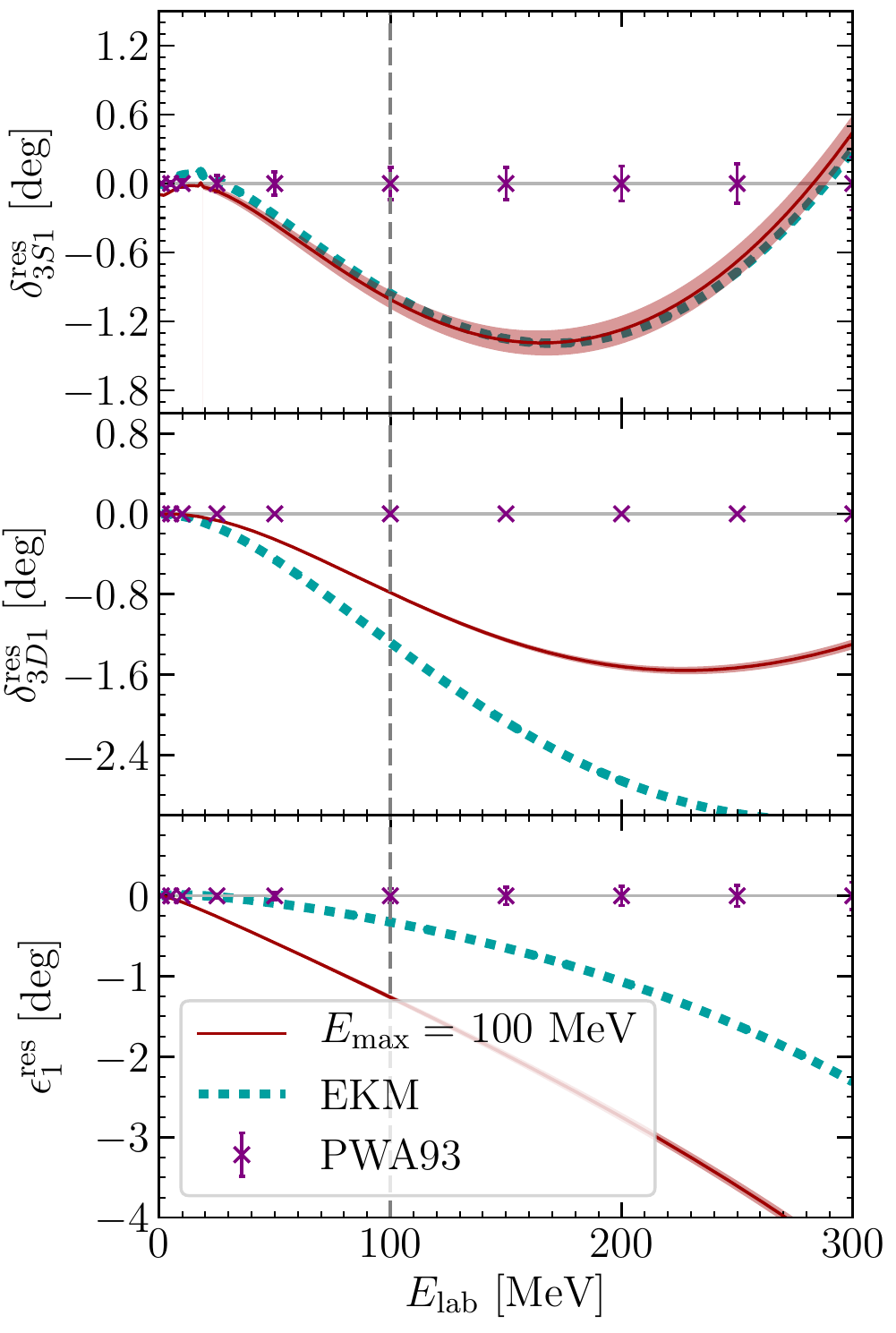}
\caption{Left: Posterior distribution functions for the NLO fit in the
$^3$S$_1$ channel to the np phase shifts with $E_{\text{max}}=100\,$MeV. The
values of the EKM NN interactions of Ref.~\cite{Epel15improved} are denoted by
the gray squares and lines. Right: The red band shows the propagated
prediction for the deviation of the phase shifts and mixing angle to the
partial-wave analysis of Ref.~\cite{Stok93PWA}. The cyan dotted line shows the
EKM result.\\
\textit{Source:} Figure taken from Ref.~\cite{Weso18NNphase}.}
\label{fig:Bayes_param_estimation}
\end{figure}

The fits of NN and 3N interactions discussed in the previous sections were all
based on some kind of $\chi^2$ minimization procedures. That means the set of
low-energy couplings of the interactions are determined such that a given
objective function, which encodes the deviation between some given input data
and their theoretical predictions, gets minimized. Recently, alternative
strategies for estimating the couplings based on Bayesian statistics were
proposed~\cite{Weso16Bayes,Ekst19Bayes}. Instead of extracting specific
values for each of the couplings, such a framework allows to determine
posterior probability distributions for each coupling (see
Figure~\ref{fig:Bayes_param_estimation}), opens ways to systematically study
correlations between different couplings and observables and to naturally
incorporate statistical and systematic uncertainties in the analysis.
Furthermore, the determination of the parameters can be guided by theoretical
expectations, such as naturalness, through the specification of Bayesian
priors, which allows to avoid problems connected to the overfitting of
parameters. In Ref.~\cite{Weso18NNphase} the framework was first applied to
the NN interactions of Ref.~\cite{Epel15improved} and two major issues could
be identified as part of the analysis. First, indications for degenerate
couplings were found, which were also validated and fixed in
Ref.~\cite{Rein17semilocal}. Furthermore, the incorporation of correlations
between observables in different kinematical regimes via Gaussian processes
was discussed (see also Ref.~\cite{Ekst19Bayes}) and the stability of the
extraction with respect to the inclusion of data at higher energies was
investigated.
\subsection{EFT uncertainty quantification}
\label{sec:EFTtruncation}

\begin{figure}[t]
\centering
\includegraphics[width=0.41\textwidth,clip]{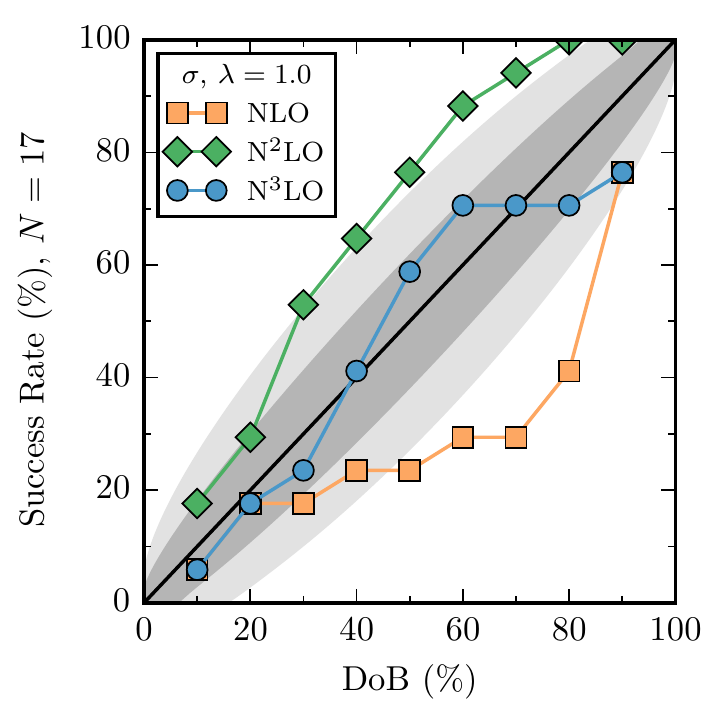}
\hspace{0.5cm}
\includegraphics[width=0.52\textwidth,keepaspectratio,clip]{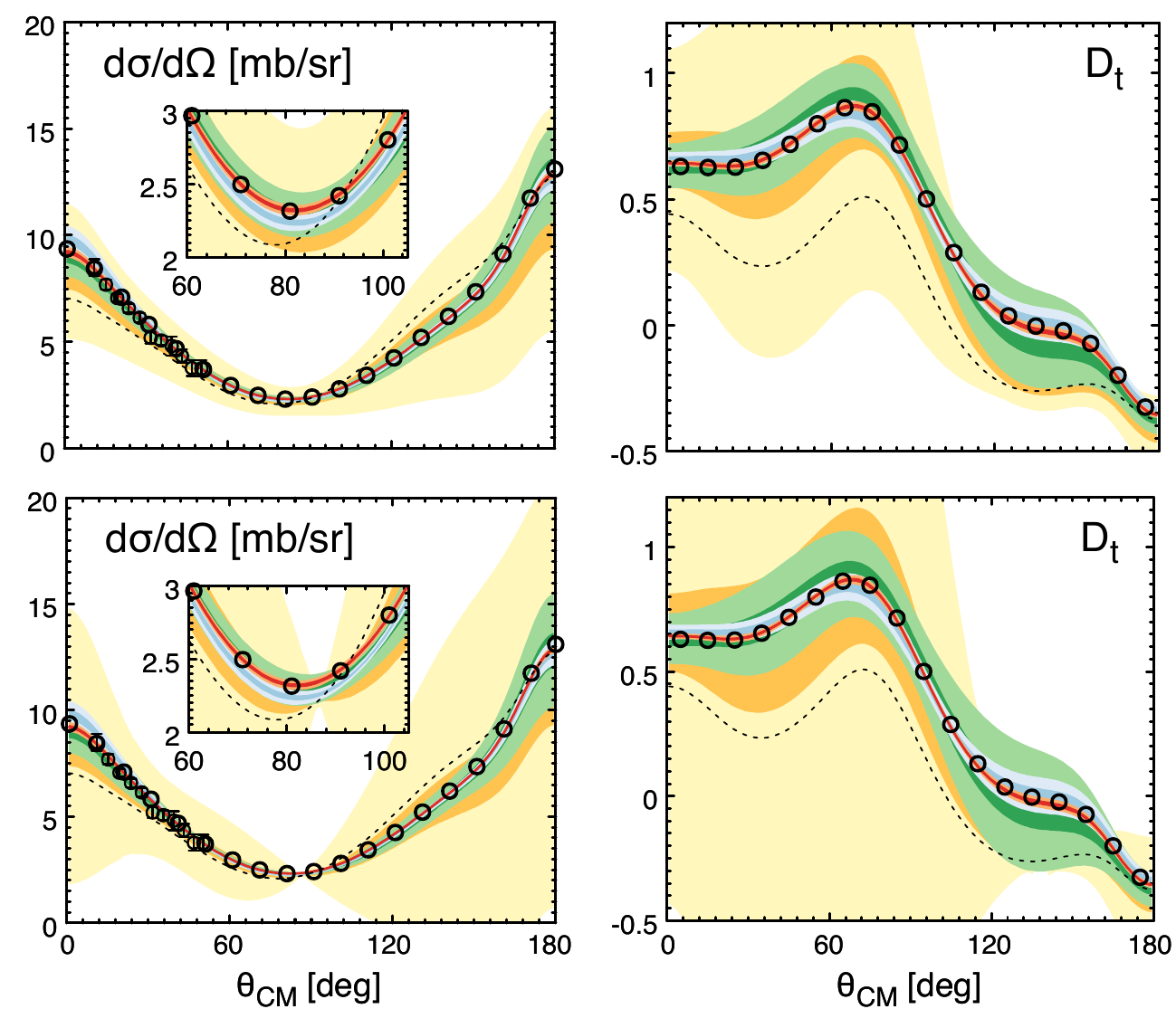}
\caption{Left: Consistency plot for the total np scattering cross
section at different orders in the chiral expansion and based on the NN
interactions of Ref.~\cite{Epel15improved}. The degree-of-beliefs (DoBs) were
determined at energies $E_{\text{lab}} =20,40,\ldots,340$\,MeV. Right: Estimated theoretical uncertainty for
the chiral EFT results for np differential cross section (left) and
polarization transfer coefficient $D_t$ (right) at laboratory energy of
$E_{\rm lab} = 143$~MeV. The top and bottom rows correspond to different
Bayesian models.  The light- (dark-) shaded yellow, green, blue and red bands
show the $95\%$ ($68\%$) DoB intervals at NLO, N$^2$LO, N$^3$LO and N$^4$LO,
respectively. Dashed lines show the LO predictions. Open circles refer to the
results of the Nijmegen partial-wave analysis~\cite{Stok93PWA}.\\
\textit{Source:} Left figure taken from Ref.~\cite{Mele17bayerror} and right figure adapted from Ref.~\cite{Epel19Bayes}.}
\label{fig:EFT_Bayes_truncation}  
\end{figure}

One of the key benefits of calculations based on nuclear forces derived within
a systematic EFT expansion is the possibility to estimate theoretical
uncertainties of results due to neglected interaction contributions at higher
orders. Calculations performed at different orders in the EFT expansion
provide expansion coefficients for observables and allow to study the
convergence of the EFT expansion, which can in turn be used to estimate the
size of the unknown higher order terms. To be specific, let us consider an
observable at a typical momentum scale $p$, for example, a scattering cross
section at a given energy. Then this observable is expected to take the
following form within the EFT expansion~\cite{Furn15uncert}:
\begin{equation}
X(p) = X_{\text{ref}} (p) \sum_{n=0}^{\infty} c_n (p) Q^n \, .
\label{eq:eft_converegnce}
\end{equation}
Here $X_{\text{ref}}$ is a natural size of the considered observable and
defines the scale of the observable. The parameter $Q$ is the expansion
coefficient with $Q=p/\Lambda$, where $\Lambda$ is the breakdown scale of the
EFT. The coefficients $c_n$ are dimensionless quantities which in general also
depend on the momentum scale $p$, but are expected to be of order one since
the scaling factor $X_{\text{ref}}$ has been factored out. Some coefficients
can also vanish due to symmetry reasons. If the series is truncated at order
$k$ then the truncation error is given by (see Ref.~\cite{Furn15uncert}):
\begin{equation}
\Delta X^{(k)} (p) = \Delta_k X_{\text{ref}} (p) = X_{\text{ref}} (p) \sum_{n=k+1}^{\infty} c_n (p) Q^n \, .
\label{eq:EFT_truncation_def}
\end{equation}
Different strategies have been developed to obtain an estimate for $\Delta
X^{(k)} (p)$ for chiral EFT given information on the size of the computed
coefficients $c_0, c_1, ..., c_k$. In Ref.~\cite{Epel15improved} a
conservative prescription for the estimate was proposed which at N$^3$LO
takes the following form:
\begin{equation}
\Delta X^{\text{N$^3$LO}} (p) = \text{max} 
\left( 
Q^5 \bigl| X^{\text{LO}} (p) \bigr|, 
Q^3 \bigl| X^{\text{LO}} (p) -  X^{\text{NLO}}(p) \bigr|, 
Q^2 \bigl| X^{\text{NLO}} (p) -  X^{\text{N$^2$LO}}(p) \bigr|, 
Q \bigl| X^{\text{N$^2$LO}}(p) -  X^{\text{N$^3$LO}}(p) \bigr|
\right) \, .
\label{eq:EKM_uncertainty}
\end{equation}
The corresponding uncertainties at other orders follow accordingly. This
estimate is based on the assumption that all coefficients $c_n$ are of the
same order of magnitude, which in turn allows to estimate the contributions at order
$n$ as the leading order result $X_{\text{ref}} c_0$ times the factors $Q^n$.
In particular, the differences of results at successive orders provide
various ways to estimate the leading order term in
Eq.~(\ref{eq:EFT_truncation_def}), for example:
\begin{equation}
Q^3 \bigl( X^{\text{NLO}} (p) -  X^{\text{LO}}(p) \bigr) = X_{\text{ref}} (p) c_2 (p) Q^5 \approx X_{\text{ref}} (p) c_5 (p) Q^5 \approx \Delta X^{\text{N$^3$LO}} (p) \, .
\end{equation}
The maximum of all available differences represents a conservative estimate
for the truncation error and leads to Eq.~(\ref{eq:EKM_uncertainty}). It
should be noted, however, that this prescription is also based on the
assumption that the leading order result $X^{\text{LO}} (p)$ already provides
a reasonable description of a given observable. If that is not the case the
first term in Eq.~(\ref{eq:EKM_uncertainty}) will typically dominate the
uncertainty bands and the resulting estimates will not necessarily be
reasonable (see, e.g.,~Ref.~\cite{Hopp19medmass}).

Another drawback of the strategy above is that it does not provide a
statistical interpretation of uncertainty estimates. A more systematic and
quantitative determination of truncation errors can be achieved by employing
Bayesian frameworks
instead~\cite{Furn15UQrecipe,Furn15uncert,Mele19corr,Epel19Bayes}. Such
Bayesian analyses allow to extract statistical degree-of-belief (DoB)
intervals based on results up to a given order (see
Figure~\ref{fig:EFT_Bayes_truncation}), to study the consistency of a chosen
breakdown scale $\Lambda$~(see, e.g. Ref.~\cite{Mele17bayerror}) and also to
explore the parameter estimation of the LECs in EFTs (see
Refs.~\cite{Weso16Bayes,Mele17bayerror,Weso18NNphase} and
Section~\ref{sec:Bayes_parameter}). First Bayesian analyses of various NN and
nucleon-deuteron scattering observables found that uncertainties based on the
Eq.~(\ref{eq:EKM_uncertainty}) were in good agreement with $68\%$ DoB
intervals for particular prior choices~\cite{Mele17bayerror,Epel19Bayes}.

The generalization and improvement of these Bayesian analyses is presently a
very active field of research. Current work aims at improving the treatment of
correlations between observables in different physical regimes via, e.g.,
Gaussian processes~\cite{Mele19corr} or the extension of the analyses to
heavier nuclei, including contributions from 3N interactions. One additional
challenge when investigating bound-state observables of atomic nuclei is the
question how to exactly determine the scale $p$ in
Eqs.~(\ref{eq:eft_converegnce}) and~(\ref{eq:EFT_truncation_def}) given that
wave functions generally contain contributions from various different momentum
scales (see, e.g., the discussion in Ref.~\cite{Bind15Fewbody}). Such global
analyses will open new ways to benchmark NN and 3N interactions more
systematically in different regimes of the nuclear chart, explore and test
different fitting strategies for the LECs, benchmark various regularization
schemes (see also Section~\ref{sec:3N_regularization}) and isolate possible
deficiencies of interactions regarding the description of specific observables
of nuclei and matter (see also Section~\ref{sec:applications}).

\clearpage
\section{Representation and calculation of 3N interactions}
\label{sect:3NF_representation}

In this section we give a comprehensive overview of fundamental techniques for
the calculation of 3N interaction matrix elements in a partial-wave basis
representation. Many state-of-the-art few- and many-body frameworks in nuclear
physics like, e.g., Faddeev(-Yakubovsky), hyperspherical harmonics, no-core
shell model, valence-shell diagonalization, in-medium similarity renormalization group,
coupled cluster, self-consistent Greens-Function and many-body perturbation
theory (see Table~\ref{tab:many_body_frameworks}) are formulated in such a
basis representation. That means the matrix elements obtained from the methods
discussed in this section can be incorporated in a straightforward way in all
these frameworks. On the other hand, frameworks based on Quantum Monte Carlo
techniques or lattice effective field theory require different representations
of 3N interactions.

Specifically, in this section we discuss in detail how to efficiently
calculate and represent 3N interactions in a three-body partial-wave momentum
representation. This basis representation has several conceptual advantages:
\begin{itemize}
\item The momentum representation does not contain an implicit infrared
cutoff scale, in contrast to, e.g, a harmonic oscillator (HO) representation
(see, e.g., Refs.~\cite{Koni14UVextrapol,Furn15IRextrapol}). This ensures
that the long-range part of the interactions is fully captured in this
representation.

\item The ultraviolet cutoff scale of the momentum representation can be
specified explicitly by appropriate choices of the discrete momentum meshes.
Note that the ultraviolet momentum cutoff scale can vary significantly,
depending of the specific choice of regularization scheme and cutoff scale
(see Section~\ref{sec:3N_regularization}).

\item The transformation from a momentum basis to a HO partial-wave
representation can be performed in a straightforward and numerically stable
way (see Section \ref{sec:HO_transf}), whereas the inverse transformation is
problematic since a large number of HO states is required to represent a
plane-wave state. This implies that a momentum representation allows to
apply the same interactions to few- and many-body frameworks that are
formulated directly in the momentum basis (like, e.g, the
Faddeev(-Yakubovsky) equations or MBPT for nuclear matter) as well as to
frameworks for medium-mass nuclei, which are typically formulated in a HO
representation (see Table~\ref{tab:many_body_frameworks}).

\end{itemize}

\subsection{Definition of coordinates}
\label{sec:3NF_coord_def}

We consider a system of three interacting point particles with mass $m_1$,
$m_2$ and $m_3$ located at the coordinates $\mathbf{x}_1$, $\mathbf{x}_2$ and
$\mathbf{x}_3$. Starting from these single-particle coordinates we can
introduce relative coordinates via the following definitions~\cite{Mess99QM,Gloe83QMFewbod}:
\begin{align}
\mathbf{r}_{\{12\}} &= \mathbf{x}_1 - \mathbf{x}_2 \nonumber \\
\mathbf{R}_{\{12\}} &= \frac{m_1 \mathbf{x}_1 + m_2 \mathbf{x}_2}{m_1 + m_2} \nonumber \\
\mathbf{s}_{\{12\}} &= \mathbf{x}_3 - \mathbf{R}_{\{12\}} \nonumber \\
\mathbf{R}_{\rm{3N}} &= \frac{1}{M} \bigl( m_1 \mathbf{x}_1 + m_2 \mathbf{x}_2 + m_3 \mathbf{x}_3 \bigr) \, ,
\label{eq:def_Jacobi_coordinates12}
\end{align}
with the total mass $M = m_1 + m_2 + m_3$. Here, $\mathbf{r}_{\{12\}}$
represents the relative distance between particle $1$ and $2$, and
$\mathbf{R}_{\{12\}}$ is the two-body center-of-mass coordinate of the
subsystem consisting of particles $1$ and $2$, indicated in the following by
the index $\{12\}$. Of course, this choice is only one possible alternative.
The choices $\{31\}$ and $\{23\}$ are equally valid and are discussed in more
detail below. The second relative coordinate $\mathbf{s}_{\{12\}}$ is defined
by the distance of the third particle to the center-of-mass coordinate of
subsystem $\{12\}$. These definitions can be generalized straightforwardly to
arbitrary particle numbers~\cite{Gloe83QMFewbod}. Parametrizing three
particles in terms of relative coordinates $\mathbf{r}_{\{12\}}$ and
$\mathbf{s}_{\{12\}}$ instead of single-particle coordinates $\mathbf{x}_1$,
$\mathbf{x}_2$ and $\mathbf{x}_3$ allows to explicitly factor out the total
center-of-mass coordinate $\mathbf{R}_{\rm{3N}}$. Thanks to this factorization
the total number of coordinates required for the description of translational
invariant system can be reduced from 3 to 2, which is of crucial importance
for the practical representation of 3N interactions.

Accordingly, in a momentum representation the three particles can be
characterized by three single-particle momenta $\mathbf{k}_1$, $\mathbf{k}_2$
and $\mathbf{k}_3$ or by the corresponding relative and total
center-of-mass momenta:
\begin{align}
\mathbf{p}_{\{12\}} &= \frac{m_2 \mathbf{k}_1 - m_1 \mathbf{k}_2}{m_1 + m_2} \nonumber \\
\mathbf{P}_{\{12\}} &= \mathbf{k}_1 + \mathbf{k}_2 \nonumber \\
\mathbf{q}_{\{12\}} &= \frac{1}{M} \bigl[ (m_1 + m_2) \mathbf{k}_3 - m_3 \mathbf{P}_{\{12\}} \bigr] \nonumber \\
\mathbf{P}_{\rm{3N}} &= \mathbf{k}_1 + \mathbf{k}_2 + \mathbf{k}_3 \, .
\label{eq:def_Jacobi_momenta12}
\end{align}

The relative momenta $\mathbf{p}$ and $\mathbf{q}$ are also called Jacobi
momenta. The kinetic energy of the system can then be expressed in terms of
single-particle momenta or Jacobi momenta:
\begin{equation}
T = \sum_{i=1}^{3} \frac{\mathbf{k}_i^2}{2 m_i} = \frac{\mathbf{P}_{\rm{3N}}^2}{2 M} + \frac{\mathbf{p}_{\{12\}}^2}{2 \mu_1} + \frac{\mathbf{q}_{\{12\}}^2}{2 \mu_2} \, ,
\label{eq:kinetic_energy_Jacobi_genmasses}
\end{equation}
with the reduced masses
\begin{equation}
\frac{1}{\mu_1} = \frac{1}{m_1} + \frac{1}{m_2}, \quad \frac{1}{\mu_2} = \frac{1}{m_1 + m_2} + \frac{1}{m_3} \, .
\end{equation}
Corresponding relations can be derived in the basis representations $\{31\}$ and
$\{23\}$ (see Figure~\ref{fig:Jacobi_representations}). Since each choice of
variables represents a complete basis to describe the relative motion of the
three particles, the different variables are all related by linear
transformations. These relations can be derived straightforwardly and are summarized
in Table~\ref{tab:Jacobi_momenta_crosstable} for the case of equal masses,
$m_1$ = $m_2$ = $m_3$.

The different basis representations are related by a cyclic (or anticyclic)
permutation of particles. Let us consider a generic three-body state $\left| a b
c \right>$ with some arbitrary single-particle quantum numbers $a$, $b$ and
$c$. Now consider the following permutation operators
\begin{equation}
P_{123} = P_{12} P_{23} \quad \text{and} \quad P_{132} = P_{13} P_{23} \, ,
\end{equation}
where $P_{ij}$ are the two-body transposition operators that exchange the
labels of particles $i$ and $j$, e.g.:
\begin{equation}
P_{12} \left| a b c \right> = \left| b a c \right>, \quad P_{13} \left| a b c \right> = \left| c b a \right>, \quad \text{etc.}
\end{equation}
It is easy to verify that the following relations hold:
\begin{align}
P_{123} \left| a b c \right> &= P_{12} P_{23} \left| a b c \right> = P_{12} \left| a c b \right> = \left| c a b \right> \, ,\\
P_{132} \left| a b c \right> &= P_{13} P_{23} \left| a b c \right> = P_{13} \left| a c b \right> = \left| b c a \right> = P_{123}^{-1} \left| a b c \right> \, .
\end{align}
That means $P_{123}$ ($P_{132}$) represent the cyclic (anticyclic) permutation
operators for three-body states. These operators play a key role for the
treatment of 3N interactions as we will discuss in detail the following
sections.

If we now consider specifically a three-body state in a momentum
representation, we can parametrize it via the single-particle momenta in the
form $\left| \mathbf{k}_1 \mathbf{k}_2 \mathbf{k}_3 \right>$ or in a relative
momentum representation $\left| \mathbf{p} \mathbf{q}
\mathbf{P}_{\text{3N}} \right>_{\{ab\}}$. Here the total three-body momentum
$\mathbf{P}_{\text{3N}}$ is identical for all basis representations $\{ab\}$
and just characterizes boosts of the three-body system. As we will discuss in
more detail in Section~\ref{sec:3NF_mom_rep}, microscopic nuclear forces do
not depend on the three-body center-of-mass momentum. This implies that the
entire structure of 3N interactions is encoded in their dependence on the
Jacobi momenta $\mathbf{p}$ and $\mathbf{q}$. For that reason, we suppress in
the following the center-of-mass quantum number $\mathbf{P}_{\text{3N}}$ in
the three-body states and write them in the form $\left| \mathbf{p} \mathbf{q}
\right>_{\{ab\}}$, while we choose the following normalization convention (see
also Appendix~\ref{sec:normalization}):
\begin{equation}
\tensor*[_{\{ab\}}]{\bigl< \mathbf{p}' \mathbf{q}' | \mathbf{p} \mathbf{q} \bigr>}{_{\{ab\}}} = (2 \pi)^6 \delta(\mathbf{p}' - \mathbf{p}) \delta(\mathbf{q}' - \mathbf{q}), 
\quad \int \frac{d \mathbf{p}}{(2 \pi)^3} \frac{d \mathbf{q}}{(2 \pi)^3}
\bigl| \mathbf{p} \mathbf{q} \bigr>_{\{ab\}} \prescript{}{\{ab\}}{\bigl<
\mathbf{p} \mathbf{q} \bigr|} = 1 \, .
\label{eq:PW_def_norm}
\end{equation}

Since the representations $\{12\}$ and $\{23\}$ are related by the
transformations (see Table~\ref{tab:Jacobi_momenta_crosstable})
\begin{equation}
\mathbf{k}_1 \rightarrow \mathbf{k}_2, \quad \mathbf{k}_2 \rightarrow
\mathbf{k}_3, \quad \mathbf{k}_3 \rightarrow \mathbf{k}_1 \, .
\label{eq:relation_ki_ab_bases}
\end{equation}

the following relations hold:
\begin{align}
\left| \mathbf{p} \mathbf{q} \right>_{\{23\}} = P_{123} \left| \mathbf{p}
\mathbf{q} \right>_{\{12\}}, \quad \left| \mathbf{p} \mathbf{q}
\right>_{\{31\}} = P_{123} \left| \mathbf{p} \mathbf{q} \right>_{\{23\}} \quad
\text{and} \quad \left| \mathbf{p} \mathbf{q} \right>_{\{12\}} = P_{123}
\left| \mathbf{p} \mathbf{q} \right>_{\{31\}} \, .
\label{eq:mombases_relations}
\end{align}
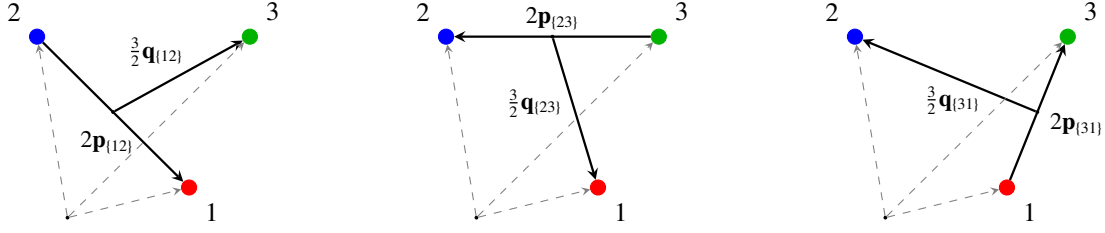
\begin{figure}[t]
\centering 
\begin{minipage}{0.32\textwidth} 
\begin{tikzpicture}
  [
    scale=0.8,
    >=stealth,
    point/.style = {draw, circle,  fill = black, inner sep = 2pt},
    dot/.style   = {draw, circle,  fill = black, inner sep = .2pt},
  ]

  \node (origin) at (0.5,1.0) [dot] {};
  \node (n1) at (2.5,1.5) [point, color = red, label = {[label distance = 0.02cm]below right:$1$}] {};
  \node (n2) at (0,4) [point, color = blue, label = {[label distance = 0.02cm]above left:$2$}] {};
  \node (n3) at (3.5,4.0) [point, color = black!30!green, label = {[label distance = 0.02cm]above right:$3$}] {};
  \node (n12) at ($ (n1)!.5!(n2)$) [dot] {};

  \draw[->, line width=0.3mm] (n2) -- node (a) {} (n1);
  \node at ($(n12) - (0.1,0.5)$) {\small $2 \mathbf{p}_{\{12\}}$};

  \draw[->, line width=0.3mm] (n12) -- node (b) {} (n3);
  \node at ($(n12) + (0.7,1.0)$) {\small $\tfrac{3}{2} \mathbf{q}_{\{12\}}$};

  \draw[->, dashed, gray] (origin) -- node (c) {} (n1);
  \draw[->, dashed, gray] (origin) -- node (d) {} (n2);
  \draw[->, dashed, gray] (origin) -- node (e) {} (n3);

\end{tikzpicture}
\end{minipage}
\begin{minipage}{0.32\textwidth} 
\begin{tikzpicture}
  [
    scale=0.8,
    >=stealth,
    point/.style = {draw, circle,  fill = black, inner sep = 2pt},
    dot/.style   = {draw, circle,  fill = black, inner sep = .2pt},
  ]

  \node (origin) at (0.5,1.0) [dot] {};
  \node (n1) at (2.5,1.5) [point, color = red, label = {[label distance = 0.02cm]below right:$1$}] {};
  \node (n2) at (0,4) [point, color = blue, label = {[label distance = 0.02cm]above left:$2$}] {};
  \node (n3) at (3.5,4.0) [point, color = black!30!green, label = {[label distance = 0.02cm]above right:$3$}] {};
  \node (n23) at ($ (n2)!.5!(n3)$) [dot] {};

  \draw[->, line width=0.3mm] (n3) -- node (a) {} (n2);
  \node at ($(n23) + (0.0,0.3)$) {\small $2 \mathbf{p}_{\{23\}}$};

  \draw[->, line width=0.3mm] (n23) -- node (b) {} (n1);
  \node at ($(n23) + (-0.3,-1.1)$) {\small $\tfrac{3}{2} \mathbf{q}_{\{23\}}$};

  \draw[->, dashed, gray] (origin) -- node (c) {} (n1);
  \draw[->, dashed, gray] (origin) -- node (d) {} (n2);
  \draw[->, dashed, gray] (origin) -- node (e) {} (n3);

\end{tikzpicture}
\end{minipage}
\begin{minipage}{0.32\textwidth} 
\begin{tikzpicture}
  [
    scale=0.8,
    >=stealth,
    point/.style = {draw, circle,  fill = black, inner sep = 2pt},
    dot/.style   = {draw, circle,  fill = black, inner sep = .2pt},
  ]

  \node (origin) at (0.5,1.0) [dot] {};
  \node (n1) at (2.5,1.5) [point, color = red, label = {[label distance = 0.02cm]below right:$1$}] {};
  \node (n2) at (0,4) [point, color = blue, label = {[label distance = 0.02cm]above left:$2$}] {};
  \node (n3) at (3.5,4.0) [point, color = black!30!green, label = {[label distance = 0.02cm]above right:$3$}] {};
  \node (n13) at ($ (n1)!.5!(n3)$) [dot] {};

  \draw[->, line width=0.3mm] (n1) -- node (a) {} (n3);
  \node at ($(n13) + (0.65,-0.2)$) {\small $2 \mathbf{p}_{\{31\}}$};

  \draw[->, line width=0.3mm] (n13) -- node (b) {} (n2);
  \node at ($(n13) + (-1.4,0.2)$) {\small $\tfrac{3}{2} \mathbf{q}_{\{31\}}$};

  \draw[->, dashed, gray] (origin) -- node (c) {} (n1);
  \draw[->, dashed, gray] (origin) -- node (d) {} (n2);
  \draw[->, dashed, gray] (origin) -- node (e) {} (n3);

\end{tikzpicture}
\end{minipage}
\caption{Definition of the three-body Jacobi momenta $\mathbf{p}$ and
$\mathbf{q}$ in the representations $\{12\}$ (left), $\{23\}$ (center) and
$\{31\}$ (right) for equal masses ($m_1 = m_2 = m_3$). The dashed arrows
denote the single-particle momenta $\mathbf{k}_1$, $\mathbf{k}_2$ and
$\mathbf{k}_3$. The relations between the momenta in the different
representations are summarized in Table~\ref{tab:Jacobi_momenta_crosstable}.}
\label{fig:Jacobi_representations}
\end{figure}

\begin{table}[t]
\centering
\begin{tabular}{c||c|c|c|c}
                            & $\mathbf{k}_i$                                    & $\{12\}$                                          & $\{23\}$                                            & $\{31\}$                                                                                          \\ \hline \hline
\parbox[0pt][1.7em][l]{0cm}{} $\mathbf{p}_{\{12\}}$   & $\frac{1}{2} (\mathbf{k}_1 - \mathbf{k}_2)$                       & $\mathbf{p}_{\{12\}}$                                     & $- \frac{1}{2} \mathbf{p}_{\{23\}} + \frac{3}{4} \mathbf{q}_{\{23\}}$               & $-\frac{1}{2} \mathbf{p}_{\{31\}} - \frac{3}{4} \mathbf{q}_{\{31\}}$                                \\ \hline
\parbox[0pt][1.7em][l]{0cm}{} $\mathbf{q}_{\{12\}}$   & $\frac{2}{3} \left[ \mathbf{k}_3 - \frac{1}{2} (\mathbf{k}_1 + \mathbf{k}_2) \right]$ & $\mathbf{q}_{\{12\}}$                                     & $-\mathbf{p}_{\{23\}} - \frac{1}{2} \mathbf{q}_{\{23\}}$                    & $\mathbf{p}_{\{31\}} - \frac{1}{2} \mathbf{q}_{\{31\}}$                                         \\ \hline
\parbox[0pt][1.7em][l]{0cm}{} $\mathbf{p}_{\{23\}}$   & $\frac{1}{2} (\mathbf{k}_2 - \mathbf{k}_3)$                       & $- \frac{1}{2} \mathbf{p}_{\{12\}} - \frac{3}{4} \mathbf{q}_{\{12\}}$             & $\mathbf{p}_{\{23\}}$                                     & $-\frac{1}{2} \mathbf{p}_{\{31\}} + \frac{3}{4} \mathbf{q}_{\{31\}}$                              \\ \hline
\parbox[0pt][1.7em][l]{0cm}{} $\mathbf{q}_{\{23\}}$   & $\frac{2}{3} \left[ \mathbf{k}_1 - \frac{1}{2} (\mathbf{k}_2 + \mathbf{k}_3) \right]$ & $\mathbf{p}_{\{12\}} - \frac{1}{2} \mathbf{q}_{\{12\}}$                     & $\mathbf{q}_{\{23\}}$                                     & $- \mathbf{p}_{\{31\}} - \frac{1}{2} \mathbf{q}_{\{31\}}$                                            \\ \hline
\parbox[0pt][1.7em][l]{0cm}{} $\mathbf{p}_{\{31\}}$   & $\frac{1}{2} (\mathbf{k}_3 - \mathbf{k}_1)$                                           & $- \frac{1}{2} \mathbf{p}_{\{12\}} + \frac{3}{4} \mathbf{q}_{\{12\}}$                       & $-\frac{1}{2} \mathbf{p}_{\{23\}} - \frac{3}{4} \mathbf{q}_{\{23\}}$                            & $\mathbf{p}_{\{31\}}$                                                                         \\ \hline
\parbox[0pt][1.7em][l]{0cm}{} $\mathbf{q}_{\{31\}}$   & $\frac{2}{3} \left[ \mathbf{k}_2 - \frac{1}{2} (\mathbf{k}_3 + \mathbf{k}_1) \right]$ & $-\mathbf{p}_{\{12\}} - \frac{1}{2} \mathbf{q}_{\{12\}}$                                    & $\mathbf{p}_{\{23\}} - \frac{1}{2} \mathbf{q}_{\{23\}}$                                         & $\mathbf{q}_{\{31\}}$                                                                         \\ \hline
\parbox[0pt][1.7em][l]{0cm}{} $\mathbf{k}_1$    & $\mathbf{k}_1$                                    & $\mathbf{p}_{\{12\}} - \frac{1}{2} \mathbf{q}_{\{12\}} + \frac{1}{3} \mathbf{P}_{\rm{3N}}$  & $\mathbf{q}_{\{23\}} + \frac{1}{3} \mathbf{P}_{\rm{3N}}$                    & $-\mathbf{p}_{\{31\}} - \frac{1}{2} \mathbf{q}_{\{31\}} + \frac{1}{3} \mathbf{P}_{\rm{3N}}$          \\ \hline
\parbox[0pt][1.7em][l]{0cm}{} $\mathbf{k}_2$    & $\mathbf{k}_2$                                    & $- \mathbf{p}_{\{12\}} - \frac{1}{2} \mathbf{q}_{\{12\}} + \frac{1}{3} \mathbf{P}_{\rm{3N}}$& $\mathbf{p}_{\{23\}} - \frac{1}{2} \mathbf{q}_{\{23\}} + \frac{1}{3} \mathbf{P}_{\rm{3N}}$    & $\mathbf{q}_{\{31\}} + \frac{1}{3} \mathbf{P}_{\rm{3N}}$                                            \\ \hline
\parbox[0pt][1.7em][l]{0cm}{} $\mathbf{k}_3$    & $\mathbf{k}_3$                                    & $\mathbf{q}_{\{12\}}  + \frac{1}{3} \mathbf{P}_{\rm{3N}}$                   & $- \mathbf{p}_{\{23\}} - \frac{1}{2} \mathbf{q}_{\{23\}}  + \frac{1}{3} \mathbf{P}_{\rm{3N}}$ & $\mathbf{p}_{\{31\}} - \frac{1}{2} \mathbf{q}_{\{31\}} + \frac{1}{3} \mathbf{P}_{\rm{3N}}$ 
\end{tabular}
\caption{Relations between the single-particle momenta $\mathbf{k}_i$ and the
Jacobi momenta $\mathbf{p}$ and $\mathbf{q}$ for a three-body system with
equal masses in the different representations \{12\}, \{31\} and \{23\} (see
Figure~\ref{fig:Jacobi_representations}).}
\label{tab:Jacobi_momenta_crosstable}
\end{table}
Hence, the momentum matrix elements of the permutation operator can be
expressed as (see Table \ref{tab:Jacobi_momenta_crosstable}):
\begin{subequations}
\begin{align}
\tensor*[_{\{12\}}]{\left< \mathbf{p}' \mathbf{q}' | P_{123} | \mathbf{p} \mathbf{q} \right>}{_{\{12\}}} &= \tensor*[_{\{12\}}]{\left< \mathbf{p}' \mathbf{q}' | \mathbf{p} \mathbf{q} \right>}{_{\{23\}}} \nonumber \\
&= (2 \pi)^6 \delta \bigl( \mathbf{p}'_{\{12\}} + \tfrac{1}{2} \mathbf{p}_{\{23\}} - \tfrac{3}{4} \mathbf{q}_{\{23\}} \bigr) \delta \bigl( \mathbf{q}'_{\{12\}} + \mathbf{p}_{\{23\}} + \tfrac{1}{2} \mathbf{q}_{\{23\}} \bigr) \label{eq:P123_momentum_def_first} \\
&= (2 \pi)^6 \delta \bigl( \mathbf{p}_{\{23\}} + \tfrac{1}{2} \mathbf{q}_{\{23\}} + \mathbf{q}'_{\{12\}} \bigr) \delta \bigl( \mathbf{p}'_{\{12\}} - \tfrac{1}{2} \mathbf{q}'_{\{12\}} - \mathbf{q}_{\{23\}} \bigr) \label{eq:P123_momentum_def_third} \\
&= (2 \pi)^6 \delta \bigl( \mathbf{p}_{\{23\}} + \tfrac{1}{2} \mathbf{p}'_{\{12\}} + \tfrac{3}{4} \mathbf{q}'_{\{12\}} \bigr) \delta \bigl( \mathbf{q}_{\{23\}} - \mathbf{p}'_{\{12\}} + \tfrac{1}{2} \mathbf{q}'_{\{12\}} \bigr) \, .
\label{eq:P123_momentum_def_fourth}
\end{align}
\end{subequations}
From Eq.~(\ref{eq:mombases_relations}) it follows immediately that
\begin{equation}
\tensor*[_{\{12\}}]{\left< \mathbf{p}' \mathbf{q}' | \mathbf{p} \mathbf{q} \right>}{_{\{23\}}} 
= \tensor*[_{\{23\}}]{\left< \mathbf{p}' \mathbf{q}' | \mathbf{p} \mathbf{q} \right>}{_{\{31\}}} 
= \tensor*[_{\{31\}}]{\left< \mathbf{p}' \mathbf{q}' | \mathbf{p} \mathbf{q} \right>}{_{\{12\}}} \, ,
\label{eq:permutation_momequiv}
\end{equation}
and, consequently, the representation of the permutation operator is identical
in all basis representations:
\begin{equation}
\left< \mathbf{p}' \mathbf{q}' | P_{123} | \mathbf{p} \mathbf{q} \right> = \tensor*[_{\{ab\}}]{\left< \mathbf{p}' \mathbf{q}' | P_{123} | \mathbf{p} \mathbf{q} \right>}{_{\{ab\}}} \, .
\label{eq:P123_rep_independent}
\end{equation}
The corresponding relations to Eq.~(\ref{eq:mombases_relations}) for the bra states can be derived by using the relations
\begin{equation}
\tensor*[_{\{12\}}]{\bigl< \mathbf{p}' \mathbf{q}' | \mathbf{p} \mathbf{q} \bigr>}{_{\{12\}}} = \tensor*[_{\{12\}}]{\bigl< \mathbf{p}' \mathbf{q}' | P_{123}^{-1} P_{123} | \mathbf{p} \mathbf{q} \bigr>}{_{\{12\}}} = \tensor*[_{\{12\}}]{\bigl< \mathbf{p}' \mathbf{q}' | P_{123}^{-1} | \mathbf{p} \mathbf{q} \bigr>}{_{\{23\}}} \, .
\end{equation}
Hence:
\begin{align}
\tensor*[_{\{23\}}]{\left< \mathbf{p} \mathbf{q} \right|}{} = \tensor*[_{\{12\}}]{\left< \mathbf{p} \mathbf{q} \right|}{} P_{123}^{-1} \, ,
\end{align}
and accordingly:
\begin{equation}
\tensor*[_{\{31\}}]{\left< \mathbf{p} \mathbf{q} \right|}{} = \tensor*[_{\{23\}}]{\left< \mathbf{p} \mathbf{q} \right|}{} P_{123}^{-1} , \quad \tensor*[_{\{12\}}]{\left< \mathbf{p} \mathbf{q} \right|}{}= \tensor*[_{\{31\}}]{\left< \mathbf{p} \mathbf{q} \right|}{} P_{123}^{-1} \, .
\label{eq:mombases_relations_bra}
\end{equation}

In Section~\ref{sec:3N_decomp_antisymmetrization} the matrix elements of the
permutation operator for particles with spin and isospin are evaluated
explicitly in a partial-wave representation. These elements are a key quantity
for the representation of 3N interactions and three-body wave functions as
well as for the implementation of the SRG flow
equations for NN and 3N interactions.

\subsection{Momentum basis representation of three-nucleon forces}
\label{sec:3NF_mom_rep}
Now we discuss how to represent 3N interactions in a momentum plane wave
basis. To this end, we first neglect spin and isospin degrees of freedom of
the nucleons in order to simplify the notation and discussion. We will discuss
the general case including all internal degrees of freedom in Section
\ref{sec:PWD_3NF_local}.

For spinless particles the matrix elements of a general three-body operator
$\hat{O}$ have the following form:
\begin{equation}
\bigl< \mathbf{k}'_1 \mathbf{k}'_2 \mathbf{k}'_3 | O | \mathbf{k}_1 \mathbf{k}_2 \mathbf{k}_3 \bigr> \, ,
\label{eq:general_3Boperator}
\end{equation}
where $\mathbf{k}_i$ ($\mathbf{k}'_i$) are the single-particle momenta in the
initial (final) state. The computation of these matrix elements is a highly
nontrivial task, given that they depend on six vector variables. For our
purposes, however, we can simplify the task considerably since microscopic
free-space 3N interactions have the following symmetry properties:
\begin{itemize}
    \item  \textit{conservation} of the center-of-mass momentum, i.e., $\mathbf{P}_{\rm{3N}} = \sum_{i=1}^3 \mathbf{k}_i = \sum_{i=1}^3 \mathbf{k}'_i$,
    \item  \textit{independence} on $\mathbf{P}_{\rm{3N}}$ (Galilean invariance),
    \item  \textit{rotational invariance}, i.e., the interaction transforms like a scalar under spatial rotations. 
\end{itemize}
The first two symmetries follow from the basic symmetries of QCD in the
non-relativistic limit, while the third symmetry holds since there is no
preferred direction in the absence of external fields. In the following we
will study the representation of such an interaction, $V_{\text{3N}}$, for the
case of equal masses, $m_1 = m_2 = m_3$. This is an excellent approximation
for practical calculations, given that the relative mass difference of the
proton and neutron is smaller than one per mille. For 3N interactions the
matrix elements in Eq.~(\ref{eq:general_3Boperator}) can be significantly
simplified by using a representation in terms of Jacobi momenta and explicitly
factorizing out the trivial dependence on the center-of-mass motion:
\begin{equation}
\tensor*[_{\{ab\}}]{\bigl< \mathbf{p}' \mathbf{q}' \mathbf{P}_{\rm{3N}}' | V_{\text{3N}} | \mathbf{p} \mathbf{q} \mathbf{P}_{\rm{3N}} \bigr>}{_{\{ab\}}} 
= \tensor*[_{\{ab\}}]{\left< \mathbf{p}' \mathbf{q}' | V_{\text{3N}} | \mathbf{p} \mathbf{q} \right>}{_{\{ab\}}} \: \delta( \mathbf{P}_{\rm{3N}}' - \mathbf{P}_{\rm{3N}}) \, ,
\label{eq:P3N_factorization}
\end{equation}
where $\{ab\}$ represents one of the three basis choices $\{12\}$, $\{31\}$ or
$\{23\}$. This factorization drastically reduces the complexity of 3N
interaction matrix elements as the number of vector variables gets reduced
from 6 to 4. Instead of computing matrix elements $\tensor*[_{\{ab\}}]{\left<
\mathbf{p}' \mathbf{q}' | V_{\text{3N}} |
\mathbf{p} \mathbf{q} \right>}{_{\{ab\}}}$ directly in the vector representation,
for practical applications it is usually most convenient and efficient to
compute the matrix elements in a partial-wave representation\footnote{In
Section~\ref{sec:no_PWD} we discuss a novel approach that allows to evaluate
matrix elements in an efficient way in this vector representation of
Eq.~(\ref{eq:general_3Boperator}) for calculations of nuclear matter.}. For
such a representation the angular dependence of the Jacobi vectors is expanded
in terms of spherical harmonics. Specifically, we define (see also Appendix
~\ref{sec:normalization}):
\begin{equation}
\tensor*[_{\{ab\}}]{\left< \mathbf{p}' \mathbf{q}' | p q (L l) \mathcal{L} \mathcal{M}_{\mathcal{L}} \right>}{_{\{ab\}}} \equiv (2 \pi)^3 \frac{\delta(p-p')}{p p'} \frac{\delta(q-q')}{q q'} \mathcal{Y}_{L l}^{\mathcal{L} \mathcal{M}_{\mathcal{L}}} (\hat{\mathbf{p}}', \hat{\mathbf{q}}') \, 
\label{eq:PW_def_nospin}
\end{equation}
with
\begin{equation}
\mathcal{Y}_{l_a l_b}^{l m} (\hat{\mathbf{a}}, \hat{\mathbf{b}}) = \sum_{m_a, m_b} \mathcal{C}_{l_a m_a l_b m_b}^{l m} Y_{l_a m_a} (\hat{\mathbf{a}}) Y_{l_b m_b} (\hat{\mathbf{b}}) \, .
\end{equation}
Here all momenta are defined in a chosen representation $\{ab\}$, while $L$
denotes the relative angular momentum corresponding to momentum $\mathbf{p}'$
and $l$ denotes the angular momentum corresponding to momentum $\mathbf{q}'$.
From Eqs.~(\ref{eq:PW_def_norm}) and (\ref{eq:PW_def_nospin}) follows:
\begin{equation}
\bigl< p' q' (L' l') \mathcal{L}' \mathcal{M}'_{\mathcal{L}} | p q (L l) \mathcal{L} \mathcal{M}_{\mathcal{L}} \bigr> = \frac{\delta (p' - p)}{p p'} \frac{\delta (q' - q)}{q q'} \delta_{L L'} \delta_{l l'} \delta_{\mathcal{L} \mathcal{L}'} \delta_{\mathcal{M}_{\mathcal{L}} \mathcal{M}'_{\mathcal{L}}} \, .
\label{eq:nospin_state_norm}
\end{equation}
In Eqs.~(\ref{eq:PW_def_nospin}) and (\ref{eq:nospin_state_norm}) we
couple both angular momenta to the total angular momentum $\mathcal{L}$, which
is a conserved quantum number in the absence of spin degrees of freedom. Since
free-space 3N interactions are invariant under spatial rotations the resulting
matrix elements are proportional to $\delta_{\mathcal{L} \mathcal{L}'}
\delta_{\mathcal{M}_{\mathcal{L}} \mathcal{M}_{\mathcal{L}'}}$ and independent
of the quantum number $\mathcal{M}_{\mathcal{L}}$. 
Using Eq.~(\ref{eq:PW_def_nospin}) we obtain the following relation for the
partial-wave matrix elements of 3N interactions for spinless particles:
\begin{equation}
\tensor*[_{\{ab\}}]{\bigl< p' q' (L' l') \mathcal{L} | V_{\text{3N}} | p q (L l) \mathcal{L} \bigr>}{_{\{ab\}}} = \frac{1}{(2 \pi)^6} \frac{1}{2 \mathcal{L} + 1} \sum_{\mathcal{M}_{\mathcal{L}}} \int d \hat{\mathbf{p}} d \hat{\mathbf{q}} d \hat{\mathbf{p}}' d \hat{\mathbf{q}}' \mathcal{Y}_{L' l'}^{* \mathcal{L} \mathcal{M}_{\mathcal{L}}} (\hat{\mathbf{p}}', \hat{\mathbf{q}}') \tensor*[_{\{ab\}}]{\left< \mathbf{p}' \mathbf{q}' | V_{\text{3N}} | \mathbf{p} \mathbf{q} \right>}{_{\{ab\}}} \mathcal{Y}_{L l}^{\mathcal{L} \mathcal{M}_{\mathcal{L}}} (\hat{\mathbf{p}}, \hat{\mathbf{q}}) \, .
\label{eq:3NF_decmomp_spinless}
\end{equation}
It is convenient to parametrize the partial-wave decomposition of 3N
interactions in the form of the following function:
\begin{equation}
F_{L l L' l'}^{m_L m_l m_{L'} m_{l'}} (p, q, p', q') = \frac{1}{(2 \pi)^6} \int d\hat{\mathbf{p}}' d\hat{\mathbf{q}}' d\hat{\mathbf{p}} d \hat{\mathbf{q}} Y_{L' m_{L'}}^{*} (\hat{\mathbf{p}}') Y_{l' m_{l'}}^{*} (\hat{\mathbf{q}}') Y_{L m_L} (\hat{\mathbf{p}}) Y_{l m_l} (\hat{\mathbf{q}}) V_{\text{3N}} (\mathbf{p},\mathbf{q},\mathbf{p}',\mathbf{q}') 
\label{eq:F_func}
\end{equation}
for fixed values of $p=|\mathbf{p}|, q = |\mathbf{q}|, p' = |\mathbf{p}'|, q'
= |\mathbf{q}'|$ and the angular momentum quantum numbers. Apart from some
additional straightforward extensions this function will also be the key
quantity for the general case of spin- and isospin-dependent interactions (see
Section~\ref{sec:general_3N_decomp}). Equation~(\ref{eq:F_func}) shows that
for each value of momenta and orbital quantum numbers an 8-dimensional angular
integral needs to be computed. By using rotational symmetries of the
interaction, it is possible to reduce the dimensionality of the angular
integrals from 8 to 5. Traditional methods have been based on an explicit
discretization and numerical evaluation of these 5
integrals~\cite{Gola10newPWD,Skib11aPWD}. Due to the large number of external
quantum numbers and momentum mesh points such algorithms require a huge amount
of computational resources for calculating all matrix elements necessary for
many-body studies. As a result, the number of matrix elements of chiral
N$^3$LO interactions have been insufficient for studies of nuclei as well as
matter and were limited to three-body
systems~\cite{Skibi113HN3LO3N,Skib13n3lo}, while scattering calculations were
limited to low energies~\cite{Wita13n3loscat}. In the following sections we
will discuss an improved framework that allows to compute matrix elements much
more efficiently, which in turn makes it possible to reach basis sizes
sufficient for large-scale many-body calculations.

\subsection{Partial-wave decomposition of local 3N interactions}
\label{sec:PWD_3NF_local}

As shown in Section~\ref{sec:3NF_mom_rep}, a general translationally invariant
3N interaction depends on the four Jacobi momenta $\mathbf{p},
\mathbf{q}, \mathbf{p}'$ and $\mathbf{q'}$. However, in addition to the two
properties discussed in Section~\ref{sec:3NF_mom_rep}, most contributions to 3N
interactions have an additional property, \textit{locality}, which can be used
to further simplify the calculations.

\begin{figure}
\centering
\begin{minipage}[c]{0.4\textwidth}
\centering
\begin{tikzpicture} 
\begin{feynman}
\vertex (a) at (0,0) {}; 
\vertex (b) at (3,0) {};
\vertex [dot] (c) at (0.75,1) {}; 
\vertex [dot] (d) at (2.25,1) {};
\vertex (e) at (0,2) {}; 
\vertex (f) at (3,2) {};
\vertex (g) at (0,1) {\(\mathbf{x}_1\)=\(\mathbf{x}'_1\)};
\vertex (h) at (3.0,1) {\(\mathbf{x}_2\)=\(\mathbf{x}'_2\)};
\vertex (h) at (1.5,0.75) {\(\mathbf{r} = \mathbf{r}'\)};
\diagram* {
(a) -- [fermion, line width=0.25mm] (c) [vertex] -- [fermion, line width=0.25mm] (e);
(b) -- [fermion, line width=0.25mm] (d) [dot] -- [fermion, line width=0.25mm] (f);
(c) -- [dashed, very thick] (d) -- [dashed, very thick];
};
\end{feynman}
\end{tikzpicture}
\end{minipage}
\hspace{0.5cm}
\begin{minipage}[c]{0.4\textwidth}
\centering
\begin{tikzpicture} 
\begin{feynman}
\vertex (a) at (0,0) {}; 
\vertex (b) at (2,0) {};
\vertex [blob, /tikz/minimum size=20pt, shape=rectangle, fill=black!20, line width=0.2mm] (c) at (1,1) {}; 
\vertex (d) at (0,2) {}; 
\vertex (e) at (2,2) {};

\vertex (i) at (0.35,0.7) {\(\mathbf{x}_1\)};
\vertex (j) at (1.7,0.7) {\(\mathbf{x}_2\)};
\vertex (k) at (0.35,1.3) {\(\mathbf{x}'_1\)};
\vertex (l) at (1.7,1.3) {\(\mathbf{x}'_2\)};
\diagram* {
(a) -- [fermion, line width=0.25mm] (c) [dot] -- [fermion, line width=0.25mm] (d);
(b) -- [fermion, line width=0.25mm] (c) [dot] -- [fermion, line width=0.25mm] (e);
};
\end{feynman}
\end{tikzpicture}
\end{minipage}
\caption{Examples of a local (left) and a nonlocal (right) contribution to an
NN interaction in coordinate representation. The dashed line in the left panel
indicates an instantaneously exchanged pion, i.e., both single-particle
coordinates are the same before and after the interaction and the interaction
is therefore local. In the right panel we show a contact interaction that is
regularized via a regulator of the form $f_{\Lambda} (\mathbf{p}^2)
f_{\Lambda} (\mathbf{p}'^2)$ (see Eq.~(\ref{eq:VNN_nonlocal}) and also
Section~\ref{sec:3N_regularization}). This regularization induces a
nonlocality, i.e., $\mathbf{x}_i \neq \mathbf{x}'_i$.}
\label{fig:VNN_local_nonlocal}
\end{figure}
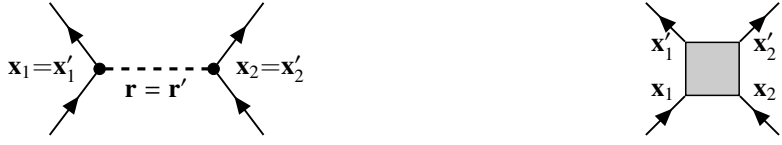

The concept of locality is illustrated most naturally in coordinate space. For
this consider first a free-space two-body nucleon-nucleon interaction.
Following the arguments of the previous sections it is easy to show that a
Galilean-invariant NN interaction can be written in the following form
\begin{equation}
\left< \mathbf{k}'_1 \mathbf{k}'_2 | V_{\rm{NN}} | \mathbf{k}_1 \mathbf{k}_2 \right> = \left< \mathbf{p}' | V_{\rm{NN}} | \mathbf{p} \right> \delta (\mathbf{P}'_{\{12\}} - \mathbf{P}_{\{12\}}) \, ,
\label{eq:VNN_Galilean_inv}
\end{equation}
with $\mathbf{p} = \mathbf{p}_{\{12\}}$ and $\mathbf{p}' =
\mathbf{p}'_{\{12\}}$. Let us analyze such interactions in coordinate space.
To this end, we consider two particles located at the coordinates $\mathbf{x}_1$
and $\mathbf{x}_2$ when they interact via the exchange of a meson. Here it is
important to note that we work in a non-relativistic limit and neglect
any retardation effects, which means that the interaction is instantaneous. In
this case the center-of-mass coordinate of the two particles before and after
the interaction process is identical, $\mathbf{R}_{\{12\}} =
\mathbf{R}'_{\{12\}}$, and the interaction can only depend on the relative
distance $\mathbf{r} \equiv \mathbf{r}_{\{12\}} =
\mathbf{r}'_{\{12\}} \equiv \mathbf{r}'$ (see also Figure~\ref{fig:VNN_local_nonlocal}):
\begin{align}
\bigl< \mathbf{x}'_1 \mathbf{x}'_2 | V_{\rm NN}^{\text{local}} \bigr| \mathbf{x}_1 \mathbf{x}_2 \bigr> &= \left< \mathbf{r}' | V_{\text{NN}} | \mathbf{r} \right> \,
\delta (\mathbf{r}' - \mathbf{r}) \, \delta (\mathbf{R}'_{\{12\}} - \mathbf{R}_{\{12\}}) = V_{\text{NN}}^{\text{local}} (\mathbf{r}) \, \delta (\mathbf{r}' - \mathbf{r}) \, \delta (\mathbf{R}'_{\{12\}} - \mathbf{R}_{\{12\}}) \, .
\end{align}
This is the definition of a \textit{local} NN interaction\footnote{We note that the
antisymmetrization of the interaction leads to an exchange term with
interchanged coordinates $\mathbf{x}'_1 \Leftrightarrow \mathbf{x}'_2$ in the final state,
i.e., $\mathbf{r}_{\{12\}} = -\mathbf{r}'_{\{12\}}$:
\begin{equation}
\bigl< \mathbf{x}'_1 \mathbf{x}'_2 | V_{\rm{NN}}^{\text{ex}} | \mathbf{x}_1 \mathbf{x}_2 \bigr> = V_{\text{NN}} (\mathbf{r}) \, \delta
(\mathbf{r}' + \mathbf{r}) \, \delta (\mathbf{R}'_{\{12\}} - \mathbf{R}'_{\{12\}}) \, .
\end{equation}}.
Fourier transform to momentum space of such an interaction leads to:
\begin{align}
\bigl< \mathbf{k}'_1 \mathbf{k}'_2 | V_{\rm{NN}}^{\text{local}} | \mathbf{k}_1 \mathbf{k}_2 \bigr> &= \int d \mathbf{x}_1 d \mathbf{x}_2 d \mathbf{x}'_1 d \mathbf{x}'_2 e^{- i (\mathbf{k}'_1 \cdot \mathbf{x}'_1 + \mathbf{k}'_2 \cdot \mathbf{x}'_2 - \mathbf{k}_1 \cdot \mathbf{x}_1 - \mathbf{k}_2 \cdot \mathbf{x}_2)} \bigl< \mathbf{x}'_1 \mathbf{x}'_2 | V_{\rm{NN}}^{\text{local}} | \mathbf{x}_1 \mathbf{x}_2 \bigr> \nonumber \\
&= \int d \mathbf{x}_1 d \mathbf{x}_2 e^{- i (\mathbf{k}'_1 - \mathbf{k}_1) \cdot \mathbf{x}_1 - i (\mathbf{k}'_2 - \mathbf{k}_2) \cdot \mathbf{x}_2} V_{\rm{NN}}^{\text{local}} (\mathbf{r}) \nonumber \\
&= \int d \mathbf{r} d \mathbf{R}_{\{12\}} e^{- i (\mathbf{p}' - \mathbf{p}) \cdot \mathbf{r} - i (\mathbf{P}'_{\{12\}} - \mathbf{P}_{\{12\}}) \cdot \mathbf{R}_{\{12\}}} V_{\rm{NN}}^{\text{local}} (\mathbf{r}) \, .
\end{align}
This implies that the interaction depends in momentum space only on
\textit{differences of Jacobi momenta}, i.e., on momentum transfers (see
Eq.~(\ref{eq:VNN_Galilean_inv})):
\begin{equation}
\bigl< \mathbf{p}' | V_{\rm{NN}}^{\text{local}} | \mathbf{p} \bigr> = \int d
\mathbf{r} e^{-i (\mathbf{p}' - \mathbf{p}) \cdot \mathbf{r}}
V_{\text{NN}}^{\text{local}} (\mathbf{r}) = V_{\rm{NN}}^{\text{local}}
(\mathbf{p}' - \mathbf{p}) \, .
\label{eq:V3N_Jacobi_difference}
\end{equation}
Examples of local interactions
are the instantaneous meson exchanges and (unregularized) contact interactions
(see left panel of Figure~\ref{fig:VNN_local_nonlocal}).

Let us consider in comparison a local interaction that is regularized via a
cutoff function $f_{\Lambda}$ in the following way:
\begin{equation}
\bigl< \mathbf{p}' | V_{\rm{NN}}^{\text{reg}} | \mathbf{p} \bigr> = f_{\Lambda} (\mathbf{p}') V_{\rm{NN}}^{\text{local}} (\mathbf{p}' - \mathbf{p}) f_{\Lambda} (\mathbf{p}) \, .
\label{eq:VNN_nonlocal}
\end{equation}
This regularization has been extensively used for chiral EFT NN interactions
(see Section~\ref{sec:nonlocal_momentum}). The resulting regularized
interaction obviously does not only depend on the momentum difference,
$\mathbf{p}' - \mathbf{p}$, but rather on the individual momenta $\mathbf{p}$
and $\mathbf{p}'$, which implies in coordinate space $\mathbf{x}_i \neq
\mathbf{x}_i'$. Such a interaction is \textit{nonlocal} (see right panel of
Figure~\ref{fig:VNN_local_nonlocal}).

The arguments above can be extended straightforwardly to 3N interactions.
Specifically, this implies that local 3N forces only depend on the difference
of the two Jacobi momenta:
\begin{equation}
V_{\rm{3N}}^{\text{local}} = V_{\rm{3N}}^{\text{local}} (\mathbf{p}'-\mathbf{p}, \mathbf{q}' - \mathbf{q}) \equiv V_{\rm{3N}}^{\text{local}} (\tilde{\mathbf{p}}, \tilde{\mathbf{q}}) \, .
\label{eq:V3N_local_form}
\end{equation}
Using the rotational symmetry of the potential $V_{\rm{3N}}^{{\rm
local}}$ we can write the 3N interaction as a function of only three independent variables:
\begin{equation}
V_{\rm{3N}}^{\text{local}} (\tilde{\mathbf{p}}, \tilde{\mathbf{q}}) = V_{\rm{3N}}^{\text{local}} (\tilde{p}, \tilde{q}, \cos \theta_{\tilde{\mathbf{p}} \tilde{\mathbf{q}}}) \, ,
\label{eq:V3N_local_form_rotational_symmetry}
\end{equation}
where $\cos \theta_{\tilde{\mathbf{p}}
\tilde{\mathbf{q}}}=\frac{\tilde{\mathbf{p}}\cdot
\tilde{\mathbf{q}}}{\tilde{p} \tilde{q}}, \tilde{p} = |\tilde{\mathbf{p}}|,
\tilde{q} = |\tilde{\mathbf{q}}|$. Note that this statement refers to
unregularized forces. The regularization is discussed in detail in
Section~\ref{sec:3N_regularization}.

The relation~(\ref{eq:V3N_local_form_rotational_symmetry}) shows that the
original eight-dimensional integral in Eq.~(\ref{eq:F_func}) actually contains
only three non-trivial integrations for local interactions, while the other
five integrations are purely geometric. In fact, after employing some integral
transformations they can all be performed analytically. The details of this
calculation are presented in Appendix~\ref{sec:eval_PWD_local}. The final
result for the function $F$ defined in Eq.~(\ref{eq:F_func}) can be written in
the following form:
\begin{align}
& F_{L l L' l'}^{m_L m_l m_{L'} m_{l'}} (p, q, p', q') \nonumber \\
&=  \delta_{m_L - m_{L'}, m_{l'} - m_l} \frac{(-1)^{m_L + m_{l'}}}{(2 \pi)^6} \frac{2 (2 \pi)^4}{p p' q q'} \sum_{\bar{l} = \text{max} (|L'-L|,|l' - l|)}^{\text{min} (L' + L, l' + l)} 
\frac{\mathcal{C}_{L' - m_{L'} L m_L}^{\bar{l} -m_{L'} + m_L} \mathcal{C}_{l' -m_{l'} l m_l}^{\bar{l} -m_{l'} + m_l}}{2 \bar{l} + 1} \nonumber \\
&\times \int_{|p'-p|}^{p'+p} d \tilde{p} \, \tilde{p} \int_{|q'-q|}^{q'+q} d \tilde{q} \, \tilde{q} \left. \mathcal{Y}_{L' L}^{\bar{l} 0} (\widehat{\tilde{p} \mathbf{e}_z + \mathbf{p}},\hat{\mathbf{p}}) \right|_{\phi_p = 0, \hat{p} \cdot \mathbf{e}_z = \frac{p'^2 - p^2 - \tilde{p}^2}{2 \tilde{p} p}} \left. \mathcal{Y}_{l' l}^{\bar{l} 0} (\widehat{\tilde{q} \mathbf{e}_z + \mathbf{q}}, \hat{\mathbf{q}}) \right|_{\phi_q = 0, \hat{q} \cdot \mathbf{e}_z = \frac{q'^2 - q^2 - \tilde{q}^2}{2 \tilde{q} q}} \nonumber \\
&\times \int_{-1}^1 d \cos \theta_{\tilde{\mathbf{p}} \tilde{\mathbf{q}}} P_{\bar{l}} (\cos \theta_{\tilde{\mathbf{p}} \tilde{\mathbf{q}}}) V_{\rm{3N}}^{\text{local}} (\tilde{q},\tilde{p},\cos \theta_{\tilde{\mathbf{p}} \tilde{\mathbf{q}}}) \, .
\label{final_result_pwd}
\end{align}
This reduction from a five-dimensional numerical integral to a
three-dimensional one represents a dramatic difference for practical
calculations. In the next chapter we discuss in more detail the corresponding
speedup factors for the calculation of matrix elements and the connected
increase in accessible basis sizes. In essence, this new framework allows to
compute matrix elements of 3N interactions at N$^2$LO about 1000 more
efficiently than the traditional approach for typical basis sizes and discrete
momentum mesh systems.

\subsection{Generalization to spin- and isospin-dependent 3N interactions}
\label{sec:general_3N_decomp}
Realistic nuclear forces also depend on the spin and isospin quantum numbers
of the nucleons. In this section we generalize the arguments of the previous
sections to a general spin- and isospin-dependent local 3N interaction. For
this we choose the standard $Jj$-coupled three-body partial-wave basis of the
form~\cite{Gloe83QMFewbod}:
\begin{align}
\bigl| p q \alpha \bigr>_{\{ab\}} \hspace{-1.6mm} \phantom \rangle &\equiv \bigl| p q; \left[ (L S) J (l s) j \right] \mathcal{J} \mathcal{M}_{\mathcal{J}} (T t) \mathcal{T} \mathcal{M}_{\mathcal{T}} \bigr>_{\{ab\}} \nonumber \\
&= \sum_{M_L, M_S, M_J, m_l, m_s, m_j} \mathcal{C}_{J M_J j m_j}^{\mathcal{J} \mathcal{M}_{\mathcal{J}}} \mathcal{C}_{L M_L S M_S}^{J M_J} \mathcal{C}_{l m_l s m_s}^{j m_j} \sum_{M_T, m_t} \mathcal{C}_{T M_T t m_t}^{\mathcal{T} \mathcal{M}_{\mathcal{T}}} \left| p q; L M_L S M_S l m_l s m_s \right>_{\{ab\}} \left| T M_T t m_t \right>_{\{ab\}} \, ,
\label{eq:Jj_bas}
\end{align}
where $L$, $S$, $J$ and $T$ denote the relative orbital angular momentum,
two-body spin, total angular momentum and total isospin of particles $a$ and
$b$ with Jacobi momentum $p$. The quantum numbers $l$, $s=\tfrac{1}{2}$, $j$ and
$t=\tfrac{1}{2}$ label the orbital angular momentum, spin, total angular momentum and
isospin of the remaining particle relative to the center-of-mass of the pair
with relative momentum $p$ (see Section~\ref{sec:3NF_coord_def}). The quantum
numbers $\mathcal{J}$ and $\mathcal{T}$ define the total three-body angular
momentum and isospin. The orbital angular momentum partial-wave states are
normalized like in Eq.~(\ref{eq:PW_def_nospin}). Hence, we immediately obtain
the following normalization for the partial-wave states (see
Appendix~\ref{sec:normalization}):
\begin{align}
\tensor*[_{\{ab\}}]{\left< p' q' \alpha' | p q \alpha \right>}{_{\{ab\}}} = \frac{\delta (p' - p)}{p p'} \frac{\delta (q' - q)}{q q'} \delta_{\alpha \alpha'}, \quad \int dp p^2 \int dq q^2 \sum_{\alpha} \left| p q \alpha \right>_{\{ab\}} \tensor*[_{\{ab\}}]{\left< p q \alpha \right|}{} = 1 \, .
\label{eq:Jj_bas_normalization}
\end{align}
Note that, even though the basis states in Eq.~(\ref{eq:Jj_bas}) contain the
quantum numbers $\mathcal{M}_{\mathcal{J}}$ and $\mathcal{M}_{\mathcal{T}}$,
the matrix elements of 3N interactions do not depend on them. Hence we omit
these quantum numbers in the basis states in the following. That means the
collective index $\alpha$ in Eq.~(\ref{eq:Jj_bas}) defines a set of six
quantum numbers $\alpha = \left\{ L,S,J,l,j,T \right\}$ for a given three-body
partial wave specified by $\left\{ \mathcal{J},\mathcal{T},\mathcal{P}
\right\}$, where $\mathcal{P}$ is the three-body parity $\mathcal{P} = (-1)^{L
+ l}$. The basis contains only states that are antisymmetric in subsystem
$\{ab\}$, i.e., we require $(-1)^{L+S+T} =  -1$. For illustration we list in
Appendix~\ref{sec:3N_config_table} all configurations for the three-body
channel with $\mathcal{J} = \tfrac{1}{2}$, $\mathcal{T} = \tfrac{1}{2}$ and $\mathcal{P} = +1$
for $J \le 7$.

In Table~\ref{tab:PW_data} we list typical basis sizes for the different
three-body channels $\left\{ \mathcal{J},\mathcal{T},\mathcal{P} \right\}$ for
practical calculations. Here we define the truncation by choosing a maximal
value of the total two-body angular momentum $J$. Note that, as we will
illustrate in Section~\ref{sec:PW_conv_matter}, usually it is sufficient to
include all partial waves up to the total three-body angular momentum
$\mathcal{J} =\tfrac{9}{2}$ and $J_{\text{max}} = 5$. Typically, matrix elements beyond
this truncation have only small effects on observables for finite nuclei and
nuclear matter (see Section~\ref{sec:PW_conv_matter}). However, the size of
these contributions depends in general on the employed NN interaction and the
regularization scheme (see Section~\ref{sec:3N_regularization}).

\begin{table}[t!]
\centering
\renewcommand{\arraystretch}{1.2}
\begin{tabular}{cc||cc|cc|cc|cc}
              &               & \multicolumn{2}{c}{$J_{\rm{max}}=5$} & \multicolumn{2}{c}{$J_{\rm{max}}=6$} & \multicolumn{2}{c}{$J_{\rm{max}}=7$} & \multicolumn{2}{c}{$J_{\rm{max}}=8$} \\
\hline \hline
$\mathcal{J}$ & $\mathcal{T}$ & $N_{\alpha}$ & $\text{dim}(V_{\rm{3N}}$) & $N_{\alpha}$ & $\text{dim}(V_{\rm{3N}}$) & $N_{\alpha}$ & $\text{dim}(V_{\rm{3N}}$) & $N_{\alpha}$ & $\text{dim}(V_{\rm{3N}}$) \\
\hline
$\tfrac{1}{2}$ & $\tfrac{1}{2}$ & 42 & $7 \times 10^8$ & 50 & $1 \times 10^9$ & 58 & $1 \times 10^9$ & 66 & $2 \times 10^9$ \\
$\tfrac{3}{2}$ & $\tfrac{1}{2}$ & 78 & $2 \times 10^9$ & 94 & $3 \times 10^9$ & 110 & $5 \times 10^9$ & 126 & $6 \times 10^9$ \\
$\tfrac{5}{2}$ & $\tfrac{1}{2}$ & 106 & $4 \times 10^9$ & 130 & $7 \times 10^9$ & 154 & $9 \times 10^9$ & 178 & $1 \times 10^{10}$ \\
$\tfrac{7}{2}$ & $\tfrac{1}{2}$ & 126 & $6 \times 10^9$ & 158 & $1 \times  10^{10}$ & 190 & $1 \times 10^{10}$ & 222 & $2 \times 10^{10}$ \\
$\tfrac{9}{2}$ & $\tfrac{1}{2}$ & 138 & $7 \times 10^9$ & 178 & $1 \times  10^{10}$ & 218 & $2 \times 10^{10}$ & 258 & $3 \times 10^{10}$ \\
$\tfrac{11}{2}$ & $\tfrac{1}{2}$ & 142 & $8 \times 10^9$ & 190 & $1 \times  10^{10}$ & 238 & $2 \times 10^{10}$ & 286 & $3 \times 10^{10}$ \\
\hline
$\tfrac{1}{2}$ & $\tfrac{3}{2}$ & 20 & $2 \times 10^8$ & 26 & $3 \times 10^8$ & 28 & $3 \times 10^8$ & 34 & $5 \times 10^8$ \\
$\tfrac{3}{2}$ & $\tfrac{3}{2}$ & 37 & $5 \times 10^8$ & 49 & $9 \times 10^8$ & 53 & $1 \times 10^9$ & 65 & $2 \times 10^9$\\
$\tfrac{5}{2}$ & $\tfrac{3}{2}$ & 50 & $1 \times 10^9$ & 68 & $2 \times 10^9$ & 74 & $2 \times 10^9$ & 92 & $3 \times 10^9$ \\
$\tfrac{7}{2}$ & $\tfrac{3}{2}$ & 59 & $1 \times 10^9$ & 83 & $3 \times 10^9$ & 91 & $3 \times 10^9$ & 115 & $5 \times 10^9$ \\
$\tfrac{9}{2}$ & $\tfrac{3}{2}$ & 64 & $2 \times 10^9$ & 94 & $3 \times 10^9$ & 104 & $4 \times 10^9$ & 134 & $7 \times 10^9$ \\
$\tfrac{11}{2}$ & $\tfrac{3}{2}$ & 65 & $2 \times 10^9$ & 101 & $4 \times 10^9$ & 113 & $5 \times 10^9$ & 149 & $9 \times 10^9$ \\
\end{tabular}
\renewcommand{\arraystretch}{1.0}
\caption{Basis sizes and dimensions of 3N interaction matrix elements for the
different three-body partial waves as a function of $J_{\rm{max}}$, the total
angular momentum corresponding to the Jacobi momentum $p$. For the dimension
estimates we have used a typical value for the number of mesh points for the
two Jacobi momenta $p$ and $q$, $N_p = N_q = 25$. $N_{\alpha}$ denotes the
number of partial-wave channels in the $Jj$-coupled partial-wave basis defined
in Eq.~(\ref{eq:Jj_bas}) for a given three-body channel $\left\{ \mathcal{J},
\mathcal{T} \right\}$. All given values apply to both three-body parities
$\mathcal{P} = (-1)^{L + l}$. Table~\ref{tab:3N_configurations} shows
explicitly the basis states for $\mathcal{J} = \mathcal{T} = 1/2$ and
$\mathcal{P} = + 1$. }

\label{tab:PW_data}
\end{table}

Due to the explicit momentum dependence of spin-momentum operators, the
relation for the fundamental function $F$ in Eq.~(\ref{eq:F_func}) needs to be
generalized. This can be achieved in a straightforward way by factorizing out
the momentum dependence of the spin operators. For illustration let us consider
a simple operator of the form $\boldsymbol{\sigma} \cdot
\mathbf{a}$, where $\mathbf{a}$ represents one of the Jacobi momenta. First,
we rewrite this scalar product in a spherical representation:
\begin{equation}
\boldsymbol{\sigma} \cdot \mathbf{a} = \sqrt{\frac{4 \pi}{3}} a \sum_{\mu=-1}^1 Y^*_{1 \mu} (\hat{\mathbf{a}}) \, \boldsymbol{\sigma} \cdot \mathbf{e}_{\mu} \, . \label{eq:spin_factorize}
\end{equation}
This factorization of the momentum dependence allows to combine the additional
spherical harmonic function with those in Eq.~(\ref{eq:F_func}) by
using~\cite{Vars88Gulag}:
\begin{align}
Y_{l m} (\hat{\mathbf{a}}) Y_{1 \mu} (\hat{\mathbf{a}}) = \sum_{\bar{L} =
|l-1|}^{l+1} \sqrt{\frac{3}{4 \pi} \frac{2 l + 1}{2 \bar{L}+1}} \mathcal{C}_{l
0 1 0}^{\bar{L} 0} \mathcal{C}_{l m 1 \mu}^{\bar{L} m + \mu} Y_{\bar{L} m +
\mu} (\hat{\mathbf{a}}) \, . \label{eq:Y_identity}
\end{align}
This strategy is completely general and can be used to reduce the expressions
for arbitrary spin-dependent interactions to the expression for
spin-independent interactions times some momentum-independent spin operators.
This step has to be performed for each momentum vector in the spin-momentum
operators. Obviously, the efficiency of the algorithm decreases with
each additional sum over the quantum numbers $\mu$ and $\bar{L}$ in
Eqs.~(\ref{eq:spin_factorize}) and (\ref{eq:Y_identity}). Note, however, that
each of these sums contains only three terms at most.

In order to factorize the momentum, spin and isospin space, it is most
convenient to perform the calculations of the matrix elements in an
$LS$-coupled basis:
\begin{equation}
\left| p q \beta \right>_{\{ab\}} \hspace{-1.6mm} \phantom \rangle \equiv \left| p q; \left[ (L l) \mathcal{L} (S s) \mathcal{S} \right] \mathcal{J} (T t)
\mathcal{T} \right>_{\{ab\}} \, , \label{eq:LS_bas}
\end{equation}
and recouple only at the end to the $Jj$-coupled basis defined in
Eq.~(\ref{eq:Jj_bas}). Here the quantum number $\mathcal{L}$ denotes the total
orbital angular momentum as in Eq.~(\ref{eq:PW_def_nospin}) and $\mathcal{S}$
is the total three-body spin. Each time the factorization in
Eq.~(\ref{eq:spin_factorize}) is applied, the spin matrix element acquires a
dependence on the quantum number $\mu$. Consequently, the matrix element in
spin space can be formally written in the form
\begin{equation}
\tensor*[_{\{ab\}}]{\bigl< (S s) \mathcal{S} \mathcal{M}_{\mathcal{S}} | \hat{O}_{\sigma} ( \{ \mu_i \} ) | (S' s') \mathcal{S}' \mathcal{M}_{\mathcal{S}'} \bigr>}{_{\{ab\}}} \, , 
\label{eq:spin_operator} 
\end{equation}
where the index $i$ counts the number of momentum vectors in the spin operator.
In the same way the function $F$ in Eq.~(\ref{eq:F_func}) becomes a function of
the  quantum numbers $\mu_i$, i.e., it takes the form $F_{L l L' l'}^{m_L m_l
m_{L'} m_{l'} \{ \mu_i \}}$. To be explicit, if we consider, the case
$\mathbf{a} = \mathbf{p}$ in Eq.~(\ref{eq:spin_factorize}), the function $F$
takes the form
\begin{equation}
F_{L l L' l'}^{m_L m_l m_{L'} m_{l'} \mu} (p, q, p', q') = p \sum_{\bar{L} = |L-1|}^{L+1} \sqrt{\frac{2 L + 1}{2 \bar{L}+1}} \mathcal{C}_{L 0 1 0}^{\bar{L} 0} \mathcal{C}_{L m_L 1 \mu}^{\bar{L} m_L + \mu} F_{\bar{L} l L' l'}^{m_L m_l m_{L'} m_{l'}} (p, q, p', q') \, , \label{eq:F_withonemu}
\end{equation} 
where we included the factor $\sqrt{\frac{4 \pi}{3}}p$ from the spin operator
factorization in Eq.~(\ref{eq:spin_factorize}) in this function. 

For an efficient implementation it is important to note that all quantities
that depend on the projection quantum numbers $m$ and $\mu$ are independent of
the momenta. Hence, it is useful to factorize this dependence in the function
$F$. Specifically, for the example shown in Eq.~(\ref{eq:F_withonemu}) we can
write
\begin{equation}
F_{L l L' l'}^{m_L m_l m_{L'} m_{l'} \mu} (p, q, p', q') \equiv \delta_{m_L - m_{L'}, m_{l'} - m_l} (-1)^{m_L + m_{l'}} \sum_{\bar{l}} \mathcal{C}_{L' -m_{L'} L m_L}^{\bar{l} -m_{L'} + m_L} \mathcal{C}_{l' -m_{l'} l m_l}^{\bar{l} -m_{l'} + m_l} \sum_{ \bar{L}} \mathcal{C}_{L 0 1 0}^{\bar{L} 0} \mathcal{C}_{L m_L 1 \mu}^{\bar{L} m_L + \mu} \tilde{F}_{L l L' l'}^{\bar{l} \bar{L} } (p,q,p',q') \, . 
\label{eq:F_func_factorize}
\end{equation}
For general interactions, the function $\tilde{F}$ depends on multiple quantum
numbers $\bar{L}_i$, hence the function takes formally the  form $\tilde{F}_{L l
L' l'}^{\bar{l} \{ \bar{L}_i \} } (p,q,p',q')$. Using this decomposition we can
first precalculate all sums over the projection  quantum numbers $m$ and $\mu_i$
and prestore the result in a function of the form $A_{\beta \beta'}^{\bar{l} \{
\bar{L}_i \}}$, where $\beta$ labels the $LS$-coupled partial-wave channels
(see Eq.~(\ref{eq:LS_bas})). Then the final matrix element in $LS$-coupling can
be calculated very efficiently via:
\begin{equation}
\tensor*[_{\{ab\}}]{\left< p' q' \beta' | V_{\rm{3N}} | p q \beta \right>}{_{\{ab\}}} = \sum_{\bar{l}} \sum_{\{ \bar{L}_i \} } A_{\beta \beta'}^{\bar{l} \{ \bar{L}_i \}} \tilde{F}_{L l L' l'}^{\bar{l} \{ \bar{L}_i \} } (p,q,p',q') \, ,
\label{eq:VLS}
\end{equation}
where values of the quantum numbers $L$, $L'$, $l$ and $l'$ are specified by
the $LS$-coupling partial-wave indices $\beta$ and $\beta'$. Similar to the
$Jj$-coupled basis, here the collective index $\beta$ defines a set of six
quantum numbers $\beta=\{ L,l,\mathcal{L},S,\mathcal{S},T\}$ for a given
three-body partial wave $\{\mathcal{J},\mathcal{T},\mathcal{P}\}.$ Finally,
the recoupling to the $Jj$-basis is achieved by applying the standard recoupling
relation~\cite{Gloe82spline,Vars88Gulag}
\begin{equation}
\left| p q \alpha \right>{_{\{ab\}}} = \sum_{\mathcal{L},\mathcal{S}} \sqrt{\hat{\mathcal{L}} \hat{\mathcal{S}} \hat{J} \hat{j}} 
\left\{
\begin{array}{ccc}
L & S & J \\
l & \tfrac{1}{2} & j \\
\mathcal{L} & \mathcal{S} & \mathcal{J}
\end{array}
\right\}
\left| p q \beta \right>{_{\{ab\}}} \, ,
\label{eq:Jj_recoupling}
\end{equation}
with $\hat{a} = 2 a + 1$ and the $9j$-symbol $\{ ... \}$. Note that by
deriving Eq.~(\ref{eq:VLS}) the original problem of numerically calculating a
5-dimensional integral for each matrix element as in Eq.~(\ref{eq:F_func}) has
been reduced to the evaluation of a few discrete sums. The calculation and
prestorage  of the matrix elements of $\tilde{F}_{L l L' l'}^{\bar{l} \{
\bar{L}_i \} } (p,q,p',q')$ can be performed relatively efficiently since only
three internal integrals have to be performed numerically. The exact speedup
factor of the present method compared to the conventional approach of
Refs.~\cite{Gola10newPWD,Skib11aPWD} depends  on the number of internal sums
over $\mu_i$ and $\bar{L}_i$, i.e., on the specific form of the interaction.
For example, the matrix elements of the chiral long-range interactions at
N$^2$LO proportional to the couplings $c_1$ and $c_3$ can be calculated with
speedup factors of greater than 1000. In practical terms, that means that it
is possible to calculate the matrix elements of all interaction terms up to
N$^3$LO on a local computer cluster for sufficiently large basis sizes needed
for converged studies of few-nucleon scattering processes, light and medium
mass nuclei and nuclear matter.

Despite the fact that the present algorithm makes explicit use of the local
nature of the 3N interaction, it is also possible to treat polynomial
nonlocal terms. This is of immediate practical importance since, e.g., the
relativistic corrections to 3N interactions at N$^3$LO have precisely this
form~\cite{Bern083Nlong,Bern113Nshort}. Consider, for example, a nonlocal
momentum structure in the center-of-mass frame of the type
\begin{equation}
(\mathbf{k}_3 + \mathbf{k}'_3) \cdot (\mathbf{k}_3 - \mathbf{k}'_3) = (\mathbf{q}'_{\{12\}} + \mathbf{q}_{\{12\}}) \cdot (\mathbf{q}'_{\{12\}} - \mathbf{q}_{\{12\}}) \, .
\end{equation}
Such terms can be treated by factorizing the momentum dependence like 
in Eq.~(\ref{eq:spin_factorize}), for example:
\begin{equation}
\mathbf{q} \cdot \mathbf{q}' = q q' \frac{4 \pi}{3} \sum_{\mu_1,\mu_2 = -1}^1  Y^*_{1 \mu_1} (\hat{\mathbf{\mathbf{q}}}) Y^*_{1 \mu_2} (\hat{\mathbf{\mathbf{q}}}') \, \mathbf{e}_{\mu_1} \cdot \mathbf{e}_{\mu_2} \, ,
\end{equation}
and then following exactly the steps after Eq.~(\ref{eq:spin_factorize}).
Obviously, the algorithm becomes less efficient for nonlocal interactions, but
this framework turns out to be still more efficient than the
conventional approach for the relativistic corrections to chiral 3N interactions at N$^3$LO.

\subsection{Decomposition of 3N interactions and antisymmetrization}
\label{sec:3N_decomp_antisymmetrization}
A local contribution to 3N interactions can generically be written in the form
\begin{equation}
V_{\text{3N}} = \sum_{i \neq j \neq k} f_Q (\mathbf{Q}_i,\mathbf{Q}_j,\mathbf{Q}_k) f_{\sigma} (\mathbf{Q}_i,\mathbf{Q}_j,\mathbf{Q}_k,\boldsymbol{\sigma}_i,\boldsymbol{\sigma}_j,\boldsymbol{\sigma}_k) f_{\tau} (\boldsymbol{\tau}_i,\boldsymbol{\tau}_j,\boldsymbol{\tau}_k) \, ,
\label{eq:generic_form_3NF}
\end{equation}
with $i,j,k \in \left\{ 1,2,3 \right\}$, the momentum transfers $\mathbf{Q}_i =
\mathbf{k}'_i - \mathbf{k}_i$ and spin (isospin) operators
$\boldsymbol{\sigma}_i$ ($\boldsymbol{\tau}_i$) of particle $i$. The function
$f_{Q}$ includes all scalar momentum dependence, $f_{\sigma}$ denotes the
spin-momentum operators and $f_{\tau}$ the isospin operators. Since the latter
one is momentum independent it can be treated straightforwardly in the
partial-wave decomposition. Here and in the following we focus only on the
dynamical degrees of freedom of the particles and suppress all physical
constants like, e.g., $g_A$ or $m_{\pi}$.

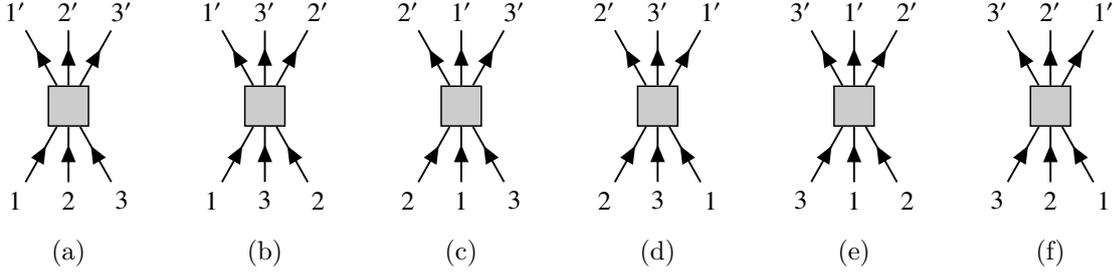
\begin{figure}
\centering
\begin{minipage}[c]{0.15\textwidth}
\begin{tikzpicture} 
\tikzfeynmanset{
  my dot/.style={
    /tikzfeynman/dot,
    /tikz/minimum size=10pt,
  },
  every vertex/.style={my dot},
}
\begin{feynman}
\vertex (a) at (0,0) {\(1\)}; 
\vertex (b) at (0.7,0) {\(2\)};
\vertex (c) at (1.4,0) {\(3\)};
\vertex [blob, /tikz/minimum size=15pt, shape=rectangle, fill=black!20, line width=0.2mm] (d) at (0.7,1.25) {}; 
\vertex (e) at (0,2.5) {\(1'\)}; 
\vertex (f) at (0.70,2.5) {\(2'\)};
\vertex (g) at (1.4,2.5) {\(3'\)};
\vertex (h) at (0.7,-0.7) {(a)};
\diagram* {
(a) -- [fermion, line width=0.25mm] (d) -- [fermion, line width=0.25mm] (e);
(b) -- [fermion, line width=0.25mm] (d) -- [fermion, line width=0.25mm] (f);
(c) -- [fermion, line width=0.25mm] (d) -- [fermion, line width=0.25mm] (g);
};
\end{feynman}
\end{tikzpicture}
\end{minipage}
\begin{minipage}[c]{0.15\textwidth}
\begin{tikzpicture} 
\tikzfeynmanset{
  my dot/.style={
    /tikzfeynman/dot,
    /tikz/minimum size=10pt,
  },
  every vertex/.style={my dot},
}
\begin{feynman}
\vertex (a) at (0,0) {\(1\)}; 
\vertex (b) at (0.7,0) {\(3\)};
\vertex (c) at (1.4,0) {\(2\)};
\vertex [blob, /tikz/minimum size=15pt, shape=rectangle, fill=black!20, line width=0.2mm] (d) at (0.7,1.25) {}; 
\vertex (e) at (0,2.5) {\(1'\)}; 
\vertex (f) at (0.70,2.5) {\(3'\)};
\vertex (g) at (1.4,2.5) {\(2'\)};
\vertex (h) at (0.7,-0.7) {(b)};
\diagram* {
(a) -- [fermion, line width=0.25mm] (d) -- [fermion, line width=0.25mm] (e);
(b) -- [fermion, line width=0.25mm] (d) -- [fermion, line width=0.25mm] (f);
(c) -- [fermion, line width=0.25mm] (d) -- [fermion, line width=0.25mm] (g);
};
\end{feynman}
\end{tikzpicture}
\end{minipage}
\begin{minipage}[c]{0.15\textwidth}
\begin{tikzpicture} 
\tikzfeynmanset{
  my dot/.style={
    /tikzfeynman/dot,
    /tikz/minimum size=10pt,
  },
  every vertex/.style={my dot},
}
\begin{feynman}
\vertex (a) at (0,0) {\(2\)}; 
\vertex (b) at (0.7,0) {\(1\)};
\vertex (c) at (1.4,0) {\(3\)};
\vertex [blob, /tikz/minimum size=15pt, shape=rectangle, fill=black!20, line width=0.2mm] (d) at (0.7,1.25) {}; 
\vertex (e) at (0,2.5) {\(2'\)}; 
\vertex (f) at (0.70,2.5) {\(1'\)};
\vertex (g) at (1.4,2.5) {\(3'\)};
\vertex (h) at (0.7,-0.7) {(c)};
\diagram* {
(a) -- [fermion, line width=0.25mm] (d) -- [fermion, line width=0.25mm] (e);
(b) -- [fermion, line width=0.25mm] (d) -- [fermion, line width=0.25mm] (f);
(c) -- [fermion, line width=0.25mm] (d) -- [fermion, line width=0.25mm] (g);
};
\end{feynman}
\end{tikzpicture}
\end{minipage}
\begin{minipage}[c]{0.15\textwidth}
\begin{tikzpicture} 
\tikzfeynmanset{
  my dot/.style={
    /tikzfeynman/dot,
    /tikz/minimum size=10pt,
  },
  every vertex/.style={my dot},
}
\begin{feynman}
\vertex (a) at (0,0) {\(2\)}; 
\vertex (b) at (0.7,0) {\(3\)};
\vertex (c) at (1.4,0) {\(1\)};
\vertex [blob, /tikz/minimum size=15pt, shape=rectangle, fill=black!20, line width=0.2mm] (d) at (0.7,1.25) {}; 
\vertex (e) at (0,2.5) {\(2'\)}; 
\vertex (f) at (0.70,2.5) {\(3'\)};
\vertex (g) at (1.4,2.5) {\(1'\)};
\vertex (h) at (0.7,-0.7) {(d)};
\diagram* {
(a) -- [fermion, line width=0.25mm] (d) -- [fermion, line width=0.25mm] (e);
(b) -- [fermion, line width=0.25mm] (d) -- [fermion, line width=0.25mm] (f);
(c) -- [fermion, line width=0.25mm] (d) -- [fermion, line width=0.25mm] (g);
};
\end{feynman}
\end{tikzpicture}
\end{minipage}
\begin{minipage}[c]{0.15\textwidth}
\begin{tikzpicture} 
\tikzfeynmanset{
  my dot/.style={
    /tikzfeynman/dot,
    /tikz/minimum size=10pt,
  },
  every vertex/.style={my dot},
}
\begin{feynman}
\vertex (a) at (0,0) {\(3\)}; 
\vertex (b) at (0.7,0) {\(1\)};
\vertex (c) at (1.4,0) {\(2\)};
\vertex [blob, /tikz/minimum size=15pt, shape=rectangle, fill=black!20, line width=0.2mm] (d) at (0.7,1.25) {}; 
\vertex (e) at (0,2.5) {\(3'\)}; 
\vertex (f) at (0.70,2.5) {\(1'\)};
\vertex (g) at (1.4,2.5) {\(2'\)};
\vertex (h) at (0.7,-0.7) {(e)};
\diagram* {
(a) -- [fermion, line width=0.25mm] (d) -- [fermion, line width=0.25mm] (e);
(b) -- [fermion, line width=0.25mm] (d) -- [fermion, line width=0.25mm] (f);
(c) -- [fermion, line width=0.25mm] (d) -- [fermion, line width=0.25mm] (g);
};
\end{feynman}
\end{tikzpicture}
\end{minipage}
\begin{minipage}[c]{0.15\textwidth}
\begin{tikzpicture} 
\tikzfeynmanset{
  my dot/.style={
    /tikzfeynman/dot,
    /tikz/minimum size=10pt,
  },
  every vertex/.style={my dot},
}
\begin{feynman}
\vertex (a) at (0,0) {\(3\)}; 
\vertex (b) at (0.7,0) {\(2\)};
\vertex (c) at (1.4,0) {\(1\)};
\vertex [blob, /tikz/minimum size=15pt, shape=rectangle, fill=black!20, line width=0.2mm] (d) at (0.7,1.25) {}; 
\vertex (e) at (0,2.5) {\(3'\)}; 
\vertex (f) at (0.70,2.5) {\(2'\)};
\vertex (g) at (1.4,2.5) {\(1'\)};
\vertex (h) at (0.7,-0.7) {(f)};
\diagram* {
(a) -- [fermion, line width=0.25mm] (d) -- [fermion, line width=0.25mm] (e);
(b) -- [fermion, line width=0.25mm] (d) -- [fermion, line width=0.25mm] (f);
(c) -- [fermion, line width=0.25mm] (d) -- [fermion, line width=0.25mm] (g);
};
\end{feynman}
\end{tikzpicture}
\end{minipage}
\caption{The six contributions to a general 3N interaction resulting from the sums over the particle indices in Eq.~(\ref{eq:generic_form_3NF}).}
\label{fig:V3N_six_particle_labels} 
\end{figure}

The interaction $V_{\text{3N}}$ in Eq.~(\ref{eq:generic_form_3NF}) is by
construction totally symmetric in all particle labels. In total there are six
contributions, which are illustrated in
Figure~\ref{fig:V3N_six_particle_labels}. It is important to note that we can
always pick a subset of two terms which are not related by a cyclic or
anticyclic permutation of the states, such that the remaining four diagrams
can then be generated by the application of the 3-body permutation operators
$P_{123}$ and $P_{132}$. In order to illustrate this point let us make a
particular choice and define $V_{\text{3N}}^{(1)}$ to be those two diagrams in
Figure~\ref{fig:V3N_six_particle_labels} in which the central fermion line
carries label ``$1$'', i.e., diagrams (c) and (e). Note that this interaction
term is by construction symmetric in particles $2$ and $3$, which follows
directly from the total symmetry of $V_{\text{3N}}$. Then it is
straightforward to show that the total interaction can be written in the form
\begin{equation}
V_{\text{3N}} = V_{\text{3N}}^{(1)} + P_{123}^{-1} V_{\text{3N}}^{(1)} P_{123} + P_{132}^{-1} V_{\text{3N}}^{(1)} P_{132} \, .
\label{eq:V3N_Fadddecomp}
\end{equation}
The quantity $V_{\text{3N}}^{(1)}$ is commonly called the \textit{Faddeev
component} of the interaction. To verify this relation note that the
permutation operator $P_{123}$ permutes the ket states cyclically and the
inverse operator $P_{123}^{-1}$ leads to a cyclic permutation of the bra
states (see discussion after Eq.~(\ref{eq:relation_ki_ab_bases}) in Section
\ref{sec:3NF_coord_def}). Obviously, we could have chosen another subset of
diagrams by selecting another particle label $i$ or another fermion line.
However, the chosen contributions will by construction always be symmetric in
two particle labels $j$ and $k$. For each possible choice we obtain a
different decomposition of the 3N interaction in
Eq.~(\ref{eq:V3N_Fadddecomp}). The practical usefulness of this decomposition
for our present purposes relies on the fact that it is sufficient to only
compute $V_{\text{3N}}^{(i)}$ explicitly for a chosen decomposition and to
generate the other terms by the application of the permutation operators.

By employing the algorithm discussed in Section~\ref{sec:PWD_3NF_local} we
obtain as the final result the partial-wave matrix elements
\begin{equation}
\tensor*[_{\{ab\}}]{\bigl< p' q' \alpha' | V_{\text{3N}}^{(i)} | p q \alpha\bigr> }{_{\{ab\}}}
\label{eq:Faddeev_components}
\end{equation}
for a given decomposition and in a specific basis representation $\{ab\}$. We
emphasize that the specific values of the matrix elements depend on
the choice of the decomposition Eq.~(\ref{eq:V3N_Fadddecomp}) and on the
chosen basis representation of the interaction. In this sense the matrix
elements in Eq.~(\ref{eq:Faddeev_components}) are scheme dependent and we
cannot expect the results of independent implementations for a given
interaction to agree. However, when computing expectation values of these
interactions with respect to wave functions the results, of course, have to be
unique and scheme independent.

In order to see that this is indeed the case consider the expectation value of
the total 3N interaction $V_{\text{3N}}$ with respect to a $3$-body
wave function $\psi$:
\begin{equation}
\left< V \right> \equiv \left< \psi \right| V_{\text{3N}} \left| \psi \right> \, ,
\label{eq:V3N_expvalue}
\end{equation}
using the normalization $\left< \psi | \psi \right> = 1$. Since we consider
nucleons as identical fermions the wave function is totally antisymmetric,
i.e., $P_{ij}
\left| \psi \right> = - \left| \psi \right>$ for two arbitrary particles $i$
and $j$, with $P_{ij}$ being the two-body transposition operators (see
Section~\ref{sec:3NF_coord_def}). Let us make the antisymmetry of the wave
function explicit by inserting the three-body antisymmetrizer~\cite{Mess99QM}:
\begin{equation}
\mathcal{A}_{123} = 1 - P_{12} - P_{13} - P_{23} + P_{123} +  P_{132} = (1 - P_{ij}) (1 + P_{123} + P_{132}) \, ,
\label{eq:A123_23decomp}
\end{equation}
where $i$ and $j$ are two arbitrary particle indices. Then the expectation
value can be written in the form
\begin{equation}
\left< V \right> = \frac{1}{36} \left< \psi \right| \mathcal{A}_{123} V_{\text{3N}} \mathcal{A}_{123} \left| \psi \right> = \frac{1}{9} \left< \psi \right| (1 + P_{123} + P_{132}) V_{\text{3N}} (1 + P_{123} + P_{132}) \left| \psi \right> \, .
\label{eq:V3N_expvalue2}
\end{equation}
The last identity follows directly from the antisymmetry of $\psi$, i.e., $(1 -
P_{ij})/2 \left| \psi \right> = \left| \psi \right>$. Now we insert the
decomposition from Eq.~(\ref{eq:V3N_Fadddecomp}) and use the relations
\begin{equation}
P_{123} (1 + P_{123} + P_{132}) = P_{132} (1 + P_{123} + P_{132}) =  (1 + P_{123} + P_{132}) \, ,
\label{eq:P123_relations}
\end{equation}
which follow directly from relations $P_{123} = P_{132}^{-1} = P_{132}^2$ and
$P_{123}^3 = P_{132}^3 = 1$. From Eq.~(\ref{eq:P123_relations}) it follows
that each term in the decomposition~(\ref{eq:V3N_Fadddecomp}) gives the same
contribution to the expectation value with the final result:
\begin{equation}
\left< V \right> = \frac{1}{3} \left< \psi \right| (1 + P_{123} + P_{132}) V_{\text{3N}}^{(i)} (1 + P_{123} + P_{132}) \left| \psi \right> \, .
\end{equation}
The arguments above are independent of the specific choice of the decomposition and also on the chosen basis
representation $\{ab\}$. In fact, the matrix elements of the \textit{antisymmetrized} 3N interaction
\begin{equation}
\bigl< p' q' \alpha' | V_{\text{3N}}^{\text{as}} | p q \alpha\bigr> = \tensor*[_{\{ab\}}]{\bigl< p' q' \alpha' | (1 + P_{123} + P_{132}) V_{\text{3N}}^{(i)} (1 + P_{123} + P_{132}) | p q \alpha\bigr> }{_{\{ab\}}}
\label{eq:Faddeev_antisymmetrized} 
\end{equation}
are unique and scheme independent. Hence, in the following we can neglect the
basis indices $\{ab\}$ for the antisymmetrized interaction.

The practical computation of antisymmetrized matrix elements involves two
steps: First, the calculation of the 3N interaction components defined in
Eq.~(\ref{eq:V3N_Fadddecomp}) and, second, the application of the permutation
operators in Eq.~(\ref{eq:Faddeev_antisymmetrized}). The latter operation is
usually performed in a partial-wave representation by inserting a complete set
of states, e.g.:
\begin{equation}
\tensor*[_{\{ab\}}]{\bigl< p' q' \alpha' | P_{123} V_{\text{3N}}^{(i)} | p q \alpha\bigr>}{_{\{ab\}}} =\int dp'' p''^2 dq'' q''^2 \sum_{\alpha''} \tensor*[_{\{ab\}}]{\bigl< p' q' \alpha' | P_{123} | p'' q'' \alpha'' \bigr>}{_{\{ab\}}} 
\tensor*[_{\{ab\}}]{\bigl< p'' q'' \alpha'' | V_{\text{3N}}^{(i)} | p q
\alpha\bigr> }{_{\{ab\}}} \, .
\label{eq:P123_times_V} 
\end{equation}
The matrix elements of the permutation operator $P_{123}$ can be derived
directly based on their definitions in momentum space,
Eqs.~(\ref{eq:P123_momentum_def_first}) to
(\ref{eq:P123_momentum_def_fourth}), and by taking into account the spin- and
isospin exchange parts. These definitions show that, in essence,
the matrix elements $\left< p' q' \alpha' | P_{123} | p q \alpha \right>$
provide two relations of the four Jacobi momenta $p$,$q$,$p'$ and $q'$, which
depend in general on all angular momentum quantum numbers. There are different
possible representations, for example:
\begin{itemize}
    \item[1.] $p' = f_1(p,q), q' = f_2(p,q)$ \quad (see Eq.~(\ref{eq:P123_momentum_def_first})),
    \item[2.] $p = f_1(q,q'), p' = f_2(q,q')$ \quad (see Eq.~(\ref{eq:P123_momentum_def_third})),
    \item[3.] $p = f_1(p',q'), q = f_2(p',q')$ \quad (see Eq.~(\ref{eq:P123_momentum_def_fourth})).
\end{itemize}
Note that we suppressed all angular, spin and isospin dependence in these
schematic relations. The preferred choice depends on the context. For our
purposes it is most convenient to choose option 3, because this choice makes
it possible to implement products like $P_{123} V_{\text{3N}}$ in
Eq.~(\ref{eq:P123_times_V}) very efficiently as a matrix product as we will
now demonstrate.

The definition of the partial-wave matrix elements of $P_{123}$ follows from a
generalization of Eqs.~(\ref{eq:permutation_momequiv}) and
(\ref{eq:P123_rep_independent}):
\begin{equation}
\left< p' q' \alpha' | P_{123} | p q \alpha \right> = \tensor*[_{\{12\}}]{\left< p' q' \alpha' | p q \alpha \right>}{_{\{23\}}} = \tensor*[_{\{23\}}]{\left< p' q' \alpha' | p q \alpha \right>}{_{\{31\}}} = \tensor*[_{\{31\}}]{\left< p' q' \alpha' | p q \alpha \right>}{_{\{12\}}} \, .
\end{equation}
The derivation of the explicit expression for the partial-wave matrix elements
of the permutation operator is straightforward but
somewhat tedious. The calculation is presented in detail
in Appendix~\ref{sec:PWD_P123}. The final result can be written in the following form:
\begin{align}
& \hspace{-1cm} \tensor*[_{\{ab\}}]{\left< p' q' \alpha' | P_{123} | p q \alpha \right>}{_{\{ab\}}} \nonumber \\
&= \sum_{\mathcal{L}, \mathcal{S}} \sqrt{\hat{J} \hat{j} \hat{J}' \hat{j}'} \hat{\mathcal{S}} \left\{
\begin{array}{ccc}
L & S & J \\
l & \tfrac{1}{2} & j \\
\mathcal{L} & \mathcal{S} & \mathcal{J}
\end{array}
\right\}
\left\{
\begin{array}{ccc}
L' & S' & J' \\
l' & \tfrac{1}{2} & j' \\
\mathcal{L} & \mathcal{S} & \mathcal{J}
\end{array}
\right\} \nonumber \\
& \times (-1)^{S} \sqrt{\hat{S} \hat{S}'} 
\left\{
\begin{array}{ccc}
\tfrac{1}{2} & \tfrac{1}{2} & S' \\
\tfrac{1}{2} & \mathcal{S} & S
\end{array}
\right\} 
(-1)^{T} \sqrt{\hat{T} \hat{T}'} 
\left\{
\begin{array}{ccc}
\tfrac{1}{2} & \tfrac{1}{2} & T' \\
\tfrac{1}{2} & \mathcal{T} & T
\end{array}
\right\} \nonumber \\
& \times 8 \pi^2 \int d \cos \theta_{\mathbf{p}' \mathbf{q}'} \frac{\delta(p - |\overline{\mathbf{p}}|)}{p^2} \frac{\delta(q - |\overline{\mathbf{q}}|)}{q^2} \sum_{\mathcal{M}_{\mathcal{L}}}  \mathcal{Y}_{L' l'}^{* \mathcal{L} \mathcal{M}_{\mathcal{L}}} \bigl( \hat{\mathbf{p}}', \hat{\mathbf{q}}' \bigr) \mathcal{Y}_{L l}^{\mathcal{L} \mathcal{M}_{\mathcal{L}}} \bigl( \hat{\overline{\mathbf{p}}},\hat{\overline{\mathbf{q}}} \bigr) \nonumber \\
&\equiv \int d \cos \theta_{\mathbf{p}' \mathbf{q}'} G_{\alpha \alpha'} (p',q',\cos \theta_{\mathbf{p}' \mathbf{q}'}) \frac{\delta(p - |\overline{\mathbf{p}}|)}{p^2} \frac{\delta(q - |\overline{\mathbf{q}}|)}{q^2} \, ,
\label{eq:P123_matrixelements}
\end{align}
with
\begin{align}
\overline{\mathbf{p}} &= \overline{\mathbf{p}} (p',q',\cos \theta_{\mathbf{p}' \mathbf{q}'}) = - \tfrac{1}{2} \mathbf{p}' - \tfrac{3}{4} \mathbf{q}' \, ,\nonumber \\ 
\overline{\mathbf{q}} &= \overline{\mathbf{q}} (p',q',\cos \theta_{\mathbf{p}' \mathbf{q}'}) = \mathbf{p}' - \tfrac{1}{2} \mathbf{q}' \, .
\label{eq:pqbar}
\end{align}
In the last step in Eq.~(\ref{eq:P123_matrixelements}) we factorized the total
matrix element into the radial delta functions and the geometric function $G_{\alpha
\alpha'} (p',q',\cos \theta_{\mathbf{p}' \mathbf{q}'})$, which contains all
the remaining terms (see also Ref.~\cite{Gloe83QMFewbod}). One of the main
differences to other expressions (see, e.g.,
Refs.~\cite{Gloe83QMFewbod,Gloe95cont}) is the fact that we directly perform
the angular integrals in Eq.~(\ref{eq:P123_matrixelements}) without
decomposing the angular dependence of the spherical harmonic functions any
further. This implementation turns out to be numerically more efficient and,
most importantly, very stable even for large values of angular momenta $L$
and $l$.

We note that the partial-wave matrix elements of the anticyclic permutation
operator $P_{132} = P_{123}^{-1}$ with respect to antisymmetric states are
also given by Eq.~(\ref{eq:P123_matrixelements}). However, applying the
operator product $P_{123} P_{132} = 1$ in this partial-wave
representation to an operator or wave function $\psi(p,q,\alpha) = \left< p q
\alpha | \psi
\right>$ will generally \textit{not} result in an identity:
\begin{equation}
\sum_{\alpha} \int dp' p'^2 dq' q'^2 dp'' p''^2 dq'' q''^2 \left< p q \alpha | P_{123} | p' q' \alpha' \right> \left< p' q' \alpha'' | P_{132} | p'' q'' \alpha'' \right> \left< p'' q'' \alpha'' | \psi \right> \neq \left< p q \alpha | \psi \right> \, .
\end{equation}
This is because the permutation operator generally couples states of different
symmetries with respect to the exchange of particles. This can be seen by
expressing the permutation operator in the following form (see
Section~\ref{sec:3NF_coord_def} and also the discussion in
Ref.~\cite{Gloe83QMFewbod}):
\begin{equation}
P_{132} = P_{13} P_{23} = P_{23} P_{12} P_{23} P_{23} = P_{23} P_{123} P_{23} \, ,
\end{equation}
and compute the following overlap matrix elements
\begin{align}
\tensor*[_{\{23\}}]{\left< p' q' \alpha' | P_{132} | p q \alpha \right>}{_{\{23\}}} &= \tensor*[_{\{23\}}]{\left< p' q' \alpha' | p q \alpha \right>}{_{\{12\}}} \nonumber \\
&= \tensor*[_{\{23\}}]{\left< p' q' \alpha' | P_{23} P_{123} P_{23} | p q \alpha \right>}{_{\{23\}}} \nonumber \\
&= \tensor*[_{\{23\}}]{\left< p' q' \alpha' | P_{123} | p q \alpha \right>}{_{\{23\}}} (-1)^{L + S + T} (-1)^{L' + S' + T'} \, .
\end{align}
If the initial and final states are antisymmetric with respect to the exchange
of particles $2$ and $3$ (like all states in the definition of the basis
in Eq.~(\ref{eq:Jj_bas})) the phase factors cancel and we obtain $\left<
p' q' \alpha' | P_{123} | p q \alpha \right> = \left< p' q' \alpha' | P_{132} | p q
\alpha \right>$. However, in general there will be contributions from couplings
to symmetric intermediate states when computing auxiliary quantities. When the
operator is eventually applied to physical states, the symmetric contributions
of the operator decouple and do not contribute to observables. However, for
intermediate steps it is key to extend the basis state by ``unphysical'' states
that are symmetric under exchange of two particles. This point will become
important for the regularization of 3N interactions (see
Section~\ref{sec:semilocal_coordinate}).

For the practical evaluation of the radial delta functions in
Eq.~(\ref{eq:P123_matrixelements}) we employ a method based on global splines,
which was originally developed for solving the Faddeev equations for
few-body systems~\cite{Gloe82spline}. This basic idea of this method is to construct
continuous spline functions $S_i (p)$ for some given mesh system $\{p_i\}$ of size
$N_p$ such that the value of a function $f$ at an arbitrary point $p$ is given
by the following global sum over all mesh points:
\begin{equation}
f(p) = \sum_{i=1}^{N_p} f(p_i) S_i (p) \, .
\label{eq:spline_sum}
\end{equation} 
The spline functions are constructed such that the interpolated function
values agrees exactly with the original values $f(p_i)$ at the mesh points.
One possible parametrization of the spline functions $S_i(p)$ is given in
Ref.~\cite{Gloe82spline}.

This interpolation method allows to express the product of the permutation
operator with another operator in an elegant and efficient way as a matrix
product. Consider as an example the product in Eq.~(\ref{eq:P123_times_V}),
which can be written in the following form ($x \equiv \cos \theta_{\mathbf{p}'
\mathbf{q}'}$):
\begin{align}
\bigl< p' q' \alpha' | P_{123} V_{\text{3N}}^{(i)} | p q \alpha \bigr> &= \sum_{\alpha''} \int dp'' p''^2 dq'' q''^2 \bigl< p' q' \alpha' | P_{123} | p'' q'' \alpha'' \bigr> \bigl< p'' q'' \alpha'' | V_{\text{3N}}^{(i)} | p q
\alpha \bigr> \nonumber \\
&= \sum_{\alpha''} \int_{-1}^1 dx G_{\alpha' \alpha''} (p',q',x) \bigl< \bar{p} \bar{q} \alpha'' | V_{\text{3N}}^{(i)} | p q \alpha \bigr>
\nonumber \\ 
&= \sum_{i,j,\alpha''} \int_{-1}^1 dx S_i (\bar{p}) S_j (\bar{q}) G_{\alpha' \alpha''} (p',q',x) \bigl< p''_i q''_j \alpha'' | V_{\text{3N}}^{(i)} | p q \alpha \bigr> \, ,
\label{eq:P123_V_product_spline}
\end{align}
where $p''_i$ and $q''_j$ are the mesh points of some chosen interpolation
grid systems, and $p = \bar{p} (p'_{i'}, q'_{j'}, x)$ and $q =
\bar{q} (p'_{i'}, q'_{j'}, x)$, given in Eq.~(\ref{eq:pqbar}). Since for
practical calculations all quantities need to be tabulated on a finite
momentum mesh system anyway it is most natural to choose the same mesh system
for the Jacobi momenta in the initial and final states in
Eq.~(\ref{eq:P123_V_product_spline}). This defines a common discrete matrix
representation for all quantities in the chosen three-body basis. In
particular, the permutation operator can be precalculated and prestored for
each three-body partial wave by defining the discrete matrix of dimension $N_p
N_q N_{\alpha}$:
\begin{equation}
\bigl< p'_{i'} q'_{j'} \alpha' | P_{123} | p_i q_j \alpha \bigr> = \int_{-1}^1 dx \: S_i (\bar{p}) S_j (\bar{q}) G_{\alpha' \alpha} (p'_{i'},q'_{j'},x) \, .
\end{equation}
This matrix can then be applied to arbitrary operators in a three-body
momentum representation via efficient BLAS (Basic Linear Algebra
Subprograms)~\cite{BLAS} matrix multiplication routines. For the application of
$P_{123}$ from the right hand side in Eq.~(\ref{eq:Faddeev_antisymmetrized})
we just apply the transposed matrix.

\subsection{Example: Calculation of two-pion exchange 3N interactions}
\label{sec:example_N2LO_calc}

In this section we illustrate the calculation of partial-wave 3N matrix
elements using the algorithm discussed above by considering as an example
the leading-order long-range 3N interactions in chiral EFT. At this point we
neglect any regulators. We will discuss the regularization in detail in the
next section.

Specifically, we consider the two-pion exchange interaction at N$^2$LO
proportional to the low-energy couplings $c_1$ and $c_3$:
\begin{equation}
V_{\text{3N}} = \frac{1}{2} \left( \frac{g_A}{2 f_{\pi}} \right)^2 \sum_{i
\neq j \neq k} \frac{(\boldsymbol{\sigma}_i \cdot \mathbf{Q}_i)
(\boldsymbol{\sigma}_j \cdot \mathbf{Q}_j)}{(\mathbf{Q}^2_i  + m_{\pi}^2)
(\mathbf{Q}^2_j  + m_{\pi}^2)} \boldsymbol{\tau}_i \cdot \boldsymbol{\tau}_j
\left[ - \frac{4 c_1 m_{\pi}^2}{f_{\pi}^2} + \frac{2 c_3}{f_{\pi}^2}
\mathbf{Q}_i \cdot \mathbf{Q}_j \right] \, ,
\label{eq:V3N_c1c3}
\end{equation}
where $\mathbf{Q}_i$ denote as usual the momentum transfers, $\mathbf{Q}_i =
\mathbf{k}_i' - \mathbf{k}_i$ (with the single-particle momenta
$\mathbf{k}_i$). This interaction has a compact form but is nevertheless
general enough to serve as an illustration of the efficient framework
discussed in Section~\ref{sec:general_3N_decomp} as this interaction contains
all the fundamental complications of more intricate interactions like those at
higher orders in chiral EFT.

As a first step we choose a particular subset of the six interaction terms
(see Figure~\ref{fig:V3N_six_particle_labels}). Here we select again, without
loss of generality, those two diagrams in which the central fermion line
caries label ``$1$'', again denoted by $V_{\text{3N}}^{(1)}$ in the following.
This uniquely defines the decomposition shown in Eq.~(\ref{eq:V3N_Fadddecomp})
of the 3N interaction, where in the present example this interaction term
takes the form
\begin{equation}
V_{\text{3N}}^{(1)} = \left( \frac{g_A}{2 f_{\pi}} \right)^2
\frac{(\boldsymbol{\sigma}_2 \cdot \mathbf{Q}_2) (\boldsymbol{\sigma}_3 \cdot
\mathbf{Q}_3)}{(\mathbf{Q}^2_2  + m_{\pi}^2) (\mathbf{Q}^2_3  + m_{\pi}^2)}
\boldsymbol{\tau}_2 \cdot \boldsymbol{\tau}_3 \left[ - \frac{4 c_1
m_{\pi}^2}{f_{\pi}^2} + \frac{2 c_3}{f_{\pi}^2} \mathbf{Q}_2 \cdot
\mathbf{Q}_3 \right] \, .
\label{eq:V3N_c1c3_1}
\end{equation} 
Note that this interaction term is, as discussed in the previous section,
symmetric in particles $2$ and $3$. In fact, in the present case both terms
are identical, so we just obtain a factor 2.

As a next step we choose one of the three basis representations for the Jacobi
momenta (see Section~\ref{sec:3NF_coord_def}). To be specific, we choose
here the basis $\{23\}$ (see Table~\ref{tab:Jacobi_momenta_crosstable}), i.e.:
\begin{align}
\mathbf{k}_1 = \mathbf{q}_{\{23\}} = \mathbf{q}, \quad \mathbf{k}_2 = \mathbf{p}_{\{23\}} - \frac{1}{2} \mathbf{q}_{\{23\}} = \mathbf{p} - \frac{1}{2} \mathbf{q}, \quad \mathbf{k}_3 = -\mathbf{p}_{\{23\}} - \frac{1}{2} \mathbf{q}_{\{23\}} = - \mathbf{p} - \frac{1}{2} \mathbf{q} \, ,
\end{align}
where we have defined $\mathbf{p}_{\{23\}} = \mathbf{p}$ and  $\mathbf{q}_{\{23\}}
= \mathbf{q}$ for the sake of simplifying the notation for the rest of this section.
Hence for the momentum transfers we obtain:
\begin{align}
\mathbf{Q}_1 = \mathbf{k}_1' - \mathbf{k}_1 = \tilde{\mathbf{q}}, \quad \mathbf{Q}_2 = \mathbf{k}_2' - \mathbf{k}_2 = \tilde{\mathbf{p}} - \frac{1}{2} \tilde{\mathbf{q}}, \quad \mathbf{Q}_3 = \mathbf{k}_3' - \mathbf{k}_3 = - \tilde{\mathbf{p}} - \frac{1}{2} \tilde{\mathbf{q}} \, ,
\end{align}
with $\tilde{\mathbf{p}}=\mathbf{p}'-\mathbf{p}$ and $\tilde{\mathbf{q}} =
\mathbf{q}' - \mathbf{q}$ like in Eq.~(\ref{eq:V3N_local_form}). Now we can
factorize the interaction into a momentum-dependent scalar function
$V_{\text{3N}}^{\text{local}} (\tilde{\mathbf{p}},\tilde{\mathbf{q}})$ times
spin- and isospin operators (see Eq.~(\ref{eq:generic_form_3NF})):
\begin{equation}
V_{\text{3N}}^{(1)} = f_Q (\tilde{\mathbf{p}},\tilde{\mathbf{q}}) \: f_{\sigma} (\mathbf{p},\mathbf{q},\mathbf{p}',\mathbf{q}') \: f_{\tau} \, ,
\end{equation}
with
\begin{equation}
f_Q (\tilde{\mathbf{p}},\tilde{\mathbf{q}}) = \frac{1}{f_{\pi}^2} \left( \frac{g_A}{2 f_{\pi}} \right)^2 \frac{- 4 c_1 m_{\pi}^2 + 2 c_3 \mathbf{Q}_2 \cdot \mathbf{Q}_3}{(\mathbf{Q}^2_2  + m_{\pi}^2) (\mathbf{Q}^2_3  + m_{\pi}^2)}, \quad f_{\sigma} (\mathbf{p},\mathbf{q},\mathbf{p}',\mathbf{q}') = (\boldsymbol{\sigma}_2 \cdot \mathbf{Q}_2) (\boldsymbol{\sigma}_3 \cdot \mathbf{Q}_3), \quad f_{\tau} = \boldsymbol{\tau}_2 \cdot \boldsymbol{\tau}_3 \, .
\end{equation}
The function $f_Q (\tilde{\mathbf{p}},\tilde{\mathbf{q}})$ enters as the
kernel in Eq.~(\ref{final_result_pwd}) after extending it by contributions
from the momentum dependence of the spin operators $f_{\sigma}$ as described
in Section \ref{sec:general_3N_decomp}. Isospin operators like $f_{\tau}$ can
be treated very easily since they do not depend on any momenta and hence do
not explicitly affect the partial-wave decomposition. For the spin operators
we first expand the momentum dependence explicitly in terms of the Jacobi
momenta
\begin{equation}
f_{\sigma} (\mathbf{p},\mathbf{q},\mathbf{p}',\mathbf{q}') = \left[
\boldsymbol{\sigma}_2 \cdot ( \mathbf{p}' - \mathbf{p} - \tfrac{1}{2} \mathbf{q}' + \tfrac{1}{2}
\mathbf{q}) \right] \left[ \boldsymbol{\sigma}_3 \cdot ( - \mathbf{p}' +
\mathbf{p} - \tfrac{1}{2} \mathbf{q}' + \tfrac{1}{2} \mathbf{q}) \right] \, ,
\label{eq:spin_operator_expanded}
\end{equation}
and then apply the factorizations shown in Eq.~(\ref{eq:spin_factorize}) and
Eq.~(\ref{eq:Y_identity}) to each of the terms. For example, for the contribution
$[\boldsymbol{\sigma}_2 \cdot (-\mathbf{p})] [\boldsymbol{\sigma}_3 \cdot
\mathbf{p}]$ the generalized function $F$ needs to be extended twice by the
quantum numbers of momentum $\mathbf{p}$ in the way shown in
Eq.~(\ref{eq:F_withonemu}). Accordingly, all the other remaining 15 terms in
Eq.~(\ref{eq:spin_operator_expanded}) can be treated. As a next step, we can
calculate the function $\tilde{F}_{L l L' l'}^{\bar{l} \{ \bar{L}_i \} }
(p,q,p',q')$ as discussed in Section \ref{sec:general_3N_decomp}. The
spherical components of the momentum-independent spin operators shown in
Eq.~(\ref{eq:spin_operator}) can either be prestored or directly computed
on the fly. Combining all these results allows the calculation of the function
$A_{\beta \beta'}^{\bar{l} \{ \bar{L}_i \}}$ and eventually of the
$LS$-coupled and $Jj$-coupled 3N interaction matrix elements (see
Eqs.~(\ref{eq:VLS}) and~(\ref{eq:Jj_recoupling})).

\subsection{Regularization of 3N interactions}
\label{sec:3N_regularization}

\begin{table}[t]
\small
\centering
\begin{tabular}{l||l|l}
            &  momentum space  &  coordinate space  \\
\hline \hline 
\textbf{nonlocal} & \underline{\textbf{nonlocal MS}} \quad \cite{Epel02fewbody} &          \\
\textit{regulators:}  &																																					& \\
long-range  &  $f^{\text{long}}_{\Lambda} (\mathbf{p},\mathbf{q}) = \exp \bigl[ - \bigl( (\mathbf{p}^2 + \tfrac{3}{4} \mathbf{q}^2)/\Lambda^2 \bigr)^n \bigr]$         &            \\
short-range &  $f^{\text{short}}_{\Lambda} (\mathbf{p},\mathbf{q}) = f^{\text{long}}_{\Lambda} (\mathbf{p},\mathbf{q}) = f_R (\mathbf{p},\mathbf{q})$         &            \\
            &                                                                                                                            &            \\
\textit{regularization:} & $\bigl< \mathbf{p}' \mathbf{q}' | V_{\text{3N}}^{\text{reg}} | \mathbf{p} \mathbf{q} \bigr> = f_R (\mathbf{p}',\mathbf{q}') \bigl< \mathbf{p}' \mathbf{q}' | V_{\text{3N}} | \mathbf{p} \mathbf{q} \bigr> f_R (\mathbf{p},\mathbf{q})$ & \\ \vspace{-0.2cm} & \\
\hline

\textbf{local} & \underline{\textbf{local MS}} \quad \cite{Navr07local3N}    &  \underline{\textbf{local CS}} \quad \cite{Lynn16QMC3N} \\
\textit{regulators:}  &																																					& \\
long-range    & $f^{\text{long}}_{\Lambda} (\mathbf{Q}_i) = \exp \bigl[ - \bigl( \mathbf{Q}_i^2/\Lambda^2 \bigr)^2 \bigr]$ & $f^{\text{long}}_R(\mathbf{r}) = 1 - \exp \bigl[ - \bigl( r^2/R^2 \bigr)^n \bigr]$ \\  
short-range    & $f^{\text{short}}_{\Lambda} (\mathbf{Q}_i) = f^{\text{long}}_{\Lambda} (\mathbf{Q}_i) = f_{\Lambda} (\mathbf{Q}_i)$ & $f^{\text{short}}_R(\mathbf{r}) = \exp \bigl[ - \bigl( r^2/R^2 \bigr)^n \bigr]$ \\  
              &                                                                                                                          &            \\
\textit{regularization:} & $\bigl< \mathbf{p}' \mathbf{q}' | V_{\text{3N}}^{\text{reg}} | \mathbf{p} \mathbf{q} \bigr> = \bigl< \mathbf{p}' \mathbf{q}' | V_{\text{3N}} | \mathbf{p} \mathbf{q} \bigr> \prod_i f_R (\mathbf{Q}_i)$ & $V^{\pi, \text{reg}}_{\text{3N}} (\mathbf{r}_{ij}) = f^{\text{long}}_R (\mathbf{r}_{ij}) V^{\pi}_{\text{3N}} (\mathbf{r}_{ij})$  \\
            &   & $V^{\delta, \text{reg}}_{\text{3N}} (\mathbf{r}_{ij}) = \alpha_n (R) f^{\text{short}}_R (\mathbf{r}_{ij})$ \\ \vspace{-0.2cm} & \\
\hline

\textbf{semilocal} & \underline{\textbf{semilocal MS}} \quad \cite{Epel19Bayes} & \underline{\textbf{semilocal CS}} \quad \cite{Epel18SCS3N} \\
\textit{regulators:}  &																																					& \\
long-range         & $f^{\text{long}}_{\Lambda} (\mathbf{Q}_i) = \exp \bigl[ - \bigl( \mathbf{Q}_i^2 + m_{\pi}^2 \bigr)/\Lambda^2 \bigr]$ & $f^{\text{long}}_R(\mathbf{r}) = \bigl( 1 - \exp \bigl[ - r^2/R^2 \bigr] \bigr)^n$ \\
short-range        & $f^{\text{short}}_{\Lambda} (\mathbf{p}) = \exp \bigl[ - \mathbf{p}^2/\Lambda^2 \bigr]$  & $f^{\text{short}}_{\Lambda} (\mathbf{p}) = \exp \bigl[ - \mathbf{p}^2 /\Lambda^2 \bigr]$ \\  
                    &   & \\                                                                                                             
\textit{regularization:}      & $\bigl< \mathbf{p}' \mathbf{q}' | V_{\text{3N}}^{\pi, \text{reg}} | \mathbf{p} \mathbf{q} \bigr> = \bigl< \mathbf{p}' \mathbf{q}' | V_{\text{3N}} | \mathbf{p} \mathbf{q} \bigr> \prod_i f^{\text{long}}_R(\mathbf{Q}_i)$ & $V^{\pi, \text{reg}}_{\text{3N}} (\mathbf{r}_{ij}) = f^{\text{long}}_R (\mathbf{r}_{ij}) V^{\pi}_{\text{3N}} (\mathbf{r}_{ij}) \overset{\text{FT}}{\rightarrow} \bigl< \mathbf{p}' \mathbf{q}' | V_{\text{3N}}^{\pi, \text{reg}} | \mathbf{p} \mathbf{q} \bigr>$ \\
     & $\bigl< \mathbf{p}' \mathbf{q}' | V_{\text{3N}}^{\delta, \text{reg}} | \mathbf{p} \mathbf{q} \bigr> = f^{\text{short}}_{\Lambda} (\mathbf{p}'_{\delta}) \bigl< \mathbf{p}' \mathbf{q}' | V_{\text{3N}}^{\delta} | \mathbf{p} \mathbf{q} \bigr> f^{\text{short}}_{\Lambda} (\mathbf{p}_{\delta}) $ & $\bigl< \mathbf{p}' \mathbf{q}' | V_{\text{3N}}^{\text{reg}} | \mathbf{p} \mathbf{q} \bigr> = f^{\text{short}}_{\Lambda} (\mathbf{p}'_{\delta}) \bigl< \mathbf{p}' \mathbf{q}' | V_{\text{3N}}^{\pi, \text{reg}} | \mathbf{p} \mathbf{q} \bigr> f^{\text{short}}_{\Lambda} (\mathbf{p}_{\delta})$ \\
\end{tabular}
\caption{Different regularization schemes for 3N interactions. We have suppressed all
spin and isospin quantum numbers for the sake of simple notation. For all
shown schemes only spin- and isospin-independent regulator functions have been
applied so far. For each choice we list the reference in which the scheme was
first proposed. We stress that a chosen regularization should be applied
consistently to NN and 3N interactions. $V_{\text{3N}}^{\pi}$ denotes the
long-range part of a given interaction contribution, like pion-exchange
interactions, whereas $V_{\text{3N}}^{\delta}$ denotes the short-range contact
contributions. The total unregularized and regularized interactions are then
given by $V_{\text{3N}} = V_{\text{3N}}^{\pi} V_{\text{3N}}^{\delta}$ and
$V_{\text{3N}}^{\text{reg}} = V_{\text{3N}}^{\pi, \text{reg}}
V_{\text{3N}}^{\delta, \text{reg}}$, respectively. $\mathbf{p}_{\delta}$
denotes the momenta of the particles that interact via the short-range force
$V_{\text{3N}}^{\delta}$ (see also Figure~\ref{fig:semilocal_N2LO_diags}),
$\alpha_n (R)$ is a normalization constant, and ``FT'' denotes the Fourier
transform to momentum space.}
\label{tab:regularization}
\end{table}

The framework discussed in Section~\ref{sec:PWD_3NF_local} allows to perform
efficiently a partial-wave decomposition in momentum basis states for 3N
interactions that are local or only contain polynomial nonlocal terms.
However, so far we have neglected the problem of regularizing the interaction.
In general, all interactions for nuclear structure calculations are
parametrized in terms effective low-energy degrees of freedom, i.e., neutrons,
protons and pions (see Section~\ref{sec:chiral_EFT}). That means the
description is, at least implicitly, based on some low-energy approximation of
the underlying quark-gluon dynamics described by QCD. However, as with any
low-energy effective theory, this description becomes inefficient beyond some
ultraviolet momentum scale, or, equivalently, below a certain interparticle
distance scale. The presence of this breakdown scale implies that the
interaction matrix elements need to be regularized in order to separate the
low-energy from the high-energy part. While the low-energy part is described
in terms of the dynamics of the effective degrees of freedom, the
contributions from high-energy physics is implicitly encoded in the low-energy
couplings. In practice, the regularization is achieved by the multiplication
of the interaction matrix elements with a regulator function $f_{\Lambda}$
($f_R$), which suppresses the contributions beyond a momentum scale $\Lambda$
(below a distance scale $R$).

There are currently active ongoing discussions about over which range of
values these scales should be varied and to which extent the effects from
these variations can be absorbed in changes of the effective low-energy
couplings. These questions are directly connected to fundamental questions
regarding the power counting of the underlying chiral EFT and
renormalizability. However, in this work we will not discuss these conceptual
questions but refer the interested reader to, e.g., Ref.~\cite{Hamm19Rev}, for
details. Instead, we will in the following discuss the practical
implementation of different type of regularization schemes for 3N
interactions. These are independent of any particular underlying power
counting scheme.

The different regularization strategies can be characterized by the nature of
the regulator function:
\begin{itemize}
\item First, we can categorize the regularization into
a \textit{momentum-space} or \textit{coordinate-space} formulation. In the
first case the regulator function is a general function of all Jacobi momenta
in some chosen basis representation $\{ab\}$:
\begin{equation}
f_{\Lambda} = f_{\Lambda} (\mathbf{p},\mathbf{q},\mathbf{p}',\mathbf{q}') \, .
\end{equation}
This regulator function is then applied as a simple multiplicative factor to
leading-order 3N contributions (see, e.g. Eq.~(\ref{eq:V3N_Fadddecomp})):
\begin{equation}
V_{\text{3N}}^{\text{reg}} = V_{\text{3N}}^{\text{reg}} (\mathbf{p},\mathbf{q},\mathbf{p}',\mathbf{q}') f_{\Lambda} (\mathbf{p},\mathbf{q},\mathbf{p}',\mathbf{q}') \, .
\end{equation}
For higher-order 3N contributions involving loop structures the regulator
functions can in general also be applied to internal loop momenta.

Accordingly, in coordinate space the regulator function depends in general on
all relative coordinates
\begin{equation}
f_R = f_R(\mathbf{r},\mathbf{s},\mathbf{r}',\mathbf{s}') \, .
\label{eq:regulator_coordinate_space}
\end{equation}
In the present work we will not discuss methods to directly apply regulator
functions in coordinate space since the calculation and decomposition of the
3N interactions is performed in momentum space. Instead, we perform a Fourier
transform of the coordinate-space regulators shown in
Eq.~(\ref{eq:regulator_coordinate_space}) to momentum space and apply them in
the basis discussed in Section~\ref{sec:general_3N_decomp} via convolution
integrals (see below for details).

\item Second, the regulator function can be categorized into \textit{local} and
\textit{nonlocal} regulator functions. According to the discussion in
Section~\ref{sec:PWD_3NF_local}, in momentum space local regulator functions
are functions of momentum transfers only, i.e., differences of Jacobi momenta:
\begin{equation}
f_{\Lambda}^{\text{local}} = f_{\Lambda} (\mathbf{p}' - \mathbf{p}, \mathbf{q}' - \mathbf{q}) = f_{\Lambda} (\tilde{\mathbf{p}}, \tilde{\mathbf{q}}) \, ,
\end{equation}
while in coordinate space the regulator function is only a function of the relative distances $\mathbf{r}' = \mathbf{r}$ and $\mathbf{s} = \mathbf{s}'$:
\begin{equation}
f_R^{\text{local}} = f_R(\mathbf{r},\mathbf{s}) \, .
\end{equation}
Nonlocal regulator functions can depend on more general combinations
of Jacobi coordinates. In practice also combinations of local and nonlocal
regulator functions within one scheme have been applied to the short- and
long-range contributions to 3N interactions (see
Table~\ref{tab:regularization}).
\end{itemize}

Table~\ref{tab:regularization} summarizes different regularization schemes for
3N interactions that have been developed and applied in recent years. Ideally,
a chosen regularization prescription should be used consistently for NN and 3N
interactions. In the following we discuss the different regularization schemes
and their practical implementation in the partial-wave momentum basis defined
in Section~\ref{sec:PWD_3NF_local} in more detail.

\subsubsection{Nonlocal momentum-space regularization}
\label{sec:nonlocal_momentum}

Nonlocal momentum regularizations were originally applied to the first
generation of ``high-precision'' NN interactions developed within chiral
EFT~\cite{Epel99nuclforc,Ente03EMN3LO,Epel05EGMN3LO}. For these interactions
the following regulator form was used:
\begin{equation}
V_{\text{NN}}^{\text{reg}} = f_{\Lambda} (\mathbf{p}') V_{\text{NN}}^{\text{reg}} (\mathbf{p}, \mathbf{p}') f_{\Lambda} (\mathbf{p}) \, , 
\end{equation}
with
\begin{equation}
f_{\Lambda} (\mathbf{p}) = \exp \bigl[ -(\mathbf{p}^2/\Lambda^2)^n \bigr] = f_{\Lambda} (p) \, .
\label{eq:reg_NN_nonlocal_mom}
\end{equation}
Here $\Lambda$ is some chosen momentum cutoff scale, $n$ some exponent and $p
= |\mathbf{p}|$. This prescription is a natural choice in the sense that the
square of the Jacobi momentum is proportional to the relative kinetic energy
of the initial and final states and hence serves as a natural measure to
characterize the high-energy part of the Hilbert space. In
Ref.~\cite{Epel02fewbody} a natural extension of this regulator to three-body
interactions was proposed. From Eq.~(\ref{eq:kinetic_energy_Jacobi_genmasses})
it follows that the three-body intrinsic kinetic energy in the center-of-mass
reference frame for $m = m_1 = m_2 = m_3$ takes the following form:
\begin{align}
T_{\text{rel}} &= \frac{1}{2 m} \left[ \sum_{i=1}^3 \mathbf{k}_i^2 - \frac{1}{3} \bigl( \sum_{i=1}^3 \mathbf{k}_i \bigr)^2 \right] = \frac{1}{6 m} \left[ (\mathbf{k}_2 - \mathbf{k}_1)^2 + (\mathbf{k}_3 - \mathbf{k}_2)^2 - (\mathbf{k}_1 - \mathbf{k}_3)^2 \right] = \frac{1}{m} \left[ \mathbf{p}_{\{ab\}}^2 + \frac{3}{4} \mathbf{q}_{\{ab\}}^2 \right] \, ,
\label{eq:kinetic_energy_equal_masses}
\end{align}
for any basis representation $\{ab\}$. This leads to the following natural extension of Eq.~(\ref{eq:reg_NN_nonlocal_mom}) to the three-body case:
\begin{equation}
f_{\Lambda} (\mathbf{p},\mathbf{q}) = \exp \bigl[ - \bigl( (\mathbf{p}^2 + \tfrac{3}{4} \mathbf{q}^2) / \Lambda^2 \bigr)^n \bigr] = f_{\Lambda} (p,q) \, ,
\label{eq:nonlocal_regulator}
\end{equation}
with $p = |\mathbf{p}|$ and $q = |\mathbf{{}q}|$. In this case, the
regularization of the initial and final states again factorizes and the
regularized interaction is given by:
\begin{equation}
V_{\text{3N}}^{\text{reg}} = f_{\Lambda} (p',q') V_{\text{3N}} f_{\Lambda} (p,q) \, .
\label{eq:nonlocal_regulator_operator}
\end{equation}
The choice in Eq.~(\ref{eq:nonlocal_regulator}) is particularly convenient
since this nonlocal regulator only depends on the absolute values of the
Jacobi momenta. Consequently, the regulator does not affect the partial-wave
decomposition and the regularized matrix elements can be obtained by a trivial
multiplicative factor to the unregularized partial-wave matrix elements shown
in Eq.~(\ref{eq:Faddeev_components}):
\begin{equation}
\tensor*[_{\{ab\}}]{\bigl< p' q' \alpha' | V_{\text{3N}}^{(i), \text{reg}} | p q \alpha \bigr>}{_{\{ab\}}} = f_{\Lambda} (p',q') \: \tensor*[_{\{ab\}}]{\bigl< p' q' \alpha' | V_{\text{3N}}^{(i)} | p q \alpha \bigr>}{_{\{ab\}}} \: f_{\Lambda} (p,q) \, .
\label{eq:regulator_nonlocal_Faddeev_components}
\end{equation}
Since the expression for the relative kinetic energy
Eq.~(\ref{eq:kinetic_energy_equal_masses}) is identical in all basis
representations $\{ab\}$, i.e., invariant under application of the permutation
operators $P_{123}$ and $P_{132}$, we can also equivalently apply the
regulator functions to the antisymmetrized interaction defined in
Eq.~(\ref{eq:Faddeev_antisymmetrized}):
\begin{equation}
\bigl< p' q' \alpha' | V_{\text{3N}}^{\text{as}, \text{reg}} | p q \alpha \bigr> = f_{\Lambda} (p',q') \: \bigl< p' q' \alpha' | V_{\text{3N}}^{\text{as}} | p q \alpha \bigr> \: f_{\Lambda} (p,q) \, ,
\label{eq:regulator_nonlocal_antisymmetrized}
\end{equation}
i.e., the regularization commutes with the antisymmetrization operation for this
type of regulator.

Applying the regulator on the operator level like in
Eq.~(\ref{eq:nonlocal_regulator_operator}) or at the level of the partial-wave
matrix elements as shown in
Eqs.~(\ref{eq:regulator_nonlocal_Faddeev_components})
and~(\ref{eq:regulator_nonlocal_antisymmetrized}) all lead to identical
results. This is a particular property of the specific regulator choice
Eq.~(\ref{eq:nonlocal_regulator}). This property has the great practical
advantage that it is sufficient to explicitly calculate only unregularized
matrix elements. The regulator functions, i.e., the values for the exponent $n$
and the cutoff scale $\Lambda$ in Eq.~(\ref{eq:nonlocal_regulator}), can be
specified at a later stage after the partial wave decomposition in a very
flexible and convenient way.

\subsubsection{Local momentum-space regularization}
\label{sec:local_momentum}

In contrast to the nonlocal regulator discussed in the previous section local
regulators depend by definition on momentum transfers and hence also on the
angles between Jacobi momenta. As a consequence, local regulators naturally
couple different partial waves in the basis defined in Eq.~(\ref{eq:Jj_bas}),
which makes it necessary to incorporate the regulators before the partial-wave
decomposition. We start from a particular choice for the Faddeev component $i$
and incorporate the regulator functions:
\begin{equation}
V_{\text{3N}}^{(i), \text{reg}} =  V^{(i)}_{\text{3N}} \prod_{j=1}^{N_j} f_{\Lambda} (\mathbf{Q}_j) \, ,
\label{eq:local_regulator_operator}
\end{equation}
where $\mathbf{Q}_j = \mathbf{k}'_j - \mathbf{k}_j$ are the momentum
transfers and the index $j$ runs over one or multiple momentum transfer
variables of a given 3N interaction. Typically, exponential forms were chosen
for $f_{\Lambda}$, similarly to the nonlocal regulator
Eq.~(\ref{eq:nonlocal_regulator}). For example, in Ref.~\cite{Navr07local3N}
the form $f_{\Lambda} (\mathbf{Q}) = \exp \bigl[ - (\mathbf{Q}^2/\Lambda^2)^2
\bigr]$ and $N_j = 2$ was used, and in Ref.~\cite{Epel19Bayes} the regulator
function $f_{\Lambda} (\mathbf{Q}) = \exp \bigl[ - (\mathbf{Q}^2
+ m_{\pi}^2)/\Lambda^2 \bigr]$ was applied to each long-range pion exchange
interaction of a given diagram, where $\mathbf{Q}_j$ is the momentum carried
by the pion.

The regularized interaction defined in
Eq.~$(\ref{eq:local_regulator_operator})$ can be straightforwardly decomposed
in a partial-wave representation using the algorithm discussed in
Section~\ref{sec:PWD_3NF_local} since the regulator preserves the local nature
of the interaction. Note, however, that the regularization in
Eq.~(\ref{eq:local_regulator_operator}) does in general not commute with the
antisymmetrization operation, in contrast to the nonlocal regulator in
Eq.~(\ref{eq:nonlocal_regulator}). This leads to ambiguities regarding the
choice of operators for the three-body interactions, which are absent for
unregularized interactions or nonlocally regularized interactions using a
function of the form Eq.~(\ref{eq:nonlocal_regulator}). Consider as an example
a pure contact interaction, which can be parametrized in terms of different
spin-isospin operators:
\begin{align}
V_{\text{3N}}^{\text{contact}} &= \sum_{i \neq j \neq k} \left[ \beta_1 + \beta_2 \boldsymbol{\sigma}_i \cdot \boldsymbol{\sigma}_j + \beta_3 \boldsymbol{\tau}_i \cdot \boldsymbol{\tau}_j + \beta_4 \boldsymbol{\sigma}_i \cdot \boldsymbol{\sigma}_j \boldsymbol{\tau}_i \cdot \boldsymbol{\tau}_j + \beta_5 \boldsymbol{\sigma}_i \cdot \boldsymbol{\sigma}_j \boldsymbol{\tau}_j \cdot \boldsymbol{\tau}_k + \beta_6 \left( (\boldsymbol{\sigma}_i \times \boldsymbol{\sigma}_j) \cdot  \boldsymbol{\sigma}_k \right) \left( (\boldsymbol{\tau}_i \times \boldsymbol{\tau}_j) \cdot \boldsymbol{\tau}_k \right) \right] \nonumber \\
&= V_{\text{3N}}^{\beta_1} + V_{\text{3N}}^{\beta_2} + V_{\text{3N}}^{\beta_3} + V_{\text{3N}}^{\beta_4} + V_{\text{3N}}^{\beta_5} + V_{\text{3N}}^{\beta_6} \, .
\label{eq:beta_i_3N_cont}
\end{align}
Antisymmetrization leads to the following relations:
\begin{align}
\mathcal{A}_{123} V_{\text{3N}}^{\beta_2} &= \mathcal{A}_{123} V_{\text{3N}}^{\beta_3} = - \mathcal{A}_{123} V_{\text{3N}}^{\beta_1} \nonumber \\ 
\mathcal{A}_{123} V_{\text{3N}}^{\beta_4} &= - \mathcal{A}_{123} V_{\text{3N}}^{\beta_5} = -3 \: \mathcal{A}_{123} V_{\text{3N}}^{\beta_1} \nonumber \\
\mathcal{A}_{123} V_{\text{3N}}^{\beta_6} &= -12 \: \mathcal{A}_{123} V_{\text{3N}}^{\beta_1} \, .
\end{align}
These relations imply that all operators are linearly dependent and it is
hence sufficient to choose just one of the operators $V_{\text{3N}}^{\beta_i}$
(see also discussion in Ref.~\cite{Epel02fewbody}). By convention, the
interaction $V_{\text{3N}}^{\beta_3}$ is usually chosen to define the
low-energy constant $c_E$ (see Eq.~(\ref{eq:V3N_expvalue})). However, for
local regulators this symmetry, usually referred to as Fierz rearrangement
freedom of Fierz symmetry, is generally not fulfilled anymore. This is in
particular the case when the local regulator is not applied in a symmetric way
to all momentum transfers in 3N interactions. In Ref.~\cite{Huth17fierz} the
effects of this ambiguity in NN interactions was studied in few-body systems
and neutron matter, whereas in Ref.~\cite{Navr07local3N} the effect of
different coordinate choices in the local regulators for the 3N interactions
was studied based on ground-state energies of $^3$H. Ref.~\cite{Lov113NNM}
studied the impact of this ambiguity for local 3N interactions for nuclear
matter. Generally, these ambiguities at a given order should be absorbable by
operators at higher orders in the chiral expansion, but more detailed studies
are needed to demonstrate this explicitly.

\begin{figure}[t!]
\centering
\includegraphics[width=0.95\textwidth]{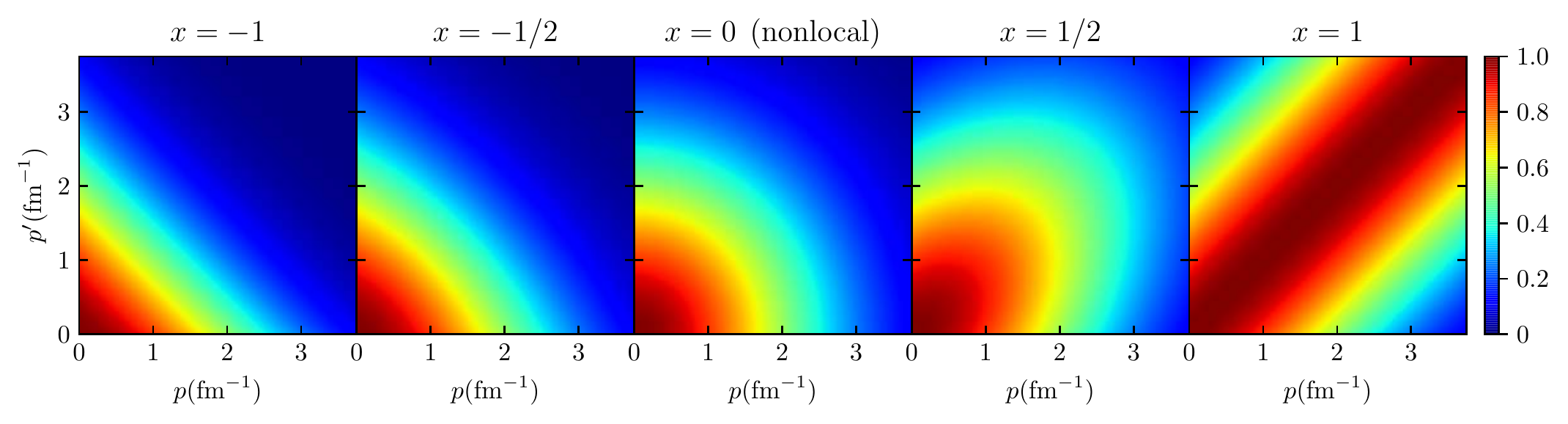}
\caption{Phase-space weights of the local regulator function $f_{\Lambda}^{\text{local}} (\mathbf{p},\mathbf{p}') =
\exp \left[ -(\mathbf{p}' - \mathbf{p})^2 /\Lambda^2 \right] = \exp \left[ -(p'^2 + p^2 - 2 p' p x) /\Lambda^2 \right]$
with ${x = \cos \theta_{\mathbf{p} \mathbf{p}'}}$ and $\Lambda = 500$ MeV. The
different panels show the phase-space contributions for different angles $x$.
The case $x=0$, i.e., $\theta_{\mathbf{p} \mathbf{p}'} = \tfrac{\pi}{2}$ (central
panel), corresponds to the nonlocal regulator.}
\label{fig:regulators_contour}
\end{figure}

The local and nonlocal regulators discussed above can differ quite
substantially, depending on the kinematical regime. We illustrate this in
Figure~\ref{fig:regulators_contour} by showing the phase space contributions
of a local regulator of form $f_{\Lambda}^{\text{local}}
(\mathbf{p},\mathbf{p}') =
\exp \bigl[ -(\mathbf{p}' - \mathbf{p})^2 /\Lambda^2 \bigr] =
f_{\Lambda}^{\text{local}} (p,p',\cos
\theta_{\mathbf{p} \mathbf{p}'})$ for different angles $\theta_{\mathbf{p}
\mathbf{p}'}$ and for the cutoff scale $\Lambda = 500 $ MeV. Obviously, for $\cos
\theta_{\mathbf{p} \mathbf{p}'} = 0$, i.e., $\theta_{\mathbf{p} \mathbf{p}'} =
\pi/2$, the local regulator agrees by construction with the nonlocal regulator
defined in Eq.~(\ref{eq:reg_NN_nonlocal_mom}) with $n=1$. Furthermore, in the
kinematical regime with small $x$ both regulators agree reasonably well.
However, when both Jacobi momenta, $\mathbf{p}$ and $\mathbf{p}'$, are getting
close to being aligned or antialigned with each other, the regulators show
significant differences. In particular for the case $x=1$, i.e., $\mathbf{p} =
\mathbf{p}'$ the local regulator exhibits a band diagonal structure and does not suppress 
contributions at large Jacobi momenta at all.

Besides such technical differences between local and nonlocal regulators,
there are also conceptual differences. In Ref.~\cite{Epel15improved} it is
argued that nonlocal cutoff functions of the form
Eq.~(\ref{eq:reg_NN_nonlocal_mom}) lead to distortions of the analytic
structure of the partial-wave scattering amplitude around the threshold since
it affects the discontinuity across the left-hand cuts (see also discussion in
Refs.~\cite{Gasp12analytic,Olle14dispersion}). Local regulators, in contrast,
can remove the short-range parts of the pion-exchange interactions and hence
make an additional spectral function regularization obsolete. In addition, it
is argued that locally regularized interactions lead to a better description
of scattering phase shifts even at relatively high energies due to reduced
finite-cutoff artifacts (see Refs.~\cite{Epel15improved,Rein17semilocal} for
details).

\subsubsection{Semilocal momentum-space regularization}
\label{sec:semilocal_momentum}

The semilocal regularization scheme first presented in
Refs.~\cite{Rein17semilocal,Epel19Bayes} combines features of the nonlocal and
local regularizations discussed in the previous two sections. Specifically, in
the semilocal regularization approach the long-range pion-exchange
contributions are regularized via local regulator functions, whereas the
short-range point couplings are regularized nonlocally (see also
Ref.~\cite{Epel19SMSreview} for details). This has the practical advantage
that the regularization of the short-range parts does not induce a coupling of
partial waves and hence the low-energy couplings in the NN interaction can be
fixed independently in different partial waves, e.g. by fitting them to
extracted scattering phase shifts in the corresponding channel.

\begin{figure}[t]
\centering
\begin{minipage}[c]{0.3\textwidth}
\centering
\begin{tikzpicture} 
\begin{feynman}
\vertex (a) at (0,0) {\(1\)}; 
\vertex (b) at (1.2,0) {\(2\)};
\vertex (c) at (2.4,0) {\(3\)};
\vertex [dot] (d) at (0,1) {}; 
\vertex [blob, /tikz/minimum size=5pt] (e) at (1.2,1) {};
\vertex [dot] (f) at (2.4,1) {};
\vertex (g) at (0,2) {\(1'\)}; 
\vertex (h) at (1.2,2) {\(2'\)};
\vertex (i) at (2.4,2) {\(3'\)};
\vertex (j) at (1.2,-0.7) {(a)};
\diagram* {
(a) -- [fermion, line width=0.25mm] (d) -- [fermion, line width=0.25mm] (g);
(b) -- [fermion, line width=0.25mm] (e) -- [fermion, line width=0.25mm] (h);
(c) -- [fermion, line width=0.25mm] (f) -- [fermion, line width=0.25mm] (i);
(e) -- [charged scalar, edge label=$\mathbf{Q}_1$, color=blue] (d);
(e) -- [charged scalar, edge label'=$\mathbf{Q}_3$, color=blue] (f);
};
\end{feynman}
\end{tikzpicture}
\end{minipage}
\hspace{0.3cm}
\begin{minipage}[c]{0.3\textwidth}
\centering
\begin{tikzpicture} 
\begin{feynman}
\vertex (a) at (0,0) {\(1\)}; 
\vertex (b) at (1.2,0) {\(2\)};
\vertex (c) at (2.4,0) {\(3\)};
\vertex [dot] (d) at (0,1) {}; 
\vertex [blob, /tikz/minimum size=6pt, color=red] (e) at (1.8,1) {};
\vertex (f) at (0,2) {\(1'\)}; 
\vertex (g) at (1.2,2) {\(2'\)};
\vertex (h) at (2.4,2) {\(3'\)};
\vertex (i) at (1.2,-0.7) {(b)};
\diagram* {
(a) -- [fermion, line width=0.25mm] (d) -- [fermion, line width=0.25mm] (f);
(b) -- [fermion, line width=0.25mm] (e) -- [fermion, line width=0.25mm] (g);
(c) -- [fermion, line width=0.25mm] (e) -- [fermion, line width=0.25mm] (h);
(e) -- [charged scalar, edge label=$\mathbf{Q}_1$, color=blue] (d);
};
\end{feynman}
\end{tikzpicture}
\end{minipage}
\hspace{0.1cm}
\begin{minipage}[c]{0.25\textwidth}
\centering
\begin{tikzpicture} 
\begin{feynman}
\vertex (a) at (0,0) {\(1\)}; 
\vertex (b) at (0.8,0) {\(2\)};
\vertex (c) at (1.6,0) {\(3\)};
\vertex [blob, /tikz/minimum size=6pt, color=red] (d) at (0.8,1) {};
\vertex (e) at (0,2) {\(1'\)}; 
\vertex (f) at (0.8,2) {\(2'\)};
\vertex (g) at (1.6,2) {\(3'\)};
\vertex (h) at (0.8,-0.7) {(c)};
\diagram* {
(a) -- [fermion, line width=0.25mm] (d) -- [fermion, line width=0.25mm] (e);
(b) -- [fermion, line width=0.25mm] (d) -- [fermion, line width=0.25mm] (f);
(c) -- [fermion, line width=0.25mm] (d) -- [fermion, line width=0.25mm] (g);
};
\end{feynman}
\end{tikzpicture}
\end{minipage}
\caption{Semilocal momentum-space regularization of the 3N interactions at
N$^2$LO. The pion exchange contributions (blue) to all interactions are
regularized locally via the long-range regulator $f_R^{\text{long}}
(\mathbf{Q}_i)$, whereas the short-range parts (red) are regularized
nonlocally by the regulator $f_{\Lambda}^{\text{short}} (\mathbf{p}^2_{\delta})$.
Here, the momentum $\mathbf{p}_{\delta}$ denotes the momentum scale related to
the relative kinetic energy in the initial and final states of those particles
that interact via the short-range coupling, i.e., specifically, for
diagram (b) $\mathbf{p}_{\delta}^2 = \left( (\mathbf{k}_2 -
\mathbf{k}_3)/2 \right)^2$ and for diagram (c) $\mathbf{p}_{\delta}^2 =
\frac{1}{6} \bigl[ (\mathbf{k}_2 -
\mathbf{k}_1)^2 + (\mathbf{k}_3 - \mathbf{k}_2)^2 + (\mathbf{k}_1 -
\mathbf{k}_3)^2\bigr] = \mathbf{p}^2 + \tfrac{3}{4} \mathbf{q}^2$.}
\label{fig:semilocal_N2LO_diags}
\end{figure}
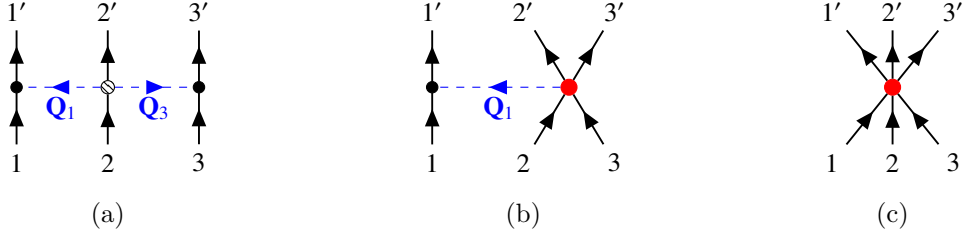

We illustrate the semilocal regularization by applying it to the 3N
contributions at N$^2$LO in chiral EFT for the interaction terms shown in
Figure~\ref{fig:semilocal_N2LO_diags}. The regularization of the purely
long-range leading-order two-pion exchange contributions is formally identical
to the local regularization of Ref.~\cite{Navr07local3N}. In diagram (a) of
Figure~\ref{fig:semilocal_N2LO_diags} the pions carry the momenta
$\mathbf{Q}_1$ and $\mathbf{Q}_3$. Hence the regularized interaction is given
by
\begin{equation}
V_{\text{3N}}^{c_i, \text{reg}} (\mathbf{Q}_1, \mathbf{Q}_3) = V_{\text{3N}}^{c_i} (\mathbf{Q}_1, \mathbf{Q}_3) f_{\Lambda}^{\text{long}} (\mathbf{Q}_1) f_{\Lambda}^{\text{long}} (\mathbf{Q}_3) \, .
\end{equation}
In Ref.~\cite{Epel19Bayes} the particular form $f_{\Lambda}^{\text{long}}
(\mathbf{Q}) = \exp \bigl[ - (\mathbf{Q}^2
+ m_{\pi}^2)/\Lambda^2 \bigr]$ was chosen (see also Table~\ref{tab:regularization}).

The intermediate-range diagram (b) in Figure~\ref{fig:semilocal_N2LO_diags},
proportional to the coupling $c_D$, consists of a long-range pion-exchange
part and short-range coupling. Here, the pion exchange is regularized via the
local long-range regulator $f_{\Lambda}^{\text{long}}$ (see above) and the
short-range two-point coupling via a nonlocal regulator of the form
$f_{\Lambda}^{\text{short}} =
\exp \bigl[ - (\mathbf{p}_{\delta}^2/\Lambda^2) \bigr]$, where the momenta 
$\mathbf{p}_{\delta}$ ($\mathbf{p}'_{\delta}$) are the initial (final) state
relative momenta of the two particles interacting via the point coupling. For
the practical calculation we choose a particular basis $\{ab\}$. In the present
case it is most convenient to choose basis $\{23\}$ since in this case the
momentum $\mathbf{p}_{\delta}$ is given by (see
Table~\ref{tab:Jacobi_momenta_crosstable})
\begin{equation}
\mathbf{p}_{\delta} = \frac{\mathbf{k}_2 - \mathbf{k}_3}{2} = \mathbf{p}_{\{23\}} \, ,
\end{equation}
i.e., the argument of the regulator functions is independent of any angles. On the other hand,
choosing another basis representation, e.g. $\{12\}$, leads to
\begin{equation}
\mathbf{p}_{\delta} = \frac{\mathbf{k}_2 - \mathbf{k}_3}{2} = - \mathbf{p}_{\{12\}} - \frac{3}{4} \mathbf{q}_{\{12\}} \, ,
\end{equation}
and the application of the nonlocal regulators obviously becomes much more
intricate due to the dependence on the angle between the vectors
$\mathbf{p}_{\{12\}}$ and $\mathbf{q}_{\{12\}}$. Of course, both choices
eventually lead to identical results for the antisymmetrized interaction.

That means in total we obtain for the regularized interaction:
\begin{equation}
V_{\text{3N}}^{c_D, \text{reg}} (\mathbf{Q}_1, \mathbf{p}_{\delta}, \mathbf{p}'_{\delta}) = V_{\text{3N}}^{c_D} (\mathbf{Q}_1) f_{\Lambda}^{\text{long}} (\mathbf{Q}_1) f_{\Lambda}^{\text{short}} (\mathbf{p}_{\delta}) f_{\Lambda}^{\text{short}} (\mathbf{p}'_{\delta}) \, .
\end{equation}
In Ref.~\cite{Epel19Bayes} the regulator form was chosen to be
$f_{\Lambda}^{\text{short}} (\mathbf{p}) = \exp \left[ -
(\mathbf{p}^2/\Lambda^2) \right]$.

The regularization of the purely short-range interaction, proportional to the
coupling $c_E$ (diagram (c) in Figure~\ref{fig:semilocal_N2LO_diags}), reduces
to the nonlocal regularization. Here all three particles participate in the
short-range interaction, i.e.,
\begin{equation}
\mathbf{p}_{\delta}^2 = \frac{1}{6} \bigl[ (\mathbf{k}_2 - \mathbf{k}_1)^2 + (\mathbf{k}_3 - \mathbf{k}_2)^2 + (\mathbf{k}_1 - \mathbf{k}_3)^2 \bigr] = \mathbf{p}^2 + \frac{3}{4} \mathbf{q}^2 \, ,
\end{equation}
and for the regularized interaction we obtain:
\begin{equation}
V_{\text{3N}}^{c_E, \text{reg}} (\mathbf{p}_{\delta}, \mathbf{p}'_{\delta}) = f_{\Lambda}^{\text{short}} (\mathbf{p}_{\delta}) V_{\text{3N}}^{c_E} f_{\Lambda}^{\text{short}} (\mathbf{p}'_{\delta}) \, .
\end{equation}
In Ref.~\cite{Epel19Bayes} a Gaussian form was chosen, i.e.:
\begin{equation}
f_{\Lambda}^{\text{short}} (\mathbf{p}_{\delta}) = \exp \bigl[ - (\mathbf{p}^2 + \tfrac{3}{4} \mathbf{q}^2) / \Lambda^2 \bigr] = f_{\Lambda}^{\text{short}} (p,q) \, .
\end{equation}
Since the argument does not depend on any angles between Jacobi momenta this
regularization can directly be applied to the partial-wave matrix elements:
\begin{equation}
\tensor*[_{\{ab\}}]{\bigl< p' q' \alpha' | V_{\text{3N}}^{c_E, \text{reg}} | p q \alpha \bigr>}{_{\{ab\}}} = f_{\Lambda}^{\text{short}} (p',q') \: \tensor*[_{\{ab\}}]{\bigl< p' q' \alpha' | V_{\text{3N}}^{c_E} | p q \alpha \bigr>}{_{\{ab\}}} \: f_{\Lambda}^{\text{short}} (p,q) \, .
\end{equation}

The regularization of the 3N contributions at N$^3$LO that involve loop
structures is more involved since the regulators can also be applied to
internal momenta.

\subsubsection{Local coordinate-space regularization}
\label{sec:local_coordinate}

Local interactions formulated in coordinate space play a key role for Quantum
Monte Carlo methods~\cite{Carl15RMP}. In recent years there has been
significant progress toward incorporating chiral EFT interactions in these
frameworks (see, e.g.,
Refs.~\cite{Geze13QMCchi,Geze14long,Lynn14QMCln,Tews16QMCPNM,Piar17LightNucl}).
Here we discuss strategies that allow to apply regulator functions defined in
coordinate space, i.e., $f_R =
f_R(\mathbf{r},\mathbf{s},\mathbf{r}',\mathbf{s}')$, to matrix elements of 3N
interactions. Generally there are two options to calculate regularized matrix
elements $\bigl< p' q' \alpha' | V_{\text{3N}}^{(i), \text{reg}} | p q
\alpha\bigr>$:
\begin{itemize}
\item[1.] Calculation of the Fourier transform of the regulator function
$f_R$ to momentum space and application of the regulator to the unregularized
momentum-space partial-wave matrix elements of Eq.~(\ref{eq:V3N_Fadddecomp}).
\item[2.] Computation of the coordinate-space partial-wave matrix elements of the
regularized 3N interaction
\begin{equation}
V_{\text{3N}}^{\text{reg}} = V_{\text{3N}} (\mathbf{r}, \mathbf{s}, \mathbf{r}', \mathbf{s}') f_R (\mathbf{r}, \mathbf{s}, \mathbf{r}', \mathbf{s}') \, ,
\end{equation}
and Fourier transform of the result to momentum space. The Fourier transform
of the chiral EFT 3N interactions at N$^2$LO can be performed to a large
extent analytically and can be expressed in terms of a few elementary
functions which can be computed numerically in a straightforward way (see,
e.g., appendix of Ref.~\cite{Tews15PhD}).
\end{itemize}

The second option has the advantage that the application of the regulator
function is just a simple multiplicative operation. On the other hand, the
expressions for various 3N contributions have so far only been derived in
momentum space (see Ref.~\cite{Bern083Nlong,Bern113Nshort}). In this sense the
first option is more general and we will hence focus on this method in the
following.

We consider a local interaction regularized by a local function
in coordinate space. For the sake of simple notation we parametrize all
quantities as a function of the interparticle distances $\mathbf{r}_{ij} =
\mathbf{x}_i - \mathbf{x}_j$ or the momentum transfers $\mathbf{Q}_i =
\mathbf{k}_i' - \mathbf{k}_i$. In coordinate space the regularized Faddeev component
is given by
\begin{equation}
V_{\text{3N}}^{(i),\text{reg}} = V^{(i)}_{\text{3N}} (\mathbf{r}_{12}, \mathbf{r}_{23})
f_R (\mathbf{r}_{12}, \mathbf{r}_{23}) \, .
\label{eq:def_Vreg_coord_space}
\end{equation}
Here we arbitrarily selected the two variables $\mathbf{r}_{12}$ and
$\mathbf{r}_{23}=-\mathbf{r}_{32}$. Since there are only two independent
relative distance variables ($\mathbf{r}_{12} + \mathbf{r}_{23} +
\mathbf{r}_{31} = 0$) we could have equally well chosen any other two
independent interparticle distance variables. The momentum-space
representation of the regulator is then given by
\begin{equation}
\tilde{f}_{R} (\mathbf{Q}_1,\mathbf{Q}_3) = \int d \mathbf{r}_{12} d \mathbf{r}_{23} e^{-i \mathbf{Q}_1 \cdot \mathbf{r}_{12}} e^{-i \mathbf{Q}_3 \cdot \mathbf{r}_{23}} f_R (\mathbf{r}_{12}, \mathbf{r}_{23}) \, ,
\label{eq:regulator_fourier_tranform}
\end{equation}
or equivalently the coordinate-space representation by
\begin{equation}
f_R (\mathbf{r}_{12},\mathbf{r}_{23}) = \int \frac{d \mathbf{Q}_{1}}{(2 \pi)^3} \frac{d \mathbf{Q}_{3}}{(2 \pi)^3} e^{i \mathbf{Q}_1 \cdot \mathbf{r}_{12}} e^{i \mathbf{Q}_3 \cdot \mathbf{r}_{23}} \tilde{f}_{R} (\mathbf{Q}_1, \mathbf{Q}_{3}) \, .
\label{eq:regulator_fourier_tranform_inverse}
\end{equation}
The role of the regulator function $f_R (\mathbf{r}_{12}, \mathbf{r}_{23})$ is
to suppress contributions to the interaction at small interparticle distances,
while leaving the interaction unchanged in the low-energy regime at large
distances. Consequently the Fourier transform cannot be directly computed in
the form given in Eq.~(\ref{eq:regulator_fourier_tranform}) since the
integration kernel is not suppressed at large interparticle distances.
Instead we first need to subtract the identity:
\begin{equation}
\tilde{f}_{R} (\mathbf{Q}_1,\mathbf{Q}_3) = \int d \mathbf{r}_{12} d \mathbf{r}_{23} e^{-i \mathbf{Q}_1 \cdot \mathbf{r}_{12}} e^{-i \mathbf{Q}_3 \cdot \mathbf{r}_{23}} \left[ f_R (\mathbf{r}_{12}, \mathbf{r}_{23}) - 1 \right] + (2 \pi)^6 \delta(\mathbf{Q}_1) \delta(\mathbf{Q}_3) = \tilde{f}_{R,\text{subtr}} (\mathbf{Q}_1,\mathbf{Q}_3) + (2 \pi)^6 \delta(\mathbf{Q}_1) \delta(\mathbf{Q}_3) \, .
\label{eq:regulator_fourier_tranform_subtraction}
\end{equation}
Inserting the Fourier representation Eq.~(\ref{eq:regulator_fourier_tranform}) and its inverse for the regulator and the interaction we obtain:
\begin{subequations}
\begin{align}
V_{\text{3N}}^{(i),\text{reg}} (\mathbf{Q}_1,\mathbf{Q}_3) &= \int d \mathbf{r}_{12} d \mathbf{r}_{23} e^{-i \mathbf{Q}_1 \cdot \mathbf{r}_{12}} e^{-i \mathbf{Q}_3 \cdot \mathbf{r}_{23}} V_{\text{3N}}^{(i)} (\mathbf{r}_{12}, \mathbf{r}_{23}) f_R (\mathbf{r}_{12}, \mathbf{r}_{23}) \nonumber \\
&= \int \frac{d \bar{\mathbf{Q}}_1}{(2 \pi)^3} \frac{d \bar{\mathbf{Q}}_3}{(2 \pi)^3} V_{\text{3N}}^{(i)} (\bar{\mathbf{Q}}_1, \bar{\mathbf{Q}}_3) \tilde{f}_{R} (\mathbf{Q}_1 - \bar{\mathbf{Q}}_1,\mathbf{Q}_3 - \bar{\mathbf{Q}}_3) 
\label{eq:regulator_no_cancallation}
\\
&= V_{\text{3N}}^{(i)} (\mathbf{Q}_1, \mathbf{Q}_3) + \int \frac{d
\bar{\mathbf{Q}}_1}{(2 \pi)^3} \frac{d \bar{\mathbf{Q}}_3}{(2 \pi)^3}
V_{\text{3N}}^{(i)} (\bar{\mathbf{Q}}_1, \bar{\mathbf{Q}}_3)
\tilde{f}_{R,\text{subtr}} (\mathbf{Q}_1 - \bar{\mathbf{Q}}_1,\mathbf{Q}_3 -
\bar{\mathbf{Q}}_3) \, ,
\label{eq:regulator_cancallation}
\end{align}
\end{subequations}
where we inserted the Fourier representation of the regulator and the
interaction (Eq.~(\ref{eq:regulator_fourier_tranform_inverse})) in the second
step. The key point of this result is the fact that the regularized
interaction is given by a sum of two terms. While the first term is just the
unregularized interaction, the second term is a well-defined integral,
which in general needs to be computed numerically. This step involves some
intricate numerical problems that are most obvious in the regime of large
momenta. Since the regulator cuts off high-energy physics at small distances
the regularized interaction in momentum space should be suppressed at large
momentum transfers. However, note that in general the unregularized
interaction $V_{\text{3N}} (\mathbf{Q}_1,
\mathbf{Q}_3)$ is not suppressed at large momenta, which means that also the
second term in Eq.~(\ref{eq:regulator_cancallation}) cannot be suppressed in
this kinematical region. Eventually, a very delicate cancellation between these
two terms is required in order to obtain a regularized interaction that has no
contributions at large momenta. In fact, the direct implementation of
Eq.~(\ref{eq:regulator_cancallation}) can lead to significant numerical noise
contaminations, in particular at large momenta.

However, there is a more clever way to apply coordinate-space local
regulators. The basic idea of this method is to insert an identity in the
definition Eq.~(\ref{eq:def_Vreg_coord_space}) of the following
form\footnote{This method was originally developed by Hermann Krebs.}
\begin{equation}
V_{\text{3N}}^{(i), \text{reg}} = V_{\text{3N}}^{(i)} (\mathbf{r}_{12}, \mathbf{r}_{23}) \frac{Q(\mathbf{r}^2_{12}) Q(\mathbf{r}_{23}^2)}{Q(\mathbf{r}^2_{12}) Q(\mathbf{r}_{23}^2)}
f_R (\mathbf{r}_{12}, \mathbf{r}_{23}) \, ,
\end{equation}
where $Q(r^2)$ is at this point an arbitrary function that depends on the
square of the interparticle distance, which we will specify further below. We
then define a pre-regularized interaction in momentum space by inserting the
functions $Q$ in the \textit{numerator} (compare
Eq.~(\ref{eq:regulator_fourier_tranform})):
\begin{align}
V_{\text{3N}}^{(i), \text{prereg}} (\mathbf{Q}_1,\mathbf{Q}_3) &= \int d \mathbf{r}_{12} d \mathbf{r}_{23} e^{-i \mathbf{Q}_1 \cdot \mathbf{r}_{12}} e^{-i \mathbf{Q}_3 \cdot \mathbf{r}_{23}} Q(\mathbf{r}^2_{12}) Q(\mathbf{r}_{23}^2) V^{(i)}_{\text{3N}} (\mathbf{r}_{12}, \mathbf{r}_{23}) = Q(- \Delta_{\mathbf{Q}_1}) Q(- \Delta_{\mathbf{Q}_3}) V_{\text{3N}}^{(i)} (\mathbf{Q}_1,\mathbf{Q}_3) \, ,
\label{eq:V3N_prereg}
\end{align}
where $\Delta_{\mathbf{Q}}$ denotes the Laplacian with respect to vector
$\mathbf{Q}$. Accordingly, we define a pre-regularized regulator in
momentum space by incorporating here the functions $Q$ in the
\textit{denominator}:
\begin{equation}
\tilde{f}_R^{\text{prereg}} (\mathbf{Q}_1,\mathbf{Q}_3) = \int d \mathbf{r}_{12} d \mathbf{r}_{23} e^{-i \mathbf{Q}_1 \cdot \mathbf{r}_{12}} e^{-i \mathbf{Q}_3 \cdot \mathbf{r}_{23}} \frac{f_R (\mathbf{r}_{12}, \mathbf{r}_{23})}{Q(\mathbf{r}^2_{12}) Q(\mathbf{r}_{23}^2)} \, .
\label{eq:regulator_preregularized}
\end{equation}
Choosing the function $Q(\mathbf{r}^2)$ in a suitable way allows to render the
integral Eq.~(\ref{eq:regulator_preregularized}) finite and well defined,
without the need to subtract the identity like in the original integral in
Eq.~(\ref{eq:regulator_fourier_tranform}). For example, the choices
$Q(\mathbf{r}^2) = r^2$ or $Q(\mathbf{r}^2) = r^4$ have the desired
properties. In addition, in cases where the original regulator function
factorizes and only depends on the absolute values of the interparticle
distances, i.e., $f_R (\mathbf{r}_{12}, \mathbf{r}_{23}) = f_R
(|\mathbf{r}_{12}|) f_R (|\mathbf{r}_{23}|)$, then also the pre-regularized
regulator factorizes and can be calculated in a particularly simple way. For
example, for $Q(r^2) = r^4$ we obtain:
\begin{equation}
\tilde{f}_R^{\text{prereg}} (\mathbf{Q}) = \int d \mathbf{r} e^{-i \mathbf{Q} \cdot \mathbf{r}} \frac{1}{r^4} f_R (r) = 4 \pi \int_0^{\infty} \frac{dr}{r^2} f_R (r) j_0 (Q r) \, ,
\label{eq:regulator_prereg_factorized}
\end{equation}
where $j_l$ are the spherical Bessel functions. However, also for
non-factorizable regulator functions the computation of the Fourier
transform poses no serious problems. Finally, the
regularized interaction can then be obtained by
\begin{equation}
V_{\text{3N}}^{(i), \text{reg}} (\mathbf{Q}_1,\mathbf{Q}_3) = \int \frac{d \bar{\mathbf{Q}}_1}{(2 \pi)^6} \frac{d \bar{\mathbf{Q}}_3}{(2 \pi)^6} V_{\text{3N}}^{(i), \text{prereg}} (\bar{\mathbf{Q}}_1, \bar{\mathbf{Q}}_3) \tilde{f}_R^{\text{prereg}} (\mathbf{Q}_1 - \bar{\mathbf{Q}}_1,\mathbf{Q}_3 - \bar{\mathbf{Q}}_3) \, .
\label{eq:V3N_regul_prereg_Q}
\end{equation}
In contrast to Eq.~(\ref{eq:regulator_no_cancallation}) this integral can be
directly evaluated and no subtractions with delicate cancellations are
required for the regularization of the interaction. By choosing a basis
representation $\{ab\}$ it is now straightforward to express
Eq.~(\ref{eq:V3N_regul_prereg_Q}) in terms of Jacobi momenta:
\begin{equation}
\tensor*[_{\{ab\}}]{\bigl< \mathbf{p}' \mathbf{q}' | V_{\text{3N}}^{(i), \text{reg}} | \mathbf{p} \mathbf{q} \bigr>}{_{\{ab\}}} = \int \frac{d \mathbf{p}''}{(2 \pi)^6} \frac{d \mathbf{q}''}{(2 \pi)^6} \tensor*[_{\{ab\}}]{\bigl< \mathbf{p}' \mathbf{q}' | V_{\text{3N}}^{(i), \text{prereg}} | \mathbf{p}'' \mathbf{q}'' \bigr>}{_{\{ab\}}} \tensor*[_{\{ab\}}]{\bigl< \mathbf{p}'' \mathbf{q}'' | \tilde{f}_R^{\text{prereg}} | \mathbf{p} \mathbf{q} \bigr>}{_{\{ab\}}} \, , 
\label{eq:V3N_regul_prereg_Q_Jacobimomenta}
\end{equation}
and after including spin and isospin degrees of freedom we obtain the
following results in the momentum partial-wave representation:
\begin{align}
& \tensor*[_{\{ab\}}]{\bigl< p' q' \alpha' | V_{\text{3N}}^{(i), \text{reg}} | p q \alpha \bigr>}{_{\{ab\}}} = \int d p'' p''^2 d q'' q''^2 \sum_{\alpha''} \tensor*[_{\{ab\}}]{\bigl< p' q' \alpha' | V_{\text{3N}}^{(i), \text{prereg}} | p'' q'' \alpha'' \bigr>}{_{\{ab\}}} \tensor*[_{\{ab\}}]{\bigl< p'' q'' \alpha'' | \tilde{f}_R^{\text{prereg}} | p q \alpha \bigr>}{_{\{ab\}}} \, .
\label{eq:V3N_regul_prereg_PW}
\end{align}
The key steps for the application of Eq.~(\ref{eq:V3N_regul_prereg_PW})
consist in the computation of the pre-regularized interaction as defined in
Eq.~(\ref{eq:V3N_prereg}) and the momentum-space partial-wave decomposition of
the interaction $V_{\text{3N}}^{\text{prereg}}$ and the regulator
$f_R^{\text{prereg}}$. We will discuss these steps in more detail in the next
section, where we apply the framework presented here to semilocally regularized
3N interactions, corresponding to the NN interactions presented in
Refs.~\cite{Epel14SCSprl,Epel15improved} and later extended by contributions from
3N interactions~\cite{Epel18SCS3N}.

\subsubsection{Semilocal coordinate-space regularization}
\label{sec:semilocal_coordinate}

\begin{figure}[t!]
\centering
\includegraphics[scale=0.48]{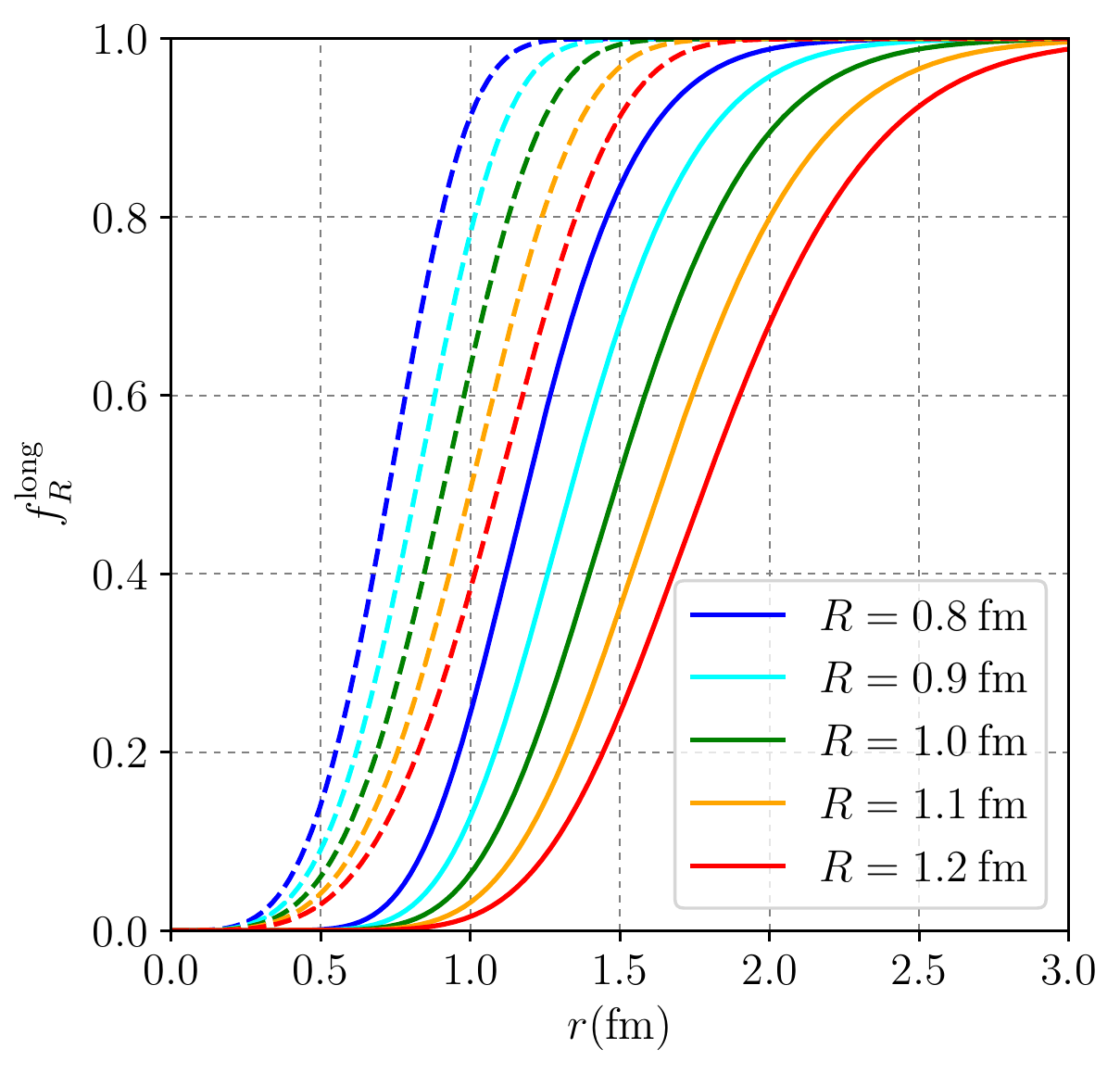}
\caption{Coordinate-space regulator functions of the ``semilocal CS''
regularization~\cite{Epel14SCSprl,Epel15improved} (solid lines, see
Eq.~(\ref{eq:semilocal_coordinate_regulator_long})) compared to those of the
``local CS'' regularization~\cite{Geze13QMCchi,Geze14long} (dashed lines)
for the coordinate-space regularization scales $R=0.8 - 1.2$ fm. Note the
different effective form and effective cutoff scales of these two sets of
regulators even when using formally the same regularization scale $R$.}
\label{fig:coord_regulators}
\end{figure}

The underlying idea of the semilocal coordinate-space regularization is the
same as for the semilocal momentum-space regularization (see
Section~\ref{sec:semilocal_momentum}), with the only difference being that the
local regulator functions for the long-range parts of the interactions are now
given in coordinate space. Specifically, in Ref.~\cite{Epel18SCS3N} the
following form was chosen:
\begin{equation}
f_R^{\text{long}} (r) = \bigl( 1 - \exp \bigl[ - r^2/R^2 \bigr] \bigr)^6 \, .
\label{eq:semilocal_coordinate_regulator_long}
\end{equation}
Here $R$ is a coordinate cutoff scale which is typically chosen to be in the range
$R=0.8 - 1.2$ fm. As shown in Figure~\ref{fig:coord_regulators}, this
regulator function is virtually vanishing at $r=0$, which implies that all
short-range coupling contributions are projected out by the local regulator
and only long range contributions from the pion exchange terms remain.
However, the figure also demonstrates that this regulator function differs
quite significantly from the regulators chosen in
Refs.~\cite{Geze13QMCchi,Geze14long} for the same values of coordinate-space
cutoff scales $R$ (see also Ref.~\cite{Hopp17WeinEVAn}). This emphasizes that
some care has to be taken when comparing interactions with the same
regularization cutoff scales but different forms of the regulators.

For interaction contributions with multiple pion-exchange interactions each
contribution is regularized via the regulator $f_R^{\text{long}}
(\mathbf{r})$. The short-range parts of the interaction are regularized in the
same way as for the semilocal momentum-space regularization by nonlocal
regulators in momentum space. In Ref.~\cite{Epel18SCS3N} a Gaussian form was
chosen:
\begin{equation}
f_{\Lambda}^{\text{short}} (\mathbf{p}) = \exp \bigl[ - p^2/\Lambda^2 \bigr] \, .
\label{eq:semilocal_coordinate_regulator_short}
\end{equation}

\begin{figure}[b!]
\centering
\begin{minipage}[c]{0.3\textwidth}
\centering
\begin{tikzpicture} 
\begin{feynman}
\vertex (a) at (0,0) {}; 
\vertex (b) at (1.2,0) {};
\vertex (c) at (2.4,0) {};
\vertex [dot] (d) at (0,1) {}; 
\vertex [blob, /tikz/minimum size=5pt] (e) at (1.2,1) {};
\vertex [dot] (f) at (2.4,1) {};
\vertex (g) at (0,2) {}; 
\vertex (h) at (1.2,2) {};
\vertex (i) at (2.4,2) {};
\vertex (j) at (1.2,-0.3) {(a)};
\vertex (k) at (-0.3,1) {\(\mathbf{x}_1\)};
\vertex (l) at (2.75,1) {\(\mathbf{x}_3\)};
\vertex (m) at (0.95,1.2) {\(\mathbf{x}_2\)};
\diagram* {
(a) -- [fermion, line width=0.25mm] (d) -- [fermion, line width=0.25mm] (g);
(b) -- [fermion, line width=0.25mm] (e) -- [fermion, line width=0.25mm] (h);
(c) -- [fermion, line width=0.25mm] (f) -- [fermion, line width=0.25mm] (i);
(e) -- [charged scalar, edge label=$\mathbf{r}_{12}$, color=blue] (d);
(e) -- [charged scalar, edge label'=$\mathbf{r}_{32}$, color=blue] (f);
};
\end{feynman}
\end{tikzpicture}
\end{minipage}
\hspace{0.3cm}
\begin{minipage}[c]{0.3\textwidth}
\centering
\begin{tikzpicture} 
\begin{feynman}
\vertex (a) at (0,0) {}; 
\vertex (b) at (1.2,0) {};
\vertex (c) at (2.4,0) {};
\vertex [dot] (d) at (0,1) {}; 
\vertex [blob, /tikz/minimum size=6pt, color=red] (e) at (1.8,1) {};
\vertex (f) at (0,2) {}; 
\vertex (g) at (1.2,2) {};
\vertex (h) at (2.4,2) {};
\vertex (i) at (1.2,-0.3) {(b)};
\vertex (k) at (-0.3,1) {\(\mathbf{x}_1\)};
\vertex (l) at (2.6,1) {\(\mathbf{x}_2 = \mathbf{x}_3\)};
\diagram* {
(a) -- [fermion, line width=0.25mm] (d) -- [fermion, line width=0.25mm] (f);
(b) -- [fermion, line width=0.25mm] (e) -- [fermion, line width=0.25mm] (g);
(c) -- [fermion, line width=0.25mm] (e) -- [fermion, line width=0.25mm] (h);
(e) -- [charged scalar, edge label=$\mathbf{r}_{12}$, color=blue] (d);
};
\end{feynman}
\end{tikzpicture}
\end{minipage}
\hspace{0.1cm}
\begin{minipage}[c]{0.25\textwidth}
\centering
\begin{tikzpicture} 
\begin{feynman}
\vertex (a) at (0,0) {}; 
\vertex (b) at (0.8,0) {};
\vertex (c) at (1.6,0) {};
\vertex [blob, /tikz/minimum size=6pt, color=red] (d) at (0.8,1) {};
\vertex (e) at (0,2) {}; 
\vertex (f) at (0.8,2) {};
\vertex (g) at (1.6,2) {};
\vertex (h) at (0.8,-0.3) {(c)};
\vertex (k) at (2.1,1) {\(\mathbf{x}_1 = \mathbf{x}_2 = \mathbf{x}_3\)};
\diagram* {
(a) -- [fermion, line width=0.25mm] (d) -- [fermion, line width=0.25mm] (e);
(b) -- [fermion, line width=0.25mm] (d) -- [fermion, line width=0.25mm] (f);
(c) -- [fermion, line width=0.25mm] (d) -- [fermion, line width=0.25mm] (g);
};
\end{feynman}
\end{tikzpicture}
\end{minipage}
\caption{Semilocal coordinate-space regularization of the 3N interactions at
N$^2$LO. The pion exchange contributions (blue) to all interactions are
regularized locally via the long-range regulator $f_R^{\text{long}}
(\mathbf{r}_{ij})$, whereas the short-range parts (red) are regularized
nonlocally in momentum space via the regulator $f_{\Lambda}^{\text{short}}
(\mathbf{p}^2_{\delta})$ (see caption of Figure
\ref{fig:semilocal_N2LO_diags}). Since all interactions terms are local,
the relation $\mathbf{x}_i = \mathbf{x}'_i$ holds.}
\label{fig:semilocal_N2LO_diags_coordinate}
\end{figure}
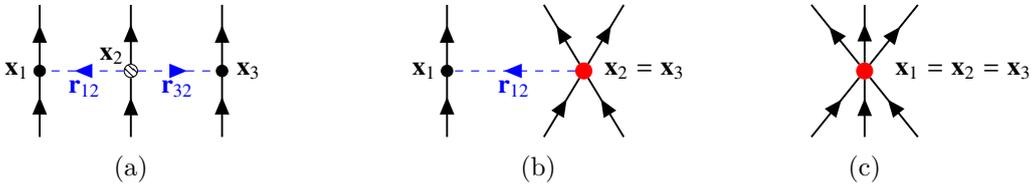

We again illustrate the regularization scheme explicitly using the 3N
interaction contributions at N$^2$LO (see Figure
\ref{fig:semilocal_N2LO_diags_coordinate}) as discussed in
Section~\ref{sec:chiral_expansion}. We start with the long-range $2\pi$ exchange
topologies, given in Eqs.~(\ref{eq:Vc}) and (\ref{eq:Vc2}). For the diagram
shown in panel (a) of Figure \ref{fig:semilocal_N2LO_diags_coordinate} we need
to apply two long-range regulators:
\begin{equation}
V_{\text{3N}}^{c_i, \text{reg}} (\mathbf{r}_{12}, \mathbf{r}_{32}) = V_{\text{3N}}^{c_i} (\mathbf{r}_{12}, \mathbf{r}_{32}) f_{R}^{\text{long}} (\mathbf{r}_{12}) f_{R}^{\text{long}} (\mathbf{r}_{32}) \, .
\end{equation}
For the practical calculation of the momentum-space elements of the
regularized interaction $\bigl< p' q' \alpha' | V_{\text{3N}}^{2\pi,
\text{reg}} | p q \alpha \bigr>$ we follow the strategy discussed in
Section~\ref{sec:local_coordinate}. As a first step we need to calculate the
pre-regularized interactions:
\begin{equation}
V_{\text{3N}}^{c_i, \text{prereg}} (\mathbf{Q}_1,\mathbf{Q}_3) = Q(- \Delta_{\mathbf{Q}_1}) Q(- \Delta_{\mathbf{Q}_3}) V_{\text{3N}}^{c_i} (\mathbf{Q}_1,\mathbf{Q}_3) \, .
\end{equation}
Starting from the unregularized expressions Eqs.~(\ref{eq:Vc}) and (\ref{eq:Vc2}) 
these can be calculated in a straightforward way. To be explicit
we give the results for the 3N contributions at N$^2$LO for $Q(-
\Delta_{\mathbf{Q}}) = - \Delta_{\mathbf{Q}}$.
\begin{align}
V_{\text{3N}}^{c_1, \text{prereg}} (\mathbf{Q}_1,\mathbf{Q}_3) &= - \frac{2 c_1 g_A^2 m_{\pi}^2}{f_{\pi}^4} \boldsymbol{\tau}_1 \cdot \boldsymbol{\tau}_3 \frac{(5 m_{\pi}^2 + \mathbf{Q}^2_1) (5 m_{\pi}^2 + \mathbf{Q}^2_3) \: \boldsymbol{\sigma}_1 \cdot \mathbf{Q}_1 \: \boldsymbol{\sigma}_3 \cdot \mathbf{Q}_3}{(\mathbf{Q}_1^2 + m_{\pi}^2)^3 (\mathbf{Q}_3^2 + m_{\pi}^2)^3} \, , \nonumber \\
V_{\text{3N}}^{c_3, \text{prereg}} (\mathbf{Q}_1,\mathbf{Q}_3) &= \frac{c_3 g_A^2}{f_{\pi}^4} \boldsymbol{\tau}_1 \cdot \boldsymbol{\tau}_3 \left[
  \frac{\boldsymbol{\sigma}_1 \cdot \boldsymbol{\sigma}_3}{(\mathbf{Q}_1^2 + m_{\pi}^2) (\mathbf{Q}_3^2 + m_{\pi}^2)} - \frac{(3 \mathbf{Q}^2_1 + 7 m_{\pi}^2) \boldsymbol{\sigma}_1 \cdot \mathbf{Q}_1 \: \boldsymbol{\sigma}_3 \cdot \mathbf{Q}_1}{(\mathbf{Q}_1^2 + m_{\pi}^2)^3 (\mathbf{Q}_3^2 + m_{\pi}^2)} \right. \nonumber \\
  &\left. - \frac{(3 \mathbf{Q}^2_3 + 7 m_{\pi}^2) \boldsymbol{\sigma}_1 \cdot \mathbf{Q}_3 \: \boldsymbol{\sigma}_3 \cdot \mathbf{Q}_3}{(\mathbf{Q}_1^2 + m_{\pi}^2) (\mathbf{Q}_3^2 + m_{\pi}^2)^3}
  +\frac{(3 \mathbf{Q}^2_1 + 7 m_{\pi}^2) (3 \mathbf{Q}^2_3 + 7 m_{\pi}^2) \mathbf{Q}_1 \cdot \mathbf{Q}_3 \: \boldsymbol{\sigma}_1 \cdot \mathbf{Q}_1 \: \boldsymbol{\sigma}_3 \cdot \mathbf{Q}_3}{(\mathbf{Q}_1^2 + m_{\pi}^2)^3 (\mathbf{Q}_3^2 + m_{\pi}^2)^3} \right] \, , \nonumber \\
V_{\text{3N}}^{c_4, \text{prereg}} (\mathbf{Q}_1,\mathbf{Q}_3) &= \frac{c_4 g_A^2}{2 f_{\pi}^4} \boldsymbol{\tau}_1 \cdot \boldsymbol{\tau}_2 \times \boldsymbol{\tau}_3 \left[
  - \frac{\boldsymbol{\sigma}_1 \cdot \boldsymbol{\sigma}_2 \times \boldsymbol{\sigma}_3}{(\mathbf{Q}_1^2 + m_{\pi}^2) (\mathbf{Q}_3^2 + m_{\pi}^2)} - \frac{(3 \mathbf{Q}^2_1 + 7 m_{\pi}^2) \boldsymbol{\sigma}_1 \cdot \mathbf{Q}_1 \: \boldsymbol{\sigma}_3 \cdot \boldsymbol{\sigma}_2 \times \mathbf{Q}_1}{(\mathbf{Q}_1^2 + m_{\pi}^2)^3 (\mathbf{Q}_3^2 + m_{\pi}^2)} \right. \nonumber \\
  &+ \left. \frac{(3 \mathbf{Q}^2_3 + 7 m_{\pi}^2) \boldsymbol{\sigma}_3 \cdot \mathbf{Q}_3 \: \boldsymbol{\sigma}_1 \cdot \boldsymbol{\sigma}_2 \times \mathbf{Q}_3}{(\mathbf{Q}_1^2 + m_{\pi}^2) (\mathbf{Q}_3^2 + m_{\pi}^2)^3} + \frac{(3 \mathbf{Q}^2_1 + 7 m_{\pi}^2) (3 \mathbf{Q}^2_3 + 7 m_{\pi}^2) \boldsymbol{\sigma}_1 \cdot \mathbf{Q}_1 \: \boldsymbol{\sigma}_3 \cdot \mathbf{Q}_3 \: \boldsymbol{\sigma}_2 \cdot \mathbf{Q}_1 \times \mathbf{Q}_3}{(\mathbf{Q}_1^2 + m_{\pi}^2)^3 (\mathbf{Q}_3^2 + m_{\pi}^2)^3} \right] \, ,
\end{align}
and the corresponding pre-regularized regulator term (cf. Eq.~(\ref{eq:regulator_prereg_factorized})):
\begin{equation}
\tilde{f}_{R}^{\text{prereg}} (Q) = 4 \pi \int_0^{\infty} dr \bigl( 1 - e^{-r^2/R^2} \bigr)^6 j_0(Q r) \, .
\end{equation}
These pre-regularized quantities can be straightforwardly decomposed in a
partial-wave representation using the framework discussed in
Section~\ref{sec:PWD_3NF_local}. The decomposition of the regulator
\begin{equation}
\tilde{f}_{R}^{\text{prereg}} (\mathbf{Q}_1, \mathbf{Q}_3) = \tilde{f}^{\text{prereg}}_R (|\mathbf{Q}_1|) \tilde{f}^{\text{prereg}}_R (|\mathbf{Q}_3|)
\end{equation} 
is particularly simple as it does not depend on any spin and isospin quantum
numbers. However, note that the regulator does not factorize anymore if the
long-range regulator is applied to pion-exchange contributions that involve
all three interparticle distances (like, e.g. for the ring contributions at
N$^3$LO) due to the relation $\mathbf{r}_{12} + \mathbf{r}_{23} +
\mathbf{r}_{31} = 0$. After the decomposition the matrix elements of the
regularized interaction can be computed via
Eq.~(\ref{eq:V3N_regul_prereg_PW}).

The regularization of the intermediate-range diagram (b) in
Figure~\ref{fig:semilocal_N2LO_diags_coordinate} is somewhat more intricate as
it involves long-range and short-range pieces. Since this interaction contains
only one pion-exchange interaction the long-range regularization takes the
following form:
\begin{equation}
V_{\text{3N}}^{1\pi, \text{reg}} (\mathbf{r}_{12}) = V_{\text{3N}}^{1\pi} (\mathbf{r}_{12}) f_{R}^{\text{long}} (\mathbf{r}_{12}) \, .
\end{equation}
Accordingly, the pre-regularized interaction reads
\begin{align}
V_{\text{3N}}^{c_D, \text{prereg}} (\mathbf{Q}_1) = - \Delta_{\mathbf{Q}_1} V_{\text{3N}}^{c_D} (\mathbf{Q}_1) = - \frac{g_A c_D}{4 f_{\pi}^4 \Lambda_{\chi}} \boldsymbol{\tau}_1 \cdot \boldsymbol{\tau}_3 \left[ - \frac{\boldsymbol{\sigma}_1 \cdot \boldsymbol{\sigma}_3}{\mathbf{Q}_1^2 + m_{\pi}^2} + \frac{(7 m_{\pi}^2 + 3 \mathbf{Q}^2_1) \boldsymbol{\sigma}_1 \cdot \mathbf{Q}_1 \boldsymbol{\sigma}_3 \cdot \mathbf{Q}_1}{(\mathbf{Q}_1^2 + m_{\pi}^2)^3} \right] \, ,
\end{align}
and the pre-regularized regulator
\begin{equation}
\tilde{f}_R^{\text{prereg}} (\mathbf{Q}_1, \mathbf{Q}_3) = \int d \mathbf{r}_{12} d \mathbf{r}_{23} e^{-i \mathbf{Q}_1 \cdot \mathbf{r}_{12}} e^{-i \mathbf{Q}_3 \cdot \mathbf{r}_{23}} \frac{f_R^{\text{long}} (r_{12})}{Q(r_{12}^2)} = (2 \pi)^3 \delta(\mathbf{Q}_3) \tilde{f}_R^{\text{prereg}} (\mathbf{Q}_1) \, .
\label{eq:regulator_prereg_delta}
\end{equation}
The partial-wave decomposition of the regulator in
Eq.~(\ref{eq:regulator_prereg_delta}) can be simplified significantly by a
clever choice of coordinates. If we choose the basis representation $\{12\}$
the momentum transfer $\mathbf{Q}_3$ is given by $\mathbf{Q}_3 =
\mathbf{q}'_{\{12\}}- \mathbf{q}_{\{12\}}$, and the argument of the delta
function does not depend on any angles. As a result, it is straightforward to
show that the partial-wave matrix elements of the regulator take the following
form:
\begin{equation}
\tensor*[_{\{12\}}]{\bigl< p' q' \alpha' | \tilde{f}_R^{\text{prereg}} | p q \alpha \bigr>}{_{\{12\}}} = \frac{\delta_{\alpha \alpha'}}{(2 \pi)^6} 2 \pi \delta(q' - q) \int_{-1}^1 dx P_{L} (x) \tilde{f}_R^{\text{prereg}} \bigl( y \bigr) \quad \text{with} \quad y^2 = p^2 + p'^2 - 2 p p' x \, ,
\label{eq:local_regulator_local_cD}
\end{equation}
and $L$ denotes the relative orbital angular momentum quantum number (see
Eq.~(\ref{eq:Jj_bas})). Any other basis choice, $\{13\}$ or $\{23\}$, leads to
nontrivial angular dependencies of the delta function and significantly
complicates the partial wave decomposition. Using the representation
Eq.~(\ref{eq:local_regulator_local_cD}) makes the application of the
long-range regulator straightforward, but complicates significantly the
application of the nonlocal regulator to the short-range coupling. To see
this, note that the nonlocal regulator takes the form
$f_{\Lambda}^{\text{short}} (\mathbf{p}_{\delta}) f_{\Lambda}^{\text{short}}
(\mathbf{p}'_{\delta})$ with $\mathbf{p}_{\delta} = (\mathbf{k}_2 -
\mathbf{k}_3)/2$ (see Section~\ref{sec:semilocal_momentum}). In
Ref.~\cite{Epel18SCS3N} the values of the local coordinate-space scale $R$ and
nonlocal scale $\Lambda$ are chosen to be connected via the relation $R =
\Lambda/2$. However, in the basis representation $\{12\}$ the momentum
$\mathbf{p}_{\delta}$ depends on angles between Jacobi momenta:
\begin{equation}
\mathbf{p}_{\delta} = \frac{1}{2} (\mathbf{k}_2 - \mathbf{k}_3) = - \frac{1}{2} \mathbf{p}_{\{12\}} - \frac{3}{4} \mathbf{q}_{\{12\}} = \mathbf{p}_{\{23\}} \, ,
\end{equation}
which makes the application of the nonlocal regulator nontrivial. On the other
hand, using basis $\{23\}$ reduces the nonlocal regularization to a simple
factor after the partial-wave decomposition.

One way out of this dilemma is to perform the regularization in the basis
$\{23\}$ and compute the corresponding matrix elements of the long-range
regulator by applying permutation operators to the expression
Eq.~(\ref{eq:local_regulator_local_cD}) (see also
Section~\ref{sec:3NF_coord_def}):
\begin{equation}
\tensor*[_{\{23\}}]{\bigl< p' q' \alpha' | \tilde{f}_R^{\text{prereg}} | p q \alpha \bigr>}{_{\{23\}}} = \tensor*[_{\{12\}}]{\bigl< p' q' \alpha' | P_{123}^{-1} \tilde{f}_R^{\text{prereg}} P_{123} | p q \alpha \bigr>}{_{\{12\}}} \, .
\label{eq:local_regulator_cD_23}
\end{equation}
It is important to note that the practical calculation of the matrix products
in Eq.~(\ref{eq:local_regulator_cD_23}) requires the calculation of the
permutation operator as well as the regulator for both the physical partial
waves with $(-1)^{L+S+T} = -1$, as well as unphysical partial waves with
$(-1)^{L+S+T} = +1$ (see also discussion in Section
\ref{sec:3N_decomp_antisymmetrization}). The unphysical partial waves
eventually decouple when the regulator is applied to the interaction via
Eq.~(\ref{eq:V3N_regul_prereg_PW}). However, for the calculation of products
of auxiliary quantities as in Eq.~(\ref{eq:local_regulator_cD_23}) it is
crucial to retain them. Finally, the regularized interaction can be computed
via
\begin{align}
& \tensor*[_{\{23\}}]{\bigl< p' q' \alpha' | V^{(i), \text{prereg}}_{\text{3N}} | p q \alpha \bigr>}{_{\{23\}}} \nonumber \\
&\quad = f_{\Lambda}^{\text{short}} (p') f_{\Lambda}^{\text{short}} (p) \int d p'' p''^2 d q'' q''^2 \sum_{\alpha''} \tensor*[_{\{23\}}]{\bigl< p' q' \alpha' | V_{\text{3N}}^{(i), \text{prereg}} | p'' q'' \alpha'' \bigr>}{_{\{23\}}} \tensor*[_{\{23\}}]{\bigl< p'' q'' \alpha'' | \tilde{f}_R^{\text{prereg}} | p q \alpha \bigr>}{_{\{23\}}} \, .
\end{align}
Obviously, the preferred basis $\{ab\}$ is determined by the particular choice
of the interaction decomposition (see discussion in
Section~\ref{sec:3N_decomp_antisymmetrization}).

Finally, the regularization of the purely short-range diagram (c) in
Figure~\ref{fig:semilocal_N2LO_diags_coordinate} again reduces to a purely
nonlocal momentum regularization, i.e., a simple multiplicative factor of the
partial-wave matrix elements:
\begin{equation}
\tensor*[_{\{ab\}}]{\bigl< p' q' \alpha' | V_{\text{3N}}^{c_E, \text{reg}} | p
q \alpha \bigr>}{_{\{ab\}}} = f_R^{\text{short}} (p',q')
\tensor*[_{\{ab\}}]{\bigl< p' q' \alpha' | V_{\text{3N}}^{c_E} | p q \alpha
\bigr>}{_{\{ab\}}} f_R^{\text{short}} (p,q) \, ,
\end{equation}
as discussed in Section~\ref{sec:semilocal_momentum}.

\subsection{Visualization and comparison of matrix elements}
\label{sec:visualization_3NF}
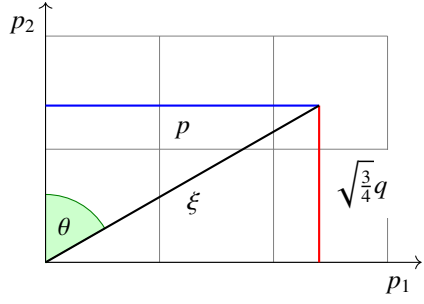
\begin{figure}[t]
\centering 
\begin{tikzpicture}[scale=3]
\draw[step=.5cm,gray,very thin] (0,0) grid (1.5,1.0);
\filldraw[fill=green!20,draw=green!50!black] (0,0) -- (0mm,3mm) arc (90:30:3mm) -- cycle;
\draw[->] (0,0) -- (1.65,0) coordinate (x axis);
\draw[->] (0,0) -- (0,1.15) coordinate (y axis);
\draw[thick,blue] (1.2,0.69282) -- node[below=2pt,fill=white] {\color{black} $p$} (0,0.69282);
\path [name path=upward line] (1.2,0) -- (1.2,1);
\path [name path=sloped line] (0,0) -- node[below=2pt,fill=white] {$\xi$} (30:1.5cm);
\draw [name intersections={of=upward line and sloped line, by=t}] [thick,red] (1.2,0) -- node [right=1pt,fill=white] {\color{black}$\sqrt{\frac{3}{4}} q$} (t);
\draw[thick] (0,0) -- (t);
\node at (0.08,0.16) {$\theta$};
\node at (1.55,-0.1) {$p_1$};
\node at (-0.1,1.05) {$p_2$};
\end{tikzpicture}
\caption{Definition of the hypermomentum $\xi^2 = p^2 + \tfrac{3}{4} q^2 =
p_1^2 + p_2^2$ and the hyperangle $\tan \theta = \frac{p}{\sqrt{3/4} q} =
\frac{p_1}{p_2}$ for a three-body state with Jacobi momenta $\left\{ p^2 = p_1^2,q^2 = \frac{4}{3} p_2^2 \right\}$.}
\label{fig:hypermomentum_illustration}
\end{figure}

In this section we illustrate and visualize the form of the partial-wave
matrix elements of different topologies within chiral EFT using the
different regularization schemes discussed in
Section~\ref{sec:3N_regularization}. Since the matrix elements effectively
depend on six variables, $p,q, \alpha, p', q'$ and $\alpha'$, it is necessary
to reduce the dimensionality of the parameter space. First, we will focus here
only on the most important partial-wave channels. Specifically, we select the
channel with $\bar{\alpha} = 0$ for the three-body channel with $\mathcal{J} =
\tfrac{1}{2}, \mathcal{T} = \tfrac{1}{2}$ and $\mathcal{P} = +1$, i.e., $L=0, S=0, J=0, T=1, l=0$
and $j=\tfrac{1}{2}$ for both initial and final states (see Appendix
\ref{sec:3N_config_table}). In this channel all orbital angular momenta are
zero, which means that it is possible to study the effect of pure $S$-wave
short-range couplings. Second, we re-parametrize the momentum dependence by
introducing the hypermomentum $\xi$ and the hyperangle $\theta$ via the
following definition (see also Figure \ref{fig:hypermomentum_illustration}):
\begin{equation}
\left\{ p,q \right\} \rightarrow \Bigl\{ \xi^2 = p^2 + \frac{3}{4} q^2, \tan \theta = \frac{p}{\sqrt{3/4} q} \Bigr\} \, .
\label{eq:hypermomentum_def}
\end{equation}
This transformation of variables is useful since the hypermomentum is directly
related to the relative three-body kinetic energy and hence serves as a
measure for the energy of the initial and final states. Since the role of the
regulators is to separate the low-energy from the high-energy part of the
Hilbert space the new variable set $\left\{ \xi,
\theta \right\}$ is particularly suited to study the nature and properties of
different regulators. In the following we visualize the matrix elements as a
function of the initial and final state momenta $\xi$ and $\xi'$ for a fixed
value of the hyperangle. In the following we arbitrarily choose $\theta =
\tfrac{\pi}{4}$, i.e., $\tan \theta = 1$ for the initial and final states. The global features 
of the results presented in the following are insensitive to this particular
choice.

\begin{figure}[t]
\includegraphics[width=0.99 \textwidth]{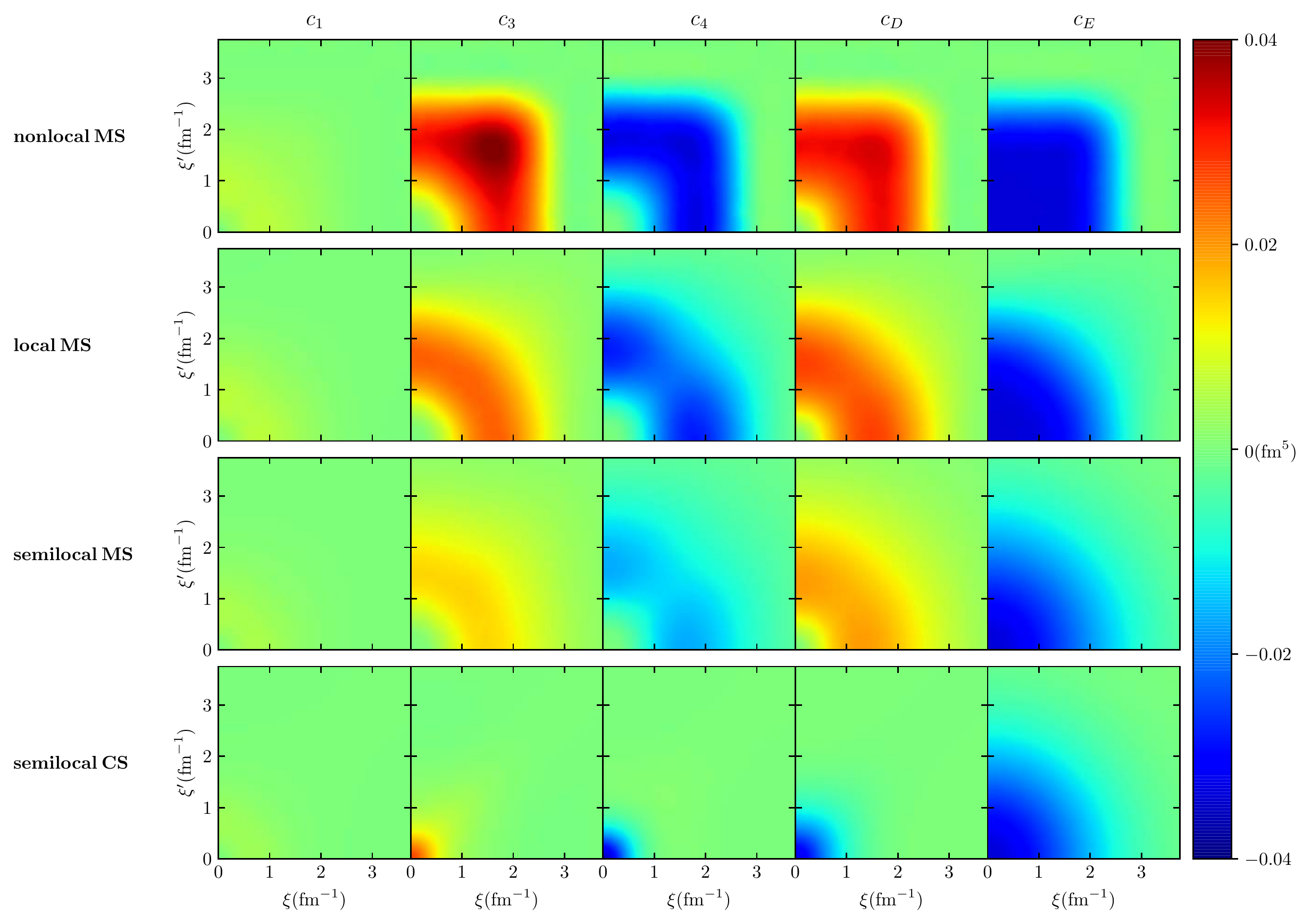}
\caption{Matrix elements of the antisymmetrized interaction $\bigl< p' q'
\bar{\alpha} | V^{\text{as}}_{\text{3N}} | p q \bar{\alpha} \bigr>$ for the
individual topologies at N$^{2}$LO (columns, indicated by the corresponding
LEC) and different regularization schemes (rows, see main text and
Table~\ref{tab:regularization}) as a function of the hypermomentum $\xi^2 =
p^2 + \tfrac{3}{4} q^2$ at the hyperangle ${\tan
\theta = p / (\sqrt{3}/2 q) = \frac{\pi}{4}}$ and in the partial wave with
$\bar{\alpha} = \left\{ L=0, S=0, J=0, T=1, l=0, j=\tfrac{1}{2} \right\}$. For
the long-range LECs, we choose the values of the Roy-Steiner analysis of
Refs.~\cite{Hofe15piNchiral,Hofe15PhysRep}: $c_1 = -0.74 \: \text{GeV}^{-1},
c_3 = -3.61 \: \text{GeV}^{-1}$ and $c_4 = 2.44 \: \text{GeV}^{-1}$. The
short-range couplings are chosen for optimized visibility: $c_D = 2.0$ and
$c_E = 0.5$. For the momentum-space regularization scales we set $\Lambda =
500 \:\text{MeV}$ and $R = 0.9 \: \text{fm}$ for the semilocal
coordinate-space regularization (corresponding to $\Lambda = 2 R \approx 355$
MeV for the nonlocal regulators).}
\label{fig:contour_N2LO_500}
\end{figure}

Here we illustrate the matrix elements of individual 3N topologies in chiral
EFT using the following four different regularization schemes (see also
Table~\ref{tab:regularization}):
\begin{itemize}
\item \textit{Nonlocal MS:} Nonlocal momentum-space regularization (see Section~\ref{sec:nonlocal_momentum}).
\item \textit{Local MS:} Local momentum-space regularization (see Section~\ref{sec:local_momentum}).
\item \textit{Semilocal MS:} Semilocal momentum-space regularization (see Section~\ref{sec:semilocal_momentum}).
\item \textit{Semilocal CS:} Semilocal coordinate-space regularization (see Section~\ref{sec:semilocal_coordinate}).
\end{itemize}

\begin{figure}[t]
\includegraphics[width=0.99 \textwidth]{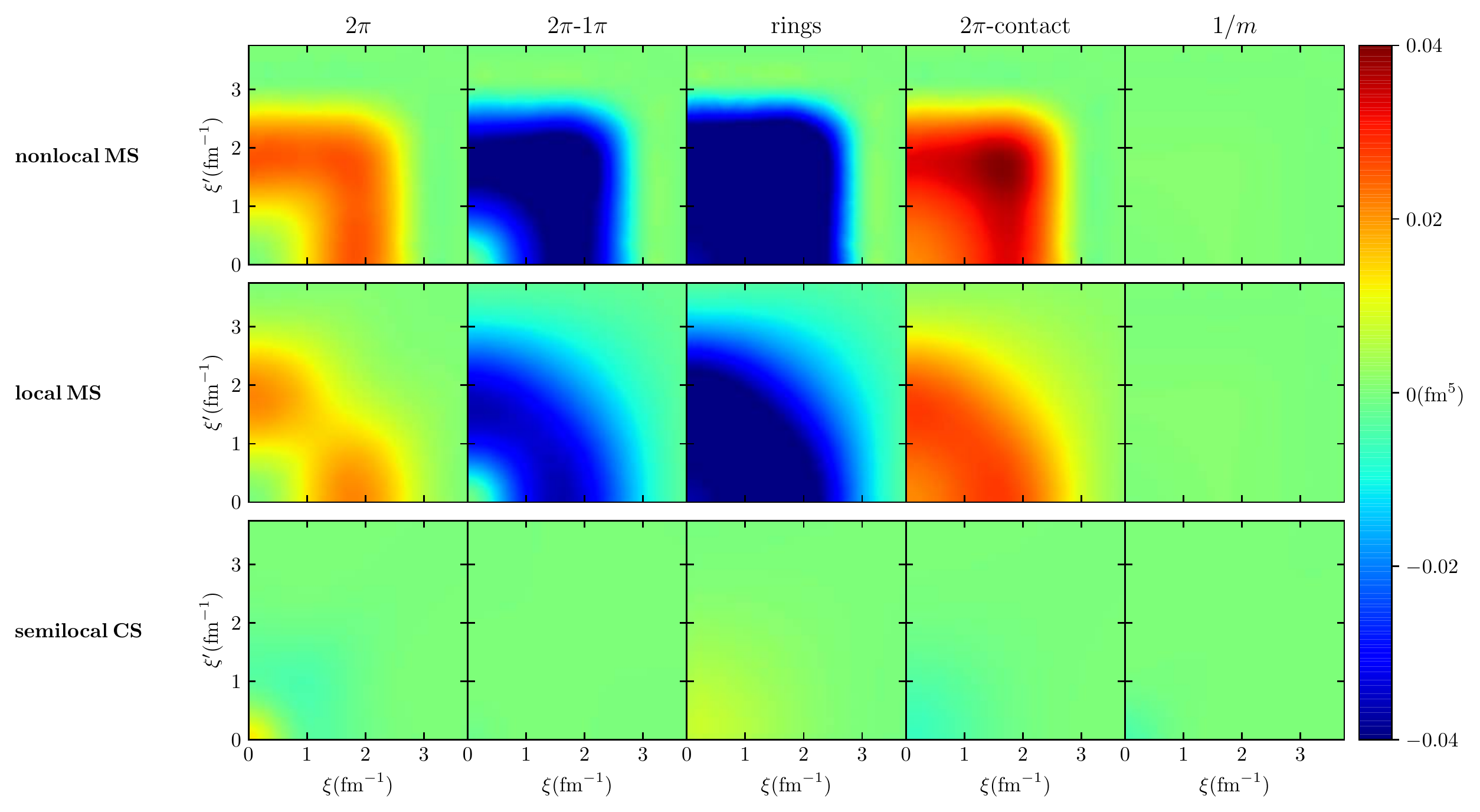}
\caption{The matrix elements for the individual N$^3$LO topologies for the
same parameters and plot ranges as in Figure~\ref{fig:contour_N2LO_500}. For
optimized visibility of the $2\pi$-contact matrix elements we set $C_T = 0.2$.
In the right column we show the relativistic corrections to the two-pion
exchange contributions. The relativistic corrections to the one-pion exchange
interactions are of the same order of magnitude and vanishingly small on the
shown plot scale. The N$^{3}$LO contributions in the ``semilocal MC''
regularization scheme are not yet fully developed. }
\label{fig:contour_N3LO_500}
\end{figure}

Based on the power counting of chiral EFT, the sum of all contributions to
observables at a given order in the chiral expansion should be suppressed
compared to contributions at lower order. However, we stress that studying the
matrix elements of individual topologies separately may be misleading due to possible
cancellation effects when computing observables. In particular, we stress that
the definition of individual topologies is in general scheme-dependent and not
unique (compare discussion in Ref.~\cite{Kreb123Nlong}).

Figure~\ref{fig:contour_N2LO_500} shows the matrix elements for the five
leading-order (N$^2$LO) 3N interaction topologies for typical cutoff values:
$\Lambda = 500$ MeV for momentum-space regulators and $R = 0.9$ fm for
coordinate-space regulators. Clearly, the values and structure of the matrix
elements are quite sensitive to the employed regularization scheme. In
general, the values are largest for the ``nonlocal MS'' (first row)
regularization, while the elements for the ``local MS'' and ``semilocal MS''
schemes look qualitatively similar, but tend to be more strongly suppressed. In
particular, the long-range topologies (first three columns) of the ``local
MS'' and ``semilocal MS'' regularized matrix elements are equal up to an
additional suppression factor $\exp \bigl[ - 2 m_{\pi}^2/\Lambda^2 \bigr]$
(see Table~\ref{tab:regularization}). The ``semilocal CS'' matrix elements
(bottom row) on the other hand show a very different qualitative form compared
to all momentum-space regularization schemes. Here the coordinate-space
regulator $f_R^{\text{long}} (\mathbf{r})$ completely suppresses all
short-range contributions of the pion exchange interactions (see discussion in
Section~\ref{sec:semilocal_coordinate}) as the regulator function virtually
vanishes at $r = 0$. The remaining contributions shown in the bottom row of
Figure \ref{fig:contour_N2LO_500} only contain long-range contributions to the
2$\pi$-exchange topologies. For the momentum-space regulators on the other
hand the pion exchange interactions still contain the short-range contact
terms. This qualitative difference in the matrix elements is also reflected in
observables as we demonstrate in Section~\ref{sec:PW_conv_matter} based on
nuclear matter properties.

In Figure~\ref{fig:contour_N3LO_500} we show the corresponding matrix elements
for the subleading topologies at N$^3$LO for the same parameters and
regularization schemes as in Figure~\ref{fig:contour_N2LO_500}. We can observe
some striking differences between the different regularization schemes. First,
the matrix elements of the individual topologies at N$^2$LO and N$^3$LO for
the shown partial-wave channel are of the same order of magnitude for both the
``nonlocal MS'' and ``local MS'' regularization scheme. This is per se not a
problem for power counting as the contributions can still add up to a
relatively small total contribution. However, the cancellation becomes more
and more fine-tuned as the size of the matrix elements for the individual
topologies increase, which might eventually lead to serious practical
problems. Second, there is a significant difference between the semilocal
regularization schemes and the nonlocal and purely local regularization
schemes. This indicates that the chosen regularization scheme could have some
significant impact on the convergence of the chiral expansion.

\begin{figure}[t]
\centering
\includegraphics[width=0.9 \textwidth]{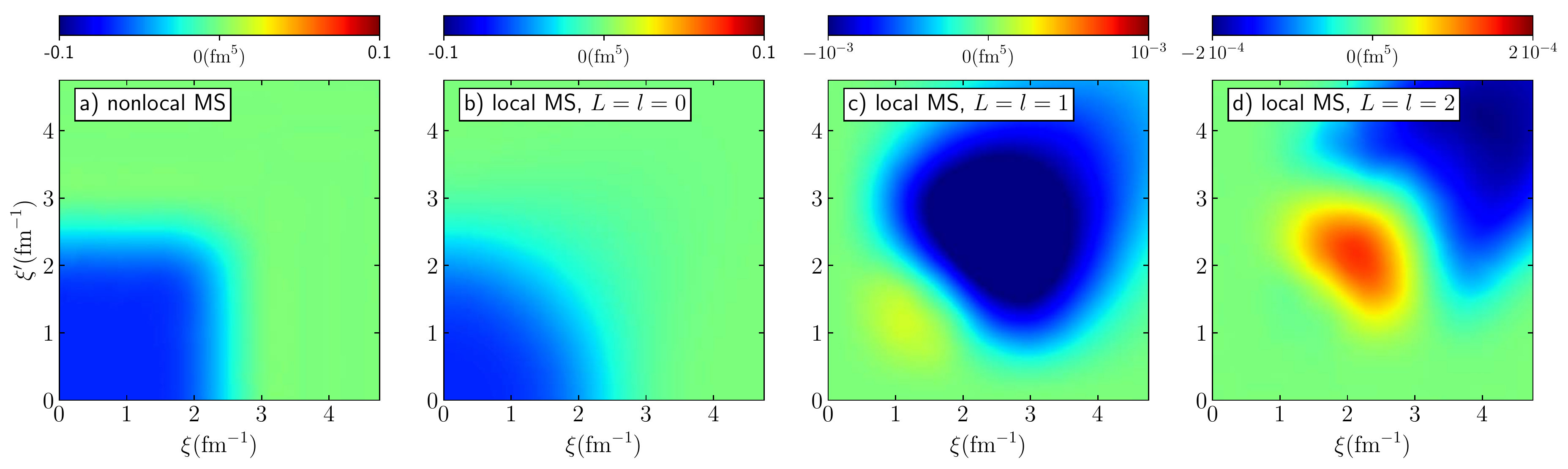}
\caption{Matrix elements of the antisymmetrized interaction $\bigl< p' q'
\bar{\alpha} | V^{\text{as}}_{\text{3N}} | p q \bar{\alpha} \bigr>$ for the
short-range interaction proportional to $c_E$ using the nonlocal
regularization (a) and the ``local MS'' regularization (b, c and
d) using $\Lambda = 500 \, \text{MeV}$ at the hyperangle $\tan \theta =
\tfrac{\pi}{4}$. In the different panels we show the partial waves with $\bar{\alpha}_a
= \bar{\alpha}_b = \left\{ L=0, S=0, J=0, T=1, l=0, j=\tfrac{1}{2} \right\}$,
$\bar{\alpha}_c = \left\{ L=1, S=0, J=1, T=0, l=1, j=\tfrac{3}{2} \right\}$ and
$\bar{\alpha}_d = \left\{ L=2, S=0, J=2, T=1, l=2, j=\tfrac{5}{2}
\right\}$ (see Appendix~\ref{sec:3N_config_table}). The matrix elements for
$L>0$ or $l>0$ vanish for the nonlocal regularization for this interaction
topology. Note the different plot scales in panels c and d.}
\label{fig:contour_cE_induced}
\end{figure}

Finally, in Figure~\ref{fig:contour_cE_induced} we show the matrix elements of
the pure contact interaction topology proportional to the LEC
$c_E$ for different configuration channels in the three-body channel with
$\mathcal{J} = \tfrac{1}{2}, \mathcal{T} = \tfrac{1}{2}$ and $\mathcal{P} = +1$. Note that for
the regularization schemes ``nonlocal MS'', ``semilocal MS'' and ``semilocal
CS'' the matrix elements always vanish for all channels with $L>0$ or
$l>0$ due to the nonlocal regularization of this topology in these schemes. In
contrast, for the ``local MS'' scheme a finite range is induced, which leads to
finite contributions in channels with nonvanishing angular momenta.

\subsection{Transformation of matrix elements to harmonic oscillator basis}
\label{sec:HO_transf}

Most many-body frameworks based on partial-wave expansion methods are developed in
harmonic oscillator (HO) bases. It is straightforward to transform momentum-space
matrix elements of the form $\left< p' q'
\alpha' | V_{\text{3N}} | p q \alpha \right>$ to the corresponding Jacobi HO basis.
Since the partial-wave expansion of the angular dependence is identical in
both bases, the transformation can be performed individually for the different
partial-wave channels:
\begin{align}
\bigl< N' n' \alpha' | V_{\text{3N}} | N n \alpha \bigr> 
&= \int dp p^2 dp' p'^2 \int dq q^2 dq' q'^2 R_{N'L'} (p',b) R_{n'l'} (q',b) \left< p' q'
\alpha' | V_{\text{3N}} | p q \alpha \right> R_{N L} (p,b) R_{n l} (q,b) \, , 
\label{eq:relHO_transf}
\end{align}
with the radial harmonic oscillator wave functions\footnote{There exist different phase conventions for the harmonic oscillator wave functions. In some conventions an additional phase $(-1)^n$ is added. This phase is either present in the coordinate-space wave function or in the momentum space representation.}
\begin{equation}
R_{n l} (p, b) = \sqrt{\frac{2 n! b^3}{\Gamma(n + l + \tfrac{3}{2})}} (p b)^l e^{- p^2 b^2/2} L_{n}^{l+\tfrac{1}{2}} (p^2 b^2) \, .
\end{equation}
Here $b$ denotes the oscillator parameter $b^{-1} = \sqrt{m \Omega}$ with the
nucleon mass $m$ and the oscillator frequency $\Omega$ and $L_{n}^k$ are the
generalized Laguerre polynomials. 

In the limit of infinite basis size many-body results become independent of
$\Omega$. For practical calculations the frequency $\Omega$ is usually chosen
such that the ground-state energy results are minimized for a given basis size
(see Figure \ref{fig:3H_hbaromega_SRG}). 

We note that some care has to be taken regarding the value of the oscillator
parameter due to different conventions of Jacobi coordinates. In the
coordinate system introduced in Section~\ref{sec:3NF_coord_def} the isotropic
oscillator Hamiltonian for three particles of mass $m$ takes the following
form:
\begin{align}
H &= \frac{\mathbf{k}_1^2 + \mathbf{k}_2^2 + \mathbf{k}_3^2}{2 m} + \frac{1}{2} m \Omega^2 (\mathbf{x}_1^2 + \mathbf{x}_2^2 + \mathbf{x}_3^2) = \frac{\mathbf{P}_{\text{3N}}^2}{2 M} + \frac{\mathbf{p}_{\{12\}}^2}{2 \mu_1} + \frac{\mathbf{q}_{\{12\}}^2}{2 \mu_2} + \frac{1}{2} M \Omega^2 \mathbf{R}_{\text{3N}}^2 + \frac{1}{2} \mu_1 \Omega^2 \mathbf{r}_{\{12\}}^2 + \frac{1}{2} \mu_2 \Omega^2 \mathbf{s}_{\{12\}}^2 \, ,
 \label{eq:H_HO_Gloeckle}
\end{align}
with $M = 3 m, \mu_1 = \frac{m}{2}, \mu_2 = \frac{2}{3} m$. An alternative
common choice of Jacobi coordinates is a more symmetric choice of factors in
momentum and coordinate space:
\begin{align}
&\tilde{\mathbf{p}}_{\{12\}} = \frac{\mathbf{k}_1 - \mathbf{k}_2}{\sqrt{2}},&  &\tilde{\mathbf{q}}_{\{12\}} = \sqrt{\frac{2}{3}} \left( \mathbf{k}_3 - \frac{1}{2} (\mathbf{k}_1 + \mathbf{k}_2) \right), & &\tilde{\mathbf{P}}_{\text{3N}} = \frac{\mathbf{k}_1 + \mathbf{k}_2 + \mathbf{k}_3}{\sqrt{3}} \nonumber \\
&\tilde{\mathbf{r}}_{\{12\}} = \frac{\mathbf{x}_1 - \mathbf{x}_2}{\sqrt{2}},&  &\tilde{\mathbf{s}}_{\{12\}} = \sqrt{\frac{2}{3}} \left( \mathbf{x}_3 - \frac{1}{2} (\mathbf{x}_1 + \mathbf{x}_2) \right), & &\tilde{\mathbf{R}}_{\text{3N}} = \frac{\mathbf{x}_1 + \mathbf{x}_2 + \mathbf{x}_3}{\sqrt{3}} \, .
\end{align}
In this representation the harmonic oscillator Hamiltonian takes the following form, as can be easily verified:
\begin{eqnarray}
 H &=& \frac{\tilde{\mathbf{P}}_{\text{3N}}^2}{2 m} + \frac{\tilde{\mathbf{p}}_{\{12\}}^2}{2 m} + \frac{\tilde{\mathbf{q}}_{\{12\}}^2}{2 m} + \frac{1}{2} m \Omega^2 \bigl( \tilde{\mathbf{R}}_{\text{3N}}^2 + \tilde{\mathbf{r}}_{\{12\}}^2 + \tilde{\mathbf{s}}_{\{12\}}^2 \bigr) \, .
 \label{eq:H_HO_Navratil}
\end{eqnarray}
By comparing of Eqs.~(\ref{eq:H_HO_Gloeckle}) and (\ref{eq:H_HO_Navratil}) we
can identify the relations between the oscillator frequencies and oscillator
parameters in both Jacobi coordinate representations for the transformation
Eq.~(\ref{eq:relHO_transf}) for the three Jacobi variables:
\begin{equation}
\tilde{b}_{cm} = \frac{1}{\sqrt{3}} b, \quad \tilde{b}_p = \sqrt{2} b, \quad \tilde{b}_q = \sqrt{\frac{3}{2}} b \, ,
\end{equation}
where $b$ is the value of the oscillator parameter in
representation~(\ref{eq:H_HO_Gloeckle}), and $\tilde{b}_p$ and $\tilde{b}_q$
are the oscillator parameters in the momentum $p$ and $q$, respectively.

\clearpage
\section{Incorporation of 3N interactions in many-body frameworks}
\label{sec:3N_incorporation}

In Section~\ref{sect:3NF_representation} we discussed the representation and
practical calculation of 3N interaction matrix elements in a partial-wave
momentum basis. These matrix elements represent the microscopic input of most
\textit{ab initio} many-body frameworks for finite nuclei as well as dense matter.
However, the step from interaction matrix elements to the extraction of
many-body observables involves several challenges, in particular since the
required many-body basis sizes grow with the particle number of the studied
nuclei. In this section we discuss different novel and established techniques
that facilitate the incorporation of 3N interactions in many-body frameworks
and help to push the reach of \textit{ab initio} methods toward heavier nuclei.
Specifically, in Section~\ref{sec:PW_conv_matter} we illustrate the
partial-wave convergence of results for nuclear-matter energies. In addition
we demonstrate that the choice of the regularization scheme has a significant
impact on the size of contributions from 3N interactions. In
Section~\ref{sec:SRG} we discuss the Similarity Renormalization Group (SRG)
evolution of NN and 3N interactions to a lower resolution scale in the
partial-wave momentum basis, which helps to significantly accelerate the
convergence of many-body calculations of matter and atomic nuclei. In
Section~\ref{sec:normal_ordering} we review recent and ongoing developments
for the normal ordering of 3N interactions. This method allows to incorporate
the main contributions from 3N interactions in many-body calculations at the
cost of NN interactions. This technique is now used in basically all
basis-expansion many-body methods that aim at studying properties of
medium-mass and heavy nuclei. Typically, SRG evolution and normal ordering are
combined in most many-body frameworks. Finally, we discuss in
Section~\ref{sec:no_PWD} a novel method to apply 3N interactions in many-body
perturbation theory without partial-wave expansion. This method is
particularly suited for calculations of nuclear matter, as it drastically
simplifies the calculation of individual diagrams in MBPT compared to
conventional approaches based on partial-wave decomposed 3N interactions. In
addition, the developed framework can also be combined with a recently
developed coupled cluster framework for nuclear matter~\cite{Hage14ccnm}.

For all these applications we will consider a general Hamiltonian in the
center-of-mass reference frame, including contributions from the kinetic
energy, NN and 3N interactions. For all the applications discussed in the
following, it is most convenient to represent all quantities in second
quantized form (see, e.g. Ref.~\cite{negele1995quantum}):
\begin{equation}
\hat{H} = \hat{T}_{\text{rel}} + \hat{V}_{\text{NN}} + \hat{V}_{\text{3N}} \, ,
\label{eq:SRG_def_H}
\end{equation}
with 
\begin{align}
\hat{T}_{\text{rel}} &= \sum_{ij} \left< i | T | j \right> \hat{a}^{\dagger}_i \hat{a}_j, \nonumber \\
\hat{V}_{\text{NN}} &= \frac{1}{(2!)^2} \sum_{ijkl} \bigl< i j | \vnn^{\text{as}} | k l \bigr> \: \hat{a}_i^{\dagger} \hat{a}_j^{\dagger} \hat{a}_l \hat{a}_k, \nonumber \\
\hat{V}_{\text{3N}} &= \frac{1}{(3!)^2} \sum_{ijklmn} \bigl< i j k | \vtn^{\text{as}} | l m n \bigr> \: \hat{a}_i^{\dagger} \hat{a}_j^{\dagger} \hat{a}_k^{\dagger} \hat{a}_n \hat{a}_m \hat{a}_l \, .
\label{eq:H_second_quantized}
\end{align}
Here all interactions are represented in terms of antisymmetrized matrix elements: 
\begin{align}
\bigl< i j | \vnn^{\text{as}} | k l \bigr> &= \bigl< i j | \mathcal{A}_{12} \vnn | k l \bigr> = \left< i j | \vnn | k l \right> - \left< j i | \vnn | k l \right> \nonumber \\
\bigl< i j k | \vtn^{\text{as}} | l m n \bigr> &= \bigl< i j k | \mathcal{A}_{123} \vtn | l m n \bigr> \nonumber \\
&= \left< i j k | \vtn | l m n \right> - \left< j i k | \vtn | l m n \right> -
\left< i k j | \vtn | l m n \right> - \left< k j i | \vtn | l m n \right> +
\left< j k i | \vtn | l m n \right> + \left< k i j | \vtn | l m n \right> \, ,
\end{align}
with the two- and three-body antisymmetrizers (see Section~\ref{sec:3N_decomp_antisymmetrization}):
\begin{align}
\mathcal{A}_{12} = 1 - P_{12}, \quad \mathcal{A}_{123} = 1 - P_{12} - P_{13} - P_{23} + P_{123} + P_{132} \, .
\end{align}
The matrix elements are assumed to fulfill the general symmetries under simultaneous interchange of particles in the initial and final states:
\begin{align}
\left< i j | \vnn | k l \right> &= \left< j i | \vnn | l k \right> \nonumber \\
\left< i j k | \vtn | l m n \right> &= \left< j i k | \vtn | m l n \right> = \left< k j i | \vtn | n m l \right>  \quad \text{etc.}
\end{align}
These relations are obviously fulfilled for 3N interactions of the form
Eq.~(\ref{eq:generic_form_3NF}). The individual contributions to the
antisymmetrized interactions $\vnn^{\text{as}}$ and $V_{\text{3N}}^{\text{as}}$ are
illustrated in Figures~\ref{fig:Vij_asymm} and ~\ref{fig:V123_asymm}.
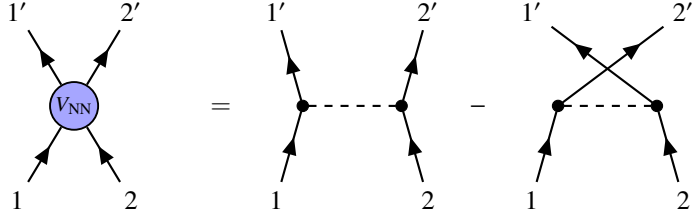
\begin{figure}[t]
\centering
\begin{minipage}[c]{0.15\textwidth}
\begin{tikzpicture} 
\tikzfeynmanset{
  my dot/.style={fill=red},
  every vertex/.style={my dot},
}
\begin{feynman}
\vertex (a) at (0,0) {\(1\)}; 
\vertex (b) at (1.5,0) {\(2\)};
\vertex [blob, /tikz/minimum size=18pt, fill=blue!35, line width=0.25mm, font=\fontsize{8}{0}\selectfont] (d) at (0.75,1.25) {\(V_{\rm NN}\)};
\vertex (e) at (0,2.5) {\(1'\)}; 
\vertex (f) at (1.5,2.5) {\(2'\)};
\diagram* {
(a) -- [fermion, thick] (d) -- [fermion, thick] (e);
(b) -- [fermion, thick] (d) -- [fermion, thick] (f);
};
\end{feynman}
\end{tikzpicture}
\end{minipage}
\hspace{0.1cm}
=
\hspace{0.1cm}
\begin{minipage}[c]{0.15\textwidth}
\begin{tikzpicture} 
\begin{feynman}
\vertex (a) at (0,0) {\(1\)}; 
\vertex (b) at (2.0,0) {\(2\)};
\vertex [dot] (c) at (0.35,1.25) {};
\vertex [dot] (d) at (1.65,1.25) {};
\vertex (e) at (0,2.5) {\(1'\)}; 
\vertex (f) at (2.0,2.5) {\(2'\)};
\diagram* {
(a) -- [fermion, thick] (c) -- [fermion, thick] (e);
(b) -- [fermion, thick] (d) -- [fermion, thick] (f);
(c) -- [dashed, thick] (d);
};
\end{feynman}
\end{tikzpicture}
\end{minipage}
\hspace{0.1cm}
$-$
\hspace{0.1cm}
\begin{minipage}[c]{0.15\textwidth}
\begin{tikzpicture} 
\tikzfeynmanset{
  my dot/.style={fill=red},
  every vertex/.style={my dot},
}
\tikzfeynmanset{
  fermion1/.style={
    /tikz/postaction={
      /tikz/decoration={
        markings,
        mark=at position 0.7 with {
          \node[
            transform shape,
            xshift=-0.5mm,
            fill,
            inner sep=1.5pt,
            draw=none,
            isosceles triangle
          ] { };
        },
      },
      /tikz/decorate=true,
    },
  }
}
\begin{feynman}
\vertex (a) at (0,0) {\(1\)}; 
\vertex (b) at (2.0,0) {\(2\)};
\vertex [dot] (c) at (0.35,1.25) {};
\vertex [dot] (d) at (1.65,1.25) {};
\vertex (e) at (0,2.5) {\(1'\)}; 
\vertex (f) at (2.0,2.5) {\(2'\)};
\diagram* {
(a) -- [fermion, thick] (c) -- [fermion1, thick] (f);
(b) -- [fermion, thick] (d) -- [fermion1, thick] (e);
(c) -- [dashed, thick] (d);
};
\end{feynman}
\end{tikzpicture}
\end{minipage}
\caption{The two contributions to the antisymmetrized NN interaction $\vnn^{\text{as}}$. On the right hand side we show for illustration a long-range $1\pi$-exchange contribution.}
\label{fig:Vij_asymm}
\end{figure}
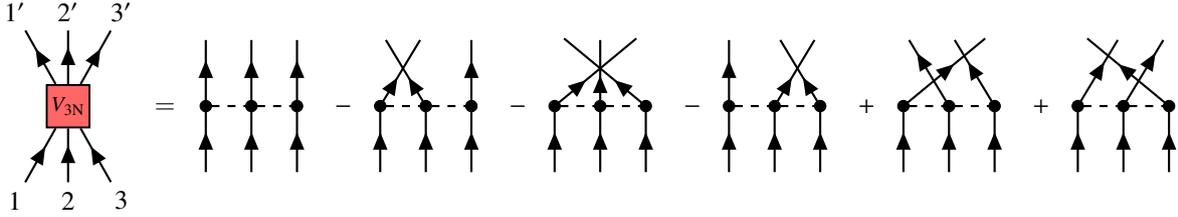
\begin{figure}[t]
\centering
\begin{minipage}[c]{0.11\textwidth}
\begin{tikzpicture} 
\begin{feynman}
\vertex (a) at (0,0) {\(1\)}; 
\vertex (b) at (0.7,0) {\(2\)};
\vertex (c) at (1.4,0) {\(3\)};
\vertex [blob, /tikz/minimum size=16pt, shape=rectangle, fill=red!60, line width=0.25mm, font=\fontsize{8}{0}\selectfont] (d) at (0.7,1.25) {\(V_{\rm 3N}\)};
\vertex (e) at (0,2.5) {\(1'\)}; 
\vertex (f) at (0.7,2.5) {\(2'\)};
\vertex (g) at (1.4,2.5) {\(3'\)};
\diagram* {
(a) -- [fermion, thick] (d) -- [fermion, thick] (e);
(b) -- [fermion, thick] (d) -- [fermion, thick] (f);
(c) -- [fermion, thick] (d) -- [fermion, thick] (g);
};
\end{feynman}
\end{tikzpicture}
\end{minipage}
\hspace{0.05cm}
=
\hspace{0.05cm}
\begin{minipage}[c]{0.095\textwidth}
\begin{tikzpicture} 
\begin{feynman}
\vertex (a) at (0,0.25) {}; 
\vertex (b) at (0.6,0.25) {};
\vertex (c) at (1.2,0.25) {};
\vertex [dot] (d) at (0.0,1.25) {};
\vertex [dot] (e) at (0.6,1.25) {};
\vertex [dot] (f) at (1.2,1.25) {};
\vertex (g) at (0,2.25) {}; 
\vertex (h) at (0.6,2.25) {};
\vertex (i) at (1.2,2.25) {};
\diagram* {
(a) -- [fermion, thick] (d) -- [fermion, thick] (g);
(b) -- [fermion, thick] (e) -- [fermion, thick] (h);
(c) -- [fermion, thick] (f) -- [fermion, thick] (i);
(d) -- [dashed, thick] (e) -- [dashed, thick] (f);
};
\end{feynman}
\end{tikzpicture}
\end{minipage}
\hspace{0.02cm}
$-$
\hspace{0.02cm}
\begin{minipage}[c]{0.095\textwidth}
\begin{tikzpicture}
\tikzfeynmanset{
  fermion1/.style={
    /tikz/postaction={
      /tikz/decoration={
        markings,
        mark=at position 0.25 with {
          \node[
            transform shape,
            xshift=-0.5mm,
            fill,
            inner sep=1.5pt,
            draw=none,
            isosceles triangle
          ] { };
        },
      },
      /tikz/decorate=true,
    },
  }
}
\begin{feynman}
\vertex (a) at (0,0.25) {}; 
\vertex (b) at (0.6,0.25) {};
\vertex (c) at (1.2,0.25) {};
\vertex [dot] (d) at (0.0,1.25) {};
\vertex [dot] (e) at (0.6,1.25) {};
\vertex [dot] (f) at (1.2,1.25) {};
\vertex (g) at (0,2.25) {}; 
\vertex (h) at (0.6,2.25) {};
\vertex (i) at (1.2,2.25) {};
\diagram* {
(a) -- [fermion, thick] (d) -- [fermion1, thick] (h);
(b) -- [fermion, thick] (e) -- [fermion1, thick] (g);
(c) -- [fermion, thick] (f) -- [fermion, thick] (i);
(d) -- [dashed, thick] (e) -- [dashed, thick] (f);
};
\end{feynman}
\end{tikzpicture}
\end{minipage}
\hspace{0.02cm}
$-$
\hspace{0.02cm}
\begin{minipage}[c]{0.095\textwidth}
\begin{tikzpicture}
\tikzfeynmanset{
  fermion1/.style={
    /tikz/postaction={
      /tikz/decoration={
        markings,
        mark=at position 0.25 with {
          \node[
            transform shape,
            xshift=-0.5mm,
            fill,
            inner sep=1.5pt,
            draw=none,
            isosceles triangle
          ] { };
        },
      },
      /tikz/decorate=true,
    },
  }
}
\begin{feynman}
\vertex (a) at (0,0.25) {}; 
\vertex (b) at (0.6,0.25) {};
\vertex (c) at (1.2,0.25) {};
\vertex [dot] (d) at (0.0,1.25) {};
\vertex [dot] (e) at (0.6,1.25) {};
\vertex [dot] (f) at (1.2,1.25) {};
\vertex (g) at (0,2.25) {}; 
\vertex (h) at (0.6,2.25) {};
\vertex (i) at (1.2,2.25) {};
\diagram* {
(a) -- [fermion, thick] (d) -- [fermion1, thick] (i);
(b) -- [fermion, thick] (e) -- [fermion1, thick] (h);
(c) -- [fermion, thick] (f) -- [fermion1, thick] (g);
(d) -- [dashed, thick] (e) -- [dashed, thick] (f);
};
\end{feynman}
\end{tikzpicture}
\end{minipage}
\hspace{0.02cm}
$-$
\hspace{0.02cm}
\begin{minipage}[c]{0.095\textwidth}
\begin{tikzpicture} 
\tikzfeynmanset{
  fermion1/.style={
    /tikz/postaction={
      /tikz/decoration={
        markings,
        mark=at position 0.25 with {
          \node[
            transform shape,
            xshift=-0.5mm,
            fill,
            inner sep=1.5pt,
            draw=none,
            isosceles triangle
          ] { };
        },
      },
      /tikz/decorate=true,
    },
  }
}
\begin{feynman}
\vertex (a) at (0,0.25) {}; 
\vertex (b) at (0.6,0.25) {};
\vertex (c) at (1.2,0.25) {};
\vertex [dot] (d) at (0.0,1.25) {};
\vertex [dot] (e) at (0.6,1.25) {};
\vertex [dot] (f) at (1.2,1.25) {};
\vertex (g) at (0,2.25) {}; 
\vertex (h) at (0.6,2.25) {};
\vertex (i) at (1.2,2.25) {};
\diagram* {
(a) -- [fermion, thick] (d) -- [fermion, thick] (g);
(b) -- [fermion, thick] (e) -- [fermion1, thick] (i);
(c) -- [fermion, thick] (f) -- [fermion1, thick] (h);
(d) -- [dashed, thick] (e) -- [dashed, thick] (f);
};
\end{feynman}
\end{tikzpicture}
\end{minipage}
\hspace{0.02cm}
$+$
\hspace{0.02cm}
\begin{minipage}[c]{0.095\textwidth}
\begin{tikzpicture}
\tikzfeynmanset{
  fermion1/.style={
    /tikz/postaction={
      /tikz/decoration={
        markings,
        mark=at position 0.65 with {
          \node[
            transform shape,
            xshift=-0.5mm,
            fill,
            inner sep=1.5pt,
            draw=none,
            isosceles triangle
          ] { };
        },
      },
      /tikz/decorate=true,
    },
  }
}
\begin{feynman}
\vertex (a) at (0,0.25) {}; 
\vertex (b) at (0.6,0.25) {};
\vertex (c) at (1.2,0.25) {};
\vertex [dot] (d) at (0.0,1.25) {};
\vertex [dot] (e) at (0.6,1.25) {};
\vertex [dot] (f) at (1.2,1.25) {};
\vertex (g) at (0,2.25) {}; 
\vertex (h) at (0.6,2.25) {};
\vertex (i) at (1.2,2.25) {};
\diagram* {
(a) -- [fermion, thick] (d) -- [fermion, thick] (i);
(b) -- [fermion, thick] (e) -- [fermion1, thick] (g);
(c) -- [fermion, thick] (f) -- [fermion, thick] (h);
(d) -- [dashed, thick] (e) -- [dashed, thick] (f);
};
\end{feynman}
\end{tikzpicture}
\end{minipage}
\hspace{0.02cm}
$+$
\hspace{0.02cm}
\begin{minipage}[c]{0.095\textwidth}
\begin{tikzpicture} 
\tikzfeynmanset{
  fermion1/.style={
    /tikz/postaction={
      /tikz/decoration={
        markings,
        mark=at position 0.65 with {
          \node[
            transform shape,
            xshift=-0.5mm,
            fill,
            inner sep=1.5pt,
            draw=none,
            isosceles triangle
          ] { };
        },
      },
      /tikz/decorate=true,
    },
  }
}
\begin{feynman}
\vertex (a) at (0,0.25) {}; 
\vertex (b) at (0.6,0.25) {};
\vertex (c) at (1.2,0.25) {};
\vertex [dot] (d) at (0.0,1.25) {};
\vertex [dot] (e) at (0.6,1.25) {};
\vertex [dot] (f) at (1.2,1.25) {};
\vertex (g) at (0,2.25) {}; 
\vertex (h) at (0.6,2.25) {};
\vertex (i) at (1.2,2.25) {};
\diagram* {
(a) -- [fermion, thick] (d) -- [fermion, thick] (h);
(b) -- [fermion, thick] (e) -- [fermion1, thick] (i);
(c) -- [fermion, thick] (f) -- [fermion, thick] (g);
(d) -- [dashed, thick] (e) -- [dashed, thick] (f);
};
\end{feynman}
\end{tikzpicture}
\end{minipage}
\caption{The six contributions to the antisymmetrized 3N interaction
$V_{\text{3N}}^{\text{as}}$. On the right hand side we choose for illustration a $2\pi$-exchange
contribution.}
\label{fig:V123_asymm}
\end{figure}

\subsection{Partial-wave convergence of 3N interaction in dense matter}
\label{sec:PW_conv_matter}

\begin{figure}[t!]
\centering
\includegraphics[scale=0.48]{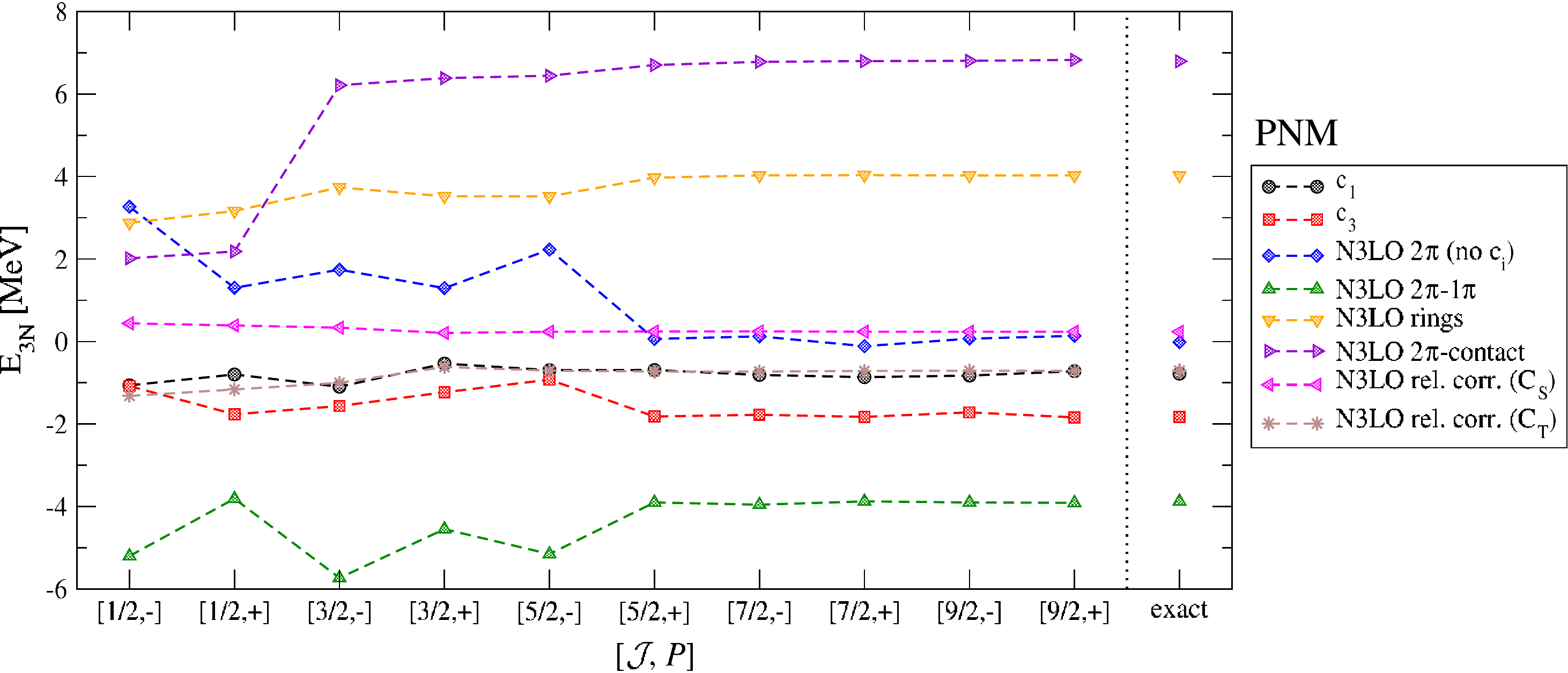}
\vspace{0.3cm} \\
\hspace{0.3cm} \includegraphics[scale=0.51]{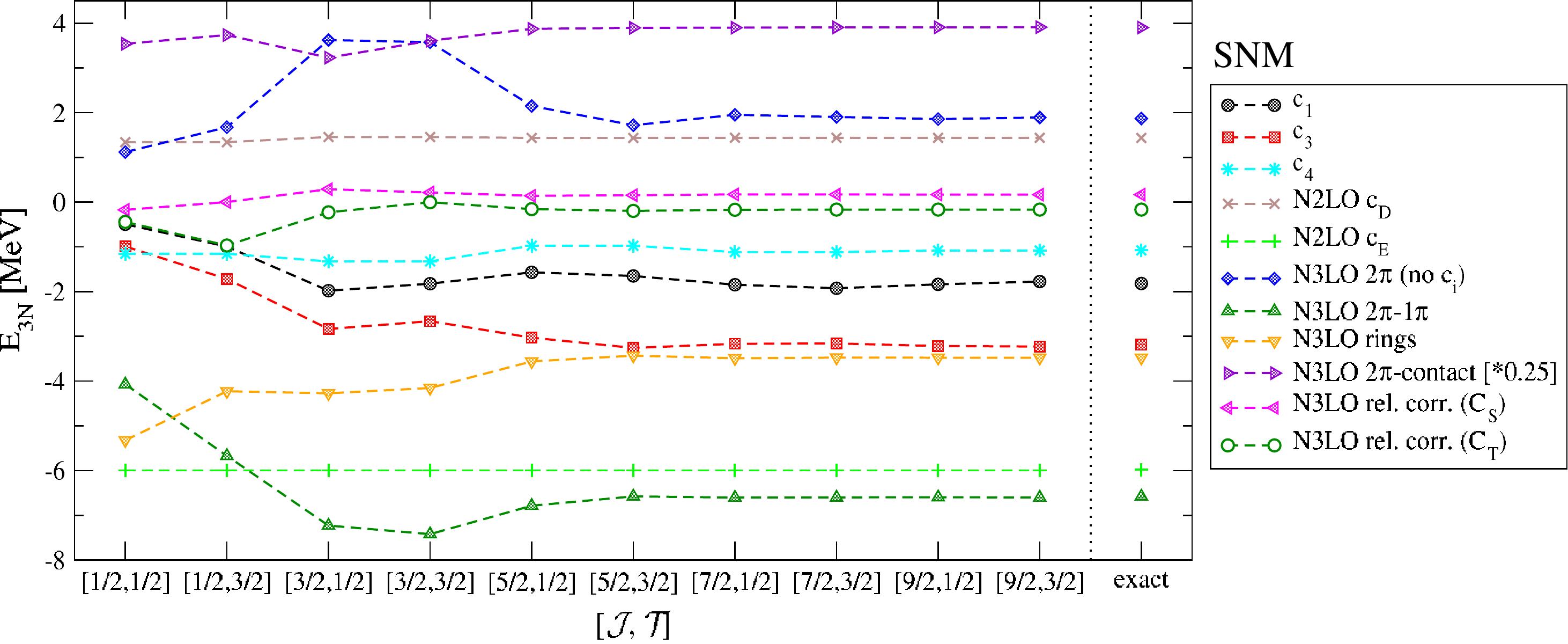}
\caption{Partial-wave contributions to the energy per particle
to pure neutron matter (top panel) and symmetric nuclear matter (lower panel)
in the Hartree-Fock approximation at nuclear saturation density for the
individual unregularized 3N interaction contributions and all topologies up to
N$^3$LO. For the shown results we use the Fermi momenta $k^n_F = 1.7 \,
\text{fm}^{-1}$ for PNM and $k^n_F = k^p_F = 1.35 \,\text{fm}^{-1}$ for SNM.
For the LECs we use the values $c_D = c_E = 1$, $C_S = C_T
= 1 \, \text{fm}^2$ and $c_i = 1 \, \text{GeV}^{-1}$. All results show the
accumulated energies including contributions up to the given partial-wave
channel using $J_{\text{max}} = 5$ for each three-body partial wave. The exact
benchmark results are calculated based on
Refs.~\cite{Tews13N3LO,Krue13N3LOlong}.\\
\textit{Source:} Figures adapted from Ref.~\cite{Hebe15N3LOpw}.}
\label{fig:EOS_3N_HF_PNM_SNM}
\end{figure}

As a first illustration we apply the 3N interactions computed in
Section~\ref{sect:3NF_representation} to calculate the energy of nuclear
matter in the Hartree-Fock approximation at zero temperature. Even though this
many-body approximation is usually not sufficient to obtain converged results,
such calculations are instructive for several reasons:
\begin{itemize}
\item[1.)] In this approximation the contributions from NN and 3N interactions
decouple completely. In addition, the contributions of the individual 3N
topologies do not mix. This allows to study the typical scale of the different
3N contributions to the energy of dense matter, independent of a particular
chosen NN interaction.
\item[2.)] Even though higher order terms in the many-body expansion
provide non-negligible contributions (depending on the regularization scheme
and properties of the NN interactions), for several recently developed chiral
NN interactions like, e.g., those of
Refs.~\cite{Ekst13optNN,Ekst15sat,Carl17UQLagrange,Ente17EMn4lo,Rein17semilocal},
the interactions behave perturbatively, at least for neutron-rich matter and
smaller regularization cutoff scales. For such interactions, calculations in
the Hartree-Fock approximation provide a reasonable estimate of the fully
converged results.
\item[3.)] In the Hartree-Fock approximation it is also possible to calculate the
3N contributions exactly without employing a partial-wave decomposition of the
interactions~\cite{Tews13N3LO,Krue13N3LOlong}. These results represent an
independent benchmark of the partial-wave matrix elements and also provide
insight into the nature of the partial-wave convergence and the required
partial-wave basis sizes (see Table~\ref{tab:PW_data}) at the mean-field
level. Typically these convergence patterns extend reasonably well to
calculations at higher orders in the many-body expansion or also for
calculations of finite nuclei such as those discussed in
Section~\ref{sec:applications}.
\end{itemize}
In the Hartree-Fock approximation, the 3N interaction contributions to the energy per
volume of nuclear matter is given by~\cite{negele1995quantum,Hebe10nmatt}:
\begin{align}
 \frac{E_{\text{HF}}}{V} &= \frac{1}{6} \prod_{i=1}^3 \text{Tr}_{\sigma_i} \text{Tr}_{\tau_i} \int \frac{d \mathbf{k}_i}{(2 \pi)^3} \bigl< 1 2 3 | V^{\text{as}}_{\text{3N}} | 1 2 3 \bigr> \: n_{\mathbf{k}_1}^{\sigma_1 \tau_1} n_{\mathbf{k}_2}^{\sigma_2 \tau_2} n_{\mathbf{k}_3}^{\sigma_3 \tau_3} \, ,
 \label{eq:HF_energy} 
\end{align}
where $n_{\mathbf{k}}^{\sigma \tau}$ are the zero-temperature occupation
numbers for particles with spin $\sigma$ and isospin $\tau$:
$n_{\mathbf{k}}^{\sigma \tau} = \Theta( k_{\text{F}}^{\sigma \tau} -
|\mathbf{k}|)$, with the Fermi momenta $k_{\text{F}}^{\sigma \tau}$. For the
matrix elements we introduced the short-hand notation $\left| 1 2 3 \right> =
\bigl| \mathbf{k}_1 \sigma_1 \tau_1 \mathbf{k}_2 \sigma_2 \tau_2 \mathbf{k}_3
\sigma_3 \tau_3 \bigr>$. In the following we consider only spin-saturated
systems, i.e., $n_{\mathbf{k}}^{\sigma \tau} = n_{\mathbf{k}}^{\tau}$. In
particular, in this section we will focus on pure neutron matter (PNM,
$n_{\mathbf{k}}^{p} = 0$) and symmetric nuclear matter (SNM,
$n_{\mathbf{k}}^{p} = n_{\mathbf{k}}^{n}$).

\begin{figure}[t]
\centering
\includegraphics[width=0.7\textwidth]{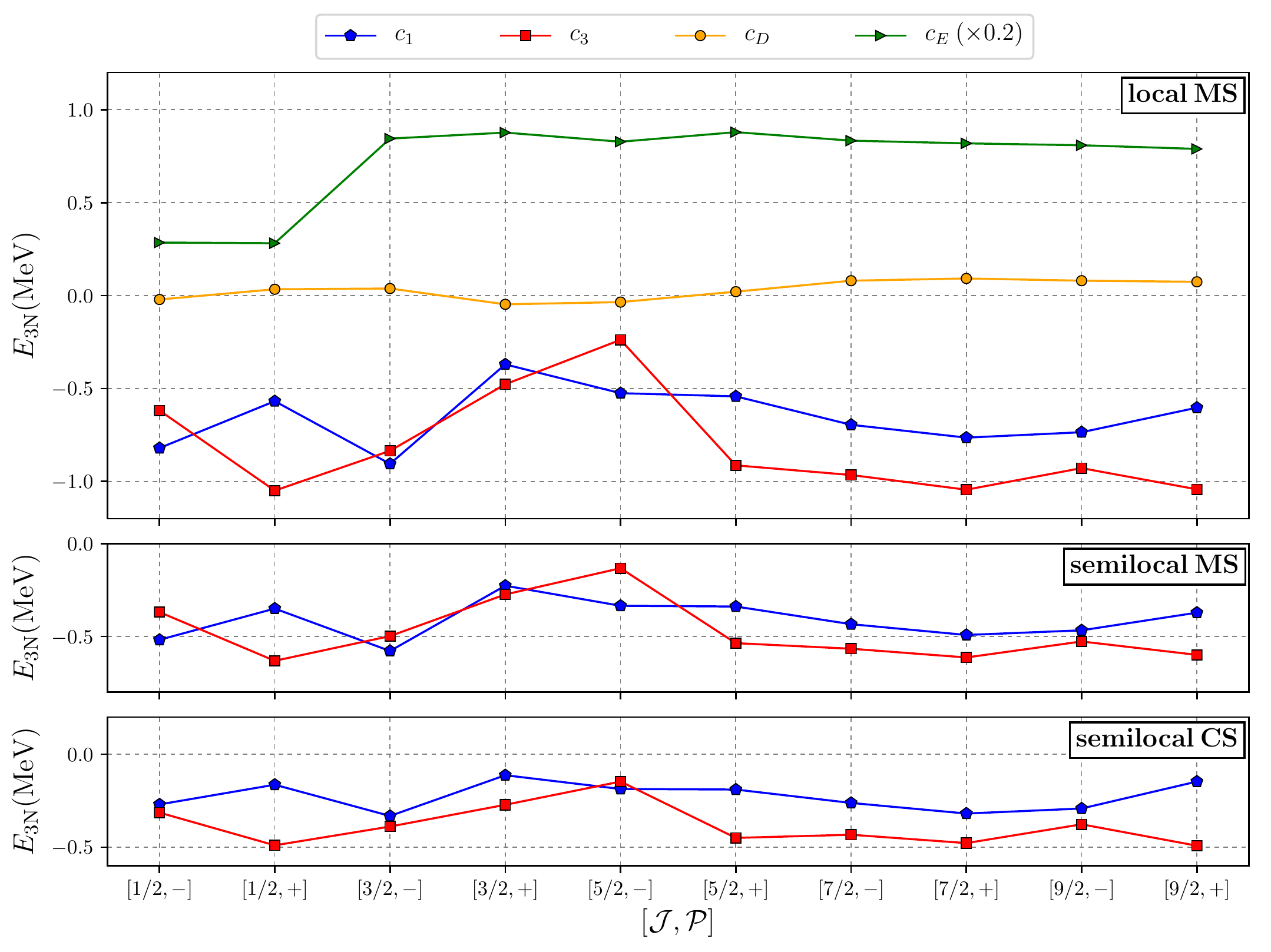}
\caption{Partial-pave contributions to the energy per particle
of PNM in the Hartree-Fock approximation at nuclear saturation density, $k^n_F
= 1.7 \, \text{fm}^{-1}$, for the individual 3N interaction contributions of
all topologies at N$^2$LO for the different regularization schemes (compare to
top panel of Figure~\ref{fig:EOS_3N_HF_PNM_SNM}). As in
Figure~\ref{fig:EOS_3N_HF_PNM_SNM} we use the LEC values $c_i = 1 \,
\text{GeV}^{-1}$ and $c_D = c_E = 1$. Note that the contributions from $c_E$
in the top panel have been scaled by a factor $\tfrac{1}{5}$ for optimized
visibility. For both semilocal regularization schemes all contributions from
the $c_4$, $c_D$ and $c_E$ topology vanish. The cutoff values $\Lambda = 500 \:
\text{MeV}$ and $R=0.9$ fm were chosen for the momentum-space and
coordinate-space regulators, respectively, and the truncation $J_{\text{max}}
= 5$ has been employed for each three-body partial wave.}
\label{fig:EOS_3N_HF_PNM_regs}
\end{figure}

The sums and integrals in Eq.~(\ref{eq:HF_energy}) can be either performed
directly based on the operator form of the interactions by performing all
momentum integrals analytically in a three-dimensional vector representation
(see Refs.~\cite{Tews13N3LO,Krue13N3LOlong}) or
numerically~\cite{Dris17MCshort} (see also Section~\ref{sec:no_PWD}). The
analytical approach is in principle straightforward, but requires some work for
antisymmetrizing the interactions. The final results can be expressed in a
rather compact way for the individual 3N topologies up to N$^3$LO (see
Ref.~\cite{Krue13N3LOlong} for details). Another alternative for evaluating
Eq.~(\ref{eq:HF_energy}) is to make use of the partial-wave representation of
the 3N interactions. Here we need to represent the spin- and isospin sums as
well as the momentum integrals in Eq.~(\ref{eq:HF_energy}) in the partial-wave
momentum basis discussed in Section~\ref{sec:general_3N_decomp}. For example,
for PNM the expression for the HF energy per volume takes the following
form~\cite{Hebe13nmattSRG}:
\begin{align}
\frac{E_{\text{HF}}^{\text{PNM}} (k_{\text{F}})}{V} &= \frac{1}{3\pi} \frac{1}{(4 \pi)^2} \int dp p^2 dq q^2 \int d\cos \theta_{\mathbf{p} \mathbf{q}} \int dP_{\text{3N}} P_{\text{3N}}^2 \int d \cos \theta_{\mathbf{P}_{\text{3N}}} \int d \phi_{\mathbf{P}_{\text{3N}}} \nonumber \\
& \times \sum_{\alpha, \alpha'} \delta_{S S'} \sum_{\bar{L},\mathcal{S}, \mathcal{L}, \mathcal{J}} \hat{\mathcal{S}} \hat{\mathcal{L}} \hat{\mathcal{J}} \sqrt{\hat{J} \hat{j} \hat{J}' \hat{j}' \hat{L} \hat{L}' \hat{l} \hat{l}'} (-1)^{l+l'+\mathcal{L}} \mathcal{C}_{l' 0 l 0}^{\bar{L} 0} \mathcal{C}_{L' 0 L 0}^{\bar{L} 0} P_{\bar{L}} (\hat{\mathbf{p}} \cdot \hat{\mathbf{q}}) \nonumber \\
& \times n_{\mathbf{p} - \mathbf{q}/2 + \mathbf{P}_{\text{3N}}/3} \: n_{\mathbf{p} + \mathbf{q}/2 - \mathbf{P}_{\text{3N}}/3} \: n_{\mathbf{q} + \mathbf{P}_{\text{3N}}/3} \nonumber \\
& \times \left\{
\begin{array}{ccc}
 L' & L & \bar{L} \\
 l & l' & \mathcal{L} 
\end{array}
\right\}
\left\{
\begin{array}{ccc}
 L & S & J \\
 l & \tfrac{1}{2} & j \\
 \mathcal{L} & \mathcal{S} & \mathcal{J}
\end{array}
\right\} 
\left\{
\begin{array}{ccc}
 L' & S & J' \\
 l' & \tfrac{1}{2} & j' \\
 \mathcal{L} & \mathcal{S} & \mathcal{J}
\end{array}
\right\} \bigl< p q \alpha' | V_{\text{3N}}^{\text{as,reg}} | p q \alpha \bigr>
  \, ,
\label{eq:spinsum_HF_PNM}
\end{align}
with $\hat{x} \equiv 2 x + 1$. The corresponding relation for SNM is identical
up to an isospin factor $2 \mathcal{T} + 1$ plus an additional sum over the
three-body isospin quantum number $\mathcal{T}$.

\begin{figure}[t]
\centering
\includegraphics[width=0.7\textwidth]{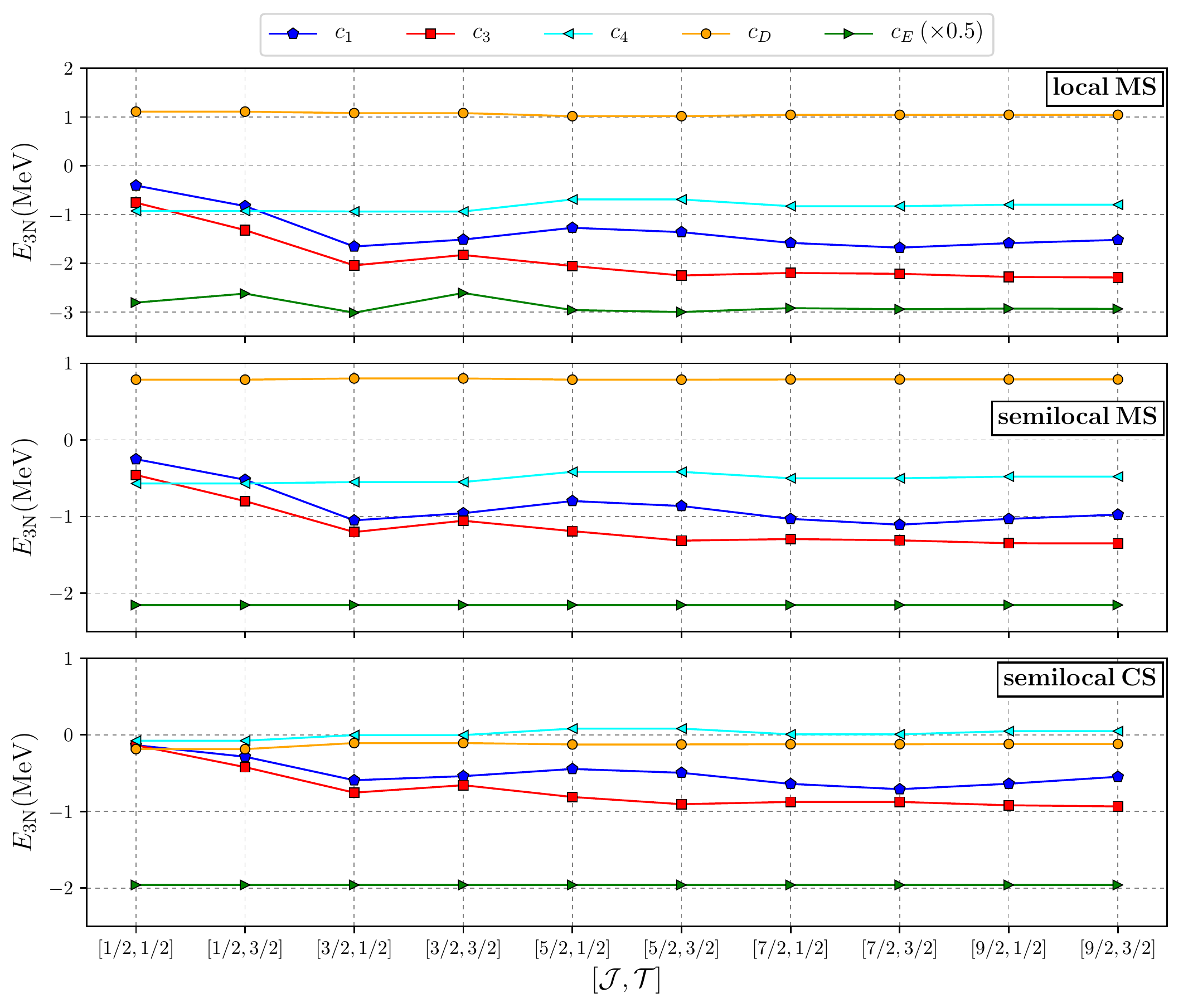}
\caption{Partial-wave contributions to the energy per particle
of SNM in the Hartree-Fock approximation at nuclear saturation density ($k^n_F
= k_F^p = 1.35 \, \text{fm}^{-1}$) for the individual 3N interaction
contributions of all topologies at N$^2$LO for the different regularization
schemes (compare with lower panel of Figure~\ref{fig:EOS_3N_HF_PNM_SNM}). The
same LEC and cutoff values as in Figure~\ref{fig:EOS_3N_HF_PNM_regs} have been
used. Note that the contributions from $c_E$ have been scaled by a factor
$\tfrac{1}{2}$ for optimized visibility.}
\label{fig:EOS_3N_HF_SNM_regs}
\end{figure}

Figure~\ref{fig:EOS_3N_HF_PNM_SNM} illustrates the partial-wave convergence of
3N contributions to the energy per particle in PNM and SNM based on
unregularized 3N interactions compared to exact results calculated in
Refs.~\cite{Tews13N3LO,Krue13N3LOlong}. Since the Fermi momenta serve as
natural ultraviolet cutoffs (see Eq.~(\ref{eq:spinsum_HF_PNM})), we do not
need to regularize the interactions for these benchmark calculations. The
plots show the accumulated 3N interaction contributions to the energy per particle of PNM
(upper panel) and SNM (lower panel) as a function of the three-body quantum
number $\mathcal{J}$ for the individual 3N interaction topologies up to N$^3$LO in chiral
EFT. Specifically, for the PNM results shown in the upper panel of
Figure~\ref{fig:EOS_3N_HF_PNM_SNM} we use $k_{\rm{F}}=1.7 \, \text{fm}^{-1}$ for the
neutron Fermi momentum, which corresponds to a neutron number density of $n
\simeq 0.166 \, \text{fm}^{-3}$. In neutron matter only matrix elements in the
three-body isospin channel $\mathcal{T} = \tfrac{3}{2}$ contribute, while the
N$^2$LO topologies that include the low-energy coupling constants $c_4$, $c_D$
and $c_E$ vanish exactly for unregularized 3N interactions~\cite{Hebe10nmatt}.
The results of Figure~\ref{fig:EOS_3N_HF_PNM_SNM} show that matrix elements up
to the partial-wave channel with $\mathcal{J} = \tfrac{5}{2}$ and both
three-body parities $\mathcal{P}=(-1)^{L+l}$ can provide significant
contributions to the energy, whereas higher partial waves give only small
corrections. Overall, we find that the results including all contributions up
to $\mathcal{J}=\tfrac{9}{2}$ are well converged and show excellent agreement
with the exact results.

For symmetric nuclear matter we find a very similar convergence pattern. In
contrast to neutron matter, here all 3N interaction topologies and also both
three-body isospin channels, $\mathcal{T}=\tfrac{1}{2}$ and $\mathcal{T}=\tfrac{3}{2}$,
contribute to the energy. For the results shown in the lower panel of
Figure~\ref{fig:EOS_3N_HF_PNM_SNM} we fix the neutron and proton Fermi momenta
to $k^n_F = k_F^p = 1.35 \, \text{fm}^{-1}$, which again corresponds to a
total number density of $n \simeq 0.166 \, \text{fm}^{-3}$. We show the
accumulated contributions to the energy for the individual partial-wave
channels, where each $[\mathcal{J},\mathcal{T}]$ channel includes
contributions from both three-body parity channels $\mathcal{P}=\pm 1$. Again,
we observe excellent partial-wave convergence and essentially perfect
agreement with the exact Hartree-Fock results.

In Figures~\ref{fig:EOS_3N_HF_PNM_regs} and ~\ref{fig:EOS_3N_HF_SNM_regs} we
show for comparison the corresponding partial-wave contributions from the 3NF
topologies at N$^2$LO to PNM and SNM for the regularization schemes ``local
MS'', ``semilocal MS'' and ``semilocal CS'' (see
Section~\ref{sec:3N_regularization}). Overall, we find that in these schemes
the overall size of the contributions in PNM tend to be suppressed compared to
the ``nonlocal MS'' or unregularized scheme. For SNM the size of the contributions
are of similar size for the ``local MS'' scheme and the unregularized case.
Moreover, for the ``local MS'' scheme we find nonzero contributions for PNM for
the short-range topologies proportional to the LECs $c_D$ and $c_E$. This is
due to the local nature of the regulator, which induces a finite range for the
short-range couplings in these interaction topologies (see also
Figure~\ref{fig:contour_cE_induced} and related discussion). In particular, we
find significant contributions from the pure short-range interaction $c_E$.
The qualitative trends of the contributions from the long-range topologies are
remarkably similar in all shown regularization schemes, except for the $c_4$
topology, which gives very small contributions in the ``semilocal CS'' scheme.

In summary, the results of this section demonstrate that the partial-wave
contributions to the Hartree-Fock energy contributions to nuclear matter show
an excellent agreement with exact mean-field results for unregularized 3N
interactions by summing up all contributions up to the total three-body
angular momentum $\mathcal{J} = \tfrac{9}{2}$. In addition, for the other
studied regularization schemes the results are reasonably well converged in
this model space. Even though certain qualitative trends appear universal, the
specific size of the individual contributions can depend quite significantly
on the employed regularization scheme.

\subsection{SRG evolution in momentum basis}
\label{sec:SRG}

\subsubsection{Flow equations}
\label{sec:SRG_flow_equations}

One of the key challenges of all basis-expansion \textit{ab initio} many-body
frameworks is achieving convergence of observables as a function of the
basis size. The total number of required basis states is mainly determined by
two factors:
\begin{itemize}
\item[1.)] The total number of particles in the system, plus possible
additional challenges like, e.g., the need to include scattering states in
the basis if the system is only loosely bound as one approaches the drip lines
of the nuclear chart (see, e.g. Refs.~\cite{Hage12Ox3N,Lang14c3N9Be}).
\item[2.)] The coupling strength of the nuclear interactions between
low-energy states to high-energy intermediate virtual states.
\end{itemize}
The first factor is predetermined by the system under investigation and there
is \textit{a priori} no freedom for optimization within a given many-body
framework, except for possibly switching to more suitable single-particle
bases (see, e.g. Ref.~\cite{Tich18natorb}). Since different methods exhibit
different scaling behaviors as a function of particles (see
Section~\ref{sec:Intro}), frameworks like IM-SRG, MBPT, CC are more suited for
studying medium-mass and heavier nuclei than conceptually exact methods like,
e.g, NCSM. The second factor, on the other hand, is determined by properties
of the employed nuclear interactions. If the NN or 3N interactions have
sizable coupling strength of low- and high-energy states, the many-body basis
needs to be large enough such that these virtual excitations induced by these
couplings can be captured. This is the case even if we are eventually only
interested in low-energy properties of nuclei, like ground-state energies,
radii, quadrupole moments or low-lying excited states~\cite{Bogn10PPNP}. The
Similarity Renormalization Group (SRG) provides a framework to systematically
decouple low- and high-energy states via a unitary transformations that
consistently renormalize all operators, including many-body forces, while
preserving low-energy observables~\cite{Glaz93SRG,Wegn94SRG,Bogn07SRG}.

The basic underlying idea of the SRG consists of
the construction of a unitary operator $U(s)$ that transforms the Hamiltonian
$\hat{H}$ (as well as other operators, see
Refs.~\cite{Ande11opSRG,Schus14opSRG,More15deep,More17sdde}):
\begin{equation}
\hat{H} (s) = \hat{U} (s) \hat{H} \hat{U}^{\dagger} (s) \equiv \hat{T}_{\text{rel}} + \hat{V}_{\text{NN}} (s) + \hat{V}_{\text{3N}} (s) \, .
\label{eq:unitary_transformation}
\end{equation}
At this point $s$ is an arbitrary parameter and will be specified further
below. Since the relative kinetic energy is effectively a two-body operator
(see Eq.~(\ref{eq:kinetic_energy_equal_masses})), the separation of
contributions from the kinetic energy and NN interactions of the transformed
Hamiltonian is in general ambiguous. The transformed NN and 3N interactions
$\hat{V}_{\text{NN}} (s)$ and $\hat{V}_{\text{3N}} (s)$ can be defined
uniquely as done in Eq.~(\ref{eq:unitary_transformation}) by choosing the
relative kinetic energy to be independent of $s$. Taking the derivative of
Eq.~(\ref{eq:unitary_transformation}) with respect to the parameter $s$ we
obtain the following flow equation
\begin{equation}
\frac{d \hat{H} (s)}{d s} = \frac{d \hat{V}_{\text{NN}} (s)}{d s} + \frac{d \hat{V}_{\text{3N}} (s)}{d s} = \bigl[ \hat{\eta} (s), \hat{H} (s) \bigr] \, , 
\label{eq:dHds}
\end{equation}
with the anti-unitary generator
\begin{equation}
\hat{\eta} (s) = \frac{d \hat{U} (s)}{ds} \hat{U}^{\dagger} (s) = - \hat{\eta}^{\dagger} (s) \, .
\label{eq:SRG_generator}
\end{equation}
Here $\hat{\eta} (s)$ specifies the unitary transformation and can be chosen
in a suitable way in order to achieve particular decoupling patterns. By far
the most common choice for practical applications of the SRG in the context of
nuclear structure has been the choice
\begin{equation}
\hat{\eta} (s) = \bigl[ \hat{T}_{\text{rel}}, \hat{H} (s) \bigr] \, , 
\label{eq:generator_T_SRG}
\end{equation}
which leads to the flow equation
\begin{equation}
\frac{d \hat{H} (s)}{d s} = \frac{d \hat{V}_{\text{NN}} (s)}{d s} + \frac{d \hat{V}_{\text{3N}} (s)}{d s} = \bigl[ \bigl[ \hat{T}_{\text{rel}}, \hat{H} (s) \bigr], \hat{H} (s) \bigr] \, .
\label{eq:dHds_Tkin}
\end{equation}
In this flow equation the parameter $s$ plays the role of a resolution
scale~\cite{Bogn10PPNP}. It is also customary to introduce an alternative
parameter $\lambda$, which has units of momenta:
\begin{equation}
\lambda = s^{-1/4} \, .
\end{equation}
This parameter serves as a measure of the degree of decoupling in NN
interactions as $\lambda^2$ is directly related to the width of the
band-diagonal structure in NN matrix elements as a function of the square of
the relative momenta~\cite{Bogn07SRG}. This implies that the evolution of NN
interactions via Eq.~(\ref{eq:dHds_Tkin}) leads to a decoupling of low and
high momenta as they are evolved to lower resolution scales $\lambda$ and to
significantly less correlated wave functions, which means that the nuclear
many-body problem becomes more perturbative~\cite{Bogn10PPNP,Furn13RPP,Tich16HFMBPT}.
The SRG flow for a given initial NN and 3N interaction is completely
determined by the generator $\hat{\eta}_s$ defined in
Eq.~(\ref{eq:SRG_generator}). We emphasize that SRG transformations generally
reshuffle strength within a given many-body sector in Fock space like NN or 3N
interactions, but also couple to other sectors like four- and higher-body
forces, even when such forces are initially absent (see discussion below).
Since SRG transformations need to be truncated in practice at a given order in
many-body operators, the chosen generator should ideally not induce strong
contributions in a regime beyond a given truncation scheme. The study of
alternative efficient generators rather than the canonical choice
Eq.~(\ref{eq:generator_T_SRG}) is presently an active field of research.
Several choices have been explored (see, e.g.,
Refs.~\cite{Ande08blockdiagSRG,Li11altgenSRG,Dica14altgenSRG}), but so far no
more powerful generator has been found, and the generator shown in
Eq.~(\ref{eq:generator_T_SRG}) still represents the preferred choice for many
state-of-the-art calculations. Hence, in the following we will stick for
illustration to the particular flow equation Eq.~(\ref{eq:dHds_Tkin}), but
stress that the framework discussed below can be straightforwardly extended to
alternative generators.

We also emphasize that Eqs.~(\ref{eq:dHds}) and (\ref{eq:dHds_Tkin}) are operator
identities and can hence be represented in arbitrary bases. In recent years
two different strategies have been employed to evolve NN and 3N interactions
via these flow equations:
\begin{itemize}
\item[(a)] Starting from realistic nuclear NN and 3N interactions it is possible to
systematically evolve the full Hamiltonian $\hat{H} (s)$ as defined in
Eq.~(\ref{eq:H_second_quantized}). This has been achieved by representing
Eq.~(\ref{eq:dHds}) in a discrete three-body harmonic oscillator
basis~\cite{Jurg08SRG3N1D,Jurg09SRG3N,Jurg09PhD,Jurg10SRG3N,Roth11SRG,Roth14SRG3N}
or hyperspherical momentum basis~\cite{Wend13msHHSRG}. The sum of all
contributions of the initial Hamiltonian at $s=0$ are represented in the
chosen three-body basis and then evolved via the flow equation
(\ref{eq:dHds_Tkin}). After the evolution the matrix elements of the evolved
NN and 3N interactions can be extracted by suitable subtractions from the full
Hamiltonian.

\item[(b)] The approach (a) involves fundamental problems for continuous bases
like the momentum basis in Eq.~(\ref{eq:Jj_bas}) due to the presence of delta
functions for two-body operators, related to spectator particles (see
Ref.~\cite{Bogn07SRG} and discussion below). Instead, for continuous bases the
evolution of NN and 3N forces needs to be separated explicitly, which allows
to avoid the need for subtractions of NN interactions from the total
Hamiltonian in a three-body basis completely. In practice that means
Eq.~(\ref{eq:dHds_Tkin}) is being reformulated in terms of two explicit flow
equations for $\hat{V}_{\text{NN}} (s)$ and $\hat{V}_{\text{3N}} (s)$ (see
Refs.~\cite{Aker11SRG1D,Hebe12msSRG}). Since SRG transformations are usually
characterized by the coupling patterns of momentum eigenstates, the momentum
basis is a well-suited basis in which to construct the SRG generator $\eta(s)$.
\end{itemize}
In the following we discuss in detail the implementation of approach (b) using
the partial-wave momentum representation discussed in detail in
Section~\ref{sect:3NF_representation}. The approach (a) is well established by
now and details can be found, e.g., in Ref.~\cite{Jurg10SRG3N}. However,
evolution in a momentum basis has several advantages compared to a discrete
oscillator basis. First, the oscillator basis has intrinsic infrared and
ultraviolet cutoffs that depend on the basis size and oscillator parameter
$\Omega$~\cite{Jurg10SRG3N}, which could lead to convergence issues for 3N
forces. This problem can be avoided by first evolving in momentum space and
then using a straightforward transformation to an oscillator basis with
\emph{any} $\Omega$ (see Section~\ref{sec:HO_transf}). Second, the
momentum-space interactions can be used directly in calculations of infinite
matter (see Section~\ref{sec:applications_matter} for first results). This
allows tests of whether consistently-evolved NN plus 3N forces, initially fit
only to few-body properties, predict empirical nuclear saturation properties
within theoretical errors, as found previously for evolved NN forces combined
with fitted 3N forces~\cite{Hebe11fits} (see Section~\ref{sec:sep_3N_fits}).
Finally, since SRG transformations are usually characterized by the decoupling
of momentum eigenstates, the momentum basis is a natural basis in which to
construct the SRG generator $\hat{\eta}_s$. In particular, momentum-diagonal
generators such as $\Trel$ (as chosen here) can be implemented very
efficiently in a momentum basis and it is straightforward to generalize to the
Hamiltonian-diagonal form advocated by Wegner~\cite{Wegn94SRG}. The
possibility of using the generator to suppress the growth of many-body forces
is also under active investigation.

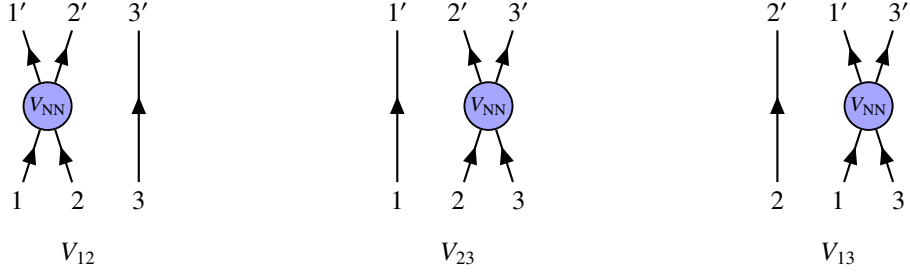
\begin{figure}
\centering
\begin{minipage}[c]{0.15\textwidth}
\begin{tikzpicture} 
\tikzfeynmanset{
  my dot/.style={fill=red},
  every vertex/.style={my dot},
}
\begin{feynman}
\vertex (a) at (0,0) {\(1\)}; 
\vertex (b) at (0.8,0) {\(2\)};
\vertex (c) at (1.6,0) {\(3\)};
\vertex (d1) at (0.3,1.25) {};
\vertex (d2) at (0.5,1.25) {};
\vertex [blob, /tikz/minimum size=18pt, fill=blue!35, line width=0.25mm, font=\fontsize{8}{0}\selectfont] (d) at (0.4,1.25) {\(V_{\rm NN}\)};
\vertex (e) at (0,2.5) {\(1'\)}; 
\vertex (f) at (0.8,2.5) {\(2'\)};
\vertex (g) at (1.6,2.5) {\(3'\)};
\vertex (h) at (0.8,-0.7) {\(V_{12}\)};
\diagram* {
(a) -- [fermion, thick] (d) -- [fermion, thick] (e);
(b) -- [fermion, thick] (d) -- [fermion, thick] (f);
(c) -- [fermion, thick] (g);
};
\end{feynman}
\end{tikzpicture}
\end{minipage}
\hspace{2.3cm}
\begin{minipage}[c]{0.15\textwidth}
\begin{tikzpicture} 
\tikzfeynmanset{
  my dot/.style={fill=red},
  every vertex/.style={my dot},
}
\begin{feynman}
\vertex (a) at (0,0) {\(1\)}; 
\vertex (b) at (0.8,0) {\(2\)};
\vertex (c) at (1.6,0) {\(3\)};
\vertex [blob, /tikz/minimum size=18pt, fill=blue!35, line width=0.25mm, font=\fontsize{8}{0}\selectfont] (d) at (1.2,1.25) {\(V_{\rm NN}\)};
\vertex (e) at (0,2.5) {\(1'\)}; 
\vertex (f) at (0.8,2.5) {\(2'\)};
\vertex (g) at (1.6,2.5) {\(3'\)};
\vertex (h) at (0.8,-0.7) {\(V_{23}\)};
\diagram* {
(a) -- [fermion, thick] (e);
(b) -- [fermion, thick] (d) -- [fermion, thick] (f);
(c) -- [fermion, thick] (d) -- [fermion, thick] (g);
};
\end{feynman}
\end{tikzpicture}
\end{minipage}
\hspace{2.3cm}
\begin{minipage}[c]{0.15\textwidth}
\begin{tikzpicture} 
\tikzfeynmanset{
  my dot/.style={fill=red},
  every vertex/.style={my dot},
}
\begin{feynman}
\vertex (a) at (0,0) {\(2\)}; 
\vertex (b) at (0.8,0) {\(1\)};
\vertex (c) at (1.6,0) {\(3\)};
\vertex [blob, /tikz/minimum size=18pt, fill=blue!35, line width=0.25mm, font=\fontsize{8}{0}\selectfont] (d) at (1.2,1.25) {\(V_{\rm NN}\)};
\vertex (e) at (0,2.5) {\(2'\)}; 
\vertex (f) at (0.8,2.5) {\(1'\)};
\vertex (g) at (1.6,2.5) {\(3'\)};
\vertex (h) at (0.8,-0.7) {\(V_{13}\)};
\diagram* {
(a) -- [fermion, thick] (e);
(b) -- [fermion, thick] (d) -- [fermion, thick] (f);
(c) -- [fermion, thick] (d) -- [fermion, thick] (g);
};
\end{feynman}
\end{tikzpicture}
\end{minipage}
\caption{The three contributions of two-body interactions $V_{ij}$ in a
three-body basis. The vertices denote antisymmetrized interactions, where we
also included the spectator particle in each diagram. The different diagrams
are related by permutation of states, specifically $V_{23} = P_{123}
V_{12} P_{123}^{-1}$ and $V_{13} = P_{132}^{-1} V_{12} P_{132}$ (see main
text).}
\label{fig:Vij_threebody_basis}
\end{figure}

For the derivation of the flow equation in momentum space we make, as a first
step, the representation of Eq.~(\ref{eq:dHds_Tkin}) more explicit by adapting
the notation of Ref.~\cite{Bogn07SRG} and representing all quantities as matrix
elements in a generic three-body basis $\left| 1 2 3 \right> =
\hat{a}_1^{\dagger} \hat{a}^{\dagger}_2 \hat{a}^{\dagger}_3 | 0 \bigr>$, $\bigl< 1' 2' 3' | = \bigl< 0 |
\hat{a}_{3'} \hat{a}_{2'} \hat{a}_{1'}$ and
$\left< 1' 2' 3' | 1 2 3
\right> = \delta_{1 1'} \delta_{2 2'} \delta_{3 3'}$. Evaluating the matrix
elements of the second-quantized operators in this basis by employing Wick's
theorem~\cite{Fett71QToMPS} results in:
\begin{align}
\bigl< 1' 2' 3' | \hat{V}_{\text{NN}} | 1 2 3 \bigr> &= \sum_{ijkl} \bigl< i j | \vnn^{\text{as}} | k l \bigr> 
\biggl[ 
\contraction[1.75ex]{\bigl< 1' 2'}{3}{' | \hat{a}}{{}_i^{\dagger}\hat{a}^{\dagger}} 
\contraction[1.0ex]{\bigl< 1'}{2}{' 3'|}{\hat{a}} 
\contraction[1.75ex]{\bigl< 1' 2' 3' | \hat{a}_i^{\dagger} \hat{a}_j^{\dagger}}{\hat{a}}{{}_l \hat{a}_k | 1 2}{3} 
\contraction[1.0ex]{\bigl< 1' 2' 3' | \hat{a}_i^{\dagger} \hat{a}_j^{\dagger} \hat{a}_l}{\hat{a}}{{}_k | 1}{2} 
\contraction[2.5ex]{\bigl<}{1}{' 2' 3'| \hat{a}_i^{\dagger} \hat{a}_j^{\dagger} \hat{a}_l \hat{a}_k |}{1} 
\bigl< 1' 2' 3' | \hat{a}_i^{\dagger} \hat{a}_j^{\dagger} \hat{a}_l \hat{a}_k | 1 2 3 \bigr> 
+ 
\contraction[1.75ex]{\bigl< 1' 2'}{3}{'| \hat{a}_i^{\dagger}}{\hat{a}} 
\contraction[1.0ex]{\bigl<}{1}{' 2' 3' |}{\hat{a}} 
\contraction[1.75ex]{\bigl< 1' 2' 3' | \hat{a}_i^{\dagger} \hat{a}_j^{\dagger}}{\hat{a}}{{}_l \hat{a}_k | 1 2}{3} 
\contraction[1.0ex]{\bigl< 1' 2' 3' | \hat{a}_i^{\dagger} \hat{a}_j^{\dagger} \hat{a}_l}{\hat{a}}{{}_k |}{1} 
\contraction[2.5ex]{\bigl< 1'}{2}{' 3' | \hat{a}_i^{\dagger} \hat{a}_j^{\dagger} \hat{a}_l \hat{a}_k | 1}{2} 
\bigl< 1' 2' 3' | \hat{a}_i^{\dagger} \hat{a}_j^{\dagger} \hat{a}_l \hat{a}_k | 1 2 3 \bigr>
+ 
\contraction[1.75ex]{\bigl< 1'}{2}{' 3'| \hat{a}_i^{\dagger}}{\hat{a}} 
\contraction[1.0 ex]{\bigl<}{1}{' 2' 3' |}{\hat{a}} 
\contraction[1.75ex]{\bigl< 1' 2' 3' | \hat{a}_i^{\dagger} \hat{a}_j^{\dagger}}{\hat{a}}{{}_l \hat{a}_k | 1}{2} 
\contraction[1.0ex]{\bigl< 1' 2' 3' | \hat{a}_i^{\dagger} \hat{a}_j^{\dagger} \hat{a}_l}{\hat{a}}{{}_k |}{1} 
\contraction[2.5ex]{\bigl< 1' 2'}{3}{' | \hat{a}_i^{\dagger} \hat{a}_j^{\dagger} \hat{a}_l \hat{a}_k | 1 2}{3} 
\bigl< 1' 2' 3' | \hat{a}_i^{\dagger} \hat{a}_j^{\dagger} \hat{a}_l \hat{a}_k | 1 2 3 \bigr> \biggr] \nonumber \\ 
&= \bigl< 2' 3' | V_{\text{NN}}^{\text{as}} | 2 3 \bigr> \, \delta_{1 1'} + \bigl< 1' 3' | V_{\text{NN}}^{\text{as}} | 1 3 \bigr> \, \delta_{2 2'} + \bigl< 1' 2' | V_{\text{NN}}^{\text{as}} | 1 2 \bigr> \, \delta_{3 3'} \nonumber \\
&\equiv V_{23} + V_{13} + V_{12} \, , 
\label{eq:NN_3N_first_quantized}
\end{align} 
where we used the fact that there are four contractions for each of the three
terms. Each of these four contractions provide identical contributions due to
the antisymmetry of the matrix elements. This combinatorial factor cancels the
factor $\tfrac{1}{4}$ in Eq.~(\ref{eq:H_second_quantized}). Note that the two-body
interactions $V_{ij}$ between particles $i$ and $j$ include an implicit delta
function $\delta_{k k'}$ with $k \neq i,j$, corresponding to the spectator
particle $k$ (see Figure~\ref{fig:Vij_threebody_basis}). For the
antisymmetrized 3N interaction we obtain 36 possible
contractions\footnote{As a first step pick one creation operator. There are three possible
single-particle final states to contract it with. For the second creation
operator there are two states left, whereas for the last one there is just a
single contraction left, giving a total factor of 6. The same factor is
obtained for the annihilation operators, giving a total factor of 36.}, each
providing the same contribution:
\begin{align}
\bigl< 1' 2' 3' | \hat{V}_{\text{3N}} | 1 2 3 \bigr> &= \sum_{ijklmn} \bigl< i j k | \vtn^{\text{as}} | l m n \bigr> 
\contraction[1.0ex]{\bigl<}{1}{' 2' 3'|}{\hat{a}} 
\contraction[1.75ex]{\bigl< 1'}{2}{' 3' | \hat{a}_i^{\dagger}}{\hat{a}}
\contraction[2.5ex]{\bigl< 1' 2'}{3}{'| \hat{a}_i^{\dagger} \hat{a}_j^{\dagger}}{\hat{a}}
\contraction[1.0ex] {\bigl< 1' 2' 3' | \hat{a}_i^{\dagger} \hat{a}_j^{\dagger} \hat{a}_k^{\dagger} \hat{a}_n \hat{a}_m}{\hat{a}}{{}_l|}{1}
\contraction[1.75ex]{\bigl< 1' 2' 3' | \hat{a}_i^{\dagger} \hat{a}_j^{\dagger} \hat{a}_k^{\dagger} \hat{a}_n}{\hat{a}}{{}_m \hat{a}_l | 1}{2}
\contraction[2.5ex]{\bigl< 1' 2' 3' | \hat{a}_i^{\dagger} \hat{a}_j^{\dagger} \hat{a}_k^{\dagger}}{\hat{a}}{{}_n \hat{a}_m \hat{a}_l | 1 2 }{3}
\bigl< 1' 2' 3' | \hat{a}_i^{\dagger} \hat{a}_j^{\dagger} \hat{a}_k^{\dagger} \hat{a}_n \hat{a}_m \hat{a}_l | 1 2 3 \bigr> = \bigl< 1' 2' 3' | \vtn^{\text{as}} | 1 2 3 \bigr> \equiv V_{123} \, .
\end{align}
The kinetic energy can be decomposed in the following way~\cite{Bogn07SRG}
(see also discussion in Section~\ref{sec:3NF_coord_def}):
\begin{equation}
\bigl< 1' 2' 3' | \hat{T}_{\text{rel}} | 1 2 3 \bigr> = T_{23} + T_1 = T_{31} + T_2 = T_{12} + T_3 = T_{\text{rel}} \, ,
\label{eq:Trel_first_quantized}
\end{equation}
where the kinetic energy terms $T_{ij}$ and $T_k$ correspond to the relative
kinetic energy between particles $i$ and $j$ and the contribution of particle
$k$, respectively. 

Note that there exist natural basis representations for each of the kinetic
energy contributions and the NN interaction terms when using the Jacobi
momentum basis defined in Eq.~(\ref{eq:Jj_bas}). For example, the NN
interaction term $V_{12}$ takes the following form in basis representation
$\{12\}$ (see Appendix~\ref{sec:normalization}):
\begin{equation}
 \tensor*[_{\{12\}}]{\left< p' q' \alpha' | V_{12} | p q \alpha \right>}{_{\{12\}}} = \left< p' \alpha_{12}' | V_{\text{NN}} | p \alpha_{12} \right> \frac{\delta(q - q')}{q q'} \delta_{\alpha_3 \alpha'_3} \, ,
 \label{eq:VN_pw_representation}
\end{equation}
where $\alpha_{12} = \{L, S, J, T\}$ represents all quantum numbers of the
subsystem consisting of particles $1$ and $2$ and accordingly $\alpha_{3} =
\{l, j\}$ those of particle $3$ (see Section~\ref{sec:general_3N_decomp}).
According representations can be obtained for the interaction contributions
$V_{23}$ and $V_{31}$ in the representations $\{23\}$ and $\{31\}$, respectively.
Using other basis representations makes the embedding of NN interactions into
the three-body basis more complicated. However, eventually we need to
represent all the terms of the flow equation~(\ref{eq:dHds_Tkin}) in a single
chosen three-body basis. This step will be discussed further below.

Similarly we obtain for the kinetic energy:
\begin{align}
 \tensor*[_{\{12\}}]{\left< p' q' \alpha' | T_{12} | p q \alpha \right>}{_{\{12\}}} = \frac{p^2}{m} \frac{\delta(p - p')}{p p'} \frac{\delta(q - q')}{q q'} \delta_{\alpha \alpha'}, \quad \tensor*[_{\{12\}}]{\left< p' q' \alpha' | T_{3} | p q \alpha \right>}{_{\{12\}}} = \frac{3}{4} \frac{q^2}{m} \frac{\delta(p - p')}{p p'} \frac{\delta(q - q')}{q q'}  \delta_{\alpha \alpha'} \, ,
\end{align}
and accordingly for the other terms in their natural basis representations. Note that from the relations above the following commutator relations follow:
\begin{equation}
\left[ V_{12}, T_3 \right] = \left[ V_{23}, T_1 \right] = \left[ V_{13}, T_2 \right] = 0 \, .
\label{eq:commutator_Vij_Tkin}
\end{equation}
The representations in Eqs.~(\ref{eq:NN_3N_first_quantized}) and
(\ref{eq:Trel_first_quantized}) allow to recast Eq.~(\ref{eq:dHds_Tkin}) as
separate SRG flow equations for the two- and three-body interactions~\cite{Bogn07SRG}:
\begin{align}
\frac{d V_{ij}}{ds} &= \bigl[ \bigl[ T_{ij}, V_{ij} \bigr], T_{ij} + V_{ij} \bigr], 
\label{eq:dVij_ds} \\
\frac{d V_{123}}{ds} &= 
\left[ \left[ T_{12}, V_{12} \right], V_{31} + V_{23} + V_{123} \right] \nonumber \\
&+\left[ \left[ T_{31}, V_{31} \right], V_{12} + V_{23} + V_{123} \right] \nonumber \\
&+\left[ \left[ T_{23}, V_{23} \right], V_{12} + V_{31} + V_{123} \right] \nonumber \\
&+\left[ \left[ \Trel, V_{123} \right], H_s \right] \, . 
\label{eq:dV123_ds}
\end{align}
Equation~(\ref{eq:dVij_ds}) follows directly from representing
Eq.~(\ref{eq:dHds_Tkin}) in a two-body basis for the subsystem consisting of
particles $i$ and $j$. In this case the kinetic energy only consists of the
term $T_{ij}$ and three-body interactions do not contribute.
Eq.~(\ref{eq:dV123_ds}) can then be derived by representing
Eq.~(\ref{eq:dHds_Tkin}) including all terms of the Hamiltonian in the
three-body basis, making use of the commutator relations
(\ref{eq:commutator_Vij_Tkin}) and the flow equation (\ref{eq:dVij_ds})
for the two-body interactions.

Equation~(\ref{eq:dV123_ds}) demonstrates explicitly that 3N forces are being
induced even if they are initially absent at $s=0$ or $\lambda = \infty$,
respectively. The same is true for all higher-body forces. That means, for
maintaining unitarity for a $N$-body system, in general $N-1$ flow
equations for the two-body to $N$-body forces need to be solved. In practice
this hierarchy of equations is typically truncated at the three-body level.
Some attempts to extend it to four-body forces have been
pursued~\cite{Schu13MSc}, but reaching sufficiently large model spaces is
currently still out of reach. A more promising and feasible approach for
dealing with higher-body forces seems to be a more suitable choice of
generators which only induce weak higher-body interactions. This is work in
progress.

Compared to Eq.~(\ref{eq:dHds_Tkin}), the system of differential equations
(\ref{eq:dVij_ds}) and (\ref{eq:dV123_ds}) has the important advantage that
terms resulting from spectator particles in two-body interaction processes
have been eliminated explicitly. That means the flow equation for $V_{ij}$
only involves particles $i$ and $j$, whereas in the evolution equation for
$V_{123}$ every term on the right hand side involves interaction processes
involving all three particles. The flow equation~\eqref{eq:dVij_ds} can be
easily represented in two-body partial-wave bases by inserting complete sets
of states. For example, for $V_{ij} = V_{12}$ (see
Eq.~(\ref{eq:VN_pw_representation})):
\begin{align}
\frac{d}{ds} \left< p' \alpha_{12}' | V_{\text{NN}} | p \alpha_{12} \right> &= - \left( \frac{p^2 - p'^2}{m} \right)^2 \left< p' \alpha'_{12} | V_{\text{NN}} | p \alpha_{12} \right> \nonumber \\
&+ \sum_{\alpha_{12}''} \int_0^{\infty} dp'' p''^2 \frac{p^2 + p'^2 - 2 p''^2}{m} \left< p' \alpha'_{12} | V_{\text{NN}} | p'' \alpha''_{12} \right> \left< p'' \alpha''_{12} | V_{\text{NN}} | p \alpha_{12} \right> \, ,
\label{eq:dVNN_ds_mombasis}
\end{align}
where we used the normalization\footnote{Note that the normalization
$\frac{2}{\pi} \sum_{\alpha_{12}} \int dp p^2 \left| p \alpha_{12} \right>
\left< p \alpha_{12} \right| = 1$ is more common in the literature
(see, e.g., Ref.~\cite{Bogn10PPNP}). The normalization of
Eq.~(\ref{eq:Jj_bas_NN_normalization}) leads to more compact expressions and
represents a natural reduction of the three-body state normalization in
Eq.~(\ref{eq:Jj_bas_normalization}), see
Appendix~\ref{sec:normalization}.\label{foot:normalization}} (see
Appendix~\ref{sec:normalization})
\begin{equation}
\left< p' \alpha'_{12} | p \alpha_{12} \right> = \frac{\delta(p - p')}{p p'} \delta_{\alpha_{12} \alpha'_{12}}, \quad \sum_{\alpha_{12}} \int dp p^2 \left| p \alpha_{12} \right> \left< p \alpha_{12} \right| = 1 \, .
\label{eq:Jj_bas_NN_normalization}
\end{equation}
For the representation of Eq.~(\ref{eq:dV123_ds}) we choose a three-body basis
representation $\{ab\}$. Let us choose without loss of generality representation
$\{12\}$. One key step represents the embedding of the NN interactions $V_{23}$ and
$V_{13}$ in this three-body basis. For this step, note that all three NN interaction terms are
related by cyclic (or anticyclic) permutations of the initial and final states,
i.e. (see Section~\ref{sec:3NF_coord_def}):
\begin{align}
\tensor*[_{\{23\}}]{\left< p' q' \alpha' | V_{23} | p q \alpha \right>}{_{\{23\}}} &= \tensor*[_{\{12\}}]{\left< p' q' \alpha' | V_{12} | p q \alpha \right>}{_{\{12\}}} \nonumber \\
&= \tensor*[_{\{12\}}]{\left< p' q' \alpha' | P_{123}^{-1} P_{123} V_{12} P_{123}^{-1} P_{123} | p q \alpha \right>}{_{\{12\}}} \nonumber \\
&= \tensor*[_{\{23\}}]{\left< p' q' \alpha' | P_{123} V_{12} P_{123}^{-1} | p q \alpha \right>}{_{\{23\}}} \, ,
\end{align}
which implies
\begin{equation}
V_{23} = P_{123} V_{12} P^{-1}_{123}, \quad \text{and accordingly} \quad V_{31} = P_{132} V_{12} P^{-1}_{132} = P_{123}^{-1} V_{12} P_{123} \, . 
\label{eq:NN_embedding}
\end{equation}
Corresponding relations hold for the kinetic energy. Hence we can rewrite Eq.~(\ref{eq:dV123_ds}) in the following form:
\begin{align}
\frac{d V_{123}}{ds} &= 
\left[ \left[ T_{12}, V_{12} \right], P^{-1}_{123} V_{12} P_{123} + P_{123} V_{12} P_{123}^{-1} + V_{123} \right] \nonumber \\
&+\left[ P_{123}^{-1} \left[ T_{12}, V_{12} \right] P_{123}, V_{12} + P_{123} V_{12} P^{-1}_{123} + V_{123} \right] \nonumber \\
&+\left[ P_{123} \left[ T_{12}, V_{12} \right] P^{-1}_{123}, V_{12} + P^{-1}_{123} V_{12} P_{123} + V_{123} \right] \nonumber \\
&+\left[ \left[ \Trel, V_{123} \right], \Trel + V_{12} + P_{123} V_{12} P_{123}^{-1} + P^{-1}_{123} V_{12} P_{123} + V_{123} \right] \, .
\label{eq:dV123_ds_embedded}
\end{align}
Equation~(\ref{eq:dV123_ds_embedded}) can be further simplified by noting that the
kinetic energy $T_{\text{rel}}$, as well as the sum of all three NN
interactions, $V_{12} + V_{23} + V_{31}$, and also the 3N interaction
$V_{123}$ are each invariant under the multiplication with the sum of
permutation operators $\bigl( 1 + P_{123} + P_{123}^{-1}
\bigr)/3$ from the left or from the right. This follows directly from the
relations (\ref{eq:NN_embedding}) and the representation
Eq.~(\ref{eq:Faddeev_antisymmetrized}) for the antisymmetrized 3N interaction.
After evaluating the products of permutation operators we finally obtain the
following flow equation in the partial-wave momentum representation:
\begin{align}
& \hspace{-0.5cm} \frac{d}{ds} \left< p' q' \alpha' | V_{123}  | p q \alpha \right> = \nonumber \\
& \frac{2}{3} \left< p' q' \alpha' | (1 + 2 P_{123}) (\left[ T_{12}, V_{12} \right] P_{123} V_{12} - V_{12} P_{123} \left[ T_{12}, V_{12} \right]) (1 + 2 P_{123}) | p q \alpha \right> \nonumber \\
& + \left< p' q' \alpha' | (1 + 2 P_{123}) \left[ T_{12}, V_{12} \right] V_{123} - V_{123} \left[ T_{12}, V_{12} \right] (1 + 2 P_{123}) | p q \alpha \right> \nonumber \\
& + \left< p' q' \alpha' | \left[ T_{\text{rel}}, V_{123} \right] V_{12} (1 + 2 P_{123}) - (1 + 2 P_{123}) V_{12} \left[ T_{\text{rel}}, V_{123} \right] | p q \alpha \right> \nonumber \\
& + \left< p' q' \alpha' | \left[ \left[ T_{\text{rel}},V_{123} \right],T_{\text{rel}} + V_{123} \right] | p q \alpha \right> \, .
\label{eq:dV123_ds_embedded_simplified}
\end{align}
For the derivation of Eq.~(\ref{eq:dV123_ds_embedded_simplified}) we used that
the permutation operators $P_{123}$ and $P_{132} = P_{123}^{-1}$ have the same
momentum-space representation (see
Section~\ref{sec:3N_decomp_antisymmetrization}). Note, again, that this
property can only be used after all products of permutation operators have
been evaluated due to the coupling to unphysical intermediate states (see
discussion in Section \ref{sec:semilocal_coordinate}). Once all products are
evaluated and each permutation operator is acting on interactions or kinetic
energy terms, contributions from these unphysical states decouple and both
operators have identical representations within the physical subset of states.
Finally, we also suppressed the basis index $\{ab\}$ in
Eq.~(\ref{eq:dV123_ds_embedded_simplified}) since the matrix elements of
antisymmetrized 3N matrix elements are invariant under this choice (see
Eq.~(\ref{eq:Faddeev_antisymmetrized})). However, for our particular choice
here it is most convenient to choose basis $\{12\}$, the natural basis for the
two-body interaction $V_{12}$. Of course, we could have equally well
represented all NN interactions in terms of $V_{23}$ or $V_{31}$. It is now
straightforward to represent the different terms in
Eq.~(\ref{eq:dV123_ds_embedded_simplified}) by inserting complete sets of
states, for example:
\begin{align}
\left< p' q' \alpha' | \left[ \left[ T_{\text{rel}},V_{123} \right], V_{123} \right] | p q \alpha \right> &= \int dp'' p''^2 dq'' q''^2 \sum_{\alpha''} \left[ \frac{p^2 + p'^2 - 2 p''^2}{m} + \frac{3}{4} \frac{q^2 + q'^2 - 2 q''^2}{m} \right] \nonumber \\
& \times \left< p' q' \alpha' | V_{123} | p'' q'' \alpha'' \right> \left< p'' q'' \alpha'' | V_{123} | p q \alpha \right> \, .
\end{align}

For a consistent evolution of NN and 3N interactions
Eqs.~(\ref{eq:dVNN_ds_mombasis}) and (\ref{eq:dV123_ds_embedded_simplified})
need to be solved simultaneously. This task is in principle straightforward,
but it involves some practical challenges~\cite{Hebe12msSRG}. First, it is
necessary to define a model space for the calculations. This step consists of
a choice for the maximal value of the three-body total angular momentum
$\mathcal{J}$ and a truncation for each three-body channel, e.g., by
choosing a maximal value for the relative angular momentum $J$ (see Section
\ref{sec:general_3N_decomp}). Since three-body interactions are diagonal in
the quantum numbers $\mathcal{J},
\mathcal{T}$ and $\mathcal{P}$, all three-body channels can be treated
separately. The basis size choices determine the dimension of the matrices as
shown in Table~\ref{tab:PW_data}, which in turn define the runtime and memory
requirements. Both can become quite substantial. In order to provide a sense
of scales, we consider as an example a typical truncation: $J_{\text{max}} =
5, N_p = N_q = 25$ and all three-body channels up to $\mathcal{J} = \tfrac{9}{2}$. This
basis is usually sufficient for well-converged many-body calculations of
heavier nuclei as well as matter (see Section~\ref{sec:PW_conv_matter}). For
$\mathcal{J}=\tfrac{9}{2}$ the matrix dimension is about $7 \times 10^9$ in this case
(see Table~\ref{tab:PW_data}), which amounts to about 50 gigabytes of memory
for the storage of a single 3N interaction matrix in double precision (64
bit). Since a typical differential equation solver requires several (up to
about 10) copies of the kernel, the total memory requirement can reach about
0.5 terabyte. In addition, the calculation of matrix products of this
dimension can become quite time consuming even when using highly optimized
BLAS routines as done for the code implementation used in
Ref.~\cite{Hebe12msSRG} and used for the results shown in the following
section.

\subsubsection{Application to state-of-the-art chiral NN and 3N interactions}
\label{sec:SRG_applications}

Apart from the many-body convergence, the model space choice also controls the
degree of unitarity of the SRG transformation when solving
Eq.~(\ref{eq:dV123_ds_embedded_simplified}). This is because a finite basis of
the form shown in Eq.~(\ref{eq:Jj_bas}) is not complete under cyclic and
anticyclic permutations of particles, as the permutation operator $P_{123}$
couples in general all partial waves (see Eq.~(\ref{eq:P123_matrixelements})).
As a consequence, non-vanishing matrix elements in all three-body partial
waves are induced when two-body operators are embedded in a three-body
momentum basis via Eq.~(\ref{eq:NN_embedding}). This problem is absent in one
dimension~\cite{Aker11SRG1D} or in a discrete oscillator
basis~\cite{Jurg08SRG3N1D}, where the permutation operator is block-diagonal
in a given model space of size $N_{\rm{max}}$. However, in practice this
violation of unitarity can be reduced to a negligible degree as we demonstrate
now.

\begin{figure}[t]
\centering
\includegraphics[scale=0.8]{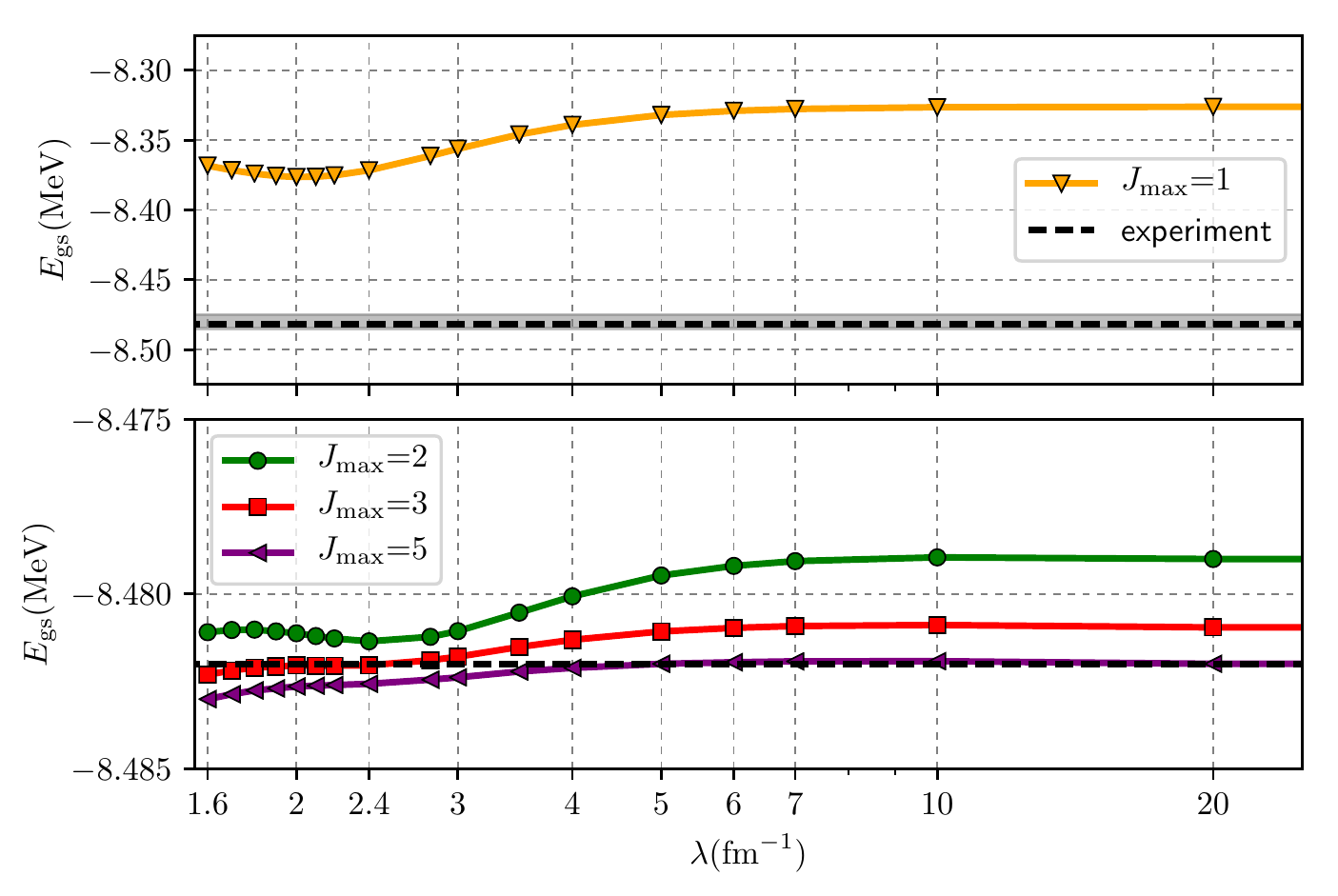}
\caption{Ground-state energy of $^3$H as a function of resolution scale
$\lambda = s^{-1/4}$ based on consistently-evolved NN and 3N interactions for
different model spaces defined by $J_{\text{max}}$ (for both NN and 3N
interactions), obtained from solutions of the Faddeev
equations~\cite{Gloe82spline}. We employed the NN interaction of
Ref.~\cite{Ente17EMn4lo} at N$^2$LO with $\Lambda = 500$ MeV plus 3N
interactions at N$^2$LO fitted to the experimental ground-state energy
$E_{\text{gs}} = - 8.482$ MeV (black dashed lines). The precise values of the
LECs and the regulator form are given in the main text. The
upper panel shows the results for $J_{\text{max}}$=1, whereas the lower panel
shows the results for higher values of $J_{\text{max}}$. Note the different
scales of the $y$ axis in both panels. The lower panel shows the very narrow
energy range around the experimental value, indicated by the gray band in the
upper panel.}
\label{fig:3H_unitarity}
\end{figure}

We illustrate the degree of unitarity using the ground-state energy of $^3$H,
i.e., we set $\mathcal{J} = \tfrac{1}{2}$, $\mathcal{T}=\tfrac{1}{2}$ and $\mathcal{P} = + 1$.
Such studies of three-body systems are particularly instructive since four-
and higher-body forces cannot contribute and it is possible to cleanly
disentangle effects from neglected higher-body forces and model space
truncations for NN and 3N interactions. In Figure~\ref{fig:3H_unitarity} we
show the ground-state energy of $^3$H for different values of $J_{\text{max}}$
as a function of the resolution scale. The calculations are based on the
nonlocal NN interaction of Ref.~\cite{Ente17EMn4lo} at N$^2$LO for the cutoff
value of $\Lambda = 500$ MeV and the ground-state energies have been obtained
from the T-matrix solutions of the momentum-space Faddeev equations. The
three-body force has been fitted to reproduce the experimental binding energy
of $^3$H with $E_{\text{gs}} = - 8.482$ MeV. Specifically, we used the values
$c_1 = -0.74$ GeV $^{-1}$, $c_3 = -3.61$ GeV $^{-1}$, $c_4 = 2.44$ GeV
$^{-1}$, $c_D = 1.0$, $c_E = -0.384$ and a regulator of the form shown in
Eq.~(\ref{eq:nonlocal_regulator}) with $n=4$. However, we stress that these
particular values are rather arbitrary and the general features shown in this
figure are independent of these specific choices. The figure demonstrates that
the results converge rapidly as a function of $J_{\text{max}}$ for the present
Hamiltonian. We find that the variation of the energy for model spaces
$J_{\text{max}} \ge 2$ is smaller than $4$ keV over the shown range of
resolution scales, while the variation decreases systematically with
increasing model space. When reaching a level of less than $1$ keV (like for
$J_{\text{max}} = 5$), results start to become sensitive to numerical
truncation effects due to the finite discretization of the momentum basis when
representing the flow equation Eq.~(\ref{eq:dV123_ds_embedded_simplified})
using $N_p = N_q = 25$.

\begin{table}[b]
\centering
\begin{tabular}{l|ccccc|l}
regularization  & $c_1$ [GeV $^{-1}$] & $c_3$ [GeV $^{-1}$] & $c_4$ [GeV $^{-1}$] & $c_D$ & $c_E$ & Ref.\\
\hline
\hline
\textbf{nonlocal MS}   & -$0.74$               & -$3.61$                & $2.44$                 & $-1.5$  & -$0.61$ & \cite{Dris17MCshort}  \\
\hline
\textbf{local MS}      & -$0.81$               & -$3.2$                & $5.4$                 & $0.83$  & -$0.052$  & \cite{Gazi08lec}  \\
\hline
\textbf{semilocal MS}  & -$0.74$               & -$3.61$               & $2.44$                & $2.0$     & $0.23$  & \\
\hline
\textbf{semilocal CS}  & -$0.81$               & -$4.69$               & $3.4$                 & $1.0$     & -$0.25$  & \cite{Epel18SCS3N}
\end{tabular}
\caption{LEC values of 3N interactions at N$^2$LO in the different
regularization schemes discussed in Section~\ref{sec:3N_regularization}. The
couplings $c_D$ and $c_E$ have been fitted to the $^3$H binding energy plus
another observable (details can be found in the given references). Except for
the ``local MS'' scheme, according NN interactions at N$^2$LO with the same
cutoff scale as for the 3N interactions have been employed. For the
momentum-space and coordinate-space cutoff scales $\Lambda = 500 \:\text{MeV}$
and $R = 0.9\: \text{fm}$ have been chosen, respectively.}
\label{tab:LEC_values_SRG_decoupling}
\end{table}

\begin{figure}[t]
\includegraphics[width=0.99 \textwidth]{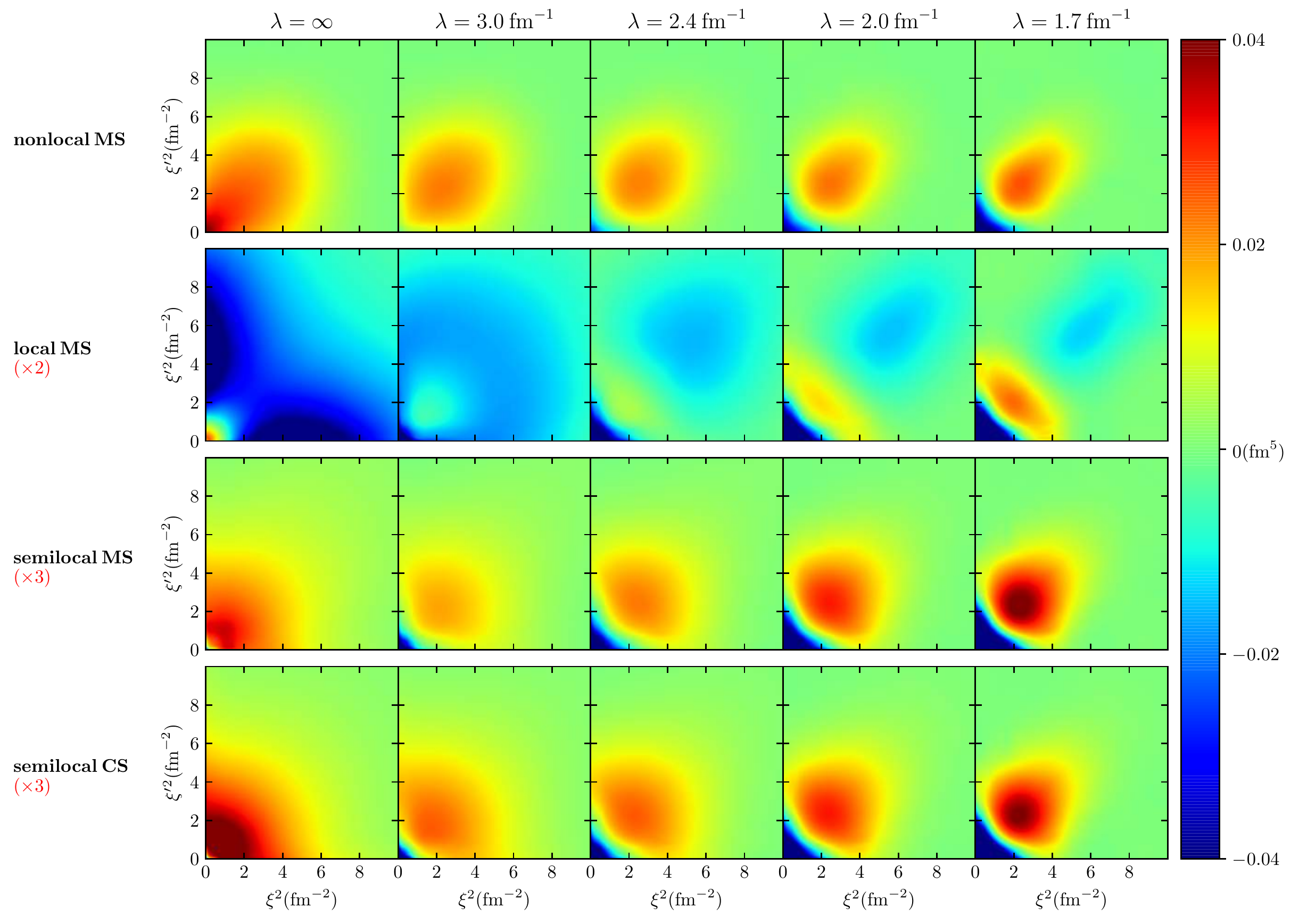}
\caption{Matrix elements of the antisymmetrized interaction $\bigl< p' q'
\bar{\alpha} | V^{\text{as}}_{\text{3N}} | p q \bar{\alpha} \bigr>$ at different resolution scales
$\lambda$ (columns) for the different regularization schemes (rows, see
Section \ref{sec:3N_regularization}). In contrast to the figures of Section
\ref{sec:visualization_3NF} we show the matrix elements as a function of the
square of the hypermomentum $\xi^2 = p^2 + \tfrac{3}{4} q^2$. As in
Figure~\ref{fig:contour_N2LO_500}, we choose the hyperangle $\tan \theta = p /
(\sqrt{3}/2 q) =
\frac{\pi}{4}$ and the partial wave with the quantum numbers $\bar{\alpha} = 0
\equiv \left\{ L=0, S=0, J=0, T=1, l=0, j=\tfrac{1}{2} \right\}$ (see
Appendix~\ref{sec:3N_config_table}). The values of the LECs for the
interactions within the different regularization schemes are shown in
Table~\ref{tab:LEC_values_SRG_decoupling}. For optimized visibility we
multiplied the matrix elements for the ``local MS'' interactions by 2 and
those of the ``semilocal MS'' and ``semilocal CS'' regularization scheme by
$3$.}
\label{fig:contour_SRG}
\end{figure}
\begin{figure}[t]
\centering
\includegraphics[width=0.9 \textwidth]{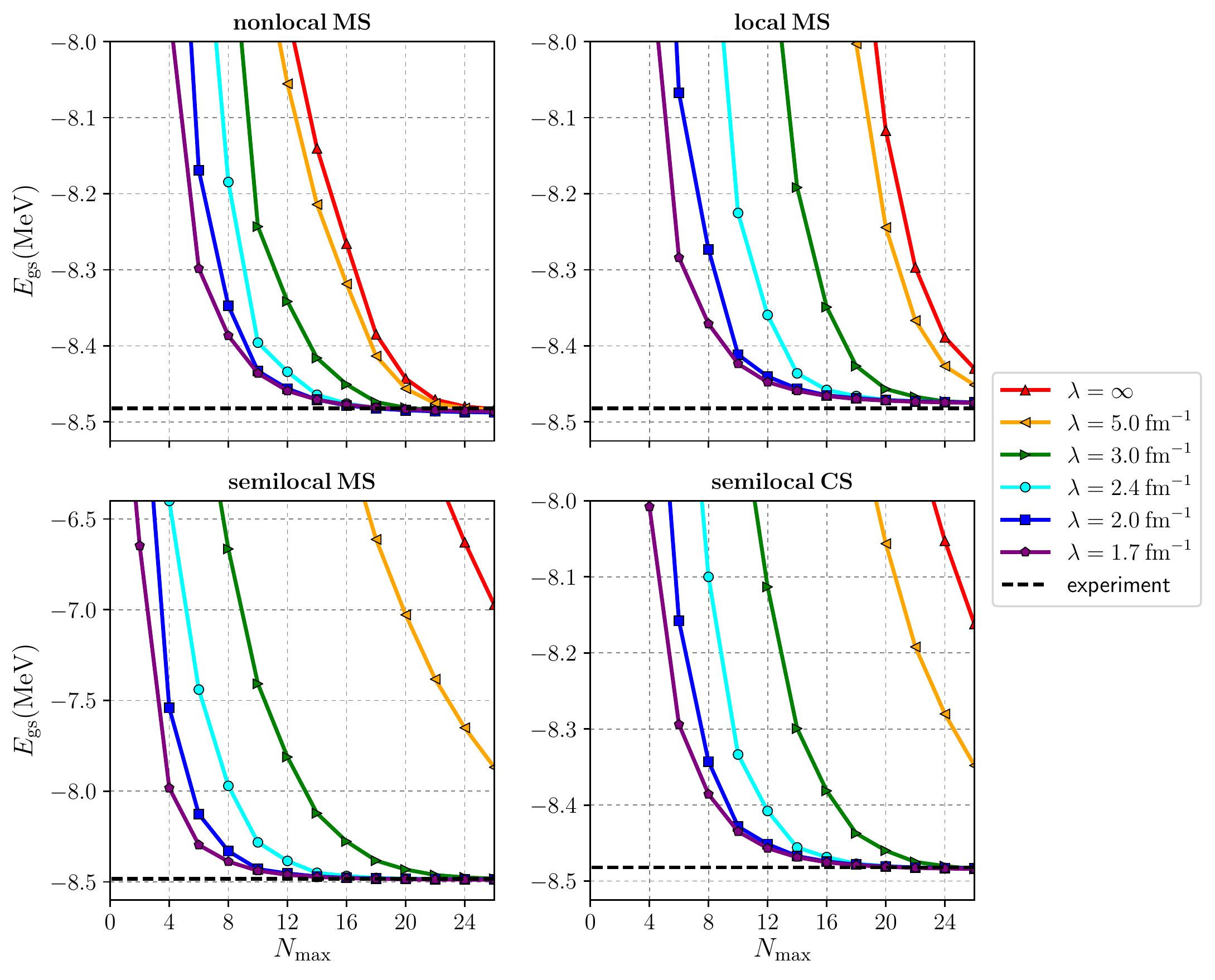}
\caption{Ground-state energies of $^3$H based on the Hamiltonians specified
in Table ~\ref{tab:LEC_values_SRG_decoupling} as a function of SRG resolution
scale $\lambda$ and the harmonic oscillator basis size $N_{\text{max}}$. For
all calculations the oscillator parameter $\Omega = 20 \: \text{MeV}$
and $J_{\text{max}} = 5$ was used. Note the different energy scale in the
panel for the ``semilocal MS'' interaction.}
\label{fig:3H_Nmax_SRG}
\end{figure}
In Figure~\ref{fig:contour_SRG} we demonstrate the decoupling of low- and
high-energy components of 3N interactions as a function of resolution scale
$\lambda$ (columns) for different regularization schemes. The panels show the
matrix elements as a function of the hypermomentum $\xi^2 = p^2
+ \tfrac{3}{4} q^2$ for a fixed hyperangle $\tan \theta = p / (\sqrt{3}/2 q) =
\frac{\pi}{4}$ (see Section~\ref{sec:visualization_3NF}) for the partial wave
with $\bar{\alpha} = 0$, i.e., $L=0, S=0, J=0, T=1, l=0$ and $j=\tfrac{1}{2}$ (see
Appendix~\ref{sec:3N_config_table}). The values of the LECs of
the 3N interactions are chosen such that the experimental binding energy of
$^3$H is reproduced. The precise values are given in
Table~\ref{tab:LEC_values_SRG_decoupling}. The figure shows that the initial
matrix elements at $\lambda = \infty$ differ quite substantially in the
different regularization schemes. However, as the resolution scale $\lambda$
is lowered, attractive components at small momenta are generated for all
interactions, while the off-diagonal contributions get systematically
suppressed. Here, the width of the diagonal band is also approximately given by
the scale $\lambda^2$, similarly to NN interactions~\cite{Bogn07SRG}. These
features are general and hold for matrix elements at different hyperangles and
partial waves. In addition, the overall effects of the SRG evolution are
stronger for initial potentials with stronger off-diagonal couplings.
\begin{figure}[t]
\centering
\includegraphics[width=0.95 \textwidth]{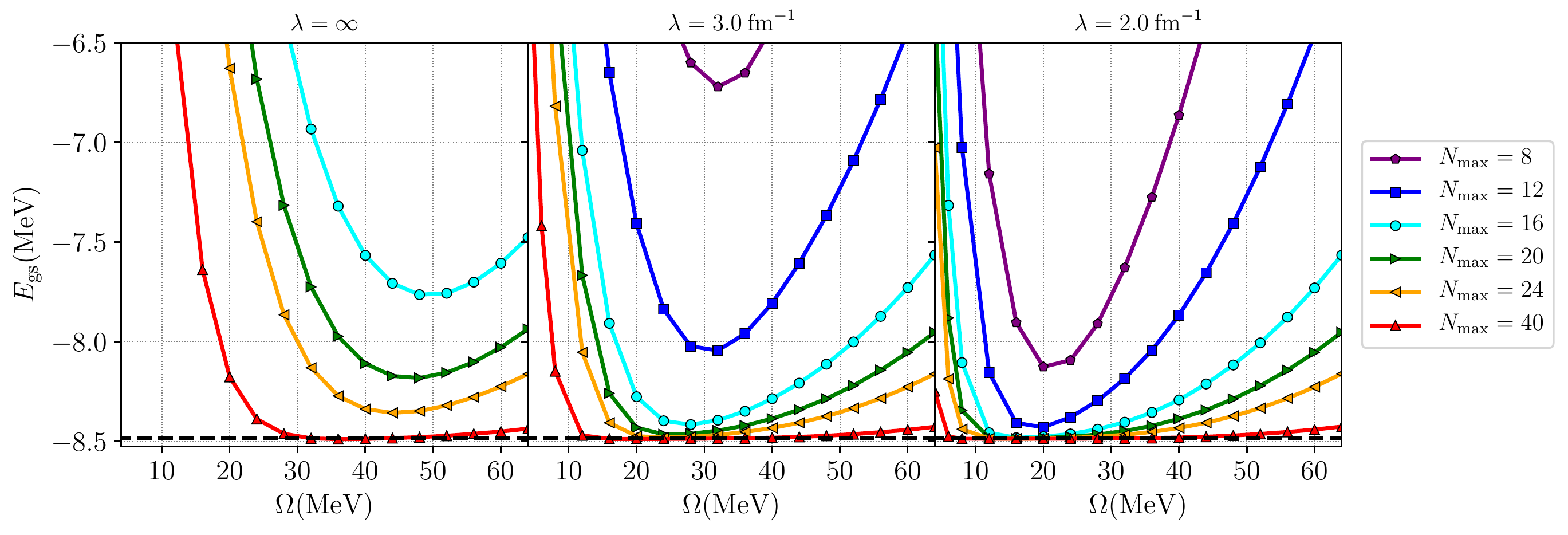}
\caption{Ground-state energies of $^3$H based on the ``semilocal MS'' NN
plus 3N interactions specified in Table ~\ref{tab:LEC_values_SRG_decoupling}
at three different resolutions scales $\lambda$ as a function of the
oscillator frequency $\Omega$ and the basis size $N_{\text{max}}$.}
\label{fig:3H_hbaromega_SRG}
\end{figure}
In addition, it is quite remarkable that at the lowest shown resolution scale
$\lambda = 1.7 \: \text{fm}^{-1}$, all interactions are very similar for the
shown hyperangle and partial-wave channel, at least qualitatively. This
property has already been observed for NN Interactions and is usually referred
to as ``universality''~\cite{Bogn07SRG,Bogn10PPNP}. This universality of NN
interactions can be attributed to common long-range pion physics and
phase-shift equivalence of all realistic potentials, which is reflected in the
matrix elements at low resolution. It is an interesting question to what
extent the same is true for 3N forces since there are important differences:
First, 3N forces up to N$^3$LO are fixed by fitting only two low-energy
constants $c_D$ and $c_E$, in contrast to numerous couplings in NN
interactions. Second, 3N forces give only subleading contributions to
observables. Since universality is only approximate in NN interactions, it is
not obvious to what extent 3N forces are constrained by long-range physics at
low resolution. The results shown in Figure~\ref{fig:contour_SRG} indicate
that universality also holds for 3N interactions at low resolution scales,
even though not to such a quantitative degree as for NN interactions (see also
Ref.~\cite{Hebe12msSRG}). However, more detailed investigations are needed to
draw more robust conclusions.

In Figure~\ref{fig:3H_Nmax_SRG} we demonstrate the improved perturbativeness and
the accelerated convergence of many-body calculations based on interactions at
lower resolution scales. The different panels show the binding energy of $^3$H
for different resolution scales $\lambda$ as a function of the model space
indicated by $N_{\text{max}}$, obtained by a diagonalization of the
Hamiltonian in a Jacobi harmonic oscillator three-body
basis~\cite{Navr99NCSM}\footnote{Credits to Andreas Ekstr\"om for providing
the diagonalization code.}. It is obvious that the convergence properties of
the initial interactions at $\lambda=\infty$ differ quite significantly for
the different regularization schemes and the shown oscillator parameter
$\Omega = 20 \: \text{MeV}$. However, note that the optimal frequency is
not the same for all interactions and also depends on the resolution scale. In
particular, for the ``semilocal MS'' and ``semilocal CS'' interactions in the bottom
row the calculations converge much faster for an oscillator parameter of about
$\Omega = 40 \: \text{MeV}$ at $\lambda = \infty$, while the optimal
frequency gets shifted systematically to lower values as we evolve to
lower resolution scales (see Figure~\ref{fig:3H_hbaromega_SRG}). This shift
represents a challenge when performing the SRG evolution in an oscillator
basis at a given frequency. In order to ensure a converged SRG evolution, a
frequency conversion method was employed~\cite{Roth14SRG3N}. In contrast, at
low scales the many-body convergence is quite similar for all shown
interactions and we are able to obtain converged results using only moderate
model space sizes of about $N_{\text{max}} = 16$.

Note that such three-body calculations can be performed without any
fundamental constraints on the resolution scale $\lambda$ since it is possible
to solve the SRG flow equations exactly up to numerical effects. For systems
with $A > 3$, however, the evolution will not be unitary anymore
since contributions from four- and higher-body interactions tend to become
stronger with decreasing resolution scale. This puts some implicit constraints
on the range of the resolution scales:
\begin{itemize}
\item First, the resolution scale $\lambda$ needs to be \textit{sufficiently
small}, such that the low- and high-momentum components are sufficiently
decoupled and it is possible to converge the many-body calculations.
\item Second, the resolution scale should be \textit{sufficiently large}, such
that contributions from induced four- and higher-body interactions remain
small.
\end{itemize}
These two conditions define a range of resolution scales that are particularly
useful for practical calculations, given the currents basis size constraints
and the typical strength of induced four- and higher-body interactions for
presently used generators and interactions. This region typically comprises
the range $\lambda \approx 1.6 - 2.4 \: \text{fm}^{-1}$ or, equivalently, $s \approx
0.03 - 0.15\:\text{fm}^{4}$ for modern interactions derived within chiral EFT
(see, e.g., Refs.~\cite{Bogn10PPNP,Roth11SRG}). A practical way to check if
both conditions are met is to perform calculations at different resolution
scales and investigate the dependence of the results on this scale and on the
many-body basis size. We will present an overview of recent state-of-the-art
calculations in Section~\ref{sec:applications}.

\subsection{Normal ordering of 3N interactions}
\label{sec:normal_ordering}

Normal ordering is a powerful and well-established method that allows to
transform the representation of a given Hamiltonian in an exact way, such that
contributions from 3N interactions can approximately be incorporated in
many-body calculations at the computational cost of two-nucleon interactions.
The underlying idea of normal ordering is to rewrite the Hamiltonian given in
Eq.~(\ref{eq:H_second_quantized}) by using Wick's
theorem~\cite{Wick50theorem,Fett71QToMPS}:
\begin{align}
\hat{A} \hat{B} \hat{C} \hat{D} \hat{E} \hat{F} ... &= N( \hat{A} \hat{B} \hat{C} \hat{D} \hat{E} \hat{F} ...) \nonumber \\
&+ \sum_{\text{singles}} 
N( \contraction[1.0ex]{}{\hat{A}}{}{\hat{B}}
\hat{A} \hat{B} \hat{C} \hat{D} \hat{E} \hat{F} ... ) + \sum_{\text{doubles}} 
N( 
\contraction[1.0ex]{}{\hat{A}}{}{\hat{B}}
\contraction[1.0ex]{\hat{A}\hat{B}}{\hat{C}}{}{\hat{D}}
\hat{A} \hat{B} \hat{C} \hat{D} \hat{E} \hat{F} ... ) + ... + N( 
\contraction[1.0ex]{}{\hat{A}}{}{\hat{B}}
\contraction[1.0ex]{\hat{A}\hat{B}}{\hat{C}}{}{\hat{D}}
\contraction[1.0ex]{\hat{A}\hat{B}\hat{C}\hat{D}}{\hat{E}}{}{\hat{F}}
\hat{A} \hat{B} \hat{C} \hat{D} \hat{E} \hat{F} ... ) \, .
\label{eq:Wick_theorem}
\end{align}
Here the operators $\hat{A}$, $\hat{B}$, etc. represent some generic creation or annihilation operators, $N(..)$ denotes the normal-ordering of the these operators with respect to a chosen reference state $\left| \psi_{\text{ref}} \right>$, such that
\begin{equation}
\bigl< \psi_{\text{ref}} | N( \hat{A} \hat{B} \hat{C} \hat{D} \hat{E} \hat{F} ...) | \psi_{\text{ref}} \bigr> = \bigl< \psi_{\text{ref}} | N( \contraction[1.0ex]{}{\hat{A}}{}{\hat{B}}
\hat{A} \hat{B} \hat{C} \hat{D} \hat{E} \hat{F} ... ) | \psi_{\text{ref}} \bigr> = 0
\end{equation}
for all non-fully-contracted strings of operators, where the contraction is defined by
\begin{equation}
\contraction[1.0ex]{}{\hat{A}}{}{\hat{B}} \hat{A} \hat{B} = \hat{A} \hat{B} - N (\hat{A} \hat{B}) \, .
\end{equation}
We emphasize that normal ordering generally depends on the definition of the
reference state $\left|
\psi_{\text{ref}} \right>$ since creation and annihilation operators are only
defined with respect to a particular state. If we consider, e.g., a
Hartree-Fock reference state $\left| \psi_{\text{HF}}
\right>$ in which all the lowest $A$ single-particle orbitals are occupied,
normal ordering involves the anticommutation of all creation operators
$\hat{a}^{\dagger}_i$ with $i \leq A$ to the left, and all annihilation
operators $\hat{a}_i$ with $i > A$ to the right. As a result, only one type of
contraction is nonvanishing and takes the simple form:
\begin{equation}
\contraction[1.5ex]{}{\hat{a}}{{}_i^{\dagger}}{\hat{a}}
\hat{a}_i^{\dagger} \hat{a}_j = \delta_{ij} - \contraction[1.5ex]{}{\hat{a}}{{}_j}{\hat{a}}
\hat{a}_j \hat{a}_i^{\dagger} = n_i \delta_{ij} \quad \text{with} \quad n_i = \left\{ 
\begin{array}{ll} 
1 \quad \text{for} \quad i \leq A \\
0 \quad \text{for} \quad i > A
\end{array}
\right. \, .
\label{eq:contractions_HF}
\end{equation}
Evaluating the contractions in Eq.~(\ref{eq:Wick_theorem}), we obtain:
\begin{equation}
\hat{H} = \Gamma^{(0)}_{\text{HF}} + \hat{\Gamma}^{(1)}_{\text{HF}} + \hat{\Gamma}^{(2)}_{\text{HF}} + \hat{\Gamma}^{(3)}_{\text{HF}} \, ,
\end{equation}
with
\begin{align}
\Gamma^{(0)}_{\text{HF}} &= \sum_i n_i \bigl< i | T | i \bigr> + \frac{1}{2} \sum_{ij} n_i n_j \bigl< i j | V_\text{NN}^{\text{as}} | i j \bigr> + \frac{1}{6} \sum_{ijk} n_i n_j n_k \bigl< i j k | V_\text{3N}^{\text{as}} | i j k \bigr>, \nonumber \\  
\hat{\Gamma}^{(1)}_{\text{HF}} &= \sum_{ij} \left[ \bigl< i | T | j \bigr> + \sum_{k} n_k \bigl< i k | V_{\text{NN}}^{\text{as}} | j k \bigr> + \frac{1}{2} \sum_{kl} n_k n_l \bigl< i k l | V_{\text{3N}}^{\text{as}} | j k l \bigr> \right] N \bigl( \hat{a}^{\dagger}_i \hat{a}_j \bigr), \nonumber \\
\hat{\Gamma}^{(2)}_{\text{HF}} &= \sum_{ijkl} \left[ \bigl< i j | V_{\text{NN}}^{\text{as}} | k l \bigr> + \sum_m n_m \bigl< i j m | V_{\text{3N}}^{\text{as}} | k l m \bigr> \right] N \bigl( \hat{a}^{\dagger}_i \hat{a}^{\dagger}_j \hat{a}_l \hat{a}_k \bigr), \nonumber \\
\hat{\Gamma}^{(3)}_{\text{HF}} &= \sum_{ijklmn} \bigl< i j k | V_{\text{3N}}^{\text{as}} | l m n \bigr> N \bigl( \hat{a}^{\dagger}_i \hat{a}^{\dagger}_j \hat{a}^{\dagger}_k \hat{a}_n \hat{a}_m \hat{a}_l \bigr) \, .
\label{eq:normord_terms_HF}
\end{align}
\begin{table}
\centering
\begin{tabular}{c|c||c|c||c|c||c}
  $n_a$ & $n_{a^{\dagger}}$ & NN & $n_{\text{NN}}$ & 3N & $n_{\text{3N}}$ & $\zeta$ \\
\hline 
\rule{0pt}{25pt} 0 & 0 & $N \bigl( \frac{1}{4} \vnn \contraction[1.25ex]{}{\hat{a}}{{}^{\dagger} \hat{a}^{\dagger} \hat{a}}{\hat{a}} \contraction[0.75ex]{\hat{a}^{\dagger}}{\hat{a}}{{}^{\dagger}}{\hat{a}} \hat{a}^{\dagger} \hat{a}^{\dagger} \hat{a} \hat{a} \bigr)$ & $2$ & $N \bigl( \frac{1}{36} \vtn \contraction[2.0ex]{}{\hat{a}}{{}^{\dagger} \hat{a}^{\dagger} \hat{a}^{\dagger} \hat{a} \hat{a}}{\hat{a}} \contraction[1.0ex]{\hat{a}}{{}^{\dagger} \hat{a}^{\dagger}}{\hat{a}^{\dagger} \hat{a}}{\hat{a}} \contraction[0.5ex]{\hat{a}^{\dagger} \hat{a}}{{}^{\dagger} \hat{a}^{\dagger}}{}{\hat{a}} \hat{a}^{\dagger} \hat{a}^{\dagger} \hat{a}^{\dagger} \hat{a} \hat{a} \hat{a} \bigr)$ & $6$ & $\frac{1}{3}$ \\
\rule{0pt}{25pt} 1 & 1 & $ N \bigl( \frac{1}{4} \vnn \contraction[0.75ex]{\hat{a}}{{}^{\dagger} \hat{a}^{\dagger}}{}{\hat{a}} \hat{a}^{\dagger} \hat{a}^{\dagger} \hat{a} \hat{a} \bigr) $ & $4$ & $N \bigl( \frac{1}{36} \vtn \contraction[1.0ex]{\hat{a}}{{}^{\dagger} \hat{a}^{\dagger}}{\hat{a}^{\dagger} \hat{a}}{\hat{a}} \contraction[0.5ex]{\hat{a}^{\dagger} \hat{a}}{{}^{\dagger} \hat{a}^{\dagger}}{}{\hat{a}} \hat{a}^{\dagger} \hat{a}^{\dagger} \hat{a}^{\dagger} \hat{a} \hat{a} \hat{a} \bigr)$ & $18$ & $\frac{1}{2}$ \\
\rule{0pt}{25pt} 2 & 2 & $N \bigl( \frac{1}{4} \vnn \hat{a}^{\dagger} \hat{a}^{\dagger} \hat{a} \hat{a} \bigr)$ & $1$ & $N \bigl( \frac{1}{36} \vtn \contraction[0.5ex]{\hat{a}^{\dagger} \hat{a}}{{}^{\dagger} \hat{a}^{\dagger}}{}{\hat{a}} \hat{a}^{\dagger} \hat{a}^{\dagger} \hat{a}^{\dagger} \hat{a} \hat{a} \hat{a} \bigr)$ & $9$ & $1$ 
\\
\rule{0pt}{1pt}  & & & & & & \\
\hline
\rule{0pt}{25pt} 0 & 0 & $N \bigl( \frac{1}{4} \vnn \contraction[0.75ex]{}{\hat{a}}{{}^{\dagger}}{\hat{a}} \contraction[0.75ex]{\hat{a}^{\dagger} \hat{a}^{\dagger}}{\hat{a}}{}{\hat{a}} 
\hat{a}^{\dagger} \hat{a}^{\dagger} \hat{a} \hat{a} \bigr)$ & $1$ & $N \bigl( \frac{1}{36} \vtn \contraction[1.0ex]{}{\hat{a}}{{}^{\dagger}}{\hat{a}} \contraction[1.0ex]{\hat{a}^{\dagger} \hat{a}^{\dagger}}{\hat{a}}{{}^{\dagger}}{\hat{a}} \contraction[1.0ex]{\hat{a}^{\dagger} \hat{a}^{\dagger} \hat{a}^{\dagger} \hat{a}}{\hat{a}}{}{\hat{a}} \hat{a}^{\dagger} \hat{a}^{\dagger} \hat{a}^{\dagger} \hat{a} \hat{a} \hat{a} \bigr)$ & $9$ & $1$ \\
\rule{0pt}{25pt} 2 & 0 & $N \bigl( \frac{1}{4} \vnn \contraction[0.75ex]{}{\hat{a}}{{}^{\dagger}}{\hat{a}} \hat{a}^{\dagger} \hat{a}^{\dagger} \hat{a} \hat{a} \bigr)$ & $1$ & $N \bigl( \frac{1}{36} \vtn \contraction[0.75ex]{}{\hat{a}}{{}^{\dagger}}{\hat{a}} \contraction[0.75ex]{\hat{a}^{\dagger} \hat{a}^{\dagger}}{\hat{a}}{{}^{\dagger}}{\hat{a}} \hat{a}^{\dagger} \hat{a}^{\dagger} \hat{a}^{\dagger} \hat{a} \hat{a} \hat{a} \bigr)$ & $9$ & $1$ \\
\rule{0pt}{25pt} 0 & 2 & $N \bigl( \frac{1}{4} \vnn \contraction[0.75ex]{\hat{a}^{\dagger} \hat{a}^{\dagger}}{\hat{a}}{}{\hat{a}} \hat{a}^{\dagger} \hat{a}^{\dagger} \hat{a} \hat{a} \bigr)$ & $1$ & $N \bigl( \frac{1}{36} \vtn \contraction[0.75ex]{\hat{a}^{\dagger} \hat{a}^{\dagger} \hat{a}^{\dagger} \hat{a}}{\hat{a}}{}{\hat{a}} \contraction[0.75ex]{\hat{a}^{\dagger} \hat{a}^{\dagger}}{\hat{a}}{{}^{\dagger}}{\hat{a}} \hat{a}^{\dagger} \hat{a}^{\dagger} \hat{a}^{\dagger} \hat{a} \hat{a} \hat{a} \bigr)$ & $9$ & $1$ \\
\end{tabular}
\caption{Combinatorial factors of the Wick contractions including normal
contractions (upper half) and anomalous contractions (lower half). For the
three-body terms only contributions including at least one normal contraction
are shown. $n_a$ denotes the number of uncontracted annihilation operators,
$n_{a^{\dagger}}$ the according number of creation operators, $n_{\text{NN}}$
the number of possible contractions for NN interactions (or number of terms if
there are no contractions) and $n_{\text{3N}}$ the corresponding number of
contractions for 3N interactions.}
\label{tab:normal_ordering_factors}
\end{table}
The non-trivial prefactors in Eq.~(\ref{eq:normord_terms_HF}) result from the
combinatorial factors associated with number of possible contractions
$n_{\text{NN}}$ and $n_{\text{3N}}$ for the zero-body term $\Gamma^{(0)}$, the
one-body term $\hat{\Gamma}^{(1)}$ and the two-body term $\hat{\Gamma}^{(2)}$.
The factors are shown in detail in Table \ref{tab:normal_ordering_factors}.
From Eq.~(\ref{eq:normord_terms_HF}) it follows that normal ordering
allows to combine contributions from free-space two-body interactions with
normal-ordered three-body contributions in the form of an effective two-body vertex
function $V_{\text{eff}}$ (see Figure~\ref{fig:Veff_diagram}):
\begin{equation}
\bigl< i j | V^{\text{as}}_{\text{eff}} | k l \bigr> = \bigl< i j | \vnn^{\text{as}} | k l \bigr> + \zeta \bigl< i j | \overline{V}_{\text{3N}} | k l \bigr> \, ,
\label{eq:effective_NN_interaction}
\end{equation}
with 
\begin{equation}
\bigl< i j | \overline{V}_{\text{3N}} | k l \bigr> \equiv \sum_m n_m \bigl< i j m | V_{\text{3N}}^{\text{as}} | k l m \bigr> \, .
\label{eq:Vbar}
\end{equation}
It is important to stress two crucial properties of the vertex function
$V_{\text{eff}}$, defined in Eq.~(\ref{eq:effective_NN_interaction}): 
\begin{itemize}
\item[a)] The value of the combinatorial factor $\zeta$ depends on the quantity of interest,
as can be seen explicitly in Eq.~(\ref{eq:normord_terms_HF}). For example, for
the calculation of Hartree-Fock energies the factor takes the value $\zeta =
\tfrac{1}{3}$ (as in $\Gamma^{(0)}_{\text{HF}}$), whereas for the calculations of self
energies $\Sigma$ in the Hartree-Fock approximation we obtain $\zeta = \tfrac{1}{2}$ (see
$\Gamma^{(1)}_{\text{HF}}$).
\item[b)] In contrast to the free-space NN interaction $\vnn^{\text{as}}$, the effective 
interaction $V^{\text{as}}_{\text{eff}}$ is in general not Galilean
invariant, i.e., it depends on the two-body center-of-mass momenta (see also
discussion in Section \ref{sec:3NF_mom_rep}). This property is a consequence
of the fact that the reference state defines a specific reference frame for
the single-particle states $m$ in Eq.~(\ref{eq:Vbar}) and hence violates
translational symmetry.
\end{itemize}
The type and number of nonvanishing contractions depends on the type of
reference state. For example, for superfluid and open-shell systems it is
convenient to choose the BCS ground state~\cite{Fett71QToMPS,Ball98QM}:
\begin{equation}
\left| \psi_{\text{BCS}} \right> = \prod_i (u_i + v_i \hat{a}^{\dagger}_{i} \hat{a}^{\dagger}_{-i}) \left| 0 \right> \, , 
\end{equation}
with $u_i^2 + v_i^2 = 1$, $u_{-i} = u_i$ and $v_{-i} = - v_{i}$. The BCS state is a vacuum state
with respect to the Bogoliubov-transformed operators:
\begin{equation}
\label{eq:BCS_bogoliubov}
\hat{\alpha}_i \left| \psi_{\text{BCS}} \right> = 0 \quad \text{with} \quad \hat{\alpha}_i = u_i \hat{a}_i - v_i \hat{a}_{-i}^{\dagger} \, .
\end{equation}
Hence, in the present case normal ordering involves the anticommutation of all
$\hat{\alpha}^{\dagger}$ operators to the left and all $\hat{\alpha}$ to the
right. By using Eq.~(\ref{eq:BCS_bogoliubov}) it follows $\hat{a}_i = u_i
\hat{\alpha}_{i} + v_i \hat{\alpha}_{-i}^{\dagger}$ and it is straightforward
to verify that the contractions take the following form:
\begin{align}
\contraction[1.5ex]{}{\hat{a}_i}{}{\hat{a}}
\hat{a}_i^{\dagger} \hat{a}_j &= v_i v_j 
\contraction[1.5ex]{}{\hat{\alpha}_{-i}}{}{\hat{\alpha}_{j}}
\hat{\alpha}_{-i} \hat{\alpha}_{-j}^{\dagger} = v^2_i \delta_{ij}, \nonumber \\
\contraction[1.5ex]{}{\hat{a}_i}{}{\hat{a}}
\hat{a}_i \hat{a}_j^{\dagger} &= u_i u_j
\contraction[1.5ex]{}{\hat{\alpha}_{i}}{}{\hat{\alpha}_{j}}
\hat{\alpha}_{i} \hat{\alpha}_{j}^{\dagger} = u^2_i \delta_{ij} = (1 - v^2_i) \delta_{ij}, \nonumber \\
\contraction[1.5ex]{}{\hat{a}_i}{}{\hat{a}}
\hat{a}_i \hat{a}_j &= u_i v_j 
\contraction[1.5ex]{}{\hat{\alpha}_{i}}{}{\hat{\alpha}_{j}}
\hat{\alpha}_{i} \hat{\alpha}_{-j}^{\dagger} = 
u_i v_j \delta_{\shortminus ij} = - u_i v_i \delta_{\shortminus ij}, \nonumber \\
\contraction[1.5ex]{}{\hat{a}_i}{}{\hat{a}}
\hat{a}_i^{\dagger} \hat{a}_j^{\dagger} &= v_i u_j 
\contraction[1.5ex]{}{\hat{\alpha}_{-i}}{}{\hat{\alpha}_{j}}
\hat{\alpha}_{\shortminus i} \hat{\alpha}_{j}^{\dagger} = 
v_i u_j \delta_{\shortminus ij} = u_i v_i \delta_{\shortminus ij} \, .
\end{align}
Clearly, in the present case the number of nonvanishing contractions is larger
than for a Hartree-Fock reference state (see Eq.~(\ref{eq:contractions_HF})).
In particular, we find nonvanishing contributions from anomalous contractions,
i.e., contractions of two creation or two annihilation operators. This is a
consequence of the fact that the BCS state is not a state with a fixed
particle number. Hence the normal-ordered Hamiltonian takes a more
complicated form than for a Hartree-Fock reference state~\cite{Sign14BogCC,Ripo19NO}:
\begin{figure}[t]
\centering
\begin{minipage}[c]{0.08\textwidth}
\begin{tikzpicture} 
\begin{feynman}
\vertex (a) at (0,0) {\(1\)}; 
\vertex (b) at (0.8,0) {\(2\)};
\vertex [blob, /tikz/minimum size=24pt, fill=green!35, line width=0.25mm, font=\fontsize{8}{0}\selectfont, shape=diamond] (d) at (0.4,1.25) {\(\overline{V}_{\rm 3N}\)};
\vertex (e) at (0,2.5) {\(1'\)}; 
\vertex (f) at (0.7,2.5) {\(2'\)};
\diagram* {
(a) -- [fermion, thick] (d) -- [fermion, thick] (e);
(b) -- [fermion, thick] (d) -- [fermion, thick] (f);
};
\end{feynman}
\end{tikzpicture}
\end{minipage}
=
\begin{minipage}[c]{0.08\textwidth}
\begin{tikzpicture} 
\begin{feynman}
\vertex (a) at (0,0) {\(1\)}; 
\vertex (b) at (0.8,0) {\(2\)};
\vertex [blob, /tikz/minimum size=16pt, shape=rectangle, fill=red!60, line width=0.25mm, font=\fontsize{8}{0}\selectfont] (d) at (0.4,1.25) {\(V_{\rm 3N}\)};
\vertex (e) at (0,2.5) {\(1'\)}; 
\vertex (f) at (0.7,2.5) {\(2'\)};
\diagram* {
(a) -- [fermion, thick] (d) -- [fermion, thick] (e);
(b) -- [fermion, thick] (d) -- [fermion, thick] (f);
(d) -- [fermion, line width=0.25mm, out=45, in=-45, loop, min distance=0.6cm] (d);
};
\end{feynman}
\end{tikzpicture}
\end{minipage}
\hspace{3cm}
\begin{minipage}[c]{0.08\textwidth}
\begin{tikzpicture} 
\tikzfeynmanset{
  my dot/.style={fill=red},
  every vertex/.style={my dot},
}
\begin{feynman}
\vertex (a) at (0,0) {\(1\)}; 
\vertex (b) at (0.8,0) {\(2\)};
\vertex (d1) at (0.3,1.25) {};
\vertex (d2) at (0.5,1.25) {};
\vertex [blob, /tikz/minimum size=18pt, fill=blue!35, line width=0.25mm, font=\fontsize{8}{0}\selectfont] (d) at (0.4,1.25) {\(V_{\rm eff}\)};
\vertex (e) at (0,2.5) {\(1'\)}; 
\vertex (f) at (0.8,2.5) {\(2'\)};
\diagram* {
(a) -- [fermion, thick] (d) -- [fermion, thick] (e);
(b) -- [fermion, thick] (d) -- [fermion, thick] (f);
};
\end{feynman}
\end{tikzpicture}
\end{minipage}
$=$
\begin{minipage}[c]{0.08\textwidth}
\begin{tikzpicture} 
\tikzfeynmanset{
  my dot/.style={fill=red},
  every vertex/.style={my dot},
}
\begin{feynman}
\vertex (a) at (0,0) {\(1\)}; 
\vertex (b) at (0.8,0) {\(2\)};
\vertex (d1) at (0.3,1.25) {};
\vertex (d2) at (0.5,1.25) {};
\vertex [blob, /tikz/minimum size=18pt, fill=blue!35, line width=0.25mm, font=\fontsize{8}{0}\selectfont] (d) at (0.4,1.25) {\(V_{\rm NN}\)};
\vertex (e) at (0,2.5) {\(1'\)}; 
\vertex (f) at (0.8,2.5) {\(2'\)};
\diagram* {
(a) -- [fermion, thick] (d) -- [fermion, thick] (e);
(b) -- [fermion, thick] (d) -- [fermion, thick] (f);
};
\end{feynman}
\end{tikzpicture}
\end{minipage}
$+ \: \zeta$
\begin{minipage}[c]{0.08\textwidth}
\begin{tikzpicture} 
\begin{feynman}
\vertex (a) at (0,0) {\(1\)}; 
\vertex (b) at (0.8,0) {\(2\)};
\vertex [blob, /tikz/minimum size=24pt, fill=green!35, line width=0.25mm, font=\fontsize{8}{0}\selectfont, shape=diamond] (d) at (0.4,1.25) {\(\overline{V}_{\rm 3N}\)};
\vertex (e) at (0,2.5) {\(1'\)}; 
\vertex (f) at (0.7,2.5) {\(2'\)};
\diagram* {
(a) -- [fermion, thick] (d) -- [fermion, thick] (e);
(b) -- [fermion, thick] (d) -- [fermion, thick] (f);
};
\end{feynman}
\end{tikzpicture}
\end{minipage}
\caption{Diagrammatic representation of the integration over single-particle
states as defined in Eq.~(\ref{eq:Vbar}) as part of the normal ordering
operation (left panel) and the definition of the resulting effective
two-nucleon interaction (see Eq.~(\ref{eq:effective_NN_interaction}). Note
the presence of the nontrivial combinatorial factor $\zeta$ (see discussion in
Section~\ref{sec:normal_ordering}).}
\label{fig:Veff_diagram}
\end{figure}
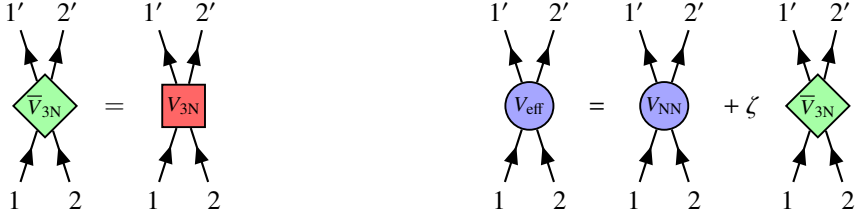
\begin{equation}
\label{eq:H_normord_BCS}
\hat{H} = \Gamma^{(0)}_{\text{BCS}} + \hat{\Gamma}^{(1)}_{1,\text{BCS}} + \hat{\Gamma}^{(1)}_{2,\text{BCS}} + \hat{\Gamma}^{(1)}_{3,\text{BCS}} + \hat{\Gamma}^{(2)}_{\text{BCS}} + \hat{\Gamma}^{(3)}_{\text{BCS}} + ... \, ,
\end{equation}
with
\begin{align}
\Gamma^{(0)}_{\text{BCS}} &= \sum_i v_i^2 \left< i | T | i \right> + \frac{1}{2} \sum_{ij} v_i^2 v_j^2 \left< i j | V_\text{NN} | i j \right> + \frac{1}{6} \sum_{ijk} v_i^2 v_j^2 v_k^2 \left< i j k | V_\text{3N} | i j k \right> \nonumber \\
& - \frac{1}{4} \sum_{ij} u_i v_i u_j v_j \left< i \shortminus i | V_\text{NN} | j \shortminus j \right> - \frac{1}{4} \sum_{ijk} u_i v_i u_j v_j v_k^2 \left< i \shortminus i k | V_\text{3N} | j \shortminus j k \right> + ... \: , \nonumber \\  
\hat{\Gamma}^{(1)}_{1,\text{BCS}} &= \sum_{ij} \left[ \left< i | T | j \right> + \sum_{k} v^2_k \left< i k | V_{\text{NN}} | j k \right> + \frac{1}{2} \sum_{kl} v_k^2 v_l^2 \left< i k l | V_{\text{3N}} | j k l \right> + ... \right] N \bigl( \hat{a}^{\dagger}_i \hat{a}_j \bigr), \nonumber \\
\hat{\Gamma}^{(1)}_{2,\text{BCS}} &= \frac{1}{4} \sum_{ij} \left[ \sum_{k} u_k v_k \left< k \shortminus k | V_{\text{NN}} | i j \right> + \sum_{kl} u_k v_k v_l^2 \left< k \shortminus k l | V_{\text{3N}} | i j l \right> + ... \right] N \bigl( \hat{a}_i \hat{a}_j \bigr), \nonumber \\
\hat{\Gamma}^{(1)}_{3,\text{BCS}} &= \frac{1}{4} \sum_{ij} \left[ \sum_{k} u_k v_k \left< i j | V_{\text{NN}} | k \shortminus k \right> - \sum_{kl} u_k v_k v_l^2 \left< i j l | V_{\text{3N}} | k \shortminus k l \right> + ... \right] N \bigl( \hat{a}^{\dagger}_i \hat{a}^{\dagger}_j \bigr), \nonumber \\
\hat{\Gamma}^{(2)}_{\text{BCS}} &= \sum_{ijkl} \left[ \left< i j | V_{\text{NN}} | k l \right> + \sum_m v_m^2 \left< i j m | V_{\text{3N}} | k l m \right> \right]  N \bigl( \hat{a}^{\dagger}_i \hat{a}^{\dagger}_j \hat{a}_l \hat{a}_k \bigr), \nonumber \\
\hat{\Gamma}^{(3)}_{\text{BCS}} &= \sum_{ijklmn} \left< i j k | V_{\text{3N}} | l m n \right> N \bigl( \hat{a}^{\dagger}_i \hat{a}^{\dagger}_j \hat{a}^{\dagger}_k \hat{a}_n \hat{a}_m \hat{a}_l \bigr) \, ,
\label{eq:normord_terms_BCS}
\end{align}
where we have neglected terms involving only anomalous contractions in the
3N interactions in Eqs.(\ref{eq:H_normord_BCS}) and
(\ref{eq:normord_terms_BCS}). Generally, anomalous contractions include only
contributions around the Fermi surface in nuclear matter and consequently
provide typically only small contributions to bulk properties of matter
compared to normal contractions. In finite nuclei such contributions might be
more relevant. In addition to Hartree-Fock and BCS reference states it is also
possible to use reference states including additional many-body correlations~(see,
e.g., Ref.~\cite{Carb14SCGFdd}).

The effectiveness of normal ordering relies on the assumption that the
reference state is sufficiently close to the exact ground state of the system,
such that the contributions from the vertex functions $\hat{\Gamma}^{(1)}$,
$\hat{\Gamma}^{(2)}$ and $\hat{\Gamma}^{(3)}$ only give small contributions
when evaluating expectation values with respect to the real ground state $\left| \psi_{0} \right>$:
\begin{equation}
\bigl< \psi_0 | \hat{H} | \psi_0 \bigr> \, .
\end{equation}
Obviously, if the normal-ordering reference state agrees with the exact ground
state, all contributions from the vertex functions $\hat{\Gamma}^{(1,2,3)}$
vanish, while the zero-body term $\Gamma^{(0)}$ takes the value of the exact
ground-state energy for a given Hamiltonian. In particular, for the practical
usefulness of normal ordering it is key to choose a reference state such that
contributions from the three-body vertex function $\hat{\Gamma}^{(3)}$ (also
called \textit{residual} 3N contributions) are small compared to the
lower-body operators and can be neglected to good approximation. In this case
the total Hamiltonian $\hat{H}$ can be written only in terms of operators up
to the two-body level, which obviously simplifies calculations tremendously.
In fact, the smallness of the contributions from $\hat{\Gamma}^{(3)}$ has been
demonstrated explicitly for calculations of ground-state energies of atomic
nuclei for specific NN and 3N interactions derived within chiral EFT in
Refs.~\cite{Hage07CC3N,Roth11normord} using a Hartree-Fock reference state and
recently also for multi-reference calculations of ground and excited
states~\cite{Gebr16MR}. However, we stress that the choice of a useful
reference state generally depends on the employed Hamiltonian and the system.
For example, interactions that are highly non-perturbative, such as, e.g., the
Argonne $v_{18}$ potential~\cite{Wiri95AV18}, induce stronger high-energy
virtual excitations, and the Hartree-Fock wave function will most likely not
be a suitable reference state for normal ordering. Instead a state with more
many-body correlations most likely needs to be chosen for such interactions.
 
\subsubsection{Normal ordering in nuclear matter}
\label{sect:normal_ordering_matter}

In this section we discuss normal ordering for nuclear matter,
i.e., for a reference state defined in terms of momentum eigenstates. If we
denote the single-particle states by momentum $\mathbf{k}_i$, spin
$\sigma_i$ and isospin $\tau_i$, $\left| i \right> = \left| \mathbf{k}_i
\sigma_i \tau_i \right>$, Eq.~(\ref{eq:Vbar}) takes the following form:
\begin{equation}
\bigl< 1' 2' | \overline{V}_{\rm{3N}} | 1 2 \bigr> = \sum_{\sigma_3} \sum_{\tau_3} \int \frac{d \mathbf{k}_3}{(2 \pi)^3}
\, n_{\mathbf{k}_3}^{\tau_3} \, \bigl< 1' 2' 3 | V^{\text{as}}_{\rm{3N}} | 1 2 3 \bigr> \, ,
\label{eq:normord_singpart}
\end{equation}
which involves sums over spin and isospin projection quantum numbers
$\sigma_3$ and $\tau_3$, as well as an integration over all momentum states,
weighted by the momentum distribution functions $n^{\tau_3}_{\bf k}$ for a
given neutron and proton density. Here we choose, without loss of generality,
to integrate over particle state $3$. Due to the antisymmetry of
$V^{\text{as}}_{\rm{3N}}$ we have the freedom to choose any other single-particle
state. In addition, we assumed for the sake of simplicity that the
distribution function does not depend on spin, i.e., we consider
spin-unpolarized matter. However, the generalization to spin-polarized systems
is straightforward~\cite{Krue15polnm}. The simplest and most common choice for
the reference state is a Hartree-Fock wave function, for which all
single-particle levels up to a Fermi momentum $\kf$ are occupied. For the
following examples, we consider for the sake of simplicity matter at zero
temperature, i.e., the distribution function takes the following form:
$n^{\tau}_{\bf k} =
\theta(\kf^{\tau} - |{\bf k}|)$. We stress, however, that the treatment can be 
extended to more general distribution functions (see e.g.
Refs.~\cite{Lov10ddNN,Carb14SCGFdd}) or finite temperatures in a
straightforward way.

Normal ordering takes the simplest form when expressed in single-particle
coordinates, like Eq.~(\ref{eq:normord_singpart}). However, since the 3N
interaction matrix elements are most efficiently expressed in a relative
coordinate basis (see Section~\ref{sec:3NF_mom_rep}) it is desirable to
perform the normal ordering in a Jacobi momentum representation. Since we
chose to integrate over particle 3 in Eq.~(\ref{eq:normord_singpart}), it is
most convenient to choose basis representation $\{12\}$ (see
Section~\ref{sec:3NF_coord_def}). By expressing all single-particle momenta in
terms of the Jacobi momenta and the two-body center-of-mass momentum
$\mathbf{P} = \mathbf{P}_{\{12\}} = \mathbf{k}_1 +
\mathbf{k}_2 = \mathbf{k}'_1 + \mathbf{k}'_2$ we obtain~\cite{Dris16asym}:
\begin{equation}
\bigl< \mathbf{p}' | \overline{V}_{\rm{3N}} (\mathbf{P}) | \mathbf{p} \bigr> =  \left( \frac{3}{2}\right)^3 \sum_{\sigma_3}
\sum_{\tau_3} \int \frac{d \mathbf{q}}{(2 \pi)^3} \, n_{(3 \mathbf{q}+\mathbf{P})/2}^{\tau_3} \, \tensor*[_{\{12\}}]{\bigl< \mathbf{p}' \mathbf{q} | V^{\text{as}}_{\rm{3N}} | \mathbf{p} \mathbf{q} \bigr>}{_{\{12\}}} \, .
\label{eq:normord_jacobi}
\end{equation}
Here we suppressed the spin and isospin quantum numbers in the matrix elements
in order to keep the notation simple and transparent. It is important to note
that, in contrast to the 3N interaction, the distribution function
$n_{\mathbf{p}}^{\tau}$ is not Galilean invariant as it depends on the
two-body center-of-mass momentum $\mathbf{P}$. This in turn leads to a
center-of-mass dependence of the effective interaction
$\overline{V}_{\text{3N}}$ (see discussion in Section
\ref{sec:normal_ordering}), which makes the calculation of normal ordering
technically quite challenging.

An exact calculation of partial-wave matrix elements for the effective NN
interaction $\overline{V}_{\text{3N}}$ in nuclear matter for general values and
angles of the center-of-mass momentum $\mathbf{P}$ has not been achieved so
far.\footnote{In Section \ref{sec:no_PWD} we present a novel method that
allows to perform the normal order operation in
Eq.~(\ref{eq:normord_singpart}) exactly, using a framework that does not rely
on a partial-wave basis.}. Note that even if such an exact calculation would
be available, the application of the resulting effective interaction in
many-body frameworks becomes significantly more involved than free-space
NN interactions in Eq.~(\ref{eq:effective_NN_interaction}) due to the
dependence on the vector $\mathbf{P}$. In most many-body frameworks for
nuclear matter the interaction matrix elements are expressed in a Jacobi
partial-wave momentum basis, consisting of the Jacobi momenta $\mathbf{p}$ and
$\mathbf{p}'$. This representation needs to be extended for general
normal-ordered 3N interactions by the center-of-mass momentum $\mathbf{P}$
plus corresponding additional angular quantum numbers, which usually requires
significant modifications of existing frameworks that have been designed only
for free-space NN interactions. 

\begin{figure}[t]
\centering
\includegraphics[width=0.95\textwidth,clip=]{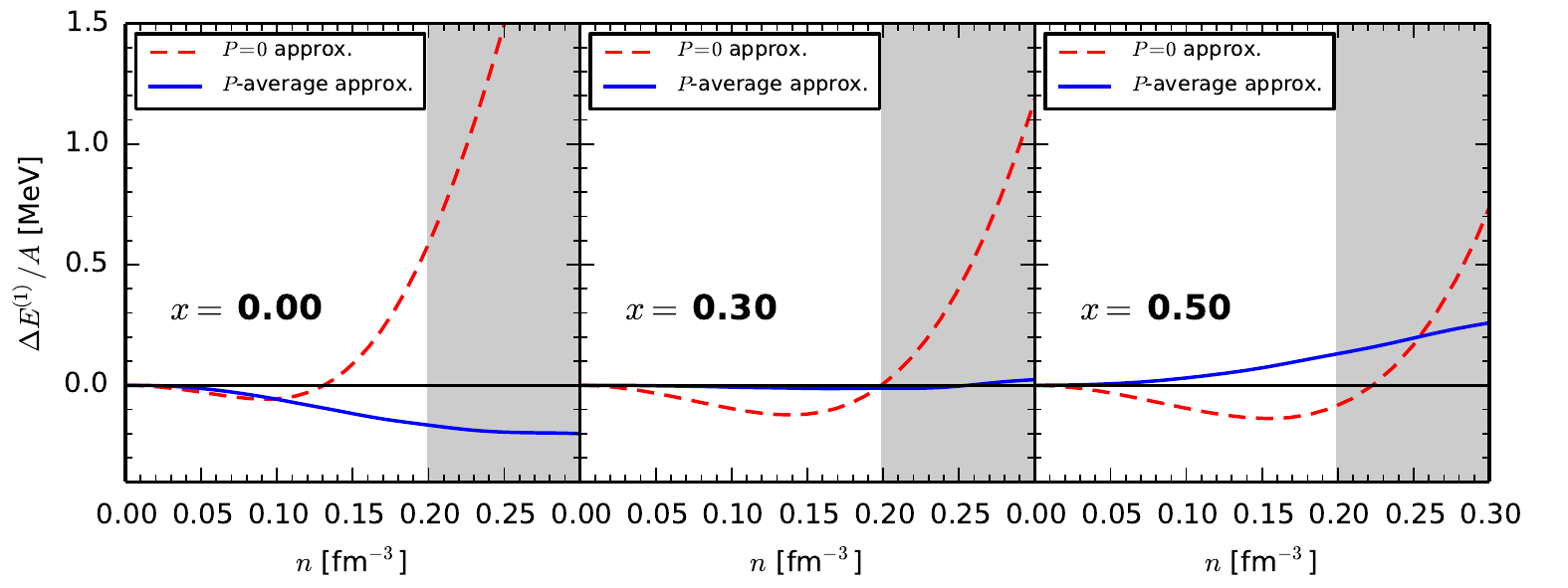}
\caption{\label{fig:comp_3N_HF}
Comparison of 3N Hartree-Fock energies for the $P = 0$~(red dashed) and
$P$-average approximation~(blue solid line) for the effective interaction
$\overline{V}_{\rm{3N}}$. Results are shown as difference to the
exact Hartree-Fock energy for three proton fractions $x=n_p/(n_p + n_n)$, with
the proton and neutron densities $n_p$ and $n_n$ respectively. The three
panels show the results for $x = 0$~(left), $x = 0.3$~(center), and $x =
0.5$~(right panel). The $P = 0$ values give larger deviations above saturation
density, whereas the $P$-average approximation behaves more systematic over
the entire density range.\\
\textit{Source:} Figure taken from Ref.~\cite{Dris16asym}.}
\label{fig:HF_matter_normalordering}
\end{figure}

Due to these complications the dependence on the momentum $\mathbf{P}$ is
usually approximated for nuclear matter calculations. So far, the following
two strategies have been employed:
\begin{itemize}
\item The simplest approximation consists of working in the center-of-mass reference
frame, i.e., setting $\mathbf{P}=0$. This approximation has been first employed in
the Refs.~\cite{Holt10ddnn,Hebe10nmatt,Hebe11fits} using a Hartree-Fock
reference state. It was shown that this approximation leads to remarkable
agreement with exact energy results at the Hartree-Fock level up to nuclear
saturation density (see Figure~\ref{fig:HF_matter_normalordering}), which
indicates that even this rather crude approximation captures the most
important contributions from 3N interactions rather well, at least at this
order in the many-body expansion. In addition, in this approximation scheme
the normal ordered 3N interaction takes exactly the same form as free-space
NN interactions, which makes the application to many-body frameworks
straightforward. However, some care has to be taken for the correct treatment
of the combinatorial factor $\zeta$ in Eq.~(\ref{eq:Vbar}).

\item In Ref.~\cite{Dris16asym} the normal ordering procedure was extended to
finite values of $\mathbf{P}$ by averaging this vector over all directions:
\begin{equation}
n_{(3 \mathbf{q}+\mathbf{P})/2}^\tau \longrightarrow \Gamma^\tau(q,P) = \frac{1}{4 \pi}\int d \hat{\mathbf{P}} \, n_{(3\mathbf{q}+\mathbf{P})/2}^\tau \, .
\label{eq:normord_Paverage}
\end{equation}
Within this approximation the effective interaction $\overline{V}_{\rm{3N}}$
acquires an additional dependence on the absolute value of $\mathbf{P}$,
while its partial-wave structure is still sufficiently simple so that it can
be combined in a straightforward way with contributions from free-space NN
interactions in many-body calculations. In addition, this approximation
reduces to the $\mathbf{P} =0$ approximation for $P = |\mathbf{P}|\rightarrow 0$ in
Eq.~(\ref{eq:normord_Paverage}).
\end{itemize}
We now discuss and compare both approximations above in more detail. For
$\mathbf{P}=0$ the analytical structure of the effective NN interaction is
sufficiently simple so that normal ordering can performed in a semi-analytical
way, at least for simple 3N interactions like those at N$^2$LO in chiral EFT.
In Refs.~\cite{Hebe10nmatt,Hebe11fits} $\overline{V}_{\rm{3N}}$ was derived
via an automated implementation of the momentum- and spin-exchange
operations by representing all spin and isospin operators in matrix form. The
antisymmetrized 3N force was then represented in this three-particle basis for
general single-particle momenta $\mathbf{k}_i$. The traces over spin and
isospin degrees of freedom can then be performed in a straightforward way. In
the last step the resulting interaction was projected on a complete set of
two-body spin and isospin operators. For completeness, we provide in Appendix
\ref{sec:Veff_nuclear_matter} the complete expressions for the effective
interaction $\overline{V}_{\rm{3N}}$ in neutron matter
($n_{\mathbf{p}}^{\text{proton}} = 0$) and symmetric nuclear matter
($n_{\mathbf{p}}^{\text{proton}} = n_{\mathbf{p}}^{\text{neutron}}$) in
operator form for the N$^2$LO 3N interactions in chiral EFT for angularly
independent regulators. The interaction includes all spin structures that are
invariant under combined rotations in spin and space in a spin-saturated
system. In addition to the central spin-independent and spin-spin
$\boldsymbol{\sigma}_1 \cdot
\boldsymbol{\sigma}_2$ interactions, it also includes tensor forces
$S_{12}$, spin-orbit interactions and additional tensor structures, which can
be expressed in terms of $S_{12}$ and quadratic spin-orbit interactions. The
resulting expression can be decomposed into partial-wave states in a
straightforward way. Obviously, for finite values of $\mathbf{P}$
the number of possible operators increases significantly, such that this
analytical treatment becomes very tedious and impractical.

In Ref.~\cite{Holt10ddnn} an alternative but related approach for the
derivation of the partial-wave matrix elements was followed. Here
semi-analytical expressions for the partial-wave momentum matrix elements for
$|\mathbf{p}| = |\mathbf{p}'|$ were derived for the N$^2$LO 3N topologies
using a Hartree-Fock reference state. These analytical approaches have the
advantage that they do not require any large input files of 3N matrix
elements, but quickly become tedious when trying to generalize them to more
complicated 3N interactions~\cite{Kais183Ndens,Kais193Ndens,Kais193Ndensint},
angularly-dependent regulators~\cite{Logo2019consistent}, more general reference
states or more sophisticated treatments of the center-of-mass momentum
dependence.

\begin{figure}[t]
\centering
\includegraphics[width=0.95\textwidth,clip=]{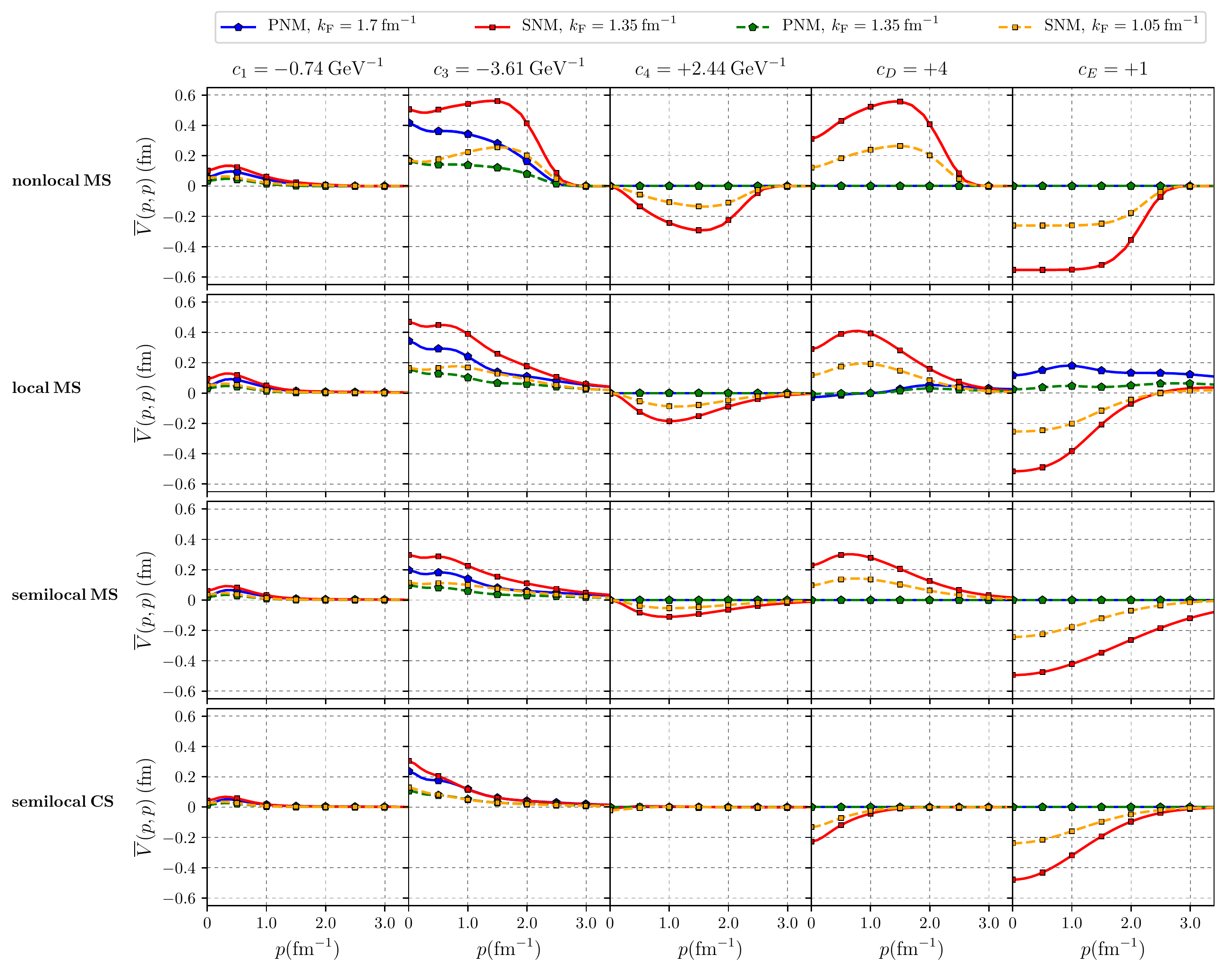}
\caption{Diagonal momentum-space matrix elements of the effective interaction
$\overline{V}_{\text{3N}} (p,p)$ in the channel $^1$S$_0$ resulting from normal ordering
of the leading-order chiral 3N interactions at N$^2$LO in pure neutron matter
(PNM) and symmetric nuclear matter (SNM) for two different particle densities.
We present the results for the individual contributions at N$^2$LO for the 4
different regularization schemes discussed in Section
\ref{sec:3N_regularization}. The used Fermi momenta correspond approximately
to saturation density and half saturation density for neutron matter
$(n_{\text{PNM}} (k_{\rm{F}}) = k_{\text{F}}^3/(3 \pi^2))$ and symmetric
matter ($n_{\text{SNM}} (k_{\rm{F}}) = 2 k_{\text{F}}^3/(3 \pi^2)$). For the
regulator cutoff scales we used $\Lambda = 500$ MeV for the momentum-space
regulators and $R=0.9$ fm for the coordinate-space regularization ``semilocal
CS''. For the shown results all partial waves up to $\mathcal{J}=\tfrac{9}{2}$ have
been included.}
\label{fig:NO_Vbar}
\end{figure}

A more general and flexible approach is to make use of partial-wave
representations of 3N interactions. In this case it is possible to derive
general relations in different approximations for any 3N interaction that is
available in a partial-wave decomposed form. In Ref.~\cite{Dris16asym} normal
ordering was first presented for general isospin-asymmetric matter using the
angular averaging prescription Eq.~(\ref{eq:normord_Paverage}) for the
$\mathbf{P}$-dependence. For a Hartree-Fock reference state at zero
temperature we can immediately simplify the angular
integrals:
\begin{equation}
\Gamma^\tau(q,P)
= \frac{1}{4 \pi}\int d \hat{\mathbf{P}} \, n_{(3\mathbf{q}+\mathbf{P})/2}^\tau = 
\begin{cases} 
1 & (3q+P) \leqslant 2 k_{F,\tau} \; , \\
0 & |3q-P| \geqslant 2 k_{F,\tau} \; ,  \\
\frac{1}{2}\int_{-1}^{\gamma} d\cos  \theta \,n_{(3\mathbf{q}+\mathbf{P})/2}^\tau  & \text{otherwise\,,}
\end{cases}
\end{equation}
with $\gamma = (4 k_{F,\tau}^2-9q^2-P^2)/(6Pq)$ and $\cos  \theta = \hat{\mathbf{q}} \cdot \hat{\mathbf{P}}$. Explicitly, we
obtain for the partial-wave matrix elements~\cite{Dris16asym}:
\begin{align} 
& \bigl< p' (L' S') J' T' m_T| \overline{V}_{\rm{3N}} (P) | p (L S) J T m_T \bigr> \nonumber \\
& \qquad = \frac{3}{(4\pi)^2} \, \left(\frac{3}{4 \pi} \right)^3 \int dq \, q^2\; \sum \limits_{\tau,\mathcal{T}}  \clebschG{T}{m_{T}}{\tfrac{1}{2}}{\tau}{\mathcal{T}}{m_{T}+\tau} \clebschG{T'}{m_{T}}{\tfrac{1}{2}}{\tau}{\mathcal{T}}{m_{T}+\tau}  \; \Gamma^\tau (q,P) \sum \limits_{\substack{l,j,\mathcal{J}}} \frac{2\mathcal{J}+1}{2 J+1} \delta_{ll'} \delta_{jj'} \delta_{JJ'} \bigl< p' q \alpha' | V^{\text{as,reg}}_{\text{3N}} |p q \alpha \bigr> \, ,
\label{eq:Veff_matter_PW}
\end{align}
where $V^{\text{as,reg}}_{\text{3N}}$ denote the regularized antisymmetrized
matrix elements as defined in Eqs.~(\ref{eq:Faddeev_antisymmetrized}) and
(\ref{eq:regulator_nonlocal_antisymmetrized}) for nonlocal regulators.
However, we stress that the relation in Eq.~(\ref{eq:Veff_matter_PW}) holds for
any type of regulator, since the regulators and their angular
dependence are already included in the matrix elements of
$V^{\text{as,reg}}_{\text{3N}}$. Note that, except for neutron and symmetric
matter, off-diagonal matrix elements in spin and isospin quantum numbers $S$
and $T$ in general contribute to the effective potential. In addition, it also
depends on the isospin projection $m_T$, as a consequence of the isospin
dependence of the occupation function $n_{k}^{\tau}$. For pure neutron matter
and symmetric nuclear matter the sum over the quantum number $\tau$ can be performed
immediately and the isospin factor simplifies to 
\begin{equation}
\sum \limits_{\tau}  \clebschG{T}{m_{T}}{\tfrac{1}{2}}{\tau}{\mathcal{T}}{m_{T}+\tau} \clebschG{T'}{m_{T}}{\tfrac{1}{2}}{\tau}{\mathcal{T}}{m_{T}+\tau} = 
\left\{
\begin{array}{ll}
\delta_{\mathcal{T}, \tfrac{3}{2}} \delta_{T, 1} \delta_{m_T, -1} & \text{for PNM} \\
\frac{2 \mathcal{T} + 1}{2 T + 1} & \text{for SNM}
\end{array}
\right. \, .
\end{equation}

\begin{figure}[t]
\centering
\includegraphics[width=0.99\textwidth,clip=]{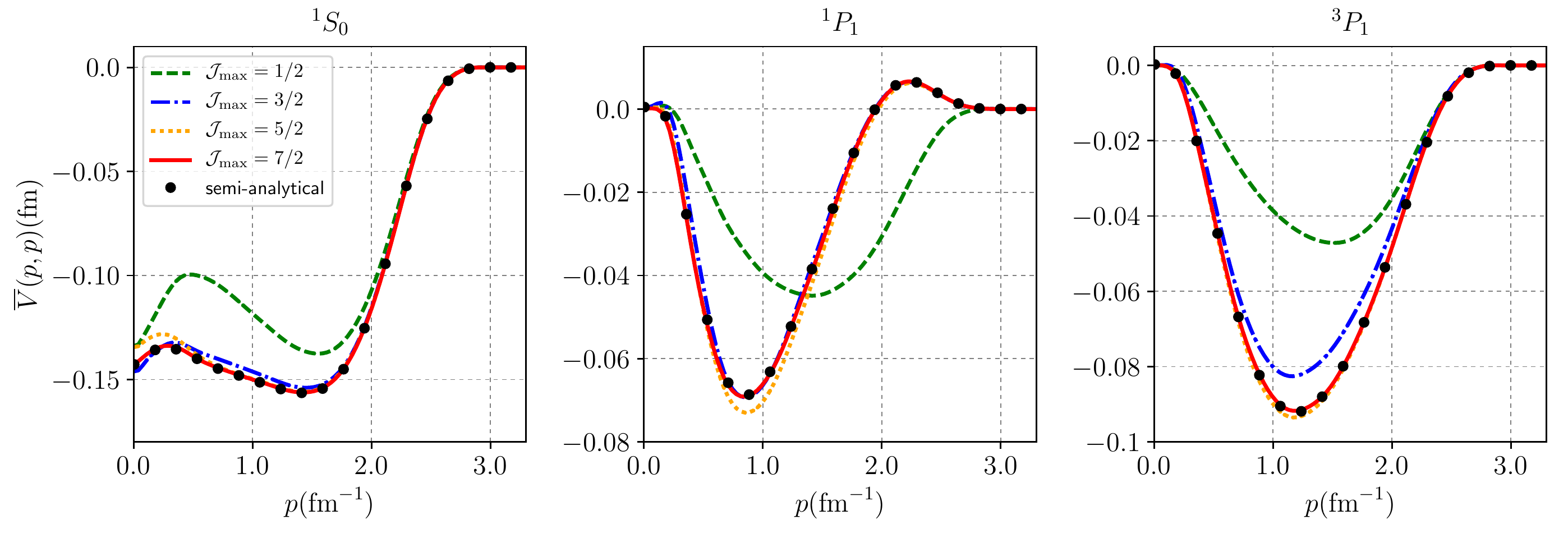}
\caption{Partial-wave convergence of the effective interaction $\overline{V}_{\text{3N}} (p,p)$ 
resulting from normal ordering of the leading order chiral 3N interaction
proportional to the LEC $c_3$ at N$^2$LO ($c_3 = 1 \: \text{GeV}^{-1}$) in symmetric
nuclear matter (SNM) at nuclear saturation density $k_{\text{F}} = 1.35 \:
\text{fm}^{-1}$ using a Hartree-Fock reference state at zero temperature and
using a ``nonlocal MS'' regularization with $n=4$ and $\Lambda=500$ MeV (see
Eq.~(\ref{eq:nonlocal_regulator})). In the different panels we show the
diagonal momentum-space matrix elements for the partial-wave channels
$^1$S$_0$ (left), $^1$P$_1$ (center) and $^3$P$_1$ (right). For comparison we
also show the results obtained from the semi-analytical approach of
Ref.~\cite{Hebe10nmatt,Hebe11fits}, i.e., by partial-wave decomposing the
effective interaction $\overline{V}{}_{\text{3N}}^{\text{SNM}}$ given in
Eq.~(\ref{Veff_pnm}) in Appendix \ref{sec:Veff_nuclear_matter}.}
\label{fig:Vbar_convergence}
\end{figure}

In Figure~\ref{fig:comp_3N_HF} we compare results for the 3N
Hartree-Fock energies based on the different approximations for the
effective NN interaction. The three panels show the energy difference
to the exact Hartree-Fock result for proton fraction $x = 0$~(left),
$x = 0.3$~(center), and $x = 0.5$~(right). The effective NN
interaction based on the $P=0$ approximation reproduces the exact
results well up to $n \simeq (0.13-0.23)$~fm$^{-3}$, depending on the
proton fraction. For higher densities the deviation systematically
increases, indicating a breakdown of the $P=0$ approximation. In
contrast, the results based on the $P$-average approximation agree
reasonably well with the exact results over the entire density range.

In Figure~\ref{fig:NO_Vbar} we present the results of some representative matrix
elements of $\overline{V}_{\text{3N}}$ in the $^1$S$_0$ channel, computed
via Eq.~(\ref{eq:Veff_matter_PW}). The normalization of the matrix
elements is chosen such that they can be combined with those of the
free-space NN interaction matrix elements for calculations in the Hartree-Fock
approximation using a combinatorial factor of $\zeta = \tfrac{1}{3}$ (see
Eq.~(\ref{eq:effective_NN_interaction}))\footnote{In the used normalization and
unit convention the NN partial-wave matrix elements the half on-shell
Lippmann-Schwinger equation takes the form (compare also Appendix~\ref{sec:normalization})
\begin{equation}
\left< p' L' | T_{\text{NN}} | p L \right> = \left< p' L' | V_{\text{NN}} | p L \right> + \frac{2}{\pi} \mathcal{P} \sum_{L''} \int dq q^2 \frac{\left< p' L' | V_{\text{NN}} | p'' L'' \right> \left< p'' L'' | T_{\text{NN}} | p L \right>}{p^2 - q^2} \, ,
\end{equation}
with the units $[V_{\text{NN}}] = [T_{\text{NN}}] = \text{fm}$. This
convention is commonly used in the literature ~\cite{Bogn10PPNP} (see also
footnote \ref{foot:normalization}).\label{foot:NN_normalization}}. Like in
Figure~\ref{fig:contour_N2LO_500}, we show the matrix elements for the
individual 3N topologies at N$^2$LO for the four different regularization
schemes discussed in Section~\ref{sec:3N_regularization} for a Hartree-Fock
reference state at zero temperature for pure neutron matter and symmetric
nuclear matter at two different densities each. The values of the long-range
LECs $c_1$, $c_3$ and $c_4$ are taken from the Roy-Steiner analysis of
Refs.~\cite{Hofe15sigmapiN,Hofe15PhysRep}, and the short-range LECs $c_D$
and $c_E$ are chosen for optimized visibility. The figure shows that the
regularization scheme has a significant impact on the values of the matrix
elements. While the overall form is quite similar for the different
regularizations, the precise values are quite sensitive to the regulator. In
particular, for pure neutron matter the effective potential is always
vanishing for nonlocal regularizations due to the Pauli
principle~\cite{Hebe10nmatt}. Local regulators, in contrast, generally induce
a finite range for this interaction and hence leads to finite
contributions in pure neutron matter (see also
Figure~\ref{fig:contour_cE_induced} and
Refs.~\cite{Tews16QMCPNM,Lynn16QMC3N,Logo2019consistent,Holt193Ndensrev}).

Finally, in Figure~\ref{fig:Vbar_convergence} we illustrate the partial-wave
convergence of the results for the effective potential for the ``nonlocal MS''
scheme. For this scheme the normal ordering has been also worked out in a
semi-analytical approach (see Appendix~\ref{sec:Veff_nuclear_matter}) and
hence it provides some independent benchmark results. We show some representative
results for the diagonal matrix elements for the partial-wave channels
$^1$S$_0$, $^1$P$_1$ and $^3$P$_1$ and the N$^2$LO topology proportional to
the LEC $c_3 = 1
\: \text{GeV}^{-1}$. It is manifest that the results obtained in a partial
wave representation via Eq.~(\ref{eq:Veff_matter_PW}) show excellent agreement
with the semi-analytical results for $\mathcal{J}_{\text{max}} = \tfrac{7}{2}$.

\subsubsection{Normal ordering in finite nuclei}
\label{sec:normal_ordering_nuclei}

Most many-body frameworks for finite nuclei based on basis expansion
representations, like, e.g., IM-SRG or CC, are formulated in a harmonic
oscillator (HO) basis. That means for these frameworks normal ordering of 3N
interactions involves summations over single-particle HO orbitals of the form
\begin{equation}
\left| a \right> = | n_a (l_a s_a) j_a m_{j_a} t_a m_{t_a} \bigr> \, ,
\label{eq:HO_single_particle}
\end{equation}
where $n_a$ is the radial quantum number, $l_a$ the single-particle orbital
angular momentum quantum number coupled with the spin $s_a = \tfrac{1}{2}$ to
the total angular momentum $j_a$. The isospin projection quantum number
$m_{t_a}$ denotes the proton $(m_{t_a} = +\tfrac{1}{2})$ and neutron orbitals
$(m_{t_a} = -\tfrac{1}{2})$. In order to distinguish the HO Orbital quantum
number from the occupation numbers $n_i$ in Eq.~(\ref{eq:normord_singpart}),
we denote the occupation number in the following by $\tilde{n}_i$.

\begin{figure}[t]
\centering
\includegraphics[scale=1.0,clip=]{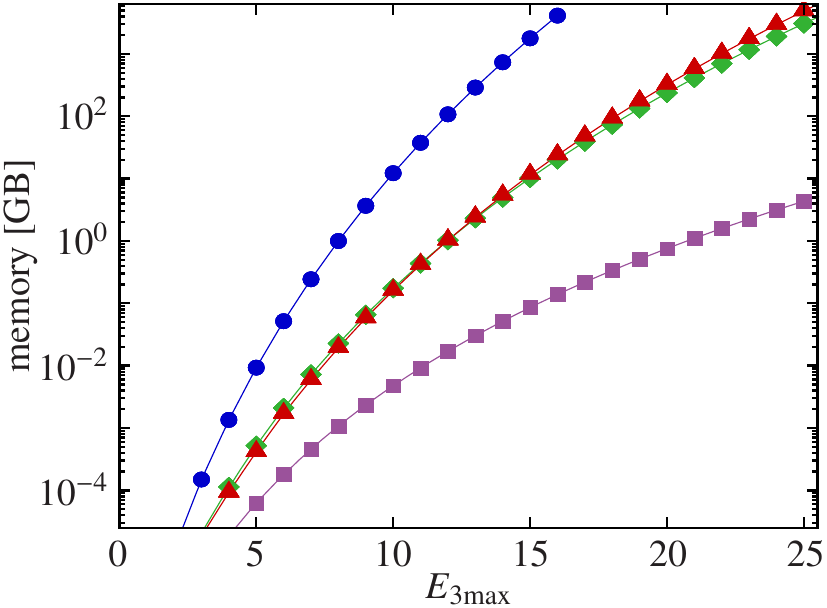}
\caption{Memory required to store the $T$-coefficients
(\symboldiamond[FGGreen]), as well as the three-body matrix elements in the
antisymmetrized-Jacobi (\symbolbox[FGViolet]), $JT$-coupled
(\symboltriangle[FGRed]) and $m$-scheme (\symbolcircle[FGBlue])
representation as function of the  maximum three-body energy quantum number
$E_{3\text{max}}$. All quantities are assumed to be single-precision floating
point numbers.\\
\textit{Source:} Figure taken from Ref.~\cite{Roth14SRG3N}.}
\label{fig:3N_sp_dimension}
\end{figure}

The traditional approach to normal ordering of 3N interactions for
applications to finite nuclei consists of the following steps (see
Ref.~\cite{Roth14SRG3N} for more details):
\begin{itemize}
\item[1.)] Transformation of 3N interaction matrix elements, calculated as
shown in Section~\ref{sect:3NF_representation} from a Jacobi momentum-space
representation to a HO Jacobi representation (see
Section~\ref{sec:HO_transf}):
\begin{align}
\bigl< p' q' \alpha' | V_{\text{3N}}^{(i), \text{reg}} | p q \alpha \bigr> &\rightarrow \bigl< N' n' \alpha' |  V_{\text{3N}}^{(i), \text{reg}} | N n \alpha \bigr> \, .
\end{align}
Of course, if the interaction matrix elements are calculated directly in HO
basis (like, e.g., in Ref.~\cite{Navr07local3N}) this step is not necessary.
\item[2.)] Antisymmetrization of matrix elements. This step can either already
be performed in the momentum basis (see
Section~\ref{sec:3N_decomp_antisymmetrization}) or in the HO
basis~\cite{Navr99NCSM}. Both choices eventually result formally in 3N matrix
elements of the form
\begin{equation}
\bigl< N' n' \alpha' |  V_{\text{3N}}^{\text{as,reg}} | N n \alpha \bigr> \, . 
\end{equation}
\item[3.)] Transformation of the Jacobi HO matrix elements to a
single-particle HO basis representation via a three-body Talmi-Moshinski
transformation~\cite{Kamu01TalmiMos,Roth14SRG3N}
\begin{equation}
\bigl< N' n' \alpha' |  V_{\text{3N}}^{\text{as,reg}} | N n \alpha \bigr> \rightarrow \bigl< 1' 2' 3' | V_{\text{3N}}^{\text{as,reg}} | 1 2 3 \bigr>  \, , 
\label{eq:3N_Moshinsky}
\end{equation}
where the single-particle states are given by
Eq.~(\ref{eq:HO_single_particle}). We note that Eq.~(\ref{eq:3N_Moshinsky}) is
only a formal representation. In practice this $m$-scheme representation of
the 3N interaction can be optimized by angular-momentum recoupling (see
Ref.~\cite{Roth14SRG3N} and Figure~\ref{fig:3N_sp_dimension}).

\item[4.)] Normal ordering of the single-particle 3N interaction matrix
elements by summing over a reference state expanded in the HO basis. As for
infinite matter, a common choice is a Hartree-Fock reference state. Then
Eq.~(\ref{eq:effective_NN_interaction}) takes the form
\begin{equation}
\bigl< 1' 2' | \overline{V}_{\text{3N}} | 1 2 \bigr> = \sum_3 \tilde{n}_3 \bigl< 1' 2' 3 | V_{\text{3N}}^{\text{as}} | 1 2 3 \bigr> \, ,
\end{equation}
where $\tilde{n}_3$ denote the HO orbital occupation numbers of the Hartree-Fock state.
\end{itemize}
The corresponding normal-ordered one- and zero-body contributions can then be
easily obtained by additional summations over the remaining single-particle
states in $\overline{V}_{\text{3N}}$, as shown in Eq.~(\ref{eq:normord_terms_HF}) and
(\ref{eq:normord_terms_BCS}). Since many-body calculations for nuclei beyond
the light sector are performed in a single-particle representation, all these
contributions can then be incorporated in a straightforward way in many-body
frameworks, under consideration of the correct combinatorial factors $\zeta$
(see discussion after Eq.~(\ref{eq:effective_NN_interaction})).

Although all steps above are straightforward conceptually, the practical
implementation involves some technical challenges and limitations. In
particular the transformation of 3N interactions to single-particle
coordinates (step 3) poses some severe challenges, in particular when
studying heavier nuclei. In this regime of the nuclear chart the required
basis sizes for storing and computing the required 3N matrix elements become
substantial (see Figure~\ref{fig:3N_sp_dimension}) and pose serious limitations.

These problems can be avoided by performing normal ordering in a Jacobi
coordinate representation of the underlying 3N interaction rather than a
single-particle representation. We now illustrate a framework that allows to
calculate the effective interaction $\overline{V}_{\text{3N}}$ in a Jacobi
momentum representation using a HO reference state. To this end, we will
neglect spin and isospin degrees of freedom for the sake of simple and
transparent notation. For details and the treatment of general 3N interactions
we refer the reader to Ref.~\cite{Dura19PhD,Dura19NO}.

As a first step we introduce a short hand notation for the HO quantum numbers:
$\left| \gamma_a \right> = \left| n_a l_a m_a \right>$, and start from the
definition of normal ordering in a single-particle HO representation:
\begin{align}
\bigl< \gamma_{1'} \gamma_{2'} | \overline{V}_{\text{3N}} | \gamma_1 \gamma_2 \bigr> = \sum_{n_3 l_3 m_3} \tilde{n}_3 \bigl< \gamma_{1'} \gamma_{2'} \gamma_3 | V_{\text{3N}}^{\text{as}} | \gamma_{1'} \gamma_{2'} \gamma_3 \bigr> \, .
\end{align}
By inserting a complete set of two-body single-particle momentum states and
projecting on these states this can be rewritten in the form
\begin{align}
\bigl< \mathbf{k}'_1 \mathbf{k}'_2 | \overline{V}_{\text{3N}} | \mathbf{k}_1 \mathbf{k}_2 \bigr> &= \sum_{n_3 l_3 m_3} \tilde{n}_3 \bigl< \mathbf{k}'_1 \mathbf{k}'_2 \gamma_3 | V_{\text{3N}}^{\text{as}} | \mathbf{k}_1 \mathbf{k}_2 \gamma_3 \bigr> \nonumber \\
&= \int \frac{d \mathbf{k}_3}{(2 \pi)^3} \frac{d \mathbf{k}'_3}{(2 \pi)^3} \bigl< \mathbf{k}'_1 \mathbf{k}'_2 \mathbf{k}'_3 | V^{\text{as}}_{\text{3N}} | \mathbf{k}_1 \mathbf{k}_2 \mathbf{k}_3 \bigr> \sum_{n_3 l_3 m_3} \tilde{n}_3 \bigl< \gamma_3 | \mathbf{k}'_3 \bigr> \bigl< \mathbf{k}_3 | \gamma_3 \bigr> \, ,
\end{align}
with $\bigl< \mathbf{k} | n l m \bigr> = R_{n l} (k) Y_{l m}
(\hat{\mathbf{k}})$. Here we used the completeness of the single-particle
momentum states $\int \frac{d \mathbf{k}_i}{(2 \pi)^3} | \mathbf{k}_i \bigr> \bigl< \mathbf{k}_i
| = 1$ and projected on the momentum states of particles $1, 2, 1'$ and $2'$
on both sides by using the orthogonality of the HO wave functions. As a next
step we rewrite the single-particle momentum representation of $\overline{V}_{\text{3N}}$
and $V_{\text{3N}}^{\text{as}}$ in a Jacobi representation by using
Eq.~(\ref{eq:P3N_factorization}):
\begin{equation}
\bigl< \mathbf{p}' \mathbf{P}' | \overline{V}_{\text{3N}} | \mathbf{p} \mathbf{P} \bigr> = \int \frac{d \mathbf{k}_3}{(2 \pi)^3} \frac{d \mathbf{k}'_3}{(2 \pi)^3} \bigl< \mathbf{p}' \mathbf{q}' | V_{\text{3N}}^{\text{as}} | \mathbf{p} \mathbf{q} \bigr> \, \delta(\mathbf{P} + \mathbf{k}_3 - \mathbf{P}' - \mathbf{k}'_{3}) \sum_{n_3 l_3 m_3 } \tilde{n}_3 \bigl< \gamma_3 | \mathbf{k}'_3 \bigr> \bigl< \mathbf{k}_3 | \gamma_3 \bigr> \, .
\end{equation}
We expressed the effective potential in terms of the Jacobi momentum
$\mathbf{p}$ and the two-body center-of-mass momentum $\mathbf{P}$, i.e.,
$\mathbf{P} = \mathbf{k}_1 + \mathbf{k}_2$ and  $\mathbf{P}' = \mathbf{k}'_1 +
\mathbf{k}'_2$. The single-particle momentum of particle 3 can be easily expressed in
terms of these momenta (see Table~\ref{tab:Jacobi_momenta_crosstable}):
$\mathbf{k}_3 = \tfrac{3}{2} \mathbf{q} + \tfrac{\mathbf{P}}{2}$. Note that the two-body
center-of-mass momentum $\mathbf{P}$ is in general not conserved since
$\mathbf{k}_3
\neq \mathbf{k}'_3$, in contrast to normal ordering with respect to a momentum
eigenstate like for nuclear matter (see
Section~\ref{sect:normal_ordering_matter}). If the orbital occupation numbers
$\tilde{n}_3$ do not depend on $m_3$ the sum can be performed
immediately:
\begin{align}
\bigl< \mathbf{p}' \mathbf{P}' | \overline{V}_{\text{3N}} | \mathbf{p} \mathbf{P} \bigr> &= \int \frac{d \mathbf{k}_3}{(2 \pi)^3} \frac{d \mathbf{k}'_3}{(2 \pi)^3} \bigl< \mathbf{p}' \mathbf{q}' | V_{\text{3N}}^{\text{as}} | \mathbf{p} \mathbf{q} \bigr> \, \delta(\mathbf{P} + \mathbf{k}_3 - \mathbf{P}' - \mathbf{k}'_{3}) \sum_{n_3 l_3} \tilde{n}_3 R_{n_3 l_3} (k_3) R_{n_3 l_3} (k'_3) \frac{2 l_3 + 1}{4 \pi} P_{l_3} (\hat{\mathbf{k}}_3 \cdot \hat{\mathbf{k}}'_3) \, .
\label{eq:Veff_Pq}
\end{align}
In the following we stick to this simplified case for illustration. However,
the generalization poses no fundamental problems. Eventually, we are
interested in the partial-wave matrix elements of the effective potential
$\overline{V}_{\text{3N}}$. Due to the non-Galilean invariance the partial-wave structure
becomes more complex compared to a free-space NN interaction (see discussion
in Section~\ref{sec:normal_ordering}). We extend the partial-wave basis by the
center-of-mass quantum numbers and project the interaction in
Eq.~(\ref{eq:Veff_Pq}) onto these states (compare Eq.~(\ref{eq:3NF_decmomp_spinless})):
\begin{align}
& \bigl< p' P' (L' L'_{cm}) \mathcal{L} | \overline{V}_{\text{3N}} | p P (L L_{cm}) \mathcal{L} \bigr> = \frac{1}{(2 \pi)^6} \sum_{\mathcal{M}_{\mathcal{L}}} \int d \hat{\mathbf{p}} d \hat{\mathbf{P}} d \hat{\mathbf{p}}' d \hat{\mathbf{P}}' \mathcal{Y}^{*\mathcal{L} \mathcal{M}_{\mathcal{L}}}_{L' L'_{cm}} (\hat{\mathbf{P}}', \hat{\mathbf{p}}') \bigl< \mathbf{p}' \mathbf{P}' | \overline{V}_{\text{3N}} | \mathbf{p} \mathbf{P} \bigr> \mathcal{Y}_{L L_{cm}}^{\mathcal{L} \mathcal{M}_{\mathcal{L}}} (\hat{\mathbf{P}}, \hat{\mathbf{p}}) \, .
\label{eq:Veff_spinless_PWD}
\end{align}
In Eq.~(\ref{eq:Veff_Pq}) we can finally make use of the partial-wave
representation of the 3N interaction such that in total the relations
(\ref{eq:Veff_Pq}) together with (\ref{eq:Veff_spinless_PWD}) give a relation
for the effective interaction. These relations can be generalized to spin- and
isospin-dependent 3N interactions~\cite{Dura19PhD,Dura19NO} and result in partial wave matrix elements of the form
\begin{equation}
\bigl< p' P' [(L' S') J' L'_{cm} ]
J_{\text{tot}} T | \overline{V}_{\text{3N}} | p P [(L
S) J L_{cm}] J_{\text{tot}} T \bigr> \, .
\end{equation}

The two final steps consist of a transformation of these relative-coordinate
momentum matrix elements of $\overline{V}_{\text{3N}}$ to a
relative-coordinate HO basis and a successive transformation to a
single-particle coordinate representation for applications to many-body
frameworks. While the first step is straightforward the latter one requires a
generalization of the well-established two-body Talmi-Moshinsky
transformation, which usually assumes a Galilean-invariant interaction, i.e.,
a trivial dependence on the center-of-mass quantum numbers. The model space in
the Jacobi representation is characterized by the total energy quantum number
$E$, which involves relative and center-of-mass quantum numbers:
\begin{equation}
E = 2 N_{\text{cm}} + L_{\text{cm}} + 2 N + L = e_1 + e_2 \, .
\end{equation}
Here $N_{\text{cm}}$ denotes the radial center-of-mass HO quantum number, $N$
the corresponding quantum number of relative excitations and $e_i$ the
according single-particle energy quantum numbers: $e_i = 2 n_i + l_i$ (see
Ref.~\cite{Roth14SRG3N}). In order to take into account all possible
recoupling contributions from the Jacobi representation to a single-particle
representation, matrix elements for sufficiently large values of
$L_{\text{cm}}$ and $L$ have to be computed for a given truncation scheme for
$e_{i}$. In practice, however, the matrix elements of
$\overline{V}_{\text{3N}}$ get suppressed as the values of $L$ and
$L_{\text{cm}}$ increase. A systematic study of these convergence patterns is
work in progress and is key for applications of this framework to realistic
many-body calculations~\cite{Dura19NO}.

\begin{figure}[t!]
\begin{center}
\vspace{1cm}
\includegraphics[width=0.8\textwidth]{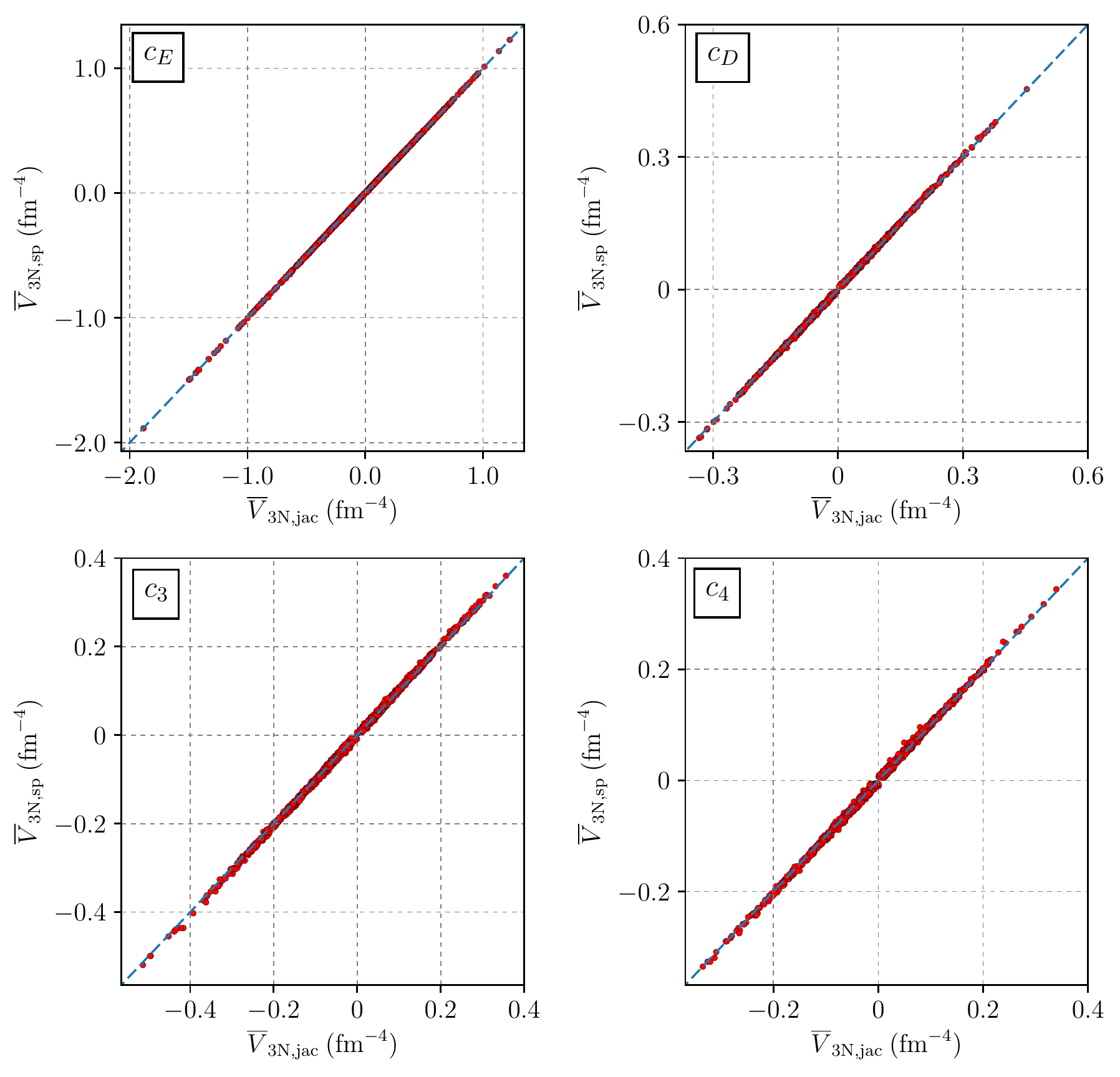}
\end{center}
\caption{Comparison of normal-ordered 3N matrix elements in the model space $e_{\text{max}} = 4$ 
using the established framework formulated in single-particle representation
($\overline{V}_{\text{3N,sp}}$) and the new framework in Jacobi momentum
representation ($\overline{V}_{\text{3N,jac}}$) for different 3N topologies at
N$^{2}$LO using $c_3 = c_4 = 1$ GeV$^{-1}$, $c_D = c_E = 1$,
$\Lambda_{\text{3N}} = \infty$ and $\Omega = 13.53$ MeV. For the calculations
a harmonic oscillator $^{16}$O reference state was used and the partial-wave
truncations $L, L_{\text{cm}} \le 4$. Credits to Johannes Simonis for
providing the reference results for $\overline{V}_{\text{3N,sp}}$. For details
see Refs.~\cite{Dura19PhD,Dura19NO}.
}
\label{fig:jacobi_normalorder}
\end{figure}

The framework above relates the partial-wave matrix elements of the effective
potential directly to the Jacobi momentum-space matrix elements of the
underlying 3N interactions. Consequently, it is not necessary to represent the
matrix elements in a single-particle basis at any point and we can hence avoid
the basis dimension problems illustrated in Figure~\ref{fig:3N_sp_dimension}.
In Figure~\ref{fig:jacobi_normalorder} we compare results for
$\overline{V}_{\text{3N}}$ obtained in the traditional normal-ordering
implementation in a single-particle basis ($\overline{V}_{\text{3N,sp}}$) and
those obtained in the novel framework formulated in a Jacobi basis
($\overline{V}_{\text{3N,jac}}$). For all the shown results we use the model
space truncations $L,L_{\text{cm}},e_{i} \le 4$ and $J_{\text{tot}} \le 2$.
Furthermore we include all three-body channels up to $\mathcal{J} =
\tfrac{5}{2}$. For simplicity we use a harmonic oscillator $^{16}$O reference
state and $\Lambda_{\text{3N}} =
\infty$ for these first proof-of-principle calculations. For the pure contact
3N interaction proportional to $c_E$ (top left panel) the results of both
approaches show perfect agreement. Here the possible angular momentum
couplings are severely restricted by the constraints $L=l=L'=l'=0$ for the
three-body states $\left| \alpha \right>$. For 3N interactions containing
long-range pion exchange contributions the incorporation of all relevant
partial-wave channels becomes more challenging. For the one-pion exchange
interaction proportional to $c_D$ the agreement is still excellent. For the
long-range topologies proportional to $c_3$ and $c_4$ (lower panels) the
coupling of channels with different angular momenta is stronger than for the
one-pion-exchange interaction. We still find good agreement for results with
$J_{\text{tot}} \le 1$, while the agreement is not as good as for the
interactions proportional to $c_D$, indicating that contributions from
channels of larger $L$ and $L_{\text{cm}}$ are more significant. One big
advantage of the new framework is the fact that the values for the radial HO
quantum numbers $N_{\text{cm}}$ and $N$ can be increased in a simple and
straightforward way since these quantum numbers are only introduced in the
transformation to HO basis after normal ordering in momentum space, which is
computationally very cheap. The main remaining challenges consist in
optimizing the normal ordering algorithm such that the accessible partial-wave
model space can be increased and in generalizing the framework to more general
reference states such that the resulting normal-ordered interactions can be
applied to state-of-the-art many-body calculations. Both these improvements
are currently work in progress~\cite{Dura19NO}.
 
\subsection{Application of 3N interactions without partial-wave decomposition}
\label{sec:no_PWD}

In Ref.~\cite{Dris17MCshort} a novel many-body framework was presented for
computing the equation of state of dense matter based on NN and 3N
interactions within many-body perturbation theory (MBPT) without employing a
partial-wave decomposition. This approach has some practical advantages
compared to traditional approaches based on partial-wave representation as we
illustrate now.

The new framework is formulated in a single-particle basis of the form
$\left| \mathbf{k}_i \sigma_i \tau_i \right>$, where $\mathbf{k}_i$ are the
single-particle momenta of particle $i$ (see
Section~\ref{sect:3NF_representation}) and $\sigma_i$ ($\tau_i$) are the
corresponding spin and isospin projections. That means NN and
3N interaction matrix elements take the following form:
\begin{align}
\bigl< 1' 2' | V_{\text{NN}} | 1 2 \bigr> &= \bigl< \mathbf{k}'_1 \sigma'_1 \tau'_1 \mathbf{k}'_2 \sigma'_2 \tau'_2 | V_{\text{NN}} | \mathbf{k}_1 \sigma_1 \tau_1 \mathbf{k}_2 \sigma_2 \tau_2 \bigr> \, , \nonumber \\
\bigl< 1' 2' 3' | V_{\text{3N}} | 1 2 3 \bigr> &= \bigl< \mathbf{k}'_1 \sigma'_1 \tau'_1 \mathbf{k}'_2 \sigma'_2 \tau'_2 \mathbf{k}'_3 \sigma'_3 \tau'_3 | V_{\text{3N}} | \mathbf{k}_1 \sigma_1 \tau_1 \mathbf{k}_2 \sigma_2 \tau_2 \mathbf{k}_3 \sigma_3 \tau_3 \bigr> \, .
\label{eq:3N_noPWD}
\end{align}
Instead of prestoring these interaction matrix elements on a discrete grid of
mesh points, they can also be represented exactly as matrices in spin-isospin
space, where the matrix elements are analytic functions of the single-particle
momenta $\mathbf{k}_i$. This allows to evaluate matrix elements efficiently on
the fly for arbitrary values of momenta, spin and isospin. This is
particularly important when using Monte-Carlo integration routines, because it
is impossible to know \textit{a priori} which mesh points are being sampled
during the many-body calculation.

For the calculations presented in Ref.~\cite{Dris17MCshort} all NN and 3N
interactions up to N$^3$LO (for nonlocal regulators) were implemented in this
vector representation, where the antisymmetrization was performed in an
automated way. For the inclusion of NN and 3N interactions whose operatorial
structure is not directly accessible, like RG-evolved potentials (see
Section~\ref{sec:SRG}), it is also possible to resum the contributions of all
partial-wave channels in order to obtain the interaction matrix elements in
momentum vector representation, e.g. for 3N interactions (see
Section~\ref{sec:general_3N_decomp}):

\begin{align}
\bigl< 1' 2' 3' | V_{\text{3N}} | 1 2 3 \bigr> &= \sum_{\mathcal{J},\mathcal{P}} \frac{1}{2 \mathcal{J} + 1} \sum_{\alpha, \alpha'} \sum_{\mathcal{M}_{\mathcal{J}}} \sum_{\substack{M_{L},m_{l},M'_{L},m'_{l}\\M_{S},m_{s},M'_{S},m'_{s}}} \sum_{M_{J},m_{j},M'_{J},m'_{j}} \mathcal{C}_{\tfrac{1}{2} \sigma_1 \tfrac{1}{2} \sigma_2}^{S M_S} \mathcal{C}_{\tfrac{1}{2} \sigma'_1 \tfrac{1}{2} \sigma'_2}^{S' M'_S} \tensor*[_{\{ab\}}]{\left< p' q' \alpha' | V_{\rm 3N} | p q \alpha \right>}{_{\{ab\}}} \nonumber \\
& \times \mathcal{C}_{J M_J j m_j}^{\mathcal{J} \mathcal{M}_{\mathcal{J}}} \mathcal{C}_{J' M'_J j' m'_j}^{\mathcal{J} \mathcal{M}_{\mathcal{J}}} \mathcal{C}_{L M_L S M_S}^{J M_J} \mathcal{C}_{L' M'_L S' M'_S}^{J' M'_J} \mathcal{C}_{l m_l \tfrac{1}{2} \sigma_3}^{j m_j} \mathcal{C}_{l' m'_l \tfrac{1}{2} \sigma'_3}^{j' m'_j} Y^*_{L' M'_L} (\hat{\mathbf{p}}_{\{ab\}}') Y^*_{l' m'_l} (\hat{\mathbf{q}}_{\{ab\}}') Y_{L M_L} (\hat{\mathbf{p}}_{\{ab\}}) Y_{l m_l} (\hat{\mathbf{q}}_{\{ab\}}) \, ,
\end{align}
where we have suppressed the isospin quantum numbers in these relations for
the sake of a more compact notation. The Jacobi momenta $\mathbf{p}$,
$\mathbf{q}$, $\mathbf{p}'$ and $\mathbf{q}'$ in the chosen representation
$\{ab\}$ are given in terms of the single particle momenta $\mathbf{k}_i$ and
$\mathbf{k}'_i$ by the relations summarized in
Table~\ref{tab:Jacobi_momenta_crosstable}. In practice all calculations are
usually performed based on antisymmetrized interactions, i.e. in terms of
Hugenholz diagrams, so that the matrix elements become independent of the
chosen basis representation $\{ab\}$ (see discussion in
Section~\ref{sec:3N_decomp_antisymmetrization}).

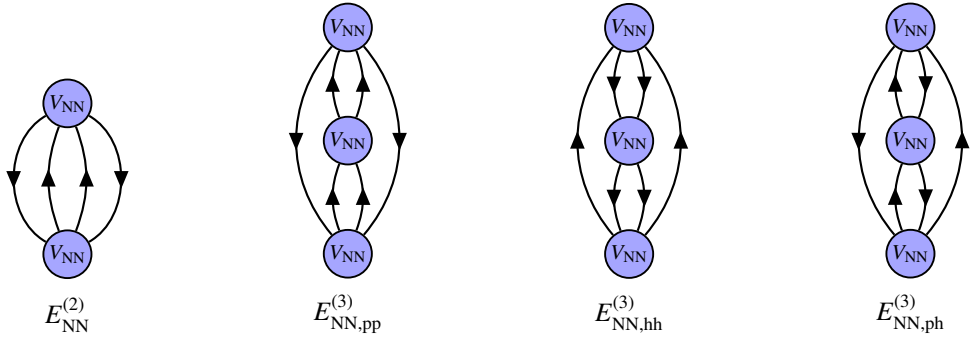
\begin{figure}[t!]
\centering
\begin{minipage}[c]{0.15\textwidth}
\begin{tikzpicture} 
\tikzfeynmanset{
  my dot/.style={fill=red},
  every vertex/.style={my dot},
}
\begin{feynman}
\vertex [blob, /tikz/minimum size=18pt, fill=blue!35, line width=0.25mm, font=\fontsize{8}{0}\selectfont] (a) at (0.0,0.0) {\(V_{\rm NN}\)};
\vertex [blob, /tikz/minimum size=18pt, fill=blue!35, line width=0.25mm, font=\fontsize{8}{0}\selectfont] (b) at (0.0,2.0) {\(V_{\rm NN}\)};
\vertex [blob, /tikz/minimum size=18pt, draw=none, fill=none] (c) at (0.0,3.0) {};
\vertex (c) at (0.0,-0.8) {\(E_{\rm NN}^{(2)}\)};
\diagram* {
(a) -- [fermion, thick, in=-70, out=70] (b);
(a) -- [fermion, thick, in=-110, out=110] (b);
(b) -- [fermion, thick, out=-30, in=30] (a);
(b) -- [fermion, thick, out=-150, in=150] (a);
};
\end{feynman}
\end{tikzpicture}
\end{minipage}
\hspace{1cm}
\begin{minipage}[c]{0.15\textwidth}
\begin{tikzpicture} 
\tikzfeynmanset{
  my dot/.style={fill=red},
  every vertex/.style={my dot},
}
\begin{feynman}
\vertex [blob, /tikz/minimum size=18pt, fill=blue!35, line width=0.25mm, font=\fontsize{8}{0}\selectfont] (a) at (0.0,0.0) {\(V_{\rm NN}\)};
\vertex [blob, /tikz/minimum size=18pt, fill=blue!35, line width=0.25mm, font=\fontsize{8}{0}\selectfont] (b) at (0.0,1.5) {\(V_{\rm NN}\)};
\vertex [blob, /tikz/minimum size=18pt, fill=blue!35, line width=0.25mm, font=\fontsize{8}{0}\selectfont] (c) at (0.0,3.0) {\(V_{\rm NN}\)};
\vertex (d) at (0.0,-0.8) {\(E_{\rm NN,pp}^{(3)}\)};
\diagram* {
(a) -- [fermion, thick, in=-70, out=70] (b);
(a) -- [fermion, thick, in=-110, out=110] (b);
(b) -- [fermion, thick, in=-70, out=70] (c);
(b) -- [fermion, thick, in=-110, out=110] (c);
(c) -- [fermion, thick, out=-50, in=50] (a);
(c) -- [fermion, thick, out=-130, in=130] (a);
};
\end{feynman}
\end{tikzpicture}
\end{minipage}
\hspace{1cm}
\begin{minipage}[c]{0.15\textwidth}
\begin{tikzpicture} 
\tikzfeynmanset{
  my dot/.style={fill=red},
  every vertex/.style={my dot},
}
\begin{feynman}
\vertex [blob, /tikz/minimum size=18pt, fill=blue!35, line width=0.25mm, font=\fontsize{8}{0}\selectfont] (a) at (0.0,0.0) {\(V_{\rm NN}\)};
\vertex [blob, /tikz/minimum size=18pt, fill=blue!35, line width=0.25mm, font=\fontsize{8}{0}\selectfont] (b) at (0.0,1.5) {\(V_{\rm NN}\)};
\vertex [blob, /tikz/minimum size=18pt, fill=blue!35, line width=0.25mm, font=\fontsize{8}{0}\selectfont] (c) at (0.0,3.0) {\(V_{\rm NN}\)};
\vertex (d) at (0.0,-0.8) {\(E_{\rm NN,hh}^{(3)}\)};
\diagram* {
(b) -- [fermion, thick, out=-70, in=70] (a);
(b) -- [fermion, thick, out=-110, in=110] (a);
(c) -- [fermion, thick, out=-70, in=70] (b);
(c) -- [fermion, thick, out=-110, in=110] (b);
(a) -- [fermion, thick, in=-50, out=50] (c);
(a) -- [fermion, thick, in=-130, out=130] (c);
};
\end{feynman}
\end{tikzpicture}
\end{minipage}
\hspace{1cm}
\begin{minipage}[c]{0.15\textwidth}
\begin{tikzpicture} 
\tikzfeynmanset{
  my dot/.style={fill=red},
  every vertex/.style={my dot},
}
\begin{feynman}
\vertex [blob, /tikz/minimum size=18pt, fill=blue!35, line width=0.25mm, font=\fontsize{8}{0}\selectfont] (a) at (0.0,0.0) {\(V_{\rm NN}\)};
\vertex [blob, /tikz/minimum size=18pt, fill=blue!35, line width=0.25mm, font=\fontsize{8}{0}\selectfont] (b) at (0.0,1.5) {\(V_{\rm NN}\)};
\vertex [blob, /tikz/minimum size=18pt, fill=blue!35, line width=0.25mm, font=\fontsize{8}{0}\selectfont] (c) at (0.0,3.0) {\(V_{\rm NN}\)};
\vertex (d) at (0.0,-0.8) {\(E_{\rm NN,ph}^{(3)}\)};
\diagram* {
(b) -- [fermion, thick, out=-70, in=70] (a);
(a) -- [fermion, thick, in=-110, out=110] (b);
(c) -- [fermion, thick, out=-70, in=70] (b);
(b) -- [fermion, thick, in=-110, out=110] (c);
(a) -- [fermion, thick, in=-50, out=50] (c);
(c) -- [fermion, thick, out=-130, in=130] (a);
};
\end{feynman}
\end{tikzpicture}
\end{minipage}
\caption{Diagrammatic representation of the second-order (left) and
third-order contributions (three right diagrams) to the energy density from NN
interactions. Arrows pointing up in the propagators indicate particle states
and arrows pointing down hole states.}
\label{fig:MBPT_NN_2nd_3rd}
\end{figure}

This representation allows to express and
implement individual diagrams in MBPT in a very compact form and opens the way
to efficient calculations in MBPT up to much higher orders than was possible
in a partial-wave representation. We illustrate the advantage of the new
method by considering as an example the second-order NN contribution in MBPT
to the energy of neutron matter (see Figure~\ref{fig:MBPT_NN_2nd_3rd}). In
order to keep the notation simple we assume without loss of generality
spin-independent occupation numbers $n_{\mathbf{k}}$ and single-particle
energies $\varepsilon_{\mathbf{k}}$. In a partial-wave representation this
contribution takes the following form (using the same normalization convention
for the NN matrix elements as for Figure~\ref{fig:NO_Vbar}, see also footnote
on page~\pageref{foot:NN_normalization})~\cite{Hebe10nmatt}:
\begin{align}
\frac{E^{(2)}_{\text{NN}}}{V} &= \frac{1}{4 \pi^5} \int dp p^2 dp' p'^{2} d \cos \theta_{\mathbf{p} \mathbf{p}'} \int d P_{\text{NN}} P_{\text{NN}}^2 \int d \cos \theta_{\mathbf{P}_{\text{NN}}} \int d \phi_{\mathbf{P}_{\text{NN}}} 
\sum_{\bar{L}} \, P_{\bar{L}} (\hat{\mathbf{p}} \cdot \hat{\mathbf{p}}') \nonumber \\
&\times \frac{n_{\mathbf{p} - \mathbf{P}_{\text{NN}}/2} n_{\mathbf{p} + \mathbf{P}_{\text{NN}}/2} (1 - n_{\mathbf{p}' - \mathbf{P}_{\text{NN}}/2}) (1 - n_{\mathbf{p}' + \mathbf{P}_{\text{NN}}/2})}{\varepsilon_{\mathbf{p} - \mathbf{P}_{\text{NN}}/2} - \varepsilon_{\mathbf{p} + \mathbf{P}_{\text{NN}}/2} - \varepsilon_{\mathbf{p}' - \mathbf{P}_{\text{NN}}/2} - \varepsilon_{\mathbf{p}' + \mathbf{P}_{\text{NN}}/2}} \nonumber \\
&\times \sum_S \sum_{J,l,l'} \, \sum_{\tj,\tl,\tlp} \, (-1)^{\tl+L'+\bar{L}} \, 
\mathcal{C}_{L 0 \tlp 0}^{\bar{L} 0} \, \mathcal{C}_{L' 0 \tl 0}^{\bar{L} 0} \bigl(1-(-1)^{\bar{L}+S+1}\bigr) \, \bigl(1-(-1)^{\tl+S+1}\bigr) \sqrt{\hat{L} \hat{L'} \hat{\tl} \hat{\tlp}} \: \hat{J} \hat{\tj} \nonumber \\
&\times \biggl\{ \begin{array}{ccc}
L & S & J \\
\tj & \bar{L} & \tlp \end{array} \biggr\}
\biggl\{ \begin{array}{ccc}
J & S & L' \\
\tl & \bar{L} & \tj \end{array} \biggr\} \,
\bigl\la p' (\tlp S) \tj | V_{\text{NN}}^{\text{as}} | p (S \tl) \tj \bigr\ra \bigl\la p (L S) J | V_{\text{NN}}^{\text{as}} | p' (L' S) J \bigr\ra \, \, ,
\label{eq:E2nd_PNM_PW}
\end{align}
where $\mathbf{P}_{\text{NN}}$ is the two-body center-of-mass momentum $\mathbf{P}_{\text{NN}} =
\mathbf{k}_1 + \mathbf{k}_2 = \mathbf{k}'_1 + \mathbf{k}'_2$. Although this
expression can be implemented in a straightforward way on a computer, it is
quite involved already at this low order in MBPT, and the analytical
derivation requires some effort. Furthermore, the implementation and
benchmarking of the corresponding expressions at third order in MBPT becomes
quite challenging and tedious~\cite{Cora14nmat,Holt16eos3pt,Kais17phring3}. In
addition, pushing this approach to even higher orders becomes eventually
impractical.

In contrast, in the single-particle vector representation given in
Eq.~(\ref{eq:3N_noPWD}) the same diagram can be expressed in a much more
compact form:
\begin{align}
\frac{E^{(2)}_{\text{NN}}}{V} &= \frac{1}{4} \prod_{i=1}^2 \text{Tr}_{\sigma_i} \int \frac{d \mathbf{k}_i}{(2\pi)^3} \frac{d \mathbf{k}'_i}{(2\pi)^3} (2 \pi)^3 \delta (\mathbf{k}_1 + \mathbf{k}_2 - \mathbf{k}'_1 - \mathbf{k}'_2) \left| \bigl< \mathbf{k}'_1 \sigma'_1 \mathbf{k}'_2 \sigma'_2 | \hat{V}_{\text{NN}} | \mathbf{k}_1 \sigma_1 \mathbf{k}_2 \sigma_2 \bigr> \right|^2 \frac{n_{\mathbf{k}_1} n_{\mathbf{k}_2} \bigl( 1 - n_{\mathbf{k}'_1} \bigr) \bigl( 1 - n_{\mathbf{k}'_2} \bigr)}{\varepsilon_{\mathbf{k}_1} + \varepsilon_{\mathbf{k}_2} - \varepsilon_{\mathbf{k}'_1} - \varepsilon_{\mathbf{k}'_2}} \nonumber \\
&= \frac{1}{4} \sum_{1,2} \sum_{1',2'} \left< 1' 2' | V_{\text{NN}} | 1 2 \right> \left< 1 2 | V_{\text{NN}} | 1' 2' \right> \delta_{1 + 2,1' + 2'} \frac{n_1 n_2 (1 - n_{1'}) (1 - n_{2'})}{\varepsilon_1 + \varepsilon_2 - \varepsilon_{1'} - \varepsilon_{2'}} \nonumber \\
&= \frac{1}{4} \sum \limits_{\substack{ij\\\alpha\beta}} \frac{\braket{\alpha \beta| V_{\text{NN}} |ij} \braket{ij| V_{\text{NN}} |\alpha \beta}}{D_{ij\alpha\beta}} \, ,
\label{eq:E_2nd_singleparticle}
\end{align}
where we suppressed the isospin projection quantum numbers $\tau_i = -\tfrac{1}{2}$. In
the last steps we introduced a compact notation via the definition
\begin{equation}
D_{ijk\ldots \alpha \beta \gamma\ldots} = \varepsilon_{i} + \varepsilon_{j} + \varepsilon_{k} + \ldots - \varepsilon_{\alpha} -\varepsilon_{\beta} -\varepsilon_{\gamma} - \ldots \, .
\end{equation}
Here, greek indices $\alpha,\beta,\gamma,\ldots$ denote particle states and
latin indices $i,j,k,\ldots$ holes, e.g.:
\begin{align}
\sum_{\beta} = \text{Tr}_{\sigma_{\beta}} \int \frac{d \mathbf{k}_{\beta}}{(2\pi)^3} \left( 1 - n_{\mathbf{k}_{\beta}} \right), \quad \sum_i = \text{Tr}_{\sigma_i} \int \frac{d \mathbf{k}_i}{(2\pi)^3} n_{\mathbf{k}_i} \, ,
\end{align}
with implicit delta functions that enforce the conservation of the
center-of-mass momentum in each interaction process. For systems with a finite
proton fraction these relations can be generalized in a straightforward way by
appropriate traces over isospin quantum numbers.

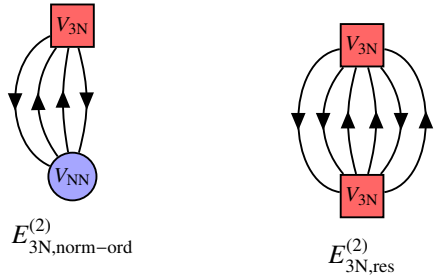
\begin{figure}[b!]
\centering
\begin{minipage}[c]{0.15\textwidth}
\begin{tikzpicture}
\tikzfeynmanset{
  my dot/.style={fill=red},
  every vertex/.style={my dot},
}
\begin{feynman}
\vertex [blob, /tikz/minimum size=18pt, fill=blue!35, line width=0.25mm, font=\fontsize{8}{0}\selectfont] (a) at (0.0,0.0) {\(V_{\rm NN}\)};
\vertex [blob, /tikz/minimum size=16pt, shape=rectangle, fill=red!60,, line width=0.25mm, font=\fontsize{8}{0}\selectfont] (b) at (0.0,2.0) {\(V_{\rm 3N}\)};
\vertex (c) at (0.0,-0.8) {\(E_{\rm 3N,norm-ord}^{(2)}\)};
\diagram* {
(a) -- [fermion, line width=0.25mm, in=-100, out=100] (b);
(a) -- [fermion, line width=0.25mm, in=-126, out=126] (b);
(b) -- [fermion, line width=0.25mm, out=-75, in=75] (a);
(b) -- [fermion, line width=0.25mm, out=45, in=-45, loop, min distance=0.6cm] (b);
(b) -- [fermion, line width=0.25mm, out=-155, in=155] (a);
};
\end{feynman}
\end{tikzpicture}
\end{minipage}
\hspace{1cm}
\begin{minipage}[c]{0.15\textwidth}
\begin{tikzpicture} 
\tikzfeynmanset{
  my dot/.style={fill=red},
  every vertex/.style={my dot},
}
\begin{feynman}
\vertex [blob, /tikz/minimum size=16pt, shape=rectangle, fill=red!60, line width=0.25mm, font=\fontsize{8}{0}\selectfont] (a) at (0.0,0.0) {\(V_{\rm 3N}\)};
\vertex [blob, /tikz/minimum size=16pt, shape=rectangle, fill=red!60, line width=0.25mm, font=\fontsize{8}{0}\selectfont] (b) at (0.0,2.0) {\(V_{\rm 3N}\)};
\vertex [blob, /tikz/minimum size=16pt, fill=none, draw = none] (ghost) at (0.0,2.53) {};
\vertex (c) at (0.0,-0.8) {\(E_{\rm 3N,res}^{(2)}\)};
\diagram* {
(a) -- [fermion, line width=0.25mm, in=-75, out=75] (b);
(a) -- [fermion, line width=0.25mm, in=-105, out=105] (b);
(b) -- [fermion, line width=0.25mm, out=-50, in=50] (a);
(b) -- [fermion, line width=0.25mm, out=-130, in=130] (a);
(a) -- [fermion, line width=0.25mm, in=-10, out=10] (b);
(b) -- [fermion, line width=0.25mm, out=-170, in=170] (a);
};
\end{feynman}
\end{tikzpicture}
\end{minipage}
\caption{Diagrammatic representation of second-order contributions to the
energy density from NN interactions and 3N interactions. Arrows pointing up in
the propagators indicate particle states and arrows pointing down hole states.
The left diagram shows a contribution that involves one normal-ordered 3N
interaction (see Figure~\ref{fig:Veff_diagram}), whereas the right diagram
shows the contributions from residual 3N interactions at this order (see
Section~\ref{sec:normal_ordering}).}
\label{fig:MBPT_3N_2nd}
\end{figure}

Equation~(\ref{eq:E_2nd_singleparticle}) shows that the computation of diagrams
effectively amounts to the evaluation of high-dimensional phase-space
integrals over the single-particle momenta, restricted by the occupation
numbers $n_{\mathbf{k}}$ and the cutoff regularization scales of the NN and 3N
interactions, plus the computation of discrete sums over spin quantum numbers.
Tracing over spin and isospin states of each particle is straightforward and
can be fully automated in the representation Eq.~(\ref{eq:3N_noPWD}). The
integrals over the momenta can be computed efficiently using adaptive
Monte-Carlo algorithms~\cite{Lepa78Vegas,Hahn05CUBA,Hahn16CUBApara}, which are
especially suitable for high-dimensional integrals.

The compact form of the expressions for each diagram makes the implementation
of arbitrary energy diagrams straightforward, even up to high orders in MBPT.
Since the number of diagrams at a given order increases rapidly, with 3, 39
and 840 at third, fourth and fifth order for NN-only interactions
\cite{OEISHugen,Stev03autgen,Shav09MBmethod}, we developed a general framework
that allows to generate the analytic expressions for each diagram in a fully
automated way~\cite{Dris17MCshort}. This framework can also be combined with
other sophisticated automated many-body diagram generation
frameworks~\cite{Artu18DiagGen} that allow to push MBPT calculations to even
higher orders.

For example, the expressions for the third order diagrams (see
Figure~\ref{fig:MBPT_NN_2nd_3rd}) are given by~\cite{Shav09MBmethod}
\begin{align}
\frac{E^{(3)}_{\text{NN,pp}}}{V} &= \frac{1}{8} \sum \limits_{\substack{ij\\\alpha \beta\gamma \delta}} \frac{\braket{ij|V_{\text{NN}}|\alpha \beta} \braket{\alpha \beta|V_{\text{NN}}|\gamma \delta} \braket{\gamma \delta|V_{\text{NN}}|ij}}{D_{ij\alpha \beta}D_{ij\gamma \delta}} \, , \nonumber \\
\frac{E^{(3)}_{\text{NN,hh}}}{V} &= \frac{1}{8} \sum \limits_{\substack{ijkl\\\alpha \beta}} \frac{\braket{\alpha \beta|V_{\text{NN}}|kl} \braket{kl|V_{\text{NN}}|ij} \braket{ij|V_{\text{NN}}|\alpha \beta}}{D_{ij\alpha\beta}D_{kl\alpha \beta}} \, , \nonumber \\
\frac{E^{(3)}_{\text{NN,ph}}}{V} &= \sum \limits_{\substack{ijk\\\alpha \beta \gamma}} \frac{\braket{ij|V_{\text{NN}}|\alpha \beta} \braket{\alpha k|V_{\text{NN}}|i\gamma} \braket{\beta \gamma|V_{\text{NN}}|jk}}{D_{ij\alpha \beta}D_{jk\beta\gamma}} \, ,
\end{align}
where the individual contributions are usually referred to particle-particle
(``pp''), hole-hole (``hh''), and particle-hole (``ph'') excitations, respectively,
according to the type of intermediate states connecting the three interaction
vertices.

\begin{figure}[t!]
\begin{center}
\includegraphics[width=0.6\textwidth]{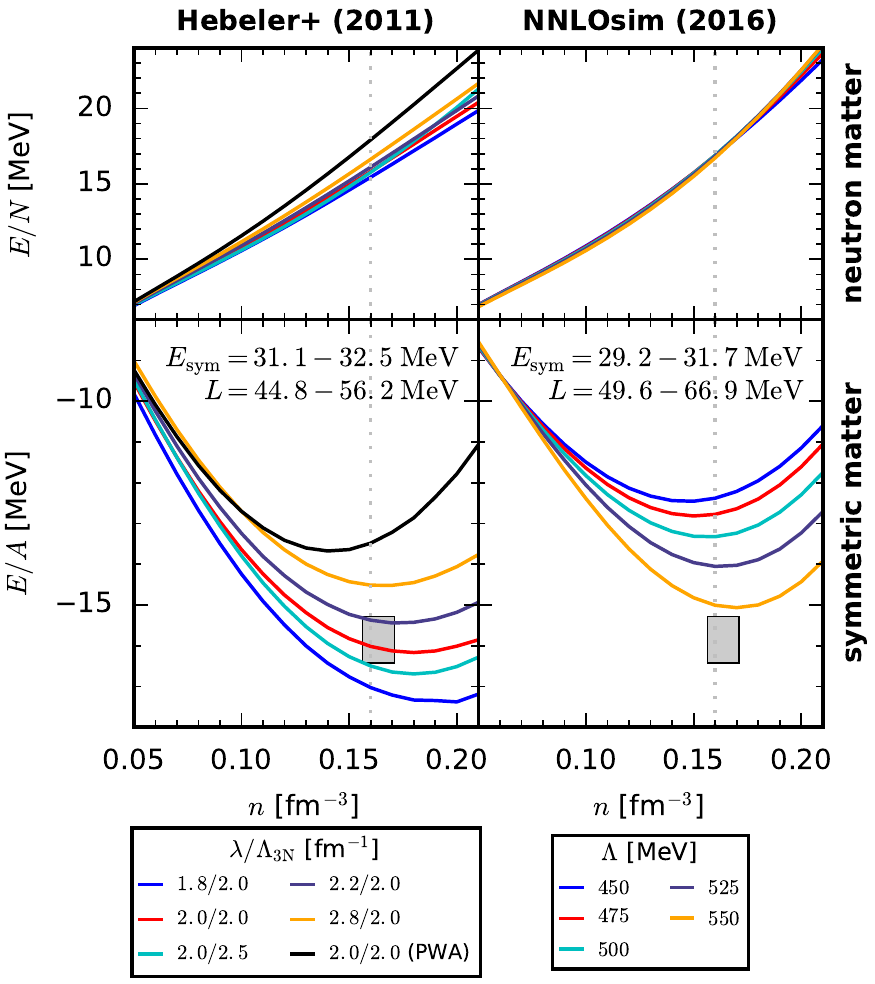}\hspace{5mm}
\end{center}
\caption{Energy per particle of neutron matter (top row) and symmetric nuclear
matter (bottom row) based on the ``Hebeler+''~\cite{Hebe11fits} and
``N$^2$LO$_{\text{sim}}$''~\cite{Carl15sim} NN and 3N interactions (columns),
computed in MBPT including contributions up to 4th order in the many-body
expansion. Results are shown for $\lambda/\Lambda_{\text{3N}}$ for the
interactions of Ref.~\cite{Hebe11fits} (see also
Section~\ref{sec:applications_fits}) and $\Lambda =
\Lambda_{\text{NN,\,3N}} $ for those of Ref.~\cite{Carl15sim}.\\
\textit{Source:} Figure taken from Ref.~\cite{Dris17MCshort}.}
\label{fig:eos_old}
\end{figure}

The contribution at second order in the MBPT expansion including one
three-body interaction (Figure~\ref{fig:MBPT_3N_2nd}) takes the following
form:
\begin{align}
\frac{E^{(2)}_{\text{3N,norm-ord}}}{V} = \frac{1}{2} \sum \limits_{\substack{ijk\\\alpha \beta}} \frac{\braket{ij|V_{\text{NN}}|\alpha \beta} \braket{\alpha \beta k|V_{\text{3N}}|ijk}}{D_{ij\alpha\beta}} \, .
\end{align}

Note that this contribution corresponds to a contribution that includes a
normal-ordered 3N interaction (see Section~\ref{sec:normal_ordering}).
Figure~\ref{fig:MBPT_3N_2nd} also shows an example of contributions from
residual 3N contributions, $E^{(2)}_{\text{3N,res}}$. This diagram is given by
the following expression:
\begin{equation}
\frac{E^{(2)}_{\text{3N,res}}}{V} = \frac{1}{36} \sum \limits_{\substack{ijk\\\alpha \beta \gamma}} \frac{\braket{ijk|V_{\text{3N}}|\alpha \beta \gamma} \braket{\alpha \beta \gamma |V_{\text{3N}}|ijk}}{D_{ijk\alpha\beta\gamma}} \, .
\end{equation}
In Ref.~\cite{Dris17MCshort} all the contributions discussed above, plus all
NN diagrams up to fourth order have been included. This new framework offers
new paths to check the normal-ordering approximation for 3N interactions in
nuclear matter and allows to push MBPT for nuclear matter to much higher
orders and hence allows to estimate uncertainties due to the many-body
expansion in a more systematic way (see also Figure~\ref{fig:coester_fits}).
Of course, the Monte-Carlo evaluation contains inherent statistical
uncertainties, which need to be checked for each diagram.

In Figure~\ref{fig:eos_old} we show the results for the energy per particle in
symmetric nuclear matter and neutron matter based on the interactions of
Ref.~\cite{Hebe11fits} (``Hebeler+'') and the ``N$^2$LO$_{\text{sim}}$''~\cite{Carl15sim} NN
and 3N interactions (see also discussion in Section~\ref{sec:sep_3N_fits}) up
to fourth order in MBPT. For symmetric matter we show the empirical saturation
region by a gray box. We also give results for the symmetry energy
$E_\text{sym} = E/N - E/A$ as well as its slope parameter $L= 3 n_0 \partial_n
E_\text{sym}$ at nuclear saturation density, $n_0=0.16\fmiq$ (dashed vertical line).

Note that the ``Hebeler+'' potentials include NN (N$^3$LO) and 3N forces
(N$^2$LO) up to different orders in the chiral expansion. Despite being fitted
to only few-body data, these interactions are able to reproduce empirical
saturation in Figure~\ref{fig:eos_old} within uncertainties given by the spread
of the individual ``Hebeler+'' interactions~\cite{Hebe11fits}. In addition,
recent calculations of medium to heavy nuclei based on some of these
interactions show remarkable agreement with
experiment~\cite{Hage16NatPhys,Ruiz16Calcium,
Hage16Ni78,Simo17SatFinNuc,Birk17dipole,Morr17Tin} and thus offer new ab
initio possibilities to investigate the nuclear chart. The second column of
Figure~\ref{fig:eos_old} shows results for the ``N$^2$LO$_{\text{sim}}$''
potentials~\cite{Carl15sim} for different cutoff values (see legend). These
interactions were obtained by a simultaneous fit of all low-energy couplings
to two-body and few-body data for $\Lambda_{\text{NN}} = \Lambda_{\text{3N}}$.
We observe a weak cutoff dependence for these potentials in neutron matter
over the entire density range and in symmetric matter up to $n \lesssim 0.08
\fmiq$. At higher densities, the variation of the energy per particle
increases up to $\approx 3 \MeV$ at $n_0=0.16\fmiq$ with a very similar density
dependence. Overall, all the ``N$^2$LO$_{\text{sim}}$'' interactions turn out to be too
repulsive compared to the empirical saturation region.

\clearpage
\section{Applications to nuclei and matter}
\label{sec:applications}

In this chapter we present recent results of ab initio calculations of light
nuclei, medium-mass nuclei as well as dense matter based on state-of-the-art
chiral NN and 3N interactions. The selection is not intended to be exhaustive,
but is rather supposed to illustrate the current status and open issues in
nuclear structure theory. The discussed results cover various observables of
nuclei in different regimes of the nuclear chart and highlight the
capabilities as well as limitations of presently used interactions and
many-body frameworks. The employed interactions include different
regularization schemes (see Section~\ref{sec:3N_regularization}) and fitting
strategies for the LECs of the NN and 3N interactions (see
Section~\ref{sec:3N_fits}).

\subsection{SRG evolution of 3N interactions versus low-resolution fits}
\label{sec:applications_fits}

\begin{table}[b!]
\centering
\begin{tabular}{c|l|c|c|c|c||c|c|c}
& \multicolumn{5}{c||}{NN SRG evolution + 3N fits} & \multicolumn{3}{c}{NN+3N SRG evolution} \\[0.2mm] \hline \hline
$\lambda_{\text{SRG}}$ (fm$^{-1}$) & $\lm_{\rm 3NF}$ (fm$^{-1}$) & $c_D$ & $c_E$ & $r_{^3\text{H}}$ (fm) & $E_{^4\text{He}}$ (MeV) & $E_{^3\text{H}}$ (MeV) & $r_{^3\text{H}}$ (fm) & $E_{^4\text{He}}$ (MeV) \\
\hline
$\infty$ & $\:\: 2.0$ & $+1.5$ & $0.114$ & $1.601$ & $-28.64(4)$ & $-8.482$ & $1.601$ & $-28.64(4)$ \\
$2.8$ & $\:\: 2.0$ \cite{Hebe11fits} & $+1.278$ & $-0.078$ & $1.604$ & $-28.75(2)$ & $-8.482$ & $1.605$ & $-28.72(2)$ \\
$2.6$ & $\:\: 2.0$& $+1.26$ & $-0.099$ & $1.605$ & $-28.77(2)$ & $-8.481$ & $1.606$ & $-28.73(2)$ \\
$2.4$ & $\:\:2.0$& $+1.265$ & $-0.115$ & $1.606$ & $-28.80(2)$ & $-8.481$ & $1.608$ & $-28.73(2)$ \\
$2.2$ & $\:\:2.0$ \cite{Hebe11fits} & $+1.214$ & $-0.137$ & $1.608$ & $-28.86(2)$ & $-8.480$& $1.611$ & $-28.74(2)$ \\
$2.0$ & $\:\:2.0$ \cite{Hebe11fits} & $+1.271$ & $-0.131$ & $1.612$ & $-28.95(2)$ & $-8.479$ & $1.615$ & $-28.75(2)$ \\
$1.8$ & $\:\:2.0$ \cite{Hebe11fits} & $+1.264$ & $-0.120$ & $1.617$ & $-29.11(2)$ & $-8.478$ & $1.622$ & $-28.76(2)$ \\ 
$1.6$ & $\:\:2.0$& $+1.25$ & $-0.075$ & $1.626$ & $-29.42(2)$ & $-8.476$ & $1.635$ & $-28.79(2)$ \\ \hline
\hline
$\infty$ & $\:\:2.5$ & $-1.45$ & $-0.633$ & $1.604$ & $-28.65(4)$ & $-8.482$ & $1.604$ & $-28.65(4)$\\
$2.8$ & $\:\:2.5$ & $-1.35$ & $-0.735$ & $1.606$ & $-28.84(2)$ & $-8.482$ & $1.608$ & $-28.75(2)$ \\
$2.6$ & $\:\:2.5$ & $-1.2$ & $-0.75$ & $1.606$ & $-28.85(2)$ & $-8.482$ & $1.609$ & $-28.76(2)$ \\
$2.4$ & $\:\:2.5$ & $-1.0$ & $-0.725$ & $1.607$ & $-28.89(2)$ & $-8.482$ & $1.610$ & $-28.77(2)$ \\
$2.2$ & $\:\:2.5$ & $-0.7$ & $-0.675$ & $1.609$ & $-28.95(2)$ & $-8.481$ & $1.613$ & $-28.77(2)$ \\
$2.0$ & $\:\:2.5$ \cite{Hebe11fits} & $-0.292$ & $-0.592$ & $1.612$ & $-29.05(2)$ & $-8.481$ & $1.617$ & $-28.77(2)$ \\
$1.8$ & $\:\:2.5$ & $0.05$ & $-0.503$ & $1.617$ & $-29.21(2)$ & $-8.480$ & $1.625$ & $-28.77(2)$ \\
$1.6$ & $\:\:2.5$ & $0.55$ & $-0.353$ & $1.626$ & $-29.48(2)$ & $-8.478$ & $1.638$ & $-28.77(2)$ \\
\end{tabular}
\caption{Results for the $c_D$ and $c_E$ couplings, fit to $E_{^3{\rm H}} 
= -8.482 \mev$ and to the point charge radius $r_{^4{\rm He}} = 1.464 \,
\text{fm}$ (based on Ref.~\cite{Sick08rmsradius4He}) for the NN/3N cutoffs and
the EM $c_i$ values ($c_1 = -0.81$ GeV$^{-1}, c_3 = -3.2$ GeV$^{-1}, c_4 = +
5.4$ GeV$^{-1}$) used, see Ref.~\cite{Hebe11fits} for details. The $^3$H
point charge radius $r_{^3\text{H}}$ is calculated from the charge form factor
solutions of the Faddeev equations and the energies $E_{^4\text{He}}$ are
computed via a Jacobi NCSM harmonic oscillator diagonalization code (credits
to Andreas Ekstr\"om for providing the code). For comparison, the experimental
$^3$H point charge radius is $1.5978 \pm 0.040$~\cite{Ange13rch}. The basis
space truncations $\mathcal{J}_{\text{max}} = \tfrac{7}{2}$ and
$J_{\text{max}} = 5$ have been used for the four-body calculations (see
Section~\ref{sec:general_3N_decomp}). The slight violation of unitarity as
seen in the $^3$H binding energy is mainly due to the treatment of the charge
dependence of the NN interaction in the SRG evolution (see main text and also
discussion in Section~\ref{sec:SRG_applications} for details).}
\label{tab:3Nmagicfits}
\end{table}

Many of the studies discussed in the following sections are based on the NN
plus 3N interactions derived in Ref.~\cite{Hebe11fits}. As already discussed
in Section~\ref{sec:sep_3N_fits}, these interactions consist of NN
interactions evolved to different SRG resolution scales $\lambda_{\text{SRG}}$
plus 3N interactions fitted to the binding energy of $^3$H and the
point-proton radius of $^4$He at each scale~(see also
Table~\ref{tab:3Nmagicfits}). Even though the interactions are only fitted to
NN and few-body observables, the interactions exhibit realistic saturation
properties of symmetric matter (see Figure~\ref{fig:eos_old}). Furthermore,
calculations based on the interaction with
$\lambda_{\text{SRG}}/\Lambda_{\text{3N}} = 1.8/2.0$ (``1.8/2.0 (EM)'') show a
remarkable agreement with experimental binding energies for medium-mass nuclei
(see also Figure~\ref{fig:magic_interactions_matter_nuclei}). In
Table~\ref{tab:3Nmagicfits} we give the specific values of the 3N couplings
$c_D$ and $c_E$ for the different values of the SRG resolution scale
$\lambda_{\text{SRG}}$ and the 3N cutoff scale $\Lambda_{\text{3N}}$. The
listed values include the results published in Ref.~\cite{Hebe11fits} as well
as results for additional resolution scales. The fits at different scales map
out a continuous trajectory for the couplings $c_D$ and $c_E$. We also provide
results for the point charge radius of $^3$H and the binding energy of $^4$He
at the different scales. Given that there exists a correlation between the
ground-state energies of three- and four-body systems (``Tjon
line'')~\cite{Tjon75tjonline,Plat04tjon} we expect that the ground state
energies for $^4$He should not change too much when varying
$\lambda_{\text{SRG}}$, since the binding energy of $^3$H is fixed by
construction in the fit. Still, the observed variation is about $800
\,\text{keV}$ over the full range of scales, while all energies are
slightly overbound compared to the experimental ground-state energy
$E_{\text{gs}} = -28.296 \, \text{MeV}$~\cite{Wang17AME16}. The point charge
radius of $^3$He changes only by about $0.025 \, \text{fm}$ for both values of
$\Lambda_{\text{3N}}$.

\begin{figure}[b!]
\begin{center}
\includegraphics[width=0.6\textwidth]{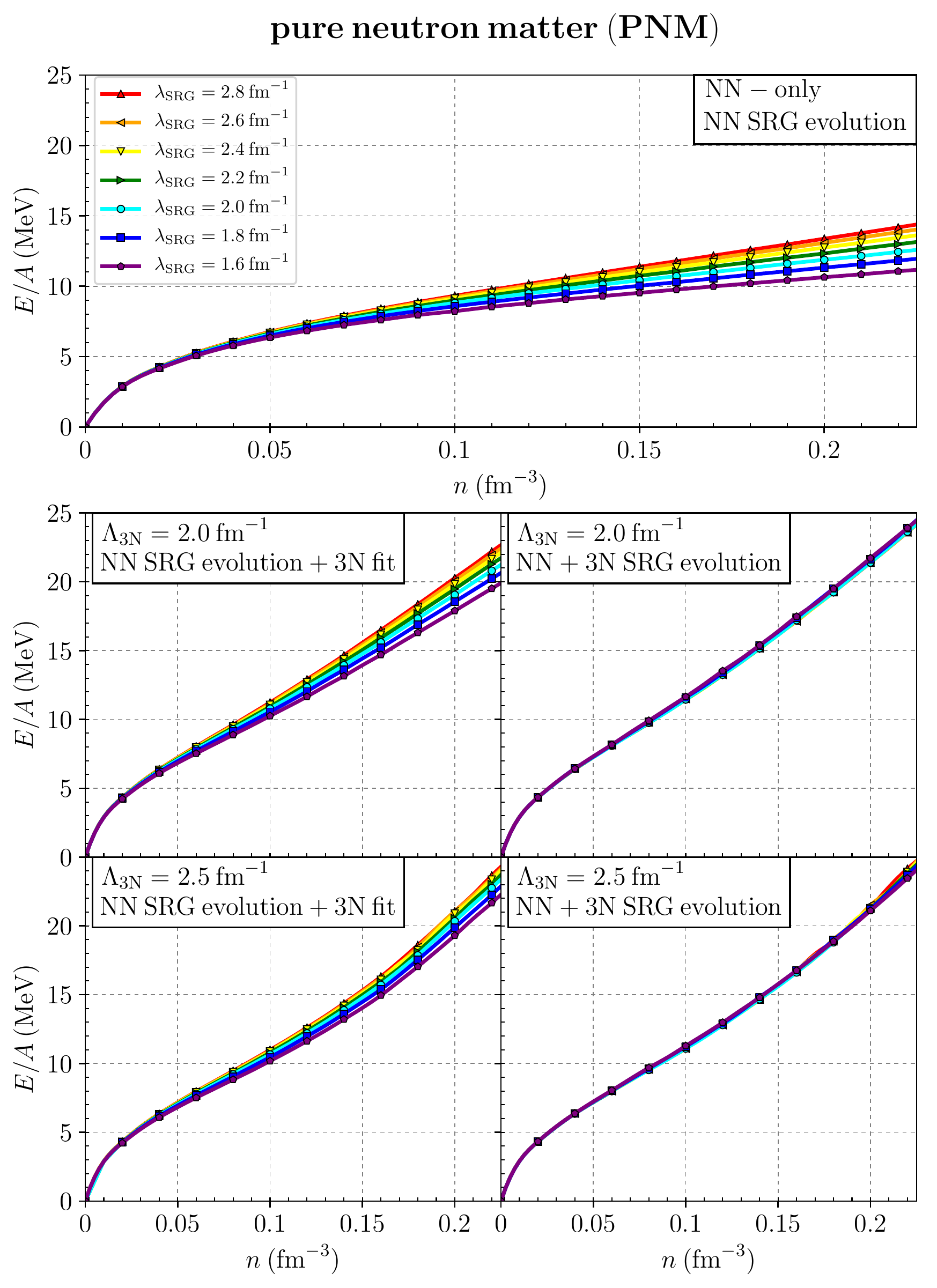}
\end{center}
\caption{The energy per particle of pure neutron matter 
for the interactions specified in Table~\ref{tab:3Nmagicfits}. The plots show
results of MBPT calculations under consideration of all NN and 3N
contributions, including residual terms up to 2nd order. The 3N contributions
at 3rd order are treated in normal-ordering approximation (using
$\mathbf{P}=0$). The top panels show the NN-only results at different
resolution scales, while the lower two rows show the results based on the
interactions defined in the left and right columns of
Table~\ref{tab:3Nmagicfits}.}
\label{fig:eos_srg_PNM}
\end{figure}

For comparison, we present the corresponding results for consistently-evolved
NN+3N interactions, using the SRG framework presented in
Section~\ref{sec:SRG_flow_equations}. The SRG evolution is performed using an
isospin-averaged NN interaction, i.e., the isospin $T=1$ channels are
treated as
\begin{equation}
V_{\text{NN}} = \frac{V_{\text{NN}}^{\text{nn}} + V_{\text{NN}}^{\text{np}} + V_{\text{NN}}^{\text{pp}}}{3} \, ,
\end{equation}
where $V_{\text{NN}}^{\text{nn}}$, $V_{\text{NN}}^{\text{np}}$ and
$V_{\text{NN}}^{\text{pp}}$ represent the neutron-neutron, neutron-proton and
proton-proton interactions, respectively. We note that this approximation
leads to a violation of unitarity for the $^3$H binding energy, which is
determined from the solutions of the Faddeev equations including the proper
treatment of the charge dependence of NN interactions~\cite{Wita89Faddeev}.
For the calculations of nuclear matter (see next section) all Coulomb
interactions are switched off in the SRG evolution, while for the few-body
results in Table~\ref{tab:3Nmagicfits} the Coulomb contributions are included
in $V_{\text{NN}}^{\text{pp}}$ and are evolved consistently. We emphasize that
for the shown results for $r_{^3 \text{H}}$ in Table~\ref{tab:3Nmagicfits} we
did not evolve the radius operator for these calculations. Due to this and due
to the isospin treatment, the radius varies by about $0.03$ fm over the shown
resolution scale range. The energy of $^4$He exhibits a significantly smaller
variation for the consistently-evolved 3N interactions for both cutoff values
$\Lambda_{\text{3N}}$ compared to the low-resolution fits shown in the left
column.

\subsection{Nuclear matter based on consistently SRG-evolved 3N interactions}
\label{sec:applications_matter}

\begin{figure}[t!]
\begin{center}
\includegraphics[width=0.6\textwidth]{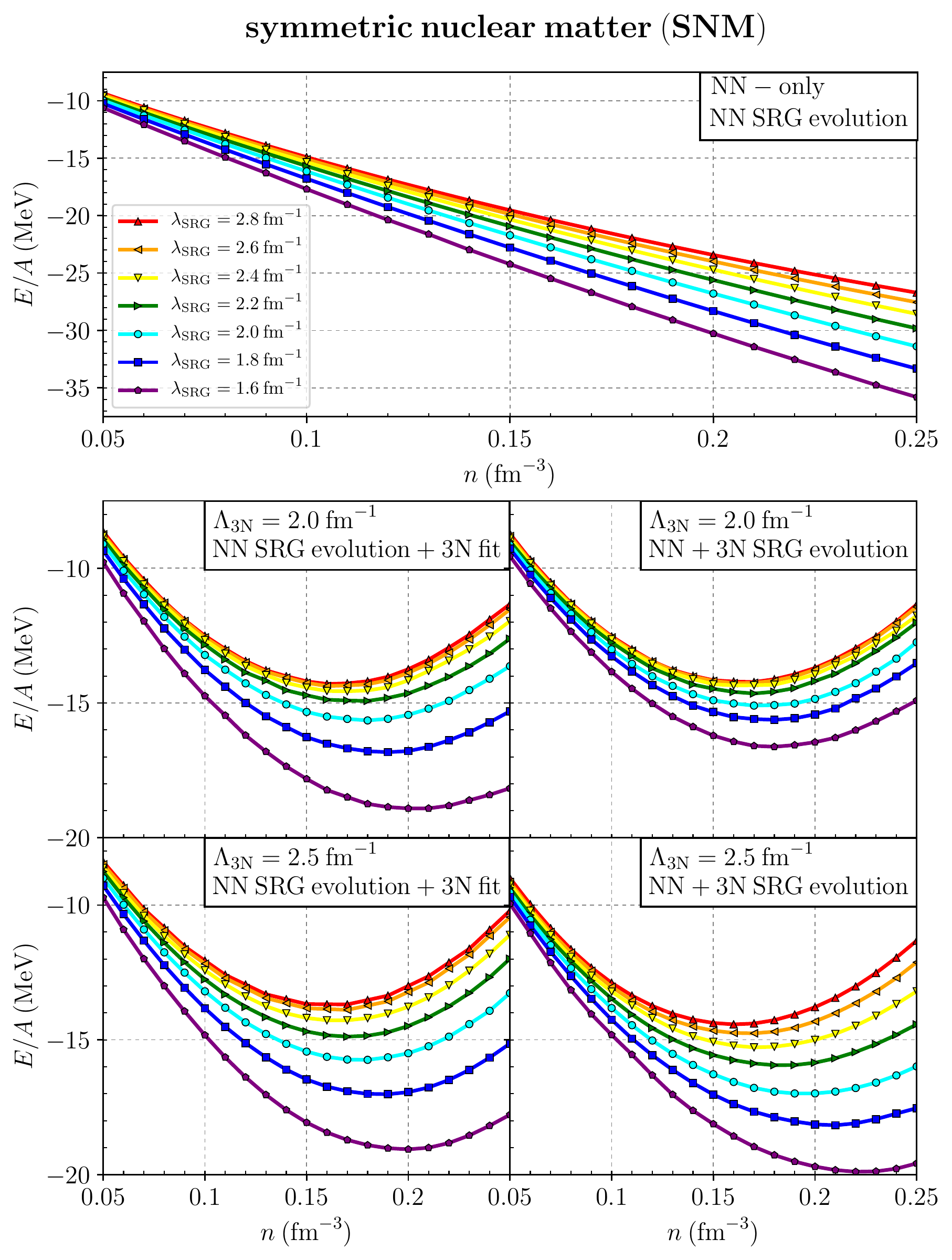}
\end{center}
\caption{The energy per particle of symmetric nuclear matter
for the interactions specified in Table~\ref{tab:3Nmagicfits}. See caption of
Figure~\ref{fig:eos_srg_PNM} for details regarding the many-body calculations
and the shown results.}
\label{fig:eos_srg_SNM}
\end{figure}

The consistent evolution of NN and 3N interactions within the SRG has opened
new avenues that allowed to push the scope of various \textit{ab initio} frameworks for
nuclei to heavier masses (see Sections~\ref{sec:Intro} and~\ref{sec:SRG}). On
the other hand, SRG-evolved NN and 3N forces have not yet been applied to
many-body frameworks for nuclear matter since the SRG evolution of 3N
interactions was always performed in a harmonic oscillator representation.
Thanks to the new developments discussed in Section~\ref{sec:SRG} it is now
possible to perform the SRG evolution in the plane-wave momentum
representation so that a given evolved interaction can now be applied to light
nuclei, medium-mass nuclei as well as nuclear matter. In this section we
present first results for pure neutron matter as well as symmetric nuclear
matter based on consistently-evolved NN plus 3N interactions. To this end, we
start from the set of interactions derived in Ref.~\cite{Hebe11fits} plus the
new fits as specified in Table~\ref{tab:3Nmagicfits}. In particular, we
perform matter calculations based on the interactions ``NN SRG evolution + 3N
fits'' and ``NN+3N SRG evolution'' at the SRG scales in the range
$\lambda_{\text{SRG}} = [1.6, 2.8]$. The many-body calculations are performed
within many-body perturbation theory (MBPT), under consideration of all
contributions from NN and 3N interactions up to 2nd order, including residual
3N contributions (see Section~\ref{sec:normal_ordering}). The third-order
diagrams are taken into account in the normal-ordering approximation setting
$\mathbf{P} = 0$ for the two-body center-of-mass momentum (see
Section~\ref{sect:normal_ordering_matter}). For the evaluation of the diagrams
we use the Monte-Carlo approach discussed in Section~\ref{sec:no_PWD}, while
all NN and 3N partial-wave contributions are resummed for each Monte-Carlo
evaluation. The results are shown in Figures~\ref{fig:eos_srg_PNM} and
~\ref{fig:eos_srg_SNM}. For the discussion of the results it is important to
note the following points:
\begin{itemize}
\item The many-body convergence is usually more rapid for PNM than
for SNM using typical chiral interactions. Specifically, the employed many-body
truncation leads to almost perfectly converged results for PNM, in
particular at small $\lambda_{\text{SRG}}$ for the consistently-evolved
interactions, whereas for SNM higher order terms in the many-body expansion
can still contribute significant contributions.
\item The relative size of 3N contributions compared to NN terms is typically
smaller in PNM than in SNM by a significant amount. This suggests that effects
of induced higher-body forces will typically be more important in SNM than in
PNM.
\end{itemize}

The results for PNM exhibit only a very mild dependence on the SRG resolution
scale for consistently-evolved interactions, whereas for the fitted 3N
interactions we observe a variation of up to about 2 MeV per particle at
saturation density (see left panels of Figure~\ref{fig:eos_srg_PNM}). The SNM
results, on the other hand, sensitively depend on the chosen regularization
scale $\Lambda_{\text{3N}}$, as shown in Figure~\ref{fig:eos_srg_SNM}. While
for $\Lambda_{\text{3N}} =2.0$ fm$^{-1}$ we find a significantly smaller
dependence on $\lambda_{\text{SRG}}$ for the consistently-evolved 3N
interactions compared to the fitted interactions, for $\Lambda_{\text{3N}}
=2.5$ fm$^{-1}$ we find a sizable dependence in both cases. This is an
indication that the strength of induced higher-body contributions increases as
the regularization cutoff scale $\Lambda_{\text{3N}}$ of the initial 3N
interactions gets larger. This trend has also been found in calculations of
atomic nuclei (see, e.g., Refs.~\cite{Roth11SRG,Herg13MR}). On the other hand,
the good invariance of the ground state energy results of $^{4}$He for
consistently-evolved NN and 3N interactions suggests that the contributions of
induced 4N interactions are rather small in light nuclei. Furthermore, it is
remarkable that for $\Lambda_{\text{3N}} = 2.0$ fm$^{-1}$ the rather narrow
uncertainty band includes the phenomenological saturation region. These
findings suggest that it could be worthwhile to apply these
consistently-evolved NN and 3N interactions also to many-body frameworks for
finite nuclei.

\subsection{Ground-state energies of nuclei}
\label{sec:gs}

\begin{figure}[t!]
\begin{center}
\includegraphics[width=0.6\textwidth]{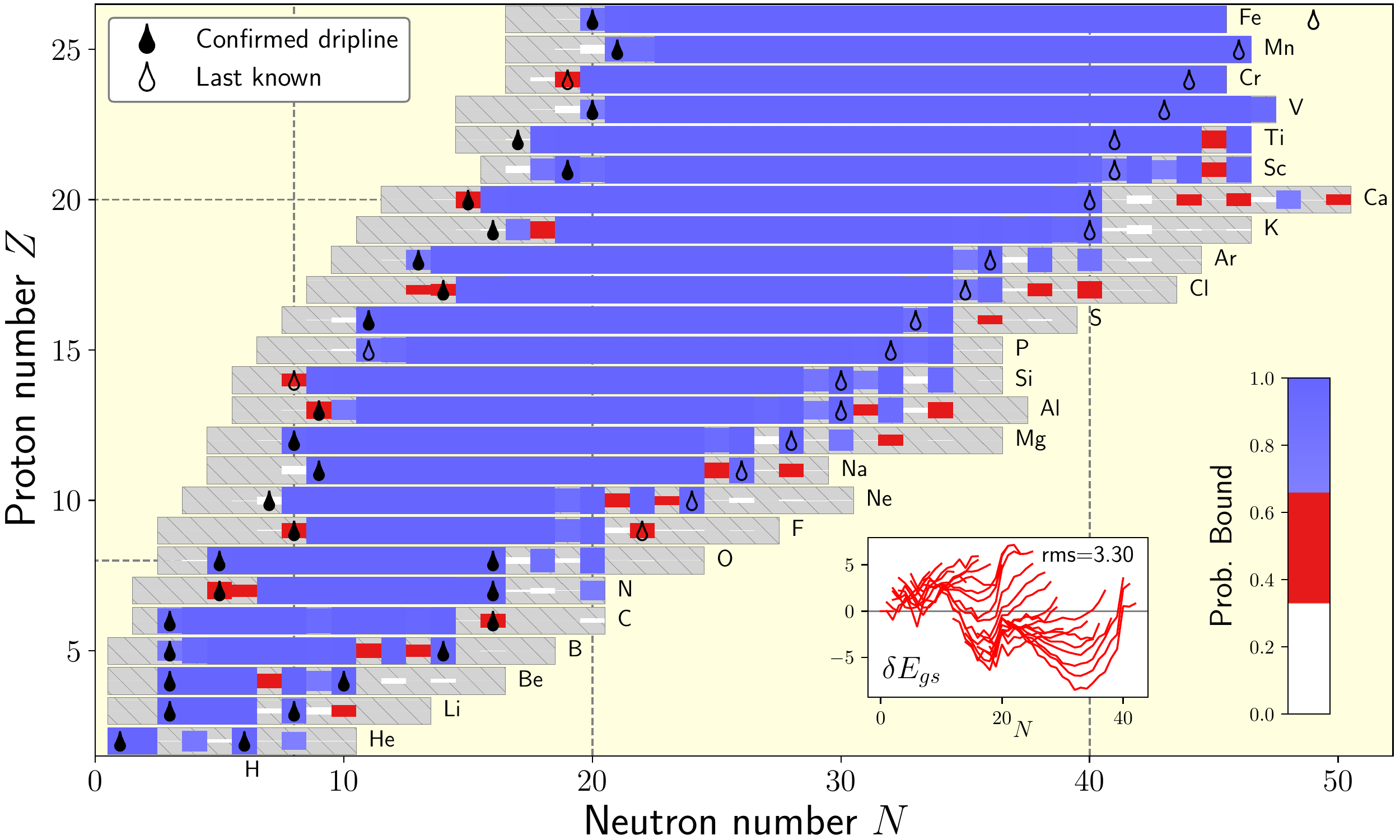}
\end{center}
\caption{Theoretical probabilities of isotopes to be bound with respect
to one- or two-nucleon separation, indicated by the size and color of the
boxes (see legend). The gray region marks all calculated nuclei within the
VS-IM-SRG framework. The inset shows the global agreement with experimental
data.\\
\textit{Source:} Figure taken from Ref.~\cite{Holt19limits}.}
\label{fig:nuclei_stability}
\end{figure}

In Ref.~\cite{Holt19limits} ground-state energies of nuclei from helium to
iron were studied within the VS-IM-SRG framework (see Section~\ref{sec:Intro})
using the interaction ``1.8/2.0 (EM)'' of Ref.~\cite{Hebe11fits} (see previous
section). Within this framework an effective interaction is computed via the
IM-SRG for a given nucleus and is then used as input for a valence-space
diagonalization, which gives access to observables of closed- and open-shell
systems within a given isotopic chain. A particular focus of this work was put
on the location of the drip line, i.e., the point in the nuclear chart beyond
which the nucleons no longer form a bound system. One indication for the drip
line is a negative one- and two-body nucleon separation energy, which involves
the decay of a nucleus via nucleon emission. Predicting the location of the drip
line plays a key role for our understanding of $r$-processes that govern the
synthesis of heavy elements in neutron star
mergers~\cite{Mump15rprocess,LIGO17NSmergers}. Given the excellent agreement
of computed ground-state energies with experimental data in the known regime
of the nuclear chart (see Figure~\ref{fig:nuclei_stability}), these calculations
are expected to provide reasonable theoretical predictions beyond the region
where data exists, including some uncertainty estimates. While this work does
not yet include, e.g., uncertainties from the chiral EFT truncation (see
Section~\ref{sec:EFTtruncation}) or many-body uncertainties from continuum
contributions, this work represents a first step toward improved global
studies of the nuclear chart including a more rigorous estimate of theoretical
uncertainties.

\begin{figure}[t!]
\begin{center}
\includegraphics[width=0.35\textwidth]{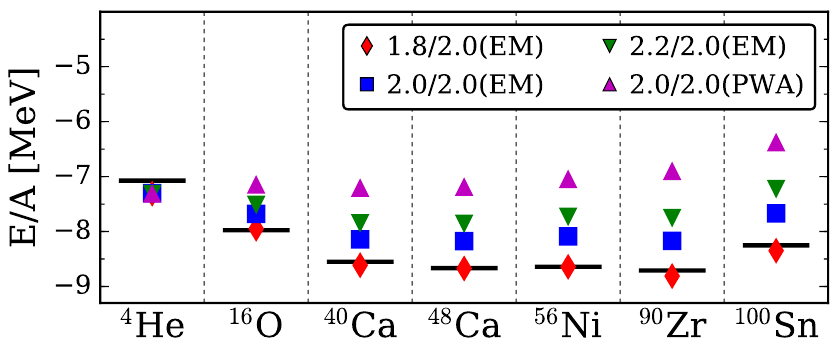}
\hspace{5mm}
\includegraphics[width=0.6\textwidth]{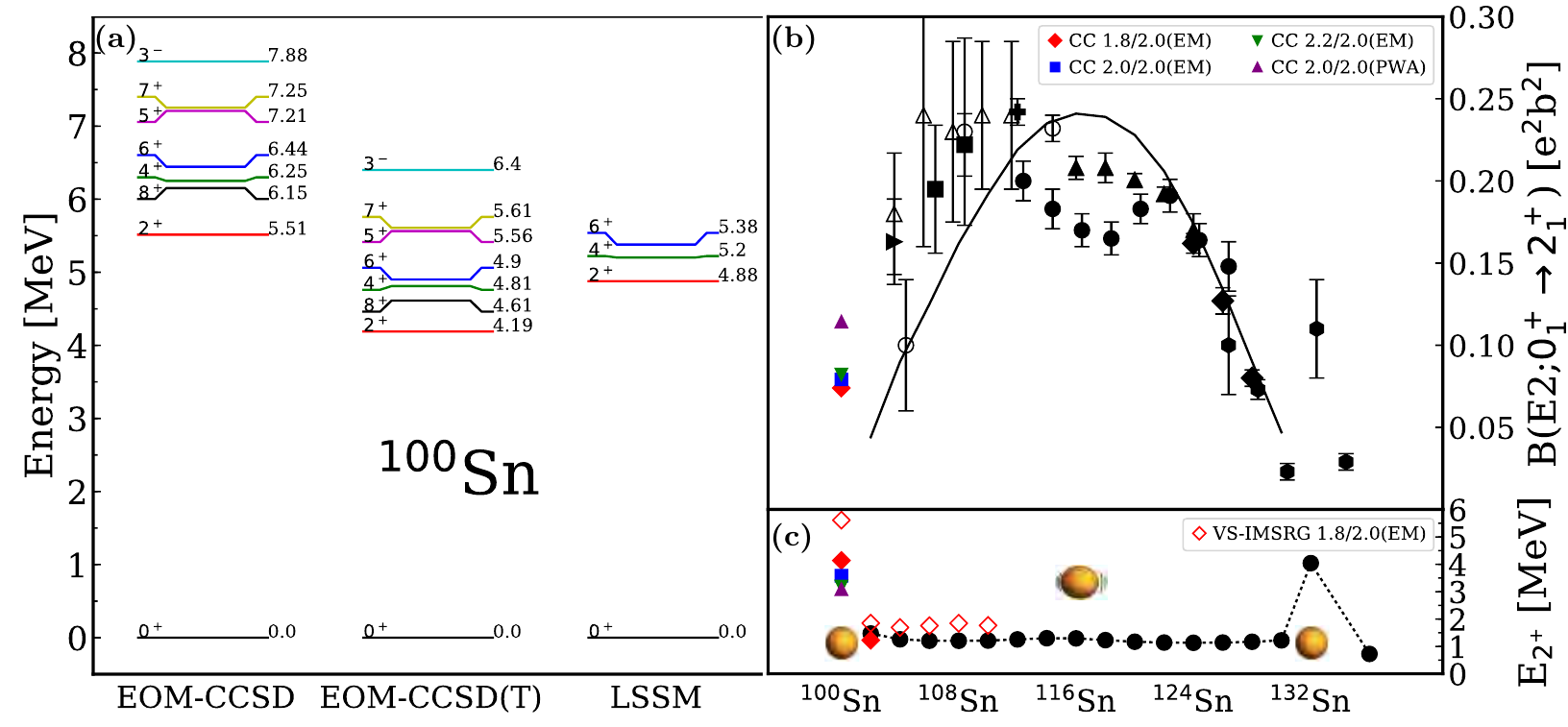}
\end{center}
\caption{Left: Ground-state energies per nucleon of closed-shell nuclei
computed with the IM-SRG based on the interactions of Ref.~\cite{Hebe11fits}
compared to experiment (black lines), Right: Panel (a) shows the
first excited states in $^{100}$Sn based on CC calculations using the interaction
``1.8/2.0 (EM)'' compared to calculations based on phenomenological interactions
(LSSM). Panel (b) shows CC results for the $B(E2)$ transition strength in
$^{100}$Sn compared to the experimental data for all other even tin isotopes,
while panel (c) shows the experimental and theoretical result for the energy of
the $J^{\pi} = 2^+$ states in even tin isotopes.\\
\textit{Source:} Figures taken from Ref.~\cite{Morr17Tin}.}
\label{fig:Sn}
\end{figure}

In Ref.~\cite{Morr17Tin} the structure of nuclei around $^{100}$Sn was studied
based on the same nuclear interactions within CC and VS-IM-SRG (see
Section~\ref{sec:Intro}). The nucleus $^{100}$Sn is currently the heaviest
known doubly-magic nucleus ($N=Z=50$), shows the largest known allowed
$\beta$ decay strength~\cite{Hink13Sn100}, and is located very close to the
proton drip line. However, beyond that not much is known about the structure
of nuclei in this regime of the nuclear chart, in particular regarding the
spectrum of these nuclei~\cite{Faes13Sn100}. These properties make the tin
isotopes a prime target of current theoretical and experimental investigations
and also a natural benchmark system for \textit{ab initio} calculations. The
computation of nuclei in this mass regime poses significant computational
challenges due to the required model spaces for the employed nuclear
interactions as well as the many-body truncation. In fact, for these studies a
new method was developed that allows to include perturbatively higher order
particle-hole excitations in CC calculations. The left panel of
Figure~\ref{fig:Sn} shows the energy per particle of doubly-magic nuclei up to
$^{100}$Sn and highlights once again the remarkable agreement with
experimental data for the interaction ``1.8/2.0 (EM)''. In the right panel
first results for the spectrum of $^{100}$Sn, the $B(E2)$ strength in
comparison with experimental data, and the theoretical and experimental
energies of the excited $J^{\pi} = 2^+$ states for different isotopes are
shown.

The results of Figures~\ref{fig:nuclei_stability} and \ref{fig:Sn} show that
calculations based on specific nuclear NN and 3N interactions are able to
reproduce ground-state energies of nuclei in different regimes of the nuclear
chart remarkably well. However, so far it is not clear why the particular
interaction ``$1.8/2.0$ (EM)'' performs so well for energies, whereas
Hamiltonians with slightly different cutoffs of the same interaction set lead
to a significant underbinding of heavier nuclei (see left panel of
Figures~\ref{fig:magic_interactions_matter_nuclei} and ~\ref{fig:Sn}). In
addition, other observables like radii turn out to be too small compared to
experiment for all these interactions (see, e.g., the right panel of
Figure~\ref{fig:magic_interactions_matter_nuclei}). In general, calculations
should ideally be performed based on a set of interactions at different orders
of the chiral expansion and also for a range of cutoff scales rather than just
for a single specific interaction. Such calculations at different orders in
the chiral expansion allow to extract systematic uncertainty bands (see
Section~\ref{sec:EFTtruncation}) and to rule out accidental agreement of
particular interactions for specific observables.

\begin{figure}[t!]
\begin{center}
\includegraphics[width=0.51\textwidth]{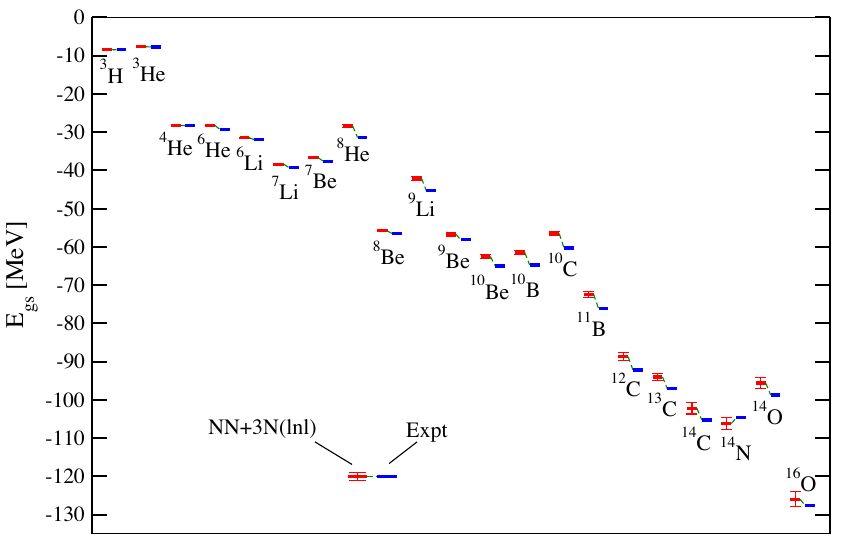}
\hspace{5mm}
\includegraphics[width=0.43\textwidth]{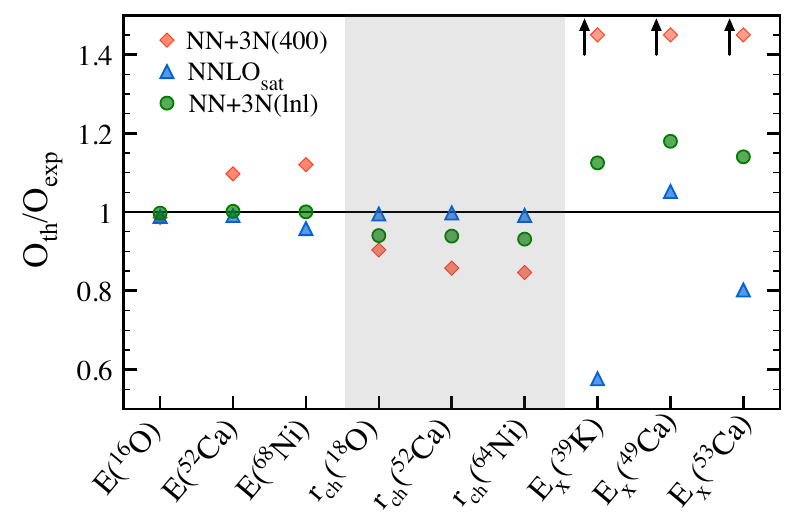}
\end{center}
\caption{Left: Ground-state energies of nuclei calculated within the NCSM based on the 
``NN+3N (lnl)'' interaction (red lines) compared to experiment (blue lines).
The interaction has been SRG-evolved to $\lambda_{\text{SRG}} = 2.0$
fm$^{-1}$. The shown error bars indicate many-body uncertainties. Right: The
ratio of theoretical and experimental results for the ground-state energies
$E$, the charge radius $r_{\text{ch}}$ and one-nucleon separation energies
$E_x$ obtained within the SCGF framework.\\
\textit{Source:} Figures taken from Ref.~\cite{Soma19novelHam}.}
\label{fig:lnl_3N_results}
\end{figure}

\begin{figure}[b!]
\begin{center}
\includegraphics[width=0.46\textwidth]{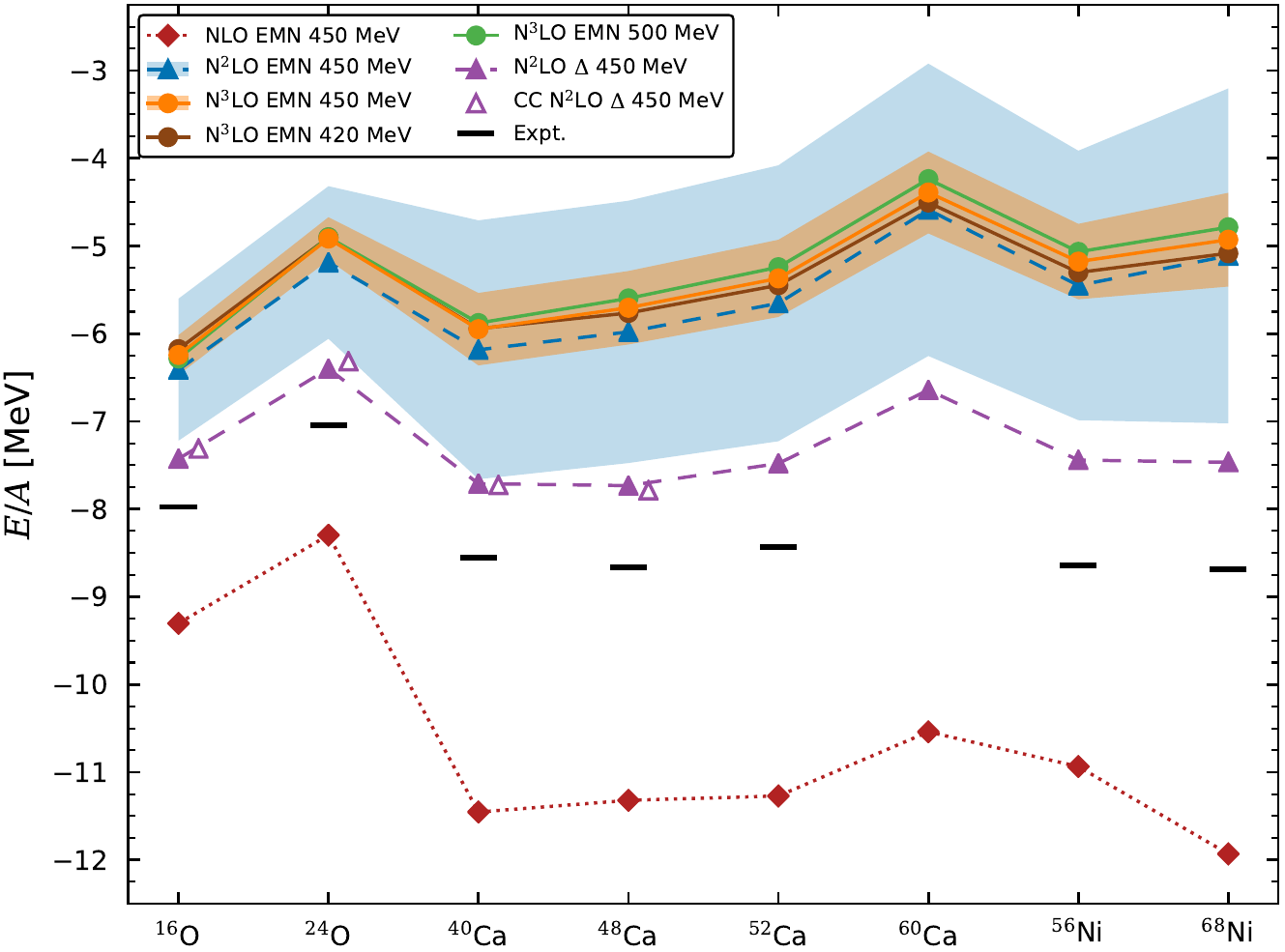}
\hspace{5mm}
\includegraphics[width=0.46\textwidth]{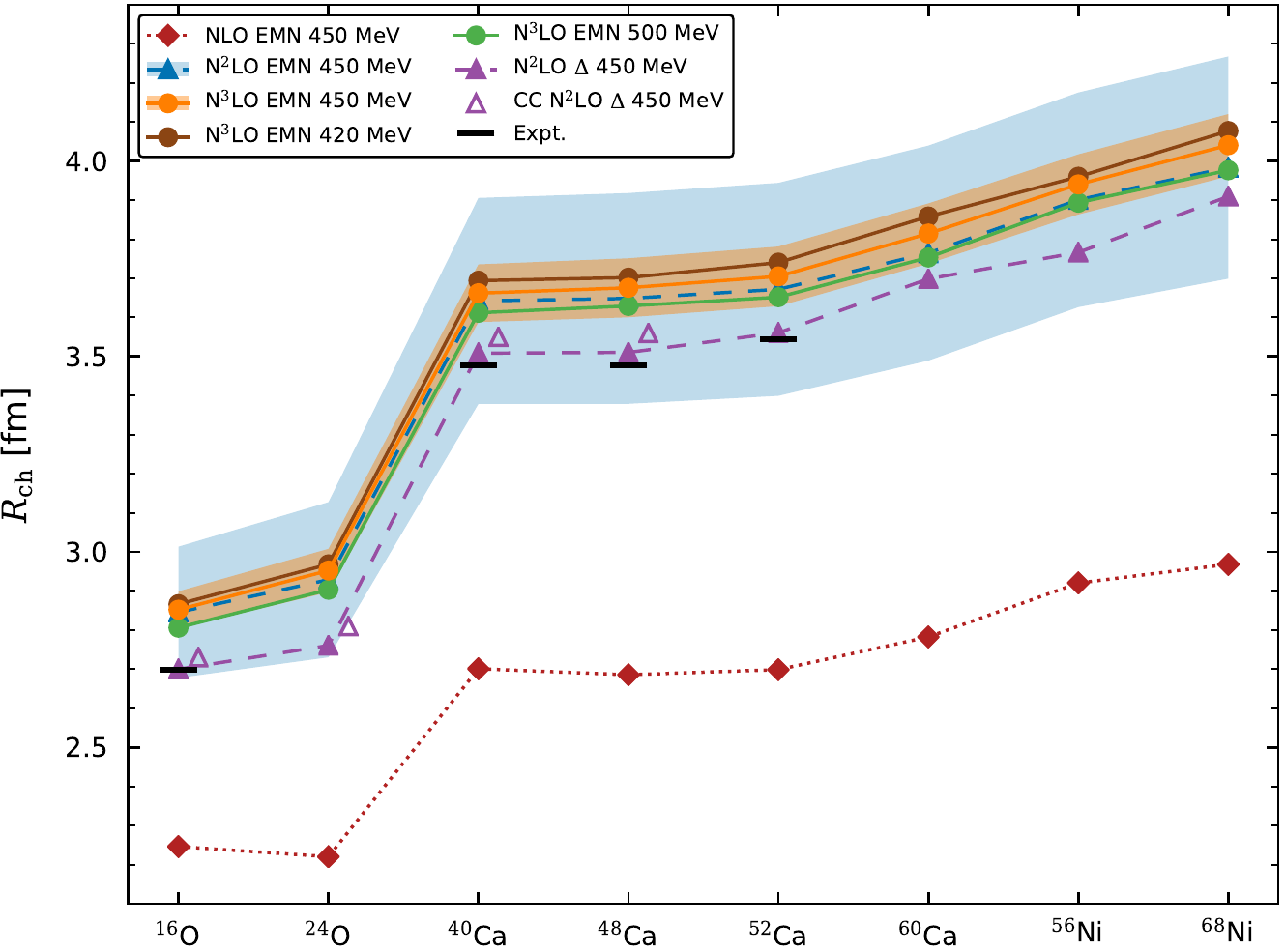}
\end{center}
\caption{Ground-state energies per nucleon (left) and charge radii (right) of
selected closed-shell nuclei. Results are shown at N$^3$LO for the EMN
potential of Ref.~\cite{Ente17EMn4lo} with cutoffs $\Lambda = 420, 450$, and
$500$ MeV depicted by the brown, orange, and green solid lines and circles,
respectively. The N$^2$LO results are given by the dashed lines for the EMN
450 MeV potential (blue line and solid up triangles) and the $\Delta$-full
interaction of Ref.~\cite{Ekst17deltasat} (purple line and solid up
triangles), while NLO results are displayed by the red-dotted line and
diamonds. The open triangles give the CC results for the $\Delta$-full
interaction from Ref.~\cite{Ekst17deltasat} for comparison. The blue and
orange bands give the N$^2$LO and N$^3$LO uncertainty estimate, respectively,
for the EMN 450 MeV interaction.\\
\textit{Source:} Figures taken from
Ref.~\cite{Hopp19medmass}.}
\label{fig:EMN_3N_results}
\end{figure}

In Ref.~\cite{Soma19novelHam} a new interaction, ``NN+3N (lnl)'' was
introduced. This interaction includes NN contributions up to N$^3$LO given by
the potential of Ref.~\cite{Ente03EMN3LO} and 3N contributions up to N$^2$LO.
Triggered by the deficiencies of the local 3N interaction derived in
Ref.~\cite{Navr07local3N} (see e.g. Ref.~\cite{Bind14CCheavy} and
Figure~\ref{fig:Tichai_compare}) an additional nonlocal regulator was
introduced, leading to a significantly improved description of medium-mass
nuclei beyond the oxygen isotopic chain, while the good agreement with
experiment for light systems was maintained by increasing the value of the
local cutoff scale to $\Lambda = 600$ MeV. The left panel of
Figure~\ref{fig:lnl_3N_results} illustrates the agreement for ground-state
energies of various s-shell and p-shell nuclei. The right panel shows the
results of SCGF calculations for different nuclei and observables (see
caption). Particularly striking is the improvement of the results based on the
novel ``NN+3N (lnl)'' interaction compared to the purely local 3N interaction
``NN+3N (400)'' for the description of heavier nuclei.

\begin{figure}[t!]
\begin{center}
\includegraphics[width=0.51\textwidth]{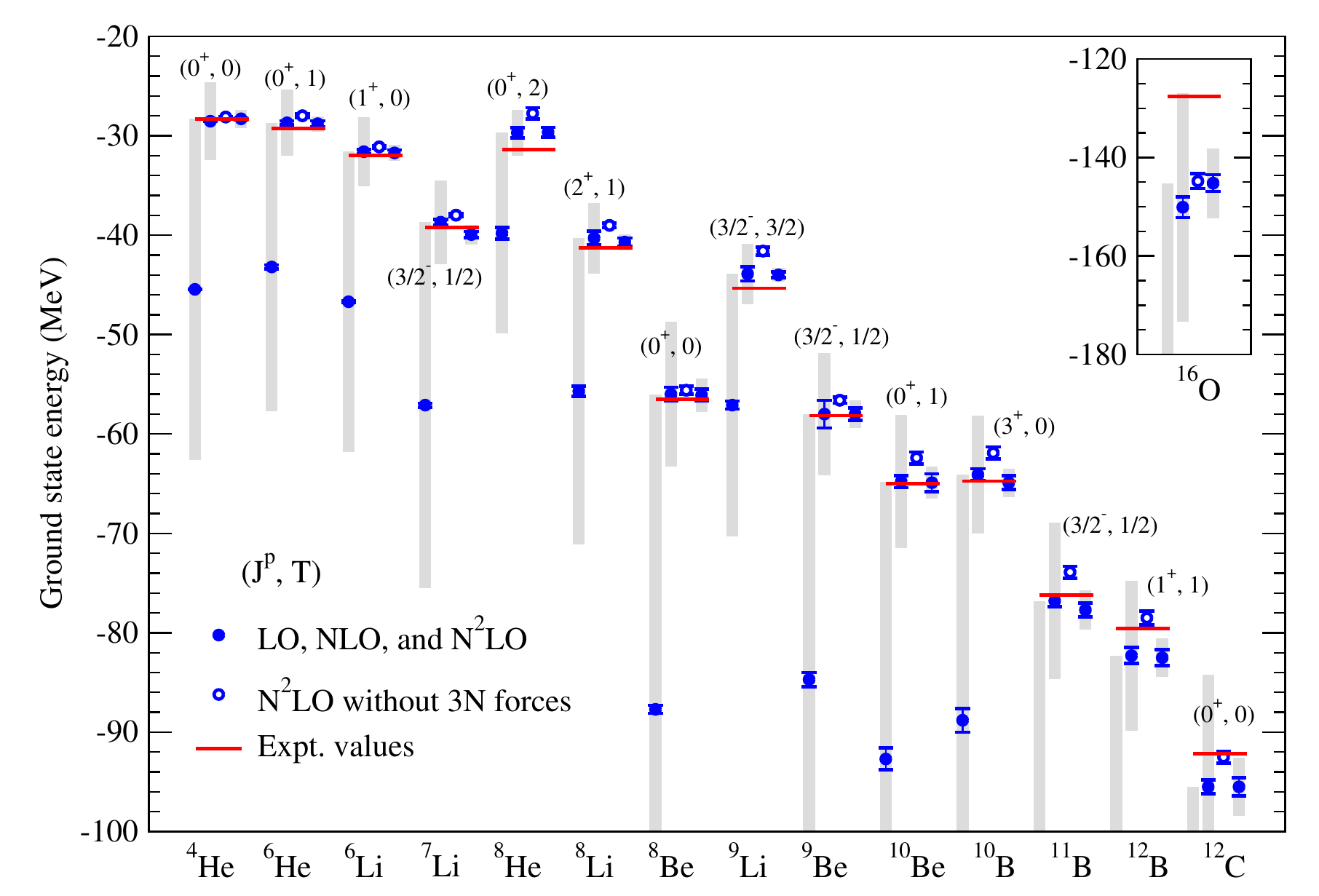}
\includegraphics[width=0.48\textwidth]{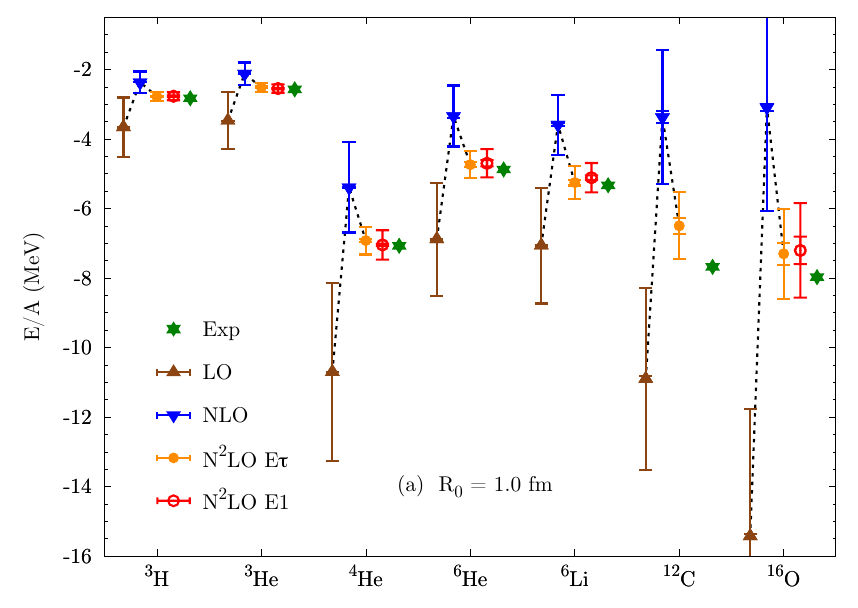}
\end{center}
\caption{Left: Ground-state energies of nuclei using semilocal chiral LO, NLO, and
N$^2$LO interactions of Refs.~\cite{Epel15improved,Bind15Fewbody,Epel18SCS3N}
at $R = 1.0$ fm (blue symbols) in comparison with experimental values (red
lines). For each nucleus the LO, NLO and N$^2$LO results are the left, middle
and right symbols and bars, respectively. The open blue symbols correspond to
incomplete calculations at N$^2$LO using NN-only interactions. Blue error bars
indicate the many-body extrapolation. The shaded bars indicate the estimated
truncation error at each chiral order following Ref.~\cite{Epel15improved}.
Right: Ground-state energies of
nuclei obtained from auxiliary-diffusion Monte-Carlo (AFDMC) results using the
local chiral interaction of Ref.~\cite{Lynn16QMC3N}. Results at different
orders of the chiral expansion and for different 3N parametrizations are
shown. Smaller error bars (barely visible up to $A = 6$) indicate the
statistical Monte Carlo uncertainties, whereas the larger error bars represent
the uncertainties from the chiral EFT truncation following
Ref.~\cite{Epel15improved}.\\
\textit{Source:} Left figure taken from Ref.~\cite{Epel18SCS3N} and right figure taken from Ref.~\cite{Lona18mediummass}.}
\label{fig:LENPIC_QMC}
\end{figure}

In Ref.~\cite{Hopp19medmass} ground-state energies and charge radii of
closed-shell medium-mass nuclei were computed in IM-SRG based on a set of
chiral NN and 3N interactions at different orders in the chiral expansion,
with a particular focus on exploring the connection between properties of
finite nuclei and nuclear matter. Specifically, the calculations were
performed using chiral interactions at NLO, N$^2$LO and
N$^3$LO~\cite{Ente17EMn4lo}, where the 3N interactions at N$^2$LO and N$^3$LO
were fit to the empirical saturation point of nuclear matter and to the $^3$H
binding energy~\cite{Dris17MCshort} (see Figures~\ref{fig:cd_ce} and
\ref{fig:coester_fits}). It is found that the results for energies and radii
of closed-shell systems at N$^2$LO and N$^3$LO overlap within uncertainties
(see Figure~\ref{fig:EMN_3N_results}) and the cutoff variation of the
interactions is within the EFT uncertainty band, which has been determined
following the prescription of Ref.~\cite{Epel15improved} (see also
Section~\ref{sec:EFTtruncation}). Overall, the ground-state energies are found
to be underbound compared to experiment, as expected from the comparison to
the empirical saturation point (see Figure~\ref{fig:coester_fits}), while the
charge radii are systematically too large.

In the left panel of Figure~\ref{fig:LENPIC_QMC} results of NCSM calculations
are shown for the ground-state energies of various nuclei up to $^{16}$O using
the semilocal NN and 3N interactions of
Refs.~\cite{Bind15Fewbody,Epel18SCS3N}. Also indicated are the chiral
truncation error estimates for these ground-state energies following
Refs.~\cite{Epel15improved}. For most of the 15 nuclei the complete results at
N$^2$LO agree with the experimental values. It is interesting to note that the
effect of the 3N interactions is noticeably larger for $^8$He and $^9$Li than
for $^8$Be and $^9$Be.  On the other hand, $^{16}$O is noticeably overbound at
N$^2$LO, with or without 3NFs, see also Ref.~\cite{Lahd13LEFT} for a related
discussion in the context of nuclear lattice simulations. This overbinding
starts at $A=12$, where, with 3N interactions, both $^{12}$B and $^{12}$C are
overbound, with the experimental value only slightly outside the chiral
truncation error estimate, and seems to be systematic for the heavier nuclei.
In the right panel of Figure~\ref{fig:LENPIC_QMC} results for ground-state
energies of AFDMC calculations are shown for nuclei up to $^{16}$O at
different orders in the chiral expansion and different parametrizations for
3N interactions, see Refs.~\cite{Lynn16QMC3N,Lona18mediummass}.  The shown
error bars include both the Monte Carlo many-body uncertainties and the
uncertainties from the truncation of the chiral expansion. For the shown
interactions at the regulator scale $R_0 = 1.0$ fm , the computed binding
energies at N$^2$LO are in good agreement with experiment. For softer
interactions ($R_0=1.2$, not shown) the agreement is also reasonable, while
the size of the uncertainties become sizable for $^{16}$O (see
Ref.~\cite{Lona18mediummass}).

In summary, the results discussed above illustrate that there are now various
different interactions available which provide ground-state energies
consistent with experimental data within theoretical uncertainties for nuclei
in a restricted sector of the nuclear chart based on order-by-order
calculations. For calculations based on specific interactions at a given
order, like, e.g., the ``1.8/2.0 (EM)'' interaction~\cite{Hebe11fits} or the
``N$^2$LO$_{\text{sat}}$'' interaction~\cite{Ekst15sat} (see
Figure~\ref{fig:N2LO_sat_nuclei}) it is even possible to reproduce known
ground-state energies over a significant part of the nuclear chart from light
systems to medium-mass nuclei up to $A\approx100$.

\begin{figure}[t!]
\begin{center}
\includegraphics[width=0.45\textwidth]{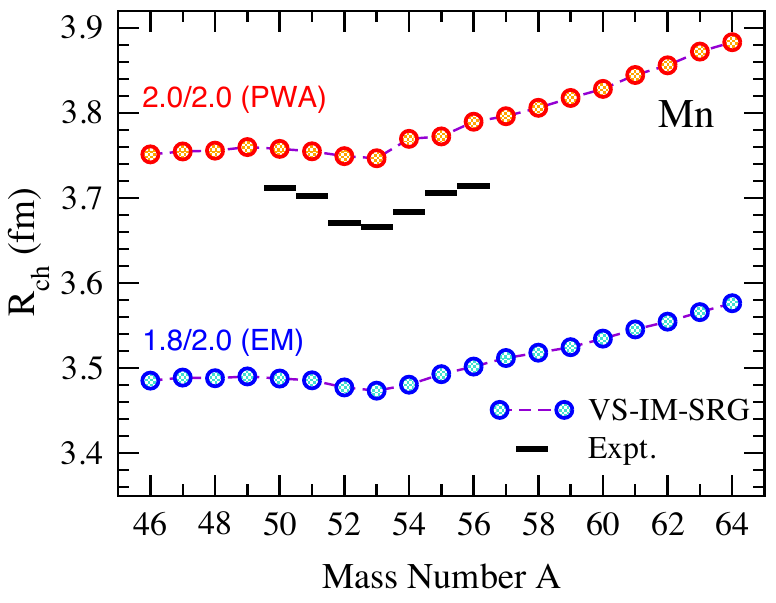}
\hspace{5mm}
\includegraphics[width=0.47\textwidth]{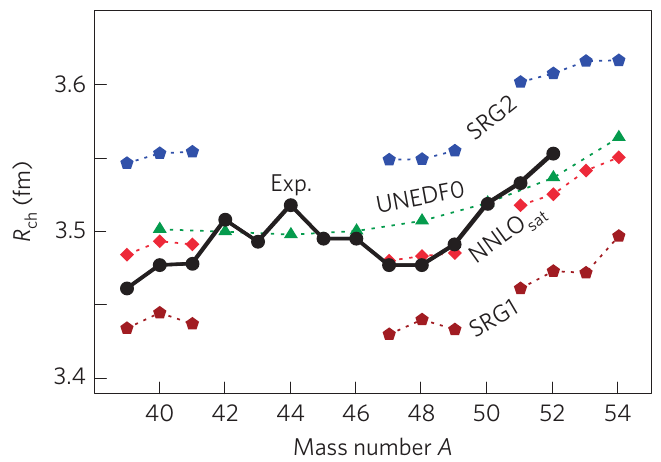}
\end{center}
\caption{
Left: Charge radii of manganese isotopes for the ``1.8/2.0 (EM)'' (blue circles)
and ``2.0/2.0 (PWA)'' (red circles) interactions of Ref.~\cite{Hebe11fits}
computed within the VS-IM-SRG. 
Right: Experimental charge radii of calcium isotopes compared to \textit{ab initio}
CC calculations with chiral EFT interactions ``N$^2$LO$_\text{sat}$''~\cite{Ekst15sat} and
``SRG1/SRG2''~\cite{Hebe11fits}, as well as DFT calculations with the UNEDF0
functional. Experimental error bars are smaller than the symbols. Note that the interaction labeled ``SRG1'' in
the right panel corresponds to the interaction usually labeled as ``2.8/2.0
(EM)'' and ``SRG2'' to ``2.0/2.0 (PWA)''.\\
\textit{Source:} Left figure taken from Ref.~\cite{Simo17SatFinNuc} and right figure taken
from Ref.~\cite{Ruiz16Calcium}.}
\label{fig:radius_magic}
\end{figure}

\subsection{Charge radii of nuclei}
\label{sec:radius_neutron_rich}

The goal of \textit{ab initio} calculations is to obtain a comprehensive understanding
of the structure of nuclei. This includes ground-state energies, as discussed
in the previous section, as well as various additional observables, like
charge and matter radii, quadrupole moments or electromagnetic response
functions. A simultaneous and realistic prediction of several or ideally all
of these quantities is a very challenging task~(see, e.g.,
Ref.~\cite{Lapo16radiiO}). As shown in Figures~\ref{fig:fits_Deltafull},
~\ref{fig:N2LO_sat_nuclei} and~~\ref{fig:EMN_3N_results} the interactions
``N$^2$LO$_{\text{sat}}$''~\cite{Ekst15sat} and
``N2LO-$\Delta$''~\cite{Ekst17deltasat} are able to provide ground state
energies and charge radii in good agreement with experiment (see also
Ref.~\cite{Maas19radii} for a recent study), while most interactions of
Ref.~\cite{Hebe11fits} give too small radii, in particular the interaction
``1.8/2.0 (EM)'' (see right panel of
Figure~\ref{fig:magic_interactions_matter_nuclei} and left panel of
Figure~\ref{fig:radius_magic}), which on the other hand provides an excellent
description of ground-state energies (see Section~\ref{sec:gs}). In
Figure~\ref{fig:radius_magic} we show some representative results for charge
radii based on the interactions of Ref.~\cite{Hebe11fits} for manganese
isotopes computed within the VS-IM-SRG (left) and calcium isotopes computed
within the coupled-cluster framework (right). Apart from the observed overall
shifts compared to experiment, the reproduction of the trends within a
isotopic chain like the steep increase in the radius toward neutron rich
systems remains a challenge for \textit{ab initio} nuclear theory~\cite{Ruiz16Calcium}.
Also shown in the right panel are nuclear DFT results obtained with the Skyrme
energy density functional UNEDF0~\cite{Kort14UNEDF2}, which fails to describe
the details of the experimental trend within the shown isotopic chain.

\begin{figure}[t!]
\begin{center}
\includegraphics[width=0.50\textwidth]{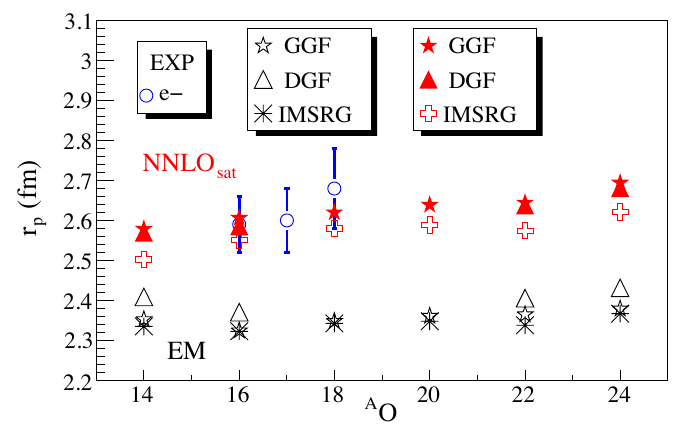}
\hspace{5mm}
\includegraphics[width=0.45\textwidth]{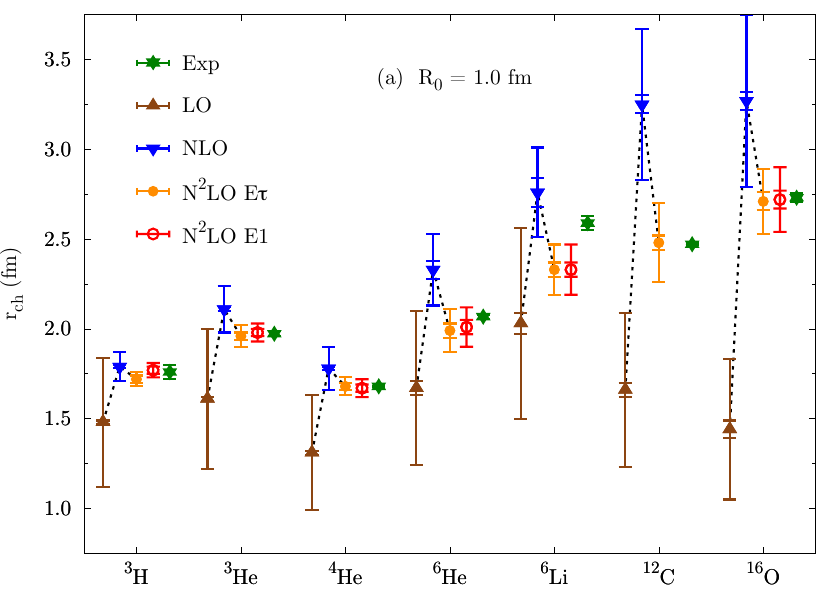}
\end{center}
\caption{Left: Proton radius of oxygen isotopes calculated with the IM-SRG and SCGF
based on the EM interaction~\cite{Ente03EMN3LO} and ``N$^2$LO$_{\text{sat}}$''
\cite{Ekst15sat}. Right: Charge
radii of nuclei obtained from auxiliary-field diffusion Monte-Carlo (AFDMC)
using the local chiral interaction of Ref.~\cite{Lynn16QMC3N}. Results at
different orders of the chiral expansion and for different 3N parametrizations
are shown (compare Figure~\ref{fig:LENPIC_QMC}).\\
\textit{Source:} Left figure taken from Ref.~\cite{Lapo16radiiO} and right figure taken from
Ref.~\cite{Lona18mediummass}. }
\label{fig:radius_QMC_O}
\end{figure}

In the left panel of Figure~\ref{fig:radius_QMC_O} we show charge radius
results for oxygen isotopes computed within IM-SRG, the Dyson-SCGF framework
(DGF)~\cite{Dick04PPNP} and the Gorkov-SCGF (GGF)~\cite{Soma11GGFform} using
the ``N$^2$LO$_{\text{sat}}$''~\cite{Ekst15sat} and the NN interaction of
Ref.~\cite{Ente03EMN3LO} plus 3N interactions~\cite{Lapo16radiiO}. Again, a
remarkable agreement between results obtained within different many-body
frameworks is observed (see discussion in Section~\ref{sec:Intro}). For a
given interaction, the uncertainties due to the many-body calculations is
smaller than the experimental uncertainties and the uncertainty coming from
the use of different interactions. On the other hand, the interaction EM
significantly underestimates the experimental radii, whereas the absolute
radii are well reproduced by ``N$^2$LO$_{\text{sat}}$''. However, we note that
the charge radius of $^{16}$O has been included in the fit of the interaction
(see Section~\ref{sec:3N_fits}). In the right panel we present the charge
radii at different orders of the chiral expansion and parametrizations of the
3N force for different nuclei up to $^{16}$O based on AFDMC
calculations~\cite{Lona18mediummass}. We emphasize that the employed local
interactions in these calculations differ quite significantly from the
interactions shown in Figure~\ref{fig:radius_magic} in terms of the softness
and the nature of the regulator. Overall, a good agreement with experiment is
found for the interaction at $R_0 = 1.0$ fm with a natural order-by-order
convergence. For a softer interaction at $R_0 = 1.2$ fm (not shown, see
Ref.~\cite{Lona18mediummass}), the results resemble those of the hard
potential up to $A = 6$, while the agreement slightly deteriorates for heavier
system and shows a stronger sensitivity to the 3N parametrization.

\begin{figure}[b!]
\begin{center}
\includegraphics[width=0.36\textwidth]{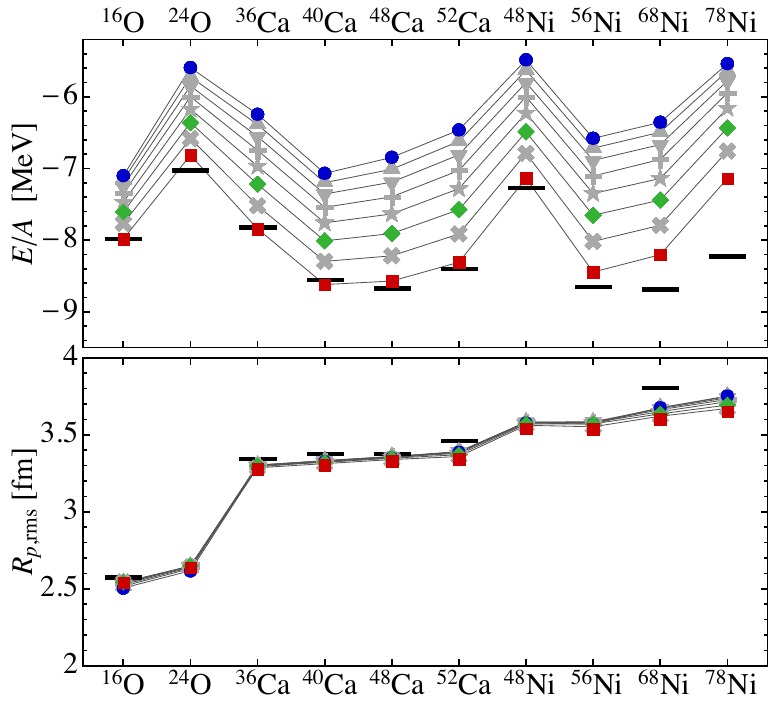}
\hspace{5mm}
\includegraphics[width=0.58\textwidth]{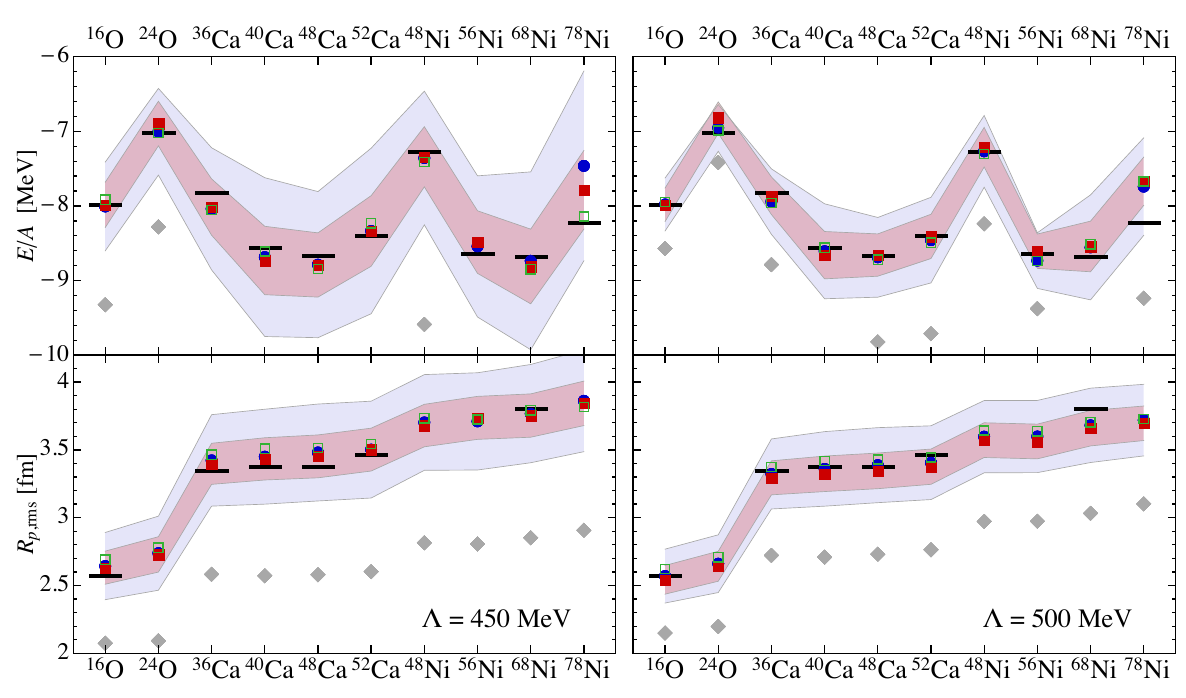}
\end{center}
\caption{Left: Ground-state energies and point-proton radii for selected
medium-mass isotopes obtained in IM-SRG based on NN+3N interaction at N$^3$LO
with $\Lambda = 500$ MeV for a range of $c_D$ parameters from $-3$ (blue) to
$+4$ (red) in steps of 1. Right: Ground-state energies (top panels) and
point-proton radii (bottom panels) based on interactions at NLO (solid
gray diamonds), N$^2$LO (blue circles), N$^3$LO (red boxes), and N$^3$LO'
(open green boxes) with $\Lambda = 450$ MeV (left) and $500$ MeV (center).
Here, N$^3$LO' refers to calculations using NN interactions at N$^3$LO
combined with 3N interactions at N$^2$LO. The uncertainty bands at N$^2$LO
(blue) and N$^3$LO (red) are obtained from the order-by-order analysis first
suggested in Ref.~\cite{Epel15improved} (see also
Section~\ref{sec:EFTtruncation}) and also include many-body
uncertainties.\\
\textit{Source:} Figures taken from Ref.~\cite{Huet19EMN3N}, right figure
modified.
}
\label{fig:radius_EMN_3N}
\end{figure}

In Figure~\ref{fig:radius_EMN_3N} we show ground-state energies and charge
radii, obtained from IM-SRG calculations performed in Ref.~\cite{Huet19EMN3N},
based on the NN interactions of Ref.~\cite{Ente17EMn4lo} plus 3N interactions
up to N$^3$LO. For these calculations the 3N interactions have been
regularized nonlocally (see Section~\ref{sec:nonlocal_momentum}) and the value
of the coupling $c_E$ as a function of $c_D$ is fixed by a fit to the $^3$H
ground state energy, i.e. following the same strategy like in
Ref.~\cite{Hopp19medmass} (see also Figure~\ref{fig:EMN_3N_results} and
related discussion). The left figure illustrates the dependence of the results
on $c_D$ (see also Figure~\ref{fig:matter_nuclei}). Remarkably, while the
ground-state energies depend quite sensitively on the value of the coupling,
the charge radius results remain almost invariant and are in very good
agreement with experimental data. This offers the opportunity to fit $c_D$ to
the ground state energies without sacrificing the agreement for the radii. In
this work, the fit was performed using $^{16}$O, which results in $c_D = +4$
for the shown cutoff scale. The same trends were found for the other studied
cutoff scales $\Lambda = 400$ and $550$ MeV. Note that the LEC values differ
quite significantly from those obtained from fits to empirical nuclear matter
properties based on the same interactions~\cite{Dris17MCshort,Hopp19medmass},
which in turn lead to a significant underbinding of nuclei. The right figure
shows the resulting uncertainty bands for the energies and radii of selected
medium-mass nuclei. Obviously, the calculations can simultaneously reproduce
both observables from p-shell nuclei up to the nickel isotopic chain and
resolves several deficiencies of previous interactions. While this new family
of interactions certainly offers new perspectives for \textit{ab initio}
studies of medium-mass and heavier nuclei including estimates of theoretical
uncertainties and also demonstrates that it is possible to derive NN and 3N
interactions that can correctly describe NN scattering phase shifts as well as
different many-body observables over a wide range of the nuclear chart, it
still exhibits deficiencies for few-body systems and hence re-emphasizes an
open question (see also Section~\ref{sec:open_questions}): How can the
incompatibility of the extracted LEC values, determined based on few-nucleon
observables, medium-mass nuclei and nuclear matter saturation properties, be
reconciled?

\subsection{Spectra of nuclei}
\label{sec:spectra}

\begin{figure}[t!]
\begin{center}
\includegraphics[width=0.45\textwidth]{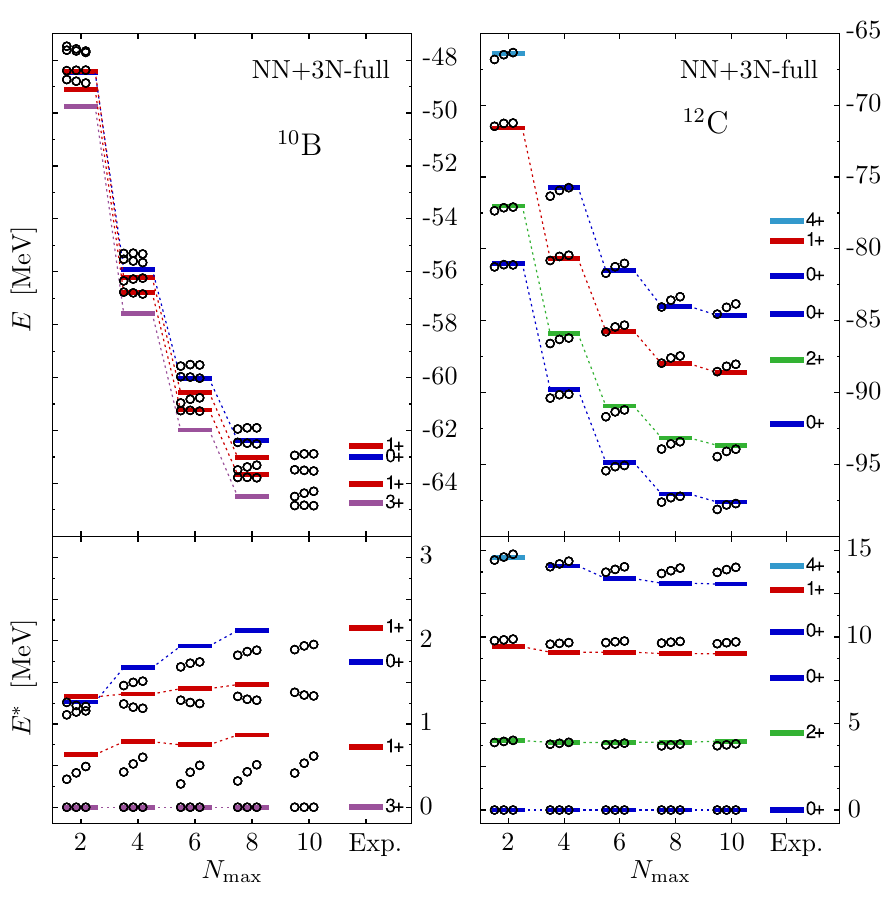}
\hspace{5mm}
\includegraphics[width=0.44\textwidth]{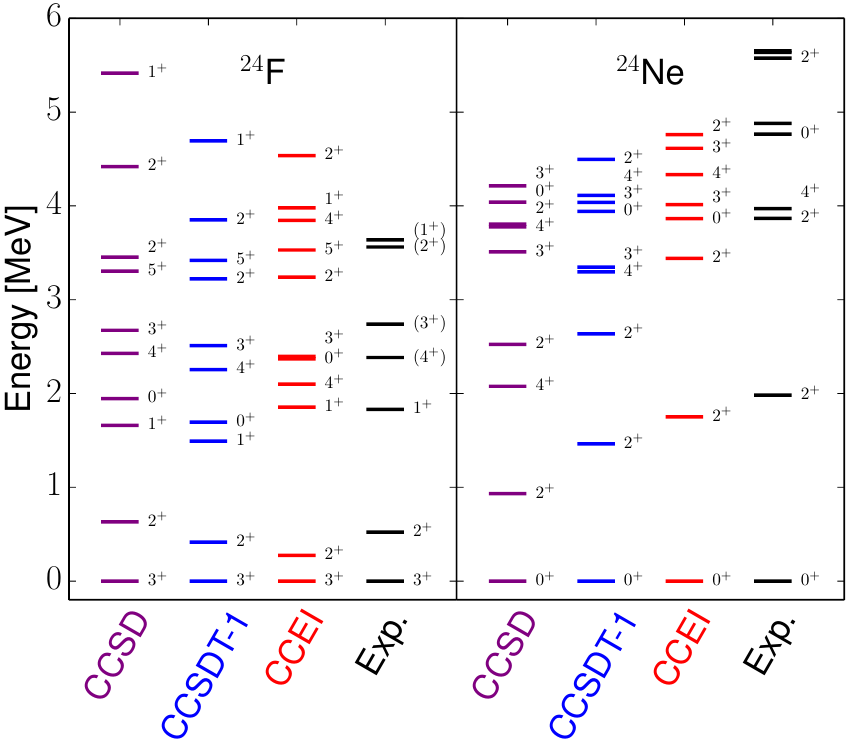}
\end{center}
\caption{Left: Absolute (top) and relative (bottom) energies of the lowest states 
of $^{10}$B and $^{12}$C as a function of basis size parameter
$N_{\text{max}}$ calculated in the importance-truncated NCSM based on
SRG-evolved NN plus 3N interactions. The solid lines correspond to the
complete 3N interaction, the circles to calculations based on normal-ordered
3N interactions. The experimental excitation energies are taken from
Ref.~\cite{nndc14ENSDF}. Right: Excited states of $^{24}$F and $^{24}$Ne computed within coupled-cluster plus
the valence-shell diagonalization.\\
\textit{Source:} Left figure adapted from Ref.~\cite{Gebr16MR} and right figure taken from Ref.~\cite{Jans16SM}.}
\label{fig:spectra}
\end{figure}

The correct description of the lowest excited states of nuclei represents
another challenge for \textit{ab initio} nuclear structure and a benchmark for
nuclear interactions. The spectrum provides important insight into the
geometric nature of a nucleus, single-particle and collective excitations, as
well as possible cluster structures. In principle, excited states can be
accessed in a straightforward way by all wave-function-based many-body
methods, like, e.g., NCSM, valence-shell diagonalization or CC (see
Section~\ref{sec:Intro}). In addition, the IMSRG framework has also been
recently extended to excited states~\cite{Parz17EOMIMSRG}. While the
valence-shell model approach was historically based on phenomenological
interactions, nowadays valence-space Hamiltonians can be computed
microscopically in \textit{ab initio} frameworks starting from chiral EFT NN
and 3N interactions using MBPT or IMSRG (see, e.g.,
Refs.~\cite{Jans16SM,Stro16TNO,Stro17ENO,Sun18CCshell,Holt19limits}). The
spectrum of light nuclei can be computed, e.g., via large scale (IT)-NCSM or
QMC methods (see right panel in Figure~\ref{fig:fits_Deltafull}).

In the left panel of Figure~\ref{fig:spectra} we show results of IT-NCSM
calculations of the lowest excited state with positive parity of
$^{10}$B and $^{12}$C as a representative set of p-shell nuclei based
on a chiral NN and 3N interaction~\cite{Gebr16MR}. The results show a natural
convergence pattern and a reasonable agreement with experiment. We note that
excited states in both nuclei have been shown to be sensitive to
contributions from 3N interactions~\cite{Roth11SRG,Navr07A1013}. Calculations
of odd-odd nuclei like $^{10}$B are particularly challenging since excitation
energies are typically smaller and the correct reproduction of the level
ordering becomes more tricky. Furthermore, we note that due to the nature of
the NCSM configuration basis, the $0^+$ resonance state in $^{12}$C at about
$7.7$ MeV excitation energy cannot be properly described in this framework due
to its pronounced cluster structure~\cite{Epel11Hoyle,Epel12Hoyle}. This
``Hoyle-state'' plays a key role for the synthesis of carbon, oxygen, nitrogen
and other elements, which form the building blocks of complex molecules of
living beings~\cite{Hoyl54state}. Finally, the shown results also illustrate
the accuracy of the normal-ordering approximation for calculations of excited
states (see Ref.~\cite{Gebr16MR} for details).

In the right panel of Figure~\ref{fig:spectra} we show the lowest excited
states of $^{24}$F and $^{24}$Ne computed within CC and the valence-shell
model. This plot compares results of CC equation-of-motion calculations
(``CCSD''), calculations including linearized contributions from
three-particle-three-hole contributions (``CCSDT-1''), and results from a
valence space diagonalization based on effective interactions computed within
CC (``CCEI'')~\cite{Jans16SM}. The results of all the different approaches are
in reasonable agreement. Furthermore, for $^{24}$F both CCEI and full-space
coupled-cluster calculations agree well with those calculated with IM-SRG
effective interactions and the experimental results from recent
measurements~\cite{Cace1524F}.

\subsection{Electromagnetic response of nuclei and neutron distributions}
\label{sec:eletric_dipole_pol}

\begin{figure}[t!]
\begin{center}
\includegraphics[width=0.99\textwidth]{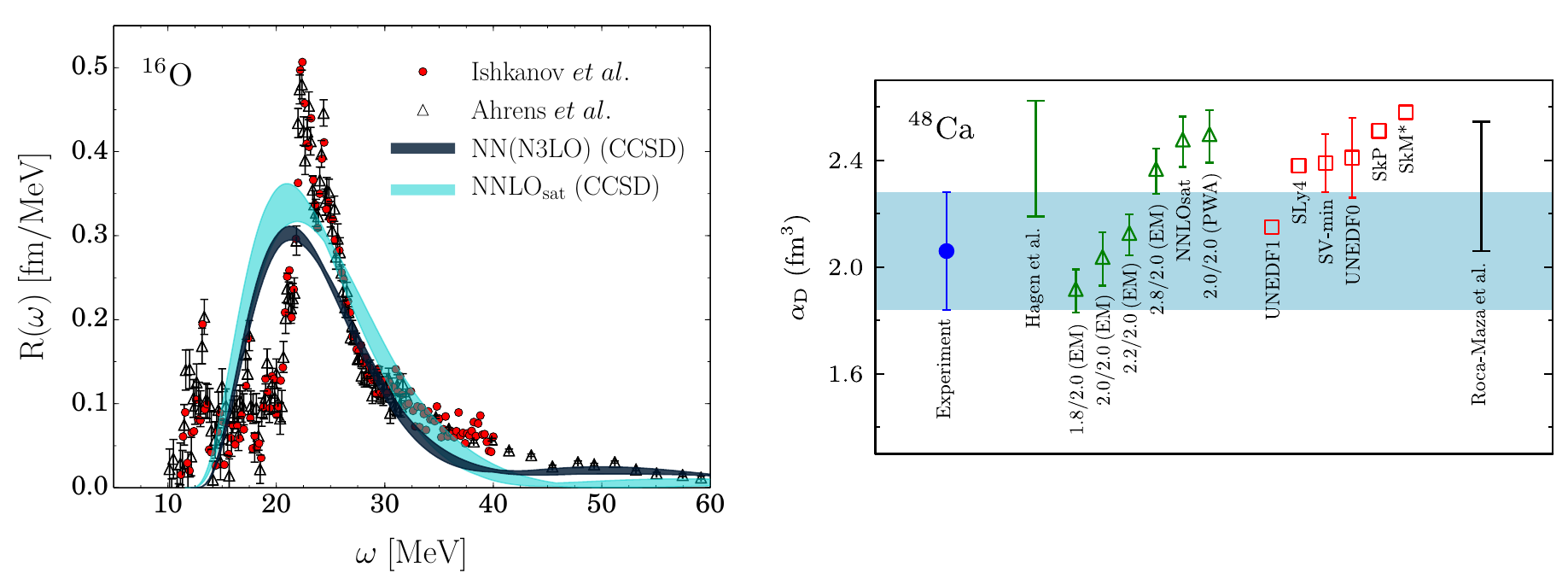}
\end{center}
\caption{
Left: Photo-absorption response function for $^{16}$O. The red circles show
the experimental data from Ref.~\cite{Ishi02crosssections}, and the white
triangles plus error bars are the results by Ahrens et
al.~\cite{Ahre75photoabsorpCS}. Calculations are performed with coupled
cluster based on the interactions of Refs.~\cite{Ekst15sat,Ente03EMN3LO}.
Right: Experimental constraints for the electric dipole polarizability of $^{48}$Ca (blue band) and
predictions from ab intio calculations based on chiral NN and 3N interactions
(green triangles) and energy density functionals (red squares).\\
\textit{Source:} Left figure taken from Ref.~\cite{Mior16dipole} and right figure taken
from Ref.~\cite{Birk17dipole}.}
\label{fig:dipole_pol_Ca48}
\end{figure}

The distribution of protons in nuclei and the resulting charge radii can be
accurately measured in electron scattering experiments via the electromagnetic
interaction~\cite{Ange13rch}. The determination of the neutron distribution on
the other hand is much more challenging. However, an accurate knowledge of
neutron distributions in atomic nuclei is key for understanding neutron-rich
systems ranging from short-lived isotopes to macroscopically large objects
such as neutron stars. The distribution of neutrons in nuclei determines the
limits of the nuclear stability~\cite{Erle12Nature}, gives rise to exotic
structures in rare isotopes~\cite{Step13Ca54} and governs basic properties of
neutron stars~\cite{Brown00radii}. Because of its fundamental importance,
experimental efforts worldwide have embarked on a program of measurements of
neutron distributions in nuclei using different probes, including hadronic
scattering~\cite{Zeni10neutrondistPb}, pion
photoproduction~\cite{Tarb14neutronskinPb208}, and parity-violating electron
scattering~\cite{Abra12prex}. Since the weak charge of the neutron, $Q^n_W
\approx -1$, is much larger than that of the proton, $Q^p_W \approx 0.07$, a
measurement of the parity-violating asymmetry $A_{\text{pv}}$ offers an
opportunity to probe the neutron distribution~\cite{Horo14CREXPREX}. In
addition, experiments focus on measuring observables that are related to the
neutron distribution, such as the dipole polarizability
$\alpha_D$~\cite{Piek12dipole}. The dipole polarizability is directly
connected to the electromagnetic response function $R(\omega)$ and
characterizes the low-energy behavior of the dipole strength. Specifically,
it is defined by (see, e.g., Ref.~\cite{Simo19response} for details)
\begin{equation}
R(\omega) = \sum_{n} \bigl| \bigl< \psi_0 | \hat{\Theta} | \psi_n \bigr> \bigr|^2 \delta( E_n - E_0 - \omega) \, ,
\label{eq:response_fct}
\end{equation}
where $\psi_0$ and $\psi_n$ are the ground- and final-state wave functions of
the nucleus with the energies $E_0$ and $E_n$, respectively, and
$\hat{\Theta}$ denotes the corresponding electromagnetic operator. Based on
this response function several different moments can be defined via: $S_n =
\int_{\omega_{\text{th}}}^{\infty} d\omega R(\omega)
\omega^n$, with some threshold energy $\omega_{\text{th}}$. The dipole
polarizability is given by
\begin{equation}
\alpha_D = 2 \alpha \int_{\omega_{\text{th}}}^{\infty} d\omega
\frac{R(\omega)}{\omega} \sim S_{-1} \, ,
\end{equation}
where $\alpha$ is the electromagnetic coupling constant. Recently, this
parameter was accurately measured for $^{208}$Pb~\cite{Tami11dipole},
$^{120}$Sn~\cite{Hash15dipole}, $^{68}$Ni~\cite{Ross13dipole} and
$^{48}$Ca~\cite{Birk17dipole} (see also right panel of
Figure~\ref{fig:dipole_pol_Ca48}). Theoretical calculations of this parameter
involve several challenges because most of the dipole strength lies in the
scattering continuum, i.e., the according final states $\psi_n$ in
Eq.~(\ref{eq:response_fct}) do not correspond to bound states. There have been
significant advances in recent years that allowed to transform this continuum
problem into a bound-state problem via the Lorenz integral transform (LIT)
method and make these observables accessible by ab initio many-body frameworks
(see, e.g.,
Refs.~\cite{Efro94LIT,Efro07LIT,Bacc13dipole,Mior16dipole,Birk17dipole,
Simo19response} and references therein). The left panel of
Figure~\ref{fig:dipole_pol_Ca48} shows experimental and theoretical results
for the $^{16}$O photo-absorption response function obtained from calculations
within CC using a N$^3$LO NN interaction~\cite{Ente03EMN3LO} (dark band) and
``N$^2$LO$_{\text{sat}}$''~\cite{Ekst15sat} (light blue band). The uncertainty
band for the ``N$^2$LO$_{\text{sat}}$'' interaction is larger since the
available model space size for these calculations was smaller than for the
NN-only calculations using the N$^3$LO potential. Generally, it is found that the
theoretical results are able to capture the bulk features of the experimental
data and that the dipole polarizability is more sensitive to the distribution
of the dipole strength at low energies rather than to the detailed structure
and shape of the response function. The right panel of
Figure~\ref{fig:dipole_pol_Ca48} shows a comparison of recent experimental
$\alpha_D$ values for $^{48}$Ca with predictions from \textit{ab initio} calculations
based on chiral NN and 3N interactions of Refs.~\cite{Hebe11fits,Ekst15sat}
and state-of-the-art energy-density functionals~\cite{Birk17dipole}. These
results show that present calculations based on NN plus 3N interactions can
provide a realistic description of the dipole polarizability even though it is
not yet possible to resolve the detailed structure of the response function
(see left panel).

\begin{figure}[t!]
\begin{center}
\includegraphics[width=0.9\textwidth]{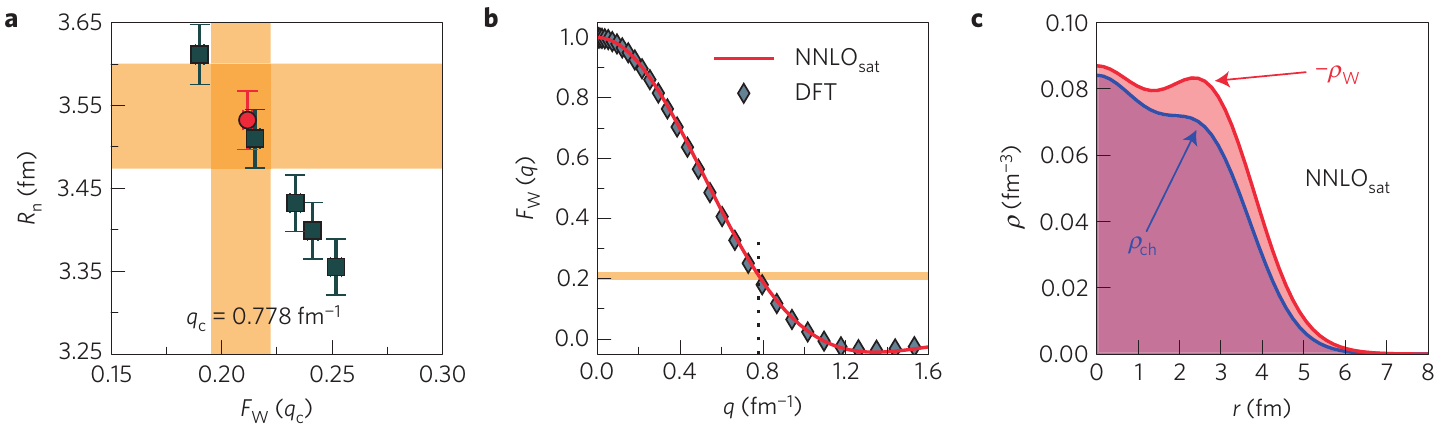}
\end{center}
\caption{Left: Point-neutron radius $R_{\text{n}}$ in $^{48}$Ca as a function
of the weak-charge form factor $F_{\text{W}} (q_{\text{c}})$ at the CREX
momentum ${q_{\text{c}} = 0.778 \text{ fm}^{-1}}$ computed in \textit{ab initio}
calculations with ``N$^2$LO$_{\text{sat}}$''~\cite{Ekst15sat} (red circle) and
chiral interactions of Ref.~\cite{Hebe11fits} (squares). The width of the
horizontal orange band shows the predicted range for $R_{\text{n}}$,
determined based on a correlation with the proton radius
$R_{\text{p}}$~\cite{Hage16NatPhys}. Center: weak-charge form factor
$F_{\text{W}}(q)$ as a function of momentum transfer $q$ with
``N$^2$LO$_{\text{sat}}$'' (red line) and DFT (diamonds). The orange
horizontal band shows $F_{\text{W}} (q_{\text{c}})$. Right: Charge
density (blue line) and (negative of) weak-charge density (red line). The
weak-charge density extends well beyond $\rho_{ch}$ as it is tightly connected
to the neutron distribution.\\
\textit{Source:} Figures taken from Ref.~\cite{Hage16NatPhys}.}
\label{fig:weak_charge}
\end{figure}

The weak-charge radius $R_{\text{W}}$ is another quantity that characterizes
the size of the nucleus. The CREX experiment will measure the parity-violating
asymmetry $A_{\text{pv}}$ in electron scattering on $^{48}$Ca at the momentum
transfer $q_c = 0.778$ fm$^{-1}$. By making some assumptions about the
weak-charge form factor, one can deduce $R_{\text{W}}$ and the neutron radius
$R_{\text{n}}$ from the single CREX data point~\cite{Horo14CREXPREX}.
Furthermore, Figure~\ref{fig:weak_charge} shows that there exists a strong
correlation between $R_{\text{n}}$ and $F_{\text{W}} (q_{\text{c}})$, which
makes it possible to extract the theoretical constraint $0.195 \le
F_{\text{W}} (q_{\text{c}}) \le 0.222$~\cite{Hage16NatPhys}. This range as
well as the momentum dependence of the weak-charge form factor (middle panel)
show good agreement with DFT results. As seen in the right panel of
Figure~\ref{fig:weak_charge}, the spatial extent of the weak-charge density
$\rho_{\text{W}} (r)$, being the Fourier transform of the weak-charge form
factor $F_{\text{W}} (q)$, is significantly greater for $^{48}$Ca than that of
the electric charge density $\rho_{\text{ch}}$. This is a reflection of the
fact that there is an excess of eight neutrons over protons in $^{48}$Ca.

\subsection{Beta decay transitions}
\label{sec:beta_decay}

Beta decay ($\beta$ decay) is the predominant decay process of atomic nuclei,
especially up to the medium-mass range. It involves the transformation of a
neutron into a proton or vice versa via the weak interaction. Generally,
$\beta$-decay transitions can be categorized into Fermi transitions and
Gamow-Teller transitions. In the case of Gamow-Teller transitions the emitted
electron-antineutrino or positron-neutrino pair form a relative
$S=1$ state, such that the total angular momentum of the nucleus changes by
$\Delta J = -1, 0$ or $1$. In contrast, for Fermi transitions, the emitted
leptons couple to an $S=0$ state and the total angular momentum $J$ of the
nucleus is hence conserved.

Specifically, the half-life $t$ for the $\beta$ decay of, e.g. $^3$H, can be
expressed in the form~\cite{Schi98weakcap,Rama78tables,Klos17triton}
\begin{equation} (1+\delta_R) t = \frac{K/G_V^2}{f_V \, \langle \text{F}
\rangle^2 + f_A \, g_A^2 \, \langle \text{GT} \rangle^2} \, ,
\label{eq:half-life}
\end{equation} 
where $\delta_R$ includes radiative corrections that originate
from virtual photon exchange between the charged particles, $f_V$ and $f_A$
are Fermi functions, which account for the deformation of the electron wave
function due to electromagnetic interactions with the nucleus, and $G_V=1$ and
$g_A=1.27$ denote the vector and axial-vector couplings. The kinematics of the
process leads to an additional constant $K= 2\pi^3\ln 2 /m_e^5$, where $m_e$
is the electron mass. The half-life depends on the nuclear matrix elements of
the vector and axial-vector currents denoted as Fermi $\langle
\text{F}\rangle$ and Gamow-Teller $\langle \text{GT}\rangle$ matrix elements,
respectively. The Fermi reduced matrix element is given by $\left< \text{F}
\right> = \langle ^3\text{He} \lVert \sum^3_{i=1} \tau_i^+ \rVert
^3\text{H}\rangle\,$, where $\tau^+=\frac{1}{2}(\tau^x+i\tau^y)$ is the
isospin-raising operator and the wave functions $|^3\text{H}\rangle$ and 
$|^3\text{He}\rangle$ denote the ground states of the mother and daughter
nuclei, $^3$H and $^3$He. The Gamow-Teller reduced matrix element contains
axial-vector one-body (1b) and two-body (2b) current contributions:
\begin{equation}
\langle \text{GT} \rangle = \frac{1}{g_A} \langle ^3\text{He} \lVert
\sum^3_{i=1}{\bf J}^+_{i,{\rm 1b}}+\sum_{i<j}{\bf J}_{ij,\text{2b}}^+ \rVert
^3\text{H}\rangle \, .
\label{eq:gt_me}
\end{equation}
The axial-vector current was derived in chiral EFT to third
order~\cite{Park02eftsolar,Hofe15powerdm} ($Q^3$), while more recent
derivations have been extended to order $Q^4$~\cite{Kreb16axial,Baro15axial}.
Since the typical Q-values of $\beta$ decays are relatively small, all
currents can be evaluated to good approximation at vanishing momentum
transfer. Therefore to order $Q^0$ and $Q^2$, only the momentum-independent
one-body current contributes: ${\bf J}^+_{i,{\rm 1b}} = g_A \tau_i^+ {\bm
\sigma}_i$. The first contributions to two-body currents enter at order $Q^3$.
The corresponding expressions can be found
in~\cite{Park02eftsolar,Hofe15powerdm,Klos17triton}.

\begin{figure}[t!]
\begin{center}
\includegraphics[width=0.43\textwidth]{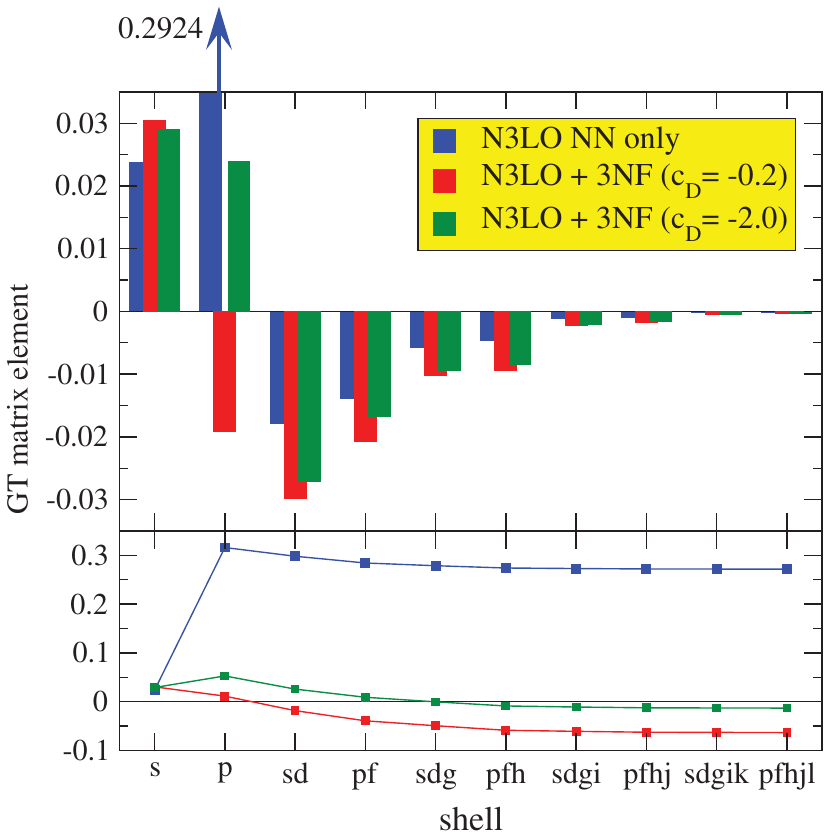}
\hspace{5mm}
\includegraphics[width=0.51\textwidth]{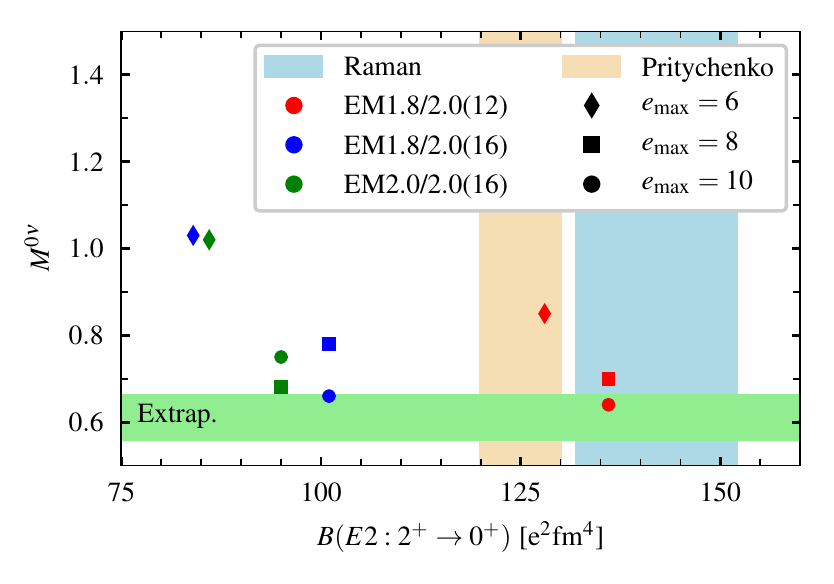}
\end{center}
\caption{Left: Contributions to the $^{14}$C Gamow-Teller matrix element for individual 
harmonic oscillator shells determined from NCSM calculations based on chiral
NN and 3N interactions. The top panel shows the contributions when only using
NN interactions (blue) and including 3N interactions (red and green). The
bottom panel displays the accumulated contributions up to the given shell. The
results show that the inclusion of 3N interactions lead to a significant
suppression of the p-shell contributions to the matrix element for the
employed interactions. Right:
$(0\nu\beta\beta$) beta decay matrix element $M^{0\nu}$ as a function of the
$B($E2:$2_1^+ \rightarrow 0^+_1$) value for $^{48}$Ti calculated within the
IM-SRG+GCM framework using different chiral EFT interactions and basis
truncations. The bands show the experimental constraints for the $B(E2)$
values of Refs.~\cite{Rama122plus,Prit132plus}.\\
\textit{Source:} Left figure taken from Ref.~\cite{Mari1114C} and right figure taken from
Ref.~\cite{Yao19beta}.}
\label{fig:beta_decay}
\end{figure}

Up to now, systematic \textit{ab initio} studies of $\beta$ decays were limited to
light nuclei~\cite{Gazi08lec,Baro16beta,Klos17triton} (see also
Figure~\ref{fig:interactions_fit_3Hdecay}) and a few selected
isotopes~\cite{Ekst14GT2bc,Morr17Tin,Gysb19beta} (see also
Figure~\ref{fig:Sn}). A particularly prominent nucleus is $^{14}$C, which is
characterized by an anomalously large half-life of about $5715$
years~\cite{Hold09t12}. This property makes this isotope a very attractive
chronometer as it is naturally present in animals and plants and hence allows
to reliably date up to about 50000 year-old samples~\cite{Libb49C14}. The left
panel of Figure~\ref{fig:beta_decay} shows results of NCSM calculations for the
Gamow-Teller matrix elements based on solutions for the ground-state wave
functions of $^{14}$C and $^{14}$N using chiral EFT NN and 3N interactions and
including only one-body current contributions
$\mathbf{J}_{i,1b}$~\cite{Mari1114C}. The results show that contributions from
3N interactions lead to a significant suppression of the matrix elements and
hence to an enhanced life time of $^{14}$C (see Eq.~(\ref{eq:half-life})). We
emphasize, however, that the shown contributions from the different harmonic
oscillator orbitals are not observable. That implies that the conclusion is
also scheme-dependent and strictly speaking only valid for the particular
employed NN and 3N interactions. Future and current work aim at including also
GT transitions to low-lying excited states of the daughter nucleus (see,
e.g.,~Refs.~\cite{Negr0614,Gysb19beta}).

On a quantitative level it is remarkable that theoretical results for
Gamow-Teller matrix elements agree reasonably well with experimental results
if the axial coupling constant $g_A$ is quenched by a factor of about
$0.75$~\cite{Wilk73quenching,Brow85beta,Chou93gamow,Mart96gA}. Different
possible sources for this deviation have been proposed, like missing
correlations in the wave functions~\cite{Town87quenching} or contributions
from higher-order nuclear currents~\cite{Mene110nbb2bc}. In fact, results from
recent \textit{ab initio} calculations indicate that the discrepancies can be
resolved when both these factors are treated properly~\cite{Gysb19beta} (see
also discussion in Section~\ref{sec:open_questions}).

Apart from first-order weak interaction processes like ordinary $\beta$
decays, also second-order processes have been observed in specific nuclei.
These decays involve a simultaneous transformation of two neutrons into
protons, or vice versa. Such decays can be observed in nuclei for which a
single $\beta$ decay is energetically forbidden, while the isobar with atomic
number two higher or lower has a larger binding energy. Given that
second-order processes are strongly suppressed, such nuclei exhibit very long
half lives~\cite{Bara15doublebeta}. All observed double-beta ($\beta\beta$)
decays correspond to the two-neutrino-$\beta\beta$ $(2\nu\beta\beta)$ decay
and need to be distinguished from an even more rare type of $\beta\beta$
decay, which involves no neutrino emission, the neutrinoless-$\beta\beta$
$(0\nu\beta\beta)$ decay. This process violates lepton number conservation and
can hence only occur if neutrinos are their own antiparticles. The observation
of such a decay would have profound implications for our understanding of the
nature of neutrinos and the Standard Model~\cite{Avig07neutrinirev} and is
thus subject of several experimental
searches~\cite{Agos17gerda,Gando16kamland,Bara15doublebeta,Albe14exo}. The
theoretical calculation of matrix elements for $\beta\beta$ decays also
involves significant challenges~\cite{Caur11beta,Verg12beta,Enge16bbrev}. As a
result, predictions from calculations using different many-body frameworks and
nuclear interactions for a given transition can vary significantly. This
situation is especially unsatisfactory since precise results are required for
the analysis and planning of present and future $0\nu\beta\beta$ decay
experiments. In the right panel of Figure~\ref{fig:beta_decay} first results
for the $0\nu\beta\beta$ matrix element $M^{0\nu}$ are shown computed with the
IM-SRG+GCM framework~\cite{Yao18GCM} including a microscopic treatment of
collective correlations based on different NN plus 3N interactions of
Ref.~\cite{Hebe11fits}. It is found that the matrix element value for the
interaction ``1.8/2.0 (EM)'', which provides ground-state energies in
excellent agreement with experiment (see Section~\ref{sec:gs}), turns out to
be around 0.6, a value significantly smaller than those of more
phenomenological approaches~\cite{Yao19beta}. It will be crucial to benchmark
these results against predictions of other many-body approaches and to
estimate the theoretical uncertainties due to EFT truncations.

\subsection{Three-body scattering cross sections}

\begin{figure}[t!]
\begin{center}
\includegraphics[width=0.55\textwidth]{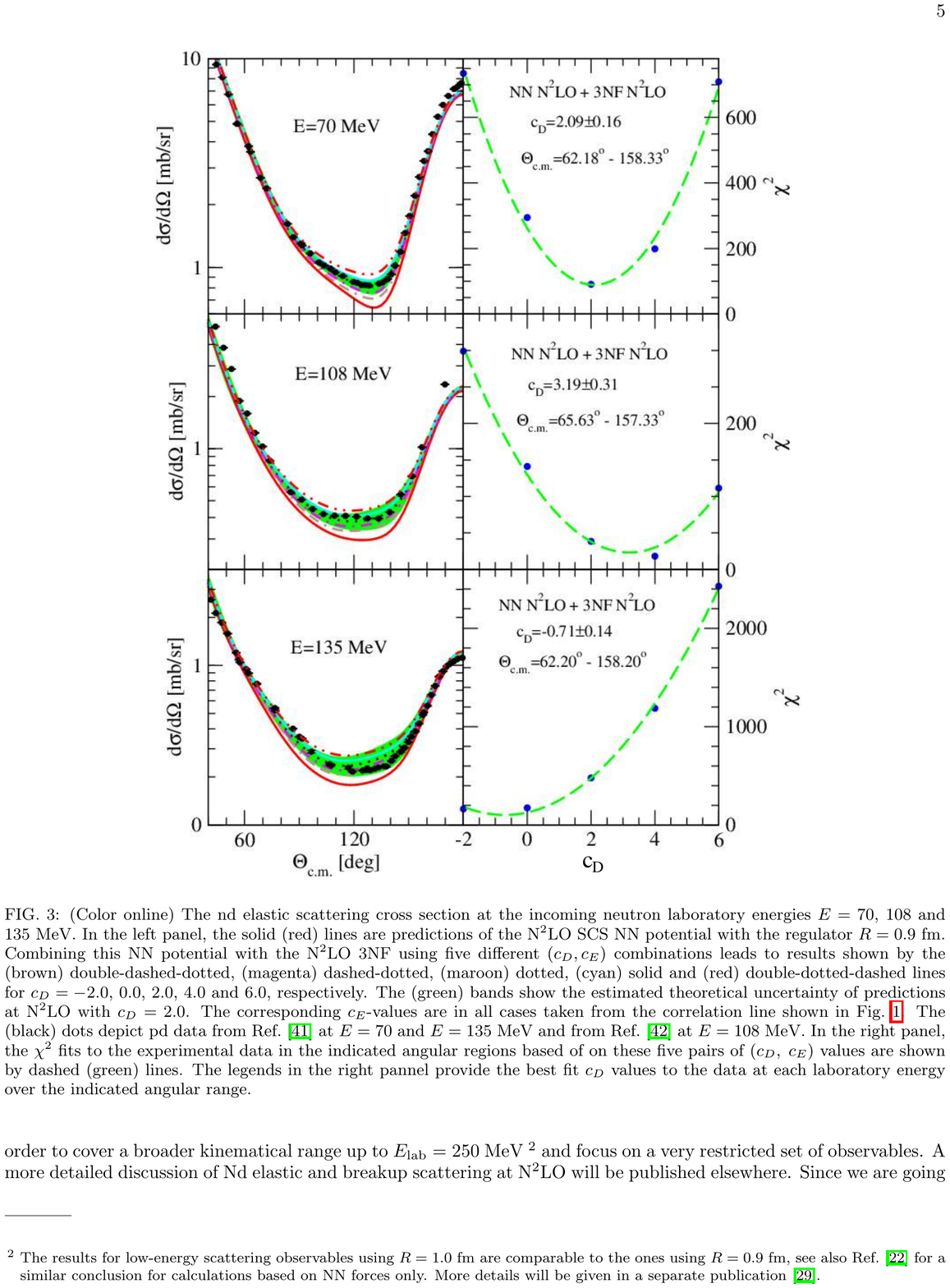}
\hspace{5mm}
\includegraphics[width=0.34\textwidth]{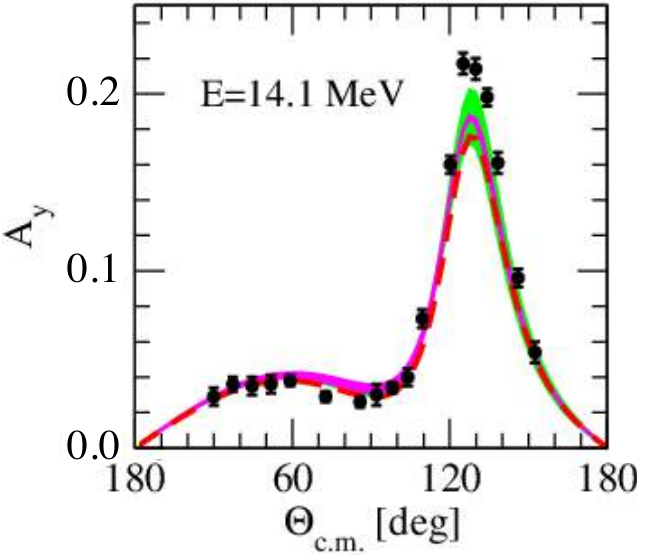}
\end{center}
\caption{Left: Neutron-deuteron (nd) elastic scattering cross section
 at different incoming neutron laboratory energies $E=70$, $108$ and $135$~MeV
 (rows). In the left column, the solid (red) lines are predictions of the
 N$^2$LO NN potential of Refs.~\cite{Epel14SCSprl,Epel15improved} with the
 regulator $R=0.9$~fm. This NN interaction is then combined with the N$^2$LO
 3N interactions using five different ($c_D,c_E$) combinations. These
 combinations lead to results shown by the brown double-dashed-dotted, magenta
 dashed-dotted, maroon dotted, cyan solid and red double-dotted-dashed lines
 for $c_D = -2.0$, $0.0$, $2.0$, $4.0$ and $6.0$, respectively. The green
 bands show the estimated theoretical uncertainty of predictions at N$^2$LO
 with $c_D=2.0$. The (black) dots depict experimental proton-deuteron data
 from Ref.~\cite{Seki02scattering} at $E=70$ and $E=135$~MeV and from
 Ref.~\cite{Ermi03scattering} at $E=108$~MeV. In the right column, the
 $\chi^2$ fits to the experimental data in the indicated angular regions based
 of on these five pairs of ($c_D,c_E$) values are shown by dashed (green)
 lines. The legends in the right column provide the best fit $c_D$ values to
 the data at each laboratory energy over the indicated angular range. Right:
 The neutron analyzing power $A_y$ in nd elastic scattering at $E_n = 14.1$
 MeV. The dashed (red) line is the prediction of the N$^2$LO NN potential of
 Refs.~\cite{Epel14SCSprl,Epel15improved} with the regulator $R = 0.9$ fm. The
 (magenta) band covers the predictions obtained with this N$^2$LO NN potential
 combined with the N$^2$LO 3N interactions using $c_D = -2.0 ... 6.0$. The
 (green) band gives the estimated theoretical uncertainty at N$^{2}$LO for the
 value of $c_D = 2.0$. The black dots depict the experimental data from
 Ref.~\cite{Torn83Ay}.\\
 \textit{Source:} Figures adapted from Ref.~\cite{Epel18SCS3N}.}
\label{fig:nd_scattering}
\end{figure}

Since their advent, numerically exact 3N continuum Faddeev
calculations of the elastic neutron-deuteron (nd) scattering and deuteron
breakup reactions have become a powerful tool to test modern nuclear
forces~\cite{Gloe95cont}. Benchmarking theoretical predictions for scattering
cross sections against precise nd elastic scattering and breakup data over a
wide range of incoming nucleon energies can help to isolate deficiencies of
present nuclear forces in a specific kinematical regime. In particular, for
three-body systems effects of four- and higher-body forces in chiral forces
can be cleanly disentangled and uncertainties from the few-body calculations
are very small since structure and reaction observables can be solved
virtually exactly.

In the left subfigure of Figure~\ref{fig:nd_scattering} we show results for
the differential cross section of elastic nd scattering (left panels) at three
different energies for different LEC values of the 3N interaction with the
corresponding $\chi^2$ values as a function of $c_D$ (right panels, see also
caption). The theoretical results are obtained by solving the Faddeev
equations in a partial-wave momentum basis as given in
Eq.~(\ref{eq:Jj_bas})~\cite{Gloe83QMFewbod,Gloe95cont}. The actual
calculations have been performed for $R = 0.9$ fm using the NN interaction of
Refs.~\cite{Epel14SCSprl,Epel15improved} for five different $c_D$ values, $c_D
= -2.0, 0.0, 2.0, 4.0$ and $6.0$ (see also Ref.~\cite{Wita19}). In all cases,
the $c_E$-value is taken from a fit to the $^3$H binding energy. The results
show that contributions from 3N interactions have a significant effect on the
agreement of the computed cross sections with the experimental data, which is
also reflected by the pronounced sensitivity of the $\chi^2$ values on $c_D$.
However, a comparison of the results at different energies also shows that a
$\chi^2$ minimization leads to different optimal $c_D$ in different regimes,
which makes it necessary to perform a global optimization of the couplings
(see also Figure~\ref{fig:cd_ce_ndscattering_length} and the discussion in
Section~\ref{sec:sep_3N_fits}).

At low energies, a particularly interesting observable is the analyzing power
$A_y$ for nd elastic scattering with polarized neutrons (see right panel in
Figure~\ref{fig:nd_scattering}, see also Section~\ref{sec:sep_3N_fits}).
Theoretical predictions of phenomenological high-precision NN potentials tend
to underestimate the experimental data for $A_y$ by up to $30
\%$~\cite{Epel18SCS3N} in the region of the maximum, which corresponds to the
center-of-mass angles of $\Theta_{\text{c.m.}} \approx 125 \degree$. Combining
these NN potentials with phenomenological 3N interactions reduces the
deviation in some cases by about a factor 2, while for other interactions the
addition has practically no effect on $A_y$~\cite{Epel18SCS3N} (see right
panel in Figure~\ref{fig:nd_scattering}). The predictions for $A_y$ based
on the chiral NN potentials appear to be similar to those of phenomenological
models, see Ref.~\cite{Bind15Fewbody} and references therein. Combining the
N$^2$LO chiral potential with the N$^2$LO 3N interactions only slightly
improves the description of $A_y$. Interestingly, the theoretical predictions
appear to be quite insensitive to the actual value of $c_D$ as visualized by a
rather narrow magenta band, which corresponds to the variation of $c_D = -2.0
... 6.0$. On the other hand, the theoretical uncertainty at N$^2$LO is rather
large and, in fact, comparable in magnitude with the observed deviation
between the predictions and experimental data. It will be interesting to see
whether the $A_y$ puzzle persists upon inclusion of higher-order corrections
to the 3N interactions.

\begin{figure}[t]
\begin{center}
\includegraphics[width=0.8\textwidth]{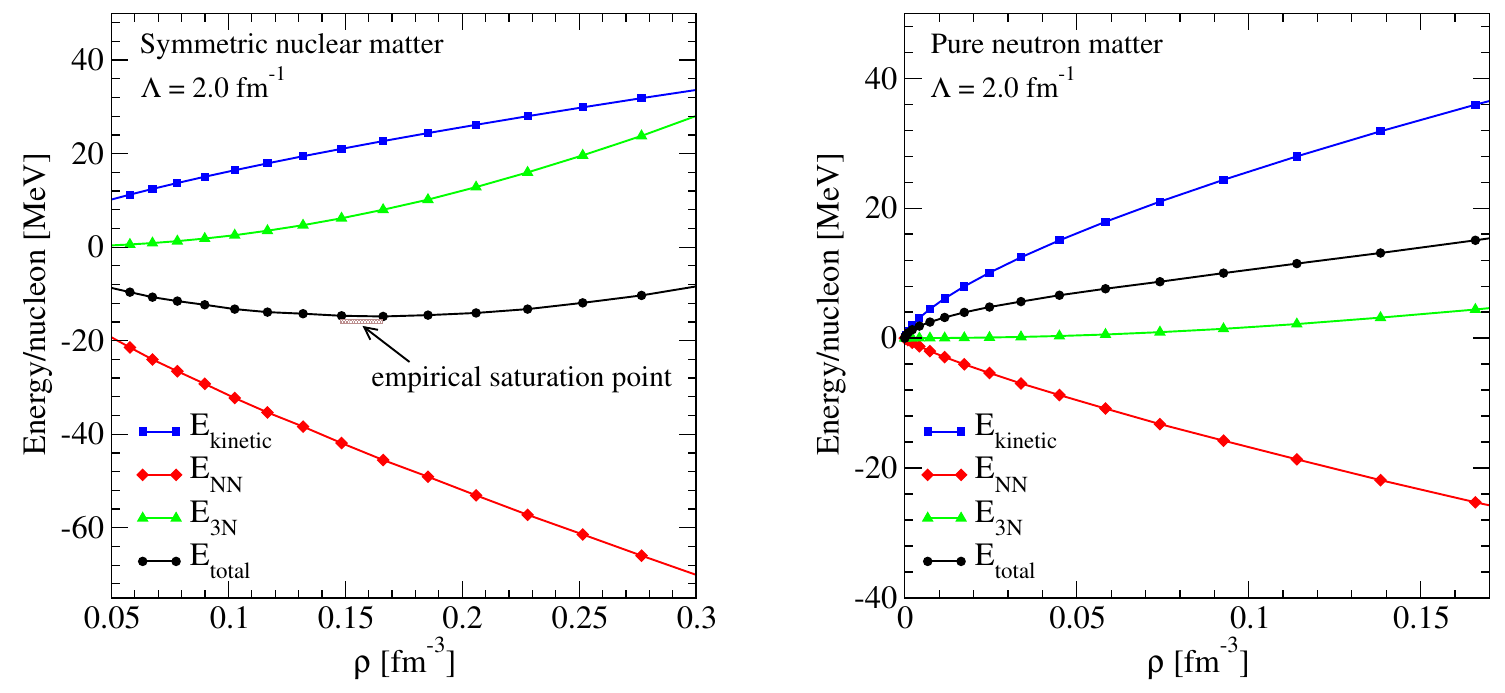}
\end{center}
\caption{Size of many-body contributions as a function of density for
SNM (left)~\cite{Hebe11fits} and PNM (right)~\cite{Hebe10nmatt}, calculated
within MBPT based on SRG-evolved NN interactions plus fitted 3N interactions
(``$2.0/2.0$(EM)'') as defined in Ref.~\cite{Hebe11fits}. $E_{\text{kinetic}}$
denotes the kinetic energy, $E_{\text{NN}}$ the energy contributions from only
NN interactions and $E_{\text{3N}}$ all contributions which include at least
one 3N interaction term.\\
\textit{Source:} Figures taken from Ref.~\cite{Furn13RPP}.
}
\label{fig:EOS_NN_3N_contribution}
\end{figure}

\subsection{Nuclear equation of state and astrophysical applications}
\label{sec:applications_astro}

The physics of neutron-rich matter covers a wide range of regimes. At very low
densities, the average interparticle distance is sufficiently large so that
details of nuclear forces are not resolved and all properties of the system
are dominated by the large s-wave scattering length. In this universal regime,
neutron matter shares many properties with cold atomic gases close to the
unitary limit~\cite{Gior08coldgas,Zwer12book}. At nuclear densities the
properties of neutron and symmetric nuclear matter are used to guide the
development of energy density functionals and to constrain the physics of
neutron-rich systems, which are key for understanding the synthesis of heavy
nuclei in the universe~(see, e.g., Ref.~\cite{Pian17rprocess}). At very high
densities, far beyond nuclear densities, the composition and properties of
nuclear matter are still unknown. Exotic states of matter containing strange
particles or quark matter may be present. Furthermore, neutron matter
constitutes a unique laboratory for chiral EFT, because only long-range 3N
interactions contribute up to N$^3$LO, at least for unregularized interactions
(see Sections~\ref{sec:chiral_expansion} and \ref{sec:3N_regularization}).
This offers the possibility to derive systematic constraints based on chiral
EFT interactions for the equation of state (EOS) of neutron-rich matter in
astrophysics, for the symmetry energy and its density dependence, and for the
structure of neutron stars, but also makes it possible to test the chiral EFT
power counting and the hierarchy of many-body forces at densities relevant for
nuclei.

\begin{figure}[t]
\begin{center}
\includegraphics[width=0.46\textwidth]{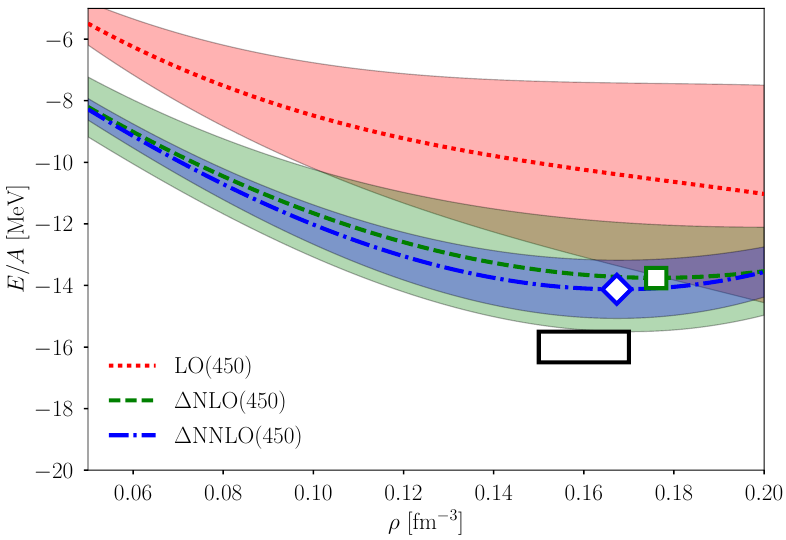}
\hspace{5mm}
\includegraphics[width=0.46\textwidth]{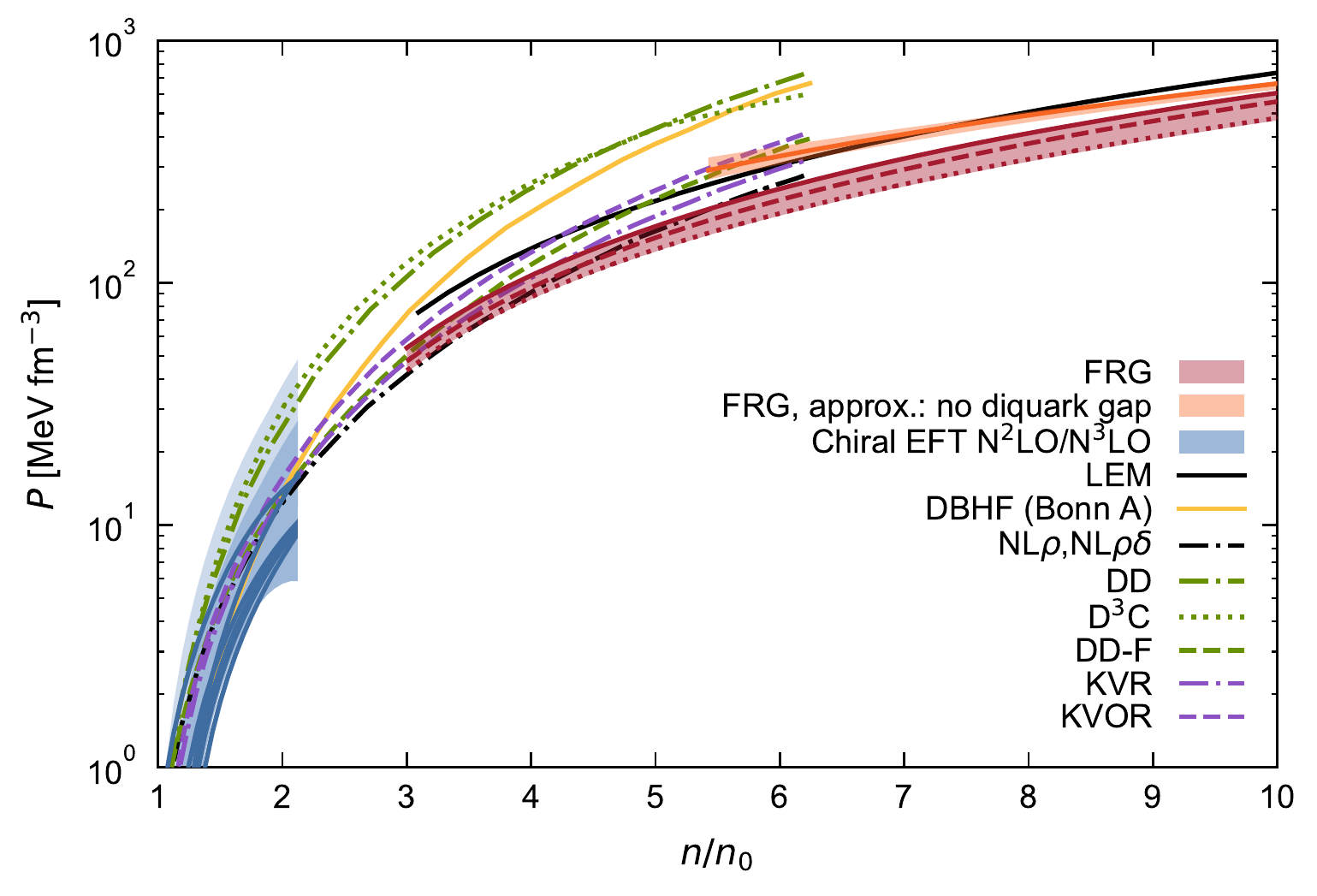}
\end{center}
\caption{Left: Energy per nucleon of symmetric nuclear matter at different
orders in $\Delta$-full chiral EFT, computed within CC. The bands show the
estimated EFT-truncation errors, and the square/diamond marks the calculated
saturation point in symmetric nuclear matter for $\Delta$NLO/$\Delta$N$^2$LO.
The black rectangle indicates the region $E/A = -16
\pm 0.5$ MeV and $\rho = 0.16 \pm 0.01$ fm$^{-3}$. Right: Pressure of symmetric nuclear matter as
obtained from chiral EFT, functional RG (FRG), and perturbative QCD (pQCD) in
comparison with different models (see main text and also
Ref.~\cite{Klah06SNM}).\\
\textit{Source:} Left figure taken from Ref.~\cite{Ekst17deltasat} and right figure taken from Ref.~\cite{Leon19SNMfRG}.}
\label{fig:SNM_delta_fRG}
\end{figure}

The importance of 3N interactions for the EOS can be clearly seen in
Figures~\ref{fig:eos_srg_PNM} and \ref{fig:eos_srg_SNM}. In fact, for
calculations based on chiral EFT interactions 3N contributions are the main
driving force of nuclear saturation due to the strong density dependence of 3N
interactions and their overall repulsive character in SNM (see also discussion
in Section~\ref{sec:sep_3N_fits}). The strong density dependence is a direct
consequence of simple phase-space arguments when computing energy
contributions for nuclear matter. Naively, since 3N interactions depend on one
additional particle state compared to NN interactions, energy contributions at
the Hartree-Fock level contain one additional power of density. Of course, for
general NN and 3N interactions this argument is only approximately correct due
to the rather complicated momentum dependence of the interactions. However,
explicit calculations show that this trend is indeed true for realistic
interactions as illustrated in Figure~\ref{fig:EOS_NN_3N_contribution}. Here
we show the individual contributions from the kinetic energy, NN interactions
and 3N interactions as a function of density, calculated within MBPT. While at
small densities the contributions from 3N interactions are very small, as
expected, they become sizable around saturation density and continue to grow
quicker than NN contributions toward higher densities, at least for SNM.

\begin{figure}[t]
\begin{center}
\includegraphics[width=0.44\textwidth]{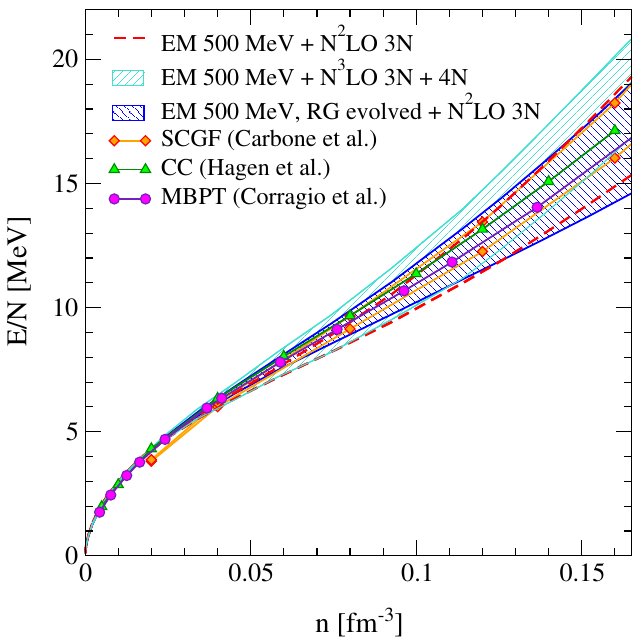}
\hspace{5mm}
\includegraphics[width=0.42\textwidth]{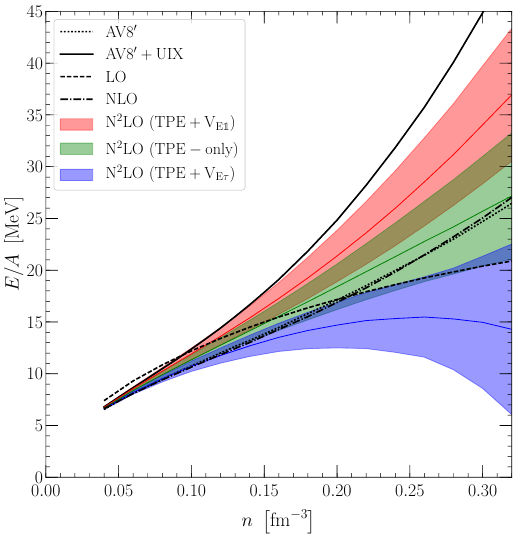}
\end{center}
\caption{Left: Energy per particle of neutron matter based on different chiral
EFT interactions and using different many-body methods. The uncertainty bands
in the left panel show the energy range based on the 500 MeV N$^3$LO NN
potential of Ref.~\cite{Ente03EMN3LO} and including N$^2$LO 3N forces in MBPT
(red lines)~\cite{Hebe13ApJ} or in the SCGF approach~\cite{Carb14SCGFdd}, as
well as including all 3N and 4N interactions to N$^3$LO (cyan
band)~\cite{Tews13N3LO,Krue13N3LOlong}. The blue band shows the results after
RG-evolution of the NN interaction~\cite{Hebe10nmatt}. In addition, we show
results obtained in CC~\cite{Hage14ccnm} and in MBPT~\cite{Cora14nmat}. Right: Auxiliary-field-diffusion Monte-Carlo
results for neutron matter based on chiral NN and 3N interactions at N$^2$LO.
The bands show the uncertainties from the EFT truncation for different choices
of the 3N short-range operator structure. For comparison, shown are also
results at LO and NLO as well as results using the phenomenological AV8' NN
and UIX 3N forces.\\
\textit{Source:} Left figure taken from Ref.~\cite{Hebe15ARNPS} and right figure taken from Ref.~\cite{Tews18CS}.}
\label{fig:PNM_results}
\end{figure}

In Figure~\ref{fig:SNM_delta_fRG} we show results for the equation of state of
symmetric nuclear matter (see also discussion in Section~\ref{sec:sep_3N_fits}
and Figures~\ref{fig:interactions_fit_4He} and \ref{fig:eos_old}). In the left
panel results are shown based on the recently developed NN and 3N interactions
within $\Delta$-full chiral EFT (see
Section~\ref{sec:sep_3N_fits_delta})~\cite{Ekst17deltasat}. Even though only
observables up to $A = 4$ were used in the fit (as in Ref.~\cite{Hebe11fits}),
the interactions provide results in remarkable agreement with the empirical
region. Furthermore, the results indicate that the inclusion of the $\Delta$
degree of freedom might lead to an accelerated convergence consistency of the
chiral expansion in the $\Delta$-full formulation. In the right panel results
are shown for the pressure of symmetric nuclear matter based on the chiral EFT
interactions of Ref.~\cite{Hebe11fits} (blue bands and lines at low density),
results from first nonperturbative functional RG calculations directly based
on QCD at higher densities (red band), as well as various relativistic
mean-field calculations (see Ref.~\cite{Leon19SNMfRG} for details).
Interestingly, the chiral EFT results show a remarkable consistency with
results obtained from the functional-RG calculations and suggest that a simple
interpolation between the two regimes might be possible in order to obtain a
comprehensive estimate of EOS uncertainties over the entire density range. In
addition, the natural emergence of a maximum in the speed of sound $c_S$ at
supranuclear densities with a value beyond the asymptotic value $c_S^2 =
\tfrac{1}{3}$ is found. The existence of such a maximum has also been
predicted for neutron-rich matter, only based on the observation of heavy
neutron stars~\cite{Beda15NS,Tews18CS,Grei18EOSsens}.

The left panel of Figure~\ref{fig:PNM_results} shows the energy per particle
of neutron matter up to saturation density. The results are obtained with
different many-body methods (see caption). For the results shown with bands,
the theoretical uncertainty of the energy is dominated by uncertainties in the
low-energy couplings $c_1$ and $c_3$, which specify the long-range
two-pion-exchange parts of 3N forces (see Section~\ref{sec:chiral_EFT}).

These results show that chiral EFT interactions provide strong constraints
for the EOS of neutron matter, which are consistent among different many-body
methods and considering variations of the Hamiltonian. The significant overlap
of the red lines and the blue band indicates that neutron matter is, to a good
approximation, perturbative for chiral NN interactions with $\Lambda = 500$
MeV. These results have also been benchmarked against first QMC calculations
with local chiral EFT interactions~\cite{Geze13QMCchi,Geze14long} (see also
right panel). In addition, neutron matter calculations using in-medium chiral
perturbation theory approaches results provide similar
results~\cite{Holt13PPNP}. In the right panel we show results for neutron
matter using local chiral interactions at LO, NLO, and N$^2$LO with three
different parametrizations for the 3N interactions specified in
Ref.~\cite{Lynn16QMC3N}. The three different choices correspond to equivalent
parametrizations in the limit of infinite momentum cutoff scale or,
equivalently, in the absence of any coordinate-space regulator functions.
However, for finite cutoff scales the different choices clearly have a
significant effect on the results due to the loss of Fierz-arrangement
symmetries for the employed local regulator functions~\cite{Tews18CS} (see
also discussion in Section~\ref{sec:local_momentum}). Furthermore, the results
show that the uncertainty estimates for the energy quickly increase for
densities beyond $n \approx n_0 = 0.16$ fm$^{-3}$.

\begin{figure}[t]
\begin{center}
\includegraphics[width=0.45\textwidth]{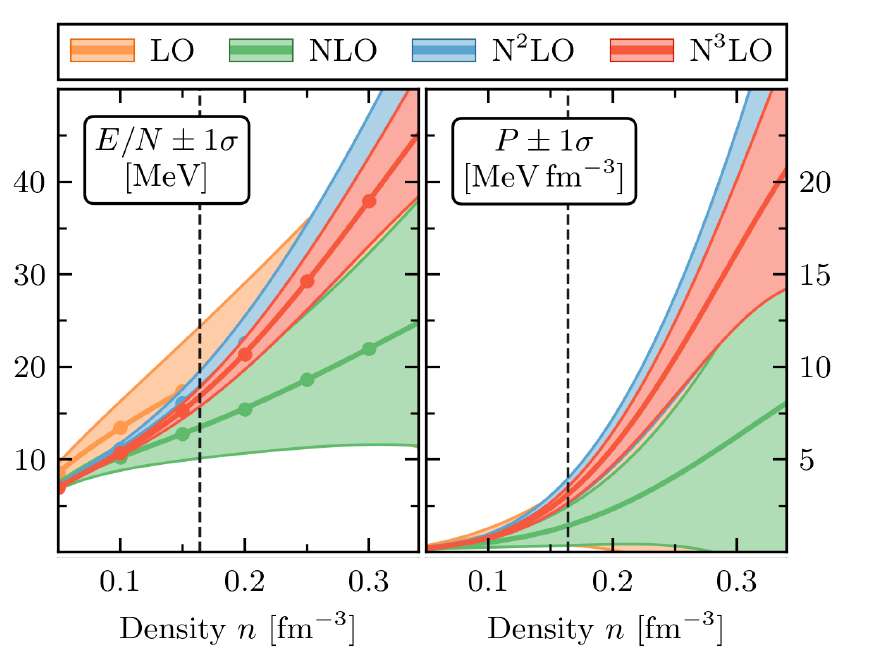} ~
\includegraphics[width=0.45\textwidth]{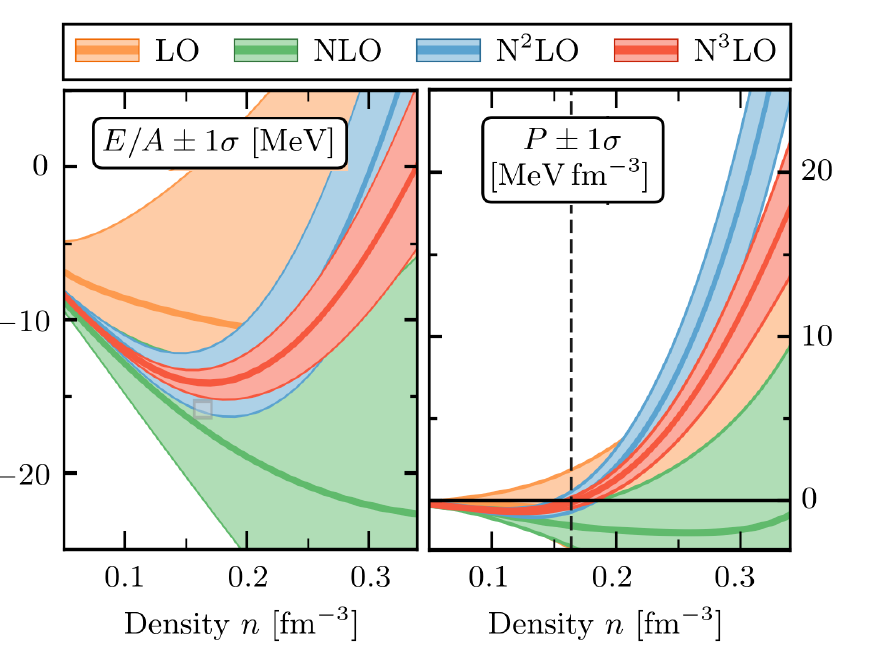}
\end{center}
\caption{68\% credible intervals for the energy per particle of PNM (left)
and SNM (right) at different orders of the chiral EFT expansion. The bands
have been determined based on the results of Ref.~\cite{Dris17MCshort}.
The gray box denotes the empirical saturation region $n_0 = 0.164 \pm 0.007
\text{fm}^{-3}$ with $E/A (n_0) = -15.86 \pm 0.57$ MeV obtained from a
selection of energy density functionals (see Ref.~\cite{Dris20Bayeslong} for details). The
vertical lines mark the density $n_0 = 0.164 \, \text{fm}^{-3}$ for guidance.\\
\textit{Source:} Left figure taken from Ref.~\cite{Dris20Bayesshort} and right figure adapted from
Ref.~\cite{Dris20Bayeslong}.}
\label{fig:EOS_Bayesian}
\end{figure}

In Refs.~\cite{Dris20Bayesshort,Dris20Bayeslong} a new framework for
quantifying theoretical uncertainties for infinite matter was presented. Based
on the calculations of Ref.~\cite{Dris17MCshort} a Bayesian framework was
employed to extract statistical uncertainties. This approach allows to account
for correlations between EOS truncation errors across different observables
and different densities~\cite{Mele19corr}. Figure~\ref{fig:EOS_Bayesian} shows
the results for the 68\% credible intervals at different orders in the chiral
EFT expansion as a function of density for PNM (left)~\cite{Dris20Bayesshort}
and SNM (right)~\cite{Dris20Bayeslong}. In agreement with the findings of
Ref.~\cite{Furn15uncert} the uncertainty bands turn out to be in reasonable
agreement with the estimates resulting from the EFT error prescription
proposed in Refs.~\cite{Epel14SCSprl,Epel15improved} as done in
Ref.~\cite{Dris17MCshort} (see also Ref.~\cite{Epel19Bayes}). The employed
framework also allows to give posterior distributions for the breakdown scale
$\Lambda_b$, which enters the in-medium EFT expansion parameter in form of the
relation $Q(k_F) = \frac{k_F}{\Lambda_{\text{b}}}$. The obtained posterior
distributions range around $\Lambda_{\text{b}} = 600$ MeV and are hence consistent with
determinations from NN scattering observables.

\begin{figure}[t]
\begin{center}
\includegraphics[width=0.45\textwidth]{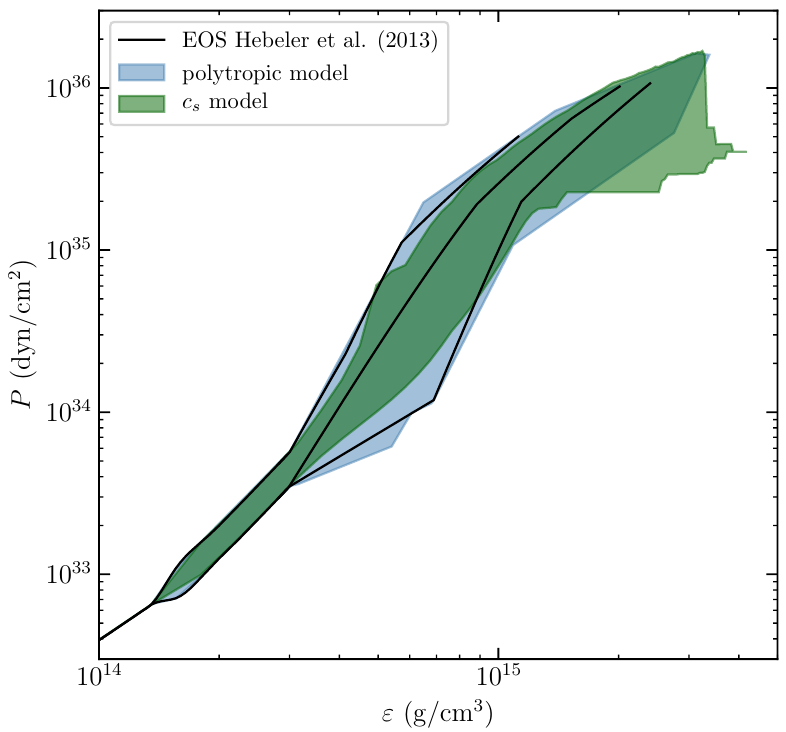}
\hspace{5mm}
\includegraphics[width=0.45\textwidth]{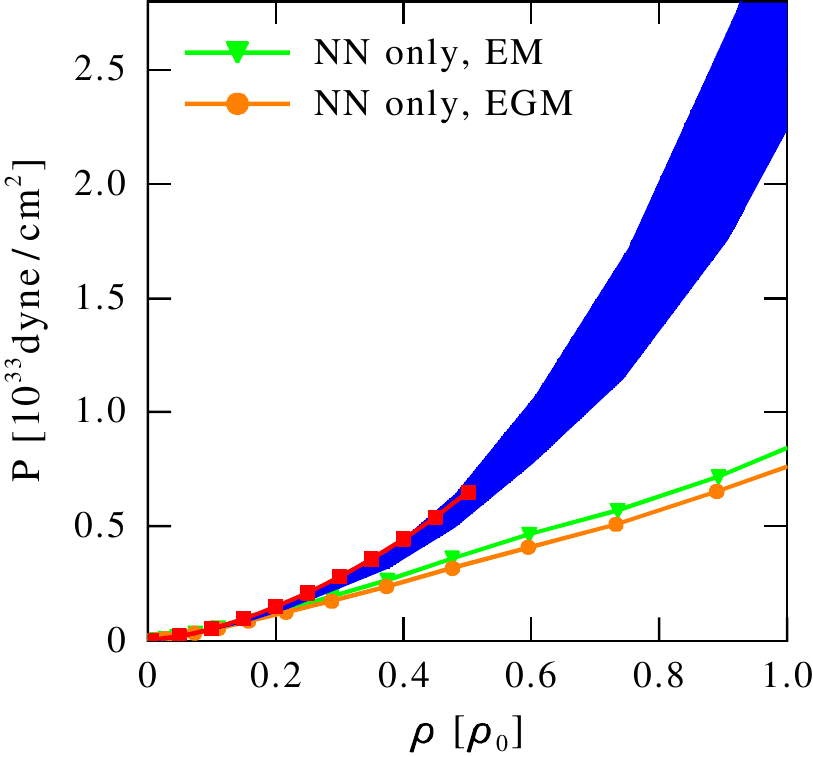}
\end{center}
\caption{Left: Comparison of the ranges for the EOS
based on the piecewise polytropic (PP) and speed-of-sound ($c_S$) high-density
parametrizations, after including constraints from causality and neutron star
masses (see Ref.~\cite{Grei18EOSsens} for details). The black lines show the
three representative equations of state of Ref.~\cite{Hebe13ApJ}, the light
blue bands show the uncertainty ranges resulting from the PP parametrization
and the darker green bands those from the $c_S$ parametrization. Right: Pressure of neutron star matter based
on chiral low-momentum interactions for densities up to nuclear saturation
density $\rho_0$. The blue band estimates the theoretical uncertainties from
many-body forces and from truncations in the many-body calculation. At low
densities, the results are compared to a standard BPS crust equation of
state~\cite{Baym71BPS,Nege73BPS} (red line), while the green and orange lines
show results based on only NN interactions.\\
\textit{Source:} Left figure taken from Ref.~\cite{Grei18EOSsens} and right figure taken from
Ref.~\cite{Hebe10PRL}.}
\label{fig:pressure_astro}
\end{figure}

The systematic increase of the uncertainty bands indicates that the expansion
parameter $Q$ becomes sizable around $n_0$ such that the chiral EFT expansion
gradually becomes inefficient with increasing density. However, for
astrophysical applications the equation of state is required over a density
regime significantly beyond the range shown in Figures~\ref{fig:PNM_results}
and \ref{fig:EOS_Bayesian}. In order to extend the EOS to densities beyond the
regime accessible by chiral EFT interactions, there are two complementary
options:
\begin{itemize}
\item Performing microscopic calculations at higher densities using
frameworks based on degrees of freedom relevant at these density scales.
Generally, this approach has the advantage providing direct insight
into the composition and properties of matter. On the other hand, the results
of such calculations are usually strongly model and scheme dependent.
\item Parametrizing the high-density part in some general way and constraining
the values of the free parameters using astrophysical observations and general
considerations like causality. This strategy assumes that the parametrization
is sufficiently complete such that all possible relevant EOSs can be
described. Generally, this approach has the advantage that it does not depend on
any assumptions regarding the degrees of freedoms and their interactions in
matter at high densities. On the downside, it does not provide any direct
insight into the microphysics of matter at supranuclear densities.
\end{itemize}

Some examples of the first approach for symmetric nuclear matter are shown in
the right panel of Figure~\ref{fig:SNM_delta_fRG}, in particular the
functional-RG results at supranuclear densities, extracted directly from the
quark-gluon dynamics described by QCD. The second approach was applied, e.g.,
in Refs.~\cite{Hebe10PRL,Hebe13ApJ,Tews18CS,Grei18EOSsens}. In these works,
two different parametrizations were developed. The first is a piecewise
polytropic form, which parametrizes the pressure as a function of baryon
density $n$ in the form $P
\sim n^{\Gamma}$~\cite{Read09polyEOS,Hebe13ApJ}, where $\Gamma$ is a free
parameter that controls the stiffness of the EOS in a given density regime.
The second parametrization is based on the speed of sound
$c_s$~\cite{Tews18CS,Grei18EOSsens}.

In the left panel of Figure~\ref{fig:pressure_astro} we show the EOS
uncertainty bands using these two high-density parametrizations based on
chiral EFT results of Refs.~\cite{Hebe10nmatt,Hebe13ApJ} up to around
saturation density. The key external constraints that determine the
uncertainty bands at higher density are the mass $M=1.97 M_{\odot}$ of the most
massive neutron star measured to date~\cite{Anto13PSRM201} and causality
considerations~\cite{Hebe13ApJ,Grei18EOSsens}. Generally, the uncertainty
bands are determined by both the low-density results that act as an anchor
point as well as the constraints from causality and neutron star
observations. We note that the results correspond to the EOS constraints of
neutron-star matter. Details on how the finite proton fraction was determined
and incorporated in the calculations can be found in Refs.~\cite{Hebe13ApJ}.
In neutron stars, matter consists of nuclei embedded in a sea of electrons at
low densities in the outer crust, while the nuclei become increasingly
neutron-rich structures in the inner crust~\cite{Baym71BPS,Nege73BPS}. The
transition to homogeneous neutron-rich matter happens around half nuclear
saturation density~\cite{Hebe13ApJ}. In the right panel of
Figure~\ref{fig:pressure_astro} the results for the pressure in this density
region are shown. Clearly, for the employed low-resolution interactions the
inclusion of 3N interactions is crucial for a continuous transition between
these two states of matter.

\begin{figure}[b!]
\begin{center}
\includegraphics[width=0.95\textwidth]{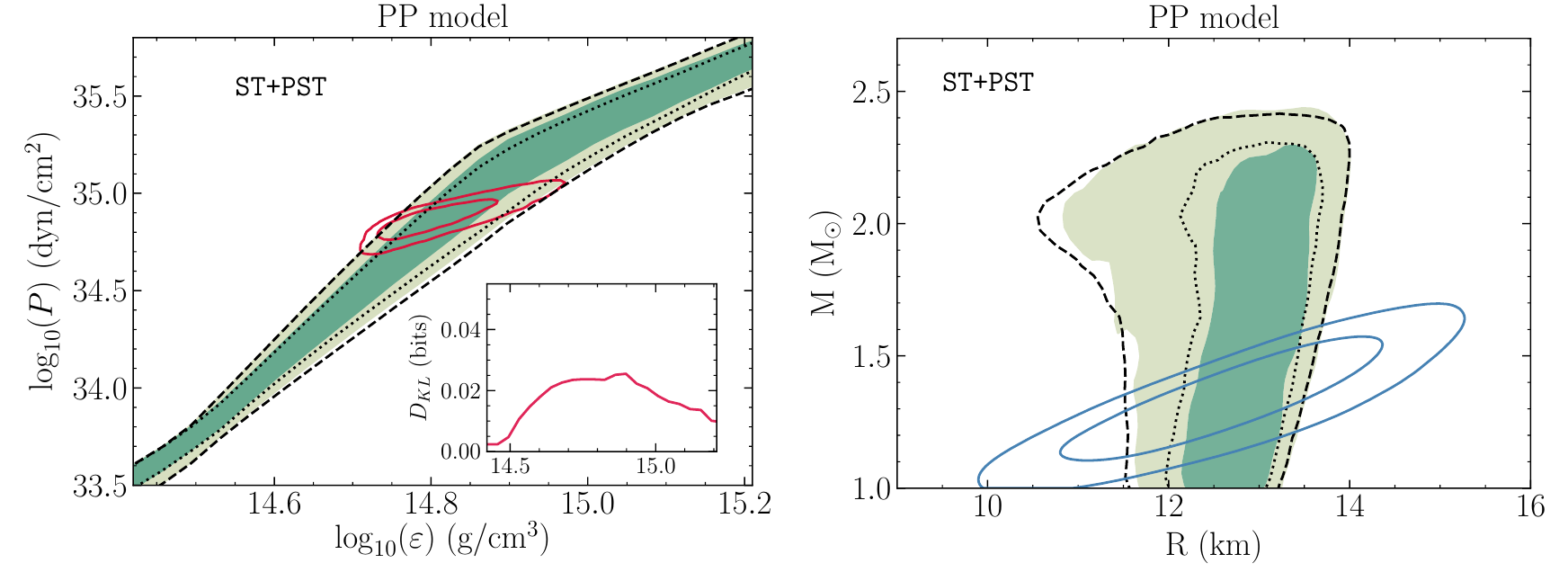}
\end{center}
\caption{Posterior distributions for the equation of state (left) and the
mass-radius relation (right) for the piecewise polytropic parametrization. The
dark/light green bands show the 68\% and 95\% posterior credible intervals,
while the black dotted and dashed lines respectively indicate the joined 68\%
and 95\% prior credible intervals. The red contours in the left panel denote
the corresponding credible regions of central energy density and central
pressure, while the blue contours in the right panel show the likelihood
functions of the PSR J0030+0451 measurement. ST+PST refers to the employed hot
region model.\\
\textit{Source:} Figures taken from Ref.~\cite{Raai19NICER}.}
\label{fig:NICER}
\end{figure}

\begin{figure}[t!]
\begin{center}
\includegraphics[width=0.9\textwidth]{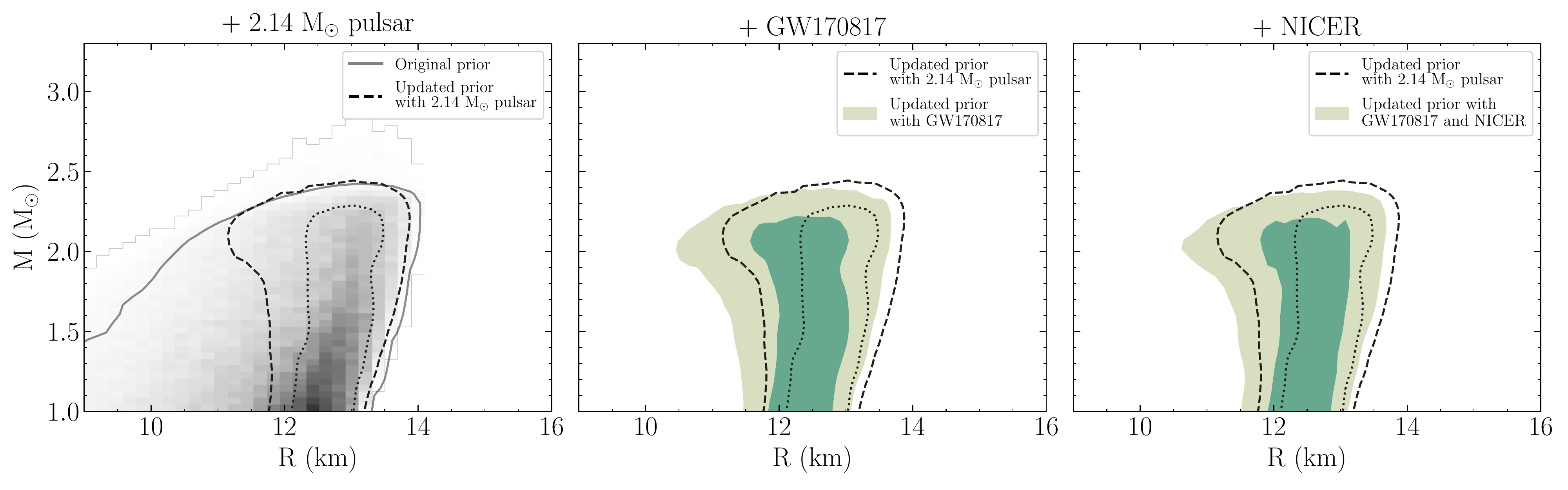} \\
\includegraphics[width=0.9\textwidth]{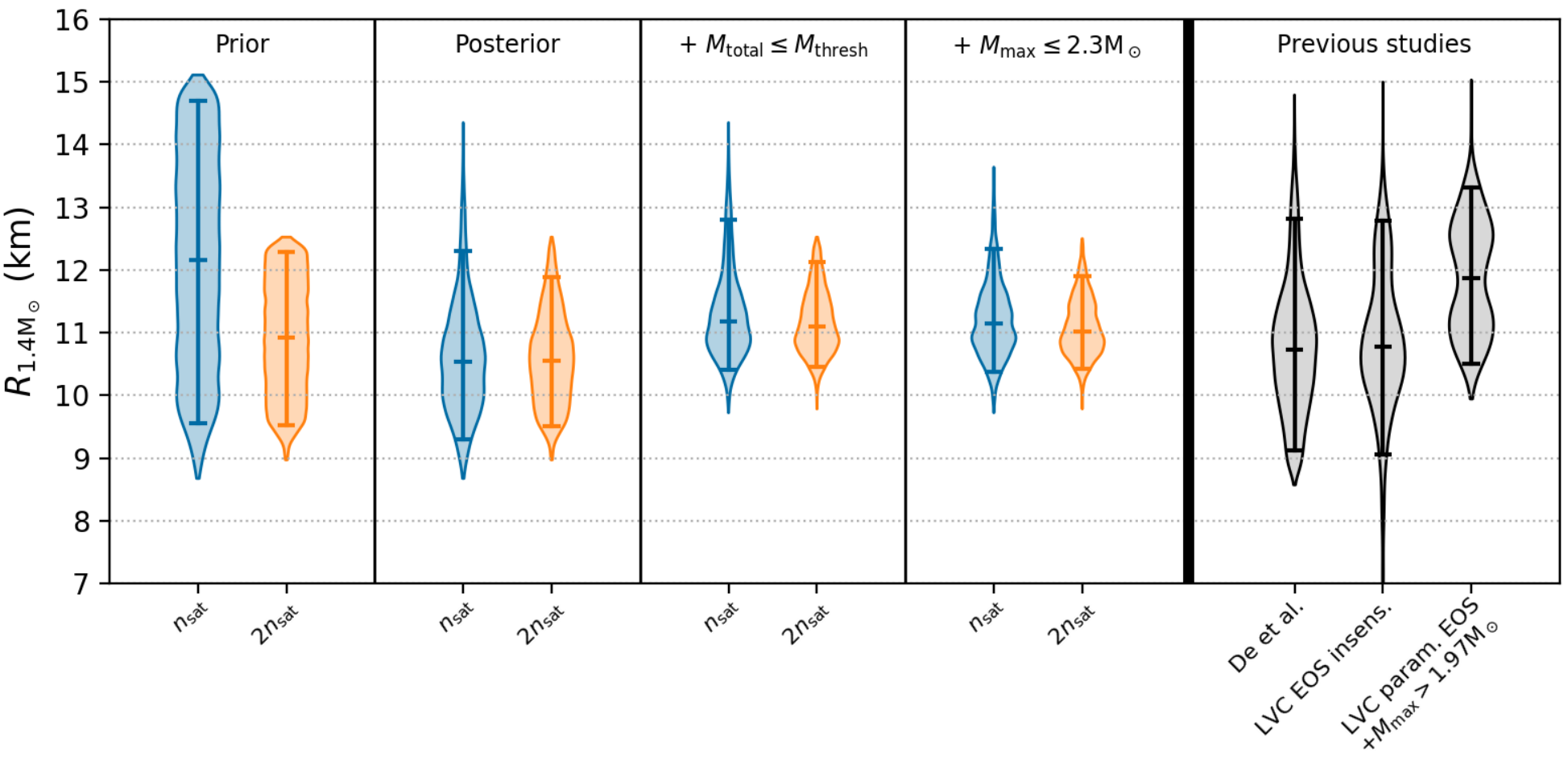} \\
\end{center}
\caption{Upper panel: Mass-radius posterior distributions conditional the PP
model (compare Figure~\ref{fig:NICER}) and given: (i) the $2.14$~M$_{\odot}$
pulsar alone (left panel); (ii) inclusion of the GW170817 measurements (middle
panel); and (iii) inclusion of the mass and radius of PSR J0030+0451 inferred
by the NICER data of Ref.~\cite{Rile19NICER} (right panel). The panels show
how the posterior distributions update the prior distributions. The contours
indicate the $68\%$ and $95\%$ credible intervals as in
Figure~\ref{fig:NICER}. Lower
panel: Posterior distributions for the radius of a $1.4 M_{\odot}$ neutron
star based on microscopic calculations (left) and after incorporating
information from gravitational-wave analyses and from constraints on the total
mass being smaller than the threshold mass $M_{\text{thresh}}$ for a prompt
collapse after merging (see Ref.~\cite{Capa19multimess} for details). The blue
bands show the results based on microscopic calculation up to nuclear
saturation density, while for the the orange bands the calculations are
extended to twice nuclear saturation density.\\
\textit{Source:} Upper figure taken from Ref.~\cite{Raai2019NICERLIGO} and lower figure taken from Ref.~\cite{Capa19multimess}.}
\label{fig:NS_combined}
\end{figure}

Recent breakthroughs like the first detection of the gravitational wave signal
of the binary neutron star merger
GW170817~\cite{LIGO17NSmergers,LIGO18NSradii} as well as ongoing missions
aiming at first direct neutron star radius measurements using x-ray
timing~\cite{Watt16RMP,NICER,NICER2} are expected to significantly enhance our
theoretical understanding of neutron-rich matter under extreme conditions.
Combining information from these ongoing efforts with existing observational
data like precise mass measurements of heavy neutron stars allows to
systematically tighten the EOS uncertainty bands, like those shown in the left
panel of Figure~\ref{fig:NICER}. Very recently, the mass and radius of the
millisecond pulsar PSR J0030+0451 have been inferred via pulse-profile
modeling of X-ray data obtained by NASA's NICER mission~\cite{Rile19NICER}.
In Ref.~\cite{Raai19NICER} the implications of this first measurement on the
EOS constraints were investigated using a Bayesian analysis (see also
Ref.~\cite{Mill19NICER}). Figure~\ref{fig:NICER} shows the posterior
probability distributions for the pressure and the mass-radius relation
obtained from a combination of constraints from chiral EFT interactions at
lower densities and information from the new NICER observations. The posterior
distributions show that not much information is gained over the prior from
these first measurements. In particular, from the distributions shown in the
left panel of Figure~\ref{fig:NICER} we observe that the changes from the
prior to the posterior are not very significant at nuclear densities, so that
from this analysis it is not possible to draw robust conclusions about further
constraints on dense matter and interactions within chiral EFT. Observations
of alternative pulsars, like PSR J$0437-4715$, which has a tight mass
constraint derived via radio-timing, promise to provide tighter constraints on
the EOS of dense matter in the near future.

The most promising approach for deriving stringent constraints on the EOS at
supranuclear densities consists in combining all available observational
constraints from neutron star mass observations, gravitational wave detections
and electromagnetic signals. The upper panel of Figure~\ref{fig:NS_combined}
shows the neutron star mass-radius uncertainty bands after incorporating the
recent detection of the heaviest know neutron star with mass $M=2.14
M_{\odot}$~\cite{Crom19heavyNS} (left), the updated posterior distributions
after incorporating the constraints from the GW170817 signal (center) and
finally the posterior after inclusion of the NICER signals
(right)~\cite{Raai2019NICERLIGO}. The results show that most information is
gained from including the $2.14 M_{\odot}$ pulsar. The binary merger GW170817
favors softer EOS than the prior, but the measured radius from PSR J0030+0451
favors stiffer EOS, resulting in a final posterior distribution very similar
to the prior (see Ref.~\cite{Raai2019NICERLIGO} for details). The lower panel
of Figure~\ref{fig:NS_combined} shows the radius posterior distributions of a
$1.4 M_{\odot}$ neutron star after including constraints from GW170817 and
from the constraint that the estimated total mass $M_{\text{total}}$ to be
less than the threshold mass for prompt collapse $M_{\text{thresh}}$ (see
Ref.~\cite{Capa19multimess} for details). The shown results in particular
highlight the importance of the value for the critical upper density up to
which the microscopic calculations based on NN and 3N interactions are being
trusted, i.e. $n_{\text{sat}}$ (blue) versus $2 n_{\text{sat}}$ (orange) (see
also discussion in Sect.~\ref{sec:open_questions}).

\begin{figure}[t!]
\begin{center}
\includegraphics[width=0.42\textwidth]{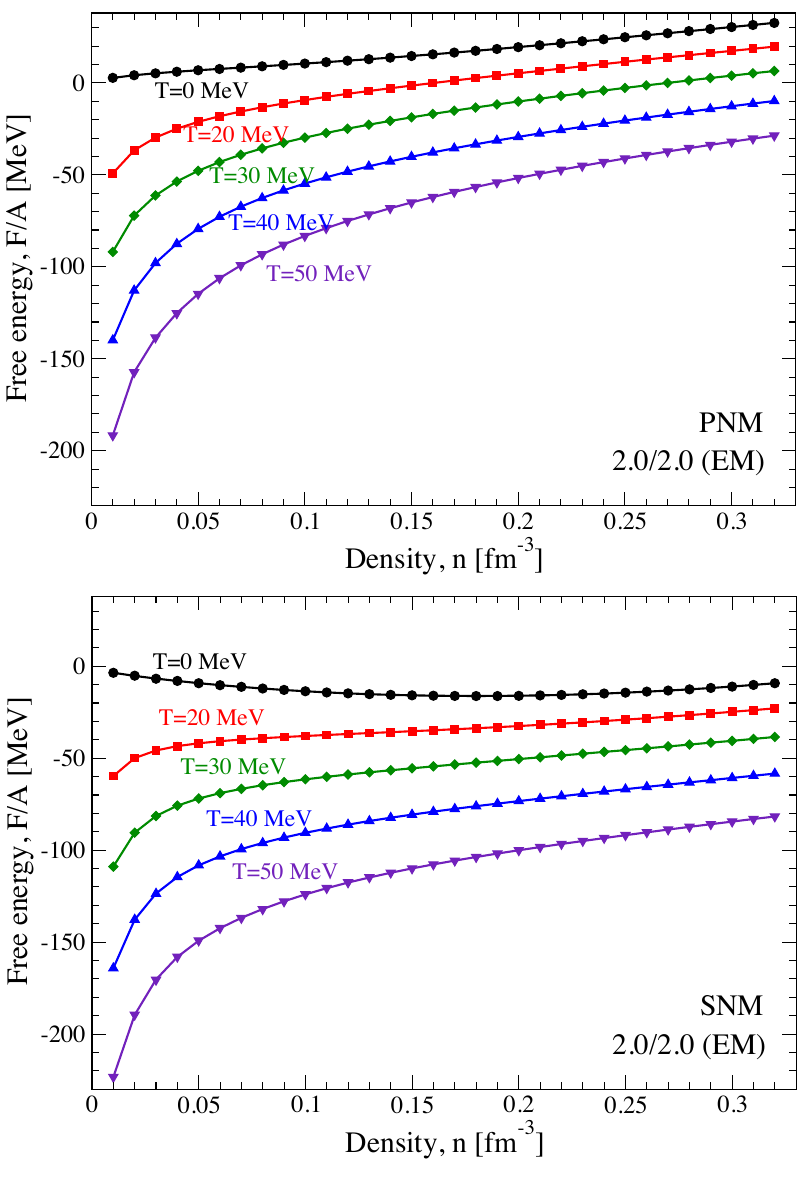} ~
\includegraphics[width=0.5\textwidth]{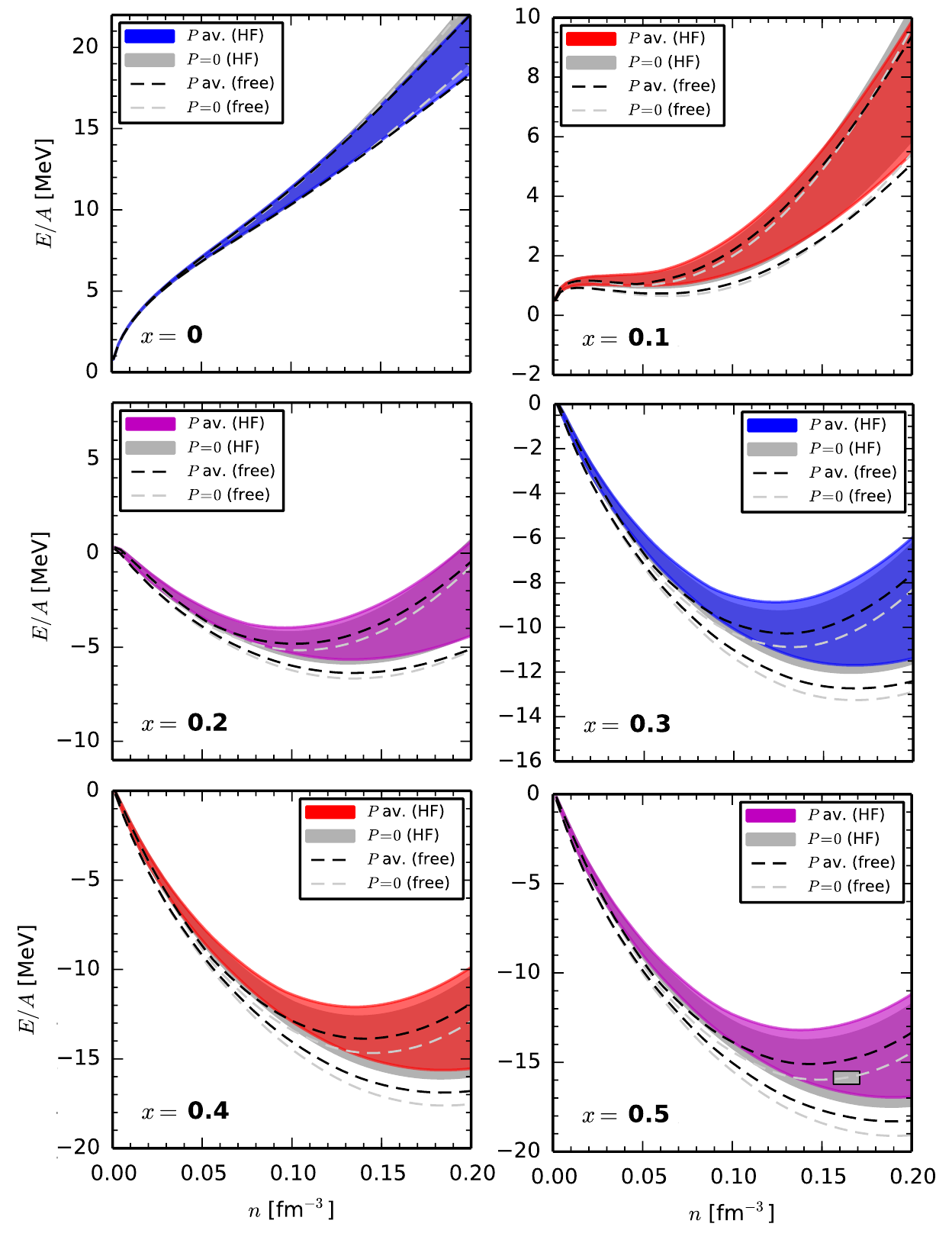}
\end{center}
\caption{Left: Free energy per nucleon for $T = 20,30,40$ and $50$ MeV as
function of density for PNM (upper panel) and SNM (lower panel). The results
are based on SCGF calculations using the ``$2.0/2.0$(EM)'' interaction of
Ref.~\cite{Hebe11fits}. Right: The energy per particle of isospin asymmetric nuclear matter as a
function density for different proton fractions $x = n_p/(n_n + n_p)$ (see
main text). The box in the figure for $x=0.5$ shows the empirical saturation
region. For details see Ref.~\cite{Dris16asym}.\\
\textit{Source:} Left figure taken from Ref.~\cite{Carb19EOSthermal} and right figure adapted from
Ref.~\cite{Dris16asym}}
\label{fig:EOS_asymm_finiteT}
\end{figure}

For the study of individual neutron stars it is sufficient to study the EOS at
zero temperature in the very neutron-rich regime (up to about 10\% protons).
Neutrino emission quickly cools new-born neutron stars to an interior
temperature of less than 1 MeV~\cite{Lattimer86birth}, which is much smaller
than the Fermi energy of a nucleon at saturation density $E_{\text{F}}
\sim 35$ MeV. However, for neutron-star merger events or supernovae explosions
the temperatures can be significantly higher such that EOS data at finite
temperature is required as the fundamental input for theoretical simulations
of such events. Conceptually, the extension of many-body calculations to
finite temperatures is straightforward, at least within MBPT or SCGF. The
particle distribution functions in momentum space become Fermi-Dirac functions
and the fundamental quantity of interest is now the free energy $F$. In the
left panel of Figure~\ref{fig:EOS_asymm_finiteT} we show the results of SCGF
calculations for the free energy of PNM and SNM at different temperatures as a
function of density based on chiral NN and 3N interactions (see
Ref.~\cite{Carb19EOSthermal} for details). Obviously, as soon as temperatures
of about $10-20$ MeV are reached, the temperature effects for the EOS become
quite significant (see also Refs.\cite{Well14nmtherm,Well15therm}). One key
challenge for the future will be to systematically extend such
finite-temperature results systematically to higher densities like at zero
temperature. However, the situation at finite temperature is somewhat more
complicated since so far no direct observational data exists that could be
used as a constraint at high densities. One promising theoretical approach
consists in extending calculations of Ref.~\cite{Leon19SNMfRG} to neutron-rich
systems and compute the EOS at finite temperature and high densities directly
based on QCD using the nonperturbative functional RG framework.

The ultimate challenge consists in deriving systematic uncertainty bands for
the EOS over the full range of densities, temperatures and proton fractions
relevant for astrophysical application. The bands should be consistent with
theoretical calculations at lower densities and with all available
astrophysical observations like, e.g., measured neutron star properties.
Extending many-body frameworks for infinite matter to general proton fractions
is in principle straightforward. However, for general proton and neutron
densities the system has fewer symmetries and the calculations hence become
somewhat more complicated and more expensive. In the right panel of
Figure~\ref{fig:EOS_asymm_finiteT} we show results of MBPT calculations for
nuclear matter at zero temperature and general isospin asymmetries based on
chiral NN and 3N interactions (see Ref.~\cite{Dris16asym} for details). The
system is characterized by the proton fraction defined by $x = n_p/(n_n +
n_p)$, where $n_p$ resp. $n_n$ denote the proton and neutron densities, i.e.
$x=0$ corresponds to PNM and $x=0.5$ to SNM. The results clearly illustrate
the increased attraction from nuclear interactions toward larger proton
fractions. Typically, this dependence on the proton fraction can be
approximated quite accurately by a simple quadratic expansion in $x$ around SNM:
\begin{equation}
\frac{E(n,x)}{A} \approx \frac{E(n,1/2)}{A} + (1-2x)^2 S_2(n) \, ,
\label{eq:Esymm_quadratic}
\end{equation}
where $S_2(n)$ is the symmetry energy. Terms beyond the quadratic
approximation based on chiral EFT interactions have been studied in
Refs.~\cite{Kais15qartic,Well16DivAsym}, which involve quartic, sextic and
even logarithmic terms. However, in practice these terms turn out to be rather
small compared to the leading quadratic coefficient such that the
approximation shown in Eq.~(\ref{eq:Esymm_quadratic}) is usually very
reasonable, at least at small temperatures. For higher temperatures it seems
more efficient and reliable to treat the isospin asymmetry of the
non-interacting part exactly and approximate only the isospin dependence of
the interaction contributions~\cite{Well15therm}.

\clearpage 

\section{Summary and outlook}
\label{sec:outlook}

In this work we discussed in detail fundamental techniques for representing and
calculating 3N interactions for \textit{ab initio} studies of nuclei and matter and
reviewed a selection of recent calculations. The presented results clearly
demonstrate the importance of contributions from 3N interactions for nuclear
observables and illustrate the capabilities of state-of-the-art many-body
frameworks. In this last section we will make an attempt to summarize the
current status of \textit{ab initio} nuclear structure theory and draw some general
conclusions based on the discussed results. Finally, we discuss some open
questions and give an overview of current and future directions.

\subsection{Present status and achievements}
\label{sec:status}

Most calculations and results discussed in this work have been made possible
thanks to major advances in two fundamental areas of ab initio nuclear
structure theory:

\begin{itemize}
\item First, the scope of \textit{ab initio} many-body frameworks with respect to the
particle number has significantly increased thanks to the development of novel
many-body expansions with a mild polynomial scaling behavior. In addition,
several existing methods have been generalized to open-shell nuclei and are
being continuously extended such that additional observables can be studied.
Presently, the scope of many-body methods formulated in a harmonic oscillator
basis representation (see Section~\ref{sec:Intro}) are limited to the regime
$A \lesssim 100$ due to the achievable basis sizes for representing 3N
interactions.

\item Second, in recent years a significant number of novel NN and 3N
interactions have been developed, derived within different regularization
schemes, using new advanced fitting strategies for the low-energy couplings,
as well as in chiral EFT formulations with and without explicit inclusion of
$\Delta$ degrees of freedom. Most of these interactions are made available at
different orders of the chiral expansion and at different cutoff scales. This
allows to perform order-by-order calculations, which can then be used to
estimate theoretical chiral EFT uncertainties. This situation is very
different from the time about 15 years ago, when only very few interactions
were available for one specific regularization scheme and at one particular
chiral order and cutoff scale. Furthermore, the methods discussed in this work
allow to compute matrix elements of 3N interactions at higher orders in the
chiral expansion and for sufficiently large basis sizes, such that the
structure of few-body systems, medium-mass nuclei and matter as well as
few-body reactions can be studied based on the same NN and 3N interactions.
\end{itemize}

The main general lessons learned from \textit{ab initio} studies in recent years can be
summarized as follows:
\begin{itemize}
\item Overall, a remarkable agreement is found between results based on
different many-body frameworks for a given low-resolution nuclear interaction
(see Figure~\ref{fig:Tichai_compare}), which implies that the many-body
uncertainties of these calculations are small. Whenever results based on such
interactions show a significant discrepancy to experiment, under consideration
of the theoretical uncertainties, they can be mainly attributed to
deficiencies of the employed nuclear interactions and operators.

\item Specific observables can be reproduced remarkably well by particular
interactions over a significant range of the nuclear landscape up to
$^{100}$Sn (see right panel of Figure~\ref{fig:N2LO_sat_nuclei} and left panel
of Figure~\ref{fig:Sn}). However, since these results are based on
interactions at a specific order of the chiral expansion, it is not possible
to assign systematic chiral EFT uncertainty estimates. Furthermore, a deeper
understanding why specific interactions perform well for some observables, but
relatively poorly for others, is still lacking. However, for all these
discussions the inherent theoretical uncertainties due to truncations in the
chiral EFT expansion of the nuclear forces need to be taken into account.

\item It is challenging to derive nuclear interactions capable of simultaneously
predicting different observables of nuclei from the light sector to the
medium-mass regime consistent with experiment. Agreement can be achieved
when information of heavier nuclei is included in the fit of the interaction.
However, in these cases, the accurate reproduction of the nucleon-nucleon phase
shifts at higher energies has to be sacrificed, at least at N$^2$LO (see,
e.g., Figure~\ref{fig:N2LO_sat_nuclei}).

\item Order-by-order calculations based on specific sets of interactions are
able to reproduce simultaneously ground-state energies and charge radii of
lighter nuclei (see right panel of Figures~\ref{fig:LENPIC_QMC} and
Figures~\ref{fig:radius_QMC_O} and \ref{fig:radius_EMN_3N}). It will be
interesting to further explore such interactions, study additional observables
of nuclei and matter and investigate in greater detail the role of
regularization schemes and scales for nuclear interactions.

\item  While theoretical predictions for systems with a significant proton
fraction, such as atomic nuclei and isospin-symmetric  matter, generally
depend more sensitively on properties of the employed interactions, the
results for pure neutron systems or very neutron-rich systems exhibit a
remarkable insensitivity to such details and are in good agreement with
present experimental constraints~\cite{Tsan12esymm,Latt12esymm,Hebe14Esymm}.
These findings suggest that, e.g., predictions for the EOS of neutron-rich
matter up to about nuclear saturation density are robust and rather well
constrained.

\end{itemize}

\subsection{Open questions and future directions}
\label{sec:open_questions}

\begin{figure}[t!]
\begin{center}
\includegraphics[width=0.50\textwidth]{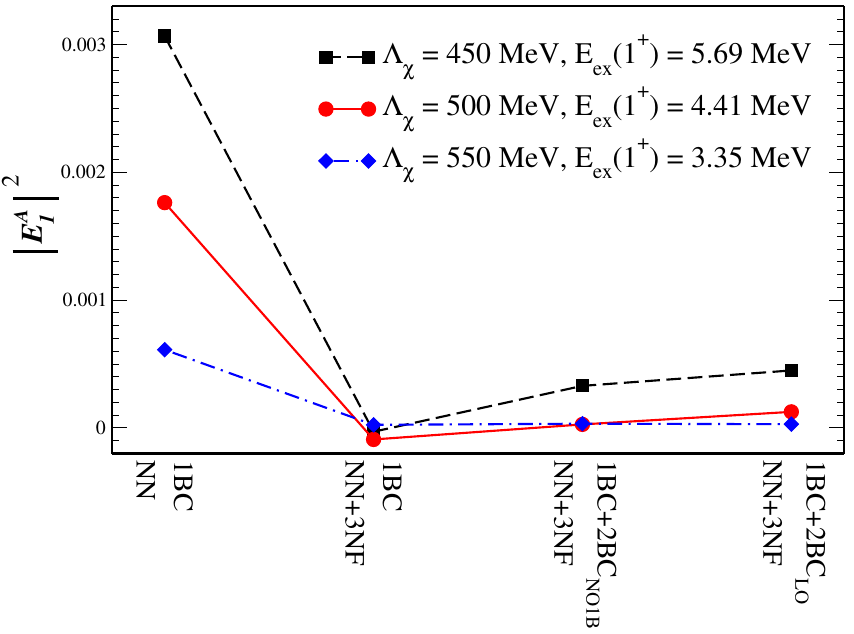}
\hspace{5mm}
\includegraphics[width=0.41\textwidth]{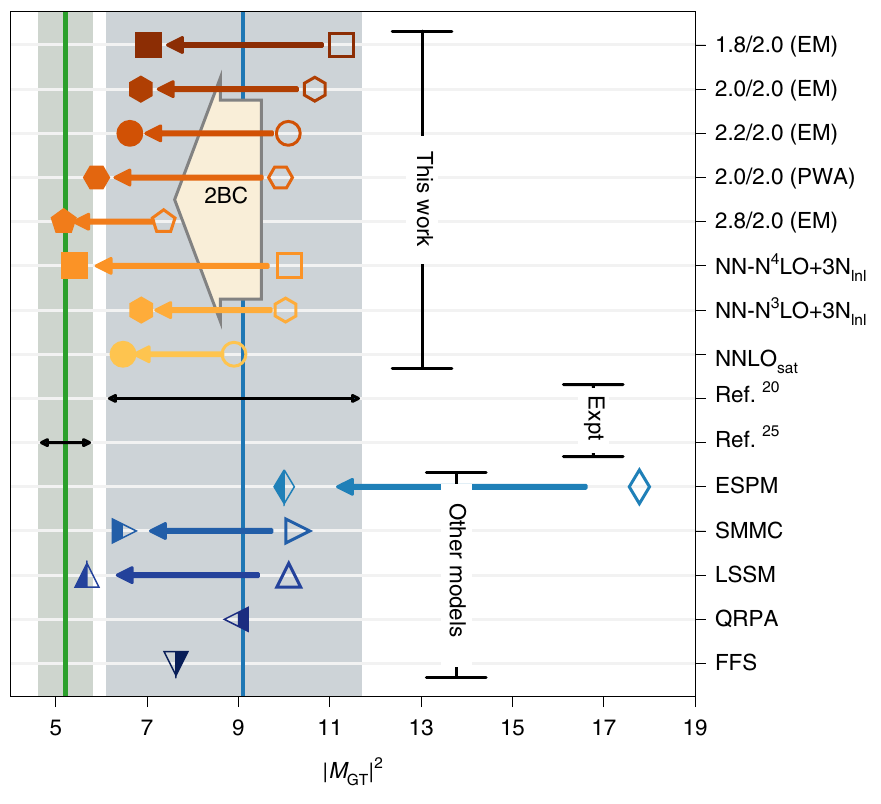}
\end{center}
\caption{Left: Squared $\beta$-decay matrix element of $^{14}$C obtained from
increasingly sophisticated calculations, ranging from NN-only calculations
including only one-body current contributions (``1BC'', left), to NN+3N
interactions including one- and leading-order two-body contributions (``2BC'',
right). Here ``2BC$_{\text{NO1B}}$'' refers to the normal ordering
approximation of the 2BC contributions, see Ref.~\cite{Ekst14GT2bc} for details. Right: Gamow-Teller
$\beta$-decay matrix elements of $^{100}$Sn calculated within coupled cluster.
Unfilled symbols show results obtained with the standard one-body Gamow-Teller
operator, while filled symbols correspond to results including two-body
currents. Partially filled symbols show the results obtained by multiplying
the computed Gamow-Teller strength in the given works (see
Ref.~\cite{Gysb19beta}) by a phenomenological quenching factor (see discussion
in Section~\ref{sec:beta_decay}).\\
\textit{Source:} Left figure taken from Ref.~\cite{Ekst14GT2bc} and right figure taken from Ref.~\cite{Gysb19beta}.}
\label{fig:gamow_2bc}
\end{figure}

Despite all the achievements discussed in the previous section there are still
fundamental open questions and challenges. Here we discuss some of them,
without any claim to completeness:

\begin{itemize}
\item So far, almost all calculations based on chiral EFT interactions have
been based on Weinberg's power counting scheme (see
Section~\ref{sec:chiral_EFT}). There are currently efforts to explore
alternative power counting formulations with an improved RG invariance, which
allow to vary the cutoff scales over a much wider range~\cite{Hamm19Rev}. The
development of these interactions involve the promotion of counter terms in
Weinberg's scheme to lower orders in the expansion. Furthermore, many-body
frameworks based on such interactions need to be modified since contributions
beyond leading order have to be treated perturbatively. The availability of
such Hamiltonians could be key for gaining a deeper understanding of the
deficiencies of presently used interactions.

\begin{figure}[t!]
\begin{center}
\includegraphics[width=0.39\textwidth]{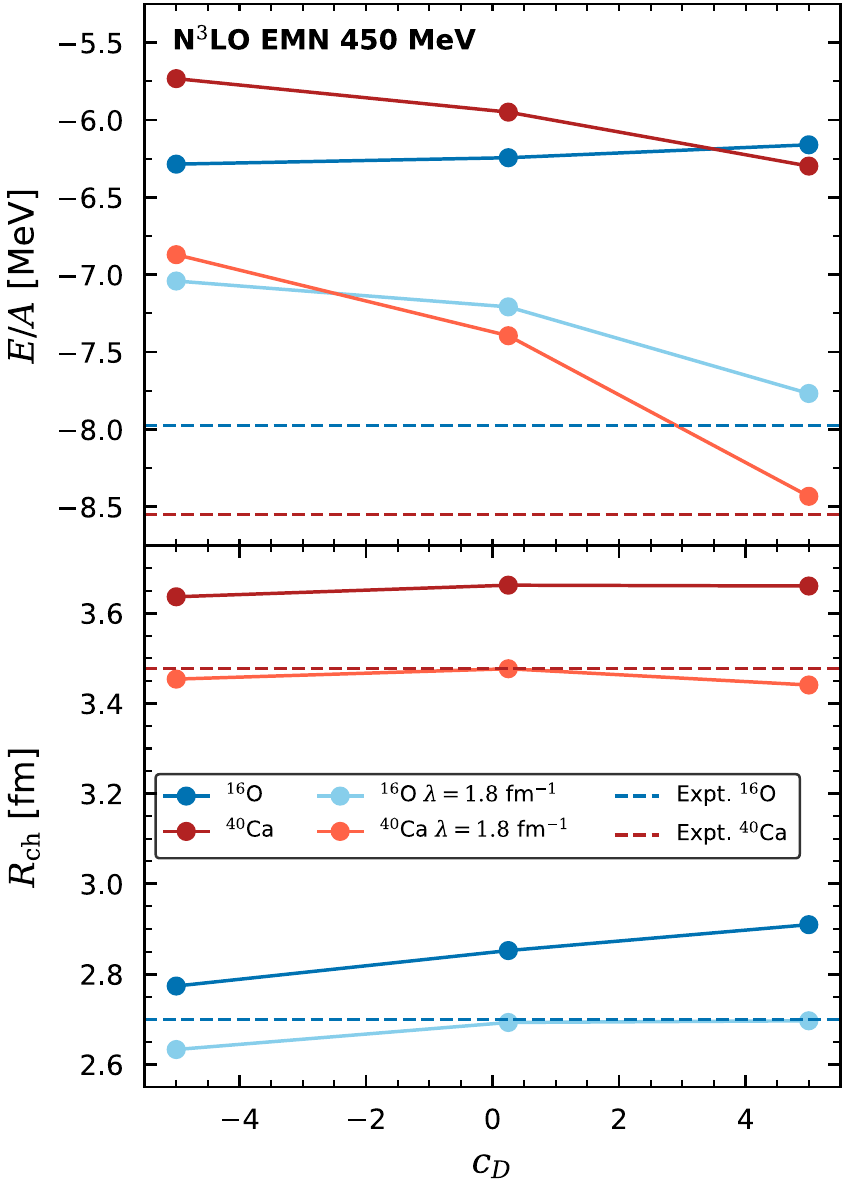}
\hspace{1cm}
\includegraphics[width=0.39\textwidth]{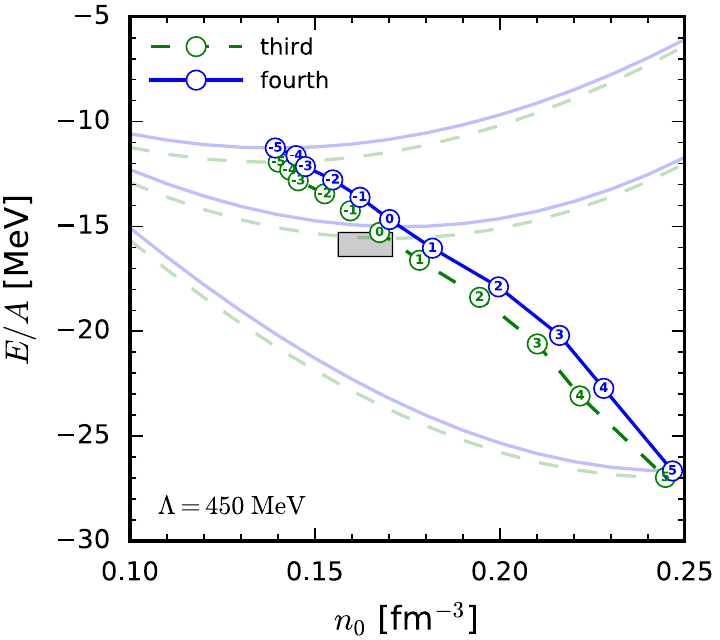}
\end{center}
\caption{Left: Ground-state energies (top panel) and charge radii (bottom panel) for
$^{16}$O and $^{40}$Ca as a function of $c_D$ (with corresponding $c_E$ value
from the $^3$H binding energy, see Figure~\ref{fig:cd_ce}). Results are shown
for unevolved and SRG-evolved potentials with $\lambda_{\text{SRG}} = 1.8 \:
\text{fm}^{-1}$ in comparison with experimental values. Right: Theoretical
saturation points of symmetric nuclear matter as a function of the coupling
$c_D$. Also shown is the energy per particle for the three representative
values, $c_D = -5, 0.25$, and $5$, at third and fourth order in MBPT.\\
\textit{Source:} Figures taken from Ref.~\cite{Hopp19medmass}.
}
\label{fig:matter_nuclei}
\end{figure}

\item In most many-body calculations the contributions to nuclear
interactions and currents in the chiral expansion are not treated on equal
footing. While there have been recent efforts to include systematically
higher-order contributions to operators like the Gamow-Teller operator (see
also Section~\ref{sec:beta_decay}), these calculations are still at their
infancy. In addition, higher-order long-range corrections to axial-vector
currents have been studied in Refs.~\cite{Klos13longSD,Hofe18WIMP} in the
context of WIMP-nucleus scattering as part of dark matter
searches~\cite{Fieg18xenon}. Figure~\ref{fig:gamow_2bc} illustrates that
contributions from higher-order two-body currents (2BC) can have significant
effects on transition strengths in nuclei. In the left panel the effects of
two-body currents on the Gamow-Teller matrix element in $^{14}$C are
illustrated, while the right panel shows the effect of 2BC on the Gamow-Teller
strength in $^{100}$Sn. The figures show that higher-order contributions can
have a significant impact on the results. While for anomalously small matrix
elements like for $^{14}$C these contributions can enhance the results by a
sizable factor (see also left panel of Figure~\ref{fig:beta_decay} and related
discussion), for $^{100}$Sn they lead to an effective suppression factor,
which is in good agreement with phenomenological quenching factors previously
introduced in order to improve the agreement with experiment (see discussion
in Section~\ref{sec:beta_decay}). Finally, in Ref.~\cite{Fili19deutradius} the
effects of two-body contributions to the charge density operator have been
investigated for the deuteron using a family of state-of-the-art NN
interactions at different orders in the chiral expansion and allowed a very
accurate extraction of the charge radius of the neutron. It will be crucial to
extend such calculations in a systematic way including estimates of
theoretical uncertainties to a larger number of nuclei and observables in the
near future.

\item The connection between properties of atomic nuclei and nuclear matter based on
a given set of interactions is still not well understood on a quantitative
level. On a qualitative level, however, some systematic trends between
observables of both systems can be found (compare, e.g., right panel of
Figure~\ref{fig:interactions_fit_4He} and
Figure~\ref{fig:magic_interactions_matter_nuclei} and related discussion in
Section~\ref{sec:sep_3N_fits}). Since nuclear forces should ideally provide a
comprehensive description of nuclei as well as nuclear matter, it will be key
to obtain a better understanding of this link. This is of particular relevance
when nuclear matter is being used as an anchor point for fits of future
nuclear forces. In Ref.~\cite{Hopp19medmass} properties of medium-mass nuclei
were studied based on 3N interactions fitted to the binding energy of $^3$H
and the saturation region of nuclear matter (see Figures~\ref{fig:cd_ce},
\ref{fig:coester_fits} and~\ref{fig:EMN_3N_results}).
Figure~\ref{fig:matter_nuclei} shows results for medium-mass nuclei and the
saturation points of symmetric nuclear matter based on the same NN and 3N
interactions as a function of the 3N coupling $c_D$. While the ground-state
energies of $^{16}$O and $^{40}$Ca change by $< 1 \MeV$ for SRG-unevolved
interactions, the change in saturation energy is $15 \MeV$ over the studied
$c_D$ range~\cite{Hopp19medmass}. In addition, for $c_D$ values that lead to a
good reproduction of the empirical saturation region, the ground-state
energies of nuclei turn out to be significantly underbound (see
Figure~\ref{fig:EMN_3N_results}). These findings indicate that nuclear matter
at lower densities might be more relevant for atomic nuclei and that the
connection between light nuclei, medium-mass nuclei and nuclear matter is
generally quite intricate and requires further investigations.

\item For calculations of dense matter it will be key to investigate more
systematically the density limit up to which chiral EFT interactions can
provide reliable predictions. Generally, the size of EOS uncertainties at
several times saturation density, determined based on calculations at low
densities and neutron star properties, crucially depends on the uncertainty
bands around and slightly above nuclear densities (see
Refs.~\cite{Hebe13ApJ,Tews18CS}). Consequently, being able to push the density
limits to slightly larger densities can significantly improve the predictive
power of the microscopic calculations for the properties of neutron stars (see
lower panel of Figure~\ref{fig:NS_combined}). Even though there are clear
indications that the uncertainty bands increase systematically with density
(see, e.g., right panel of Figure~\ref{fig:PNM_results}), these studies so far
depend on particular choices for the value of the expansion parameter $Q$ and
the chosen prescription for extracting uncertainty bands. Bayesian frameworks
might offer a powerful tool to address this question more systematically. Work
along these lines is in progress.

\item Generally, estimates of theoretical uncertainties should be assigned 
to all results of \textit{ab initio} calculations. Ideally, these uncertainties should
have a well-defined statistical interpretation. In recent years Bayesian
frameworks have been developed to estimate EFT truncations errors for two- and
three-body calculations (see, e.g., Refs.~\cite{Mele17bayerror,Epel19Bayes}).
It will be crucial to extend these analyses to many-body systems based on NN
and 3N interactions formulated in different regularization schemes and for
different Bayesian prior choices. Given the availability of various different
NN and 3N interactions, it is of central importance to obtain a deeper
understanding of the capabilities and limits of particular interactions
regarding the correct description of different observables of nuclei.
Differences due to possible regulator choices should generally be effects of
higher order in the chiral expansion and results should eventually become
independent at sufficiently high orders. Strictly, this is true as long as
all contributions to interactions and currents are properly taken into account
up to a given order. Presently, it has not yet been demonstrated to what
extent results based on different interactions are consistent with each other.

\begin{figure}[t!]
\begin{center}
\includegraphics[width=0.31\textwidth]{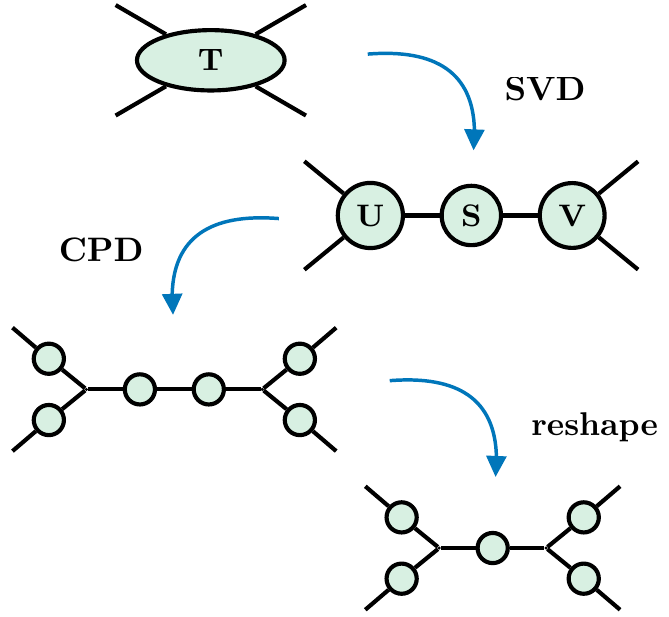}
\hspace{1cm}
\includegraphics[width=0.59\textwidth]{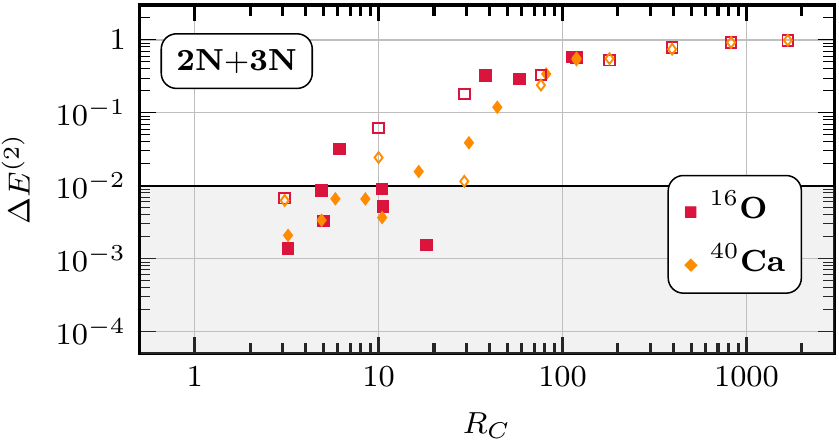}
\end{center}
\caption{Left: Diagrammatic representation of the tensor hypercontraction
(THC) decomposition of a rank-4 tensor. It involves a singular value
decomposition (SVD), a canonical polyadic decomposition (CPD) and a final
reshaping of the decomposition. Right: Relative energy difference $\Delta
E(2)$ tensor-decomposed ground-state energy correction in second-order
many-body perturbation theory as a function of the data compression factor
$R_C$ for $^{16}$O and $^{40}$Ca. Full (empty) symbols correspond to different
THC approximations. The gray region denotes the region of sub-percent
accuracy.\\
\textit{Source:} Figures taken from Ref.~\cite{Tich18tensdecomp}, left figure modified.}
\label{fig:tensor_network}
\end{figure}

Bayesian frameworks offer a promising tool for analyzing the parameter space
of existing interactions or for constructing improved nuclear interactions in
the future~\cite{Ekst19ideas}. In Ref.~\cite{Weso18NNphase} it was
demonstrated that such frameworks can be used to study the robustness of the
parameter estimation of given NN interactions and to isolate issues connected
to redundant and correlated couplings based on phase shift data (see also
Section~\ref{sec:Bayes_parameter}). However, the generalization of these
studies to many-body systems and to NN plus 3N interactions for different
regularization schemes involves significant computational challenges. Both,
parameter fitting and obtaining posterior distributions for calculated
observables involve sampling a large number of points in a high-dimensional
parameter space, where the required large number of exact calculations would
be prohibitively expensive. Recently, in Ref.~\cite{Koni19EC} it was shown
that a new method called eigenvector continuation (EC) can be used as a fast
and accurate emulator, thereby making uncertainty quantification in
multi-nucleon systems practical (see also
Refs.~\cite{Ekst19Bayes,Ekst19sens}). On the basis of calculations for $^4$He,
it was demonstrated that EC is superior to established methods like Gaussian
processes or polynomial interpolation. For nuclear physics, and statistical
analysis of the nuclear many-body problem in particular, the EC method can be
the key ingredient that enables the large-scale Markov-Chain Monte Carlo
evaluation of Bayesian posteriors of parameters in effective field theory
models of the nuclear interaction. Applications to this and related studies
are already under way.

\item Storing and handling three- and higher-body operators represents
presently one of the fundamental limitations of many-body frameworks for
medium-mass nuclei. Given the complexity of this problem, there is urgent need
to develop efficient methods that allow to reduce the computational costs of
such calculations and allow to push the scope of these frameworks to larger
mass numbers. Inspired by methods used in quantum chemistry and solid-state
physics in Ref.~\cite{Tich18tensdecomp}, a tensor decomposition technique for
\textit{ab initio} nuclear structure calculations was presented. The underlying idea of
this method is to represent a given tensor approximately in terms of products
of lower-rank tensors. If we consider, e.g., a rank-four tensor (i.e.,
a two-body interaction), which depends on four generic single-particle
quantum numbers $i_a$, it can be written via a so-called tensor
hypercontraction
(THC)~\cite{Hohe12tensor1,Hohe12tensor3,Schut17tensor,Tich18tensdecomp}
\begin{equation}
T_{i_1 i_2 i_3 i_4} = \sum_{\alpha,\beta} X_{i_1 \alpha}^{1} X_{i_2 \alpha}^{2} W_{\alpha \beta} X_{i_3 \alpha}^{3} X_{i_4 \alpha}^{4} \, .
\end{equation}
In practice, the THC decomposition can be performed via a multistep process,
which involves a singular value decomposition (SVD) and a canonical polyadic
decomposition (CPD) (see left panel of Figure~\ref{fig:tensor_network}). For
details we refer the reader to Ref.~\cite{Tich18tensdecomp}. In this work it
was also demonstrated that a significant reduction in the required memory
space can be achieved while maintaining a remarkable accuracy for energies
compared to the full result for MBPT calculations (see right panel of
Figure~\ref{fig:tensor_network}). In Ref.~\cite{Tich19preprocess} this method
was applied to open-shell systems. In the near future this strategy promises
to be a powerful tool that allows to significantly scale down the
computational complexity of state-of-the-art many-body frameworks and might
also be a path toward the inclusion of four- and higher-body interactions in
such calculations. Even though four-nucleon (4N) interactions can in principle
be treated by a straightforward extension of the methods presented in
Section~\ref{sect:3NF_representation}, the practical implementation of a
partial-wave decomposition is extremely challenging and the required basis
sizes for a complete representation are presently beyond the limit of
feasibility (see, e.g., Ref.~\cite{Schu18PhD}). Even though first studies
suggest that effects from 4N interactions in nuclei and matter are
small~\cite{Nogg104N,Krue12MSc,Krue13N3LOlong,Schu18PhD}, chiral EFT dictates
the inclusion of these interactions at N$^3$LO and beyond (see
Section~\ref{sec:chiral_EFT}). The utilization of a clever tensor
decomposition, possibly in combination with importance truncation methods (see
Ref.~\cite{Tich19preprocess}), might allow to include contributions from such
many-body interactions approximately at a much lower computational cost.

\item The field of quantum computing is rapidly evolving these days with more
and more powerful devices becoming available every year. Some inherently hard
problems for ``classical'' computers have been shown to be solvable on quantum
computers very efficiently (see, e.g.,
Refs.~\cite{Shor97qcalgorithms,Niel10Quantum} for an overview and
Ref.~\cite{Arut19quantum} for a very recent application). In recent years, it
has been demonstrated that problems in quantum chemistry and solid state
physics can already be solved via quantum
computers~\cite{Lany10Quantum,Paru14Quantum,Malle16Quantum,Kand17Quantum}.
\begin{figure}[t!]
\begin{center}
\vspace{1cm}
\includegraphics[width=0.46\textwidth]{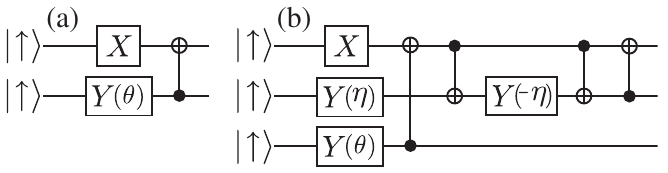}
\end{center}
\caption{Quantum circuits for generating unitary rotations for one basis state
and the corresponding angle $\theta$ (left) and two states characterized by
the angles $\theta$ and $\eta$ (right). $X$ represents a Pauli spin matrix and
$Y(\theta)$ rotation around the $Y$ axis, see also Ref.~\cite{Niel10Quantum}.\\
\textit{Source:} Figure taken from Ref.~\cite{Dumi18cloud}.}
\label{fig:quantum_computing}
\end{figure}
The fundamental units of quantum computers are circuits of entangled qubits.
Theoretically, the manipulation of qubits can be represented in terms of Pauli
matrices~\cite{Niel10Quantum}. Present quantum devices developed by Google,
IonQ and IBM are limited to about 50 to 79 non-error corrected qubits, while
full scale many-body problems in nuclear physics would require at least 100
error-corrected bits~\cite{Dumi18cloud}. One fundamental challenge for the
development of algorithms for quantum computers lies in the reduction of the
required circuit depths and number of entangling operations, such that all
operations can be performed within the device's decoherence
time~\cite{Babu17Quantum}. In Ref.~\cite{Dumi18cloud} a first
proof-of-principle quantum computation of the deuteron binding energy via two
different quantum devices was presented. For the calculations a discrete
variable representation was employed to match the connectivity of the
available hardware, while the ground-state energy of the system was determined
using a variational approach using 2 or 3 basis states. The corresponding
qubit gate structures for these deuteron calculations are shown in the left
and right panel of Figure~\ref{fig:quantum_computing}, respectively. Results of
these calculations were in good agreement with exact solutions. However, it
was also demonstrated that noise effects connected to measurement errors
increase significantly with circuit depth. This highlights the necessity for
improved error correction mechanisms for applications of quantum computers to
full-scale problems in \textit{ab initio} nuclear
physics~\cite{Lu18quantum,Klco18quantum}.

\end{itemize}

\section*{Acknowledgements}

\noindent A major part of the presented work in this review is the result of
very fruitful long-term collaborations I am very happy to be part of.
Specifically, I would like to thank Sonia Bacca, Andreas Bauswein, Sven
Binder, Scott Bogner, Jens Braun, Angelo Calci, Arianna Carbone, Christian
Drischler, Thomas Duguet, Victoria Durant, Alex Dyhdalo, Andreas Ekstr\"om,
Evgeny Epelbaum, Christian Forss\'en, Bengt Friman, Dick Furnstahl, Alex
Gezerlis, Jacek Golak, Svenja Greif, Gaute Hagen, Hans-Werner Hammer, Heiko
Hergert, Javier Hernandez, Matthias Heinz, Martin Hoferichter, Jason Holt, Jan
Hoppe, Thomas H{\"u}ther, Jonas Keller, Philipp Klos, Sebastian K{\"o}nig,
Hermann Krebs, Thomas Kr{\"u}ger, Jim Lattimer, Dean Lee, Marc Leonhardt, Joel
Lynn, Pieter Maris, Ulf-G. Mei{\ss}ner, Javier Men{\'e}ndez, Sushant More, Andreas
Nogga, Thomas Papenbrock, Chris Pethick, Geert Raaijmakers, Robert Roth, Achim
Schwenk, Rodric Seutin, Johannes Simonis, Roman Skibi{\'{n}}ski, Vittorio
Som{\`a}, Ragnar Stroberg, Ingo Tews, Alex Tichai, James Vary, and Anna Watts
for numerous discussions on various topics discussed in this work over the
years. In addition, I thank Matthias Heinz, Jan Hoppe, Rodric Seutin and Alex
Tichai for carefully reading the manuscript and providing most helpful
comments. Finally, I thank the entire Theoriezentrum at the TU Darmstadt for
the very stimulating and enjoyable atmosphere. The presence of so many bright
young researchers makes this truly a special place.

This work was supported by the Deutsche Forschungsgemeinschaft (DFG, German
Research Foundation) -- Projektnummer 279384907 -- SFB 1245 and the BMBF under
Contract No. 05P15RDFN1. Computational resources have been provided by the
J\"ulich Supercomputing Center and by the Lichtenberg high performance
computer of the TU Darmstadt.

\clearpage

\appendix 

\renewcommand*{\thesection}{\hspace{-0.3cm}\Alph{section}}

\section{Normalization convention of momentum basis states}
\label{sec:normalization}
A general Galilean-invariant NN interaction takes the following form in
vector representation:
\begin{equation}
\left< \mathbf{p}' S M_{S'} T M_T | V_{\text{NN}} | \mathbf{p} S M_{S} T M_T \right> \, . \label{eq:NN_vector}
\end{equation}
Here, the momenta $\mathbf{p}$ and $\mathbf{p}'$ denote the Jacobi momenta
of the initial and final states (see Section~\ref{sec:3NF_coord_def}), the
quantum numbers $S$, $M_S$ and $M_{S'}$ the two-body spin channel and
corresponding spin projections of the initial and final state. $T$ and $M_T$
denote the corresponding two-body isospin quantum numbers. Due to the presence
of tensor interactions, the interaction has in general finite contributions
for different spin projection $M_S \neq M_{S'}$, but is diagonal in the
quantum numbers $S$, $T$ and $M_T$. In the following we will suppress the
isospin quantum numbers for two-body quantities for the sake of compact
notation. Since all these quantities are diagonal in both isospin quantum
numbers, they can be added in a trivial way at any stage below.

The Lippmann-Schwinger equation for the scattering $T$-matrix for the energy
$E$ is given in vector form by~\cite{LandauQM}
\begin{equation}
\left< \mathbf{p}' S M_{S'} | T_{\text{NN}} (E) | \mathbf{p} S M_S \right> = \left< \mathbf{p}' S M_{S'} | V_{\text{NN}} | \mathbf{p} S M_{S} \right> + \sum_{M_{S''}} \int \frac{d\mathbf{q}}{(2 \pi)^3} \frac{\left< \mathbf{p}' S M_{S'} | V_{\text{NN}} | \mathbf{q} S M_{S''} \right> \left< \mathbf{q} S M_{S''} | T_{\text{NN}} (E) | \mathbf{p} S M_{S'} \right>}{E - \mathbf{q}^2/(2 \mu) \pm i \epsilon} \, , \label{eq:LSE_vec}
\end{equation}
with 
\begin{equation}
\left< \mathbf{p}' S' M_{S'} | \mathbf{p} S M_S \right> = (2\pi)^3 \delta(\mathbf{p} - \mathbf{p}') \: \delta_{S S'} \delta_{M_S M_{S'}}
\end{equation}
and the reduced mass $\mu$ of the interacting particles\footnote{It is common
to include the mass factor in the definition of the
potential.}. We introduce two-body partial-wave states via:
\begin{equation}
\left| \mathbf{p} S M_S \right> = A_{\text{NN}} \sum_{L,M_L,J,M_J} \mathcal{C}_{L M_L S M_S}^{J M_J} Y^*_{L M_L} (\hat{\mathbf{p}}) \left| p (L S) J M_J \right> \, , \label{eq:pw_states_NN}
\end{equation}
with $\left< \hat{\mathbf{p}} | L M \right> = Y_{L M} (\hat{\mathbf{p}})$.
Here we couple the orbital relative angular momentum quantum number $L$ with
the two-body spin $S$ to a total relative angular momentum $J$. The constant
$A_{\text{NN}}$ defines the normalization convention for the partial-wave
states (see below) and can take different values. The
definition~(\ref{eq:pw_states_NN}) leads to the following decomposition of the
two-body quantities, e.g. for NN interactions:
\begin{equation}
\left< \mathbf{p}' S M_{S'} | V_{\text{NN}} | \mathbf{p} S M_S \right> = A_{\text{NN}}^2 \sum_{L,L',M_L, M_L'} \sum_{J, M_J} \mathcal{C}_{L M S M_S}^{J M_J} \mathcal{C}_{L' M_{L'} S M_{S'}}^{J M_J} Y_{L'
 M_{L'}} (\hat{\mathbf{p}}') \left< p' (L' S) J | V_{\text{NN}} | p (L S) J \right> Y^*_{L M_L} (\hat{\mathbf{p}}) \, . \label{eq:VNN_pwd}
\end{equation}
Since the interaction transforms like a scalar under rotations, the
partial-wave matrix elements $ \left< p' (L' S) J | V_{\text{NN}} | p (L S) J
\right>$ are diagonal in $J$ and $M_J$ and do not depend on the
projection quantum number $M_J$. Hence we will suppress the $M_J$ quantum
number in the following for the matrix elements. Inverting relation
(\ref{eq:VNN_pwd}) gives:
\begin{align}
\left< p' (L' S) J | V_{\text{NN}} | p (L S) J \right> = \frac{1}{A_{\text{NN}}^2} \sum_{\substack{M_L, M_S\\M_{L'}, M_{S'}}} \int d \hat{\mathbf{p}} d \hat{\mathbf{p}}' \mathcal{C}_{L M_L S M_S}^{J M_J} \mathcal{C}_{L' M_{L'} S M_{S'}}^{J M_J} Y^*_{L' M_{L'}} (\hat{\mathbf{p}}') \left< \mathbf{p}' S M_{S'} | V_{\text{NN}} | \mathbf{p} S M_S \right> Y_{L M_{L}} (\hat{\mathbf{p}}) \, .
\end{align}
Inserting the expansion Eq.~(\ref{eq:VNN_pwd}) in Eq.~(\ref{eq:LSE_vec}) and
projecting onto the individual partial waves leads to
\begin{equation}
\left< p' (L' S) J | T_{\text{NN}} | p (L S) J \right> = \left< p' (L' S) J | V_{\text{NN}} | p (L S) J \right> + \frac{A_{\text{NN}}^2}{(2\pi)^3} \sum_{L''} \int dq q^2 \frac{\left< p' (L' S) J | V_{\text{NN}} | q (L'' S) J \right> \left< q (L'' S) J | T_{\text{NN}} | p (L S) J \right>}{E - q^2/(2 \mu) \pm i \epsilon} \, .
\end{equation}
Possible and common choices for the normalization constant $A_{\text{NN}}$ are
$A_{\text{NN}}^2 = (4 \pi)^2$ (see, e.g., Ref.~\cite{Epel05EGMN3LO}) and
$A_{\text{NN}}^2 = (2 \pi)^3$ (see, e.g., Ref.~\cite{Ente03EMN3LO}). The
convention $A_{\text{NN}}^2 = (2 \pi)^3$ minimizes the number of $\pi$ factors
in equations represented in partial-wave basis. We will adopt this convention
in this work as it allows to generalize relations for partial-wave matrix
elements in a very simple and natural way to three-body operators (see
Table~\ref{tab:states_normalization}).

\begin{table}[t!]
\centering
\resizebox{\columnwidth}{!}{
$\begin{array}{|l|}
\multicolumn{1}{l}{\textit{two-body states}} \\
\hline
\begin{aligned}
\addlinespace
&\left< \mathbf{p}' S' M_{S'} T' M_{T'} | \mathbf{p} S M_S T M_T \right> = (2 \pi)^3 \delta(\mathbf{p} - \mathbf{p}') \, \delta_{SS'} \delta_{M_S M_{S'}} \delta_{TT'} \delta_{M_T M_{T'}}\\
&\sum_{S, M_S, T, M_T} \int \frac{d\mathbf{p}}{(2 \pi)^3} \left| \mathbf{p} S M_S T M_T \right> \left< \mathbf{p} S M_S T M_T \right| = 1\\
\addlinespace
\end{aligned}\\
\hline
\begin{aligned}
\addlinespace
&\left< p' (L' S') J' M_{J'} T' M_{T'} | p (L S) J M_J T M_{T} \right> = \frac{\delta (p - p')}{p p'} \delta_{L L'} \delta_{S S'} \delta_{J J'} \delta_{M_J M_{J'}} \delta_{T T'} \delta_{M_T M_{T'}} \\
&\sum_{L,S,J,M_J, T, M_T} \int dp p^2 \left| p (L S) J M_J T M_T \right> \left< p (L S) J M_J T M_T \right| = 1\\
\addlinespace
\end{aligned}\\
\hline
\begin{aligned}
\addlinespace
&\left< \mathbf{p}' S' M_{S'} T' M_{T'} | p (L S) J M_J T M_T \right> = (2 \pi)^{3/2} \frac{\delta(p -p')}{p p'} \delta_{S S'} \delta_{T T'} \delta_{M_T M_{T'}} \sum_{M_L} \mathcal{C}_{L M_L S M_{S'}}^{J M_J} Y_{L M_L} (\hat{\mathbf{p}}') \\
\addlinespace
\end{aligned}\\
\hline
\addlinespace
\addlinespace
\multicolumn{1}{l}{\textit{three-body states}} \\
\hline
\begin{aligned}
\addlinespace
&\left< \mathbf{p}' \mathbf{q}' S' M_{S'} m_{s'} T' M_{T'} m_{t'} | \mathbf{p} \mathbf{q} S M_S m_s T M_T m_t \right> = (2 \pi)^6 \delta(\mathbf{p} - \mathbf{p}') \, \delta(\mathbf{q} - \mathbf{q}') \delta_{SS'} \delta_{M_S M_{S'}} \delta_{m_s m_{s'}} \delta_{TT'} \delta_{M_T M_{T'}} \delta_{m_t m_{t'}}\\
&\sum_{S, M_S, m_s} \sum_{T, M_T, m_t} \int \frac{d\mathbf{p}}{(2 \pi)^3} \frac{d\mathbf{q}}{(2 \pi)^3} \left| \mathbf{p} \mathbf{q} S M_S m_s T M_T m_t \right> \left< \mathbf{p} \mathbf{q} S M_S m_s T M_T m_t \right| = 1 \\
\addlinespace
\end{aligned}\\
\hline
\begin{aligned}
\addlinespace
&\bigl< p' q' [ (L' S') J' (l' \tfrac{1}{2}) j' ] \mathcal{J}' M_{\mathcal{J}'} (T' \tfrac{1}{2}) \mathcal{T}' M_{\mathcal{T}'} | p q [ (L S) J (l \tfrac{1}{2}) j ] \mathcal{J} M_{\mathcal{J}} (T \tfrac{1}{2}) \mathcal{T} M_{\mathcal{T}} \bigr> \\
&\hspace{4cm} = \frac{\delta (p - p')}{p p'} \frac{\delta (q - q')}{q q'} \delta_{L L'} \delta_{S S'} \delta_{J J'} \delta_{T T'} \delta_{l l'} \delta_{j j'} \delta_{m_j m_{j'}} \delta_{\mathcal{J} \mathcal{J}'} \delta_{M_{\mathcal{J}} M_{\mathcal{J}'}} \delta_{\mathcal{T} \mathcal{T}'} \delta_{M_{\mathcal{T}} M_{\mathcal{T}'}} \\
&\sum_{L,S,J,T,l,j} \sum_{\mathcal{J},M_{\mathcal{J}},\mathcal{T},M_{\mathcal{T}}} \int dp p^2 \int dq q^2 | p q [ (L S) J (l \tfrac{1}{2}) j ] \mathcal{J} M_{\mathcal{J}} (T \tfrac{1}{2}) \mathcal{T} M_{\mathcal{T}} \bigr> \bigl< p q [ (L S) J (l \tfrac{1}{2}) j ] \mathcal{J} M_{\mathcal{J}} (T \tfrac{1}{2}) \mathcal{T} M_{\mathcal{T}} | = 1 \\
\addlinespace
\end{aligned}\\
\hline
\begin{aligned}
\addlinespace
&\bigl< \mathbf{p}' \mathbf{q}' S' M_{S'} m_{s'} T' M_{T'} m_{t'} | p q [ (L S) J (l \tfrac{1}{2}) j ] \mathcal{J} M_{\mathcal{J}} (T \tfrac{1}{2}) \mathcal{T} M_{\mathcal{T}} \bigr> \\
& \hspace{1cm} = (2 \pi)^3 \frac{\delta(p -p')}{p p'} \frac{\delta(q -q')}{q q'} \delta_{S S'} \delta_{T T'} 
\sum_{\substack{M_L, m_l\\M_J, m_j}} \mathcal{C}_{J M_J j m_j}^{\mathcal{J} M_{\mathcal{J}}} \mathcal{C}_{L M_L S M_{S'}}^{J M_J} \mathcal{C}_{l m_l \tfrac{1}{2} m_{s'}}^{j m_j} Y_{L M_L} (\hat{\mathbf{p}}') Y_{l m_l} (\hat{\mathbf{q}}') \: \mathcal{C}_{T M_{T'} \tfrac{1}{2} m_t'}^{\mathcal{T} M_{\mathcal{T}}}  \\
\addlinespace
\end{aligned}\\
\hline
\end{array}$}
\caption{Normalization convention for two-body and three-body momentum states
in vector and partial-wave representation. The spin and isospin quantum
numbers $S$, $M_S$,$T$ and $M_T$ denote the two-body spin and isospin states
of the two particles with relative momentum $\mathbf{p}$, while
$m_s$ and $m_t$ denote the spin and isospin projections of the particle with
Jacobi momentum $\mathbf{q}$.}
\label{tab:states_normalization}
\end{table}

Following Eq.~(\ref{eq:NN_vector}), a general Galilean-invariant 3N
interaction can be written in the form (see
Section~\ref{sect:3NF_representation})
\begin{equation}
\left< \mathbf{p}' \mathbf{q}' S' M_{S'} m_{s'} T' M_{T'} m_{t'} | V_{\text{3N}} | \mathbf{p} \mathbf{q} S M_{S} m_s T M_T m_t \right> \, , \label{eq:3N_vector}
\end{equation}
where $m_s$ and $m_t$ are the spin- and isospin projections of the third
particle. Note that in contrast to NN interactions, 3N interactions are
generally off-diagonal in the isospin quantum numbers $T$ and $M_T$. We adopt
the same normalization convention as for NN interactions. The specific
relations are given in Table~\ref{tab:states_normalization}. 

We define three-body partial-wave states by generalizing
Eq.~(\ref{eq:pw_states_NN}):
\begin{align}
& \left| \mathbf{p} \mathbf{q} S M_S m_s T M_T m_t \right> \nonumber \\
& \hspace{0.5cm} = A_{\text{3N}} \sum_{\substack{L,M_L,l,m_l\\J,M_J}} \sum_{\substack{\mathcal{J},M_\mathcal{J}\\\mathcal{T},M_\mathcal{T}}} \mathcal{C}_{J M_J j m_j}^{\mathcal{J} M_{\mathcal{J}}} \mathcal{C}_{L M_L S M_S}^{J M_J} \mathcal{C}_{l m_l \tfrac{1}{2} m_s}^{j m_j} Y^*_{L M_L} (\hat{\mathbf{p}}) Y^*_{l m_l} (\hat{\mathbf{q}}) \: \mathcal{C}_{T M_{T} \tfrac{1}{2} m_t}^{\mathcal{T} M_{\mathcal{T}}} | p [ (L S) J (l \tfrac{1}{2}) j ] \mathcal{J} M_{\mathcal{J}} (T \tfrac{1}{2}) \mathcal{T} M_{\mathcal{T}} \bigr> \, . \label{eq:pw_states_3N}
\end{align}
In this work we choose $A_{\text{3N}} = (2 \pi)^3$, which leads to a
particularly compact form of equations in partial-wave representation and allows
to generalize relations for two-body operators in a simple way for the choice
$A_{\text{NN}} = \sqrt{(2 \pi)^3}$. The specific normalization conventions for
two-body and three-body states adopted in this work are summarized in
Table~\ref{tab:states_normalization}.

\section{Integral for partial-wave decomposition of local 3N interactions}
\label{sec:eval_PWD_local}

In this Appendix we describe the integral transformations that allow for a
decoupling of the three non-trivial integrations over spherical harmonics in
the partial-wave decomposition of 3N interactions (see Eq.~(\ref{eq:F_func}))
from the other five, which can be performed analytically. We start with
Eq.~(\ref{eq:F_func}) and add a radial integration over $\tilde{p}^\prime$ and
$\tilde{q}^\prime$ in order to obtain a translationally-invariant measure. We
can achieve this by introducing additional integrations with delta-functions
via
\begin{equation}
F_{L l L' l'}^{m_L m_l m_{L'} m_{l'}} (p, q, p', q') = \int \frac{d\tilde{\mathbf{p}}' d\tilde{\mathbf{q}}' d\hat{\mathbf{p}} d \hat{\mathbf{q}}}{(2 \pi)^6} \frac{\delta(p' - \tilde{p}')}{p'^2} \frac{\delta(q' - \tilde{q}')}{q'^2} Y_{L' m_{L'}}^{*} (\hat{\tilde{\mathbf{p}}}') Y_{l' m_{l'}}^{*} (\hat{\tilde{\mathbf{q}}}') Y_{L m_L} (\hat{\mathbf{p}}) Y_{l m_l} (\hat{\mathbf{q}}) V^{\text{local}}_{\text{3N}} (\tilde{\mathbf{p}}' -\mathbf{p},\tilde{\mathbf{q}}' -\mathbf{q}) \, ,
\end{equation}
where we renamed the angular parts of the vectors, $\hat{\mathbf{p}}'
\rightarrow \hat{\tilde{\mathbf{p}}}'$ and $\hat{\mathbf{q}}' \rightarrow
\hat{\tilde{\mathbf{q}}}'$, and made the local nature of the 3N interaction
explicit (see Section~\ref{sec:PWD_3NF_local}). Now we can make the
substitutions:
\begin{equation}
\tilde{\mathbf{p}}^\prime \rightarrow \tilde{\mathbf{p}} + \mathbf{p} \quad {\rm and}\quad \tilde{\mathbf{q}}^\prime \rightarrow \tilde{\mathbf{q}} + \mathbf{q} \, , 
\end{equation}
which leads to:
\begin{align}
F_{L l L' l'}^{m_L m_l m_{L'} m_{l'}} (p, q, p', q') &= \frac{1}{(2 \pi)^6} \int d \tilde{\mathbf{p}} d \tilde{\mathbf{q}}  d\hat{\mathbf{p}} d\hat{\mathbf{q}} \frac{\delta(p^\prime - |\tilde{\mathbf{p}}+\mathbf{p}|\,)}{p^{\prime\,2}} \frac{\delta(q^\prime - |\tilde{\mathbf{q}} + \mathbf{q}|\,)}{q^{\prime\,2}} V^{\text{local}}_{\text{3N}} (\tilde{\mathbf{p}},\tilde{\mathbf{q}}) \nonumber\\
& \times Y_{L' m_{L'}}^{*} (\widehat{\tilde{\mathbf{p}}+\mathbf{p}}\,) Y_{l' m_{l'}}^{*} (\widehat{\tilde{\mathbf{q}} + \mathbf{q}}\,) Y_{L m_L} (\hat{\mathbf{p}}) Y_{l m_l} (\hat{\mathbf{q}}) \, . 
\end{align}
To evaluate these integrals, we perform as a first step a rotation
$R(\tilde{\mathbf{p}})$ of the vectors $\tilde{\mathbf{q}}$ and $\mathbf{p}$,
\begin{equation}
\tilde{\mathbf{q}} \rightarrow R(\tilde{\mathbf{p}}) \tilde{\mathbf{q}} \quad \text{and} \quad \mathbf{p} \rightarrow R (\tilde{\mathbf{p}}) \mathbf{p} \quad \text{with} \quad R^{-1} \tilde{\mathbf{p}} = \tilde{p}\mathbf{e}_z \, .
\end{equation}

As a second step, we perform a rotation of the vector $\mathbf{q}$,
\begin{equation}
\mathbf{q} \rightarrow R(\tilde{\mathbf{p}}) Q(\tilde{\mathbf{q}}) \mathbf{q} \quad \text{with} \quad Q^{-1} \tilde{\mathbf{q}} = \tilde{q} \mathbf{e}_z \, .
\end{equation}
This results in:
\begin{align}
& \hspace{-1cm} F_{L l L' l'}^{m_L m_l m_{L'} m_{l'}} (p, q, p', q') = \nonumber \\
&= \frac{1}{(2 \pi)^6} \frac{1}{p^{\prime\,2} q^{\prime\,2}}\int d \tilde{\mathbf{p}} d \tilde{\mathbf{q}} d\hat{\mathbf{p}}  d\hat{\mathbf{q}} \delta(p^\prime - |\tilde{p} \mathbf{e}_z + \mathbf{p}|\,) \delta(q^\prime - |\tilde{q} \mathbf{e}_z + \mathbf{q}|\,)
V^{\text{local}}_{\text{3N}} (\tilde{p},\tilde{q}, \hat{\tilde{\mathbf{q}}} \cdot \mathbf{e}_z)\nonumber\\
&\times Y_{L' m_{L'}}^{*} (R (\widehat{\tilde{p} \mathbf{e}_z +\mathbf{p}})) Y_{l' m_{l'}}^{*} (RQ (\widehat{\tilde{q} \mathbf{e}_z + \mathbf{q}})\,) Y_{L m_L} (R \hat{\mathbf{p}}) Y_{l m_l} (R Q \hat{\mathbf{q}}) \nonumber \\
&= \frac{1}{(2 \pi)^6} \frac{1}{p^{\prime\,2} q^{\prime\,2}}\int d \tilde{\mathbf{p}} d \tilde{\mathbf{q}} d\hat{\mathbf{p}} d\hat{\mathbf{q}}
\delta(p^\prime - |\tilde{p} \mathbf{e}_z + \mathbf{p}|\,) \delta(q^\prime - |\tilde{q} \mathbf{e}_z + \mathbf{q}|\,)
V^{\text{local}}_{\text{3N}} (\tilde{p},\tilde{q}, \hat{\tilde{\mathbf{q}}} \cdot \mathbf{e}_z)\nonumber\\
&\times \sum_{\bar{m}_{L},\bar{m}_{L'}, \bar{m}_{l},\bar{m}_{l'}} Y_{L' \bar{m}_L'}^{*} (\widehat{\bar{p} \mathbf{e}_z +\mathbf{p}}) Y_{l' \bar{m}_{l'}}^{*} (\widehat{\tilde{q} \mathbf{e}_z + \mathbf{q}}) Y_{L \bar{m}_L} (\hat{\mathbf{p}}) Y_{l \bar{m}_l} (\hat{\mathbf{q}}) D_{\bar{m}_{L'} m_{L'}}^{* L'} (R) D_{\bar{m}_{l'} m_{l'}}^{* l'} (RQ) D_{\bar{m}_{L} m_{L}}^{L} (R) D_{\bar{m}_{l} m_{l}}^{l} (RQ) \, , \label{eq:3NPWD_F}
\end{align}
where $D$ denote the Wigner $D$-functions. The product of $D$-functions can be expanded:
\begin{align}
D_{\bar{m}_{L'} m_{L'}}^{* L'} (R) D_{\bar{m}_{l'} m_{l'}}^{* l'} (RQ) D_{\bar{m}_{L} m_{L}}^{L} (R) D_{\bar{m}_{l} m_{l}}^{l} (RQ) = \sum_{\bar{\bar{m}}_l, \bar{\bar{m}}_{l'}} D_{\bar{m}_{L'} m_{L'}}^{* L'} (R) D_{\bar{m}_{l'} \bar{\bar{m}}_{l'}}^{* l'} (Q) D_{\bar{\bar{m}}_{l'} m_{l'}}^{* l'} (R) D_{\bar{m}_{L} m_{L}}^{L} (R) D_{\bar{m}_{l} \bar{\bar{m}}_{l}}^{l} (Q) D_{\bar{\bar{m}}_{l} m_l}^{l} (R) \, . \label{eq:Dfunc_id}
\end{align}

For the evaluation of Eq.~(\ref{eq:3NPWD_F}) it is important to note that the
integration over the azimuthal angles $\phi_p$ and $\phi_q$ results in the
constraints $\bar{m}_L = \bar{m}_{L'}$ and $\bar{m}_l = \bar{m}_{l'}$ since
the addition of a vector in the $z$-direction does not affect the azimuthal angles.
Furthermore we can make use of the identities
\begin{align}
D_{\bar{m} m'}^{* l'} (R) D_{\bar{m} m}^l (R) = (-1)^{\bar{m} - m} \sum_{\bar{l}} \mathcal{C}_{l' -\bar{m} l \bar{m}}^{\bar{l} 0} \mathcal{C}_{l' -m' l m}^{\bar{l} m-m'} \sqrt{\frac{4 \pi}{2 \bar{l} + 1}} Y_{\bar{l} \, m - m'} (\hat{\tilde{\mathbf{p}}})
\end{align}
and the according relation for the rotation $Q$. The integration of the latter
relation over $\phi_{\tilde{q}}$ implies $\bar{\bar{m}}_{l} =
\bar{\bar{m}}_{l'}$in Eq.~(\ref{eq:Dfunc_id}), specifically:
\begin{equation}
\int_0^{2\pi} d \phi_{\tilde{q}} D_{\bar{m}_{l} \bar{\bar{m}}_{l'}}^{* l'} (Q) D_{\bar{m}_{l} \bar{\bar{m}}_{l}}^{l} (Q) = 2\pi \, (-1)^{\bar{m}_l- \bar{\bar{m}}_l} \delta_{\bar{\bar{m}}_l, \bar{\bar{m}}_{l'}} \sum_{\bar{l}} 
\mathcal{C}_{l' -\bar{m}_l l \bar{m}_l}^{\bar{l} 0} 
\mathcal{C}_{l' -\bar{\bar{m}}_{l'} l \bar{\bar{m}}_l}^{\bar{l} \bar{\bar{m}}_l-\bar{\bar{m}}_{l'}} P_{\bar{l}} (\hat{\tilde{\mathbf{q}}} \cdot \mathbf{e}_z) \, .
\end{equation}
Using these relations we can perform also the integral over
$\hat{\tilde{\mathbf{p}}}$:
\begin{align}
&\hspace{-1cm} \int d \hat{\tilde{\mathbf{p}}} D_{\bar{m}_{L} m_{L'}}^{* L'} (R) D_{\bar{m}_{L} m_{L}}^{L} (R) D_{\bar{\bar{m}}_{l} m_{l'}}^{* l'} (R) D_{\bar{\bar{m}}_{l} m_{l}}^{l} (R) \nonumber \\
& = (-1)^{\bar{m}_L + \bar{\bar{m}}_l + m_L + m_{l'}} \delta_{m_{L} - m_{L'}, m_{l'} - m_{l}} \sum_{\bar{l} = \text{max} (|L' - L|, |l' - l|)}^{\text{min} (|L' + L|, |l' + l|)} \frac{4 \pi}{2 \bar{l} + 1}
\mathcal{C}_{L' -\bar{m}_L L \bar{m}_L}^{\bar{l} 0} 
\mathcal{C}_{L' -m_{L'} L m_L}^{\bar{l} m_L - m_{L'}}
\mathcal{C}_{l' -\bar{\bar{m}}_L l \bar{\bar{m}}_L}^{\bar{l} 0} 
\mathcal{C}_{l' -m_{l'} l m_l}^{\bar{l} m_l - m_{l'}} \, .
\end{align}
Finally, the integrals with respect to $\phi_{\mathbf{p}}$ and
$\phi_{\mathbf{q}}$ in Eq.~(\ref{eq:3NPWD_F}) are trivial and just provide a
factor $2 \pi$ each. The integrals with respect to $\theta_{\mathbf{p}}$ and
$\theta_{\mathbf{q}}$ are performed via the delta functions, providing the
relations
\begin{equation}
\hat{\mathbf{p}} \cdot \mathbf{e}_z = \frac{p'^2 - p^2 - \tilde{p}^2}{2 \tilde{p} p} \quad \text{and} \quad \hat{\mathbf{q}} \cdot \mathbf{e}_z = \frac{q'^2 - q^2 - \tilde{q}^2}{2 \tilde{q} q} \, ,
\end{equation}
and also specifying the limits of the $\tilde{p}$ and $\tilde{q}$ integrations:
\begin{equation}
|p'-p| \le \tilde{p} \le p'+p, \quad \text{and} \quad |q'-q| \le \tilde{q} \le q'+ q \, .
\end{equation}
In the end only three non-trivial integrations remain in
Eq.~(\ref{eq:3NPWD_F}), leading finally to Eq.~(\ref{final_result_pwd}):
\begin{align}
F_{L l L' l'}^{m_L m_l m_{L'} m_{l'}} (p, q, p', q') &=  \delta_{m_L - m_{L'}, m_{l'} - m_l} \frac{(-1)^{m_L + m_{l'}}}{(2 \pi)^6} \frac{2 (2 \pi)^4}{p p' q q'} \sum_{\bar{l} = \text{max} (|L'-L|,|l' - l|)}^{\text{min} (L' + L, l' + l)} 
\frac{\mathcal{C}_{L' - m_{L'} L m_L}^{\bar{l} -m_{L'} + m_L} \mathcal{C}_{l' -m_{l'} l m_l}^{\bar{l} -m_{l'} + m_l}}{2 \bar{l} + 1} \nonumber \\
&\times \int_{|p'-p|}^{p'+p} d \tilde{p} \, \tilde{p} \int_{|q'-q|}^{q'+q} d \tilde{q} \, \tilde{q} \left. \mathcal{Y}_{L' L}^{\bar{l} 0} (\widehat{\tilde{p} \mathbf{e}_z + \mathbf{p}},\hat{\mathbf{p}}) \right|_{\phi_{\mathbf{p}} = 0, \widehat{\mathbf{p}} \cdot \mathbf{e}_z = \frac{p'^2 - p^2 - \tilde{p}^2}{2 \tilde{p} p}} \left. \mathcal{Y}_{l' l}^{\bar{l} 0} (\widehat{\tilde{q} \mathbf{e}_z + \mathbf{q}}, \hat{\mathbf{q}}) \right|_{\phi_{\mathbf{q}} = 0, \widehat{\mathbf{q}} \cdot \mathbf{e}_z = \frac{q'^2 - q^2 - \tilde{q}^2}{2 \tilde{q} q}} \nonumber \\
&\times \int_{-1}^1 d \cos \theta_{\tilde{\mathbf{p}}\tilde{\mathbf{q}}} P_{\bar{l}} (\cos \theta_{\tilde{\mathbf{p}}\tilde{\mathbf{q}}}) V^{\text{local}}_{\text{3N}} (\tilde{p},\tilde{q},\cos \theta_{\tilde{\mathbf{p}}\tilde{\mathbf{q}}}) \, .
\end{align}

\section{Partial-wave matrix elements of permutation operator $P_{123}$}
\label{sec:PWD_P123}
Here we derive the partial-wave matrix elements of the three-body cyclic
permutation operator $P_{123}$. In Section~\ref{sec:3NF_coord_def} the
momentum-space structure of the operator was already discussed. For the
derivation of the full matrix elements we start from
Eq.~(\ref{eq:P123_momentum_def_fourth}) and add spin and isospin degrees of
freedom to the particle states. To this end, it is most convenient to work in
$LS$-coupling scheme since then we can immediately factorize the momentum
space part from the spin and isospin part:
\begin{align}
&\tensor*[_{\{12\}}]{\left< \mathbf{p}' \mathbf{q}' (S' s) \mathcal{S}' \mathcal{M}_{\mathcal{S}'} (T' t) \mathcal{T}' \mathcal{M}_{\mathcal{T}'} | P_{123} | \mathbf{p} \mathbf{q} (S s) \mathcal{S} \mathcal{M}_{\mathcal{S}} (T t) \mathcal{T} \mathcal{M}_{\mathcal{T}} \right>}{_{\{12\}}} \nonumber \\
& = \tensor*[_{\{12\}}]{\left< \mathbf{p}' \mathbf{q}' (S' s) \mathcal{S}' \mathcal{M}_{\mathcal{S}'} (T' t) \mathcal{T}' \mathcal{M}_{\mathcal{T}'} | \mathbf{p} \mathbf{q} (S s) \mathcal{S} \mathcal{M}_{\mathcal{S}} (T t) \mathcal{T} \mathcal{M}_{\mathcal{T}} \right>}{_{\{23\}}} \nonumber \\
& = \tensor*[_{\{12\}}]{\left< \mathbf{p}' \mathbf{q}' | \mathbf{p} \mathbf{q} \right>}{_{\{23\}}} \tensor*[_{\{12\}}]{\left< (S' s) \mathcal{S}' \mathcal{M}_{\mathcal{S}'} | (S s) \mathcal{S} \mathcal{M}_{\mathcal{S}} \right>}{_{\{23\}}} \tensor*[_{\{12\}}]{\left< (T' t) \mathcal{T}' \mathcal{M}_{\mathcal{T}'}  | (T t) \mathcal{T} \mathcal{M}_{\mathcal{T}} \right>}{_{\{23\}}} \, .
\end{align}
Now we can discuss the individual factors separately. The spin and isospin
terms can be easily evaluated by standard recoupling
techniques~\cite{Vars88Gulag}, e.g. for the spin:
\begin{equation}
\tensor*[_{\{12\}}]{\left< (S' s) \mathcal{S}' \mathcal{M}_{\mathcal{S'}} | (S s) \mathcal{S} \mathcal{M}_{\mathcal{S}} \right>}{_{\{23\}}} = \delta_{\mathcal{S} \mathcal{S}'} \delta_{\mathcal{M}_{\mathcal{S}} \mathcal{M}_{\mathcal{S'}}} (-1)^S \sqrt{\hat{S} \hat{S}'}
\left\{
\begin{array}{ccc}
\tfrac{1}{2} & \tfrac{1}{2} & S' \\
\tfrac{1}{2} & \mathcal{S} & S
\end{array}
\right\} \, ,
\end{equation}
with the 6j symbol $\{ .. \}$. We obtain identical results for the overlap matrix elements
$\tensor*[_{\{23\}}]{\left< (S' s) \mathcal{S}' \mathcal{M}_{\mathcal{S'}} | (S
s) \mathcal{S} \mathcal{M}_{\mathcal{S}} \right>}{_{[31]}}$ and
$\tensor*[_{\{31\}}]{\left< (S' s) \mathcal{S}' \mathcal{M}_{\mathcal{S'}} | (S
s) \mathcal{S} \mathcal{M}_{\mathcal{S}} \right>}{_{\{12\}}}$.
The momentum part can also be evaluated in a straightforward way. Making use
of the partial-wave states defined in Appendix~\ref{sec:normalization} and
Eq.~(\ref{eq:P123_momentum_def_fourth}) for the momentum exchange operator, we
obtain:
\begin{align}
&\tensor*[_{\{12\}}]{\left< p' q' (L' l') \mathcal{L}' \mathcal{M}_{\mathcal{L}'} | p q (L l) \mathcal{L} \mathcal{M}_{\mathcal{L}} \right>}{_{\{23\}}} \nonumber \\
&= \int \frac{d \mathbf{p}''}{(2 \pi)^3}  \frac{d \mathbf{q}''}{(2 \pi)^3} \frac{d \mathbf{p}'''}{(2 \pi)^3} \frac{d \mathbf{q}'''}{(2 \pi)^3} \tensor*[_{\{12\}}]{\left< p' q' (L' l') \mathcal{L}' \mathcal{M}_{\mathcal{L}'} | \mathbf{p}'' \mathbf{q}'' \right>}{_{\{12\}}} 
\tensor*[_{\{12\}}]{\left< \mathbf{p}'' \mathbf{q}'' | \mathbf{p}''' \mathbf{q}''' \right>}{_{\{23\}}} 
\tensor*[_{\{23\}}]{\left< \mathbf{p}''' \mathbf{q}''' | p q (L l) \mathcal{L} \mathcal{M}_{\mathcal{L}} \right>}{_{\{23\}}} \nonumber \\
&= \int d \hat{\mathbf{p}}' d \hat{\mathbf{q}}' \mathcal{Y}_{L' l'}^{* \mathcal{L}' \mathcal{M}_{\mathcal{L}'}} (\hat{\mathbf{p}}', \hat{\mathbf{q}}')
\mathcal{Y}_{L l}^{\mathcal{L} \mathcal{M}_{\mathcal{L}}}  (\hat{\bar{\mathbf{p}}},\hat{\bar{\mathbf{q}}}) \frac{\delta(p - \bar{p})}{p^2} \frac{\delta(q - \bar{q})}{q^2} \, ,
\label{eq:P123_mompart_1}
\end{align}
where we renamed the angular integration variables in the second step,
$\hat{\mathbf{p}}'' \rightarrow \hat{\mathbf{p}}'$ and $\hat{\mathbf{q}}''
\rightarrow \hat{\mathbf{q}}'$, and introduced the momenta
\begin{equation}
\bar{\mathbf{p}} = - \tfrac{1}{2} \mathbf{p}' - \tfrac{3}{4} \mathbf{q}', \quad \bar{\mathbf{q}} = \mathbf{p}' - \tfrac{1}{2} \mathbf{q}' \, .
\end{equation}
As a next step we can make use of the fact
that this term is a scalar under rotations and hence has to be proportional to
$\delta_{\mathcal{L} \mathcal{L}'}
\delta_{\mathcal{M}_{\mathcal{L}} \mathcal{M}_{\mathcal{L}'}}$ and independent
of $\mathcal{M}_{\mathcal{L}}$. By summing over $\mathcal{M}_{\mathcal{L}}$
and dividing by $2 \mathcal{L} + 1$ we can then arbitrarily fix the direction
of our coordinate system. For the practical implementation it is most
convenient to choose the $z$-axis along the direction of the $\mathbf{p}'$
vector and fix the polar angle of $\mathbf{q}'$ to zero. In this case the
expressions for the spherical harmonics simplify considerably since we can use:
$Y_{LM} (\hat{\mathbf{p}}') = \sqrt{\frac{2 L + 1}{4 \pi}} \delta_{M0}$.

We can immediately perform three of the four angular integrations in
Eq.~(\ref{eq:P123_mompart_1}), leading to a trivial factor $8 \pi^2$:
\begin{align}
\tensor*[_{\{12\}}]{\left< p' q' (L' l') \mathcal{L}' | p q (L l) \mathcal{L} \right>}{_{\{23\}}} = 8 \pi^2 \int d \cos \theta_{\mathbf{p}' \mathbf{q}'} \frac{\delta_{\mathcal{L} \mathcal{L}'}}{2 \mathcal{L} + 1} \sum_{\mathcal{M}_{\mathcal{L}}} \mathcal{Y}_{L' l'}^{* \mathcal{L} \mathcal{M}_{\mathcal{L}}} (\hat{\mathbf{p}}', \hat{\mathbf{q}}')
\mathcal{Y}_{L l}^{\mathcal{L} \mathcal{M}_{\mathcal{L}}}  (\hat{\bar{\mathbf{p}}},\hat{\bar{\mathbf{q}}}) \frac{\delta(p - \bar{p})}{p^2} \frac{\delta(q - \bar{q})}{q^2} \, .
\end{align}
Now we can combine all results and compute the matrix elements $\left< p' q' \alpha' | P_{123} | p q \alpha \right>$ to
obtain the matrix elements in $LS$-coupling scheme:
\begin{align}
\left< p' q' \beta' | P_{123} | p q \beta \right> 
&= \sum_{\mathcal{M}_{\mathcal{L}}, \mathcal{M}_{\mathcal{S}}} \sum_{\mathcal{M}_{\mathcal{L}'}, \mathcal{M}_{\mathcal{S}'}} \mathcal{C}_{\mathcal{L} \mathcal{M}_{\mathcal{L}} \mathcal{S} \mathcal{M}_{\mathcal{S}}}^{\mathcal{J} \mathcal{M}_{\mathcal{J}}} \mathcal{C}_{\mathcal{L}' \mathcal{M}_{\mathcal{L}'} \mathcal{S}' \mathcal{M}_{\mathcal{S}'}}^{\mathcal{J} \mathcal{M}_{\mathcal{J}}} \nonumber \\
& \times \tensor*[_{\{12\}}]{\left< p' q' (L' l') \mathcal{L}' | p q (L l) \mathcal{L} \right>}{_{\{23\}}} \tensor*[_{\{12\}}]{\left< (S' s) \mathcal{S}' \mathcal{M}_{\mathcal{S}'} | (S s) \mathcal{S} \mathcal{M}_{\mathcal{S}} \right>}{_{\{23\}}} \tensor*[_{\{12\}}]{\left< (T' t) \mathcal{T}' \mathcal{M}_{\mathcal{T}'}  | (T t) \mathcal{T} \mathcal{M}_{\mathcal{T}} \right>}{_{\{23\}}} \nonumber \\
&= \tensor*[_{\{12\}}]{\left< p' q' (L' l') \mathcal{L} | p q (L l) \mathcal{L} \right>}{_{\{23\}}} \tensor*[_{\{12\}}]{\left< (S' s) \mathcal{S} | (S s) \mathcal{S} \right>}{_{\{23\}}} \tensor*[_{\{12\}}]{\left< (T' t) \mathcal{T} | (T t) \mathcal{T} \right>}{_{\{23\}}} \delta_{\mathcal{L} \mathcal{L}'} \delta_{\mathcal{S} \mathcal{S}'} \delta_{\mathcal{T} \mathcal{T}'} \, .
\end{align}
In the last step we  used the independence of all overlap relations on the
projection quantum numbers and the diagonal structure in the total orbital
angular momentum, total spin and total isospin quantum numbers. The
application of the standard recoupling relation~\cite{Vars88Gulag}
\begin{align}
\bigl< p' q' \alpha' | P_{123} | p q \alpha \bigr> =  
\sum_{\mathcal{L}, \mathcal{S}} \sum_{\mathcal{L}', \mathcal{S}'} \sqrt{\hat{J} \hat{j} \hat{\mathcal{L}} \hat{\mathcal{S}}} \sqrt{\hat{J}' \hat{j}' \hat{\mathcal{L}'} \hat{\mathcal{S}}'} 
\left\{
\begin{array}{ccc}
L & S & J \\
l & \tfrac{1}{2} & j \\
\mathcal{L} & \mathcal{S} & \mathcal{J}
\end{array}
\right\}
\left\{
\begin{array}{ccc}
L' & S' & J' \\
l' & \tfrac{1}{2} & j' \\
\mathcal{L}' & \mathcal{S}' & \mathcal{J}
\end{array}
\right\} 
\bigl< p' q' \beta' | P_{123} | p q \beta \bigr>
\end{align}
leads finally to the result shown in Eq.~(\ref{eq:P123_matrixelements}).

\section{Normal-ordered effective interactions for nuclear matter}
\label{sec:Veff_nuclear_matter}
In this Appendix we provide the explicit results for the effective interaction
given by Eqs.~(\ref{eq:normord_singpart}) and (\ref{eq:normord_jacobi}) in
the approximation $\mathbf{P} = 0$ for neutron matter and symmetric nuclear
matter. For angularly-independent regulators only the long-range contributions
proportional to the LECs $c_1$ and $c_3$ contribute in neutron
matter, and the effective interaction takes the following form~\cite{Hebe10nmatt}:
\be
\overline{V}{}^{\text{PNM}}_{\text{3N}} (\mathbf{p}, \mathbf{p}') = \frac{g^2_A}{4 f^4_\pi} \biggl[ - 2 c_1 m_\pi^2 \,
A_{\text{PNM}} ({\bf p},{\bf p}') + c_3 \, B_{\text{PNM}} ({\bf p},{\bf p}') \biggr] \,,
\label{eq:Veff_pnm}
\ee
where $\mathbf{p}$ and $\mathbf{p}'$ are the Jacobi momenta of the initial
and final states, $\mathbf{p} = \tfrac{1}{2} (\mathbf{k}_1 - \mathbf{k}_2)$, and the
functions $A_{\text{PNM}} ({\bf p},{\bf p}')$ and $B_{\text{PNM}} ({\bf p},
{\bf p}')$ include all spin dependences:
\begin{align}
A_{\text{PNM}} ({\bf p},{\bf p}') &= 2 \, \Bigl[ \rho_+^2({\bf p},{\bf p}')
+ 2 a^1({\bf p},{\bf p}') - a^1({\bf p},-{\bf p}')
- \overline{b}^1({\bf p},{\bf p}') \Bigr] \nonumber \\[1mm]
&- \frac{2}{3} \, {\bm \sigma}_1 \cdot {\bm \sigma}_2 \,
\Bigr[ 2 \rho_-^2({\bf p},{\bf p}') + \rho_+^2({\bf p},{\bf p}')
+ 3 a^1({\bf p},-{\bf p}') - \overline{b}^1({\bf p},{\bf p}')
- 2 \overline{b}^1({\bf p},-{\bf p}') \Bigr] \nonumber \\[1mm]
&+ 4 \, \Bigl[ S_{12}({\bf p} + {\bf p}') \, \rho_+^0({\bf p},{\bf p}')
- S_{12}({\bf p} - {\bf p}') \, \rho_-^0({\bf p},{\bf p}') \Bigr] \nonumber \\
&- 4 \, {\bm \sigma}_1^a {\bm \sigma}_2^b \,
\Bigl[ \overline{d}^0_{ab}({\bf p},{\bf p}') 
- \overline{d}^0_{ab}({\bf p},-{\bf p}') \Bigr] \nonumber \\[1mm]
&- 2 i \, ({\bm \sigma}_1 + {\bm \sigma}_2)^a \,
\Bigl[ c^0_a({\bf p},{\bf p}') - c^0_a({\bf p},-{\bf p}') \Bigr]
\,, \label{eq:Veff_A_pnm} \\
B_{\text{PNM}} ({\bf p},{\bf p}') &= - 2 \, \Big[ \rho^4_+({\bf p},{\bf p}')
+ 2 a^2({\bf p},{\bf p}') - a^2({\bf p},-{\bf p}')
- \overline{b}^2({\bf p},{\bf p}') \Big] \nonumber \\[1mm]
&+ \frac{2}{3} \, {\bm \sigma}_1 \cdot {\bm \sigma}_2 \,
\Bigr[ 2 \rho_-^4({\bf p},{\bf p}') + \rho_+^4({\bf p},{\bf p}')
+ 3 a^2({\bf p},-{\bf p}') - \overline{b}^2({\bf p},{\bf p}')
- 2 \overline{b}^2({\bf p},-{\bf p}') \Big] \nonumber \\[1mm]
&- 4 \, \Bigl[ S_{12}({\bf p} + {\bf p}') \, \rho_+^2({\bf p},{\bf p}')
- S_{12}({\bf p} - {\bf p}') \, \rho_-^2 ({\bf p},{\bf p}') \bigr] \nonumber \\
&+ 4 \, {\bm \sigma}_1^a {\bm \sigma}_2^b \, 
\Big[ \overline{d}^1_{ab}({\bf p},{\bf p}') 
- \overline{d}^1_{ab}({\bf p},-{\bf p}') \Big] \nonumber \\[1mm]
&+ 2 i \, ({\bm \sigma}_1 + {\bm \sigma}_2)^a \,
\Bigl[ c^1_a({\bf p},{\bf p}') - c^1_a({\bf p},-{\bf p}') \Big] \,,
\label{eq:Veff_B_pnm}
\end{align}
and the basic integral functions are defined by
\begin{align}
\rho_{\pm}^n({\bf p},{\bf p}') &=
\frac{({\bf p} \pm {\bf p}')^n}{\bigl(({\bf p} \pm {\bf p}')^2
+ m_\pi^2\bigr)^2} \int_{{\bf k}_3} 1 \,, \nonumber \\[1mm]
a^n({\bf p},{\bf p}') &= \int_{{\bf k}_3}
\frac{\bigl( ({\bf p} + {\bf k}_3) \cdot ({\bf p}' + {\bf k}_3)
\bigr)^n}{\bigl( ({\bf p} + {\bf k}_3)^2 + m_{\pi}^2 \bigr) \bigl(
({\bf p}' + {\bf k}_3)^2 + m_{\pi}^2 \bigr)} \,, \nonumber  \\[1mm]
b^n({\bf p},{\bf p}') &= \int_{{\bf k}_3} 
\frac{\bigl( ({\bf p} + {\bf p}') \cdot ({\bf p} + {\bf k}_3) 
\bigr)^n}{\bigl( ({\bf p} + {\bf p}')^2 + m_{\pi}^2 \bigr) \bigl(
({\bf p} + {\bf k}_3)^2 + m_{\pi}^2 \bigr)} \,, \nonumber  \\[1mm]
c^n_a({\bf p},{\bf p}') &= \int_{{\bf k}_3} 
\frac{\bigl( ({\bf p} + {\bf k}_3) \cdot ({\bf p}' + {\bf k}_3) 
\bigr)^{n} \bigl( ({\bf p} + {\bf k}_3) \times ({\bf p}' + {\bf k}_3)
\bigr)_a}{\bigl( ({\bf p} + {\bf k}_3)^2 + m_{\pi}^2 \bigr)
\bigl( ({\bf p}' + {\bf k}_3)^2 + m_{\pi}^2 \bigr)}
\,, \nonumber \\[1mm]
d^n_{ab}({\bf p},{\bf p}') &= \int_{{\bf k}_3}
\bigl( ({\bf p} + {\bf p}') \cdot ({\bf p} + {\bf k}_3) \bigr)^{n} \, \frac{({\bf p} + {\bf p}')_a ({\bf p} + {\bf k}_3)_b 
+ ({\bf p} + {\bf p}')_b ({\bf p} + {\bf k}_3)_a - \frac{2}{3}
\, \delta_{ab} \, ({\bf p} + {\bf p}') \cdot 
({\bf p} + {\bf k}_3)}{2 \, \bigl( ({\bf p} + {\bf p}')^2 + m_{\pi}^2
\bigr) \bigl( ({\bf p} + {\bf k}_3)^2 + m_{\pi}^2 \bigr)} \,.
\label{eq:PNMfunction}
\end{align}

In Eqs.~(\ref{eq:Veff_A_pnm})--(\ref{eq:PNMfunction}), the indices $a, b$ run over
the three Cartesian components of the spin operators, the tensor operator is
given by $S_{12}({\bf p}) = ({\bm \sigma}_1 \cdot {\bf p}) ({\bm
\sigma}_2 \cdot {\bf p}) - \tfrac{1}{3} p^2 \, {\bm \sigma}_1 \cdot {\bm
\sigma}_2$ and the overline denotes a symmetrization in
the relative momentum variables, $\overline{x}({\bf p}, {\bf p}') =
x({\bf p}, {\bf p}') + x({\bf p}', {\bf p})$. In addition, we have
introduced the short-hand notation
\be
\int_{{\bf k}_3} = \int \frac{d{\bf k}_3}{(2\pi)^3} \: n_{{\bf k}_3}
\, \widetilde{f}_{\text{R}}(k',k_3) \, \widetilde{f}_{\text{R}}(k,k_3) \, .
\ee
For example, for a nonlocal regulator given in Eq.~(\ref{eq:nonlocal_regulator}),
expressed in terms of the relative and third-particle momenta for $P=0$, is
given by $\widetilde{f}_{\text{R}}(k,k_3) = \exp[- ((k^2 + k_3^2/3)/
\lm^2)^n]$.

In symmetric nuclear matter the effective interaction takes the following form:
\begin{align}
\overline{V}{}^{\text{SNM}}_{\text{3N}}  (\mathbf{p},\mathbf{p}') &= \frac{1}{2} \frac{g^2_A}{4 f_{\pi}^2} \biggl[ - 4 c_1 m_\pi^2 \, A_{\text{SNM}} ({\bf p},{\bf p}') + 2 c_3 \, B_{\text{SNM}} ({\bf p},{\bf p}') + c_4 \, C_{\text{SNM}} ({\bf p},{\bf p}') \biggl] \nonumber \\
&- \frac{g_A}{8 f_{\pi}^4 \Lambda_{\chi}} c_D \, D_{\text{SNM}} ({\bf p},{\bf p}') + \frac{1}{2} \frac{c_E}{f_{\pi}^4 \Lambda_{\chi}} \, E_{\text{SNM}} ({\bf p},{\bf p}') \,,
\label{Veff_pnm}
\end{align}
where the functions $A_{\text{SNM}} ({\bf p},{\bf p}')$, $B_{\text{SNM}} ({\bf
p}, {\bf p}'), C_{\text{SNM}} ({\bf p},{\bf p}'), D_{\text{SNM}} ({\bf p},{\bf
p}')$ and $E_{\text{SNM}} ({\bf p},{\bf p}')$ are given by:
\begin{align}
A_{\text{SNM}} ({\bf p},{\bf p}') &= 3 \, \Bigl[ 2 \rho_+^2({\bf p},{\bf p}')
+ 4 a^1({\bf p},{\bf p}') - a^1({\bf p},-{\bf p}')
- \overline{b}^1({\bf p},{\bf p}') \Bigr] \nonumber \\[1mm]
&- ({\bm \sigma}_1 \cdot {\bm \sigma}_2 + {\bm \tau}_1 \cdot {\bm \tau}_2) \,
\Bigr[ 2 \rho_+^2({\bf p},{\bf p}') + 3 a^1({\bf p},-{\bf p}') - \overline{b}^1({\bf p},{\bf p}')  \Bigr] \nonumber \\[1mm]
&+ \frac{1}{3} ({\bm \sigma}_1 \cdot {\bm \sigma}_2) ({\bm \tau}_1 \cdot {\bm \tau}_2) \,
\Bigr[ 2 \, \rho_+^2({\bf p},{\bf p}') - 8 \, \rho_-^2({\bf p},{\bf p}') - 9 \, a^1({\bf p},-{\bf p}') - \overline{b}^1({\bf p},{\bf p}') + 4 \, \overline{b}^1({\bf p},-{\bf p}')  \Bigr] \nonumber \\[1mm]
&+ 12 \, S_{12}({\bf p} + {\bf p}') \, \rho_+^0({\bf p},{\bf p}')
- 6 \, {\bm \sigma}_1^a {\bm \sigma}_2^b \, \overline{d}^0_{ab}({\bf p},{\bf p}') - 3 \, i \, ({\bm \sigma}_1 + {\bm \sigma}_2)^a \,
\Bigl[ 2 c^0_a({\bf p},{\bf p}') - (1 + ({\bm \tau}_1 \cdot {\bm \tau}_2)) \, c^0_a({\bf p},-{\bf p}') \Bigr] \nonumber \\[1mm]
&- 2 \, ({\bm \tau}_1 \cdot {\bm \tau}_2) \Bigr[ 4 \, S_{12}({\bf p} - {\bf p}') \, \rho_-^0({\bf p},{\bf p}') + 2 \, S_{12}({\bf p} + {\bf p}') \, \rho_+^0({\bf p},{\bf p}') - {\bm \sigma}_1^a {\bm \sigma}_2^b \, ( \overline{d}^0_{ab}({\bf p},{\bf p}') + 2 \, \overline{d}^0_{ab}({\bf p},-{\bf p}')) \Bigr] \, , \\
B_{\text{SNM}} ({\bf p},{\bf p}') &= -3 \, \Bigl[ 2 \rho_+^4({\bf p},{\bf p}')
+ 4 a^2({\bf p},{\bf p}') - a^2({\bf p},-{\bf p}')
- \overline{b}^2({\bf p},{\bf p}') \Bigr] \nonumber \\[1mm]
&+ ({\bm \sigma}_1 \cdot {\bm \sigma}_2 + {\bm \tau}_1 \cdot {\bm \tau}_2) \,
\Bigr[ 2 \rho_+^4({\bf p},{\bf p}') + 3 a^2({\bf p},-{\bf p}') - \overline{b}^2({\bf p},{\bf p}')  \Bigr] \nonumber \\[1mm]
&- \frac{1}{3} ({\bm \sigma}_1 \cdot {\bm \sigma}_2) ({\bm \tau}_1 \cdot {\bm \tau}_2) \,
\Bigr[ 2 \, \rho_+^4({\bf p},{\bf p}') - 8 \, \rho_-^4({\bf p},{\bf p}') - 9 \, a^2({\bf p},-{\bf p}') - \overline{b}^2({\bf p},{\bf p}') + 4 \, \overline{b}^2({\bf p},-{\bf p}')  \Bigr] \nonumber \\[1mm]
&- 12 \, S_{12}({\bf p} + {\bf p}') \, \rho_+^2({\bf p},{\bf p}')
+ 6 \, {\bm \sigma}_1^a {\bm \sigma}_2^b \, \overline{d}^1_{ab}({\bf p},{\bf p}') + 3 \, i \, ({\bm \sigma}_1 + {\bm \sigma}_2)^a \,
\Bigl[ 2 c^1_a({\bf p},{\bf p}') - (1 + ({\bm \tau}_1 \cdot {\bm \tau}_2)) \, c^1_a({\bf p},-{\bf p}') \Bigr] \nonumber \\[1mm]
&+ 2 \, ({\bm \tau}_1 \cdot {\bm \tau}_2) \Bigr[ 4 \, S_{12}({\bf p} - {\bf p}') \, \rho_-^2({\bf p},{\bf p}') + 2 \, S_{12}({\bf p} + {\bf p}') \, \rho_+^2({\bf p},{\bf p}') - {\bm \sigma}_1^a {\bm \sigma}_2^b \, ( \overline{d}^1_{ab}({\bf p},{\bf p}') + 2 \, \overline{d}^1_{ab}({\bf p},-{\bf p}')) \Bigr] \, , \\
C_{\text{SNM}} ({\bf p},{\bf p}') &= 2 \Bigl[ ({\bm \sigma}_1 \cdot {\bm \sigma}_2) + ({\bm \tau}_1 \cdot {\bm \tau}_2) - 3 \Bigr] \Bigl[ \tilde{a}^2 ({\bf p},-{\bf p}') + \overline{\tilde{b}}^2 ({\bf p}, {\bf p}') \Bigr] \nonumber \\
&+ \frac{8}{3} ({\bm \sigma}_1 \cdot {\bm \sigma}_2) ({\bm \tau}_1 \cdot {\bm \tau}_2) \Bigl[ \overline{\tilde{b}}^2 ({\bf p}, -{\bf p}') + \tilde{a}^2 ({\bf p},{\bf p}') - \frac{1}{4} \overline{\tilde{b}}^2 ({\bf p}, {\bf p}') - \frac{1}{4} \tilde{a}^2 ({\bf p},-{\bf p}') \Bigr] \nonumber \\
&+ 6 \, {\bm \sigma}_1^a {\bm \sigma}_2^b \Bigl[ \overline{e}_{ab} ({\bf p}, {\bf p}') - 2 f_{ab} ({\bf p}, -{\bf p}') \Bigr] \nonumber \\
&- 2 ({\bm \tau}_1 \cdot {\bm \tau}_2) {\bm \sigma}_1^a {\bm \sigma}_2^b \Bigl[ 2 \overline{e}_{ab} ({\bf p}, -{\bf p}') - 2 f_{ab} ({\bf p}, -{\bf p}') + \overline{e}_{ab} ({\bf p}, {\bf p}') -4 f_{ab} ({\bf p}, {\bf p}') \Bigr] \nonumber \\
&- 12 \, {\bm \sigma}_1^a {\bm \sigma}_2^b \overline{g}_{ab} ({\bf p},{\bf p'}) + 4 \, ({\bm \tau}_1 \cdot {\bm \tau}_2) \, {\bm \sigma}_1^a {\bm \sigma}_2^b \Bigl[ 2\, \overline{g}_{ab} ({\bf p},-{\bf p'}) + \overline{g}_{ab} ({\bf p},{\bf p'}) \Bigr] \nonumber \\
&+ \,i \, ({\bm \sigma}_1 + {\bm \sigma}_2)^a \, \Bigl[ ( 2 ({\bm \tau}_1 \cdot {\bm \tau}_2) - 6) c^1_a({\bf p},-{\bf p}') + 4 ({\bm \tau}_1 \cdot {\bm \tau}_2) \, c^1_a({\bf p},{\bf p}') \Bigr] \, , \\
D_{\text{SNM}} ({\bf p},{\bf p}') &= \Bigl[ ({\bm \sigma}_1 \cdot {\bm \sigma}_2) + ({\bm \tau}_1 \cdot {\bm \tau}_2) - 3 \Bigr] \Bigl[ \tilde{\rho}_+^2({\bf p},{\bf p}')
+ i ({\bf p}) + i ({\bf p}') \Bigr] \nonumber \\[1mm]
&+ 2 \, ({\bm \sigma}_1 \cdot {\bm \sigma}_2) ({\bm \tau}_1 \cdot {\bm \tau}_2) \,
\Bigr[ \frac{2}{3} \, \tilde{\rho}_-^2({\bf p},{\bf p}') - \frac{1}{6} \, \tilde{\rho}_+^2({\bf p},{\bf p}') + \frac{i({\bf p})}{2} + \frac{i({\bf p}')}{2} \Bigr] \nonumber \\[1mm]
&- 6 \, S_{12}({\bf p} + {\bf p}') \, \tilde{\rho}_+^0({\bf p},{\bf p}')
+ 3 \, {\bm \sigma}_1^a {\bm \sigma}_2^b \, \Bigl[ h_{ab} ({\bf p}) + h_{ab} ({\bf p}') \Bigr] \nonumber \\[1mm]
&+ 6\, ({\bm \tau}_1 \cdot {\bm \tau}_2) \Bigr[ \frac{2}{3} \, S_{12}({\bf p} - {\bf p}') \, \tilde{\rho}_-^0({\bf p},{\bf p}') + \frac{1}{3} \, S_{12}({\bf p} + {\bf p}') \, \tilde{\rho}_+^0({\bf p},{\bf p}') - \frac{1}{2} {\bm \sigma}_1^a {\bm \sigma}_2^b \, \Bigl( h_{ab} ({\bf p}) + h_{ab} ({\bf p}') \Bigr) \Bigr] \, , \\
E_{\text{SNM}} ({\bf p},{\bf p}') &= \Bigl[ 9 - 3 ({\bm \sigma}_1 \cdot {\bm \sigma}_2) - 3 ({\bm \tau}_1 \cdot {\bm \tau}_2) + ({\bm \sigma}_1 \cdot {\bm \sigma}_2) \, ({\bm \tau}_1 \cdot {\bm \tau}_2) \Bigr] \, \int_{{\bf k}_3} 1 \, ,
\end{align}
and the basic integral functions are defined by
\begin{align}
\rho_{\pm}^n({\bf p},{\bf p}') &=
\frac{({\bf p} \pm {\bf p}')^n}{\bigl(({\bf p} \pm {\bf p}')^2
+ m_\pi^2\bigr)^2} \int_{{\bf k}_3} 1 \,, \nonumber \\[1mm]
\tilde{\rho}_{\pm}^n({\bf p},{\bf p}') &=
\frac{({\bf p} \pm {\bf p}')^n}{\bigl(({\bf p} \pm {\bf p}')^2
+ m_\pi^2\bigr)} \int_{{\bf k}_3} 1 \,, \nonumber  \\[1mm]
a^n({\bf p},{\bf p}') &= \int_{{\bf k}_3}
\frac{\bigl( ({\bf p} + {\bf k}_3) \cdot ({\bf p}' + {\bf k}_3)
\bigr)^n}{\bigl( ({\bf p} + {\bf k}_3)^2 + m_{\pi}^2 \bigr) \bigl(
({\bf p}' + {\bf k}_3)^2 + m_{\pi}^2 \bigr)} \,, \nonumber \\[1mm]
\tilde{a}^n({\bf p},{\bf p}') &= \int_{{\bf k}_3}
\frac{\bigl( ({\bf p} + {\bf k}_3) \times ({\bf p}' + {\bf k}_3)
\bigr)^n}{\bigl( ({\bf p} + {\bf k}_3)^2 + m_{\pi}^2 \bigr) \bigl(
({\bf p}' + {\bf k}_3)^2 + m_{\pi}^2 \bigr)} \,, \nonumber \\[1mm]
b^n({\bf p},{\bf p}') &= \int_{{\bf k}_3} 
\frac{\bigl( ({\bf p} + {\bf p}') \cdot ({\bf p} + {\bf k}_3) 
\bigr)^n}{\bigl( ({\bf p} + {\bf p}')^2 + m_{\pi}^2 \bigr) \bigl(
({\bf p} + {\bf k}_3)^2 + m_{\pi}^2 \bigr)} \,, \nonumber \\[1mm]
\tilde{b}^n({\bf p},{\bf p}') &= \int_{{\bf k}_3} 
\frac{\bigl( ({\bf p} + {\bf p}') \times ({\bf p} + {\bf k}_3) 
\bigr)^n}{\bigl( ({\bf p} + {\bf p}')^2 + m_{\pi}^2 \bigr) \bigl(
({\bf p} + {\bf k}_3)^2 + m_{\pi}^2 \bigr)} \,, \nonumber \\[1mm]
c^n_a({\bf p},{\bf p}') &= \int_{{\bf k}_3} 
\frac{\bigl( ({\bf p} + {\bf k}_3) \cdot ({\bf p}' + {\bf k}_3) 
\bigr)^{n} \bigl( ({\bf p} + {\bf k}_3) \times ({\bf p}' + {\bf k}_3)
\bigr)_a}{\bigl( ({\bf p} + {\bf k}_3)^2 + m_{\pi}^2 \bigr)
\bigl( ({\bf p}' + {\bf k}_3)^2 + m_{\pi}^2 \bigr)}
\,, \nonumber \\[1mm]
d^n_{ab}({\bf p},{\bf p}') &= \int_{{\bf k}_3}
\bigl( ({\bf p} + {\bf p}') \cdot ({\bf p} + {\bf k}_3) \bigr)^{n} \,
\frac{({\bf p} + {\bf p}')_a ({\bf p} + {\bf k}_3)_b 
+ ({\bf p} + {\bf p}')_b ({\bf p} + {\bf k}_3)_a - \frac{2}{3}
\, \delta_{ab} \, ({\bf p} + {\bf p}') \cdot 
({\bf p} + {\bf k}_3)}{2 \, \bigl( ({\bf p} + {\bf p}')^2 + m_{\pi}^2
\bigr) \bigl( ({\bf p} + {\bf k}_3)^2 + m_{\pi}^2 \bigr)} \,, \nonumber \\[1mm]
e_{ab}({\bf p},{\bf p}') &= \int_{{\bf k}_3}
\frac{(({\bf p} + {\bf p}') \times ({\bf p} + {\bf k}_3))_a (({\bf p} + {\bf p}') \times ({\bf p} + {\bf k}_3))_b - \frac{1}{3} \delta_{ab} (({\bf p} + {\bf p}') \times ({\bf p} + {\bf k}_3))^2}{\bigl( ({\bf p} + {\bf p}')^2 + m_{\pi}^2
\bigr) \bigl( ({\bf p} + {\bf k}_3)^2 + m_{\pi}^2 \bigr)} \,, \nonumber \\[1mm]
f_{ab}({\bf p},{\bf p}') &= \int_{{\bf k}_3}
\frac{(({\bf p} + {\bf k}_3) \times ({\bf p}' + {\bf k}_3))_a (({\bf p} + {\bf k}_3) \times ({\bf p}' + {\bf k}_3))_b - \frac{1}{3} \delta_{ab} (({\bf p} + {\bf k}_3) \times ({\bf p}' + {\bf k}_3))^2}{\bigl( ({\bf p} + {\bf k}_3)^2 + m_{\pi}^2
\bigr) \bigl( ({\bf p}' + {\bf k}_3)^2 + m_{\pi}^2 \bigr)} \,, \nonumber \\[1mm]
g_{ab}({\bf p},{\bf p}') &= \int_{{\bf k}_3}
\frac{({\bf p} + {\bf p}')^2 ({\bf p} + {\bf k}_3)_a ({\bf p} + {\bf k}_3)_b - ({\bf p} + {\bf k}_3)^2 ({\bf p} + {\bf p}')_a ({\bf p} + {\bf p}')_b}{2 \bigl( ({\bf p} + {\bf p}')^2 + m_{\pi}^2
\bigr) \bigl( ({\bf p} + {\bf k}_3)^2 + m_{\pi}^2 \bigr)} \,, \nonumber \\[1mm]
h_{ab}({\bf p}) &= \int_{{\bf k}_3}
\frac{({\bf p} + {\bf k}_3)_a ({\bf p} + {\bf k}_3)_b - \frac{1}{3} ({\bf p} + {\bf k}_3)^2 \delta_{ab}}{\bigl( ({\bf p} + {\bf k}_3)^2 + m_{\pi}^2
\bigr)} \,, \nonumber  \\[1mm]
i ({\bf p}) &= \int_{{\bf k}_3}
\frac{({\bf p} + {\bf k}_3)^2}{\bigl( ({\bf p} + {\bf k}_3)^2 + m_{\pi}^2
\bigr)} \,.
\label{dfunction}
\end{align}

\clearpage 
\section{List of three-body configurations}
\label{sec:3N_config_table}

\begin{table}[h!]
\centering
\begin{minipage}{.3\linewidth}
\centering
\renewcommand{\arraystretch}{1.2}
\begin{tabular}{c|c|c|c|c|c|c}
$\alpha$ & $L$ & $S$ & $J$ & $T$ & $l$ & $j$    \\
\hline
0   & 0   &  0   &  0   &  1   &  0   &  $\tfrac{1}{2}$    \\
1   & 1   &  1   &  0   &  1   &  1   &  $\tfrac{1}{2}$    \\
2   & 1   &  0   &  1   &  0   &  1   &  $\tfrac{1}{2}$    \\
3   & 1   &  0   &  1   &  0   &  1   &  $\tfrac{3}{2}$    \\
4   & 0   &  1   &  1   &  0   &  0   &  $\tfrac{1}{2}$    \\
5   & 0   &  1   &  1   &  0   &  2   &  $\tfrac{3}{2}$    \\
6   & 1   &  1   &  1   &  1   &  1   &  $\tfrac{1}{2}$    \\
7   & 1   &  1   &  1   &  1   &  1   &  $\tfrac{3}{2}$    \\
8   & 2   &  1   &  1   &  0   &  0   &  $\tfrac{1}{2}$    \\
9   & 2   &  1   &  1   &  0   &  2   &  $\tfrac{3}{2}$    \\
10  &  2  &   0  &   2  &   1  &   2  &   $\tfrac{3}{2}$ \\
11  &  2  &   0  &   2  &   1  &   2  &   $\tfrac{5}{2}$ \\
12  &  1  &   1  &   2  &   1  &   1  &   $\tfrac{3}{2}$ \\
13  &  1  &   1  &   2  &   1  &   3  &   $\tfrac{5}{2}$ \\
14  &  2  &   1  &   2  &   0  &   2  &   $\tfrac{3}{2}$ \\
15  &  2  &   1  &   2  &   0  &   2  &   $\tfrac{5}{2}$ \\
16  &  3  &   1  &   2  &   1  &   1  &   $\tfrac{3}{2}$ \\
17  &  3  &   1  &   2  &   1  &   3  &   $\tfrac{5}{2}$ \\
18  &  3  &   0  &   3  &   0  &   3  &   $\tfrac{5}{2}$ \\
19  &  3  &   0  &   3  &   0  &   3  &   $\tfrac{7}{2}$ \\
20  &  2  &   1  &   3  &   0  &   2  &   $\tfrac{5}{2}$ \\
21  &  2  &   1  &   3  &   0  &   4  &   $\tfrac{7}{2}$ \\
22  &  3  &   1  &   3  &   1  &   3  &   $\tfrac{5}{2}$ \\
23  &  3  &   1  &   3  &   1  &   3  &   $\tfrac{7}{2}$ \\
24  &  4  &   1  &   3  &   0  &   2  &   $\tfrac{5}{2}$ \\
25  &  4  &   1  &   3  &   0  &   4  &   $\tfrac{7}{2}$ \\
26  &  4  &   0  &   4  &   1  &   4  &   $\tfrac{7}{2}$ \\
27  &  4  &   0  &   4  &   1  &   4  &   $\tfrac{9}{2}$ \\
28  &  3  &   1  &   4  &   1  &   3  &   $\tfrac{7}{2}$
\end{tabular}
\end{minipage}
\begin{minipage}{.4\linewidth}
\centering
\renewcommand{\arraystretch}{1.2}
\begin{tabular}{c|c|c|c|c|c|c}
$\alpha$ & $L$ & $S$ & $J$ & $T$ & $l$ & $j$    \\
\hline
29  &  3  &   1  &   4  &   1  &   5  &   $\tfrac{9}{2}$ \\
30  &  4  &   1  &   4  &   0  &   4  &   $\tfrac{7}{2}$ \\
31  &  4  &   1  &   4  &   0  &   4  &   $\tfrac{9}{2}$ \\
32  &  5  &   1  &   4  &   1  &   3  &   $\tfrac{7}{2}$ \\
33  &  5  &   1  &   4  &   1  &   5  &   $\tfrac{9}{2}$ \\    
34  &  5  &   0  &   5  &   0  &   5  &   $\tfrac{9}{2}$ \\
35  &  5  &   0  &   5  &   0  &   5  &   $\tfrac{11}{2}$  \\
36  &  4  &   1  &   5  &   0  &   4  &   $\tfrac{9}{2}$ \\
37  &  4  &   1  &   5  &   0  &   6  &   $\tfrac{11}{2}$  \\
38  &  5  &   1  &   5  &   1  &   5  &   $\tfrac{9}{2}$ \\
39  &  5  &   1  &   5  &   1  &   5  &   $\tfrac{11}{2}$  \\
40  &  6  &   1  &   5  &   0  &   4  &   $\tfrac{9}{2}$ \\
41  &  6  &   1  &   5  &   0  &   6  &   $\tfrac{11}{2}$  \\
42  &  6  &   0  &   6  &   1  &   6  &   $\tfrac{11}{2}$  \\
43  &  6  &   0  &   6  &   1  &   6  &   $\tfrac{13}{2}$  \\
44  &  5  &   1  &   6  &   1  &   5  &   $\tfrac{11}{2}$  \\
45  &  5  &   1  &   6  &   1  &   7  &   $\tfrac{13}{2}$  \\
46  &  6  &   1  &   6  &   0  &   6  &   $\tfrac{11}{2}$  \\
47  &  6  &   1  &   6  &   0  &   6  &   $\tfrac{13}{2}$  \\
48  &  7  &   1  &   6  &   1  &   5  &   $\tfrac{11}{2}$  \\
49  &  7  &   1  &   6  &   1  &   7  &   $\tfrac{13}{2}$  \\
50  &  7  &   0  &   7  &   0  &   7  &   $\tfrac{13}{2}$  \\
51  &  7  &   0  &   7  &   0  &   7  &   $\tfrac{15}{2}$  \\
52  &  6  &   1  &   7  &   0  &   6  &   $\tfrac{13}{2}$  \\
53  &  6  &   1  &   7  &   0  &   8  &   $\tfrac{15}{2}$  \\
54  &  7  &   1  &   7  &   1  &   7  &   $\tfrac{13}{2}$  \\
55  &  7  &   1  &   7  &   1  &   7  &   $\tfrac{15}{2}$  \\
56  &  8  &   1  &   7  &   0  &   6  &   $\tfrac{13}{2}$  \\
57  &  8  &   1  &   7  &   0  &   8  &   $\tfrac{15}{2}$
\end{tabular}
\end{minipage}
\caption{List of three-body configurations for the three-body channel with $\mathcal{J} = \tfrac{1}{2}$, $\mathcal{T}=\tfrac{1}{2}$ and $\mathcal{P} = +1$ using $J_{\text{max}} = 7$.}
\label{tab:3N_configurations}
\end{table}

\clearpage 

\renewcommand{\arraystretch}{1.0}

\renewcommand{\bibname}{References}
\addcontentsline{toc}{section}{\protect\numberline{}\bibname}
\biboptions{numbers,sort&compress}
\bibliographystyle{elsarticle-num}
\bibliography{./strongint}

\end{document}